\documentclass[11pt]{report} 

\usepackage[papersize={8.5 in, 11 in}, nohead, includeheadfoot, left=1.5 in, right = 1 in, vmargin= 1 in]{geometry}

\usepackage[doublespacing]{setspace} 
\usepackage{etoolbox}
\BeforeBeginEnvironment{table}{\begin{singlespace}}
\AfterEndEnvironment{table}{\end{singlespace}}

\usepackage{calc}
\usepackage{array} 
\usepackage{booktabs}
\usepackage{latexsym}
\usepackage{amsmath,amssymb,amsfonts,graphicx,slashed,color,amsthm,mathtools,upgreek, enumerate,tensor}
\usepackage{caption,subcaption,multicol,longtable,float,numprint,colortbl,xcolor}
\npdecimalsign{.}
\nprounddigits{3}
\usepackage{lscape,makecell,tabularx,rotating}
\usepackage[export]{adjustbox}
\usepackage{arydshln}
\usepackage{tabu}

\usepackage{hyperref}
\usepackage[utf8]{inputenc}
\usepackage{enumitem}
\usepackage[titletoc]{appendix}
\numberwithin{equation}{section} 

\usepackage{tocloft}
\renewcommand{\cftchappresnum}{CHAPTER }
\renewcommand{\cftchapaftersnum}{:}
\newlength{\mylen}   
\renewcommand{\cfttabpresnum}{TABLE }

\newlength{\mylent}   
\renewcommand{\cftfigpresnum}{FIGURE }

\newlength{\mylenf}   
\usepackage[compact]{titlesec}

\usepackage{parskip} 
\usepackage{url} 

\usepackage{upgreek}
\usepackage[percent]{overpic}
\usepackage{xfrac}
\usepackage{lmodern} 
\allowdisplaybreaks

\newcommand{\N}{\mathcal{N}}
\newcommand{\reflambdaetaPL}{Hoare:2015gda,Sfetsos:2015nya,Klimcik:2015gba}
\newcommand{\reflongDFTPLlist}{Hassler:2017yza,Demulder:2018lmj,Sakatani:2019jgu,Catal-Ozer:2019hxw,Hassler:2019wvn}
\newcommand{\refonelooprenorm}{Valent:2009nv,Sfetsos:2009dj,Sfetsos:2009vt,Avramis:2009xi,Severa:2016lwc,Pulmann:2020omk}
\newcommand{\refPLalphaprime}{Hassler:2020tvz,Borsato:2020wwk,Codina:2020yma}
\newcommand{\refoneLRGintegr}{Delduc:2020vxy,Hoare:2020fye,Georgiou:2018gpe,Georgiou:2017jfi}
\renewcommand{\b}{\beta}
\newcommand{\DD}{\mathcal{D}}
\newcommand{\mfd}{\mathfrak{d}}
\newcommand{\dd}{\mathrm{d}}

\newcommand{\Pb}{\overline{P}}
\newcommand{\dg}{\delta g}
\newcommand{\dB}{\delta B}
\newcommand{\dphi}{\delta \phi}
\newcommand{\bh}{\widehat{\beta}}
\newcommand{\bb}{\overline{\beta}}
\newcommand{\B}{\upbeta}
\newcommand{\Bb}{\overline{\upbeta}}
\newcommand{\Bh}{\widehat{\upbeta}}
\newcommand{\Eh}{\widehat{E}}
\newcommand{\dEh}{\delta\Eh}
\newcommand{\dEgf}{\delta\Eh^{\mathrm{gf}}}
\newcommand{\Fh}{\widehat{F}}
\newcommand{\Dh}{\widehat{D}}
\newcommand{\A}{A}
\newcommand{\Al}{A_\lambda}
\newcommand{\intvol}{\int \dd^D x e^{-2d}}
\newcommand{\BhEgf}{\Bh^{E\mathrm{gf}}}
\newcommand{\AlE}{\Al^{(1)} \begin{pmatrix} \widehat E & d \end{pmatrix}}
\newcommand{\DL}{\Delta_\Lambda}
\newcommand{\Lh}{\widehat{L}}
\newcommand{\Sh}{\widehat{S}}
\newcommand{\Kh}{\widehat{K}}

\newcommand{\scR}{\mathcal{R}}
\newcommand{\scRh}{\widehat{\mathcal{R}}}
\newcommand{\scGh}{\widehat{\mathcal{G}}}
\newcommand{\qh}{\widehat{q}}
\newcommand{\jb}{\overline{j}}
\newcommand{\kh}{\widehat{k}}
\newcommand{\zb}{\overline{z}}
\newcommand{\wb}{\overline{w}}
\newcommand{\ci}{\mathrm{i}}
\def\bZ{\mathbb{Z}}

\newcommand{\cN}{\mathcal{N}}
\newcommand{\cO}{\mathcal{O}}

\newcommand{\mf}{\mathfrak}
\newcommand{\ksu}{\mathfrak{su}}
\newcommand{\kso}{\mathfrak{so}}
\DeclareMathOperator{\SU}{\mathit{SU}}
\DeclareMathOperator{\SO}{\mathit{SO}}

\DeclareMathOperator{\su}{\mathit{su}}

\newcommand{\tr}{\, {\rm tr}}

\newcommand{\op}{\oplus }

\newcommand{\ba}{\begin{eqnarray}}
\newcommand{\ea}{\end{eqnarray}}
\newcommand{\be}{\begin{equation}}
\newcommand{\ee}{\end{equation}}
\newcommand{\al}[1]{\begin{align}#1\end{align}}
\newcommand{\co}[1]{\parbox{1.7em}{\centering$\displaystyle #1$}}
\newcommand{\pb}[1] {\overline{#1}}
\newcommand{\p}[1] {\underline{#1}}

\definecolor{LightCyan}{rgb}{0.88,1,1}
\usepackage{tikz}
\usepackage{tikz-cd}
\usetikzlibrary{decorations.pathreplacing}
\usetikzlibrary{calc}
\usetikzlibrary{arrows}
\usetikzlibrary{decorations.markings}
\newcommand{\yshift}{.6cm}
\tikzstyle{small}=[scale=0.6, every node/.style={scale=0.6}]
\tikzstyle{resize}=[scale=0.8, every node/.style={scale=0.8}]
\tikzstyle{A}=[fill=black, draw=black, shape=circle, inner sep=2pt,anchor=center]
\tikzstyle{BETA}=[draw=gray!60!black,shape=circle,inner sep=10pt,{label={center:{\huge $\beta$}}}]
\tikzstyle{FA}=[fill=black, draw=black, shape=rectangle, inner sep=3pt,anchor=center]
\tikzstyle{none}=[]
\tikzstyle{arrow}=[style={-,decoration={markings,mark=at position 0.6 with {\arrow[scale=1.,black]{>}}}, postaction={decorate}, line width=1pt, shorten <=0, shorten >=0}]
\tikzstyle{dashed}=[dash pattern=on 3pt off 3pt, -, draw=black]
\tikzstyle{der}=[-triangle 60]
\tikzstyle{dashder}=[dash pattern=on 3pt off 3pt, -triangle 60]
\tikzstyle{derI}=[triangle 60-]
\tikzstyle{dashderI}=[dash pattern=on 3pt off 3pt, triangle 60-]
\tikzstyle{derder}=[triangle 60-triangle 60]
\tikzstyle{dashderder}=[dash pattern=on 3pt off 3pt, triangle 60-triangle 60]
\tikzstyle{derloop}=[loop, -triangle 60, looseness=20]
\tikzstyle{derloopback}=[style={-,decoration={markings,mark=at position 0.9 with {\arrow[scale=1.,black]{triangle 60}}}, postaction={decorate}, loop, looseness=20}]
\tikzstyle{dashderloop}=[dash pattern=on 3pt off 3pt, -triangle 60, looseness=20]
\tikzstyle{dashderbackI}=[style={-,decoration={markings,mark=at position 0.2 with {\arrow[scale=-1.,black]{triangle 60}}}, postaction={decorate}, looseness=20, dash pattern=on 3pt off 3pt}]
\tikzstyle{derbackI}=[style={-,decoration={markings,mark=at position 0.2 with {\arrow[scale=-1.,black]{triangle 60}}}, postaction={decorate}, looseness=20}]
\tikzstyle{dashderloopback}=[style={-,decoration={markings,mark=at position 0.9 with {\arrow[scale=1.,black]{triangle 60}}}, postaction={decorate}, loop, looseness=20, dash pattern=on 3pt off 3pt}]
\tikzstyle{derback}=[style={-,decoration={markings,mark=at position 0.98 with {\arrow[scale=1.,black]{triangle 60}}}, postaction={decorate}, loop, looseness=20}]
\tikzstyle{dashderback}=[style={-,decoration={markings,mark=at position 0.98 with {\arrow[scale=1.,black]{triangle 60}}}, postaction={decorate}, loop, looseness=20, dash pattern=on 3pt off 3pt}]
\pgfdeclarelayer{nodelayer}
\pgfdeclarelayer{edgelayer}
\pgfsetlayers{nodelayer,edgelayer,main}
\newcommand{\width}{.7}
\newcommand{\height}{.7}
\newcommand{\FAFA}[1]
{\scalebox{\width}[\height]{
\begin{tikzpicture}[baseline={([yshift=-.5ex]current bounding box.center)},small]
	\begin{pgfonlayer}{nodelayer}
		\node [style=FA] (0) at (-1.75, 0) {};
		\node [style=FA] (1) at (1.75, 0) {};
	\end{pgfonlayer}
	\begin{pgfonlayer}{edgelayer}
          \ifnum #1 >0 {\draw [style=dashed] (0) to (1);} \else {\draw (0) to (1);}\fi
	\end{pgfonlayer}
\end{tikzpicture}
}}
\newcommand{\FADA}[1]
{\scalebox{\width}[\height]{
\begin{tikzpicture}[baseline={([yshift=-.5ex-\yshift]current bounding box.center)},small]
	\begin{pgfonlayer}{nodelayer}
		\node [style=FA] (0) at (0, 0) {};
	\end{pgfonlayer}
	\begin{pgfonlayer}{edgelayer}
	  \ifnum #1 >0 {\draw [style=dashderback] (0) arc (-90:270:1cm);}
          \else {\draw [style=derback] (0) arc (-90:270:1cm);}\fi
	\end{pgfonlayer}
\end{tikzpicture}
}}
\newcommand{\Fsquare}[3]
{\scalebox{\width}[\height]{
\begin{tikzpicture}[baseline={([yshift=-.5ex]current bounding box.center)},small]
	\begin{pgfonlayer}{nodelayer}
		\node [style=A] (0) at (-2, 0) {};
		\node [style=A] (1) at (2, 0) {};
	\end{pgfonlayer}
	\begin{pgfonlayer}{edgelayer}
		\ifnum #1 >0 {\draw [style=dashed] (0) arc (180:0:2);}\else {\draw (0) arc (180:0:2);}\fi
		\ifnum #2 >0 {\draw [style=dashed] (0) to (1);}\else {\draw (0) to (1);}\fi
		\ifnum #3 >0 {\draw [style=dashed] (0) arc (-180:0:2);}\else {\draw (0) arc (-180:0:2);}\fi
	\end{pgfonlayer}
\end{tikzpicture}
}}
\newcommand{\dDAFA}
{\scalebox{\width}[\height]{
\begin{tikzpicture}[baseline={([yshift=-.5ex]current bounding box.center)},small]
	\begin{pgfonlayer}{nodelayer}
		\node [style=FA] (0) at (0, 0) {};
		\node [style=none] (1) at (2, 0) {};
		\node [style=none] (2) at (-2, 0) {};
	\end{pgfonlayer}
	\begin{pgfonlayer}{edgelayer}
		\draw (2.center) to (0);
		\draw [style=dashder] (1.center) to (0);
	\end{pgfonlayer}
\end{tikzpicture}
}}
\newcommand{\dFsquare}[2]
{\scalebox{\width}[\height]{
\begin{tikzpicture}[baseline={([yshift=-.5ex]current bounding box.center)},small]
  \begin{pgfonlayer}{nodelayer}
    \node [style=A] (0) at (0.75, 0) {};
    \node [style=A] (1) at (-.75, 0) {};
    \node [style=none] (2) at (2, 0) {};
    \node [style=none] (3) at (-2, 0) {};
  \end{pgfonlayer}
  \begin{pgfonlayer}{edgelayer}
    \draw (3.center) to (1);
    \draw [style=dashed] (2.center) to (0);
    \ifnum #1 >0 {\draw [style=dashed] (0) arc (0:-180:0.75);}\else {\draw (0) arc (0:-180:0.75);}\fi
    \ifnum #2 >0 {\draw [style=dashed] (0) arc (0:180:0.75);} \else {\draw (0) arc (0:180:0.75);}\fi
  \end{pgfonlayer}
\end{tikzpicture}
}}
\newcommand{\dFADA}[1]
{\scalebox{\width}[\height]{
\begin{tikzpicture}[baseline={([yshift=-.5ex-\yshift]current bounding box.center)},small]
  \begin{pgfonlayer}{nodelayer}
    \node [style=A] (0) at (0, 0) {};
    \node [style=none] (1) at (2, 0) {};
    \node [style=none] (2) at (-2, 0) {};
  \end{pgfonlayer}
  \begin{pgfonlayer}{edgelayer}
    \draw (2.center) to (0);
    \draw [style=dashed] (1.center) to (0);
    \ifnum #1 >0 {\draw [style=dashderloopback] (0) arc (-90:270:1cm);}\else {\draw [style=derloopback] (0) arc (-90:270:1cm);}\fi
  \end{pgfonlayer}
\end{tikzpicture}
}}
\newcommand{\dFFA}[1]
{\scalebox{\width}[\height]{
\begin{tikzpicture}[baseline={([yshift=-\yshift]current bounding box.center)},small]
  \begin{pgfonlayer}{nodelayer}
    \node [style=A] (0) at (0, 0) {};
    \node [style=none] (1) at (2, 0) {};
    \node [style=none] (2) at (-2, 0) {};
    \node [style=FA] (3) at (0, 1.75) {};
  \end{pgfonlayer}
  \begin{pgfonlayer}{edgelayer}
    \draw (2.center) to (0);
    \draw [style=dashed] (1.center) to (0);
    \ifnum #1 >0 {\draw [style=dashed] (0) to (3);}\else {\draw (0) to (3);}\fi
  \end{pgfonlayer}
\end{tikzpicture}
}}
\newcommand{\HCNaT}
{\scalebox{\width}[\height]{
  \begin{tikzpicture}[baseline={([yshift=-.5ex]current bounding box.center)},small]
    \begin{pgfonlayer}{nodelayer}
      \node [style=none] (0) at (-2, 0) {};
      \node [style=A] (1) at (-1.25, 0) {};
      \node [style=A] (2) at (-0.375, 0) {};
      \node [style=none] (3) at (2, 0) {};
      \node [style=A] (4) at (0.375, 0) {};
      \node [style=A] (5) at (1.25, 0) {};
    \end{pgfonlayer}
    \begin{pgfonlayer}{edgelayer}
      \draw (0.center) to (1);
      \draw (2) arc (0:180:0.875/2);
      \draw (1) arc (-180:0:0.875/2);
      \draw (2) to (4);
      \draw (5) arc (0:180:0.875/2);
      \draw (4) arc (-180:0:0.875/2);
      \draw (5) to (3.center);
    \end{pgfonlayer}
\end{tikzpicture}
}}
\newcommand{\HCNa}[5]
{\scalebox{\width}[\height]{
  \begin{tikzpicture}[baseline={([yshift=-.5ex]current bounding box.center)},small]
    \begin{pgfonlayer}{nodelayer}
      \node [style=none] (0) at (-2, 0) {};
      \node [style=A] (1) at (-1.25, 0) {};
      \node [style=A] (2) at (-0.375, 0) {};
      \node [style=none] (3) at (2, 0) {};
      \node [style=A] (4) at (0.375, 0) {};
      \node [style=A] (5) at (1.25, 0) {};
    \end{pgfonlayer}
    \begin{pgfonlayer}{edgelayer}
      \draw (0.center) to (1);
      \ifnum #1 >0 {\draw [style=dashed] (2) arc (0:180:0.875/2);}\else {\draw (2) arc (0:180:0.875/2);}\fi
      \ifnum #2 >0 {\draw [style=dashed] (1) arc (-180:0:0.875/2);}\else {\draw (1) arc (-180:0:0.875/2);}\fi
      \ifnum #3 >0 {\draw [style=dashed] (2) to (4);}\else {\draw (2) to (4);}\fi
      \ifnum #4 >0 {\draw [style=dashed] (5) arc (0:180:0.875/2);}\else {\draw (5) arc (0:180:0.875/2);}\fi
      \ifnum #5 >0 {\draw [style=dashed] (4) arc (-180:0:0.875/2);}\else {\draw (4) arc (-180:0:0.875/2);}\fi
      \draw [style=dashed] (5) to (3.center);
    \end{pgfonlayer}
\end{tikzpicture}
}}
\newcommand{\HCNbT}
{\scalebox{\width}[\height]{
  \begin{tikzpicture}[baseline={([yshift=-.5ex]current bounding box.center)},small]
	\begin{pgfonlayer}{nodelayer}
		\node [style=none] (0) at (-2, 0) {};
		\node [style=A] (1) at (-1, 0) {};
		\node [style=A] (2) at (0, 1) {};
		\node [style=A] (3) at (1, 0) {};
		\node [style=none] (4) at (2, 0) {};
		\node [style=A] (5) at (0, -1) {};
	\end{pgfonlayer}
	\begin{pgfonlayer}{edgelayer}
		\draw (0.center) to (1);
		\draw (3) to (4.center);
                \draw (1) arc (0:-90:-1);
                \draw (1) arc (0:90:-1);
                \draw (2) to (5);
                \draw (3) arc (0:90:1);
                \draw (3) arc (0:-90:1);
	\end{pgfonlayer}
  \end{tikzpicture}
}  }
\newcommand{\HCNb}[5]
{\scalebox{\width}[\height]{
  \begin{tikzpicture}[baseline={([yshift=-.5ex]current bounding box.center)},small]
	\begin{pgfonlayer}{nodelayer}
		\node [style=none] (0) at (-2, 0) {};
		\node [style=A] (1) at (-1, 0) {};
		\node [style=A] (2) at (0, 1) {};
		\node [style=A] (3) at (1, 0) {};
		\node [style=none] (4) at (2, 0) {};
		\node [style=A] (5) at (0, -1) {};
	\end{pgfonlayer}
	\begin{pgfonlayer}{edgelayer}
		\draw (0.center) to (1);
		\draw [style=dashed] (3) to (4.center);
                \ifnum #1 >0 {\draw [style=dashed] (1) arc (0:-90:-1);}\else {\draw (1) arc (0:-90:-1);}\fi
                \ifnum #2 >0 {\draw [style=dashed] (1) arc (0:90:-1);}\else {\draw (1) arc (0:90:-1);}\fi
                \ifnum #3 >0 {\draw [style=dashed](2) to (5);}\else {\draw (2) to (5);}\fi
                \ifnum #4 >0 {\draw [style=dashed] (3) arc (0:90:1);}\else {\draw (3) arc (0:90:1);}\fi
                \ifnum #5 >0 {\draw [style=dashed] (3) arc (0:-90:1);}\else {\draw (3) arc (0:-90:1);}\fi
	\end{pgfonlayer}
  \end{tikzpicture}
}  }
\newcommand{\HCNcT}
{\scalebox{\width}[\height]{
  \begin{tikzpicture}[baseline={([yshift=-.5ex]current bounding box.center)},small]
	\begin{pgfonlayer}{nodelayer}
		\node [style=none] (0) at (-2, 0) {};
		\node [style=A] (1) at (-1, 0) {};
		\node [style=A,anchor=north] (2) at (-0.5, 1) {};
		\node [style=A] (3) at (1, 0) {};
		\node [style=none] (4) at (2, 0) {};
		\node [style=A,anchor=north] (5) at (0.5, 1) {};
	\end{pgfonlayer}
	\begin{pgfonlayer}{edgelayer}
		\draw (0.center) to (1);
		\draw (3) to (4.center);
                \draw (1) arc (0:-atan2(1,.5):-1);
                \draw (3) arc (0:-180:1);
                \draw (5) arc (0:180:.5);
                \draw (5) arc (0:-180:.5);
                \draw (3) arc (0:atan2(1,.5):1);
	\end{pgfonlayer}
  \end{tikzpicture}
}  }
\newcommand{\HCNc}[5]
{\scalebox{\width}[\height]{
  \begin{tikzpicture}[baseline={([yshift=-.5ex]current bounding box.center)},small]
	\begin{pgfonlayer}{nodelayer}
		\node [style=none] (0) at (-2, 0) {};
		\node [style=A] (1) at (-1, 0) {};
		\node [style=A,anchor=north] (2) at (-0.5, 1) {};
		\node [style=A] (3) at (1, 0) {};
		\node [style=none] (4) at (2, 0) {};
		\node [style=A,anchor=north] (5) at (0.5, 1) {};
	\end{pgfonlayer}
	\begin{pgfonlayer}{edgelayer}
		\draw (0.center) to (1);
		\draw [style=dashed] (3) to (4.center);
                \ifnum #1 >0 {\draw [style=dashed] (1) arc (0:-atan2(1,.5):-1);}\else {\draw (1) arc (0:-atan2(1,.5):-1);}\fi
                \ifnum #2 >0 {\draw [style=dashed] (3) arc (0:-180:1);}\else {\draw (3) arc (0:-180:1);}\fi
                \ifnum #3 >0 {\draw [style=dashed] (5) arc (0:180:.5);}\else {\draw (5) arc (0:180:.5);}\fi
                \ifnum #4 >0 {\draw [style=dashed] (5) arc (0:-180:.5);}\else {\draw (5) arc (0:-180:.5);}\fi
                \ifnum #5 >0 {\draw [style=dashed] (3) arc (0:atan2(1,.5):1);}\else {\draw (3) arc (0:atan2(1,.5):1);}\fi
	\end{pgfonlayer}
  \end{tikzpicture}
}   }         
\newcommand{\betaLoop}[1]{(#1.north east) let \p1 = ($(#1.north)-(#1)$) in arc (-45:225:{veclen(\x1,\y1)})}
\newcommand{\DADA}[1]
{\scalebox{\width}[\height]{
  \begin{tikzpicture}[baseline={([yshift=-14pt]current bounding box.center)},resize]
    \begin{pgfonlayer}{nodelayer}
      \node [style=none] (0) at (2, 0) {};
      \node [style=BETA] (1) at (0, 0) {};
      \node [style=none] (2) at (-2, 0) {};
      \node [style=none] (3) at (1.center) {};
      \node [style=none] (4) at (1.north east) {};
    \end{pgfonlayer}
    \begin{pgfonlayer}{edgelayer}
      \ifnum #1 >0 {\draw [style=dashderder] \betaLoop1;}
      \else {\draw [style=derder] \betaLoop1;}\fi
      \draw [style=dashed] (0.center) to (1);
      \draw (2.center) to (1);
    \end{pgfonlayer}
  \end{tikzpicture}
}}
\newcommand{\DDl}[1]
{\scalebox{\width}[\height]{
  \begin{tikzpicture}[baseline={([yshift=-14pt]current bounding box.center)},resize]
    \begin{pgfonlayer}{nodelayer}
	\node [style=none] (0) at (2, 0) {};
	\node [style=BETA] (1) at (0, 0) {};
	\node [style=none] (2) at (-2, 0) {};
    \end{pgfonlayer}
    \begin{pgfonlayer}{edgelayer}
	\draw [style=dashed] (0.center) to (1);
	\draw [style=der] (2.center) to (1);
        \ifnum #1 >0 {\draw [style=dashder] \betaLoop1 ;}
        \else {\draw [style=derloop] \betaLoop1 ;}\fi
	\end{pgfonlayer}
  \end{tikzpicture}
}   }
\newcommand{\FDla}[2]
{\scalebox{\width}[\height]{
  \begin{tikzpicture}[baseline={([yshift=1pt]current bounding box.south)},resize]
    \begin{pgfonlayer}{nodelayer}
      \node [style=none] (0) at (2, 0) {};
      \node [style=A] (1) at (0, 0) {};
      \node [style=none] (2) at (-2, 0) {};
      \node [style=BETA] (3) at (0, 1) {};
    \end{pgfonlayer}
    \begin{pgfonlayer}{edgelayer}
      \draw [style=dashed] (0.center) to (1);
      \draw (2.center) to (1);
      \ifnum #1 >0 {\draw [style=dashed] (1) to (3);} \else {\draw (1) to (3);}\fi
      \ifnum #2 >0 {\draw [style=dashder] \betaLoop3;} \else {\draw [style=der] \betaLoop3 ;}\fi
    \end{pgfonlayer}
  \end{tikzpicture}
}       }
\newcommand{\FDlb}[2]
{\scalebox{\width}[\height]{
  \begin{tikzpicture}[baseline={([yshift=-10pt]current bounding box.center)},resize]
    \begin{pgfonlayer}{nodelayer}
      \node [style=none] (0) at (2, 0) {};
      \node [style=none] (1) at (-2, 0) {};
      \node [style=BETA] (2) at (.5, 0) {};
      \node [style=A] (3) at (-1, 0) {};
    \end{pgfonlayer}
    \begin{pgfonlayer}{edgelayer}
      \draw [style=dashed] (0.center) to (2);
      \draw (1.center) to (3);
      \ifnum #1 >0 {\draw [style=dashed] (3) to (2);} \else {\draw (3) to (2);}\fi
      \ifnum #2 >0 {\draw [style=dashderloopback] (3) arc (-90:270:.5);} \else {\draw [style=derloopback] (3) arc (-90:270:.5);}\fi
    \end{pgfonlayer}
  \end{tikzpicture}
}  }                       
\newcommand{\DFa}[2]
{\scalebox{\width}[\height]{
  \begin{tikzpicture}[baseline={([yshift=1pt]current bounding box.south)},resize]
	\begin{pgfonlayer}{nodelayer}
		\node [style=none] (0) at (2, 0) {};
		\node [style=A] (1) at (0, 0) {};
		\node [style=none] (2) at (-2, 0) {};
		\node [style=BETA] (3) at (0, 1.5) {};
	\end{pgfonlayer}
	\begin{pgfonlayer}{edgelayer}
		\draw [style=dashed] (0.center) to (1);
		\draw [style=der] (2.center) to (1);
                \ifnum #1 >0 {\draw [style=dashed] (1) arc (-90:-230:.75);}\else {\draw (1) arc (-90:-230:.75);}\fi
                \ifnum #2 >0 {\draw [style=dashed] (1) arc (-90:50:.75);}\else {\draw (1) arc (-90:50:.75);}\fi
	\end{pgfonlayer}
  \end{tikzpicture}
}  }
\newcommand{\DFb}[2]
{\scalebox{\width}[\height]{
  \begin{tikzpicture}[baseline={([yshift=1pt]current bounding box.south)},resize]
	\begin{pgfonlayer}{nodelayer}
		\node [style=none] (0) at (2, 0) {};
		\node [style=A] (1) at (0, 0) {};
		\node [style=none] (2) at (-2, 0) {};
		\node [style=BETA] (3) at (0,1.5) {};
	\end{pgfonlayer}
	\begin{pgfonlayer}{edgelayer}
		\draw [style=dashed] (0.center) to (1);
		\draw  (2.center) to (1);
                \ifnum #1 >0 {\draw [style=dashderbackI] (1) arc (-90:-230:.75);}\else {\draw [style=derbackI] (1) arc (-90:-230:.75);}\fi
                \ifnum #2 >0 {\draw [style=dashed] (1) arc (-90:50:.75);}\else {\draw (1) arc (-90:50:.75);}\fi
	\end{pgfonlayer}
  \end{tikzpicture}
}}
\newcommand{\DFc}[2]
{\scalebox{\width}[\height]{
  \begin{tikzpicture}[baseline={([yshift=-2pt]current bounding box.center)},resize]
    \begin{pgfonlayer}{nodelayer}
      \node [style=none] (0) at (2, 0) {};
      \node [style=A] (1) at (1, 0) {};
      \node [style=none] (2) at (-2, 0) {};
      \node [style=BETA] (3) at (-.5, 0) {};
    \end{pgfonlayer}
    \begin{pgfonlayer}{edgelayer}
      \draw [style=dashed] (0.center) to (1);
      \draw [style=der] (2.center) to (3);
      \ifnum #1 >0 {\draw [style=dashed] (3.south east) arc (30:180:-.6);}\else {\draw (3.south east) arc (30:180:-.6);}\fi
      \ifnum #2 >0 {\draw [style=dashed] (3.north east) arc (-30:-180:-.6);}\else {\draw (3.north east) arc (-30:-180:-.6);}\fi
    \end{pgfonlayer}
  \end{tikzpicture}
}}
\newcommand{\DFd}[2]
{\scalebox{\width}[\height]{
  \begin{tikzpicture}[baseline={([yshift=-3pt]current bounding box.center)},resize]
    \begin{pgfonlayer}{nodelayer}
      \node [style=none] (0) at (2, 0) {};
      \node [style=A] (1) at (1, 0) {};
      \node [style=none] (2) at (-2, 0) {};
      \node [style=BETA] (3) at (-.5, 0) {};
    \end{pgfonlayer}
    \begin{pgfonlayer}{edgelayer}
      \draw [style=dashed] (0.center) to (1);
      \draw [style=none] (2.center) to (3);
      \ifnum #1 >0 {\draw [style=dashderI] (3.south east) arc (30:180:-.6);}\else {\draw [style=derI] (3.south east) arc (30:180:-.6);}\fi
      \ifnum #2 >0 {\draw [style=dashed] (3.north east) arc (-30:-180:-.6);}\else {\draw (3.north east) arc (-30:-180:-.6);}\fi
    \end{pgfonlayer}
  \end{tikzpicture}
}}
\newcommand{\FADla}[1]
{\scalebox{\width}[\height]{
  \begin{tikzpicture}[baseline={([yshift=11pt]current bounding box.south)},resize]
    \begin{pgfonlayer}{nodelayer}
      \node [style=none] (0) at (2, 0) {};
      \node [style=BETA] (1) at (0, 0) {};
      \node [style=none] (2) at (-2, 0) {};
      \node [style=FA] (3) at (-1, 0) {};
    \end{pgfonlayer}
    \begin{pgfonlayer}{edgelayer}
      \draw [style=dashed] (0.center) to (1);
      \ifnum #1 >0 {\draw [style=dashderloop] \betaLoop1;}\else {\draw [style=derloop] \betaLoop1;}\fi
      \draw (2.center) to (3);
    \end{pgfonlayer}
  \end{tikzpicture}
}}
\newcommand{\FADlb}[1]
{\scalebox{\width}[\height]{
  \begin{tikzpicture}[baseline={([yshift=11pt]current bounding box.south)},resize]
    \begin{pgfonlayer}{nodelayer}
      \node [style=none] (0) at (2, 0) {};
      \node [style=BETA] (1) at (0, 0) {};
      \node [style=none] (2) at (-2, 0) {};
      \node [style=FA] (3) at (0, 1.75) {};
    \end{pgfonlayer}
    \begin{pgfonlayer}{edgelayer}
      \draw [style=dashed] (0.center) to (1);
      \draw (2.center) to (1);
      \ifnum #1 >0 {\draw [style=derloopback] (3) arc (-270:90:.5);}\else {\draw [style=derloopback] (3) arc (-270:90:.5);}\fi
    \end{pgfonlayer}
  \end{tikzpicture}
}}
\newcommand{\DFAa}[1]
{\scalebox{\width}[\height]{
  \begin{tikzpicture}[baseline={([yshift=11pt]current bounding box.south)},resize]
	\begin{pgfonlayer}{nodelayer}
		\node [style=none] (0) at (2, 0) {};
		\node [style=BETA] (1) at (0, 0) {};
		\node [style=none] (2) at (-2, 0) {};
		\node [style=FA] (3) at (0, 1.25) {};
	\end{pgfonlayer}
	\begin{pgfonlayer}{edgelayer}
		\draw [style=dashed] (0.center) to (1);
		\draw [style=der] (2.center) to (1);
		\ifnum #1 >0 {\draw [style=dashed] (1) to (3);}\else {\draw (1) to (3);}\fi
	\end{pgfonlayer}
  \end{tikzpicture}
}}
\newcommand{\DFAb}[1]
{\scalebox{\width}[\height]{
  \begin{tikzpicture}[baseline={([yshift=-.5ex]current bounding box.center)},resize]
	\begin{pgfonlayer}{nodelayer}
		\node [style=none] (0) at (2, 0) {};
		\node [style=none] (1) at (-2, 0) {};
		\node [style=BETA] (2) at (.5, 0) {};
		\node [style=FA] (3) at (-1, 0) {};
	\end{pgfonlayer}
	\begin{pgfonlayer}{edgelayer}
		\draw [style=dashed] (0.center) to (2);
		\draw (1.center) to (3);
		\ifnum #1 >0 {\draw [style=dashder] (2) to (3);}\else {\draw [style=der] (2) to (3);}\fi
	\end{pgfonlayer}
  \end{tikzpicture}
}}
\newcommand{\DFAc}[1]
{\scalebox{\width}[\height]{
  \begin{tikzpicture}[baseline={([yshift=-.5ex]current bounding box.center)},resize]
	\begin{pgfonlayer}{nodelayer}
		\node [style=none] (0) at (2, 0) {};
		\node [style=none] (1) at (-2, 0) {};
		\node [style=BETA] (2) at (0.5, 0) {};
		\node [style=FA] (3) at (-1, 0) {};
	\end{pgfonlayer}
	\begin{pgfonlayer}{edgelayer}
		\draw [style=dashed] (0.center) to (2);
		\draw [style=der] (1.center) to (3);
		\ifnum #1 >0 {\draw [style=dash] (2) to (3);}\else {\draw (2) to (3);}\fi
	\end{pgfonlayer}
  \end{tikzpicture}
}}
\newcommand{\DFAd}[1]
{\scalebox{\width}[\height]{
  \begin{tikzpicture}[baseline={([yshift=11pt]current bounding box.south)},resize]
	\begin{pgfonlayer}{nodelayer}
		\node [style=none] (0) at (2, 0) {};
		\node [style=BETA] (1) at (0, 0) {};
		\node [style=none] (2) at (-2, 0) {};
		\node [style=FA] (3) at (0, 1.75) {};
	\end{pgfonlayer}
	\begin{pgfonlayer}{edgelayer}
		\draw [style=dashed] (0.center) to (1);
		\draw (2.center) to (1);
                \ifnum #1 >0 {\draw [style=dashder] (3) to (1);}\else {\draw [style=der] (3) to (1);}\fi
	\end{pgfonlayer}
  \end{tikzpicture}
}}
\newcommand{\FFAa}[2]
{\scalebox{\width}[\height]{
  \begin{tikzpicture}[baseline={([yshift=-.5ex]current bounding box.center)},resize]
	\begin{pgfonlayer}{nodelayer}
		\node [style=none] (0) at (2, 0) {};
		\node [style=FA] (1) at (1.25, 0) {};
		\node [style=none] (2) at (-2, 0) {};
		\node [style=BETA] (3) at (0.25, 0) {};
		\node [style=A] (4) at (-1.25, 0) {};
	\end{pgfonlayer}
	\begin{pgfonlayer}{edgelayer}
		\draw [style=dashed] (0.center) to (1);
		\draw (2.center) to (4);
                \ifnum #1 >0 {\draw [style=dashed] (4) arc (0:-141:-.75);}\else {\draw (4) arc (0:-141:-.75);}\fi
                \ifnum #2 >0 {\draw [style=dashed] (4) arc (0:141:-.75);}\else {\draw (4) arc (0:141:-.75);}\fi
	\end{pgfonlayer}
  \end{tikzpicture}
}}
\newcommand{\FFAb}[2]
{\scalebox{\width}[\height]{
  \begin{tikzpicture}[baseline={([yshift=1pt]current bounding box.south)},resize]
	\begin{pgfonlayer}{nodelayer}
		\node [style=none] (0) at (2, 0) {};
		\node [style=A] (1) at (0, 0) {};
		\node [style=none] (2) at (-2, 0) {};
		\node [style=BETA] (3) at (0,1) {};
		\node [style=FA] (4) at (0,2) {};
	\end{pgfonlayer}
	\begin{pgfonlayer}{edgelayer}
		\draw [style=dashed] (0.center) to (1);
		\draw (2.center) to (1);
                \ifnum #1 >0 {\draw [style=dashed] (1) to (3);}\else {\draw (1) to (3);}\fi
                \ifnum #2 >0 {\draw [style=dashed] (3) to (4);}\else {\draw (3) to (4);}\fi
	\end{pgfonlayer}
  \end{tikzpicture}
}}
\newcommand{\FFAc}[2]
{\scalebox{\width}[\height]{
  \begin{tikzpicture}[baseline={([yshift=8pt]current bounding box.south)},resize]
	\begin{pgfonlayer}{nodelayer}
		\node [style=none] (0) at (2, 0) {};
		\node [style=none] (1) at (-2, 0) {};
		\node [style=BETA] (2) at (.5, 0) {};
		\node [style=A] (3) at (-1, 0) {};
		\node [style=FA] (4) at (-1, 1.25) {};
	\end{pgfonlayer}
	\begin{pgfonlayer}{edgelayer}
		\draw [style=dashed] (0.center) to (2);
		\draw (1.center) to (3);
                \ifnum #1 >0 {\draw [style=dashed] (3) to (4);}\else {\draw (3) to (4);}\fi
                \ifnum #2 >0 {\draw [style=dashed] (2) to (3);}\else {\draw (2) to (3);}\fi
	\end{pgfonlayer}
  \end{tikzpicture}
}}
\newcommand{\FFa}[3]
{\scalebox{\width}[\height]{
  \begin{tikzpicture}[baseline={([yshift=1pt]current bounding box.south)},resize]
    \begin{pgfonlayer}{nodelayer}
      \node [style=none] (0) at (2, 0) {};
      \node [style=A] (1) at (0, .75) {};
      \node [style=none] (2) at (-2, 0) {};
      \node [style=BETA] (3) at (0, 1.75) {};
      \node [style=A] (4) at (0, 0) {};
    \end{pgfonlayer}
    \begin{pgfonlayer}{edgelayer}
      \draw (2.center) to (4);
      \draw [style=dashed] (0.center) to (4);
      \ifnum #1 >0 {\draw [style=dashed] (4) to (1)=;}\else {\draw (4) to (1);}\fi
      \ifnum #2 >0 {\draw [style=dashed] (1) arc (-90:-210:.5);}\else {\draw (1) arc (-90:-210:.5);}\fi
      \ifnum #3 >0 {\draw [style=dashed] (1) arc (-90:30:.5);}\else {\draw (1) arc (-90:30:.5);}\fi
    \end{pgfonlayer}
  \end{tikzpicture}
}}
\newcommand{\FFb}[3]
{\scalebox{\width}[\height]{
  \begin{tikzpicture}[baseline={([yshift=-.5ex]current bounding box.center)},resize]
    \begin{pgfonlayer}{nodelayer}
      \node [style=none] (0) at (2, 0) {};
      \node [style=none] (1) at (-2, 0) {};
      \node [style=A] (2) at (0, 0) {};
      \node [style=A] (3) at (-1, 0) {};
      \node [style=BETA] (4) at (1, 0) {};
    \end{pgfonlayer}
    \begin{pgfonlayer}{edgelayer}
      \draw [style=dashed] (0.center) to (4);
      \draw (1.center) to (3);
      \ifnum #1 >0 {\draw [style=dashed] (2) arc (0:180:.5);}\else {\draw (2) arc (0:180:.5);}\fi
      \ifnum #2 >0 {\draw [style=dashed] (2) arc (0:-180:.5);}\else {\draw (2) arc (0:-180:.5);}\fi
      \ifnum #3 >0 {\draw [style=dashed](2) to (4);}\else {\draw (2) to (4);}\fi
    \end{pgfonlayer}
  \end{tikzpicture}
}}
\newcommand{\FFc}[3]
{\scalebox{\width}[\height]{
  \begin{tikzpicture}[baseline={([yshift=-2.1ex]current bounding box.center)},resize]
    \begin{pgfonlayer}{nodelayer}
      \node [style=none] (0) at (2, 0) {};
      \node [style=A] (1) at (1, 0) {};
      \node [style=none] (2) at (-2, 0) {};
      \node [style=BETA] (3) at (0, 1) {};
      \node [style=A] (4) at (-1, 0) {};
    \end{pgfonlayer}
    \begin{pgfonlayer}{edgelayer}
      \draw [style=dashed] (0.center) to (1);
      \draw (2.center) to (4);
      \ifnum #1 >0 {\draw [style=dashed] (1) arc (0:-180:1);}\else {\draw (1) arc (0:-180:1);}\fi
      \ifnum #2 >0 {\draw [style=dashed] (1) arc (0:61:1);}\else {\draw (1) arc (0:61:1);}\fi
      \ifnum #3 >0 {\draw [style=dashed] (4) arc (0:-61:-1);}\else {\draw (4) arc (0:-61:-1);}\fi
    \end{pgfonlayer}
  \end{tikzpicture}
}}
\newcommand{\DDFF}[4]
{\scalebox{\width}[\height]{
  \begin{tikzpicture}[baseline={([yshift=-.5ex]current bounding box.center)},small]
    \begin{pgfonlayer}{nodelayer}
      \node [style=A] (0) at (-2, 0) {};
      \node [style=A] (1) at (2, 0) {};
    \end{pgfonlayer}
    \begin{pgfonlayer}{edgelayer}
      \ifnum #1 >0 {\draw [style=dashed] (1) arc (0:180:2);}\else {\draw (1) arc (0:180:2);}\fi
      \ifnum #2 >0 {\draw [style=dashed,bend left=35] (0) to (1);}\else {\draw [bend left=35] (0) to (1);}\fi
      \ifnum #3 >0 {\draw [style=dashed,bend right=35] (0) to (1);}\else {\draw [bend right=35] (0) to (1);}\fi
      \ifnum #4 >0 {\draw [style=dashderder] (1.south) arc (-1:-179:2);}\else {\draw [style=derder] (1.south) arc (-1:-179:2);}\fi
    \end{pgfonlayer}
  \end{tikzpicture}
}}
\newcommand{\FFDDl}[4]
{\scalebox{\width}[\height]{
  \begin{tikzpicture}[baseline={([yshift=-.5ex]current bounding box.center)},small]
    \begin{pgfonlayer}{nodelayer}
      \node [style=A] (0) at (0, 0) {};
      \node [style=A] (1) at (2, 0) {};
    \end{pgfonlayer}
    \begin{pgfonlayer}{edgelayer}
      \ifnum #1 >0 {\draw [style=dashderloopback] (0) arc (0:360:1);}\else {\draw [style=derloopback] (0) arc (0:360:1);}\fi
      \ifnum #2 >0 {\draw [style=dashed] (1) arc (0:180:1);}\else {\draw (1) arc (0:180:1);}\fi
      \ifnum #3 >0 {\draw [style=dashder] (1) to (0);}\else {\draw [style=der] (1) to (0);}\fi
      \ifnum #4 >0 {\draw [style=dashed] (1) arc (0:-180:1);}\else {\draw (1) arc (0:-180:1);}\fi
    \end{pgfonlayer}
  \end{tikzpicture}
}}
\newcommand{\FFDDlprime}[4]
{\scalebox{\width}[\height]{
  \begin{tikzpicture}[baseline={([yshift=-.5ex]current bounding box.center)},small]
    \begin{pgfonlayer}{nodelayer}
      \node [style=A] (0) at (0, 0) {};
      \node [style=A] (1) at (2, 0) {};
    \end{pgfonlayer}
    \begin{pgfonlayer}{edgelayer}
      \ifnum #1 >0 {\draw [style=dashderloopback] (0) arc (0:-360:1);}\else {\draw [style=derloopback] (0) arc (0:-360:1);}\fi
      \ifnum #2 >0 {\draw [style=dashed] (1) arc (0:180:1);}\else {\draw (1) arc (0:180:1);}\fi
      \ifnum #3 >0 {\draw [style=dashder] (1) to (0);}\else {\draw [style=der] (1) to (0);}\fi
      \ifnum #4 >0 {\draw [style=dashed] (1) arc (0:-180:1);}\else {\draw (1) arc (0:-180:1);}\fi
    \end{pgfonlayer}
  \end{tikzpicture}
}}
\newcommand{\FFDlDl}[4]
{\scalebox{\width}[\height]{
  \begin{tikzpicture}[baseline={([yshift=-.5ex]current bounding box.center)},small]
    \begin{pgfonlayer}{nodelayer}
      \node [style=A] (0) at (-1, 0) {};
      \node [style=A] (1) at (1, 0) {};
    \end{pgfonlayer}
    \begin{pgfonlayer}{edgelayer}
      \ifnum #1 >0 {\draw [style=dashderloopback] (0) arc (0:360:.5);}\else {\draw [style=derloopback] (0) arc (0:360:.5);}\fi
      \ifnum #2 >0 {\draw [style=dashed] (1) arc (0:180:1);}\else {\draw (1) arc (0:180:1);}\fi
      \ifnum #3 >0 {\draw [style=dashed] (1) arc (0:-180:1);}\else {\draw (1) arc (0:-180:1);}\fi
      \ifnum #4 >0 {\draw [style=dashderloopback] (1) arc (0:360:-.5);}\else {\draw [style=derloopback] (1) arc (0:360:-.5);}\fi
    \end{pgfonlayer}
  \end{tikzpicture}
}}
\newcommand{\FFFD}[5]
{\scalebox{\width}[\height]{
  \begin{tikzpicture}[baseline={([yshift=-.5ex]current bounding box.center)},small]
    \begin{pgfonlayer}{nodelayer}
      \node [style=A] (0) at (-2, 0) {};
      \node [style=A] (1) at (2, 0) {};
      \node [style=A] (2) at (0, 2) {};
    \end{pgfonlayer}
    \begin{pgfonlayer}{edgelayer}
      \ifnum #1 >0 {\draw [style=dashderback] (2) arc (90:178:2);}\else {\draw [style=derback] (2) arc (90:178:2);}\fi
      \ifnum #2 >0 {\draw [style=dashed] (0) to (2);}\else {\draw (0) to (2);}\fi
      \ifnum #3 >0 {\draw [style=dashed] (1) arc (0:90:2);}\else {\draw (1) arc (0:90:2);}\fi
      \ifnum #4 >0 {\draw [style=dashed] (0) to (1);}\else {\draw (0) to (1);}\fi
      \ifnum #5 >0 {\draw [style=dashed] (1) arc (0:-180:2);}\else {\draw (1) arc (0:-180:2);}\fi
    \end{pgfonlayer}
  \end{tikzpicture}
}}
\newcommand{\FFFADa}[4]
{\scalebox{\width}[\height]{
  \begin{tikzpicture}[baseline={([yshift=-.5ex]current bounding box.center)},small]
     \begin{pgfonlayer}{nodelayer}
	\node [style=A] (0) at (-2, 0) {};
	\node [style=A] (1) at (2, 0) {};
	\node [style=FA] (2) at (0, 2) {};
     \end{pgfonlayer}
     \begin{pgfonlayer}{edgelayer}
	\ifnum #1 >0 {\draw [style=dashderback] (0) arc (0:-88:-2);}\else {\draw [style=derback] (0) arc (0:-88:-2);}\fi
	\ifnum #2 >0 {\draw [style=dashed] (1) arc (0:90:2);}\else {\draw (2) arc (0:90:2);}\fi
	\ifnum #3 >0 {\draw [style=dashed] (0) to (1);}\else {\draw (0) to (1);}\fi
	\ifnum #4 >0 {\draw [style=dashed] (1) arc (0:-180:2);}\else {\draw (1) arc (0:-180:2);}\fi
	\end{pgfonlayer}
  \end{tikzpicture}
}}
\newcommand{\FFFADb}[4]
{\scalebox{\width}[\height]{
  \begin{tikzpicture}[baseline={([yshift=-.5ex]current bounding box.center)},small]
    \begin{pgfonlayer}{nodelayer}
      \node [style=A] (0) at (-1, 0) {};
      \node [style=A] (1) at (2, 0) {};
      \node [style=FA] (2) at (-2, 0) {};
    \end{pgfonlayer}
    \begin{pgfonlayer}{edgelayer}
      \ifnum #1 >0 {\draw [style=dashed] (0) to (2);}\else {\draw (0) to (2);}\fi
      \ifnum #2 >0 {\draw [style=dashed] (1) arc (0:180:1.5);}\else {\draw (1) arc (0:180:1.5);}\fi
      \ifnum #3 >0 {\draw [style=dashder] (1) to (0);}\else {\draw [style=der] (1) to (0);}\fi
      \ifnum #4 >0 {\draw [style=dashed] (1) arc (0:-180:1.5);}\else {\draw (1) arc (0:-180:1.5);}\fi
    \end{pgfonlayer}
  \end{tikzpicture}
}}
\newcommand{\FFFADl}[4]
{\scalebox{\width}[\height]{
  \begin{tikzpicture}[baseline={([yshift=-.5ex]current bounding box.center)},small]
    \begin{pgfonlayer}{nodelayer}
      \node [style=A] (0) at (-1, 0) {};
      \node [style=A] (1) at (1, 0) {};
      \node [style=FA] (2) at (-2, 0) {};
    \end{pgfonlayer}
    \begin{pgfonlayer}{edgelayer}
      \ifnum #1 >0 {\draw [style=dashed] (0) to (2);}\else {\draw (0) to (2);}\fi
      \ifnum #2 >0 {\draw [style=dashed] (1) arc (0:180:1);}\else {\draw (1) arc (0:180:1);}\fi
      \ifnum #3 >0 {\draw [style=dashed] (1) arc (0:-180:1);}\else {\draw(1) arc (0:-180:1);}\fi
      \ifnum #4 >0 {\draw [style=dashderloopback] (1) arc (0:360:-.5);}\else {\draw [style=derloopback] (1) arc (0:360:-.5);}\fi
    \end{pgfonlayer}
  \end{tikzpicture}
}}
\newcommand{\FFFAFA}[4]
{\scalebox{\width}[\height]{
  \begin{tikzpicture}[baseline={([yshift=-.5ex]current bounding box.center)},small]
    \begin{pgfonlayer}{nodelayer}
      \node [style=A] (0) at (-1, 0) {};
      \node [style=A] (1) at (1, 0) {};
      \node [style=FA] (2) at (-2, 0) {};
      \node [style=FA] (3) at (2, 0) {};
    \end{pgfonlayer}
    \begin{pgfonlayer}{edgelayer}
      \ifnum #1 >0 {\draw [style=dashed] (0) to (2);}\else {\draw (0) to (2);}\fi
      \ifnum #2 >0 {\draw [style=dashed] (1) arc (0:180:1);}\else {\draw (1) arc (0:180:1);}\fi
      \ifnum #3 >0 {\draw [style=dashed] (1) arc (0:-180:1);}\else {\draw (1) arc (0:-180:1);}\fi
      \ifnum #4 >0 {\draw [style=dashed] (1) to (3);}\else {\draw (1) to (3);}\fi
    \end{pgfonlayer}
  \end{tikzpicture}
}}
\newcommand{\FFFFa}[6]
{\scalebox{\width}[\height]{
  \begin{tikzpicture}[baseline={([yshift=-.5ex]current bounding box.center)},small]
    \begin{pgfonlayer}{nodelayer}
      \node [style=A] (0) at (-1.75, 0) {};
      \node [style=A] (1) at (1.75, 0) {};
      \node [style=A,anchor=north] (2) at (-0.75, 1.65) {};
      \node [style=A,anchor=north] (3) at (0.75, 1.65) {};
    \end{pgfonlayer}
    \begin{pgfonlayer}{edgelayer}
      \ifnum #1 >0 {\draw [style=dashed] (0) arc (0:-atan2(1,.5):-1.75);}\else {\draw (0) arc (0:-atan2(1,.5):-1.75);}\fi
      \ifnum #2 >0 {\draw [style=dashed] (3) arc (0:180:.75);}\else {\draw (3) arc (0:180:.75);}\fi
      \ifnum #3 >0 {\draw [style=dashed] (3) arc (0:-180:.75);}\else {\draw (3) arc (0:-180:.75);}\fi
      \ifnum #4 >0 {\draw [style=dashed] (1) arc (0:atan2(1,.5):1.75);}\else {\draw (1) arc (0:atan2(1,.5):1.75);}\fi
      \ifnum #5 >0 {\draw [style=dashed] (0) to (1);}\else {\draw (0) to (1);}\fi
      \ifnum #6 >0 {\draw [style=dashed] (1) (1) arc (0:-180:1.75);}\else {\draw (1) (1) arc (0:-180:1.75);}\fi
    \end{pgfonlayer}
  \end{tikzpicture}
}}
\newcommand{\FFFFaT}
{\scalebox{\width}[\height]{
  \begin{tikzpicture}[baseline={([yshift=-.5ex]current bounding box.center)},small]
    \begin{pgfonlayer}{nodelayer}
      \node [style=A] (0) at (-1.75, 0) {};
      \node [style=A] (1) at (1.75, 0) {};
      \node [style=A,anchor=north] (2) at (-.75, 1.65) {};
      \node [style=A,anchor=north] (3) at (.75, 1.65) {};
    \end{pgfonlayer}
    \begin{pgfonlayer}{edgelayer}
      \draw (0) arc (0:-atan2(1,.5):-1.75);
      \draw (3) arc (0:180:.75);
      \draw (3) arc (0:-180:.75);
      \draw (1) arc (0:atan2(1,.5):1.75);
      \draw (0) to (1);
      \draw (1) arc (0:-180:1.75);
    \end{pgfonlayer}
  \end{tikzpicture}
}}
\newcommand{\FFFFb}[6]
{\scalebox{\width}[\height]{
  \begin{tikzpicture}[baseline={([yshift=-.5ex]current bounding box.center)},small]
    \begin{pgfonlayer}{nodelayer}
      \node [style=A] (0) at (-2, 0) {};
      \node [style=A] (1) at (0, 0) {};
      \node [style=A] (2) at (2, 0) {};
      \node [style=A] (3) at (0, -2) {};
    \end{pgfonlayer}
    \begin{pgfonlayer}{edgelayer}
      \ifnum #1 >0 {\draw [style=dashed](2) arc (0:180:2);}\else {\draw (2) arc (0:180:2);}\fi
      \ifnum #2 >0 {\draw [style=dashed](0) to (1);}\else {\draw (0) to (1);}\fi
      \ifnum #3 >0 {\draw [style=dashed](1) to (2);}\else {\draw (1) to (2);}\fi
      \ifnum #4 >0 {\draw [style=dashed](3) arc (-90:-180:2);}\else {\draw (3) arc (-90:-180:2);}\fi
      \ifnum #5 >0 {\draw [style=dashed](1) to (3);}\else {\draw (1) to (3);}\fi
      \ifnum #6 >0 {\draw [style=dashed](3) arc (-90:90:2);}\else {\draw (3) arc (-90:90:2);}\fi
    \end{pgfonlayer}
  \end{tikzpicture}
}}
\newcommand{\FFFFbT}
{\scalebox{\width}[\height]{
  \begin{tikzpicture}[baseline={([yshift=-.5ex]current bounding box.center)},small]
    \begin{pgfonlayer}{nodelayer}
      \node [style=A] (0) at (-2, 0) {};
      \node [style=A] (1) at (0, 0) {};
      \node [style=A] (2) at (2, 0) {};
      \node [style=A] (3) at (0, -2) {};
    \end{pgfonlayer}
    \begin{pgfonlayer}{edgelayer}
      \draw (2) arc (0:180:2);
      \draw (0) to (1);
      \draw (1) to (2);
      \draw (3) arc (-90:-180:2);
      \draw (1) to (3);
      \draw (3) arc (-90:90:2);
    \end{pgfonlayer}
  \end{tikzpicture}
}}
\newcommand{\dFAFFFaT}
{\scalebox{\width}[\height]{
  \begin{tikzpicture}[baseline={([yshift=-.5ex-\yshift]current bounding box.center)},small]
	\begin{pgfonlayer}{nodelayer}
		\node [style=none] (0) at (2, 0) {};
		\node [style=none] (1) at (-2, 0) {};
		\node [style=A] (2) at (0, 0) {};
		\node [style=A] (3) at (0, 1) {};
		\node [style=A] (4) at (0, 2) {};
		\node [style=FA] (5) at (1.75, 2) {};
	\end{pgfonlayer}
	\begin{pgfonlayer}{edgelayer}
		\draw [style=none] (0.center) to (2);
		\draw (1.center) to (2);
		\draw (2) to (3);
		\draw (3) arc (-90:-270:.5);
		\draw (3) arc (-90:90:.5);
		\draw  (4) to (5);
	\end{pgfonlayer}
  \end{tikzpicture}
}}
\newcommand{\dFAFFFa}[4]
{\scalebox{\width}[\height]{
  \begin{tikzpicture}[baseline={([yshift=-.5ex-\yshift]current bounding box.center)},small]
	\begin{pgfonlayer}{nodelayer}
		\node [style=none] (0) at (2, 0) {};
		\node [style=none] (1) at (-2, 0) {};
		\node [style=A] (2) at (0, 0) {};
		\node [style=A] (3) at (0, 1) {};
		\node [style=A] (4) at (0, 2) {};
		\node [style=FA] (5) at (1.75, 2) {};
	\end{pgfonlayer}
	\begin{pgfonlayer}{edgelayer}
		\draw [style=dashed] (0.center) to (2);
		\draw (1.center) to (2);
		\ifnum #1 >0 {\draw [style=dashed](2) to (3);}\else {\draw (2) to (3);}\fi
		\ifnum #2 >0 {\draw [style=dashed] (3) arc (-90:-270:.5);}\else {\draw (3) arc (-90:-270:.5);}\fi
		\ifnum #3 >0 {\draw [style=dashed] (3) arc (-90:90:.5);}\else {\draw (3) arc (-90:90:.5);}\fi
		\ifnum #4 >0 {\draw [style=dashed] (4) to (5);}\else {\draw  (4) to (5);}\fi
	\end{pgfonlayer}
  \end{tikzpicture}
}}
\newcommand{\dFAFFFbprime}[4]
{\scalebox{\width}[\height]{
  \begin{tikzpicture}[baseline={([yshift=.5ex-\yshift]current bounding box.center)},small]
	\begin{pgfonlayer}{nodelayer}
		\node [style=none] (0) at (2,0) {};
		\node [style=none] (1) at (-2,0) {};
		\node [style=A] (2) at (0,0) {};
		\node [style=A] (4) at (-1,0) {};
		\node [style=FA] (5) at (1,1.75) {};
		\node [style=A] (6) at (1,0) {};
	\end{pgfonlayer}
	\begin{pgfonlayer}{edgelayer}
		\draw (1.center) to (4);
		\draw [style=dashed] (0.center) to (6);
		\ifnum #1 >0 {\draw [style=dashed] (6) to (5);}\else {\draw (6) to (5);}\fi
		\ifnum #2 >0 {\draw [style=dashed] (6) to (2);}\else {\draw (6) to (2);}\fi
		\ifnum #3 >0 {\draw [style=dashed] (2) arc (0:180:.5);}\else {\draw (2) arc (0:180:.5);}\fi
		\ifnum #4 >0 {\draw [style=dashed] (2) arc (0:-180:.5);}\else {\draw (2) arc (0:-180:.5);}\fi
	\end{pgfonlayer}
  \end{tikzpicture}
}}
\newcommand{\dFAFFFb}[4]
{\scalebox{\width}[\height]{
  \begin{tikzpicture}[baseline={([yshift=.5ex-\yshift]current bounding box.center)},small]
	\begin{pgfonlayer}{nodelayer}
		\node [style=none] (0) at (2,0) {};
		\node [style=none] (1) at (-2,0) {};
		\node [style=A] (2) at (1,0) {};
		\node [style=A] (3) at (-1,0) {};
		\node [style=A] (4) at (0,0) {};
		\node [style=FA] (5) at (-1,1.75) {};
	\end{pgfonlayer}
	\begin{pgfonlayer}{edgelayer}
		\draw [style=dashed] (0.center) to (2);
		\draw (1.center) to (3);
		\ifnum #1 >0 {\draw [style=dashed] (2) arc (0:-180:.5);}\else {\draw (2) arc (0:-180:.5);}\fi
		\ifnum #2 >0 {\draw [style=dashed] (2) arc (0:180:.5);}\else {\draw (2) arc (0:180:.5);}\fi
		\ifnum #3 >0 {\draw [style=dashed](4) to (3);}\else {\draw (4) to (3);}\fi
		\ifnum #4 >0 {\draw [style=dashed] (3) to (5);}\else {\draw (3) to (5);}\fi
	\end{pgfonlayer}
  \end{tikzpicture}
}}
\newcommand{\dFAFFFbT}
{\scalebox{\width}[\height]{
  \begin{tikzpicture}[baseline={([yshift=-.5ex]current bounding box.center)},small]
	\begin{pgfonlayer}{nodelayer}
		\node [style=none] (0) at (2,0) {};
		\node [style=none] (1) at (-2,0) {};
		\node [style=A] (2) at (1,0) {};
		\node [style=A] (3) at (-1,0) {};
		\node [style=A] (4) at (0,0) {};
		\node [style=FA] (5) at (-1,1.75) {};
	\end{pgfonlayer}
	\begin{pgfonlayer}{edgelayer}
		\draw [style=none] (0.center) to (2);
		\draw (1.center) to (3);
		\draw (2) arc (0:-180:.5);
		\draw (2) arc (0:180:.5);
		\draw (4) to (3);
		\draw (3) to (5);
	\end{pgfonlayer}
  \end{tikzpicture}
}}
\newcommand{\dFAFFFc}[4]
{\scalebox{\width}[\height]{
  \begin{tikzpicture}[baseline={([yshift=-.75ex-\yshift/2]current bounding box.center)},small]
	\begin{pgfonlayer}{nodelayer}
		\node [style=none] (0) at (-2, 0) {};
		\node [style=none] (1) at (2, 0) {};
		\node [style=A] (2) at (-1, 0) {};
		\node [style=A] (3) at (1, 0) {};
		\node [style=A] (4) at (0, 1) {};
		\node [style=FA] (5) at (0, 2) {};
	\end{pgfonlayer}
	\begin{pgfonlayer}{edgelayer}
		\draw [style=dashed] (1) to (3);
		\draw (0) to (2);
		\ifnum #1 >0 {\draw [style=dashed] (3) arc (0:-180:1);}\else {\draw  (3) arc (0:-180:1);}\fi
		\ifnum #2 >0 {\draw [style=dashed] (3) arc (0:90:1);}\else {\draw  (3) arc (0:90:1);}\fi
		\ifnum #3 >0 {\draw [style=dashed] (4) arc (90:180:1);}\else {\draw  (4) arc (90:180:1);}\fi
		\ifnum #4 >0 {\draw [style=dashed] (4) to (5);}\else {\draw  (4) to (5);}\fi
	\end{pgfonlayer}
  \end{tikzpicture}
}}
\newcommand{\dFAFFFcT}
{\scalebox{\width}[\height]{
  \begin{tikzpicture}[baseline={([yshift=-.5ex]current bounding box.center)},small]
	\begin{pgfonlayer}{nodelayer}
		\node [style=none] (0) at (-2, 0) {};
		\node [style=none] (1) at (2, 0) {};
		\node [style=A] (2) at (-1, 0) {};
		\node [style=A] (3) at (1, 0) {};
		\node [style=A] (4) at (0, 1) {};
		\node [style=FA] (5) at (0,2) {};
	\end{pgfonlayer}
	\begin{pgfonlayer}{edgelayer}
		\draw [style=none] (1) to (3);
		\draw (0) to (2);
		\draw (3) arc (0:-180:1);
		\draw  (3) arc (0:90:1);
		\draw  (4) arc (90:180:1);
		\draw  (4) to (5);
	\end{pgfonlayer}
  \end{tikzpicture}
}}

\begin{document}
\makeatletter
\@removefromreset{table}{chapter}
\makeatother
\renewcommand{\thetable}{\arabic{table}}
\makeatletter
\@removefromreset{figure}{chapter}
\makeatother
\renewcommand{\thefigure}{\arabic{figure}}

\def\sym#1{\ifmmode^{#1}\else\(^{#1}\)\fi} 

\def\mytitle{GEOMETRIC APPROACHES TO QUANTUM FIELDS AND STRINGS AT STRONG COUPLINGS} 
\def\myauthor{Thomas Rochais}
\def\myauthorfull{Thomas Bernard Bruno Rochais}
\def\mysupervisorname{Jonathan J. Heckman}
\def\mysupervisortitle{Associate Professor of Physics and Astronomy}
\newlength{\superlen}   
\settowidth{\superlen}{\mysupervisorname, \mysupervisortitle} 
\def\gradchairname{Mirjam Cvetic }
\def\gradchairtitle{Professor of Physics and Astronomy}
\newlength{\chairlen}   
\settowidth{\chairlen}{\gradchairname, \gradchairtitle} 
\newlength{\maxlen}
\setlength{\maxlen}{\maxof{\superlen}{\chairlen}}
\def\mydepartment{Physics and Astronomy}
\def\myyear{2021}
\def\signatures{35 pt} 


\pagenumbering{roman}
\pagestyle{plain}


\begin{titlepage}
\thispagestyle{empty} 
\begin{center}

\onehalfspacing

\mytitle

\myauthor

A DISSERTATION

in 

\mydepartment 

Presented to the Faculties of the University of Pennsylvania

in 

Partial Fulfillment of the Requirements for the

Degree of Doctor of Philosophy

\myyear

\end{center}


\begin{flushleft}

Supervisor of Dissertation\\[\signatures] 

\renewcommand{\tabcolsep}{0 pt}
\begin{table}[h]
\begin{tabularx}{\maxlen}{l}
\toprule
\mysupervisorname, \mysupervisortitle\\ 
\end{tabularx}
\end{table}

Graduate Group Chairperson\\[\signatures] 

\begin{table}[h]
\begin{tabularx}{\maxlen}{l}
\toprule
\gradchairname, \gradchairtitle\\ 
\end{tabularx}
\end{table}
\singlespacing

Dissertation Committee 

Mirjam Cvetic, Professor of Physics and Astronomy

Jonathan Heckman, Associate Professor of Physics and Astronomy

Justin Khoury, Professor of Physics and Astronomy

Burt Ovrut, Professor of Physics and Astronomy

Evelyn Thomson, Associate Professor of Physics and Astronomy

\end{flushleft}

\end{titlepage}


\doublespacing

\thispagestyle{empty} 

\vspace*{\fill}

\begin{flushleft}
\mytitle

 \copyright \space COPYRIGHT
 
\myyear

\myauthorfull\\[24 pt] 




\end{flushleft}
\pagebreak 

\newenvironment{preliminary}{}{}
\titleformat{\chapter}[hang]{\large\center}{\thechapter}{0 pt}{}
\titlespacing*{\chapter}{0pt}{-33 pt}{0 pt} 
\begin{preliminary}

\setcounter{page}{3}  
\begin{center}
\textit{This thesis is dedicated to my mom and my two sisters.}
\end{center}


\clearpage
\chapter*{ACKNOWLEDGMENT}
\addcontentsline{toc}{chapter}{ACKNOWLEDGMENT} 

First of all I would like to thank my advisor Jonathan Heckman for giving me the opportunity to explore the fascinating field of string theory and quantum field theories. I greatly benefited from his diverse knowledge and physical intuition. I would also like to thank Mirjam Cvetic for guidance, especially during my early days in graduate school.

Furthermore, special thanks go to Falk Hassler for patiently listening to all of my questions and guiding me consistently through many projects. His insights and detailed explanations were of great help throughout the years. I would also like to thank Fabio Apruzzi, Markus Dierigl, Craig Lawrie, Tom Rudelius, and Gianluca Zoccarato for their friendly advice and collaborations.

Thank you to my office mate Muyang Liu for the support and interesting discussions. And many thanks to my fellow graduate students Hao Zhang and Ethan Torres for many enjoyable collaborations.


Moreover, as a graduate student, I had the pleasure of attending various summer schools and workshops. I am thankful to the University of Pennsylvania and the many institutions that have hosted me over the past five years: the Simons Center for Geometry and Physics at Stony Brook, the University of Oxford, CERN, and Harvard University. I give thanks for the chance I was given to meet with so many great researchers there.

Finally, and most importantly, this work would not have been possible without all the support and love from my friends and family who helped me keep my research in perspective and make my everyday life so enjoyable. Thank you!


\clearpage
\chapter*{ABSTRACT}
\addcontentsline{toc}{chapter}{ABSTRACT} 
\begin{center}
\mytitle

\myauthor

\mysupervisorname

\end{center}

Geometric structures and dualities arise naturally in quantum field theories and string theory. In fact, these tools become very useful when studying strong coupling effects, where standard perturbative techniques can no longer be used. In this thesis we look at several conformal field theories in various dimensions. We first discuss the structure of the nilpotent networks stemming from T-brane deformations in 4D $\N=1$ theories and then go to the stringy origins of 6D superconformal field theories to realize deformations associated with T-branes in terms of simple combinatorial data. We then analyze non-perturbative generalizations of orientifold 3-planes (i.e. S-folds) in order to produce different 4D $\N=2$ theories. 
Afterwards, we turn our attention towards a few dualities found at strong coupling. For instance, abelian T-duality is known to be a full duality in string theory between type IIA and type IIB. Its nonabelian generalization, Poisson-Lie T-duality, has only been conjectured to be so. We show that Poisson-Lie symmetric $\sigma$-models are at least two-loop renormalizable and their $\beta$-functions are invariant under Poisson-Lie T-duality. 
Finally, we review recent progress leading to phenomenologically relevant dualities between M-theory on local $G_2$ spaces and F-theory on locally elliptically fibered Calabi-Yau fourfolds. In particular, we find that the 3D $\N=1$ effective field theory defined by M-theory on a local $Spin(7)$ space unifies the Higgs bundle data associated with 4D $\N=1$ M-theory and F-theory vacua. We finish with some comments on 3D interfaces at strong coupling.


\clearpage
\tableofcontents


\clearpage
\phantomsection
\listoftables
\addcontentsline{toc}{chapter}{LIST OF TABLES}


\clearpage
\phantomsection
\listoffigures
\addcontentsline{toc}{chapter}{LIST OF ILLUSTRATIONS}
\end{preliminary}

\newenvironment{mainf}{}{}
\titleformat{\chapter}[hang]{\large\center}{CHAPTER \thechapter}{0 pt}{: }
\titlespacing*{\chapter}{0pt}{-29 pt}{0 pt} 
\begin{mainf}

\newpage
\pagenumbering{arabic}
\pagestyle{plain} 

\setlength{\parskip}{10 pt} 
\setlength{\parindent}{0pt}

\titleformat{\chapter}[hang]{\large\center}{\thechapter}{0 pt}{}
\chapter*{INTRODUCTION}
\addcontentsline{toc}{chapter}{INTRODUCTION} 
\titleformat{\chapter}[hang]{\large\center}{CHAPTER \thechapter}{0 pt}{: }
Our universe presents itself with many mysteries that have fascinated humankind for millennia. By studying the making of the universe, whether it be atoms, or yet smaller elementary particles, a whole zoo of particles has been unearthed. Those are divided in two classes: fermions, which make up the known matter (and antimatter) of the universe, and bosons which mediate interactions between fermions. The discovery of those force carriers has allowed the so-called Standard Model to explain three of the four known fundamental forces. First we have the electromagnetic force, initially explained by Maxwell's equations, which is carried by photons. Then, the weak interaction is mediated by both the W and Z bosons, and finally the strong nuclear force is mediated by the three family of gluons. Paradoxically, this leaves gravity -- the one force we are most acquainted with -- largely unexplained. Newton first came up with an inverse-square law relation which was later improved upon by Einstein's theory of general relativity. However, this framework only applies at large scales; and to this day, there does not exist any complete theory of gravity which can work at the quantum scale. As a result, the scientific community is left with the task of unifying Einstein theory, which describes gravity on large scales, with quantum field theory and the Standard Model, which already unifies the other three fundamental forces. To that end, string theory appears to be the most likely candidate for a ``theory of everything". 

The postulate of string theory is rather simple: the fundamental building blocks of the universe are not point-like particles such as electrons, or photons, but instead they are tiny vibrating strings, around twenty orders of magnitude smaller. Strings then come in two flavors: either open strings, with two ends free to move and attach to larger objects called ``branes", or closed strings forming loops. One such closed string turns out to be the graviton, i.e. the particle mediating the gravitational force. In fact, while the theory of general relativity is modified at very short distance scales, string theory presents itself in exactly the form proposed by Einstein at ordinary distances. Furthermore, it solves many issues found in ordinary quantum field theory such as UV divergences, it has no free parameters to be artificially fixed by hand, and it is, so far, the only consistent theory of quantum gravity we know sufficiently well. 

However, string theory also comes with its fair share of surprises. Not least of all is the prediction of extra dimensions: ten space-time dimensions in its original formulation, but eleven in M-theory, or even twelve dimensions for F-theory. So, how does one recover the four space-time dimensions we observe? It is possible that the extra dimensions that are needed to define a consistent string theory in $D=10$ are actually compact and so small that they have avoided detection at the energy scales accessible by current particle accelerators. Intuitively, this is similar to an observer standing far away from a very long cylinder of very small radius. Such a person would only see a one-dimensional line instead of a two-dimensional object. This idea predates string theory. It finds its origin in the proposal of Kaluza who attempted to unify electromagnetism with gravity by introducing a fifth dimension. Klein then gave a quantum interpretation to this classical extension of general relativity by hypothesizing that the fifth dimension was curled up like a circle and was microscopic. We can now generalize this process by taking the long dimension of the cylinder to be our four-dimensional space-time while the small circle dimension is replaced by an appropriate $d$-dimensional compact manifold. This Kaluza-Klein procedure is now commonly known as compactification. It is important to note that even though the small internal manifold is invisible to current experiments, it plays an important role in determining the particle content and structure of the four-dimensional theory. Different choices of topologies, for instance, will lead to vastly different theories. So, it is natural to study various configurations to learn more about the properties of string theory as well as which setups would be more likely to yield a phenomenologically relevant model. For that purpose, Calabi-Yau manifolds were first considered for compactifying six extra dimensions. Those are K\"ahler manifolds in $n$ complex dimensions with $SU(n)$ holonomy. Similarly, manifolds of special holonomy such as $Spin(7)$ and $G_2$ have garnered much recent interest.

Another important feature of string theory is that it requires the existence of supersymmetry, which is a symmetry that relates bosons to fermions. It is possible to formulate a bosonic string theory, but the complete absence of fermions makes it unrealistic as a proper description of our universe. Instead, bosonic fields must be paired up with fermionic partners. It turns out to be possible to have more than one kind of supersymmetry transformation, controlled by the spinor generators. For instance, in four dimensions, a spinor has four degrees of freedom and thus for the minimal amount of supersymmetry, i.e. $\N=1$, we have four supersymmetry generators. On the other hand, having eight copies of supersymmetry, i.e. $\N=8$, yields 32 supersymmetry generators. The existence of supersymmetry is a very strong prediction of string theory, which has yet to be observed at the energy scales being probed by current particle accelerators. It is thus necessary to find mechanisms that can explain the breaking of this supersymmetry as one moves to lower energy scales. One path is to carefully choose to compactify string theory on manifolds that break the initial supersymmetry.  Unlike the circle initially considered by Kaluza-Klein the purpose of these manifolds is to break symmetries rather than make them. However the exact choice of compact manifold can feel somewhat arbitrary and the number of possible low-energy effective theories compatible with string theory could be on the order of $10^{500}$ or more. As a result, it is crucial to classify the various resulting theories to properly identify which configurations can or cannot yield a model that realistically describes the observed world.

One prominent tool to study string theory is its many dualities and symmetries. For instance, two different versions of string theory, called type IIA and type IIB, compactified on different Calabi-Yau manifolds can turn out to be equivalent in a non-trivial way. This particular situation is known as mirror symmetry. In fact, internal constraints imply the existence of exactly five consistent string theories, all related by a web of dualities as illustrated in figure \ref{fig:dualityweb}. In the superstring formalism we observe both left-moving modes and right-moving modes. It can be shown that the supersymmetries associated with left-movers and right-movers can have either opposite handedness or the same handedness. We thus get two options called type IIA and type IIB superstring theories. By performing an orientifold projection it is possible to mod-out the left-right symmetry of the type IIB theory, thus yielding what is known as the type I superstring theory. The last two options are known as ``heterotic" due to their construction which involves formalism from the 26-dimensional bosonic string and the 10-dimensional superstring. They can either have $SO(32)$ or $E_8 \times E_8$ gauge groups.

Among the dualities relating these theories, one is called T-duality. This duality implies that in many cases two different geometries for the extra dimensions are physically equivalent. In its simplest form it relates to the fact that a circle of radius $R$ is equivalent to a circle of radius $l_\mathrm{s}^2/R$, where $l_\mathrm{s}$ is the fundamental string length scale. In particular it relates the two type II and the two heterotic theories. A second kind of duality, called S-duality, exists. Instead of relating different geometries it relates the string coupling constant $g_\mathrm{s}$ to $1/g_\mathrm{s}$. Thus, once we know the perturbative behavior for these theories we can immediately deduce their behavior when  $g_\mathrm{s} \gg 1$. This however leaves out the strongly coupled phases where the string coupling is of order one. Finally, when S-duality is applied to the type IIA and the $E_8 \times E_8$ heterotic theories, that is in the regime where  $g_\mathrm{s}$ becomes large, they grow an eleventh dimension of size $g_\mathrm{s}l_\mathrm{s}$. In this large coupling limit we are outside the regime of perturbative string theory and a new type of quantum theory in 11 dimensions, called M-theory, emerges. It is also possible to realize a non-perturbative version of type IIB by observing that it has an $SL(2,\mathbb{Z})$ symmetry, the modular group of a torus. It further contains a complex scalar field $\tau$ which transforms under that $SL(2,\mathbb{Z})$ as the complex structure of a torus. Geometrically, the type IIB theory is then interpreted as having an auxiliary two-torus. This framework then leads to what is known as F-theory.

\begin{figure}[t]
    \centering
    \includegraphics[width=0.75\textwidth]{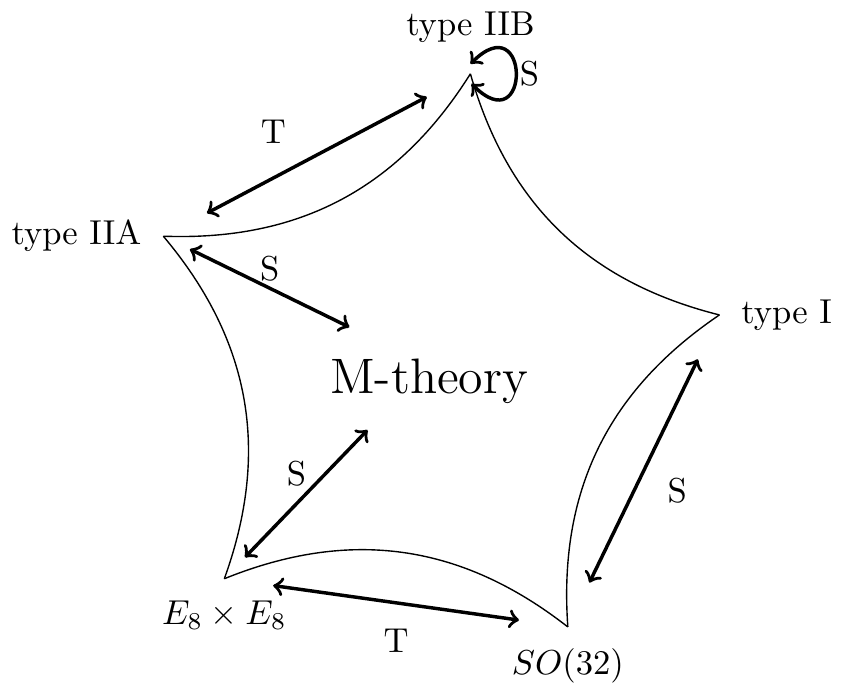}
    \caption{Duality web of the five string theories in 10 dimensions, all interconnected via the 11 dimensional M-theory. Furthermore, T-duality relates the two type II as well as the two heterotic theories, while S-duality relates the $SO(32)$ heterotic theory with type I string theory. S-duality maps type IIB back to itself, so it is actually a symmetry there.}
    \label{fig:dualityweb}
\end{figure}

On top of all these geometrical tools, there is one particular symmetry which turns out to be a very important feature of string theory: conformal symmetry. We can see this symmetry already from the study of bosonic string theory. There we begin with the free motion of a string in space-time which can be described using the principle of minimal action. The string sweeps out a two-dimensional surface as it moves through space-time. This surface is referred to as the worldsheet. The points on the worldsheet are parametrized by the two coordinates $\tau$, which is time-like, and $\sigma$, which is space-like. A closed string is then obtained by taking $\sigma$ to be periodic, while an open string requires $\sigma$ to cover a finite interval. The action is then given in analogy to that of a point particle. For the classical motion of a point particle the action is proportional to the invariant length of the particle's path. Similarly, the string's action is proportional to the area it sweeps out. This action, called the Nambu-Goto action takes the form: 
\begin{equation}
	S_{\mathrm{NG}} = -T \int d\sigma d\tau \sqrt{\left( \dot{X} \cdot X'\right) ^2 - \dot{X}^2 X^{\prime 2}}\,,
\end{equation}
where $T$ is the tension and
\begin{equation}
	\dot{X}^\mu = \frac{\partial X^\mu}{\partial \tau}\,, \quad X^{\mu \prime} = \frac{\partial X^\mu}{\partial \sigma}\,.
\end{equation}
Thus we see that the classical string motion extremizes the worldsheet area. After Wick rotation of the worldsheet to a Riemann surface $\Sigma$, string theory can be stated in terms of a two-dimensional conformal field theory on the worldsheet. 
In the conformal gauge the bosonic string action is then given by:
\begin{equation}
	S = -\frac{1}{2\pi} \int d\sigma d\tau \partial_\alpha X_\mu \partial^\alpha X^\mu \,.
\end{equation}
To generalize this action, and incorporate supersymmetry, we must introduce additional internal degrees of freedom describing fermions on the worldsheet. Explicitly, we add $D$ Majorana fermions so that the action now reads:
\begin{equation}
	S = -\frac{1}{2\pi} \int d\sigma d\tau \left( \partial_\alpha X_\mu \partial^\alpha X^\mu + \overline{\psi}^\mu \rho^\alpha \partial_\alpha \psi_\mu \right) \,,
\end{equation}
where $\rho^\alpha$, with $\alpha = 0,1$ represents the two-dimensional Dirac matrices. 

The fact that the worldsheet is described by a two-dimensional conformal field theory is especially interesting for the study of another duality known as AdS/CFT correspondence. This holographic duality is a conjectured relationship between anti-de Sitter (AdS) spaces -- which are used in quantum gravities such as string theory or M-theory -- and CFTs. This is another strong-weak duality: when the fields of the QFT are strongly interacting, the ones in the gravitational theory are weakly interacting and vice-versa. Thus, it can give a non-perturbative formulation of quantum gravity, at least in AdS space. However, a non-perturbative understanding of conformal field theories would be necessary.

As a matter of fact, a realistic phenomenological model would most likely arise in a non-perturbative regime. Thus, an important objective is to make precise the various effects one can have from interacting strings. Naturally, we then ask: \textit{What do strongly coupled string theory and quantum field theories look like?} To answer this question we can use the rich geometric structure and various dualities of string theory and quantum field theory. In particular, string dualities tell us that seemingly different string compactifications may nevertheless describe aspects of the same physical system, simply in different regimes of validity. So, by linking together various strongly coupled theories we can obtain a more complete approach to constructing and studying string vacua of phenomenological relevance.

On the QFT side, we have superconformal field theories (SCFTs) which are fixed points of the renormalization group (RG). As a result, CFTs are scale invariant, which means their physics does not change with scale, implying that they are fixed points of the renormalization group. By introducing small deformations, RG flows can take us from one fixed point to the next. In between we can have other strongly coupled QFTs. Thus, by mapping the fixed points of RG flows in various dimensions it is possible to probe various strong coupling effects in QFTs and get a more precise idea of the structure and geometry of quantum field theories in general. One important area of research has been to better understand the structure of all possible 6D RG flows obtained from deformations of different conformal fixed points, with the ultimate goal being to obtain a full classification of such RG flows. To attack this problem, we can use several geometrical tools once again. Indeed, M5-branes probing an ADE singularity lead to 6D SCFTs with $\N=(1,0)$ supersymmetry. Also, the geometry of F-theory can be used to extract various data as the M5-branes are moved off the singularity. 


This thesis is divided into three parts. In part I we look at SCFTs in both four and six dimensions. First, by starting from a general $\N=2$ SCFT, we can study the network of $\N=1$ SCFTs obtained from relevant deformations by nilpotent mass parameters. Those are associated with T-branes (for ``Triangular branes"), which are non-abelian bound states of branes characterized by the condition that, on some loci, their matrix of normal deformation, or Higgs field is upper triangular. It turns out that nilpotent elements of semi-simple algebras admit a partial ordering connected by a corresponding directed graph. We find strong evidence that the resulting fixed points are connected by a similar network of 4D RG flows. To illustrate these general concepts, we also present a full list of nilpotent deformations in the case of explicit $\N=2$ SCFTs, including the case of a single D3-brane probing a $D$- or $E$-type F-theory 7-brane, and 6D $(G,G)$ conformal matter compactified on a torus, as described by a single M5-brane probing a $D$- or $E$-type singularity.

The next chapter then returns to the stringy origin of six dimensional SCFTs. There is an intricate correspondence between certain Higgs branch deformations and nilpotent orbits of flavor symmetry algebras associated with T-branes. We show that many aspects of these deformations can be understood in terms of simple combinatorial data associated with multi-pronged strings stretched between stacks of intersecting 7-branes in F-theory. This data lets us determine the full structure of the nilpotent cone for each semi-simple flavor symmetry algebra, and it further allows us to characterize symmetry breaking patterns in quiver-like theories with classical gauge groups.

In the third chapter we turn to S-folds, which are a non-perturbative generalization of orientifold 3-planes which figure prominently in the construction of 4D $\N=2$ SCFTs. There we develop a general procedure for reading off the flavor symmetry experienced by D3-branes probing 7-branes in the presence of an S-fold. We develop an S-fold generalization of orientifold projection which applies to non-perturbative string junctions. This procedure leads to a different 4D flavor symmetry algebra depending on whether the S-fold supports discrete torsion. We also show that this same procedure allows us to read off admissible representations of the flavor symmetry in the associated 4D $\N=2$ SCFTs. Furthermore, this provides a prescription for how to define F-theory in the presence of S-folds with discrete torsion.

In Part II, we turn our focus more specifically towards T-duality. Abelian T-duality is known to be a full duality in string theory between type IIA and type IIB. On the other hand, its nonabelian generalization -- Poisson-Lie (PL) T-duality -- has only been conjectured to be so. At first we show that, to leading order in $\alpha'$ (the inverse of the string tension) PL T-duality is a proper map between CFTs. A very powerful tool to make the duality manifest has been Double Field Theory (DFT). Indeed, PL symmetric target spaces can look very complicated but their underlying structure becomes much simpler in the framework of DFT, where they are expressed in the language of generalized geometry. Thus, we actually start from the doubled (unifying) description and extract both PL T-dual target spaces according to the diagram of figure \ref{fig:diags}.

To finish part II, we show that the one-loop and two-loop $\beta$-functions of the closed, bosonic string can be written in a manifestly $O(D,D)$-covariant form. Based on this result, we prove that 1) Poisson-Lie symmetric $\sigma$-models are two-loop renormalizable, and 2) their $\beta$-functions are invariant under Poisson-Lie T-duality. Moreover, we identify a distinguished scheme in which Poisson-Lie symmetry is manifest. It simplifies the calculation of two-loop $\beta$-functions significantly and thereby provides a powerful new tool to advance into the quantum regime of integrable $\sigma$-models and generalized T-dualities.

We end in part III by exploring situations which could have promising phenomenological applications. First, we have F-theory which is a strongly coupled formulation of type IIB string theory. Upon compactification on a Calabi-Yau ($CY$) fourfold, it leads to a 4D $\N=1$ theory. On the other hand, M-theory compactified on a $G_2$ manifold also yields 4D $\N=1$. As a result, it is natural to expect the existence of some duality between the two. Indeed, we show the existence of a geometric unification of the Higgs bundle data associated with 4D $\N=1$ M-theory and F-theory vacua. While finding a map between M-theory on a $G_2$ and F-theory on a $CY_4$ would be hard to obtain directly, we lift up the problem by studying M-theory on a local $Spin(7)$ space and then show how it reduces to $G_2$ or $SU(4)$. This is schematically shown in the diagram of figure \ref{fig:diags}. As a result, we are able to go back and forth between two sides of this duality, even though it was not immediately manifest. This technique turns out to be very useful as, for instance, it would allow one to use F-theory to gain some insight into $G_2$ manifolds, which are notoriously difficult to understand. 

\begin{figure}[t]
  \centering
  \hspace{4em}
     \begin{subfigure}[b]{0.3\textwidth}
         \centering
         \includegraphics[width=\textwidth]{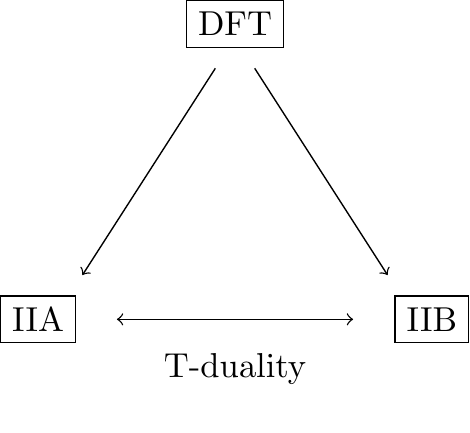}\vspace{-3em}
         \caption*{}\label{fig:d1}
     \end{subfigure}
     \hfill
     \begin{subfigure}[b]{0.3\textwidth}
         \centering
         \includegraphics[width=\textwidth]{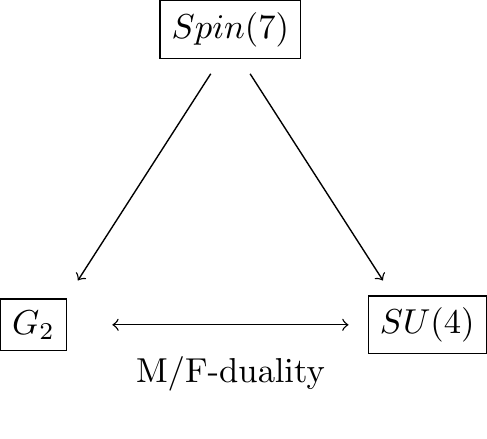}\vspace{-3em}
         \caption*{}\label{fig:d2}
     \end{subfigure}
     \hspace{4em}
     \caption{On the right, we have a duality between two 4D $\mathcal{N}=1$ M- and F-theory vacua made manifest by embedding $G_2$ and $SU(4)$ inside $Spin(7)$. On the left, nonabelian T-duality between IIA and IIB is made apparent by lifting to DFT.}\label{fig:diags}
\end{figure}

In the final chapter, we study 4D systems in which parameters of the theory have position dependence in one spatial direction. In the limit where these parameters jump, this can lead to 3D interfaces supporting localized degrees of freedom. A priori, this sort of position dependence can occur at either weak or strong coupling. Demanding time-reversal invariance for $U(1)$ gauge theories with a duality group $\Gamma \in SL(2,\mathbb{Z})$ leads to interfaces at strong coupling which are characterized by the real component of a modular curve specified by $\Gamma$. This provides a geometric method for extracting the electric and magnetic charges of possible localized states. We illustrate these general considerations by analyzing some 4D $\N=2$ theories with 3D interfaces. These 4D systems can also be interpreted as descending from a six-dimensional theory compactified on a three-manifold generated by a family of Riemann surfaces fibered over the real line. We show more generally that 6D superconformal field theories compactified on such spaces also produce trapped matter by using the known structure of anomalies in the resulting 4D bulk theories.

\part{Superconformal Field Theories}

\chapter{Nilpotent Network and 4D RG flows}\label{chapter1}

\section{Introduction \label{sec:INTRO1}}

Conformal field theories (CFTs)\ play a central role in physics. Deformations
which drive one fixed point to another also provide important insights into
more general quantum field theories.

Even so, it is often difficult to establish the existence of fixed points, let alone
determine deformations to new ones. Common techniques include combinations of methods
from supersymmetry, string compactification, holography, and / or the conformal bootstrap.

Part of the issue with understanding relevant perturbations of CFTs is that
(by definition)\ they grow deep in the infrared. From this
perspective, it is perhaps not surprising that comparatively short flows where
there is only a small drop in the number of degrees of freedom (as measured by
various anomalies)\ are often easier to study.

One way to understand long flows is to break them up into a sequence of nearby
short flows. This strategy has recently been used to make surprisingly sharp
statements in the study of 6D\ supersymmetric
RG\ flows \cite{Heckman:2015ola, Cordova:2015fha, Heckman:2015axa,
Cremonesi:2015bld, Heckman:2016ssk, Mekareeya:2016yal, Heckman:2018pqx}. In particular, the
mathematical partial ordering of nilpotent orbits in flavor symmetry algebras
automatically defines a hierarchy of 6D\ RG\ flows \cite{Heckman:2016ssk,
Mekareeya:2016yal, Heckman:2018pqx}. For a recent review of 6D superconformal field theories,
see reference \cite{Heckman:2018jxk}.

In this chapter we ask whether the same mathematical
structure leads to an improved understanding of RG\ flows in lower-dimensional
systems. The specific class of theories we study are $\mathcal{N}=1$
deformations of 4D $\mathcal{N}=2$ SCFTs. For the UV\ theories under
consideration, we assume the existence of a flavor symmetry algebra 
$\mathfrak{g}_{\text{flav}}$, which a priori could be composed of 
several simple factors:%
\begin{equation}
\mathfrak{g}_{\text{flav}}=\mathfrak{g}_{\text{flav}}^{(1)}\times
...\times\mathfrak{g}_{\text{flav}}^{(n)}%
\end{equation}
for $\mathfrak{g}_{\text{flav}}^{(i)}$ a simple Lie algebra. Associated with
this flavor symmetry are a collection of mass parameters $m_{\text{adj}}$, and 
corresponding dimension two mesonic operators $\mathcal{O}_{\text{adj}}$ 
transforming in the adjoint representation\footnote{More canonically, one can
view the mass parameters as elements in the dual $\mathfrak{g}_{\text{flav}}^{\ast}$.},
which can be used to activate relevant deformations to new conformal fixed
points in the IR via superpotential deformations:%
\begin{equation}
\delta W=\text{Tr}_{\mathfrak{g}_{\text{flav}}}\left(  m_{\text{adj}}%
\cdot\mathcal{O}_{\text{adj}}\right)  . \label{plainmass}%
\end{equation}
Promoting the mass parameters to a chiral superfield $M_{\text{adj}}$
transforming in the adjoint representation of $\mathfrak{g}_{\text{flav}}$, we can
consider the related deformations associated with expanding around background vacuum expectation values (vevs) for these ``flipper fields:''
\begin{equation}
\delta W=\text{Tr}_{\mathfrak{g}_{\text{flav}}}\left(  (m_{\text{adj}%
}+M_{\text{adj}})\cdot\mathcal{O}_{\text{adj}}\right)  , \label{MSdef}%
\end{equation}
where now, we interpret the mass deformation $m_{\text{adj}}=\left\langle
M_{\text{adj}}\right\rangle $ as a background vev.

The key point we shall be exploiting in this work is that given a flavor
symmetry Lie algebra $\mathfrak{g}_{\text{flav}}$, there is a partial ordering
available for nilpotent elements, as defined by the orbit of an element under
the adjoint action of the algebra. Given nilpotent elements $\mu,\nu
\in\mathfrak{g}_{\text{flav}}$, we say that $\mu\prec\nu$ when Orbit$(\mu
)\subset\overline{\text{Orbit}(\nu)}$. Since the mass parameters
$m_{\text{adj}}$ transform in the adjoint, this sets up a conjectural relation
between relevant deformations, as in lines (\ref{plainmass}) and (\ref{MSdef})
and 4D\ RG\ flows. Intuitively, as the size of the orbit increases, the number
of degrees of freedom which pick up a mass also increases, leading to a longer
flow into the infrared.

Another quite interesting feature of nilpotent mass deformations is that at
least in the case where we have a plain mass deformation as in line
(\ref{plainmass}), the Seiberg-Witten curve of the UV\ $\mathcal{N}=2$ theory
descends to an $\mathcal{N} = 1$ curve of the deformed $\mathcal{N}=1$ theory which
fixes the relative scaling dimensions of various operators \cite{Heckman:2010qv}.
The fact that it is still singular provides evidence of an $\mathcal{N} = 1$ fixed point.

One of our aims in this work will be to provide substantial evidence that this network
of nilpotent orbits defines a corresponding hierarchy of 4D\ RG\ flows. For the most part,
this involves a mild generalization of the procedure proposed in \cite{Heckman:2010fh}, studied in detail in
\cite{Heckman:2010qv} (see also \cite{Cecotti:2010bp}) and further extended in references \cite{Gadde:2013fma,
Agarwal:2013uga, Agarwal:2014rua, Agarwal:2015vla, Maruyoshi:2016tqk, Maruyoshi:2016aim, Agarwal:2016pjo},
and applied in various model building contexts in references \cite{Heckman:2011hu, Heckman:2011bb, Heckman:2012nt, Heckman:2012jm, Heckman:2015kqk, DelZotto:2016fju}.

The appearance of a nilpotent element $\mu$ implies the existence of
an $\mathfrak{su}(2)\subset\mathfrak{g}_{\text{flav}}$ subalgebra, with
generators $\mu$, $\mu^{\dag}$ and $[\mu,\mu^{\dag}]$. Labeling the
associated generator of the Cartan subalgebra for this$\mathfrak{\ su}(2)$
subalgebra as $T_{3}$, the infrared R-symmetry is given by a linear combination of the form
(see e.g. \cite{Heckman:2010qv}):%
\begin{equation}
R_{\mathrm{IR}}=R_{\mathrm{UV}}+\left(  \frac{t}{2}-\frac{1}{3}\right)  J_{\mathcal{N}=2}%
-tT_{3}+\underset{i}{\sum}t_{i}F_{i},
\end{equation}
where $R_{\mathrm{UV}}$ and $R_{\mathrm{IR}}$ respectively denote the UV\ and IR\ R-symmetry
(treated as an $\mathcal{N}=1$ theory), $J_{\mathcal{N}=2}$ is an additional
$U(1)$ symmetry which is always present in an $\mathcal{N}=2$ SCFT when
interpreted as an $\mathcal{N}=1$ theory. The last set of terms refers to the
possibility of additional $U(1)$'s, including those which emerge in the
infrared. The IR\ R-symmetry is then fixed via the procedure of a-maximization
over the parameters $t$ and $t_{i}$, as in reference
\cite{Intriligator:2003jj}.\footnote{In practice it is
often necessary to make additional assumptions about
these emergent symmetries to actually carry out concrete calculations.}

In the absence of these emergent $U(1)$'s, we find strong evidence that the
partially ordered set defined by the nilpotent elements of a Lie algebra
exactly aligns with the corresponding hierarchy of 4D\ RG\ flows. For example,
the conformal anomalies $a_{\mathrm{IR}}$ and $c_{\mathrm{IR}}$ decrease along such
trajectories, and anomalies involving flavor currents (with generators
suitably normalized) also decrease along such flows.

Far more non-trivial is that \textit{even in the presence} of emergent $U(1)$'s, there
is still such a partial ordering of 4D theories, as dictated by the nilpotent
cone of the Lie algebra. This is considerably more subtle and requires a case
by case analysis. For this reason, we focus on explicit examples.

One class of theories already studied in \cite{Heckman:2010qv} for plain mass deformations,
and with some masses promoted to chiral superfields in \cite{Maruyoshi:2016tqk, Maruyoshi:2016aim}
involves nilpotent mass deformations of the $\mathcal{N}=2$ theories defined by a D3-brane
probing an F-theory 7-brane with constant axio-dilaton. This includes the
$H_{0}$, $H_{1}$, $H_{2}$ Argyres-Douglas theories \cite{Argyres:1995jj, Argyres:1995xn},
the $E_{6}$, $E_{7}$, $E_{8}$ Minahan Nemeschansky theories \cite{Minahan:1996fg, Minahan:1996cj},
and $\mathcal{N}=2$ $SU(2)$ gauge theory with four flavors and corresponding $SO(8)$ flavor symmetry
(namely $D_{4}$) \cite{Seiberg:1994aj}. The string theory interpretation of nilpotent deformations
is also quite interesting, as they are associated with T-brane configurations
of 7-branes (see e.g. \cite{Aspinwall:1998he, Donagi:2003hh, Cecotti:2010bp, Anderson:2013rka, Collinucci:2014taa, Collinucci:2014qfa, Bena:2016oqr, Marchesano:2016cqg, Anderson:2017rpr, Bena:2017jhm, Marchesano:2017kke, Cvetic:2018xaq}),
namely they leave intact the Weierstrass model of the associated F-theory geometry,
but nevertheless deform the physical theory.

Here, we systematically study all possible nilpotent deformations for the $D$-
and $E$-series theories, systematically sweeping out the corresponding network
of 4D\ RG\ flows (we do not consider the $H$-series in any detail since they have only a few
nilpotent deformations). An interesting feature of these examples is that only the
Coulomb branch operator sometimes appears to drop below the unitarity bound,
and even this happens only for the largest nilpotent orbits. In such cases, we
see no evidence that the fixed point does not exist (since the underlying
geometry is still singular), and instead find it most plausible that the Coulomb
branch operator decouples as a free field, with a corresponding emergent $U(1)$ acting on only
this operator, as per the procedure advocated in \cite{Kutasov:2003iy, Intriligator:2003mi}.

We also study nilpotent mass deformations of
4D\ $\mathcal{N}=2$ conformal matter, namely the compactification of
6D\ conformal matter \cite{DelZotto:2014hpa, Heckman:2014qba} on a $T^{2}$. Here, we
consider the case where there is a
$G_{L}\times G_{R}$ flavor symmetry with $G_{L}=G_{R} = G$ given by $SO(8)$, $E_{6}$,
$E_{7}$, or $E_{8}$. The 4D\ anomaly polynomials for
these theories were computed in \cite{Ohmori:2015pua, Ohmori:2015pia}.
The Seiberg-Witten and Gaiotto curves for these models are known, both via mirror symmetry
\cite{DelZotto:2015rca}, and via its relation to compactifications of class $\mathcal{S}$
theories \cite{Ohmori:2015pua, Ohmori:2015pia}.

Nilpotent mass deformations of 4D conformal matter involve specifying a pair
of nilpotent elements, one for each flavor symmetry factor. In this case, the
string theory interpretation involves a pair of 7-branes intersecting along
the common $T^{2}$. Such nilpotent deformations involve activating background
values for gauge fields of the corresponding 7-branes.

This already leads to many new $\mathcal{N}=1$ fixed
points and the partial ordering for the product Lie algebra predicts a
corresponding hierarchy of 4D\ fixed points. We present strong evidence that
this is the case, again sweeping over all pairs of nilpotent orbits, and for
each one computing the corresponding values of various IR\ anomalies, checking
there is a corresponding decrease along a given trajectory in the nilpotent cone.

One issue which shows up in these cases is that in sufficiently long flows,
mesonic operators often decouple. This in turn signals that such operators
cannot be used to trigger further flows.
A priori, this could mean that the network of connections in the
nilpotent cone may have links which do not produce 4D RG flows.
Even though we have not found a single example where this actually occurs, we leave
a systematic analysis of this possibility for future work.

\begin{figure}[t!]
\begin{center}
\includegraphics[trim={10cm 0cm 10cm 1cm},clip,scale=.5]{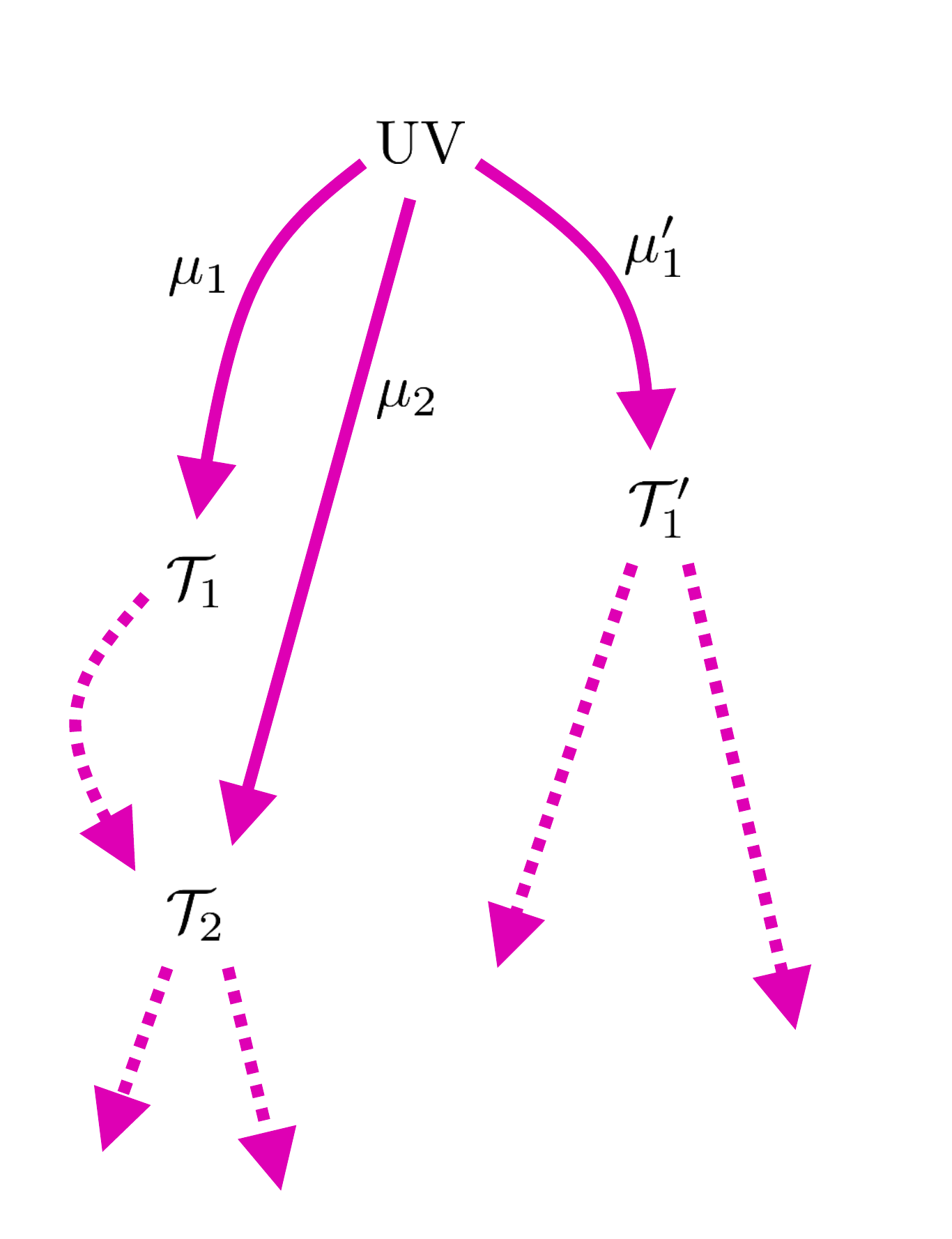}
\end{center}
\caption{Depiction of the network of 4D RG flows generated by elements
of the nilpotent cone. Starting from a UV $\mathcal{N} = 2$ fixed point,
each nilpotent orbit in the flavor symmetry algebra
determines a candidate $\mathcal{N} = 1$ fixed point. Additionally, the network of
connections between nilpotent orbits also motivates the existence of additional
flows between these $\mathcal{N} = 1$ fixed points.}
\label{fig:flowsketch}
\end{figure}

With this set of theories in hand, additional numerical studies are amenable
to treatment, though the list of theories is so large that we have chosen to
collect the full data set in an accompanying \texttt{Mathematica} package
available for download with the \texttt{arXiv} submission of \cite{Apruzzi:2018xkw}. For example, by
sweeping over all theories, we find several examples of theories where the
conformal anomalies $a_{\mathrm{IR}}$ and $c_{\mathrm{IR}}$ are rational numbers. In some cases
such as reference \cite{Maruyoshi:2016tqk, Maruyoshi:2016aim},
this was interpreted as evidence for an emergent $\mathcal{N}=2$
supersymmetry in the infrared, and we find another example of this type for a
deformation of the $E_{7}$ Minahan-Nemeschansky theory. It is not
clear to us whether there is $\mathcal{N}=2$ enhancement in all cases,
but certainly the list of such rational theories we find suggests additional
structure is present. Another numerical curiosity we observe is that for a given
choice of UV $\mathcal{N} = 2$ SCFT, the value
of the ratio:%
\begin{equation}
\frac{a_{\mathrm{IR}}}{c_{\mathrm{IR}}}\simeq \text{constant} \pm O(1\% - 5\%)
\end{equation}
is nearly constant over all nilpotent deformations, in line with the
observation made in reference \cite{Maruyoshi:2018nod} for a different set of theories.

The rest of this chapter is organized as follows. First, in section \ref{sec:GENERAL}
we analyze for a general $\mathcal{N}=2$ theory with flavor
symmetries, the structure of the $\mathcal{N}=1$ theories obtained via both
plain mass deformations and their extension to flipper field deformations. In particular,
we analyze the network of 4D RG flows predicted by the nilpotent cone.
Section \ref{sec:INHERIT} discusses the structure of IR fixed points assuming no
operators decouple, and section \ref{sec:EMERGE} discusses the structure of theories
in the presence of emergent IR symmetries. In section \ref{sec:D3} we discuss nilpotent
deformations of D3-brane probes of $D$- and $E$-type 7-branes and in section \ref{sec:CM} we
discuss nilpotent deformations of 4D $\mathcal{N} = 2$ conformal matter. We
conclude in section \ref{sec:CONC1}. Some additional review material, as well as technical details and instructions on
how to use the companion \texttt{Mathematica} files are presented in the Appendices.

\section{Nilpotent Deformations: Generalities \label{sec:GENERAL}}

In this section we discuss some general features of nilpotent mass
deformations of $\mathcal{N}=2$ SCFTs. Throughout, we assume the existence of
a continuous flavor symmetry algebra which may consist of several simple
factors:%
\begin{equation}
\mathfrak{g}_{\mathrm{UV}} \equiv \mathfrak{g}_{\text{flav}}=\mathfrak{g}_{\text{flav}}^{(1)}\times
...\times\mathfrak{g}_{\text{flav}}^{(n)}. \label{flavaflav}%
\end{equation}
We assume either that there are no abelian factors in the UV, or more
generally, that the only non-vanishing anomalies involving flavor symmetry
currents involve precisely two insertions of the same kind (which is automatic
in the traceless non-abelian case). Note that we can then also allow the
appearance of abelian symmetry factors, provided they satisfy this condition.

We assume adjoint valued mass parameters $m_{\text{adj}}$, 
and corresponding dimension two mesonic operators $\mathcal{O}_{\text{adj}}$ 
which serve as coordinates on the Higgs branch of moduli space. 
Note that there could be non-trivial chiral ring relations for these operators,
as can often happen when there is more than one simple Lie algebra factor for $\mathfrak{g}_{\mathrm{UV}}$. Since we
will couch our analysis in terms of basic properties of symmetry breaking patterns,
our analysis will not depend on such detailed knowledge of the UV theory.

It will prove useful to view our $\mathcal{N} = 2$ SCFT as an $\mathcal{N} = 1$ SCFT 
with additional symmetries. Along these lines, we recall that the $\mathcal{N} = 2$ SCFT 
has an $SU(2) \times U(1)$ R-symmetry. Labeling the generator
of the Cartan subalgebra for the $SU(2)$ factor by $I_{3}$ with eigenvalues
$\pm1/2$ in the fundamental representation, and $R_{\mathcal{N} = 2}$ for the $U(1)$
factor normalized so that the complex scalar of a free $\mathcal{N}=2$ vector
multiplet has charge $+2$, the $\mathcal{N}=1$ R-symmetry is given by the
linear combination (see e.g \cite{Tachikawa:2009tt, Heckman:2010qv}):%
\begin{equation}
R_{\mathrm{UV}}=\frac{1}{3}R_{\mathcal{N}=2}+\frac{4}{3}I_{3}.
\end{equation}
There is another linear combination which we can form which is a global
symmetry of the UV\ theory. We label this as:%
\begin{equation}
J_{\mathcal{N}=2}=R_{\mathcal{N}=2}-2I_{3}.
\end{equation}
See table \ref{tab:N2} for the charge assignments of Coulomb branch operators and 
mesonic operators which serve as coordinates on the Higgs branch.
\begin{table}[t]
  \centering
  \begin{tabular}{ |c||c|c|c|c| }
 \hline
 & $\cO_{\text{adj}}$ & $Z_i$  \\  \hline 
 $R_{\mathrm{UV}}$ & $4/3$ & $2/3$ $\Delta_{\mathrm{UV}}(Z_i)$\\  \hline
 $J_{\cN=2}$ & $-2$ & $2$ $\Delta_{\mathrm{UV}}(Z_i)$  \\  \hline
 $R_{\cN=2}$ & $0$ & $2$ $\Delta_{\mathrm{UV}}(Z_i)$\\  \hline
 $I_{3}$ & $1$ & $0$  \\  \hline
  \end{tabular}
  \caption{Charge assignments for the mesons $\cO_{\text{adj}}$ and Coulomb branch parameters $Z_i$ in the UV theory.}
  \label{tab:N2}
\end{table}

The Higgs branch is parameterized by dimension two operators transforming in
the adjoint representation of $\mathfrak{g}_{\text{flav}} \equiv \mathfrak{g}_{\mathrm{UV}}$,
which we denote by $\mathcal{O}_{\text{adj}}$. The mass parameters $m_{\text{adj}}$ which pair
with these operators transform in the adjoint representation of $\mathfrak{g}%
_{\text{flav}}$. 

We consider both the case of a plain mass deformation:%
\begin{equation}
\delta W_{\text{plain}}=\text{Tr}_{\mathfrak{g}_{\text{flav}}}\left(  m_{\text{adj}}%
\cdot\mathcal{O}_{\text{adj}}\right)  , \label{plainagain}%
\end{equation}
as well as the flipper field deformations
associated with promoting the mass parameters to a dynamical chiral superfield in the adjoint
of the flavor symmetry which mixes with the original interacting theory:%
\begin{equation}
\delta W_{\text{flip}}=\text{Tr}_{\mathfrak{g}_{\text{flav}}}\left(  (m_{\text{adj}%
}+M_{\text{adj}})\cdot\mathcal{O}_{\text{adj}}\right)  .
\end{equation}
We shall often first deal with the case of plain mass deformations, since
flipper field deformations are a mild extension of this case (though the
resulting IR\ physics can be quite different, see e.g. \cite{Gadde:2013fma, 
Agarwal:2015vla, Maruyoshi:2016tqk, Maruyoshi:2016aim}).
An important feature of our analysis is that the general structure of symmetries and anomalies enables us
to give a uniform analysis of RG\ flows for many such relevant deformations.

Though it may be difficult to explicitly construct, we know that the IR\ physics
on the Coulomb branch is controlled by a Seiberg-Witten curve
\cite{Seiberg:1994rs, Seiberg:1994aj}, and
mass deformations enter as flavor symmetry neutral combinations constructed
from the holomorphic Casimir invariants of $\mathfrak{g}_{\text{flav}}$. In
the special case of an $\mathcal{N}=2$ SCFT, all mass deformations have been
switched off and this curve will exhibit singularities, as required to have
massless degrees of freedom at the origin of the Coulomb branch.

We will in particular be interested in nilpotent deformations. For the
classical algebras, these can always be presented in terms of an explicit
nilpotent matrix, which upon conjugation by a complexified symmetry generator
can always be taken to be proportional to a matrix in Jordan normal form. For example,
in $\mathfrak{su}(4)$ we have:%
\begin{equation}
\left[
\begin{array}
[c]{cccc}%
0 & m_{1\overline{2}} & 0 & 0\\
0 & 0 & m_{2\overline{3}} & 0\\
0 & 0 & 0 & m_{3\overline{4}}\\
0 & 0 & 0 & 0
\end{array}
\right]  \sim m \times \left[
\begin{array}
[c]{cccc}%
0 & 1 & 0 & 0\\
0 & 0 & 1 & 0\\
0 & 0 & 0 & 1\\
0 & 0 & 0 & 0
\end{array}
\right]  .
\end{equation}
The labeling scheme for the classical $\mathfrak{su}$, $\mathfrak{sp}$ and
$\mathfrak{so}$ algebras are dictated by its presentation as a direct sum of
nilpotent Jordan blocks. These blocks in turn define a partition of an integer
which we write as $[\mu_{1}^{a_{1}},...,\mu_{k}^{a_{k}}]$ with $\mu
_{1}>...>\mu_{k}>0$ and $a_{i}$ the multiplicity. In the case of
$\mathfrak{su(}N)$, each partition of the integer $N$ defines a nilpotent
orbit. In the case of $\mathfrak{so}(2N)$, there are some additional
restrictions on partitions of $2N$, namely we require every even number in a partition to appear an
even number of times. Similar considerations hold for $\mathfrak{sp}(N)$ and $\mathfrak{so}(2N + 1)$.
In the case of the exceptional algebras, we instead
label the nilpotent orbit by its embedding in some subalgebra of the
larger parent algebra, which is known as the Bala-Carter label.

Now, one of the very interesting features of nilpotent mass deformations is
that all holomorphic Casimir invariants (by definition)\ must vanish, and so
the presentation of the singular geometry is exactly the same as the
$\mathcal{N}=2$ theory. In contrast to the $\mathcal{N}=2$ case, however, this
does not mean it is possible to read absolute scaling dimensions of operators
from the curve (see reference \cite{Argyres:1995xn} for the analysis of $\mathcal{N}=2$
theories), but instead only the relative scaling dimensions of operators \cite{Heckman:2010qv}.
Nevertheless, the appearance of a singular curve provides one indication
that we are still dealing with a conformal field theory, albeit one with
reduced supersymmetry.

Assuming the existence of such a fixed point, there is a partial ordering of
nilpotent orbits which suggests a physical ordering of theories. Given a pair
of nilpotent elements $\mu$ and $\nu$, we say that $\mu\prec\nu$ when
Orbit$(\mu)\subset\overline{\text{Orbit}(\nu)}$, where the overline denotes
the Zariski closure of the orbit in $\mathfrak{g}_{\text{flav}}$.

Physically, the bigger the orbit, the more degrees of freedom have picked up a
mass. So, it is natural to expect bigger orbits to be deeper in
the infrared. Moreover, for each of the simple Lie algebras, there is a
classification of all possible nilpotent orbits, and the associated
containment relations for these choices. This partially ordered set and its
interconnections defines a directed graph, namely the Hasse diagram of the
nilpotent cone. Returning to our example of explicit nilpotent matrices in $\mathfrak{su}(4)$,
for example, we can see a clear hierarchy:%
\begin{equation}
\left[
\begin{array}
[c]{cccc}%
0 & 0 & 0 & 0\\
0 & 0 & 0 & 0\\
0 & 0 & 0 & 0\\
0 & 0 & 0 & 0
\end{array}
\right]  \prec\left[
\begin{array}
[c]{cccc}%
0 & m_{1\overline{2}} & 0 & 0\\
0 & 0 & 0 & 0\\
0 & 0 & 0 & 0\\
0 & 0 & 0 & 0
\end{array}
\right]  \prec\left[
\begin{array}
[c]{cccc}%
0 & m_{1\overline{2}} & 0 & 0\\
0 & 0 & m_{2\overline{3}} & 0\\
0 & 0 & 0 & m_{3\overline{4}}\\
0 & 0 & 0 & 0
\end{array}
\right]  .
\end{equation}

It is tempting to also interpret this diagram as a
collection of candidate RG\ flows between $\mathcal{N}=1$ fixed points.
Given a sequence of theories $\mathcal{T}_{\mathrm{UV}}\rightarrow
...\rightarrow\mathcal{T}_{i}\rightarrow\mathcal{T}_{i+1}\rightarrow...$, and
associated nilpotent orbits $\varnothing\prec ... \prec \mu_{i}\prec\mu_{i+1} \prec ...$, we can ask
whether there is a flow directly from the intermediate $\mathcal{N}=1$ fixed
point $\mathcal{T}_{i}$ to $\mathcal{T}_{i+1}$. Indeed, we can
subtract the two deformations of the original parent theory:%
\begin{equation}
\delta W_{i \rightarrow i+1}=\text{Tr}_{\mathfrak{g}_{\text{flav}}}\left(  \left(  \mu
_{i+1}-\mu_{i}\right)  \cdot\mathcal{O}_{\text{adj}}\right)  , \label{defdef}%
\end{equation}
which is itself a relevant deformation of the UV\ fixed point theory. Assuming
that the operators necessary to perform such a deformation do not decouple in
theory $\mathcal{T}_{i}$, this strongly indicates that each link in the
directed graph defined by the Hasse diagram also defines a flow between
$\mathcal{N}=1$ fixed points. Carrying out a systematic analysis of this is
somewhat subtle, especially when operators start to decouple in long flows,
but this at least shows that the structure of the nilpotent cone leads to a
rich network of 4D\ RG\ flows.
See figure \ref{fig:slicesketch} for a depiction of the flows generated by these
mesonic operators.

\begin{figure}[t!]
\begin{center}
\includegraphics[trim={8cm 5cm 8cm 5cm},clip,scale=0.5]{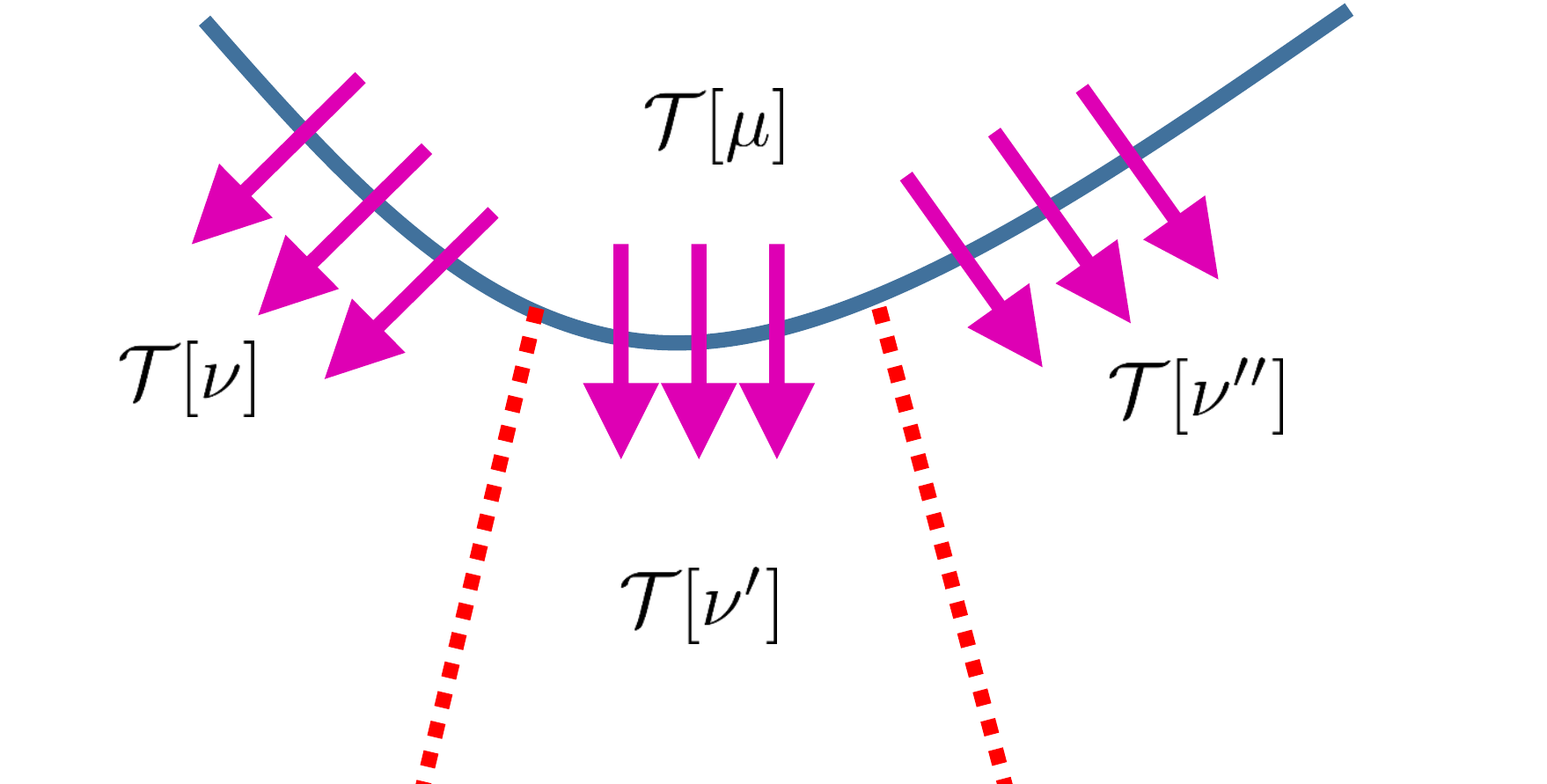}
\end{center}
\caption{Depiction of the deformations from one nilpotent orbit to another. Here, we label a theory
by a choice of nilpotent orbit $\mathcal{T}[\mu]$, and subsequent deformations
deeper down in the nilpotent cone to theories $\mathcal{T}[\nu]$, $\mathcal{T}[\nu^{\prime}]$
and $\mathcal{T}[\nu^{\prime \prime}]$. These physical paths to new orbits are parameterized
by the remnants of the original mesonic operators. An important subtlety with this picture is that
as we proceed from the UV to the IR, various mesonic operators may decouple, severing some of the
candidate links between theories. In explicit examples, however, we have not observed this
pathological behavior.}
\label{fig:slicesketch}
\end{figure}

Let us now make more precise the sense in which operator deformations such as
those of line (\ref{defdef}) lead to perturbations of one fixed point to
another. Along these lines, we start in some theory $\mathcal{T}[\mu]$, as
characterized by Orbit$(\mu)$. Given a nilpotent element, the Jacobson-Morozov
theorem guarantees the existence of a homomorphism $\mathfrak{su}%
(2)\rightarrow\mathfrak{g}_{\mathrm{UV}}$, and we label the generators of this algebra
by $T_{3}$, $T_{+}$ and $T_{-}$ in the obvious notation. Decomposing the
adjoint representation into irreducible representations of this $\mathfrak{su}%
(2)$ subalgebra, we get:%
\begin{equation}
V_{\text{adj}}=\underset{j}{%
{\displaystyle\bigoplus}
}(j),
\end{equation}
where we allow each spin $j$ to come with some multiplicity. The highest spin
states of each representation specify the deformations of the nilpotent orbit.
Indeed, a convenient way to compute the dimension of the orbit is via the
formula:%
\begin{equation}
\dim\text{Orbit}(\mu)=\dim V_{\text{adj}}-\dim V_{0}-\dim V_{1/2},
\end{equation}
where here, we have decomposed the states of the adjoint representation under
the $T_{3}$ grading:%
\begin{equation}
V_{\text{adj}}=\underset{s}{%
{\displaystyle\bigoplus}
}V_{s}.
\end{equation}

In the physical theory, these top spin states are distinguished by their role
in the breaking pattern of the flavor symmetry. More formally, we begin with
the $\mathcal{N}=1$ current supermultiplet for the flavor symmetry of the
original theory $\mathcal{J}_{A}$, with $A$ an index in the adjoint
representation. In the unbroken phase, we have the conservation rule:%
\begin{equation}
\overline{D}^{2}\mathcal{J}_{A}=0\text{.}%
\end{equation}
We can also track what becomes of this relation in the broken phase (after the
mass deformation has been switched on). Since $\mathcal{J}_{A}$
transforms in the adjoint representation of $\mathfrak{g}_{\mathrm{UV}}$, we can decompose it into
representations of this $\mathfrak{su}(2)$ subalgebra, so we
label it by a choice of spin $j$, and $T_{3}$ charge $s$, namely
$\mathcal{J}_{j,s}$. In the broken phase, the current is not conserved, since
it is explicitly broken by our mass deformation. We can follow the standard
Noether procedure to see the source of the current non-conservation.
Introducing a \textquotedblleft pion\textquotedblright\ chiral superfield
$\Lambda$ which parameterizes the flavor symmetry generators, we can send:%
\begin{equation}
\mathcal{O}_{\text{adj}}\rightarrow e^{i\Lambda}\mathcal{O}_{\text{adj}%
}e^{-i\Lambda}.
\end{equation}
Then, the superpotential deformation transforms as:%
\begin{equation}
\delta W \rightarrow\text{Tr}_{\mathfrak{g}_{\mathrm{UV}}}(m_{\text{adj}}\cdot
e^{i\Lambda}\mathcal{O}_{\text{adj}}e^{-i\Lambda}),
\end{equation}
so since $m_{\text{adj}}$ can, without loss of generality, be taken to be the
raising operator of the $\mathfrak{su}(2)_{D}$ subalgebra, we learn
that we instead have (see e.g. \cite{Xie:2016hny, Maruyoshi:2016tqk}):%
\begin{equation}
-\frac{1}{4}\overline{D}^{2}\mathcal{J}_{j,s}=\mathcal{O}_{j,s-1}\text{.}
\label{D2J}%
\end{equation}
Note in particular the relative shift in the $T_{3}$ charge $s$.

As explained in \cite{Xie:2016hny, Maruyoshi:2016tqk},
this relation tells us that in the perturbed chiral ring
relations, operators which are not the highest spin states can pair with
components of the current multiplet, forming a long multiplet. Said
differently, in the chiral ring, the operators appearing on the right-hand side
of equation (\ref{D2J}) are automatically set to zero (since they appear as $\overline{D}^2$ of something else),
and do not parameterize vacua of the deformed theory. This leaves us with just the highest spin
states, namely $\mathcal{O}_{j,j}$ for the various spin $j$ representations.
Indeed, all other mesons with $\mathcal{O}_{j,s}$ for $s<j$ can be expressed
in terms of the $\mathcal{O}_{j,j}$ using the field equations \cite{Xie:2016hny,
Maruyoshi:2016tqk, Maruyoshi:2016aim, Benvenuti:2017lle}.

In particular, we see that any further deformations of the nilpotent orbit,
namely a candidate flow from theory $\mathcal{T}_{i}$ to a theory
$\mathcal{T}_{i+1}$, will involve precisely these directions.
Provided no such operators decouple as we flow from the
UV\ to the IR, this shows that the directed graph defined by
the Hasse diagram is also a network of RG\ flows. The caveat to this statement
is that it could indeed happen that some operators decouple as we flow from
the UV\ to the IR. Indeed, as we will shortly explain, for a given
$\mathfrak{su}(2)$ representation, the highest spin states
have lowest scaling dimension.

To study this and related issues in more detail, it is of course helpful to
have an explicit example where the underlying
theory is described by a Lagrangian. In subsequent sections we will present a
more general analysis which does not rely on the existence of a Lagrangian.

\subsection{Illustrative Lagrangian Example}

We now illustrate some of the above considerations for a UV $\mathcal{N} = 2$
SCFT which has a Lagrangian description. Most of the other examples
we consider do not admit a convenient presentation of this sort, and so we will instead need to
rely on more general abstract considerations.

The example we consider is $\mathcal{N}=2$ $SU(2)$ gauge theory
with four flavors in the fundamental representation. Some nilpotent mass deformations for this theory
were considered previously in \cite{Heckman:2010qv}, so we refer the interested reader there for
additional background. Our main interest here will be to characterize every
possible nilpotent orbit of the parent $\mathfrak{so}(8)$ flavor symmetry
algebra, and to discuss the explicit structure of the broken symmetry generators.

\sloppy From the definition of the theory, there is a manifest $\mathfrak{su}(4)$
flavor symmetry which rotates the fields. In $\mathcal{N}=1$ language, we
specify four chiral superfields $q$ in the
$(\mathbf{2},\mathbf{4)}$ of $\mathfrak{su}(2)_{\text{gauge}}\times\mathfrak{su}(4)_{\text{flav}}$, and
four chiral superfields $\widetilde{q}$ in the $(\mathbf{2},\overline
{\mathbf{4}})$ of $\mathfrak{su}(2)_{\text{gauge}}\times\mathfrak{su}(4)_{\text{flav}}$.
There is also a coupling to the adjoint valued chiral superfield
associated with the $\mathfrak{su}(2)_{\text{gauge}}$ $\mathcal{N}=2$ vector multiplet:%
\begin{equation}
W_{\mathcal{N}=2}=\sqrt{2}\widetilde{q}_{\overline{f}}\varphi q^{f},
\end{equation}
where the sum on $f=1,...,4$ runs over the flavors of the model, and
we suppress $\mathfrak{su}(2)_{\text{gauge}}$ indices. This
presentation allows us to explicitly track nilpotent mass deformations
associated with the $\mathfrak{su}(4)$ symmetry algebra, as in reference \cite{Heckman:2010qv}.

Though convenient, this presentation obscures the fact that there is actually
an $\mathfrak{so}(8)$ flavor symmetry. We can assemble the $q$ and
$\widetilde{q}$ into an eight-dimensional representation of $SO(8)$, and
instead treat our field content as a half hypermultiplet transforming in the
$(\mathbf{2},\mathbf{8}_{s})$ of $\mathfrak{su}(2)_{\text{gauge}}\times\mathfrak{so}(8)_{\text{flav}}$.
Labeling the associated holomorphic chiral superfield by $Q^{i}$ with
$i=1,...,8$, we introduce a conjugate spinor of $SO(8)$ $Q_{i}^{c}$ which
canonically pairs with this field so that the superpotential can then be written
as:%
\begin{equation}
W_{\mathcal{N}=2}=\sqrt{2}Q_{i}^{c}\varphi Q^{i},
\end{equation}
where again, we suppress the $\mathfrak{su}(2)_{\text{gauge}}$ indices.
The associated mesons can be written as:%
\begin{equation}
\mathcal{O}^{A}=\left(  \rho^{A}\right)  _{j}^{i}Q_{i}^{c}Q^{j},
\end{equation}
with $\rho^{A}$ the explicit matrix representatives acting on the
$\mathbf{8}_{s}$, and $A$ an adjoint index of $SO(8)$. In this language,
nilpotent mass deformations can be viewed as specific choices for the
$\rho^{A}$ (upon complexification of the flavor symmetry algebra).
\begin{figure}[ptb]
\centering
\includegraphics[scale=1]{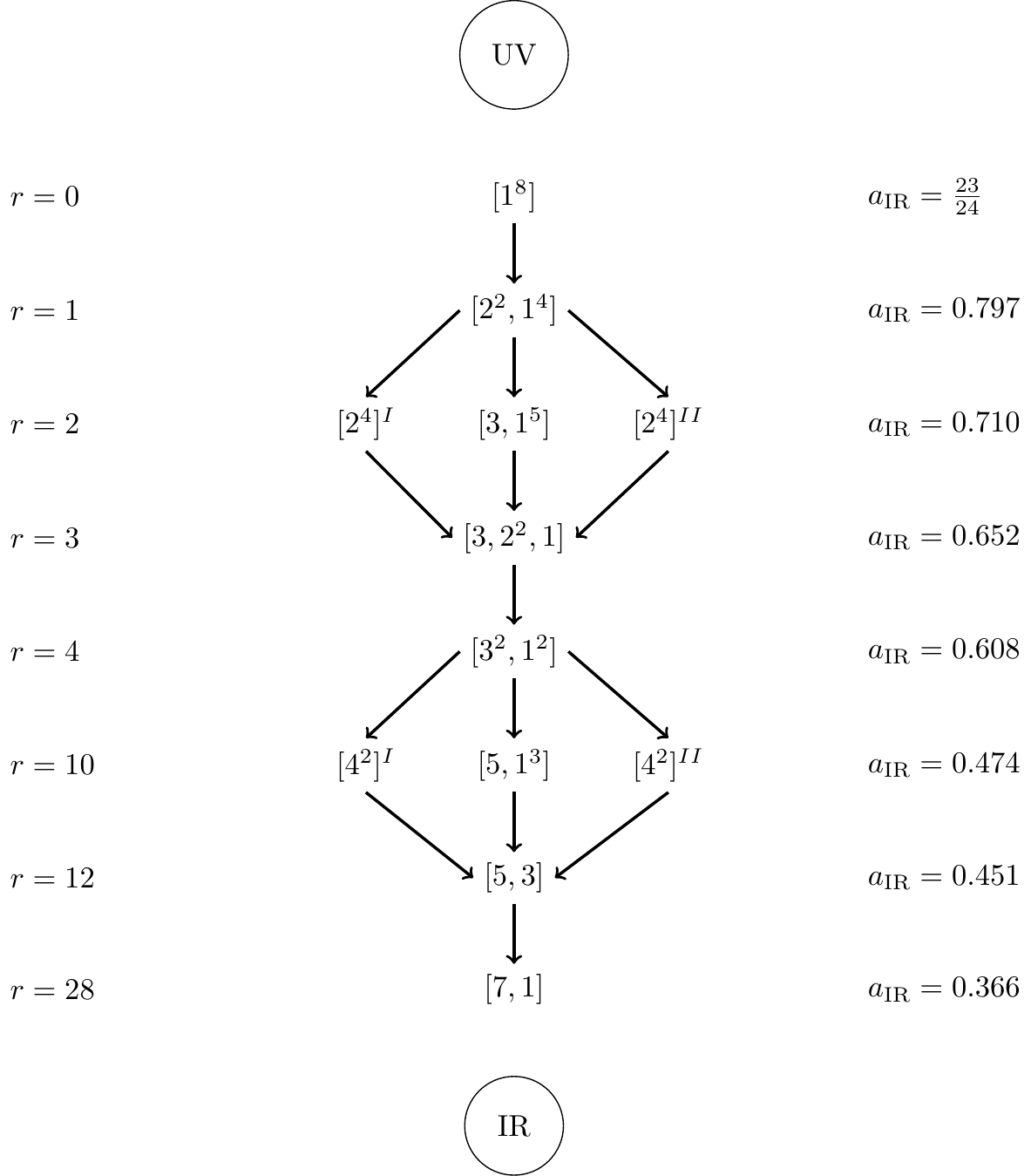}
\caption{The network of RG flows induced by nilpotent plain mass
deformations for $\mathcal{N} = 2$ Super Yang-Mills with $SU(2)$ gauge group and four flavors.
This theory has an $SO(8)$ flavor symmetry in the UV. This network is identical to the Hasse diagram
of the Lie algebra $\mathfrak{so}(8)$. The parameter $r = 2 \mathrm{Tr}_{\mathfrak{so}(8)} (T_3 T_3)$
is the embedding index for the homomorphism $\mathfrak{su}(2) \rightarrow \mathfrak{so}(8)$ defined
by a nilpotent orbit. The value of the
conformal anomaly $a_{\mathrm{IR}}$ decreases, as expected. These flows are determined using the method
described in sections \ref{sec:INHERIT} and \ref{sec:EMERGE}.}
\label{RGflow}%
\end{figure}

Figure \ref{RGflow} illustrates the resulting network of nilpotent orbits and
RG\ flows in this specific case. We also display the value of $a_{\mathrm{IR}}$ as we
pass from the UV\ to the IR. The specific method used to calculate the
IR\ R-charges is essentially the same as in reference \cite{Heckman:2010qv}, and we will
discuss it in greater detail in sections \ref{sec:INHERIT} and
\ref{sec:EMERGE}.

Another important aspect of this example is that we can also explicitly track
the structure of the broken symmetry currents. To do so, we observe
that the Lagrangian density for the $SO(8)$ theory is, in $\mathcal{N}=1$
language, given by:%
\begin{equation}
\mathcal{L}_{\mathcal{N}=2}=\mathcal{L}_{\text{gauge}}+\int d^{2}\theta
\,d^{2}\bar{\theta}\ Q_{i}^{\dagger}e^{V}Q^{i}+\int d^{2}\theta
\ W_{\mathcal{N}=2}+h.c.\,,
\end{equation}
with $V$ the $SU(2)$ $\mathcal{N} = 1$ vector multiplet. Here, $\mathcal{L}_{\text{gauge}}$
includes the remaining contributions to the $\mathcal{N}=2$ vector multiplet, namely
the kinetic terms for the vector multiplet and adjoint valued chiral superfield.

By varying the action with respect to $Q^{i}$, we obtain the following
equation of motion:
\begin{equation}
- \frac{1}{4} \overline{D}^{2} Q_{i}^{\dagger}e^{V} + 2 \sqrt{2}(Q^{c})_{i}\varphi\, = 0.
\end{equation}
For the theory with no mass deformations, we have the on-shell F-term constraint:
\begin{equation}
(Q^{c})_{i}\varphi=0.
\end{equation}
Using the on shell equations of motion, we observe that the flavor current in the UV:
\begin{equation}
\mathcal{J}_{A}=(\rho_{A})^{j}{}_{i}(Q^{c})_{j}^{\dagger}%
e^{V}Q^{i}\,,
\end{equation}
is actually conserved, namely $\overline{D}^{2}\mathcal{J}_{A}=0$.

Next, we add the superpotential deformation:
\begin{equation}
W_{D}=m^{j}{}_{i}(Q^{c})_{j}Q^{i}.
\end{equation}
The current $\mathcal{J}_{A}$ is no longer conserved, because of this explicit breaking term.
To see what happens, consider following the Noether procedure with flavor transformation:
\begin{equation}
\delta_{\text{flav}}Q^{i}=\epsilon_{A}(\rho^{A})^{i}{}_{j}Q^{j}.
\end{equation}
This yields:
\begin{equation}
-\frac{1}{4}\overline{D}^{2}\mathcal{J}_{A}= (Q^{c})_{i}m^{i}{}_{j}(\rho_{A})^{j}{}_{l}%
Q^{l}\,.
\end{equation}
$m^{i}_{j}$ is the raising operator of the $\mathfrak{su}(2)_{D}$ subalgebra 
and expressing the adjoint index $A$ in terms of spin $j$ and $T_3$ eigenvalue 
results exactly in equation (\ref{D2J}). As already mentioned, an analogous 
procedure also works for non-Lagrangian theories (see
e.g. \cite{Xie:2016hny, Maruyoshi:2016tqk, Maruyoshi:2016aim}).

\section{Inherited Infrared Symmetries\label{sec:INHERIT}}

In this section we turn to an analysis of the 4D $\mathcal{N}=1$ fixed points
generated by nilpotent mass deformations, focusing on the structure of the
symmetries inherited from the original UV $\mathcal{N}=2$ SCFT. Our aim will
be to understand both the structure of the infrared R-symmetry, as well as
global symmetries preserved by a nilpotent mass deformation. Additionally, we
compute the anomalies associated with these symmetries.

One technical assumption we make in this section is that there are no emergent
abelian symmetries. When emergent symmetries are present, as necessarily
occurs when some operators decouple, it is necessary to track which operators
have dimension coming close to the unitarity bound. This requires a more case
by case treatment of the nilpotent deformation in question, and is best
handled by way of explicit cases.

We begin by treating the case of plain mass deformations and then turn to the
case of flipper field deformations. After this, we show that under mild assumptions on the
values of $a_{\mathrm{UV}}$ and $c_{\mathrm{UV}}$ that various numerical quantities are
strictly monotonic along directed paths through the Hasse diagram of nilpotent orbits.

\subsection{Plain Mass Deformations}

Suppose, then, that we introduce a nilpotent mass deformation of a 4D
$\mathcal{N}=2$ SCFT. This initiates an explicit breaking
pattern of the $SU(2) \times U(1)$ R-symmetry of the UV theory, 
as well as well as the flavor symmetries
$\mathfrak{g}_{\mathrm{UV}}$. By definition, there is
a generator $T_{3}$ in the Cartan subalgebra such that the operator
Tr$_{\mathfrak{g}_{\mathrm{UV}}}\left(  \mu\cdot\mathcal{O}_{\text{adj}%
}\right)$ has $T_{3}$ charge $-1$. What this means is that a linear
combination of $T_{3}$ and $J_{\mathcal{N}=2}$ will remain unbroken along the
entire flow to the infrared.

In addition to these symmetries, there are of course all the generators of
$\mathfrak{g}_{\mathrm{UV}}$ which commute with our nilpotent orbit. This defines
another flavor symmetry algebra $\mathfrak{g}_{\mathrm{IR}}$ which may also include
various abelian symmetry factors.

Assuming that we indeed flow to a new fixed point in the infrared with
$\mathcal{N}=1$ supersymmetry, the infrared R-symmetry will be a linear
combination of all available abelian symmetries:%
\begin{equation}
R_{\mathrm{IR}}=R_{\mathrm{UV}}+t_{J}J_{\mathcal{N}=2}-tT_{3}+t_{\text{other}}T_{\text{other}},
\end{equation}
where $T_{\text{other}}$ is shorthand for all other abelian symmetries
inherited from the UV.

Now, for our plain mass deformation to be a relevant perturbation, it follows
that the IR R-charge of this operator deformation is fixed to be $+2$. Since
Tr$_{\mathfrak{g}_{\text{flav}}}\left(  \mu\cdot\mathcal{O}_{\text{adj}%
}\right)  $ has charges $R_{\mathrm{UV}}=+4/3$, $J_{\mathcal{N} = 2}=-2$, $T_{3}=-1$ and is neutral
under $T_{\text{other}}$, we learn that the IR\ R-symmetry is actually
constrained to be:%
\begin{equation}
R_{\mathrm{IR}}=R_{\mathrm{UV}}+\left(  \frac{t}{2}-\frac{1}{3}\right)  J_{\mathcal{N}=2}%
-tT_{3}+t_{\text{other}}T_{\text{other}},
\end{equation}
where to fix the remaining parameters $t$ and $t_{\text{other}}$, we must
resort to a-maximization \cite{Intriligator:2003jj}, namely we calculate the trial value of the
conformal anomaly $a_{\text{trial}}(t,t_{\text{other}})$ as a function of $t$
and $t_{\text{other}}$:%
\begin{equation}
a_{\text{trial}}(t,t_{\text{other}})=\frac{3}{32}\left(  3\text{Tr}R_{\mathrm{IR}}%
^{3}(t,t_{\text{other}})-\text{Tr}R_{\mathrm{IR}}(t,t_{\text{other}})\right)  ,
\end{equation}
and find the local maximum with respect to these parameters.

Since we are assuming the absence of emergent symmetries in the infrared, we
can use anomaly matching to express various IR\ quantities in terms of
UV\ data. In particular, we shall have need to reference the anomalies:%
\begin{align}
a_{\mathrm{UV}}  &  =\frac{3}{32}\left(  3\text{Tr}R_{\mathrm{UV}}^{3}-\text{Tr}R_{\mathrm{UV}}\right) \\
c_{\mathrm{UV}}  &  =\frac{1}{32}\left(  9\text{Tr}R_{\mathrm{UV}}^{3}-5\text{Tr}R_{\mathrm{UV}}\right)
\\
k_{\mathrm{UV}}\times\delta^{AB}  &  =-6\text{Tr}\left(  R_{\mathrm{UV}}J_{\text{flav}}%
^{A}J_{\text{flav}}^{B}\right)  ,
\end{align}
in the obvious notation.

Let us first establish that $t_{\text{other}}$ actually vanishes. To this end,
we note that since we have assumed below line (\ref{flavaflav}) that the
anomalies involving the UV flavor symmetries always involve precisely two
insertions of the \textit{same} flavor symmetry,\footnote{Indeed, recall that
the \textquotedblleft other\textquotedblright\ in $t_{\text{other}}$ is
shorthand for labeling possibly multiple abelian symmetry factors. This means
there could be mixed terms between these factors. If all these abelian factors
descend from a non-abelian symmetry, such mixed anomalies automatically
vanish, but it could a priori still be present for abelian symmetries
inherited from the UV\ theory. This is the main reason the assumption below
line (\ref{flavaflav}) is required.} the only way for $t_{\text{other}}$ to
make an appearance in $a_{\text{trial}}$ is through a mixed anomaly with a
symmetry generator of the $SU(2)\times U(1)$ R-symmetry of the $\mathcal{N}=2$
SCFT. Since the dependence on $t_{\text{other}}$ has only quadratic
dependence, the local maximum necessarily has $t_{\text{other}}=0$. Hence, the
infrared R-symmetry is actually given by the linear combination:%
\begin{equation}
R_{\mathrm{IR}}=R_{\mathrm{UV}}+\left(  \frac{t}{2}-\frac{1}{3}\right)  J_{\mathcal{N}=2}%
-tT_{3},
\end{equation}
with $t$ to be fixed by a-maximization.

This analysis was already carried out in reference \cite{Heckman:2010qv} for a specific class of
deformations, but the generalization to our case follows formally the same
steps. The only change is that now, we need to pay attention to the appearance
of possibly multiple UV symmetry factors\ in:%
\begin{equation}
\mathfrak{g}_{\mathrm{UV}}=\mathfrak{g}_{\mathrm{UV}}^{(1)}\times...\times\mathfrak{g}%
_{\mathrm{UV}}^{(n)},
\end{equation}
so we need to label the $RFF$ anomaly for each such factor:%
\begin{equation}
\text{Tr}\left(  R_{\mathrm{UV}}J_{A_i}^{(i)}J_{B_i}^{(i)}\right)  =-\frac{k_{\mathrm{UV}}^{(i)}}%
{6}\delta_{A_i B_i}.
\end{equation}
Since we can decompose our $T_{3}$ generator as a direct sum for each simple
factor:%
\begin{equation}
T_{3}=T_{3}^{(1)}\oplus...\oplus T_{3}^{(n)}.
\end{equation}
The value of $a_{\text{trial}}(t)$ is given by:%
\begin{equation}
\begin{split}
a_{\text{trial}}(t)=\frac{3}{32}&\Bigg[  \left(  36a_{\mathrm{UV}}-27c_{\mathrm{UV}}-\frac{9}%
{4}\underset{i=1}{\overset{n}{\sum}}k_{\mathrm{UV}}^{(i)}r^{(i)}\right)  t^{3} \\
&+(-72a_{\mathrm{UV}}+36c_{\mathrm{UV}})t^{2}+(48a_{\mathrm{UV}}-12c_{\mathrm{UV}})t\Bigg]  ,
\end{split}
 \label{atrialplain}%
\end{equation}
where in obtaining this formula we have used the structure of anomalies as
dictated by the UV$\ \mathcal{N}=2$ theory.
Here, $r^{(i)}$ refers to the embedding index for the generator
$T_{3}^{(i)}$ in $\mathfrak{g}_{\mathrm{UV}}^{(i)}$:%
\begin{equation}
r^{(i)}\equiv 2 \text{Tr}_{\mathfrak{g}_{\mathrm{UV}}^{(i)}}\left(  T_{3}^{(i)}%
T_{3}^{(i)}\right)  ,
\end{equation}
see Appendix \ref{app:EMBED} for details.

The local maximum of $a_{\text{trial}}(t)$ is then given by the critical
point:%
\begin{equation}
t_{\ast}=\frac{4}{3}\times\frac{8a_{\mathrm{UV}}-4c_{\mathrm{UV}}-\sqrt{4c_{\mathrm{UV}}^{2}%
+(4a_{\mathrm{UV}}-c_{\mathrm{UV}})\underset{i=1}{\overset{n}{\sum}}k_{\mathrm{UV}}^{(i)}r^{(i)}}%
}{16a_{\mathrm{UV}}-12c_{\mathrm{UV}}-\underset{i=1}{\overset{n}{\sum}}k_{\mathrm{UV}}^{(i)}r^{(i)}}.
\label{tstarplain}%
\end{equation}
With this in hand, we can evaluate the anomalies of our candidate infrared
fixed point. In the case of the flavor symmetry anomalies, the structure
depends on the remaining flavor symmetry generators associated with each
semi-simple factor, and we denote these unbroken symmetry currents by
$J_{A_{i}}^{(i)}$. In terms of the parameter $t_{\ast}$, the IR\ values of
these anomalies are:%
\begin{align}\label{IRanomalies}
a_{\mathrm{IR}}  &  =\frac{3}{32} \Bigg[  \left(  36a_{\mathrm{UV}}-27c_{\mathrm{UV}}-\frac{9}%
{4}\underset{i=1}{\overset{n}{\sum}}k_{\mathrm{UV}}^{(i)}r^{(i)}\right)  t_{\ast}
^{3} +(-72a_{\mathrm{UV}}+36c_{\mathrm{UV}})t_{\ast}^{2} \nonumber \\ 
& +(48a_{\mathrm{UV}}-12c_{\mathrm{UV}})t_{\ast}\Bigg] \\
c_{\mathrm{IR}}  &  =\frac{1}{32}\Bigg[  \left(  108a_{\mathrm{UV}}-81c_{\mathrm{UV}}-\frac{27}{4}\underset{i=1}{\overset{n}{\sum}}k_{\mathrm{UV}}^{(i)}r^{(i)}\right)  t_{\ast}^{3} (-216a_{\mathrm{UV}}+108c_{\mathrm{UV}})t_{\ast}^{2} \nonumber \\
&+(96a_{\mathrm{UV}}+12c_{\mathrm{UV}})t_{\ast}\Bigg]
\end{align}
and:
\begin{equation}
K_{\mathrm{IR}}^{(i)} = \frac{3}{2}k_{\mathrm{UV}}^{(i)}\times t_{\ast},
\end{equation}
In the above, we have introduced the anomaly coefficient $K_{\mathrm{IR}}^{(i)}$:%
\begin{equation}
\text{Tr}\left(  R_{\mathrm{IR}}J_{A_i}^{(i)}J_{B_i}^{(i)}\right)  =-\frac{K_{\mathrm{IR}}^{(i)}}%
{6}\delta_{A_i B_i},
\end{equation}
where we take the same normalization of all Lie algebra generators as
inherited from the parent UV\ symmetry. In a given simple factor in the IR,
there could be several subalgebras:%
\begin{equation}
\mathfrak{h}_{1}^{(i)}\times...\times\mathfrak{h}_{m_{i}}^{(i)}\subset
\mathfrak{g}_{\mathrm{IR}}^{(i)}\subset\mathfrak{g}_{\mathrm{UV}}^{(i)}\text{,}%
\end{equation}
each with a different embedding index. We can of course take generators
normalized with respect to these unbroken flavor symmetries to define the more
standard quantity via the embedding index:%
\begin{equation}
k_{l_{i} , IR}^{(i)}=\text{Ind}(\mathfrak{h}_{l_{i}}^{(i)}\rightarrow\mathfrak{g}%
_{\mathrm{UV}}^{(i)})\times K_{\mathrm{IR}}^{(i)}\text{.} \label{kvsK}%
\end{equation}
The physically more meaningful quantity is $k_{\mathrm{IR}}^{(i)}$, though it is often
more straightforward to evaluate $K_{\mathrm{IR}}^{(i)}$.

\subsubsection{Operator Scaling Dimensions}

Having determined the infrared R-symmetry, we can now extract the scaling
dimensions for a number of operators. It is helpful to organize this analysis
according to the representation content of the subalgebra $\mathfrak{g}%
_{\mathrm{IR}}\times\mathfrak{su}(2)_{D}$, where $\mathfrak{su}(2)_{D}$ is the 
subalgebra implicitly defined by a choice of nilpotent
orbit. For example, since the mesons transform in the adjoint representation
of $\mathfrak{g}_{\mathrm{UV}}$, there is a corresponding decomposition into
representations:%
\begin{align}
\mathfrak{g}_{\mathrm{UV}}  &  \supset\mathfrak{g}_{\mathrm{IR}}\times\mathfrak{su}%
(2)_{D}\\
\text{adj}(\mathfrak{g}_{\mathrm{UV}})  &  \rightarrow\underset{a}{%
{\displaystyle\bigoplus}
}\left(  R_{(a)},j_{(a)}\right)  ,
\end{align}
where on the right-hand side we implicitly sum over irreducible representations
of $\mathfrak{g}_{\mathrm{IR}}\times\mathfrak{su}(2)_{D}$ which appear in the
decomposition of the adjoint. More generally, given operators in some
representation of $\mathfrak{g}_{\mathrm{UV}}$, we can always decompose into
irreducible representations of $\mathfrak{g}_{\mathrm{IR}}\times\mathfrak{su}%
(2)_{D}$.

Supposing then that we have a UV operator transforming in a spin $j$
representation of $\mathfrak{su}(2)_{D}$, we get operators of $T_{3}$ charge
$j,j-1,...,-j$, and we can calculate their scaling dimension in the IR\ theory using
our infrared R-symmetry:%
\begin{equation}
\Delta_{\mathrm{IR}}=\frac{3}{2}\left(  R_{\mathrm{UV}}+\left(  \frac{t_{\ast}}{2}-\frac{1}%
{3}\right)  J_{\mathcal{N}=2}-t_{\ast}T_{3}\right)  .
\end{equation}
In the specific case of a Coulomb branch scalar $Z$, we know that since it has
vanishing $I_{3}$ charge, we have $3R_{\mathrm{UV}}(Z)=J_{\mathcal{N}=2}\left(
Z\right)  $, and $T_{3}(Z)=0$ (as it is neutral under all of $\mathfrak{g}%
_{\mathrm{UV}}$), so we immediately obtain:%
\begin{equation}
\Delta_{\mathrm{IR}}\mathcal{(}Z)=\frac{3}{2}t_{\ast}\times\Delta_{\mathrm{UV}}(Z).
\label{opcoul}%
\end{equation}
In the case of a mesonic operator $\mathcal{O}_{j,s}$ transforming in a spin
$j$ representation of $\mathfrak{su}\left(  2\right)  _{D}$, with
$T_{3}$ charge $s$, the scaling dimension in the IR\ is:%
\begin{equation}
\Delta_{\mathrm{IR}}\left(  \mathcal{O}_{j,s}\right)  =3-\frac{3}{2}t_{\ast}(1+s).
\label{opmes}%
\end{equation}

\subsubsection{Monotonicity}

With these results in place, we now show that various numerical quantities are
indeed monotonic as we proceed to larger orbits in the nilpotent cone. We will
also establish this numerically by \textquotedblleft brute
force\textquotedblright\ when we turn to an analysis of explicit
$\mathcal{N}=2$ theories.

To begin, we recall from reference \cite{Hofman:2008ar, Hofman:2016awc} that there is the
Hofman-Maldacena bound on the ratio $a_{\mathrm{UV}}/c_{\mathrm{UV}}$ for any $\mathcal{N} = 2$ SCFT:
\begin{equation}
\frac{1}{2}\leq\frac{a_{\mathrm{UV}}}{c_{\mathrm{UV}}}\leq\frac{5}{4}. \label{HofMalda}%
\end{equation}
We now use this general bound to establish some monotonicity results for
nilpotent mass deformations.

Now, as we proceed to larger orbits, the size of the corresponding embedding
indices necessarily increases. Introducing the parameter:%
\begin{equation}
\mathcal{K}\equiv\underset{i=1}{\overset{n}{\sum}}k_{\mathrm{UV}}^{(i)}r^{(i)},
\end{equation}
we observe that this quantity always increases as we proceed down a directed
path in the Hasse diagram. To establish various monotonicity results, it thus
suffices to evaluate their response as we vary $\mathcal{K}$.

First of all, we can consider the parameter $t_{\ast}$ given by equation (\ref{tstarplain}),
treated as a function of $\mathcal{K}$. If we introduce the Hofman-Maldacena
bounds, as well as the constraints from unitarity 
$a_{\mathrm{UV}},c_{\mathrm{UV}},k_{\mathrm{UV}}^{(i)}>0$, we immediately find (as can be checked
explicitly using \texttt{Mathematica}) that the derivative:%
\begin{equation}
\frac{\partial t_{\ast}}{\partial\mathcal{K}}<0,
\end{equation}
so in particular, $t_{\ast}$ always decreases along a flow. Moreover, since
the Coulomb branch operators are all proportional to $t_{\ast}$, we also learn
that these dimensions are also always strictly decreasing.

One can also perform a similar analysis for the parameter $a_{\mathrm{IR}}$ as a
function of $\mathcal{K}$. In addition to the numerical bounds already introduced,
we also require $t_{\ast}>0$, which in turn requires $16a_{\mathrm{UV}}-12c_{\mathrm{UV}}%
-\mathcal{K}>0$. Curiously enough, we find that in order for this quantity to
decrease monotonically, we need to impose a slightly stronger condition than
that of line (\ref{HofMalda}) for the lower bound:%
\begin{equation}
\frac{3}{4}\leq\frac{a_{\mathrm{UV}}}{c_{\mathrm{UV}}}\leq\frac{5}{4}. \label{sharpee}%
\end{equation}

The most conservative interpretation of this sharper requirement is that as we
pass to larger orbits, we should expect some operators to decouple, in
which case the expressions used for $t_{\ast}$ and $a_{\mathrm{IR}}$ would need to be
modified anyway. We will indeed see examples of this type, though we hasten to
add that in the explicit models we consider, the sharper condition of line
(\ref{sharpee}) is actually satisfied.

\subsection{Flipper Field Deformations}

Having dealt with the case of plain mass deformations, we now turn to
flipper field deformations of an $\mathcal{N}=2$ SCFT. Recall that this
involves promoting the mass parameters of the $\mathcal{N}=2$ theory to an
adjoint valued chiral superfield, and switching on a background vev:%
\begin{equation}
\delta W=\text{Tr}_{\mathfrak{g}_{\text{flav}}}\left(  (m_{\text{adj}%
}+M_{\text{adj}})\cdot\mathcal{O}_{\text{adj}}\right)  .
\end{equation}
Again, we confine our analysis to the case where this vev is a nilpotent mass deformation.

Since we are activating a breaking pattern which is identical to the case of
the plain mass deformation, much of the analysis of the previous section will
carry over unchanged. The primary issue is that now, we need to track the
additional modifications to the infrared R-symmetry which come from having
these additional fields transforming in the adjoint representation.

From the perspective of the UV theory, we have two decoupled SCFTs, namely the
original $\mathcal{N}=2$ fixed point, and a decoupled free chiral multiplet.
Consequently, there is a $U(1)$ flavor symmetry with generator $T_{\text{flip}}$ which
acts on each flipper field, so that it has charge $+1$.
The trial infrared R-symmetry is then a general linear combination of the
form:
\begin{equation}
R_{\mathrm{IR}}^{\text{flip}}(t) = R_{\mathrm{IR}}^{\text{plain}}(t)+t_{\text{flip}}T_{\text{flip}}
\end{equation}
where we have also left implicit the sum over all flippers.
Here, the trial infrared R-symmetry in the case of a plain mass deformation
is:
\begin{equation}
R_{\mathrm{IR}}^{\text{plain}}(t)=R_{\mathrm{UV}}+\left(  \frac{t}{2}-\frac{1}{3}\right)
J_{\mathcal{N}=2}-tT_{3}.
\end{equation}

Now, upon decomposing into representations of $\mathfrak{su}(2)_{D}$, we see that
all flipper fields will deform the theory via operators such as $M_{j,-s}%
\mathcal{O}_{j,s}$. If we first activate the plain mass deformation, and then
couple to the flipper fields, we see that since the operators $\mathcal{O}_{j,j}$
with the highest spin have the lowest scaling dimension, then these are the
operators which actually drive a new flow \cite{Maruyoshi:2016tqk, Maruyoshi:2016aim}. 
For this to be so, we require a constraint on the infrared R-charge assignments 
(see e.g. \cite{Gadde:2013fma, Benvenuti:2017lle}):%
\begin{equation}
R_{\mathrm{IR}}(M_{j,-j})+R_{\mathrm{IR}}(\mathcal{O}_{j,j})=2, \label{flippar}%
\end{equation}
so the new trial IR\ R-symmetry is:%
\begin{equation}
R_{\mathrm{IR}}^{\text{flip}}(t)=R_{\mathrm{IR}}^{\text{plain}}(t)+\left(  t-\frac{2}{3}\right)
T_{\text{flip}}.
\end{equation}

We can also calculate the new trial $a_{\text{trial}}^{\text{flip}}(t)$ by
breaking up the trace over states into those coming from the original
$\mathcal{N}=2$ theory, and those coming from the flipper fields which actually
participate in the flow. Doing so, we get:%
\begin{equation}
a_{\mathrm{trial}}^{\mathrm{flip}}\left(  t\right)  =a_{\mathrm{trial}%
}^{\text{plain}}(t)+\sum_{j_{(a)}}{\left[  \frac{3}{32}\left(  3\left(
R_{\mathrm{IR}}^{\text{flip}}(M_{j_{(a)},-j_{(a)}})-1\right)  ^{3}-\left(
R_{\mathrm{IR}}^{\text{flip}}(M_{j_{(a)},-j_{(a)}})-1)\right)  \right)
\right]  ,}%
\end{equation}
where in the first term, $a_{\text{\textrm{trial}}}^{\text{plain}}(t)$ is
the same quantity as in line (\ref{atrialplain}), and in the second set of
terms, we sum over all highest spin states which appear in the branching rules
for the $\mathfrak{su}(2)_{D}$ subalgebra. The R-charge for each
such flipper field is evaluated with respect to the original R-symmetry of the
plain mass deformation case, namely:%
\begin{equation}
R_{\mathrm{IR}}^{\text{plain}}(M_{j_{(a)},-j_{(a)}})=\frac{2}{3}%
+j_{(a)}\times t.
\end{equation}

Maximizing over the parameter $t$ appearing in $a_{\mathrm{trial}%
}^{\mathrm{flip}}\left(  t\right)  $, we again obtain the infrared R-symmetry,
and can read off the scaling dimensions of operators, much as before. By a
similar token, we can also read off the new value of the conformal anomaly
$c_{\mathrm{IR}}^{\mathrm{flip}}$. Collecting these expressions here, we have \cite{Giacomelli:2017ckh}:%
\begin{align}
a_{\mathrm{IR}}^{\mathrm{flip}}  &  =a_{\mathrm{IR}}^{\text{plain}}(t_{\ast
})+\sum_{j_{(a)}}{\left[  \frac{3}{32}\left(  3\left(  R_{\mathrm{IR}%
}^{\text{flip}}(M_{j_{(a)},-j_{(a)}})-1\right)  ^{3}-\left(  R_{\mathrm{IR}%
}^{\text{flip}}(M_{j_{(a)},-j_{(a)}})-1)\right)  \right)  \right]  }\\
c_{\mathrm{IR}}^{\mathrm{flip}}  &  =a_{\mathrm{IR}}^{\text{plain}}(t_{\ast
})+\sum_{j_{(a)}}{\left[  \frac{1}{32}\left(  9\left(  R_{\mathrm{IR}%
}^{\text{flip}}(M_{j_{(a)},-j_{(a)}})-1\right)  ^{3}-5\left(
R_{\mathrm{IR}}^{\text{flip}}(M_{j_{(a)},-j_{(a)}})-1)\right)  \right)
\right]  ,}%
\end{align}
in the obvious notation.

With the infrared R-symmetry in hand, we can also evaluate the new anomalies
involving the flavor symmetry. Since the flipper fields also transform in
irreducible representations of $\mathfrak{g}_{\mathrm{IR}}$, the IR\ flavor symmetry, we need to
take into account the specific branching rules associated with the
decomposition of the adjoint representation. With notation as in line
(\ref{kvsK}), we have:
\begin{equation}
k_{\mathrm{IR}}(\mathfrak{h}_{l_i}^{(i)})=\text{Ind}(\mathfrak{h}_{l_i}%
^{(i)}\rightarrow\mathfrak{g}_{\mathrm{UV}}^{(i)})\times K_{\mathrm{IR}}(\mathfrak{g}%
_{\mathrm{UV}})+6\sum_{j_{(a)}}(1-(1 - T_{3}(M_{j_{(a)},-j_{(a)}}))t_{\ast
})\mathrm{Ind}(\rho_{a}(\mathfrak{h}_{l_i}^{(i)})).
\end{equation}
Here, $\mathrm{Ind}(\rho_{a}(\mathfrak{h}_{l}^{(i)}))$ indicates the
index of the representation associated with a given flipper field for the
flavor symmetry algebra $\mathfrak{h}_{l_i}^{(i)}$.

Much as in the case of the plain mass deformations, we can read off the
scaling dimensions of our operators. The operator scaling dimensions for the
Coulomb branch scalars and mesonic operators are basically the same as in
lines (\ref{opcoul}) and (\ref{opmes}) except that now we use a modified value
for $t_{\ast}$ due to the coupling to flipper fields. In the case of the
flipper fields, we can read off the scaling dimensions of those that actually
participate in a flow via equation (\ref{flippar}). For those flipper fields which
do not actually participate in a flow, we instead have a collection of
decoupled free fields. In what follows, we shall ignore these contributions,
focusing exclusively on the interacting fixed point.

\section{Emergent Symmetries and Operator Decoupling\label{sec:EMERGE}}

In our analysis so far, we have assumed that there are no emergent symmetries
in the infrared. Our aim in this section will be to discuss some general
features of when to expect emergent symmetries in the case of nilpotent mass
deformations. We turn to specific UV\ theories in the following sections.
Turning the discussion around, the mathematical ordering of nilpotent orbits
provides some helpful clues on the nature of these candidate fixed points.

Now, one way such emergent symmetries can show up is when various operators
start to decouple. Assuming that a fixed point is
really present, if we assume the absence of emergent symmetries and find
the pathological behavior that some operator has dimension below the
unitarity bound, then it is an indication
that this operator has actually decoupled.
The minimal procedure of reference \cite{Kutasov:2003iy}
prescribes that we introduce an additional $U(1)$ flavor symmetry which only
acts on the offending operator. From our starting point of an $\mathcal{N}=2$
theory, the main thing we will be able to check is the scaling dimension of
the Coulomb branch and mesonic operators of the UV\ parent theory.

Another related possibility is that the IR\ theory actually enhances to an
$\mathcal{N}=2$ supersymmetric theory in the infrared. This can occur, for
example, in the case of flipper field deformations \cite{Maruyoshi:2016tqk, Maruyoshi:2016aim},
and recently a set of general sufficient conditions for such behavior to occur were proposed in
\cite{Giacomelli:2018ziv}. A necessary (but insufficient) condition to have such an enhancement is that the various anomalies of the IR fixed point all become
rational numbers rather than the algebraic numbers present for a more general
nilpotent mass deformation.
There are however known counter-examples that have rational anomalies but no SUSY enhancement to $\cN=2$ \cite{Evtikhiev:2017heo}.

Our plan in this section will be to setup some general diagnostics for
symmetry enhancement in the case of nilpotent mass deformations. First, we
consider the decoupling of Coulomb branch operators, and then we turn to the
decoupling of mesonic operators. After this we discuss some special cases
associated with rational values for the anomalies.
Finally, we discuss some preliminary aspects of how the partial ordering
implied by a Hasse diagram lines up with the physical RG\ flows.

\subsection{Decoupling of Coulomb Branch Operators}

Suppose then, that we perform our initial a-maximization procedure, and,
assuming the absence of any emergent $U(1)$'s, we calculate the scaling
dimension of a Coulomb branch operator $Z$. According to our general formula
from line (\ref{opcoul}), we have:%
\begin{equation}
\Delta_{\mathrm{IR}}\mathcal{(}Z)=\frac{3}{2}t_{\ast}\times\Delta_{\mathrm{UV}}(Z).
\end{equation}
If this yields a value less than one, but we still expect the presence of an
IR\ fixed point, this is a strong indication that this operator has actually
decoupled (and so has dimension exactly one). 
By inspection of our expression for the parameter $t_{\ast}$ we see
that this occurs whenever the embedding index becomes sufficiently large.

Assuming this is the only operator to decouple, it is also straightforward to
calculate the new infrared R-symmetry. Following Appendix B of
\cite{Intriligator:2003mi}, we have:
\begin{align}
a_{\mathrm{IR}}^{\mathrm{new}}(t)  &  =a_{\mathrm{IR}}^{\mathrm{old}}%
(t)+\frac{3}{32}\big[\left(  3\left(  R_{\mathrm{old}}(Z)+t_{Z}-1\right)
^{3}-3\left(  R_{\mathrm{old}}(Z)-1\right)  ^{3}\right) \nonumber\\
&  -\left(  \left(  R_{\mathrm{old}}(Z)+t_{Z}-1\right)  -\left(
R_{\mathrm{old}}(Z)-1\right)  \right)  \big]
\end{align}
for $a$ in the IR. Here, $t_{Z}$ denotes the charge of $Z$ under the emergent
$U(1)$ which only acts on this operator. Performing $a$-maximization with respect to
$t_{Z}$ then yields
\begin{equation}
R_{\mathrm{new}}(Z)\equiv R_{\mathrm{old}}(Z)+t_{Z}=\frac{2}{3}\,.
\end{equation}
At this point, we see that adding the emergent $U(1)$ indeed corrects the
scaling dimension of the offending operator to one, and it decouples.
Substituting in this result, along with the fact that
$R_{\mathrm{old}}(Z)=t\times\Delta_{\mathrm{UV}}(Z)$ implies
\begin{equation}
a_{\mathrm{IR}}^{\mathrm{new}}(t)=a_{\mathrm{IR}}^{\mathrm{old}}(t)-\frac
{3}{32}\left[  3\left(  \Delta_{\mathrm{UV}}(Z)t-1\right)  ^{3}-\left(
\Delta_{\mathrm{UV}}(Z)t-1\right)  \right]  +\frac{1}{48}\,.
\end{equation}
Now, we perform the second part of $a$-maximization by taking the partial
derivative of $a_{\mathrm{IR}}^{\mathrm{new}}(t)$ with respect to $t$ and
setting it equal to zero. For the new value
\begin{align}
  t^{\mathrm{new}}_* &= -\frac{4}{3 \left(48 a_{\mathrm{UV}} - 36 c_{\mathrm{UV}} - 3 k_{\mathrm{UV}} r - 4 \Delta_{\mathrm{UV}}^3\right) } \Big(
  -24 a_{\mathrm{UV}} + 12 c_{\mathrm{UV}} + 3 \Delta_{\mathrm{UV}}^2 \nonumber \\
  &+ \Big\{36 c_{\mathrm{UV}}^2 + 36 a_{\mathrm{UV}} k_{\mathrm{UV}} r - 6 k_{\mathrm{UV}} r \Delta_{\mathrm{UV}} + 48 a_{\mathrm{UV}} \left(-2 + \Delta_{\mathrm{UV}}\right) \left(-1 + \Delta_{\mathrm{UV}}\right) \Delta_{\mathrm{UV}} \nonumber \\
  & \ \ + \Delta_{\mathrm{UV}}^4 
   - 3 c_{\mathrm{UV}} \left(3 k_{\mathrm{UV}} r + 4 \Delta_{\mathrm{UV}} \left(6 + \left(-6 + \Delta_{\mathrm{UV}}\right) \Delta_{\mathrm{UV}}\right)\right) \Big\}^{1/2} \Big)
\end{align}
we find a maximum of $a_{\mathrm{IR}}^{\mathrm{new}}$. Note that we use the
abbreviation $\Delta_{\mathrm{UV}}$ for $\Delta_{\mathrm{UV}}(Z)$ in this
equation to increase the brevity. One can check that the second derivative of the trial
$a_{\mathrm{IR}}^{\text{new}}(t)$ is indeed negative definite at the critical point, so we do get a local
maximum.

Let us summarize the central charges after decoupling the offending operator:
\begin{align}
a_{\mathrm{IR}}^{\mathrm{new}}  &  =a_{\mathrm{IR}}^{\mathrm{old}%
}(t^{\mathrm{new}}_{\ast})-\frac{3}{32}\left[  3\left(  \Delta_{\mathrm{UV}%
}(Z)t^{\mathrm{new}}_{\ast}-1\right)  ^{3}-\left(  \Delta_{\mathrm{UV}%
}(Z)t^{\mathrm{new}}_{\ast}-1\right)  \right]  +\frac{1}{48}\\
c_{\mathrm{IR}}^{\mathrm{new}}  &  =c_{\mathrm{IR}}^{\mathrm{old}%
}(t^{\mathrm{new}}_{\ast})-\frac{1}{32}\left[  9\left(  \Delta_{\mathrm{UV}%
}(Z)t^{\mathrm{new}}_{\ast}-1\right)  ^{3}-5\left(  \Delta_{\mathrm{UV}%
}(Z)t^{\mathrm{new}}_{\ast}-1\right)  \right]  +\frac{1}{24}\\
K_{\mathrm{IR}}^{\mathrm{new}}  &  =K_{\mathrm{IR}}^{\mathrm{old}%
}(t^{\mathrm{new}}_{\ast})\,,
\end{align}
where $a_{\mathrm{IR}}^{\mathrm{old}}$, $c_{\mathrm{IR}}^{\mathrm{old}}$, and
$K_{\mathrm{IR}}^{\mathrm{old}}$ are the central charges which were computed
without the emergent $U(1)$. We emphasize that $K_{\mathrm{IR}}$ does not receive
any additional contributions besides $K_{\mathrm{IR}}^{\mathrm{old}%
}(t^{\mathrm{new}}_{\ast})$ due to the fact that $Z$ is not charged under the
flavor symmetry. Thus, removing the contribution from such operators does not
directly affect the flavor central charge, just indirectly by modifying the
value of $t_{\ast}$.

\subsection{Decoupling of Mesonic Operators}

Let us now turn to the possible decoupling of mesonic operators. When we turn
to specific examples, we find that this does not occur for the probe D3-brane theories,
but does occur for 4D conformal matter theories.

We first treat the case of plain mass deformations, and then turn to the case
of flipper field deformations. Returning to our general formula for the
operator scaling dimensions (in the absence of emergent $U(1)$'s), we see from
equation (\ref{opmes}) that the scaling dimension of an operator
$\mathcal{O}_{j,s}$ is:%
\begin{equation}
\Delta_{\mathrm{IR}}\left(  \mathcal{O}_{j,s}\right)  =3-\frac{3}{2}t_{\ast}(1+s).
\end{equation}
So, the bigger the spin of the operator under the $\mathfrak{su}%
(2)_{D}$ subalgebra, the smaller the scaling dimension.
This is counteracted to some extent by the decreasing value of $t_{\ast}%
$, though in practice, it is still true that as we descend to larger nilpotent
orbits, more mesonic operators start to decouple. For a given spin $j$
representation of $\mathfrak{su}(2)_{D}$, it is hopefully clear that
the highest spin state with $s=j$ will have lowest candidate scaling
dimension, so if this operator has scaling dimension above the unitarity
bound, the remaining operators in the same $\mathfrak{su}(2)_{D}$
multiplet will also be above the bound.

On the other hand, if the highest spin operator falls below the unitarity bound, we can again posit that
it decouples, with a single emergent $U(1)$ which acts
only on this operator. Now, in
addition to the highest spin operator $\mathcal{O}_{j,j}$, there are
often other values of $s$ in the same multiplet which might also appear to violate
the unitarity bound. Note, however, that via our previous discussion of the
broken flavor symmetry generators and the relation of equation (\ref{D2J}):%
\begin{equation}
-\frac{1}{4}\overline{D}^{2}\mathcal{J}_{j,s}= \mathcal{O}%
_{j,s-1}\text{,}%
\end{equation}
we know that components of the flavor current and the mesons pair up in
long multiplets. As a result, we again only need to apply our procedure
for the \textquotedblleft top spin\textquotedblright\ operators of a given
$\mathfrak{su}(2)_{D}$ multiplet.

Once again, reference \cite{Intriligator:2003mi} tells us that all we need to do is
remove the contribution from the offending operator $\mathcal{O}_{i}$ as follows:
\begin{align}
a_{\mathrm{IR}}^{\mathrm{new}}(t)  &  =a_{\mathrm{IR}}^{\mathrm{old}}%
(t)+\sum_{i}\frac{3}{32}\big[\left(  3\left(  R_{\mathrm{old}}(\mathcal{O}%
_{i})+t_{\mathcal{O}_{i}}-1\right)  ^{3}-3\left(  R_{\mathrm{old}}%
(\mathcal{O}_{i})-1\right)  ^{3}\right) \nonumber\\
&  -\left(  \left(  R_{\mathrm{old}}(\mathcal{O}_{i})+t_{\mathcal{O}_{i}%
}-1\right)  -\left(  R_{\mathrm{old}}(\mathcal{O}_{i})-1\right)  \right)
\big]\,.
\end{align}
Naively, one would take the index $i$ in this equation to run over all mesons which
appear to have dimension below the unitarity bound. However, our discussion of the deformed
symmetry current near line (\ref{D2J}) shows that only the highest spin component of each
$\mathfrak{su}(2)_{D}$ multiplet actually participates in the chiral ring of the IR fixed point.

The procedure of $a$-maximization with respect to
$t_{\mathcal{O}_{i}}$ then yields
\[
R_{\mathrm{new}}(\mathcal{O}_{i})\equiv R_{\mathrm{old}}(\mathcal{O}%
_{i})+t_{\mathcal{O}_{i}}=\frac{2}{3}\,.
\]
Again, we see that all bad $\mathcal{O}_{i}$ decouple. The value of $t_{\ast}$
is determined by $a$-maximization of $a_{\mathrm{IR}}^{\mathrm{new}}(t)$
and the corresponding anomalies are:
\begin{align}
a_{\mathrm{IR}}^{\mathrm{new}}  &  =a_{\mathrm{IR}}^{\mathrm{old}}(t)-\sum
_{i}\frac{3}{32}\left[  3\left(  R_{\mathrm{old}}(\mathcal{O}_{i})-1\right)
^{3}-\left(  R_{\mathrm{old}}(\mathcal{O}_{i})-1\right)  \right]  +\frac
{1}{48}\\
c_{\mathrm{IR}}^{\mathrm{new}}  &  =c_{\mathrm{IR}}^{\mathrm{old}}(t)-\sum
_{i}\frac{1}{32}\left[  9\left(  R_{\mathrm{old}}(\mathcal{O}_{i})-1\right)
^{3}-5\left(  R_{\mathrm{old}}(\mathcal{O}_{i})-1\right)  \right]  +\frac
{1}{24}\,.
\end{align}
We can also give a general formula for the
new $k_{\mathrm{IR}}(\mathfrak{h}_{l_i}^{(i)})$ after we 
decouple all the offending mesons:
\begin{equation}
k_{\mathrm{IR}}(\mathfrak{h}_{l_i}^{(i)})=\text{Ind}(\mathfrak{h}_{l_i}%
^{(i)}\rightarrow\mathfrak{g}_{\mathrm{UV}}^{(i)})\times K_{\mathrm{IR}}(\mathfrak{g}_{\mathrm{UV}}^{(i)}) - 6\sum_{a}(1-(1 + T_{3}(\mathcal{O}_{a}))t_{\ast
})\mathrm{Ind}(\rho_{a}(\mathfrak{h}_{l_i}^{(i)})),
\end{equation}
where $\mathrm{Ind}(\rho_{a}(\mathfrak{h}_{l}^{(i)}))$ is the index of the irreducible
representation under which $\mathcal{O}_{i}$ transforms,
and $t_{\ast}$ is the fixed value of the maximization parameter at the last step when there are no
unitarity bound violations anymore.

Consider next the case of mesonic operators which decouple in the
flipper field deformations. As noted in \cite{Benvenuti:2017lle}, when an operator decouples,
one can introduce an additional \textquotedblleft flipping
field\textquotedblright\ which couples to this field. Doing this is equivalent
to the standard procedure of introducing an additional $U(1)$ anyway. Let us
see how this works in detail.

With each $M$, there comes an additional $U(1)$ symmetry in the UV theory.
Coupling the mesons to the $M$'s protects them from dropping below the
unitarity bound in the IR. From another point of view, the process of removing one
of the previously offending $\mathcal{O}$'s is equivalent to adding a coupling to
$M$, as explained in \cite{Benvenuti:2017lle}. Compared to the plain mass
deformation the new UV $U(1)$ is equivalent to the emergent $U(1)$ that we
would have to introduce by hand, once a meson drops below the unitarity bound.
Hence, for all flipper field deformations we do not need to worry about any
of the mesons decoupling or how it might affect the anomalies. This is
automatically being taken care of by the $M$'s. In fact as explained in
\cite{Benvenuti:2017lle}, the mesons $\mathcal{O}$ are zero in the chiral
ring, and therefore there are no unitarity violations associated to them. In
the following, we describe this intriguing mechanism in more detail from
another point of view.

The analysis involves essentially the same equations
as already presented in section \ref{sec:INHERIT}, which we
present here for convenience of the reader.
Recall that with flipper field deformations, we have a free chiral superfield $M$ in the
adjoint of $\mathfrak{g}_{\mathrm{UV}}$ coupled to $\mathcal{O}_{\text{adj}}$ via $\delta W =
\mathrm{Tr}_{\mathfrak{g}_{\mathrm{UV}}} (M_{\text{adj}} \cdot\mathcal{O}_{\text{adj}})$,
with a background value $\langle M_{\text{adj}} \rangle = m_{\text{adj}}$ our nilpotent mass term.
There is automatically an extra $U(1)$ symmetry
for each $M_{j_{(a)},-j_{(a)}}$ in the UV. The first part of the
trial IR R-charge is fixed by the plain mass deformation term $\mathrm{Tr}%
_{\mathfrak{g}_{\mathrm{UV}}}(m_{adj} \cdot \mathcal{O}_{\text{adj}})$. In the UV the $M_{j_{(a)},-j_{(a)}}$ are free multiplets
and they are charged under an extra $U(1)$. We call the generator
corresponding the this extra $U(1)$ $T_{\text{flip}}$. The charge of the
fluctuation of $M$ is normalized to $T_{\text{flip}}(M) = 1$, and nothing else is
charged under it. Moreover we know that $T_{3}(M_{j_{(a)},-j_{(a)}})=-
T_{3}(\mathcal{O}_{(j_{(a)},j_{(a)})})=-j_{(a)}$. Now, we have to take this
additional symmetry into account while computing the trial IR R-charge
\begin{equation}
\label{RirU1}R_{\mathrm{IR}}=R_{\mathrm{UV}}+\left(  \frac{t}{2}-\frac{1}%
{3}\right)  J_{\mathcal{N}=2} - t T_{3} + t_{\text{flip}} T_{\text{flip}}\,.
\end{equation}
Applying this relation to the superpotential deformation $\delta W$, we find
\begin{equation}
R^{\mathrm{new}}_{\mathrm{IR}}(\delta W) =R^{\mathrm{old}}_{\mathrm{IR}%
}(\mathcal{O}_{j_{(a)},j_{(a)}})+ t_{\text{flip}} + \frac{2}{3} - t T_{3}(M_{(j_{(a)}%
,-j_{(a)})})
\end{equation}
So, we have:
\begin{align}
\label{RirU1s} &  R^{\mathrm{old}}_{\mathrm{IR}}(\mathcal{O})+ t_{\text{flip}} + \frac{2}{3}
- t T_{3}(M_{j_{(a)},-j_{(a)}})= 2\\
\implies &  t_{\text{flip}} =\frac{4}{3}-R^{\mathrm{old}}_{\mathrm{IR}}(\mathcal{O})+ t
T_{3}(M_{j_{(a)},-j_{(a)}})=t-\frac{2}{3}%
\end{align}

This implies an additional contribution to $a_{\mathrm{IR}}=a_{\mathrm{IR}%
}^{old}+\delta a_{\mathrm{IR}}$ as follows: {\small
\begin{align}
\label{dair}\delta a_{\mathrm{IR}}  &  = \frac{3}{32} \left[  3\left(
t_{\text{flip}} T_{\text{flip}}(M) - t_{\ast} T_{3}(M) + R_{\mathrm{UV}}(M) - 1\right)
^{3}-\left(  t_{\text{flip}} T_{\text{flip}}-t_{\ast} T_{3}(M) + R_{\mathrm{UV}(M) - 1}\right)  \right] \nonumber\\
&  =\frac{3}{32} \left[  3(t_{\text{flip}} T_{\text{flip}}(M)- t_{\ast} T_{3}(M))^{3}-3(t_{\text{flip}} T_{\text{flip}}(M) - t_{\ast} T_{3}(M))^{2}+\frac{2}{9} \right]
\nonumber\\
&  = \frac{3}{32}\left[  3\left(  \frac{4}{3}-R_{\mathrm{IR}%
}(\mathcal{O})\right)  ^{3}-3\left(  \frac{4}%
{3}-R_{\mathrm{IR}}(\mathcal{O})\right)
^{2}+\frac{2}{9} \right] \nonumber\\
&  = -\frac{3}{32}\left[  3\left(  R_{\mathrm{IR}}%
(\mathcal{O})-1\right)  ^{3}-\left(  R_{\mathrm{IR}}(\mathcal{O})-1\right)  \right]  .
\end{align}
} As a result we can see that adding an additional $U(1)$ through the above
coupling is equivalent to removing the contribution from the ``bad'' operators
directly. This is why the flipper fields automatically rescue
the mesons whenever they would naively drop below the unitarity bound had this
coupling not been there. These additional coupling terms are identical to the ones that we were forced
to add whenever one of the mesons dropped below the unitarity bound before adding flipper fields.

Another quicker approach which builds upon equation \eqref{dair} is to make
use of the fact that $R(M)+R(\mathcal{O})=2$ so that we get:
\begin{align}
\label{dairquick}\delta a_{\mathrm{IR}}  &  = -\frac{3}{32}\left[  3\left(
2-R_{\mathrm{IR}}(M)-1\right)  ^{3}-\left(  2-R_{\mathrm{IR}}(M)-1\right)
\right] \nonumber\\
&  = -\frac{3}{32}\left[  3\left(  -R_{\mathrm{IR}}(M)+1\right)  ^{3}-\left(
-R_{\mathrm{IR}}(M)+1\right)  \right] \nonumber\\
&  = \frac{3}{32}\left[  3\left(  R_{\mathrm{IR}}(M)-1\right)  ^{3}-\left(
R_{\mathrm{IR}}(M)-1\right)  \right]  .
\end{align}
Therefore, adding directly the contribution from the $M$'s is equivalent to
removing the contribution from the ``bad'' $\mathcal{O}$'s. This recovers
our expressions for $a_{\mathrm{IR}}$ and $c_{\mathrm{IR}}$ up to the presence of
free chiral multiplets that do not couple.

As a result, none of the mesons in the flipper deformed theories can drop below
the unitarity bound because they are all automatically rescued by the
$M$'s to which they couple.

\subsection{Rational Theories}

One of the interesting features of the ``brute force'' sweeps we perform in later sections
reveals that in some cases, the anomalies are all rational numbers, even though
a priori, we should only expect algebraic numbers as per the procedure of a-maximization.
We refer to such IR fixed points as rational theories.
Clearly, this suggests some additional emergent structure in the infrared, and in some
favorable circumstances, this can also be identified with the appearance of enhanced $\mathcal{N} = 2$
supersymmetry, as in the case of the Maruyoshi-Song deformations \cite{Maruyoshi:2016tqk, Maruyoshi:2016aim}.
In the specific examples we consider, we find that this can happen both with and without operators decoupling,
and both for plain mass deformations and flipper field deformations, see Appendix \ref{completeTables} for details.

There has very recently been some progress in
understanding some additional sufficient criteria for $\mathcal{N}=2$
enhancements \cite{Giacomelli:2018ziv}. The main idea in this analysis is that whenever we
encounter a flavor singlet operator of the IR\ theory, we need to be able to
interpret as a scalar operator parameterizing a direction of the Coulomb
branch. This is not the case in our rational theories, but it is also
unclear whether there is any additional supersymmetry enhancement. We leave a full treatment
of possible enhancements in these theories for future work.

\subsection{Ordering of RG Flows}

As we can see, there is no clean expression that describes
$a_{\mathrm{IR}}$ as a function of the embedding index, once we take into
account operators that decouple in the IR. One might rightfully worry that $a_{\mathrm{IR}}$
would not necessarily be a simple monotonically decreasing function of $r$
anymore. However, we observe empirically that the RG flows continue to follow
the trajectory of paths through the Hasse diagram, even
after introducing emergent $U(1)$'s and flipper field operators. This is
explicitly shown in the explicit examples we consider.

We close this section with two important remarks:

\begin{enumerate}
\item If no operator drops below the unitarity bound, the theories are
guaranteed to follow the flow pattern specified by the Hasse diagrams.

\item In all of the other cases studied in this chapter, even when operators
decouple, we \textit{still} observe that the RG flows respect the partial ordering of
nilpotent orbits. So, while the RG flows could have a weaker ordering than the
mathematical ordering (if the wrong mesons hit the unitarity bound) we see
that they do not appear to violate the partial ordering of nilpotent orbits.
\end{enumerate}

\section{D3-Brane Probe Theories \label{sec:D3}}

In the previous sections we introduced a general procedure for treating
nilpotent mass deformations. In this section, we turn to a systematic analysis
of all such deformations for the $\mathcal{N}=2$ theories defined by a
D3-brane probing a 7-brane with $D_{4}$, $E_{6}$, $E_{7}$ or $E_{8}$ flavor
symmetry. In what follows we do not include the contribution
from the decoupled free hypermultiplet with scalars parameterizing motion of the D3-brane
parallel to the 7-brane.

Some examples of nilpotent mass deformations for these theories were analyzed
in \cite{Heckman:2010qv}, as well as \cite{Maruyoshi:2016tqk}.
In the F-theory interpretation where we wrap
the 7-brane on a surface $\mathcal{S}_{\text{GUT}}$, we have a partially
twisted gauge theory with a $(0,1)$-connection and an adjoint valued $(2,0)$
form $\Phi_{(2,0)}$ \cite{Beasley:2008dc} (see also \cite{Bershadsky:1997zs, Donagi:2008ca}).
In terms of the associated F-theory geometry,
deformations of $\Phi_{(2,0)}$ with non-vanishing Casimir invariant
translate to complex structure deformations of the associated elliptically
fibered Calabi-Yau fourfold. The nilpotent case is especially interesting
because it is essentially \textquotedblleft invisible\textquotedblright\ to
the complex geometry of the model. We can then view the mass parameters
$m_{\text{adj}}$ as background values for $\Phi_{(2,0)}$
\cite{Heckman:2010fh, Heckman:2010qv}, and
the particular case of a nilpotent mass deformation
defines a T-brane configuration \cite{Aspinwall:1998he, Donagi:2003hh, Cecotti:2010bp, Anderson:2013rka, Collinucci:2014taa, Collinucci:2014qfa, Bena:2016oqr, Marchesano:2016cqg, Anderson:2017rpr, Bena:2017jhm, Marchesano:2017kke, Cvetic:2018xaq}.

From this perspective, it is also natural to view the flipper field deformation
as promoting the zero mode of $\Phi_{(2,0)}$ to a dynamical field. This is actually
somewhat subtle in the context of a full F-theory compactification, because
making $\Phi_{(2,0)}$ dynamical requires us to wrap the 7-brane on a compact
K\"{a}hler surface, which also introduces dynamical gauge fields (zero modes
from the (0,1) connection can be eliminated by choosing a suitable surface and
background vector bundle). However, by introducing a sufficiently large number
of additional spectator fields which also interact with this gauge field, we
can always take a limit where this gauge theory is infrared free (in contrast
to the case typically assumed in decoupling limits from gravity).

In both the case of plain mass deformations as well as its extension to
flipper field deformations, we see that the IR\ fixed points defined by the
D3-brane provide additional insight into the structure of T-brane
configurations in F-theory.

Let us now turn to an analysis of the fixed points in these theories. Much as
in the earlier sections of this chapter, it is helpful to split our analysis up
into the cases of plain mass deformations and flipper field deformations. We
also discuss in detail the special case of rational theories, which suggest
additional structure in the IR. This includes all the previous $\mathcal{N}=2$
enhancement theories found in \cite{Maruyoshi:2016tqk}, as well as another one which comes about
from deformations of the $E_{7}$ Minahan-Nemeschansky theory (see also \cite{Carta:2018qke}).

\subsection{Summary of UV $\mathcal{N}=2$ Fixed Points}

In this section we briefly summarize some aspects of the $\mathcal{N}=2$
theories. We first list the anomalies and scaling
dimensions of the Coulomb branch operator $Z$. These
values can be found in \cite{Aharony:2007dj} and are
summarized in table \ref{UVscaling} for later convenience:\footnote{While it
is entirely possible to study nilpotent deformations of the Argyres-Douglas
theories they are too simple to be of interest. However, for convenience we do
list their UV values in table \ref{UVscaling}.}
\begin{table}[H]
  \centering
  {\renewcommand{\arraystretch}{1.2}
\begin{tabular}{ |c||c|c|c|c|c|c|c| }
 \hline
 $G$ & $H_0$ & $H_1$ & $H_2$ & $D_4$ & $E_6$  & $E_7$ & $E_8$ \\  \hline 
 $\Delta_{\mathrm{UV}}$($Z$) & $\frac{6}{5}$ &$\frac{4}{3}$ &$\frac{3}{2}$ &2 & 3 & 4 & 6 \\  \hline
 $a_{\mathrm{UV}}$ &  $\frac{43}{120}$ &$\frac{11}{24}$ &$\frac{7}{12}$ &$\frac{23}{24}$ & $\frac{41}{24}$  & $\frac{59}{24}$ & $\frac{95}{24}$ \\  \hline
 $c_{\mathrm{UV}}$ &  $\frac{11}{30}$ &$\frac{1}{2}$ &$\frac{2}{3}$ &$\frac{7}{6}$ & $\frac{13}{6}$  & $\frac{19}{6}$ & $\frac{31}{6}$ \\  \hline
 $k_{\mathrm{UV}}$ &  $\frac{12}{5}$ &$\frac{8}{3}$ &3 &4 & 6  & 8 & 12 \\  \hline
\end{tabular}}
\caption{Scaling dimensions and anomalies of rank 1 4D $\cN=2$ SCFTs.}
\label{UVscaling}
\end{table}

From there the anomalies and scaling dimensions in the IR can directly be
computed from the previously derived equations. The only necessary information
is the embedding index of the $\mathfrak{su}(2)_D$ subalgebra defined by the 
nilpotent orbit. Since we only
have one flavor symmetry factor, the Cartan matrix is uniquely
specified by the nilpotent orbit one wants to consider. Then it is only a
matter of evaluating the formulae of sections \ref{sec:INHERIT} and \ref{sec:EMERGE}.

\subsection{Plain Mass Deformations}

It is noteworthy that for all of the rank one probe D3-brane theories,
the mesons never appear to decouple. However,
$\Delta_{\mathrm{IR}}(Z)$ sometimes does decouple when
the value of $r$ becomes too large. In general the unitarity bound for the
operator $Z$ is violated whenever:
\begin{align}
r  &  \geq5 &  &  \text{for }\SO(8)\nonumber\\
r  &  \geq19 &  &  \text{for }E_{6}\nonumber\\
r  &  \geq40 &  &  \text{for }E_{7}\nonumber\\
r  &  \geq107 &  &  \text{for }E_{8}\,.
\end{align}

There are a large number of possible nilpotent deformations. Due to the size
of the resulting tables we only list our results for flavor symmetry $D_{4}$
and all rational results for the exceptional groups. Rational coefficients are
of particular interest as they suggest additional structure present in the IR.
When comparing our results with the subset of cases studied
in \cite{Heckman:2010qv} we find perfect agreement aside from the last column of table 5
which contains the correct value of $t_{\ast}$ but a minor typo
for the values of $a_{\mathrm{IR}}$ and $c_{\mathrm{IR}}$.

The complete list of all the possible deformations can be accessed via a
\texttt{Mathematica} routine summarized in Appendix~\ref{completeTables}. Due to
the very large amount of data we only list here the rational results for the
exceptional groups in Appendix \ref{completeTables}.

\begin{figure}[ptb]
\centering
\begin{subfigure}[b]{0.45\textwidth}
\includegraphics[width=\textwidth]{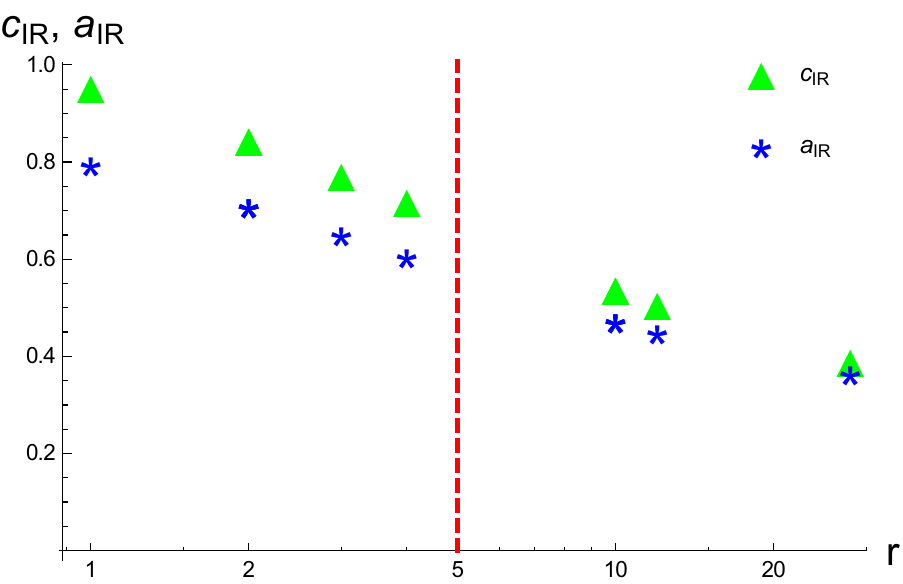}
\caption{$D_4$}
\label{MNcD4}
\end{subfigure}
\hspace{12pt} \begin{subfigure}[b]{0.45\textwidth}
\includegraphics[width=\textwidth]{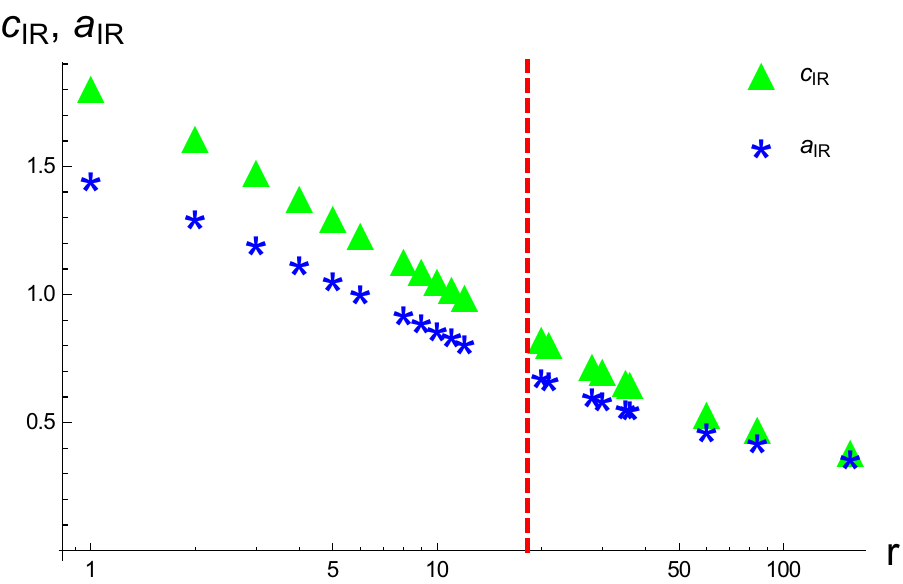}
\caption{$E_6$}
\label{MNcE6}
\end{subfigure}
\begin{subfigure}[b]{0.45\textwidth}
\includegraphics[width=\textwidth]{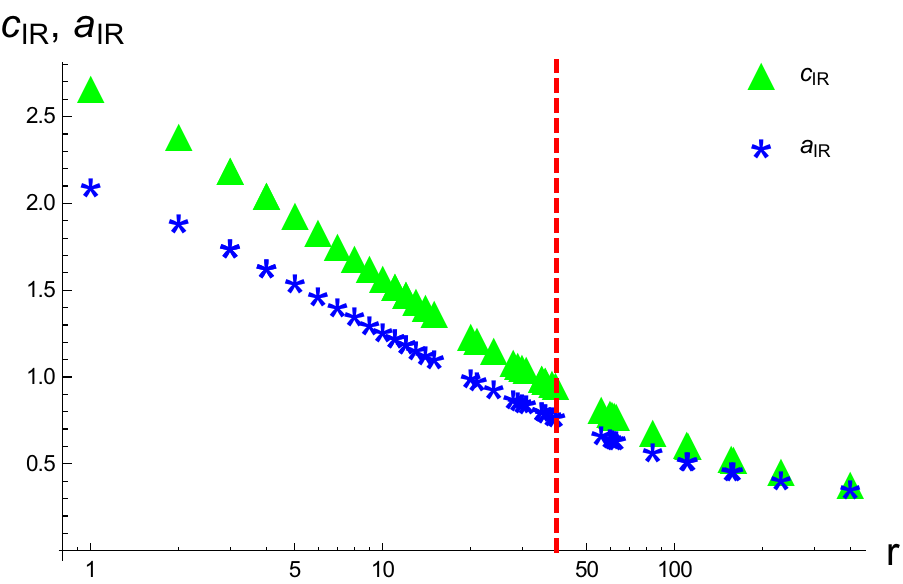}
\caption{$E_7$}
\label{MNcE7}
\end{subfigure}
\hspace{12pt} \begin{subfigure}[b]{0.45\textwidth}
\includegraphics[width=\textwidth]{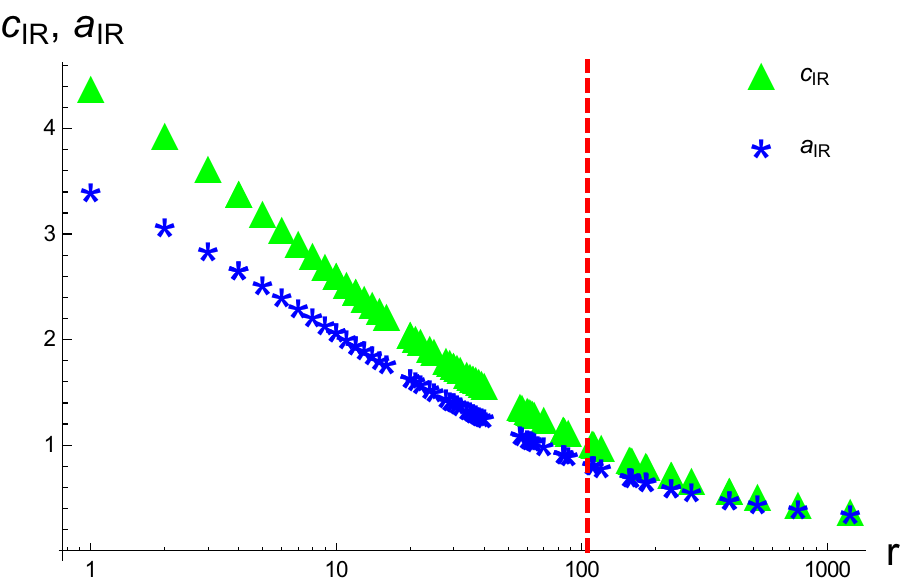}
\caption{$E_8$}
\label{MNcE8}
\end{subfigure}
\caption{Plots of $a_{\mathrm{IR}}$ (blue stars) and $c_{\mathrm{IR}}$ (green
triangles) vs embedding index $r$ for the different probe D3-brane
theories. The red vertical dashed line denotes the largest value of $r$ before
the Coulomb branch operator $Z$ decouples.
Anything to the right of this line has a single emergent $U(1)$
to rescue the Coulomb branch operator. The plots are log-scaled on the x-axis
for presentation purposes due to the fact that the region of deformed theories
is denser around lower values of $r$ and becomes more sparse as $r$
increases.}%
\label{MNc}%
\end{figure}

The tables are organized as follows. For the top tables, first we list
the Bala-Carter label of the deformation, or simply the partition of the
fundamental representation's splitting in the case of $\SO(8)$. The second
column gives the value of the embedding index $r$. The following three columns
give the anomalies $a_{\mathrm{IR}}$ and $c_{\mathrm{IR}}$, as well as the
value of the parameter $t$ after re-doing any $a$-maximization if necessary.
Whenever fields decouple (because they first hit the unitarity bound and are
rescued by emergent $U(1)$'s) then we can look at the interacting part versus the
complete contribution to $a_{\mathrm{IR}}$ and $c_{\mathrm{IR}}$. Indeed,
whenever an operator decouples it contributes a factor of $1/48$ or $1/24$ to
$a_{\mathrm{IR}}$ and $c_{\mathrm{IR}}$ respectively, and we separately 
report these values in our tables. The first number in
columns 3 and 4 is only the interacting piece, while the second number also
includes the contribution from any free multiplets that decoupled. Thus those
numbers only differ by an integer $n$ times $1/48$ (or $1/24$), where $n$ is
equal to the number of multiplets generators that have decoupled and become
free. If there is no emergent $U(1)$ introduced and no field decouples then
there is only an interacting piece and only the first number makes sense and
is listed. Finally, the last two columns give the scaling dimension of the
Coulomb branch parameter $Z$ and the lowest scaling dimension of the mesons
$\mathcal{O}$'s.

For the bottom tables we first list the Bala-Carter label of the deformation,
followed by the residual flavor symmetry. The following four columns
correspond to the flavor central charges $k_{\mathrm{IR}}$ taken with
respect to the residual flavor symmetry. For each we list their value with only the
interacting part of the theory or including the free fields which decoupled in
separate columns. Finally, we note that there are separate values for each of
the subgroups in the product decomposition of the residual flavor, hence the
multiple values listed in each column. For the theories with exceptional flavor
symmetry we only list values that have rational anomalies.

Furthermore, as it is impractical to list all the other values in a single
table we provide plots of $a_{\mathrm{IR}}$ and $c_{\mathrm{IR}}$ as functions
of the embedding index $r$:

\begin{figure}[ptb]
\centering
\begin{subfigure}[b]{0.45\textwidth}
\includegraphics[width=\textwidth]{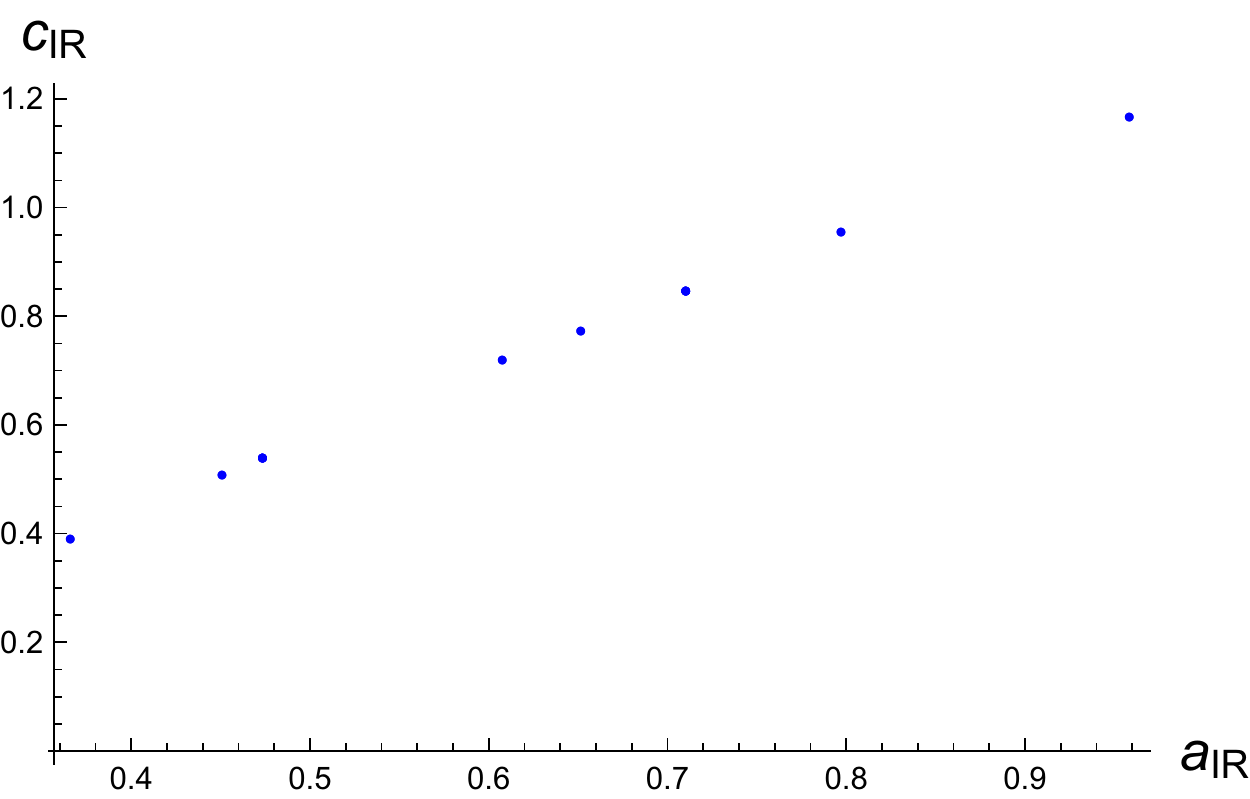}
\caption{$D_4$}
\label{MNratioD4}
\end{subfigure}
\hspace{12pt} \begin{subfigure}[b]{0.45\textwidth}
\includegraphics[width=\textwidth]{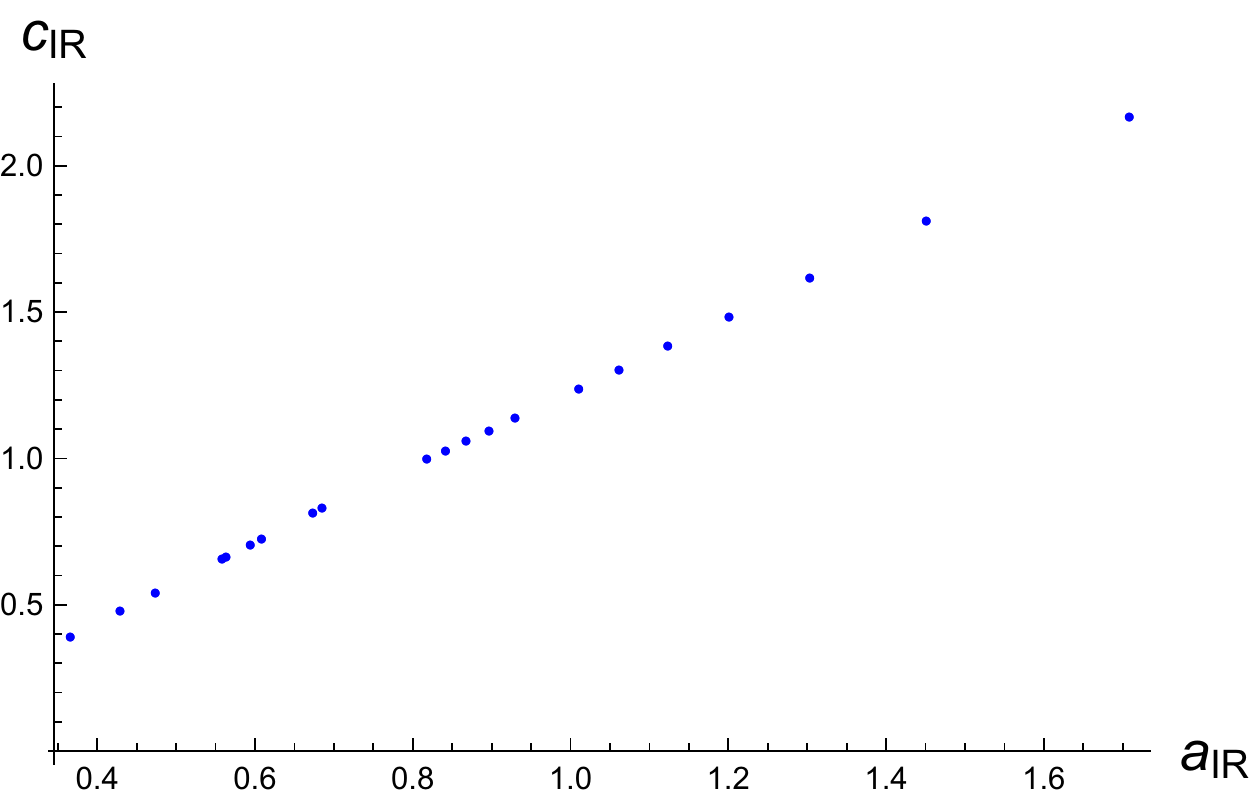}
\caption{$E_6$}
\label{MNratioE6}
\end{subfigure}
\begin{subfigure}[b]{0.45\textwidth}
\includegraphics[width=\textwidth]{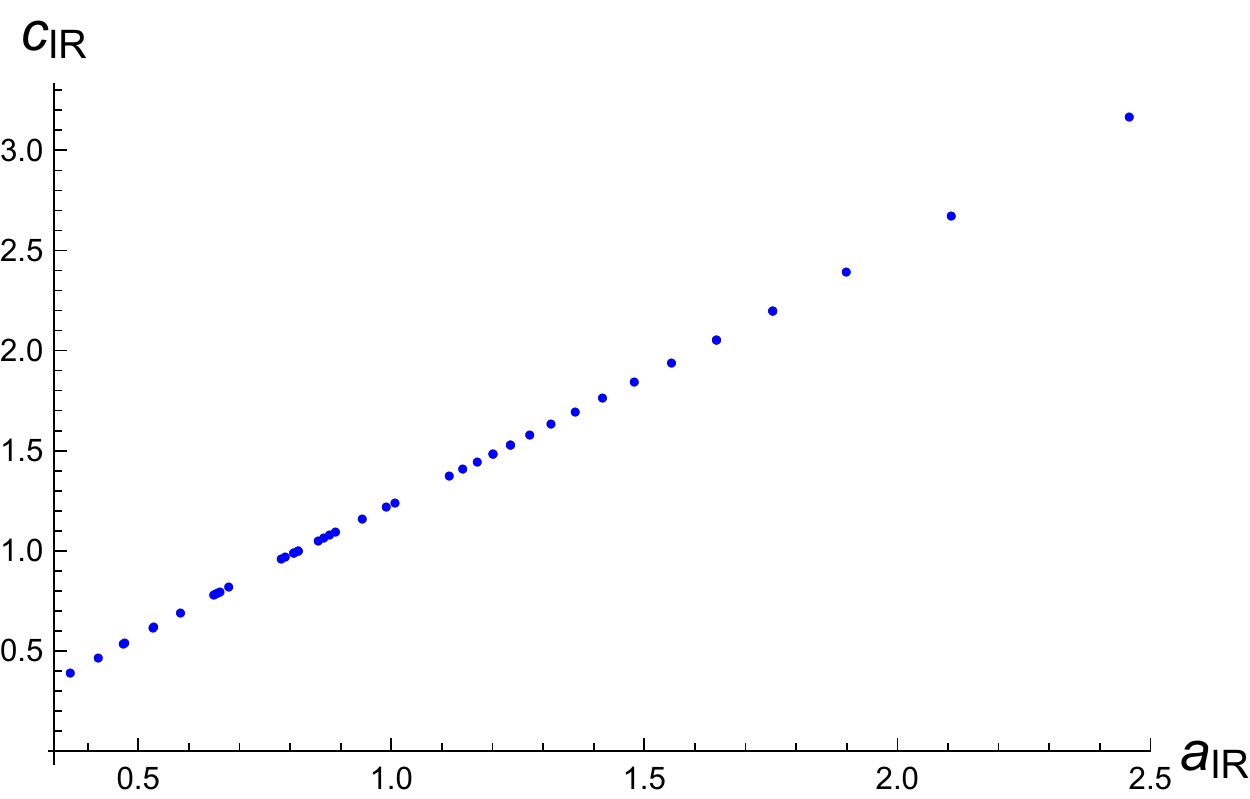}
\caption{$E_7$}
\label{MNratioE7}
\end{subfigure}
\hspace{12pt} \begin{subfigure}[b]{0.45\textwidth}
\includegraphics[width=\textwidth]{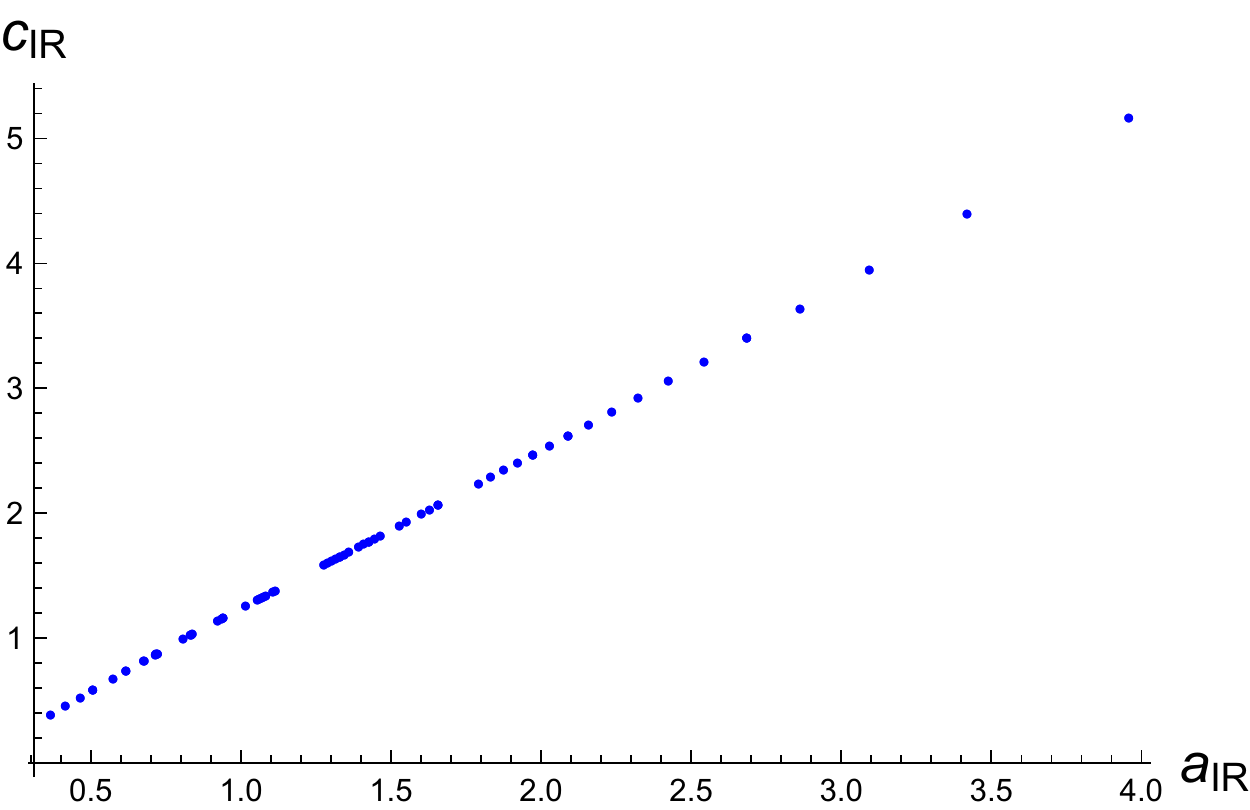}
\caption{$E_8$}
\label{MNratioE8}
\end{subfigure}
\caption{Plots of $c_{\mathrm{IR}}$ vs. $a_{\mathrm{IR}}$ for plain nilpotent
mass deformations of the different probe D3-brane theories.}%
\label{MNratio}%
\end{figure}

As we can see, as $r$ increases, the anomalies
decrease. Whenever an additional deformation is introduced the embedding index
increases. Physically, this translates in a flow to a lower IR theory down the
Hasse diagram of possible RG flows. As a result we expect the degrees of
freedom to decrease, that is $a_{\mathrm{IR}}$ should decrease along this
Hasse diagram. The fact that $a_{\mathrm{IR}}$ is a monotonically decreasing
function of $r$ is an easy consistency check. We also note that the
interacting piece of the anomaly (first value of columns 3) also decreases the
same way.

It is also interesting to note that for a given UV $\mathcal{N} = 2$ fixed point,
the ratio of anomalies $a_{\mathrm{IR}}%
$/$c_{\mathrm{IR}}$ remains roughly constant over the entire nilpotent network. Reference
\cite{Maruyoshi:2018nod} noticed a similar effect.
We also determine the overall statistical spread in the value of the ratio
$a_{\mathrm{IR}}/c_{\mathrm{IR}}$ for plain mass deformations of the probe D3-brane theories.
By inspection of the plots in figure \ref{MNc},
we see that there is a roughly constant value for each theory.
We also calculate the mean and standard deviation by sweeping
over all such theories, the results of which
are shown in table \ref{tab:nilpD3}. Quite remarkably, the standard deviation
is on the order of $1\%$ to $5\%$, indicating a remarkably stable value across
the entire network of flows. Another curious feature is that the mean value of
$a_{\mathrm{IR}}/c_{\mathrm{IR}}$ decreases as we increase to larger flavor symmetries.
Precisely the opposite behavior is observed in the nilpotent networks of
4D\ conformal matter.

\begin{table}[H]
\centering
\begin{tabular}
[c]{|c|c|c|c|c|}\hline
  & $D_{4}$ & $E_{6}$ & $E_{7}$ & $E_{8}$\\\hline
Mean & $0.86$ & $0.83$ & $0.82$ & $0.81$\\\hline
Std. Dev. & $0.03$ & $0.03$ & $0.03$ & $0.03$\\\hline
Max & $0.94$ & $0.94$ & $0.94$ & $0.94$\\\hline
Min & $0.82$ & $0.79$ & $0.78$ & $0.77$\\\hline
\end{tabular}
\caption{Table of means and standard deviations for the ratio $a_{\mathrm{IR}}/c_{\mathrm{IR}}$ across
the entire nilpotent network defined by plain mass deformations of probe
D3-brane theories. We also display the maximum and minimum values.}
\label{tab:nilpD3}
\end{table}

\subsection{Flipper Field Deformations}

Consider next flipper field deformations of the probe D3-brane theories.
As one would expect, we recover the results from \cite{Agarwal:2016pjo}.
In Appendix \ref{completeTables} we present all our results for $D_{4}$
flavor symmetry and only list the values with rational anomalies for the
exceptional flavors $E_{6,7,8}$. Furthermore, we highlight cases where we obtain
known enhancements to $\mathcal{N}=2$ theories such as $H_{0}$, $H_{1}$, and
$H_{2}$ (as already pointed out in \cite{Agarwal:2016pjo}), and we find
an enhancement of the $E_{7}$ Minahan-Nemeschansky
theory to the Argyres-Douglas theory $H_{1}$, in agreement with \cite{Giacomelli:2017ckh, Giacomelli:2018ziv, Carta:2018qke}. 
It is associated with the Bala-Carter label $E_6$ which has embedding index $r = 156$.
In such cases we can compute the embedding index $r_{F}$ of
the residual flavor symmetry and see that not only $a_{\mathrm{IR}}$ and
$c_{\mathrm{IR}}$ match the known values but $k_{\mathrm{IR}} r_{F}$ also
yields the proper value for the flavor central charge of these theories. It is
noteworthy that in those particular cases, the chiral multiplets,
$M_{j_{(a)},-j_{(a)}}$, that survive transform trivially under the residual
flavor symmetry and therefore do not introduce any additional contributions to
the flavor central charge. This is however not true in general.

We also again plot $a_{\mathrm{IR}}$ and $c_{\mathrm{IR}}$ as functions of
the embedding index $r$ for each of the above cases.
\begin{figure}[ptb]
\centering
\begin{subfigure}[b]{0.45\textwidth}
\includegraphics[width=\textwidth]{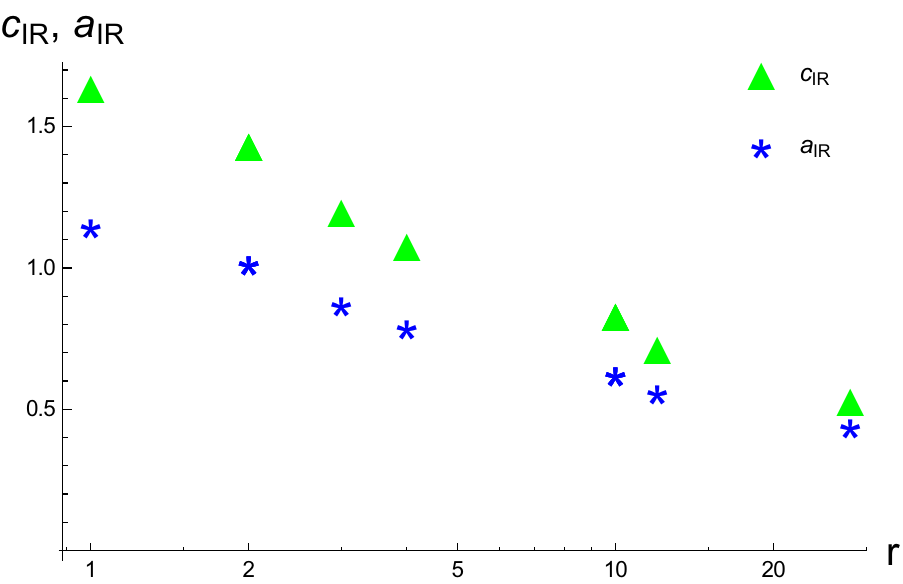}
\caption{$D_4$}
\label{MSMNcD4}
\end{subfigure}
\hspace{12pt} \begin{subfigure}[b]{0.45\textwidth}
\includegraphics[width=\textwidth]{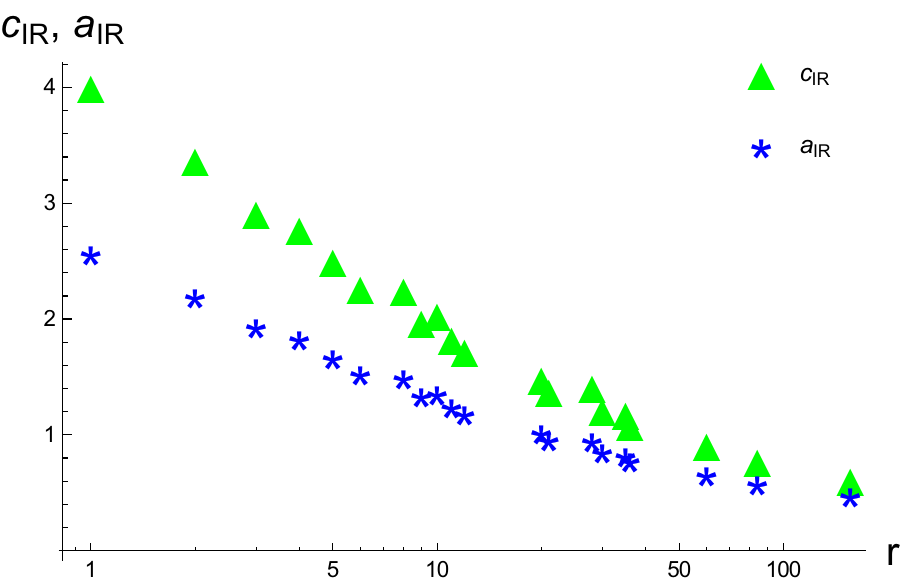}
\caption{$E_6$}
\label{MSMNcE6}
\end{subfigure}
\begin{subfigure}[b]{0.45\textwidth}
\includegraphics[width=\textwidth]{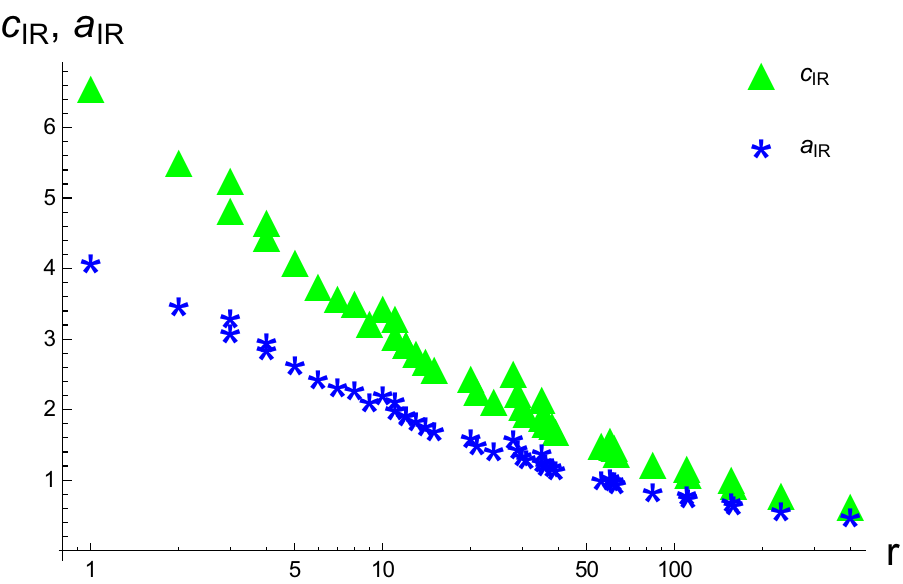}
\caption{$E_7$}
\label{MSMNcE7}
\end{subfigure}
\hspace{12pt} \begin{subfigure}[b]{0.45\textwidth}
\includegraphics[width=\textwidth]{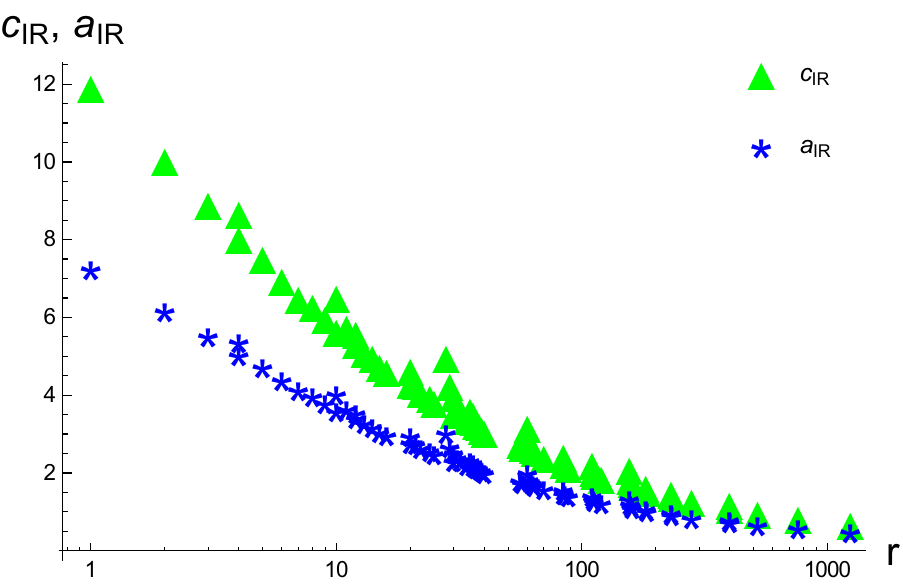}
\caption{$E_8$}
\label{MSMNcE8}
\end{subfigure}
\caption{Plots of $a_{\mathrm{IR}}$ (blue stars) and $c_{\mathrm{IR}}$ (green
triangles) vs embedding index $r$ for the different flipper field
deformations of probe D3-brane theories.}
\label{MSMNc}%
\end{figure}

This time we see that the central charges do not exactly decrease
as the embedding index $r$ increases. However, they do decrease along the
flows defined by the Hasse-diagrams, as expected. Another interesting feature of
these Hasse diagram flows is that the number of flipper field deformations which actually participate in
a flow can vary wildly from orbit to orbit (since the number of $\mathfrak{su}(2)_D$ irreducible representations
also jumps a fair amount). Of course, such fields must be included in computing
various anomalies, even if they serve to decouple mesonic operators which drop below the unitarity bound.
Doing so, we find that $a_{\mathrm{IR}}$ indeed decreases monotonically along a flow.

This raises the question of alternative numerical invariants instead of the embedding index which might be used to order RG~flows in this class of theories. We have chosen the embedding index because this is the quantity which naturally appears in the construction of the infrared R-symmetry (see equations \ref{IRanomalies}). Additionally, it is numerically simple to obtain and often a useful proxy for the ordering of the RG~flows. We are not aware of any other quantity which could provide a better trade off between accuracy and the complexity to compute it. Looking at the Hasse-diagram of the corresponding nilpotent orbits, one would expect that a more accurate description requires more parameters than just one. This would turn the presented plots into higher dimensional ones. For instance, the $x$-axis would need to be replaced by a series of branches corresponding to the full Hasse diagrams. The resulting plots would be much more complex than they need to be. Especially given how closely the embedding index gets to properly ordering the RG~flows. Hence, we continue to rely on this physical parameter rather than try and introduce a less natural quantity.

Finally, another interesting feature of our analysis is that the ratio $a_{\mathrm{IR}}$%
/$c_{\mathrm{IR}}$ is roughly constant for a fixed deformation, given a flavor
symmetry $\mathfrak{g}_{\mathrm{UV}}$ in the UV (see figure \ref{MNMSratio}).
Much as for the plain nilpotent mass deformations,
the overall statistical spread in the value of the ratio
$a_{\mathrm{IR}}/c_{\mathrm{IR}}$ is also remarkably small, and is
on the order of $1\%$ to $5\%$, indicating a remarkably stable value across
the entire network of flows. Another curious feature is that the mean value of
$a_{\mathrm{IR}}/c_{\mathrm{IR}}$ decreases as we increase to larger flavor symmetries.
Precisely the opposite behavior is observed in the nilpotent networks of
4D\ conformal matter. See table \ref{tab:flipD3} for the specific values.

\begin{table}[H]
\centering
\begin{tabular}
[c]{|c|c|c|c|c|}\hline
& $D_{4}$ & $E_{6}$ & $E_{7}$ & $E_{8}$\\\hline
Mean & $0.73$ & $0.69$ & $0.67$ & $0.65$\\\hline
Std. Dev. & $0.04$ & $0.04$ & $0.03$ & $0.03$\\\hline
Max & $0.83$ & $0.78$ & $0.77$ & $0.75$\\\hline
Min & $0.66$ & $0.62$ & $0.6$ & $0.59$\\\hline
\end{tabular}
\caption{Table of means and standard deviations for the ratio $a_{\mathrm{IR}}/c_{\mathrm{IR}}$ across
the entire nilpotent network defined by flipper field deformations of probe
D3-brane theories. We also display the maximum and minimum values.}
\label{tab:flipD3}
\end{table}

\begin{figure}[ptb]
\centering
\begin{subfigure}[b]{0.45\textwidth}
\includegraphics[width=\textwidth]{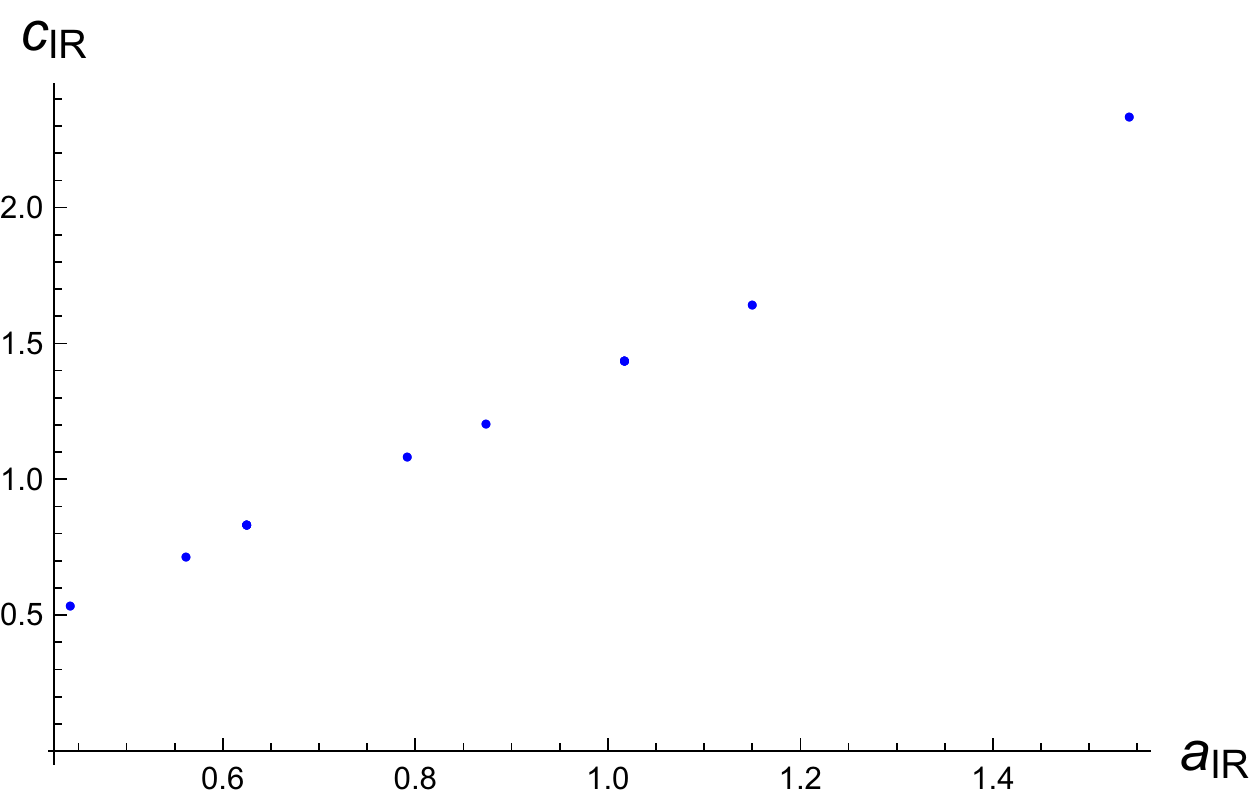}
\caption{$D_4$}
\label{MNMSratioD4}
\end{subfigure}
\hspace{12pt} \begin{subfigure}[b]{0.45\textwidth}
\includegraphics[width=\textwidth]{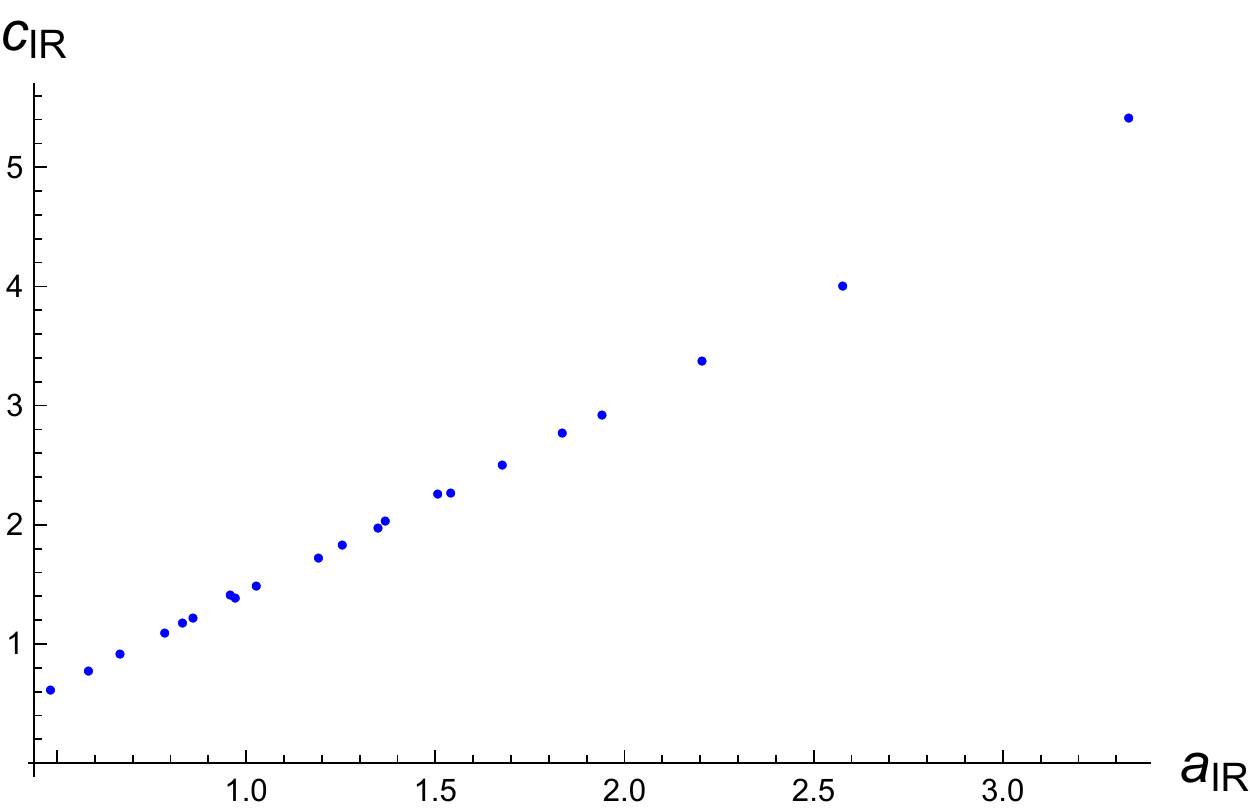}
\caption{$E_6$}
\label{MNMSratioE6}
\end{subfigure}
\begin{subfigure}[b]{0.45\textwidth}
\includegraphics[width=\textwidth]{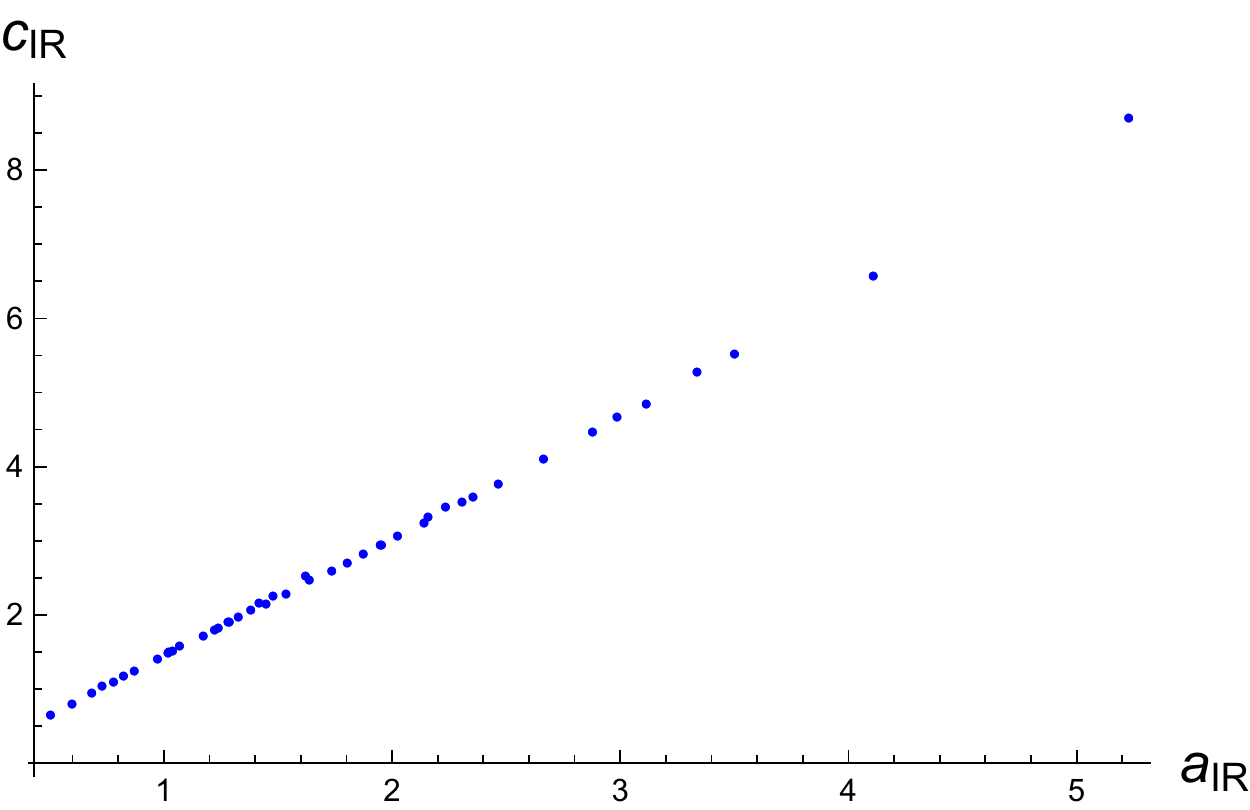}
\caption{$E_7$}
\label{MNMSratioE7}
\end{subfigure}
\hspace{12pt} \begin{subfigure}[b]{0.45\textwidth}
\includegraphics[width=\textwidth]{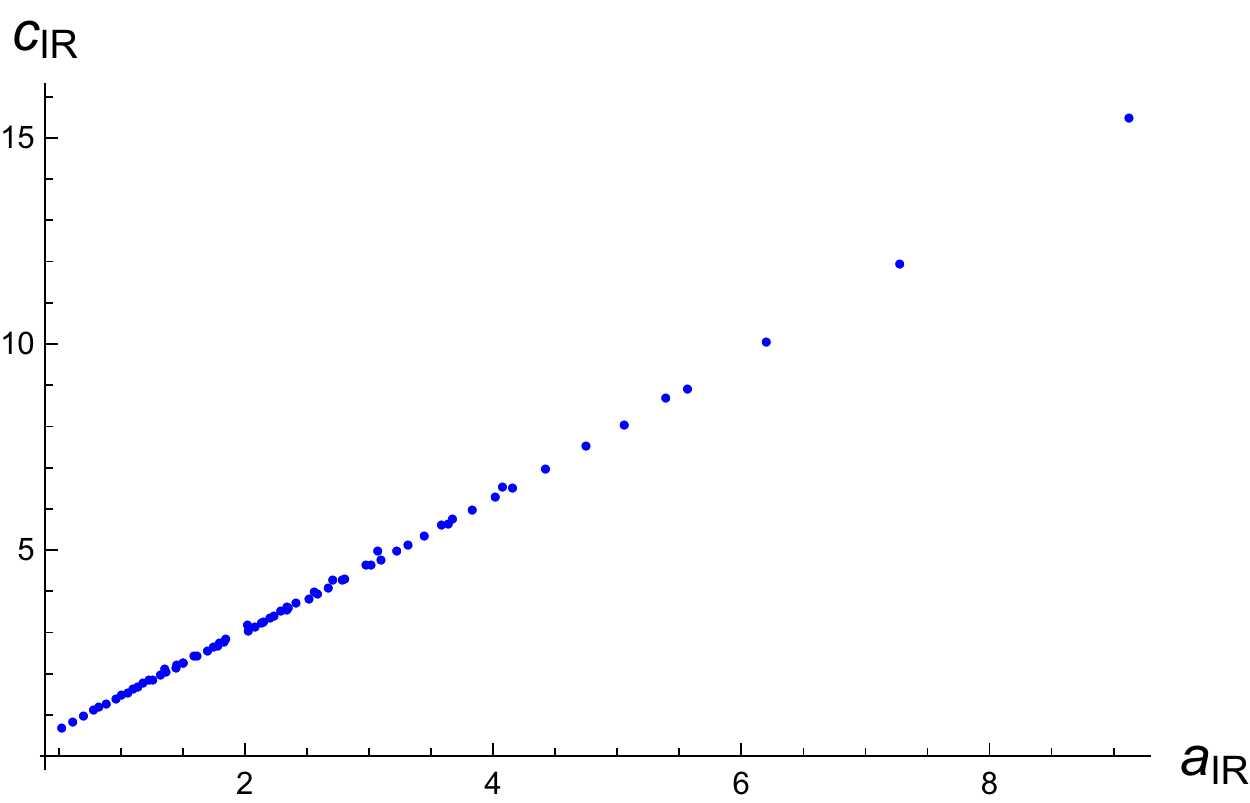}
\caption{$E_8$}
\label{MNMSratioE8}
\end{subfigure}
\caption{Plots of $c_{\mathrm{IR}}$ vs. $a_{\mathrm{IR}}$ for the different
flipper field deformations of probe D3-brane theories.}
\label{MNMSratio}%
\end{figure}

\section{4D Conformal Matter Theories \label{sec:CM}}

In this section we turn to the case of 4D conformal matter theories. In
F-theory terms, these are obtained from a pair of intersecting 7-branes each
with gauge group $G$ which intersect along a common $T^{2}$, namely we have
the compactification of 6D\ conformal matter to an $\mathcal{N}=2$ theory.
Some properties of these theories such as the anomaly polynomial were
determined in \cite{Ohmori:2015pua, Ohmori:2015pia}, and their role as building blocks in
generalized quiver gauge theories was studied in \cite{Apruzzi:2017iqe, Apruzzi:2018oge}.

Now, in this case, the interpretation of the mass parameters is somewhat
different from the D3-brane case. The reason is that the 4D conformal matter
defines a current which couples to the gauge fields of the 7-brane.
More precisely, from the $(0,1)$ connection and the adjoint valued $(2,0)$-form, it
is now the pullback of the $(0,1)$ connection $\mathbb{A}_{(0,1)}$ onto the
$T^{2}$ which actually couples to the 4D conformal matter. A mass deformation
then corresponds to switching on a zero mode for this connection along the
curve. Now in the case where the associated Wilson loop is not unipotent (so
that the zero mode is not nilpotent), this would be an element of the Deligne cohomology
$\mathcal{D}_{2,2}(CY_{4})$ for the associated elliptically fibered Calabi-Yau fourfold
of the F-theory model (see \cite{Aspinwall:1998he} as well as
\cite{Anderson:2013rka}). This can also be viewed as a T-brane deformation of
sorts, because in the limit where the mass parameter is nilpotent, this
deformation is \textquotedblleft invisible\textquotedblright\ in
the associated moduli space problem.\footnote{More precisely, the moduli space
can develop singularities, and as explained in \cite{Anderson:2013rka}, the gauge
theory on the 7-brane serves to complete the moduli space in these singular
limits.} Clearly, it is also natural to promote these background parameters to
a dynamical field, as will happen if we wrap these 7-branes on compact
K\"{a}hler surfaces, and some examples of weakly gauging flavor symmetries in
this way were studied in \cite{Apruzzi:2018oge}. To get a stringy
embedding of the flipper fields, however, we must take a suitable limit
where the gauge fields become IR\ free, but the chiral superfields remain dynamical.

Our plan in the remainder of this section will be to discuss some further
aspects of these conformal matter theories. We begin by reviewing some aspects
of the original $\mathcal{N}=2$ theories, and then turn to an analysis of the
resulting nilpotent network of $\mathcal{N} = 1$ fixed points. When we turn 
to the plots and statistics for these networks, we treat the nilpotent orbit with 
$G_L \leftrightarrow G_R$ interchanged as distinct.

\subsection{Summary of UV $\mathcal{N}=2$ Fixed Points}

We now review some aspects of $\mathcal{N} = 2$ $(G,G)$
4D conformal matter obtained from compactification of
$(G,G)$ 6D conformal matter on a $T^2$. We present in
table \ref{CMUVscaling} the values for the central
charges and flavor symmetries, together with the dimensions and multiplicities
of the Coulomb branch operators. We give further details on how those results
are obtained in Appendix~\ref{CMappendix}.

\begin{table}[h]
\resizebox{\textwidth}{!}{
\centering
{\renewcommand{\arraystretch}{1.2}
\begin{tabular}
[c]{|c||c|c|c|c|}\hline
$(G_L,G_R)$ & $(D_k , D_k)$ & $(E_{6},E_6)$ & $(E_{7},E_7)$ & $(E_{8},E_8)$\\\hline\hline
$a_{\mathrm{UV}}$ & $\frac{1}{24}\left(  k(14k-19)-53\right)  $ & $\frac
{613}{24}$ & $\frac{817}{12}$ & $\frac{1745}{8}$\\\hline
$c_{\mathrm{UV}}$ & $\frac{1}{6}\left(  k(4k-5)-13\right)  $ & $\frac{173}{6}$
& $\frac{442}{6}$ & $\frac{457}{2}$\\\hline
$k^{flav}_{L},k^{flav}_{R}$ & $4k-4$ & 24 & 36 & 60\\\hline
$\Delta(Z_{i})$ & $\{6_{1},...,(2k-2)_{1}\}$ &
\parbox{2cm}{$\{6_1,8_1,9_1,$ $12_2\}$} &
\parbox{2.5cm}{$\{6_1,8_1,10_1,$ $12_2,14_2,18_3\}$} &
\parbox{3cm}{$\{6_1,8_1,12_2,14_2,$ $18_3,20_3,24_4,30_5\}$}\\\hline
\end{tabular}
}}\caption{Anomalies and scaling dimensions for 4D $\mathcal{N} = 2$ $(G,G)$
conformal matter. In the last row, the subscripts are the multiplicities, i.e. the number of Coulomb
branch operators with that specific scaling dimension.}%
\label{CMUVscaling}%
\end{table}

The dimension of the Coulomb branch for the different conformal matter
theories on $T^{2}$ are
\begin{align}
\mathrm{dim}_{\mathbb{C}}\left(  \mathrm{Coul}\left[  (D_k,D_k) \right]  \right)   &  =k-3,\\
\mathrm{dim}_{\mathbb{C}}\left(  \mathrm{Coul}\left[  (E_{6},E_6) \right]  \right)
&  =5,\\
\mathrm{dim}_{\mathbb{C}}\left(  \mathrm{Coul}\left[  (E_{7},E_7)\right]  \right)
&  =10,\\
\mathrm{dim}_{\mathbb{C}}\left(  \mathrm{Coul}\left[  (E_{8},E_8)\right]  \right)
&  =21.
\end{align}
which matches the expectation from 6D \cite{DelZotto:2014hpa}:
\begin{equation}
\mathrm{dim}_{\mathbb{C}}\left(  \mathrm{Coul}\left[  G\right]  \right)
)=h_{G}^{\vee}-r_{G}-1\,,
\end{equation}
where $r_{G}$ is the rank of $G$ and $h_{G}^{\vee}$ is
the dual Coxeter number of $G$. In order to extract the dimensions of the
Coulomb branch operators for the different conformal matter theories, we read
off the scaling dimension of the deformations from the mirror geometries of the
elliptic threefold of the F-theory geometry. The mirror
geometries for $(E_n,E_n)$ theories were provided in
\cite{DelZotto:2015rca} and the $(D_k, D_k)$ case can be obtained from
the curve in equation (5.4) of reference \cite{Ohmori:2015pua}.

\subsection{Plain Mass Deformations}

The computations for conformal matter follow the
general procedure outlined in previous sections.
We now have two flavor groups, so two nilpotent orbits labeled
by corresponding Bala-Carter labels. Each one comes with an embedding index
$r_{L}$ and $r_{R}$.

We have actually already encountered the ($D_{4}$, $D_{4}$) 4D conformal matter theory: it is
simply the rank one $E_8$ Minahan-Nemeschansky theory (it can still be
accessed with the code described in Appendix~\ref{completeTables}). It mainly serves as a
cross-check on the general procedure, and we find perfect agreement for those
deformations which live in an $\mathfrak{so}(8) \times \mathfrak{so}(8)$ subalgebra.
Thus, we simply list in Appendix \ref{completeTables} the rational theories
in the case where the parent 4D conformal matter theory has exceptional flavor symmetry.
Due to their large size the tables are also split
in their length. The top half contains the Bala-Carter labels, embedding
indices, anomalies and $t_{*}$. The bottom half repeats the Bala-Carter labels
and $t_{*}$ before providing scaling dimensions. Finally, the tables for the
flavor central charges are too large to include here. So, we refer the reader
to the companion \texttt{Mathematica} code for those results.

We also provide contour plots of $a_{\mathrm{IR}}$ vs. the embedding indices
of the right and left flavors. We hasten to add that while the partial
ordering of nilpotent orbits enforces a corresponding ordering for the associated
embedding indices, the converse is not true (the Hasse diagram has more fine structure).
This is an unfortunate artifact of displaying all of our data with respect to a two-dimensional
contour plot.
\begin{figure}[ptb]
\centering
\begin{subfigure}[b]{0.495\textwidth}
\includegraphics[width=\textwidth]{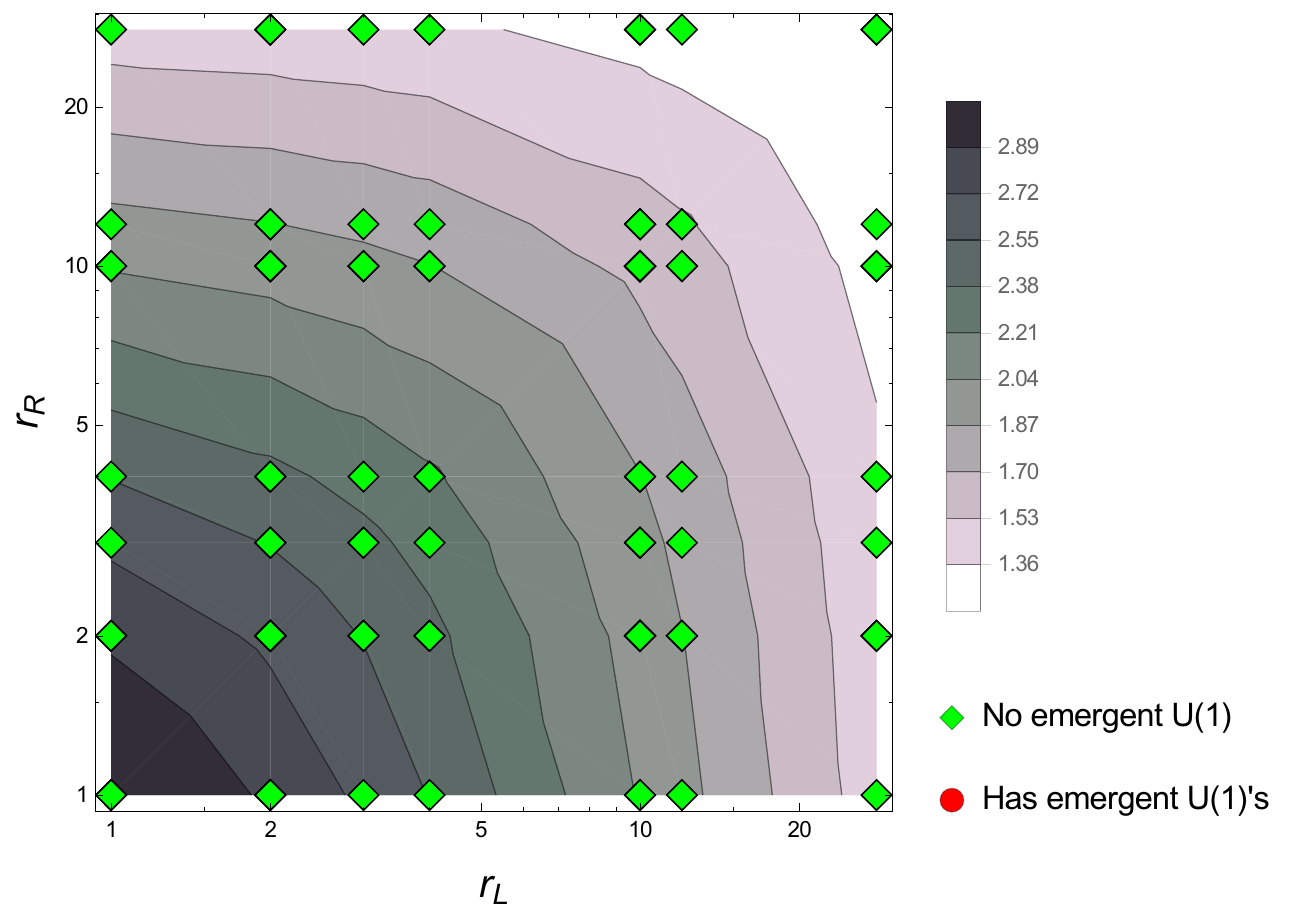}
\caption{($D_4$, $D_4$)}
\label{CMcontourD4}
\end{subfigure}
\begin{subfigure}[b]{0.495\textwidth}
\includegraphics[width=\textwidth]{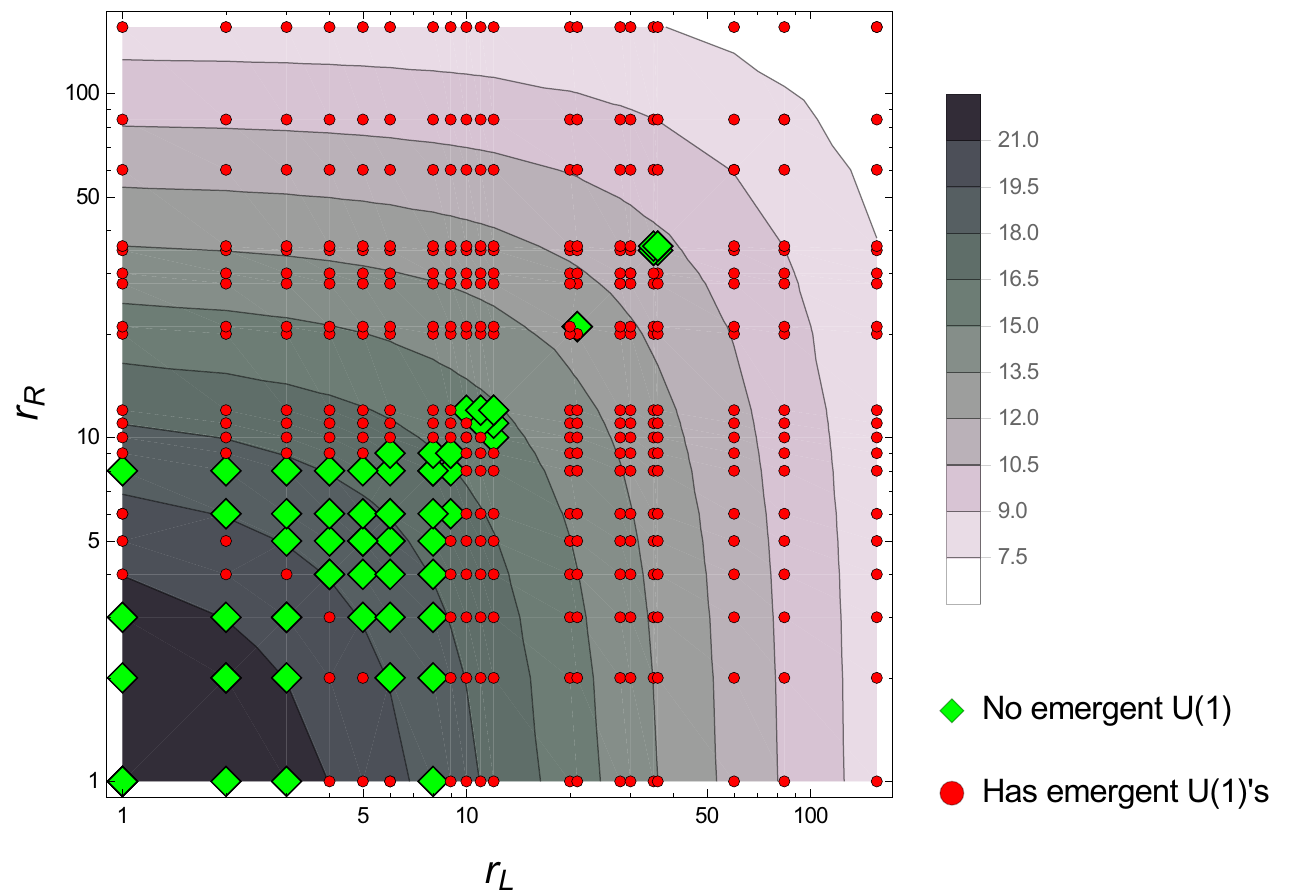}
\caption{($E_6$, $E_6$)}
\label{CMcontourE6}
\end{subfigure}
\begin{subfigure}[b]{0.495\textwidth}
\includegraphics[width=\textwidth]{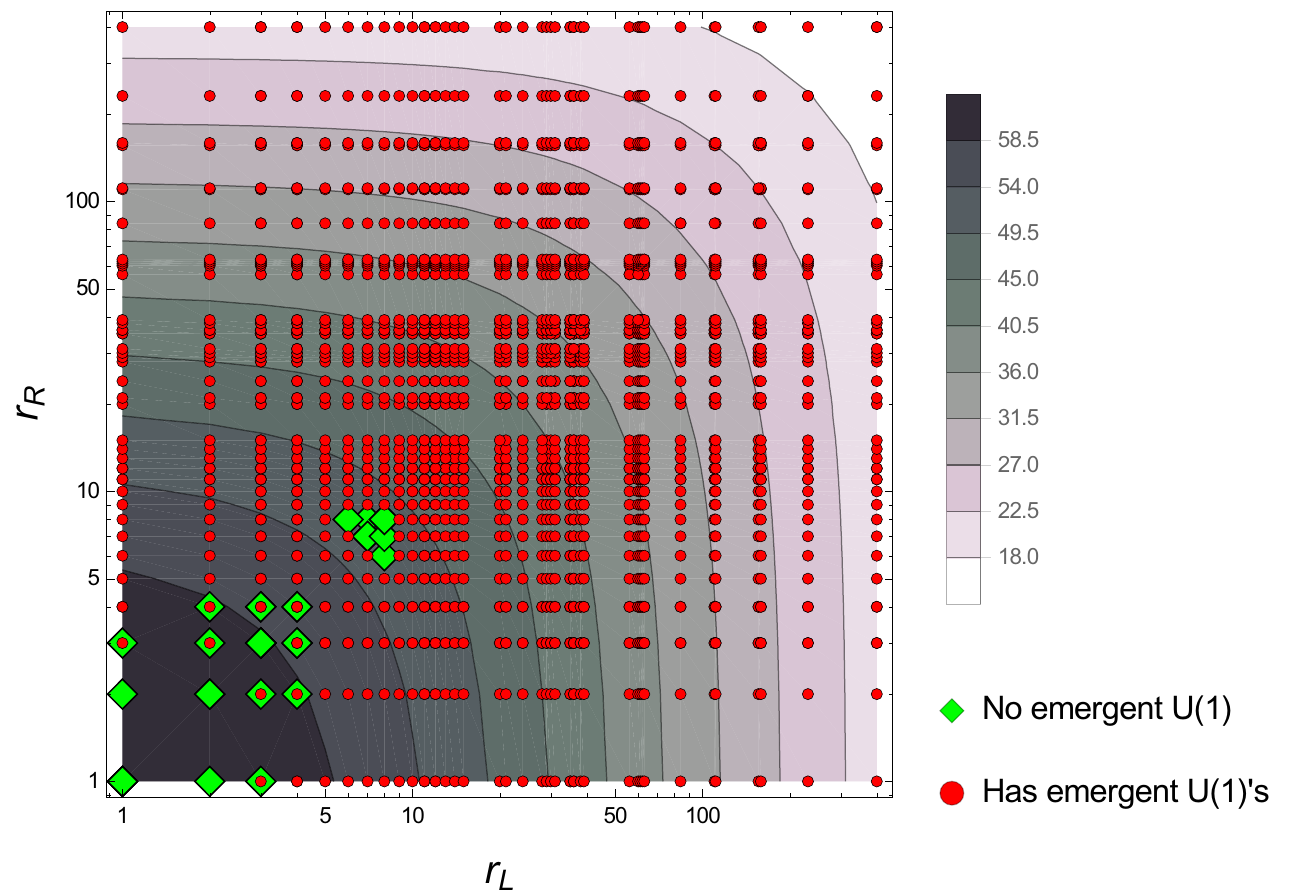}
\caption{($E_7$, $E_7$)}
\label{CMcontourE7}
\end{subfigure}
\begin{subfigure}[b]{0.495\textwidth}
\includegraphics[width=\textwidth]{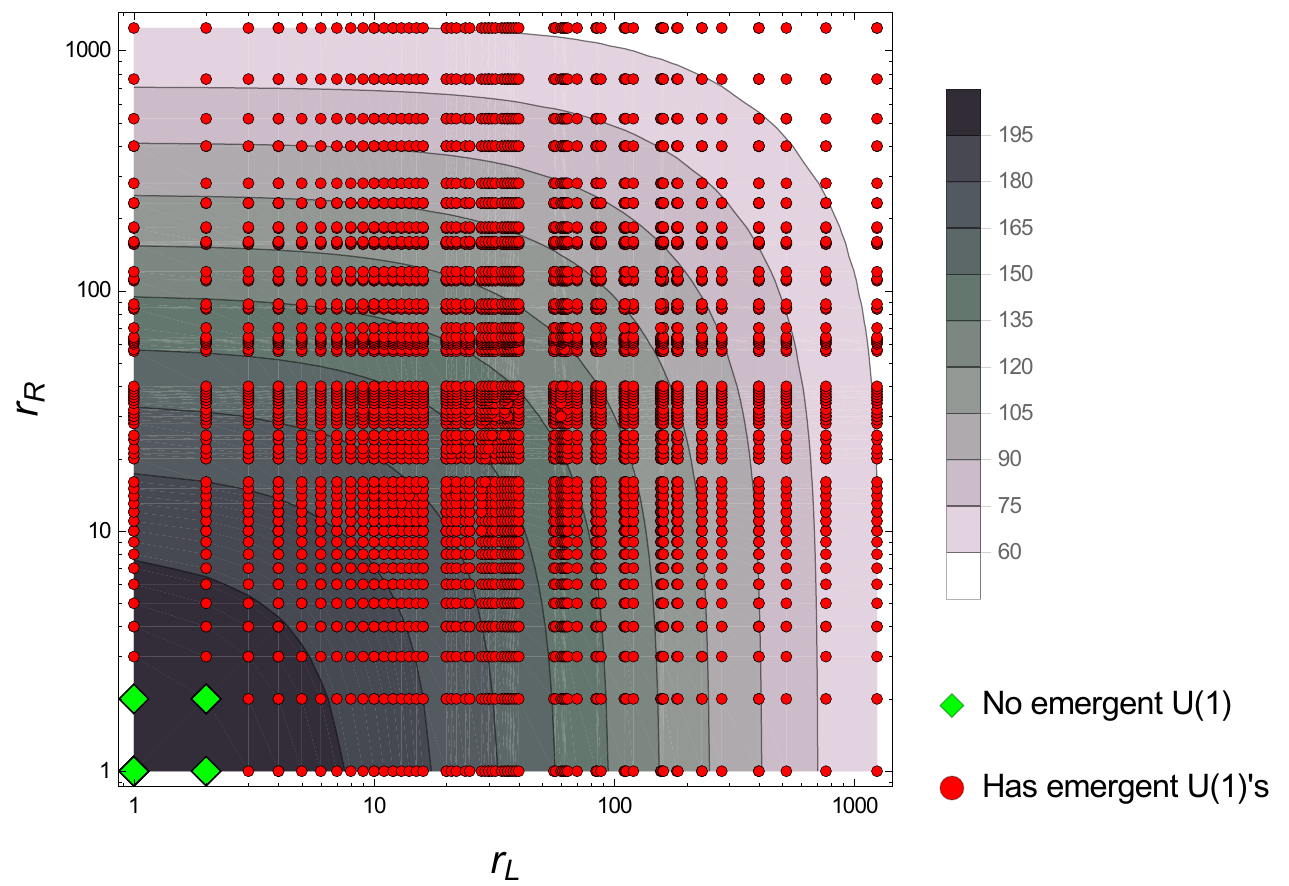}
\caption{($E_8$, $E_8$)}
\label{CMcontourE8}
\end{subfigure}
\caption{Plots of $a_{\mathrm{IR}}$ vs left and right embedding indices for the
different plain mass nilpotent
deformations of 4D conformal matter theories. The contour plots are obtained by
extrapolating between the actual data points which are labeled in green diamonds and
red circles. The green diamonds correspond to deformations where all operators remain
above the unitarity bound and no emergent $U(1)$ appears. The red circles
correspond to deformations where some operators hit the unitarity bound and
emergent $U(1)$'s are present. We emphasize that sometimes different nilpotent orbits can have the
same embedding index. A log-scale is used to spread the dense region
at low values of the embedding indices.}%
\label{CMcontour}%
\end{figure}
Of course, the plots (just like the tables) are symmetric under
the interchange of $r_{L}$ with $r_{R}$. We also see that for any fixed value
of $r_{L}$ the value of $a_{\mathrm{IR}}$ decreases as the deformation on the
right increases (along the Hasse diagram) when the interacting piece plus free
decoupled fields are considered, as well as when central charges of only the
interacting piece are analyzed (the plots for only the interacting piece would
look very similar).

Furthermore, if we simultaneously increase both $r_{L}$ and $r_{R}$ while
keeping $r_{L}=r_{R}$ (along the Hasse diagram), then $a_{\mathrm{IR}}$ monotonically decreases. This is
again consistent with the expectation that the number of degrees of freedom should
decrease as the deformations becomes larger along the RG flows.

Another interesting feature of our numerical sweep is that we sometimes encounter theories where an operator
decouples, but further down the Hasse diagram, we see no apparent unitarity bound violations. This does
not contradict the general structure implied by the nilpotent cone, since deeper down in the Hasse diagram it
often happens that the top spin operator of $\mathfrak{su}(2)_D$ may not be a top-spin operator deeper down in the nilpotent
cone. As we have already explained, the lower spin operators are trivial in the chiral ring of the IR fixed point, so
it is neither here nor there to see a jump in the number of emergent $U(1)$'s as we proceed deeper into the IR.

We also determine the overall statistical spread in the value of the ratio
$a_{\mathrm{IR}}/c_{\mathrm{IR}}$ for plain mass deformations of the probe D3-brane theories.
By inspection of the plots in figure \ref{CMratio},
we see that there is a roughly constant value for each theory. We
also calculate the mean and standard deviation by sweeping
over all such theories. Just as in the case of the probe D3-brane theories,
we find that the standard deviation
is on the order of $1\%$ to $5\%$, indicating a remarkably stable value across
the entire network of flows. The specific values are displayed in
table \ref{tab:plainCM}. Another curious feature is that the mean value of
$a_{\mathrm{IR}}/c_{\mathrm{IR}}$ increases as we go to larger UV flavor symmetries.
Precisely the opposite behavior is observed in the nilpotent networks of
probe D3-brane theories.

\begin{table}[H]
\centering
\begin{tabular}
[c]{|c|c|c|c|c|}\hline
& $(D_{4},D_{4})$ & $(E_{6},E_{6})$ & $(E_{7},E_{7})$ & $(E_{8},E_{8}%
)$\\\hline
Mean & $0.80$ & $0.91$ & $0.94$ & $0.97$\\\hline
Std. Dev. & $0.01$ & $0.01$ & $0.004$ & $0.003$\\\hline
Max & $0.81$ & $0.91$ & $0.95$ & $0.98$\\\hline
Min & $0.77$ & $0.89$ & $0.92$ & $0.96$\\\hline
\end{tabular}
\caption{Table of means and standard deviations for the ratio $a_{\mathrm{IR}}/c_{\mathrm{IR}}$ across
the entire nilpotent network defined by plain mass deformations of 4D conformal matter.
We also display the maximum and minimum values.}
\label{tab:plainCM}
\end{table}

\begin{figure}[ptb]
\centering
\begin{subfigure}[b]{0.45\textwidth}
\includegraphics[width=\textwidth]{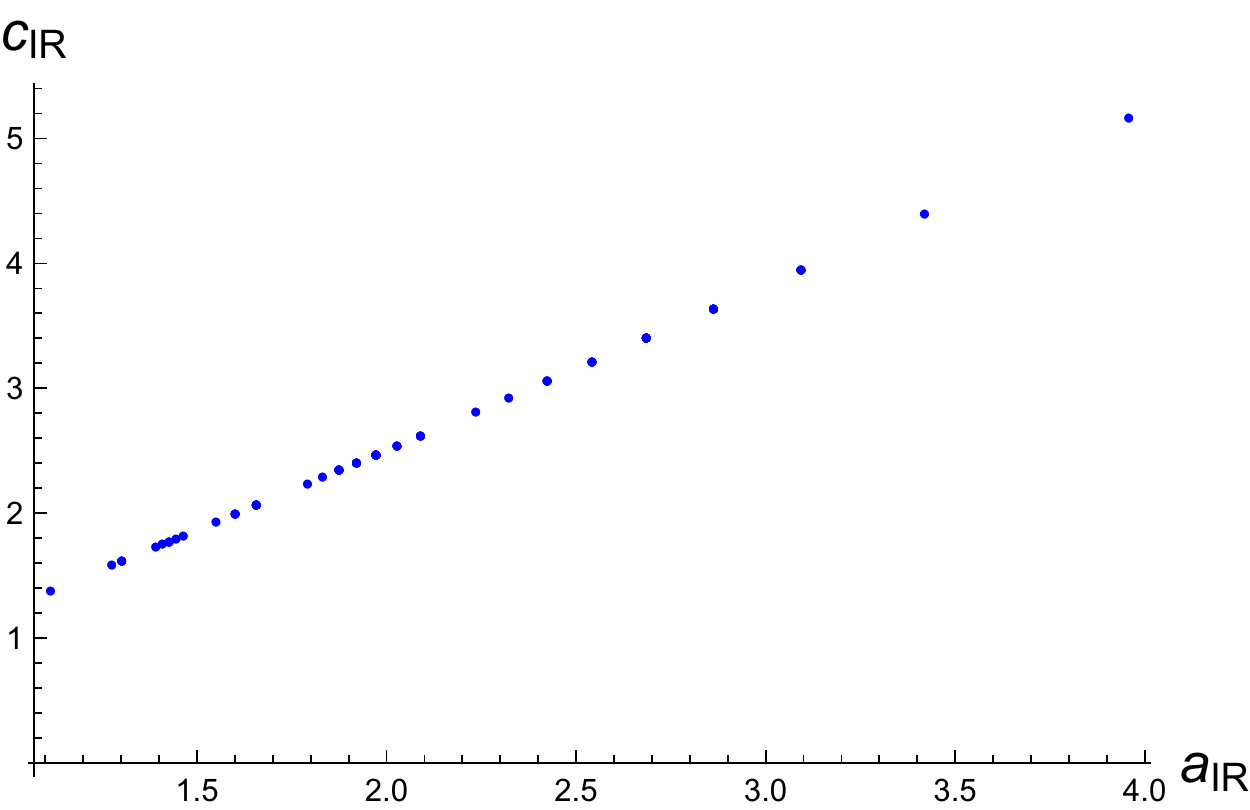}
\caption{$(D_4 , D_4)$}
\label{CMratioD4}
\end{subfigure}
\hspace{12pt} \begin{subfigure}[b]{0.45\textwidth}
\includegraphics[width=\textwidth]{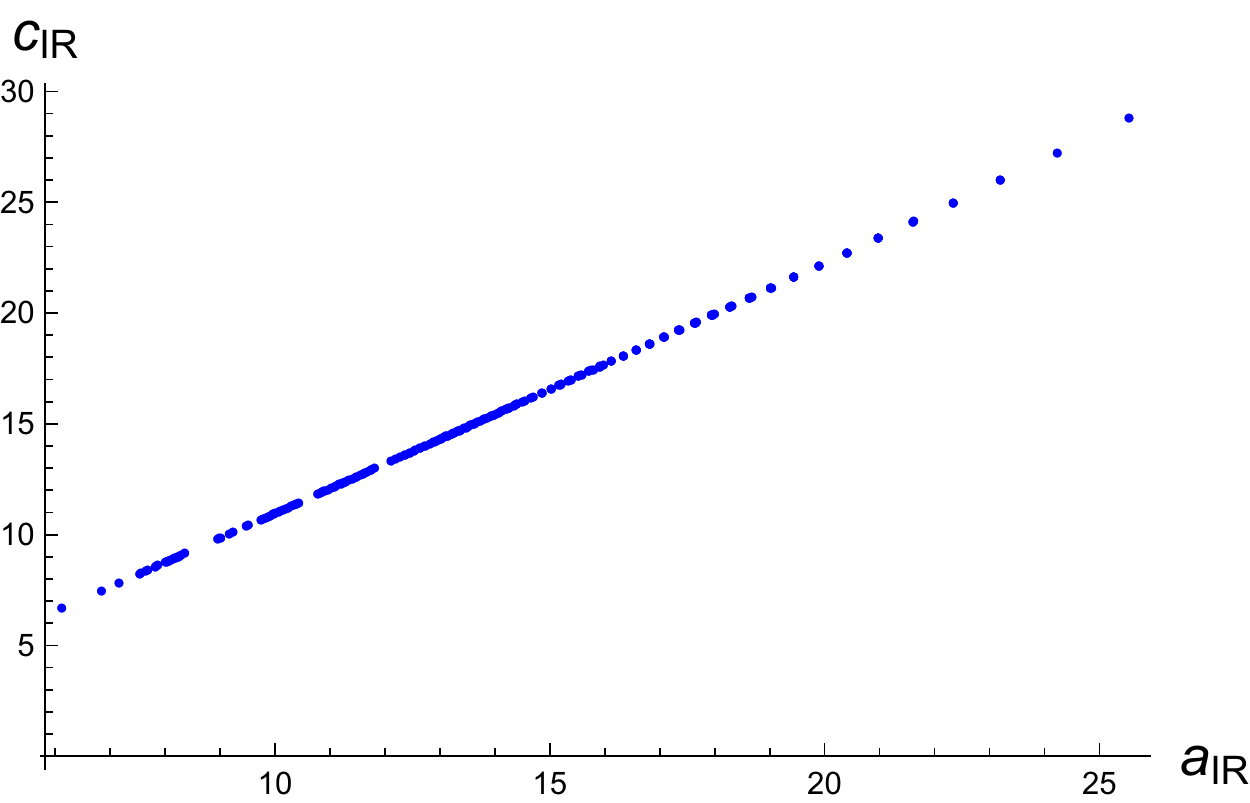}
\caption{$(E_6 , E_6)$}
\label{CMratioE6}
\end{subfigure}
\begin{subfigure}[b]{0.45\textwidth}
\includegraphics[width=\textwidth]{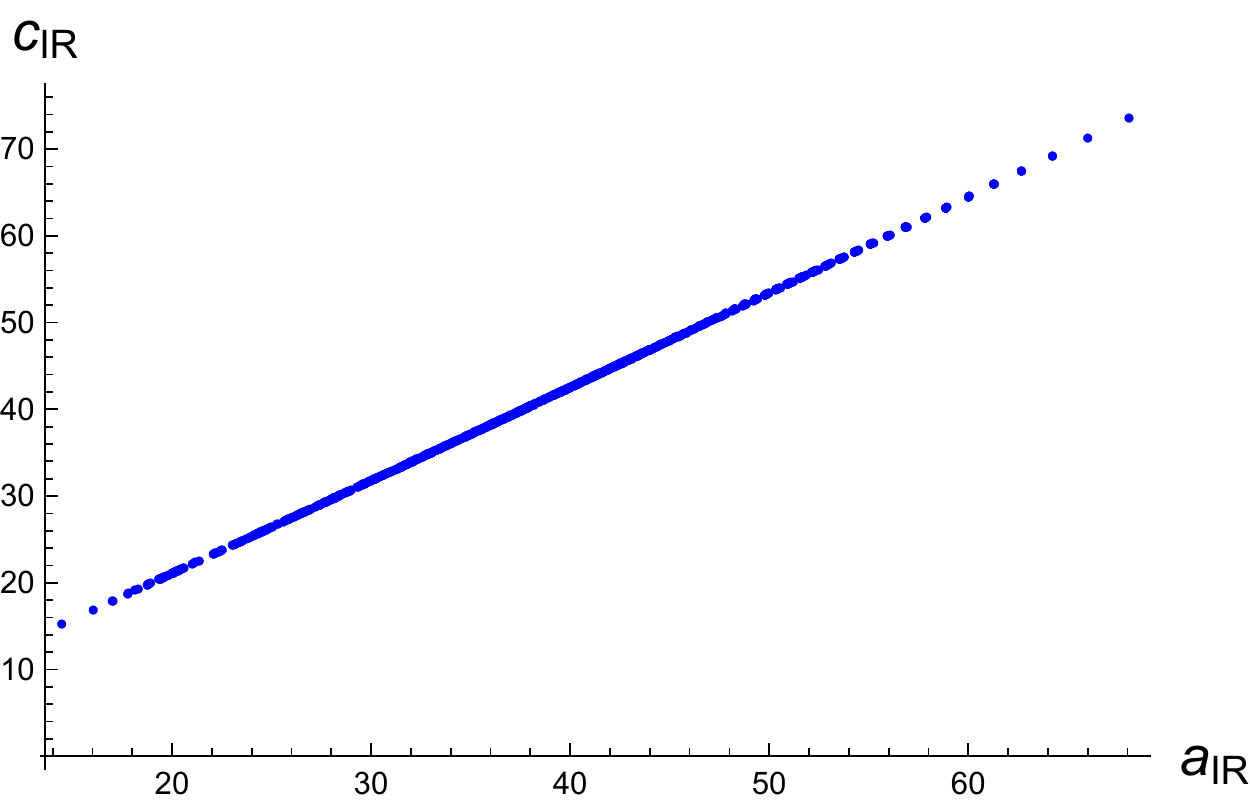}
\caption{$(E_7 , E_7)$}
\label{CMratioE7}
\end{subfigure}
\hspace{12pt} \begin{subfigure}[b]{0.45\textwidth}
\includegraphics[width=\textwidth]{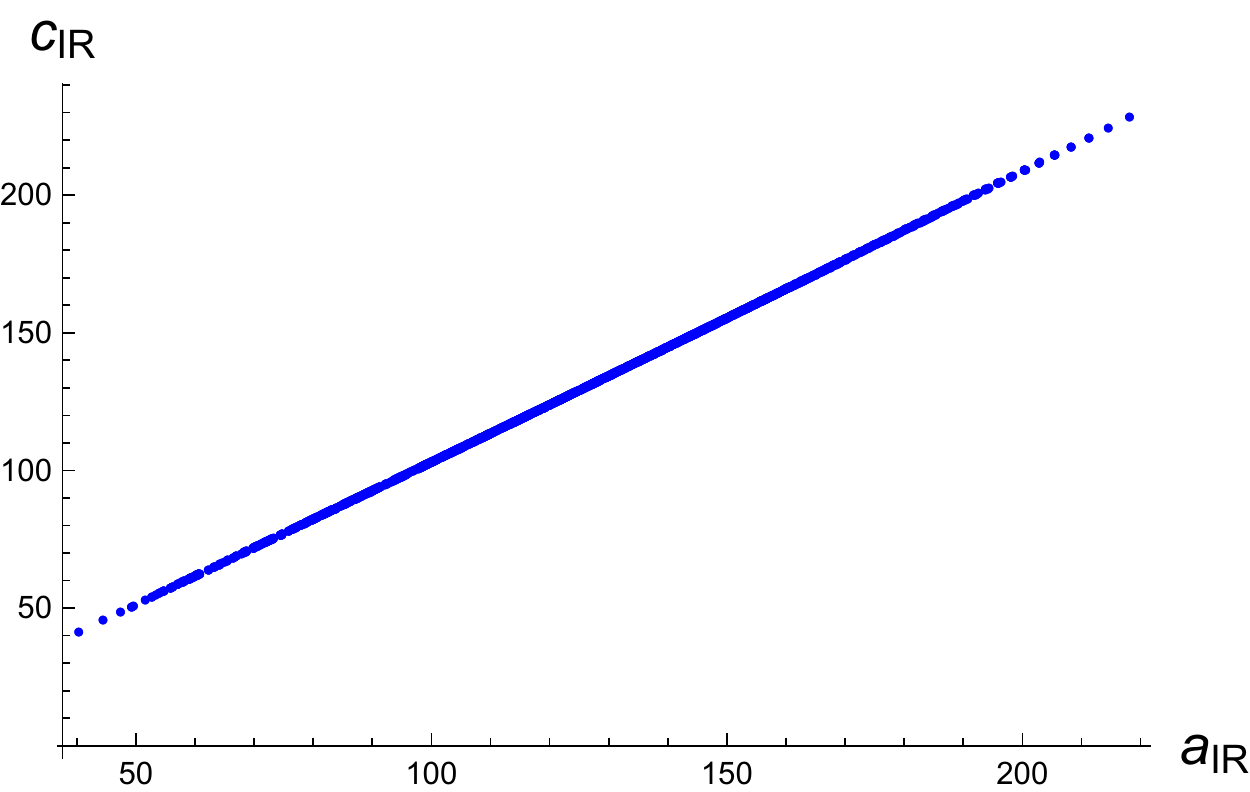}
\caption{$(E_8 , E_8)$}
\label{CMratioE8}
\end{subfigure}
\caption{Plots of $c_{\mathrm{IR}}$ vs $a_{\mathrm{IR}}$ for the different plain mass nilpotent
deformations of 4D conformal matter theories.}%
\label{CMratio}%
\end{figure}

\subsection{Flipper Field Deformations}

Finally, we come to flipper field deformations of conformal matter. The analysis is
simplified by the fact that we do not need to worry about mesons decoupling since they are automatically
rescued (if they drop below the unitarity bound) by the flipper fields $M$ to which they couple.

As before, the results with rational values are tabulated in Appendix \ref{completeTables},
and more general deformations can be accessed via the \texttt{Mathematica} code.

Finally, we provide contour plots of $a_{\mathrm{IR}}$ vs. the left and
right embedding indices $r_{L}$ and $r_{R}$. Again, we emphasize that what
really needs to be monotonic is the flow down the Hasse diagram, which in most cases (though not all)
aligns with the increase of the embedding indices $r_L$ and $r_R$. Quite remarkably,
even this coarse data based on the embedding indices (though there are a few exceptions)
usually is enough to establish monotonicity.

\begin{figure}[ptb]
\centering
\begin{subfigure}[b]{0.475\textwidth}
\includegraphics[width=\textwidth]{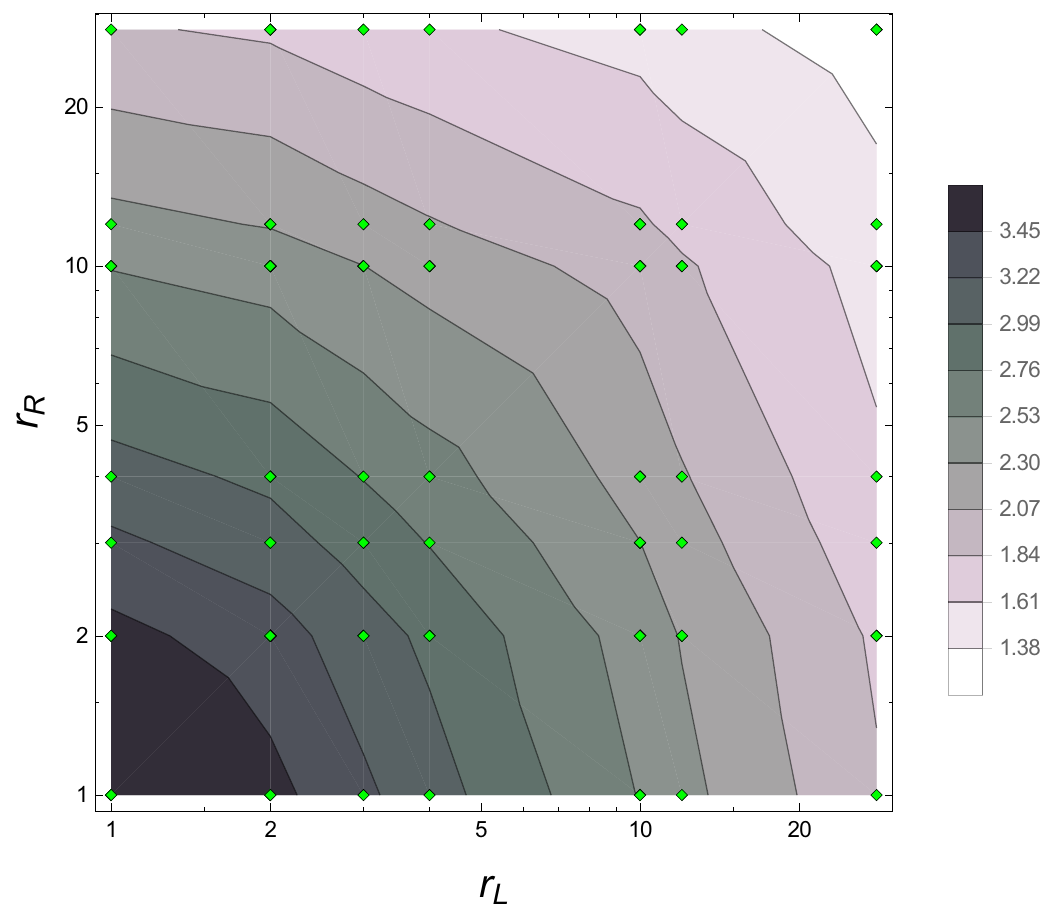}
\caption{($D_4$, $D_4$)}
\label{MSCMcontourD4}
\end{subfigure}
\hspace{12pt} \begin{subfigure}[b]{0.475\textwidth}
\includegraphics[width=\textwidth]{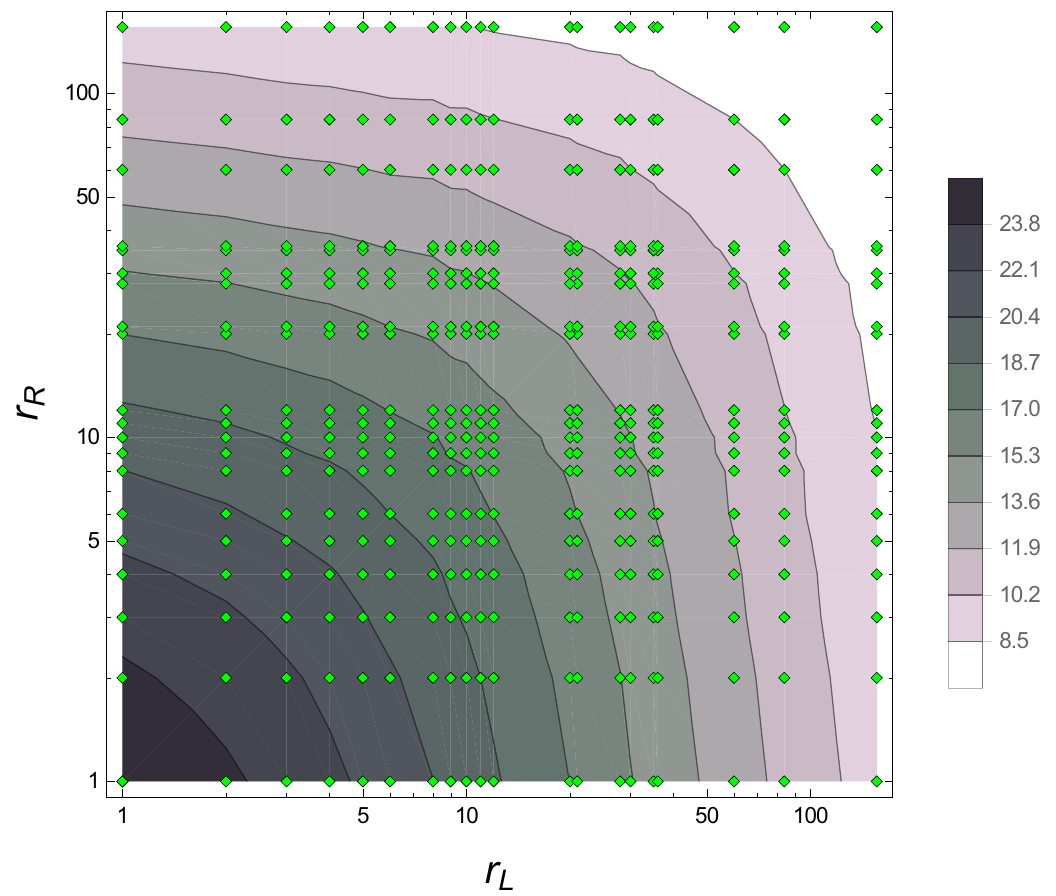}
\caption{($E_6$, $E_6$)}
\label{MSCMcontourE6}
\end{subfigure}
\par
\bigskip
\par
\bigskip\begin{subfigure}[b]{0.475\textwidth}
\includegraphics[width=\textwidth]{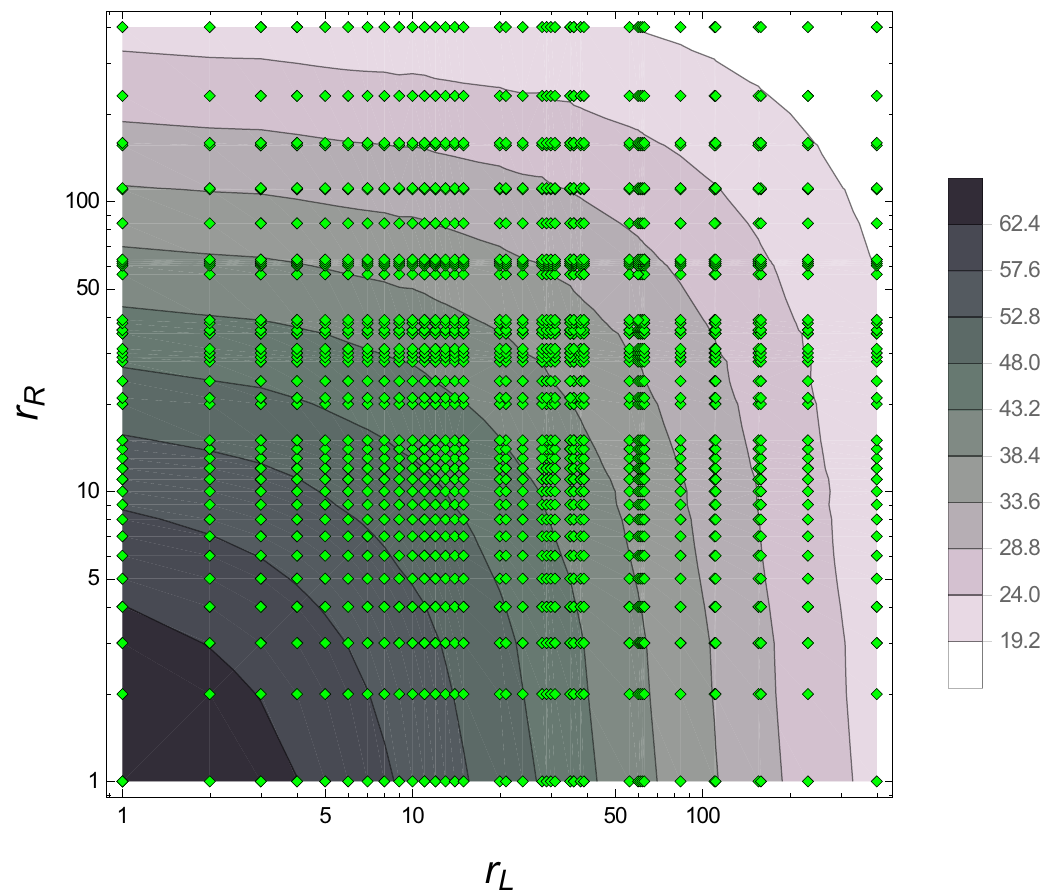}
\caption{($E_7$, $E_7$)}
\label{MSCMcontourE7}
\end{subfigure}
\hspace{12pt} \begin{subfigure}[b]{0.475\textwidth}
\includegraphics[width=\textwidth]{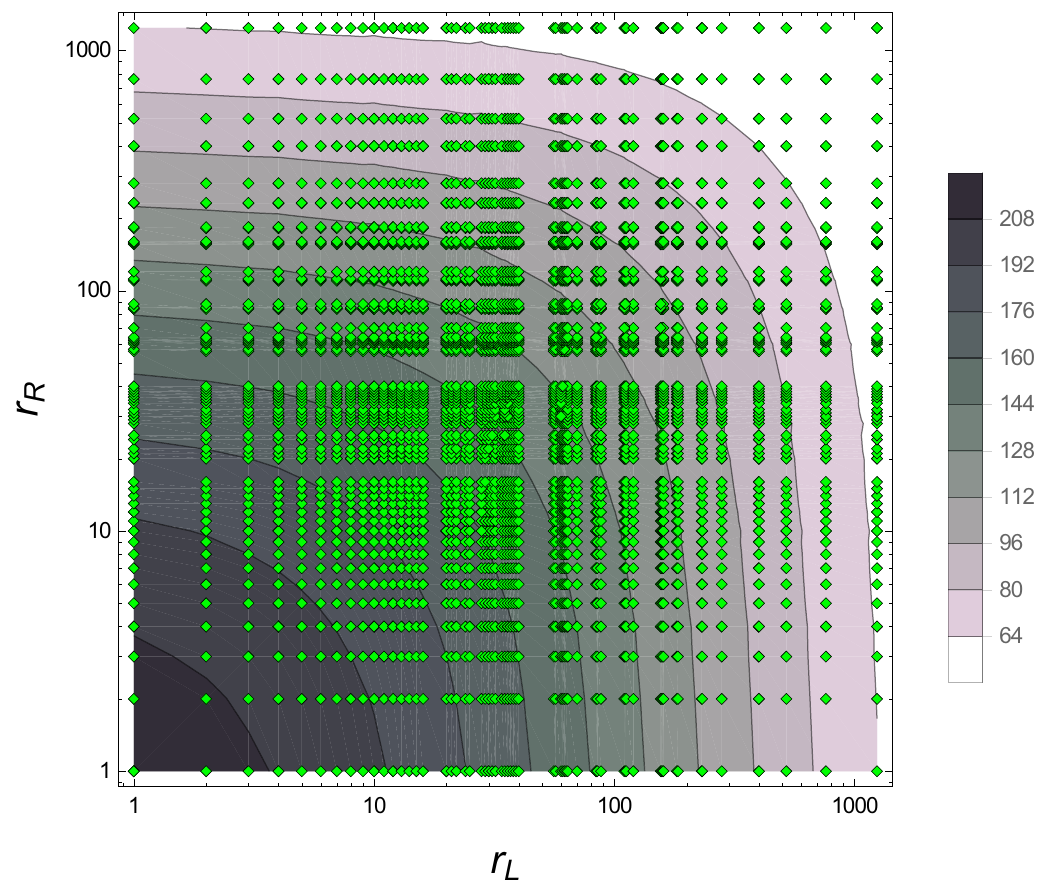}
\caption{($E_8$, $E_8$)}
\label{MSCMcontourE8}
\end{subfigure}
\caption{Plots of $a_{\mathrm{IR}}$ vs left and right embedding indices for the
different conformal matter theories, with flipper field deformations. The
contour plots are obtain by extrapolating between the actual data points which
are labeled in green.}%
\label{MSCMcontour}%
\end{figure}

We also determine the overall statistical spread in the value of the ratio
$a_{\mathrm{IR}}/c_{\mathrm{IR}}$ for flipper field deformations of 4D conformal matter.
By inspection of the plots in figure \ref{CMMSratio},
we see that there is a roughly constant value for each theory. We also
calculate the mean and standard deviation by sweeping
over all such theories, displaying the results
in table \ref{tab:flipCM}. As in all the other cases we have considered, the standard deviation
is on the order of $1\%$ to $5\%$, indicating a remarkably stable value across
the entire network of flows. Another curious feature is that the mean value of
$a_{\mathrm{IR}}/c_{\mathrm{IR}}$ increases as we increase to larger flavor symmetries.
Precisely the opposite behavior is observed in the nilpotent networks of
probe D3-brane theories.

\begin{table}[H]
\centering
\begin{tabular}
[c]{|c|c|c|c|c|}\hline
& $(D_{4},D_{4})$ & $(E_{6},E_{6})$ & $(E_{7},E_{7})$ & $(E_{8},E_{8}%
)$\\\hline
Mean & $0.73$ & $0.87$ & $0.92$ & $0.96$\\\hline
Std. Dev. & $0.01$ & $0.02$ & $0.01$ & $0.01$\\\hline
Max & $0.80$ & $0.91$ & $0.95$ & $0.98$\\\hline
Min & $0.68$ & $0.81$ & $0.86$ & $0.91$\\\hline
\end{tabular}
\caption{Table of means and standard deviations for the ratio $a_{\mathrm{IR}}/c_{\mathrm{IR}}$ across
the entire nilpotent network defined by flipper field deformations of 4D conformal matter.
We also display the maximum and minimum values.}
\label{tab:flipCM}
\end{table}

\begin{figure}[ptb]
\centering
\begin{subfigure}[b]{0.45\textwidth}
\includegraphics[width=\textwidth]{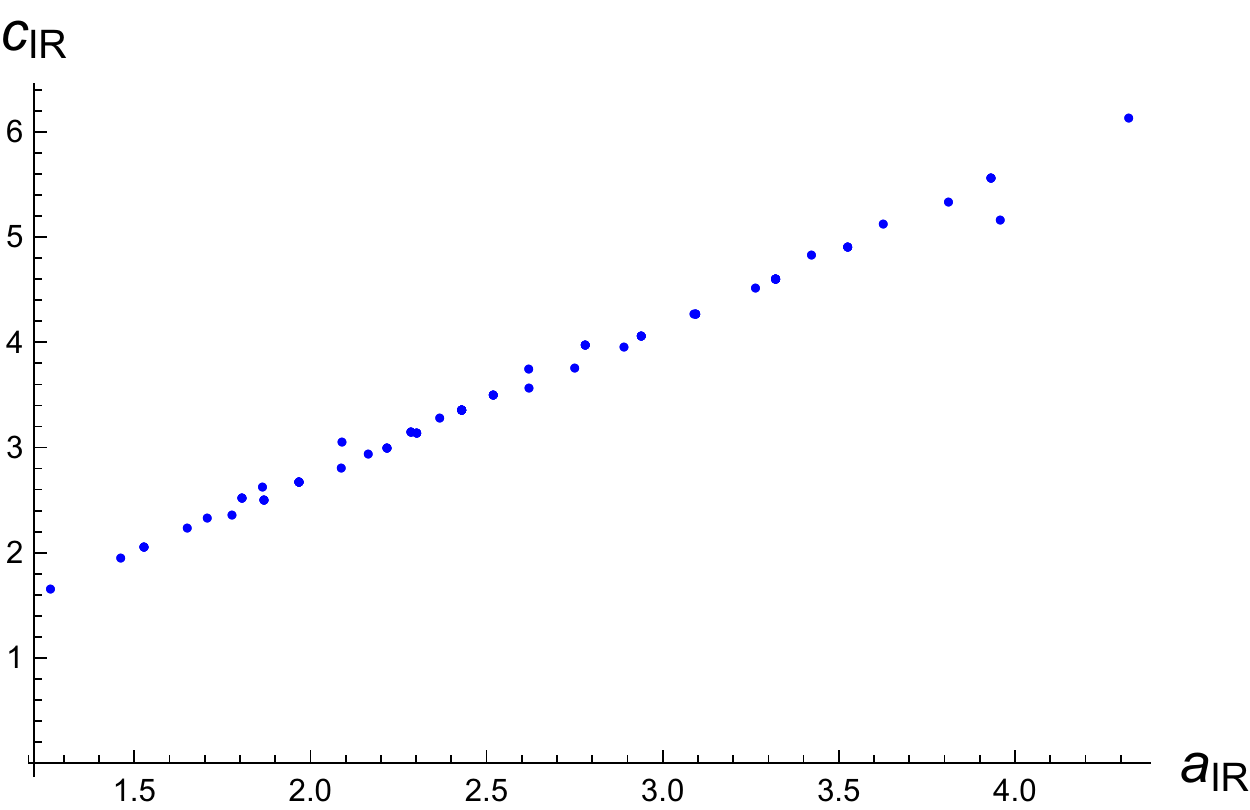}
\caption{$(D_4,D_4)$}
\label{CMMSratioD4}
\end{subfigure}
\hspace{12pt} \begin{subfigure}[b]{0.45\textwidth}
\includegraphics[width=\textwidth]{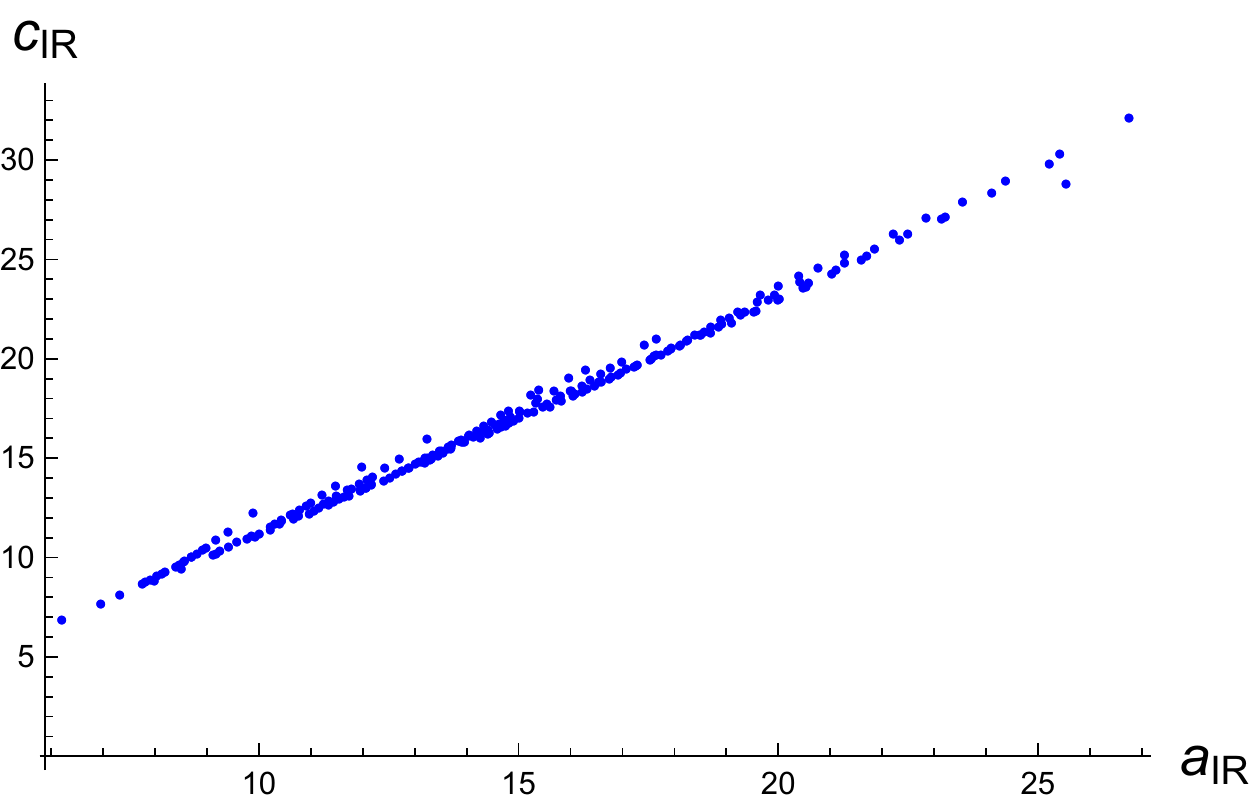}
\caption{$(E_6,E_6)$}
\label{CMMSratioE6}
\end{subfigure}
\begin{subfigure}[b]{0.45\textwidth}
\includegraphics[width=\textwidth]{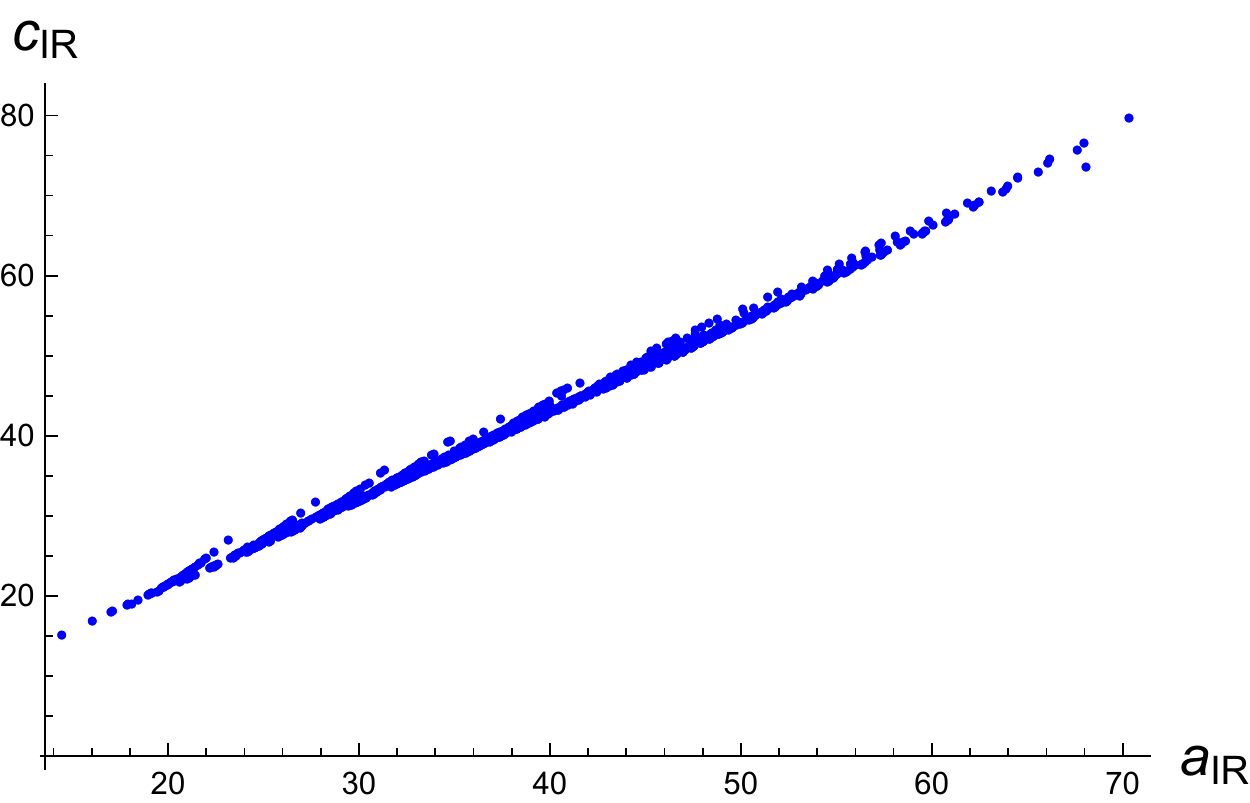}
\caption{$(E_7,E_7)$}
\label{CMMSratioE7}
\end{subfigure}
\hspace{12pt} \begin{subfigure}[b]{0.45\textwidth}
\includegraphics[width=\textwidth]{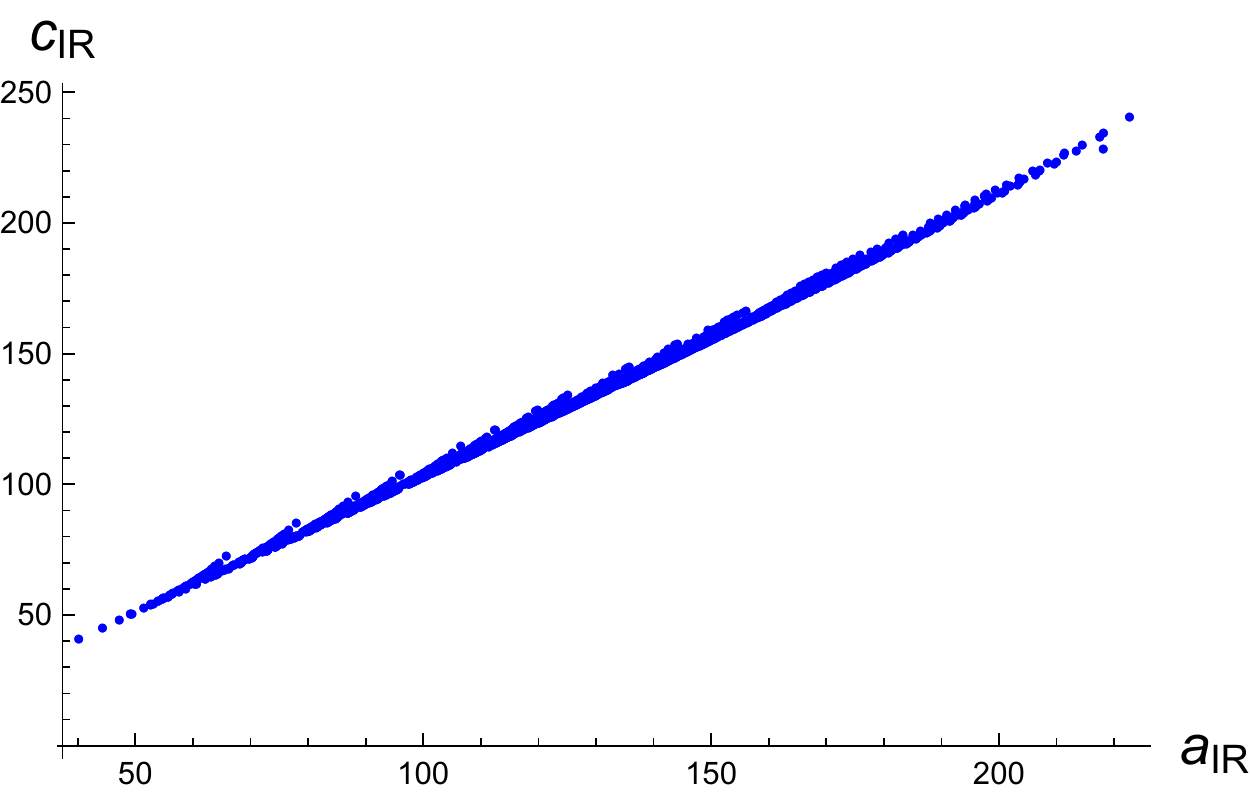}
\caption{$(E_8,E_8)$}
\label{CMMSratioE8}
\end{subfigure}
\caption{Plots of $c_{\mathrm{IR}}$ vs. $a_{\mathrm{IR}}$ for flipper field
deformations of 4D conformal matter.}%
\label{CMMSratio}%
\end{figure}

\newpage

\section{Conclusions \label{sec:CONC1}}

One of the important open issues in the study of conformal field theories is
to better understand the totality of fixed points, and their
network of flows under deformations. In this chapter we have
shown that a great deal of information on the structure of RG\ flows for
4D\ SCFTs can be extracted in the special case of nilpotent mass deformations.
Starting from a UV\ $\mathcal{N}=2$ SCFT, we have presented a general analysis of the
resulting $\mathcal{N}=1$ fixed points, both in the case of plain mass
deformations, as well as in the generalization to flipper field deformations,
where these parameters are treated as background vevs for a dynamical adjoint
valued $\mathcal{N}=1$ chiral superfield of the parent theory. In addition to
presenting a general analysis of the resulting fixed points, we have performed
an explicit sweep over all possible nilpotent deformations for the
$\mathcal{N}=2$ theories defined by D3-branes probing a $D$- or $E$-type 7-brane,
as well as the nilpotent deformations of 4D $(G,G)$ conformal matter. In both
cases, we have found strong evidence that the mathematical partial ordering
defined by the nilpotent cone of the associated Lie algebras is obeyed in the
physical theories as well. Moreover, the directed graph of this partially
ordered set also lines up with the possible relevant deformations of the
physical theory, providing a very detailed picture of the possible RG\ flows
from one fixed point to another.
The structure of the Hasse diagrams obtained provides a partially ordered set, which cleanly matches to physical 4D RG~flows. We can then take advantage of this fact (even in a more general setting) whenever there is a flavor symmetry present and we activate a breaking pattern generated by a nilpotent orbit.
In addition to presenting the full sweep
over theories in a companion \texttt{Mathematica} program, we have also observed a
number of intriguing \textquotedblleft phenomenological\textquotedblright%
\ features, including the appearance of several theories with
rational anomalies. We have also seen that for a given UV $\mathcal{N} = 2$ fixed point,
the ratio $a_{\mathrm{IR}}/c_{\mathrm{IR}}$ is roughly constant over the entire nilpotent network.
In the remainder of this section we discuss some avenues of further investigation.

One item left open by our analysis is a full treatment of the full network of
RG\ flows in cases where mesonic operators decouple from the new IR\ fixed
point. As we have already explained, such mesonic operators are often
necessary to perform further perturbations deeper down in the Hasse diagram,
so the absence of these operators could a priori pose some issues in the
context of matching the full network defined by the Hasse diagram to corresponding RG\ flows.
Even so, we have not found an explicit example which demonstrates that any
links are in fact \textquotedblleft broken.\textquotedblright\ It would be
most illuminating to further understand this class of theories.

Even within the class of theories considered here, there are some additional
relevant deformations we could contemplate switching on. This includes the
possibility of mass deformations which are semi-simple, namely their matrix
representatives are diagonalizable. Since such diagonal elements can also be
presented as the sum of two nilpotent elements, it is quite likely that the
analysis presented here may implicitly cover such cases as well, and may
actually help to \textquotedblleft explain\textquotedblright\ the appearance
of our rational theories. It would be interesting to analyze this issue further.

The bulk of this chapter has focused on determining various properties of the
new infrared fixed points generated by nilpotent mass deformations, including
the operator scaling dimensions of various operators. Another tractable
quantity to potentially extract is the superconformal index.
This could shed additional light on the IR\ properties
of these theories. Additionally, it would be quite interesting to see whether
there is a corresponding partial ordering for these indices, as induced by the
partial ordering on nilpotent orbits.

Much of our analysis has focused on the case of a single D3-brane probing an
F-theory 7-brane, as well as the case of \textquotedblleft rank one 6D
conformal matter,\textquotedblright\ namely (in M-theory terms) a single
M5-brane probing an ADE singularity. It would be quite natural to extend the
analysis presented here to the case of additional branes. While the anomalies
for the case of multiple D3-branes have already been determined \cite{Aharony:2007dj},
the corresponding statements for multiple M5-branes probing an
ADE\ singularity, and the resulting 4D\ anomaly polynomial are apparently
unknown. With this result in hand, it would then be possible to study
nilpotent mass deformations for this class of theories as well.

Another natural class of theories involves the compactification of
6D\ conformal matter on more general Riemann surfaces in the presence of
background fluxes and punctures. In this case, even before switching on
nilpotent mass deformations, we expect from the general procedure outlined in
\cite{Benini:2009mz} to get a 4D $\mathcal{N}=1$ SCFT, as in
references \cite{Gaiotto:2015usa,
Ohmori:2015pua, DelZotto:2015rca, Franco:2015jna, Ohmori:2015pia, Coman:2015bqq,
Morrison:2016nrt, Heckman:2016xdl, Razamat:2016dpl, Bah:2017gph, Apruzzi:2017iqe,
Kim:2017toz, Hassler:2017arf, Bourton:2017pee, Kim:2018bpg, Apruzzi:2018oge}.
Many of these theories admit a weakly coupled Lagrangian description \cite{Kim:2018bpg, Razamat:2018gro}, so
studying the possible nilpotent deformation and comparing the central charges
with the class of theories studied here might lead to
Lagrangian descriptions for some of the resulting IR\ fixed points.

Moreover, we have also seen a number of numerical coincidences, including the
appearance of rational theories, as well as a relatively constant value for
$a_{\mathrm{IR}}/c_{\mathrm{IR}}$ over an entire nilpotent network.
It would be very interesting to understand whether these
coincidences have a simple top down interpretation.

Finally, it is interesting to look at SCFTs in higher dimensions and ask how they are affected by nilpotent deformations. This is particularly useful given that many higher dimensional theories can be compactified to four dimensions. In fact, many 6D SCFTs serve as the ``master theories" for understanding a wide variety of lower-dimensional, strongly coupled phenomena. A typical example is the flat $T^2$ compactification of 6D SCFTs with 16 real supercharges which yields the 4D $\N = 4$ super Yang-Mills theory. We will thus probe several 6D SCFTs by introducing nilpotent deformations from strings stretching between stacks of intersecting 7-branes. The aim of the next chapter is to leverage the power of string junctions to better understand the Higgs branch deformations and nilpotent orbits of these theories.

\chapter{T-Branes, String Junctions, and 6D SCFTs}\label{chapter2}

\section{Introduction}

\label{sec:INTRO2}

One of the surprises from string theory is the
prediction of whole new classes of quantum field theories decoupled from
gravity. While the previous chapter dealt with 4D superconformal field theories (SCFTs), we will now look to other central examples of this sort: 6D SCFTs. The only known way to reliably engineer examples of such theories is to start with a
background geometry in string / M- / F-theory, and to consider a singular
limit in which all length scales are sent to zero or infinity (for early work in this
direction see e.g. \cite{Witten:1995zh, Strominger:1995ac, Seiberg:1996qx}).
Since small deformations away from these scaling limits have a sensible coupling to
higher-dimensional gravity, there is strong evidence that this leads to an
interacting conformal fixed point.

The most flexible method known for constructing such theories is via F-theory on a non-compact, elliptically-fibered Calabi-Yau threefold. SCFTs are generated by simultaneously contracting a configuration of curves in the base geometry. There is now a classification of
all elliptic threefolds which can generate a 6D SCFT, and in fact, each known
6D SCFT can be associated with some such threefold \cite{Heckman:2013pva, Heckman:2015bfa} (see also \cite{Bhardwaj:2015xxa, Bhardwaj:2019hhd}).\footnote{The caveat to
this statement is that in all known constructions, there is a non-trivial tensor branch. Additionally,
in F-theory there can be ``frozen'' singularities \cite{Witten:1997bs, Tachikawa:2015wka, Bhardwaj:2018jgp}.
We note that all such models still are described by elliptic threefolds with
collapsing curves in the base.} For a recent review,
see reference \cite{Heckman:2018jxk}.

In these sorts of constructions, one
begins away from the fixed point of interest and then tunes to zero various
operator vevs in the low energy effective field theory. In this UV limit, the
effective field theory description breaks down, but the stringy description
still remains well-behaved. From this perspective, the main question is to
better understand the microscopic structure of these 6D SCFTs.

The F-theory realization of 6D SCFTs provides insight into the corresponding
structure of these theories as well as their moduli spaces (see \cite{Heckman:2018jxk}).
Perhaps surprisingly, all known 6D SCFTs resemble generalizations of quiver gauge
theories in which (on a partial tensor branch) the theory involves ADE gauge
groups linked together by 6D conformal matter \cite{DelZotto:2014hpa, Heckman:2014qba}.
The topology of these quivers is rather simple, and consists of a single spine of such gauge groups. The
space of tensor branch deformations translates in the geometry to the moduli
space of volumes for the contractible curves in the base of the elliptic
threefolds. Additionally, Higgs branch deformations translate to complex
structure deformations of the corresponding elliptic threefolds.

The quiver-like description of 6D SCFTs also suggests that
Higgs branch deformations can be understood in terms of breaking patterns
associated with the flavor symmetries of these theories. For example, in the
$7$-brane gauge theory, nilpotent
elements of the flavor symmetry algebra correspond to ``T-brane
configurations'' of $7$-branes. For a partial list of references to the T-brane literature, see references
\cite{Aspinwall:1998he, Donagi:2003hh,Cecotti:2009zf,Cecotti:2010bp,Donagi:2011jy,Anderson:2013rka,
Collinucci:2014qfa,Cicoli:2015ylx,Heckman:2016ssk,Collinucci:2016hpz,Bena:2016oqr,
Marchesano:2016cqg,Anderson:2017rpr,Collinucci:2017bwv,Cicoli:2017shd,Marchesano:2017kke,
Heckman:2018pqx, Apruzzi:2018xkw, Cvetic:2018xaq, Carta:2018qke, Marchesano:2019azf, Bena:2019rth, Barbosa:2019bgh}.

A pleasant aspect of nilpotent elements is that they
come equipped with a partial ordering, as dictated by the symmetry breaking
pattern in the original UV theory. Indeed, the orbit of each nilpotent element
under the adjoint action specifies (under Zariski closure) a partially ordered set.
This partial ordering determines fine-grained structure for Higgs branch
flows between different 6D SCFTs \cite{Heckman:2016ssk, Mekareeya:2016yal} and points the way to a possible
classification of RG flows between 6D SCFTs \cite{Heckman:2018pqx}.\footnote{See also references
\cite{Heckman:2010qv, Apruzzi:2018xkw} for a related discussion of partial ordering in
the case of certain 4D SCFTs.}

This has been established in the case of 6D SCFTs with a sufficient number of gauge group factors in the quiver-like description,
i.e., ``long quivers,'' where Higgsing of the different flavor symmetries is uncorrelated,
and there are also hints that it extends to the case of ``short quivers'' in which the structure of
Higgsing is correlated.

One feature which is somewhat obscure in this characterization of Higgs
branch flows is the actual breaking pattern taking place in the quiver-like
gauge theory. Indeed, in the case of a weakly-coupled quiver gauge theory, the
appearance of matter transforming in representations of different gauge groups
means that the corresponding D-flatness conditions for one vector multiplet
will automatically be correlated with those of neighboring gauge group nodes.
This means that each breaking pattern defined on the exterior of a quiver will
necessarily propagate towards the interior of the quiver. Even in the case of
quiver gauge theories with classical algebras, the resulting combinatorics for
tracking the breaking pattern of a Higgs branch deformation can be quite intricate.

To address these issues, in this chapter we use the physics of
brane recombination to extract the combinatorics of Higgs branch flows in 6D SCFTs. In stringy
terms, brane recombination is associated with the condensation of strings
stretched between different branes. In the context of F-theory, strings can be
bound states of F1- and D1- strings, and they can have multiple ends. Our task,
then, will be to show how such multi-pronged strings attach between different
stacks of branes, and moreover, how this leads to a natural characterization
of brane recombination for Higgs branch flows in 6D SCFTs.

Since we will be primarily interested in flows driven by nilpotent orbits,
we first spell out how a given configuration of multi-pronged strings attached to bound states of $[p,q]$
$7$-branes maps on to the breaking pattern associated with a particular nilpotent orbit of an algebra. Separating
these branes from one another corresponds to a choice of Cartan subalgebra,
and strings stretched between these separated branes correspond to Lie algebra
elements associated with roots of the Lie algebra, defining a directed graph
between the nodes spanned by these branes. In particular, we show that we can
always generate a nilpotent element of the (complexified) Lie algebra by
working in terms of a directed graph which points in one direction. We also
show that, starting from such a directed graph, appending additional strings
always leads to a nilpotent element with a strictly larger nilpotent orbit.
We thus construct the entire nilpotent cone of each Lie algebra of type ABCDEFG using such multi-pronged string junctions.

With this result in place, we next turn to an analysis of Higgs branch flows
in quiver-like 6D SCFTs, as generated by T-brane deformations. We primarily
focus on 6D SCFTs generated by M5-branes probing an ADE singularity
with flavor symmetry $G_{ADE} \times G_{ADE}$, as well as tensor branch
deformations of these cases to non-simply laced flavor symmetry algebras. As found in
\cite{Heckman:2018pqx}, these are progenitor theories for many
6D SCFTs (the other being E-string probes of ADE singularities \cite{Heckman:2014qba, Heckman:2015bfa, Mekareeya:2017jgc, Frey:2018vpw, Cabrera:2019izd}). The partial
tensor branch of these parent UV theories are all of the form:
\begin{equation}
\label{quivo}[G_{0}] - G_{1} - ... - G_{k} - [G_{k+1}]
\end{equation}
with $G_{0}, G_{k+1}$ flavor symmetries and $G_1,...,G_k$ gauge symmetries. We show that Higgs
branch flows are determined by a system of coupled D-term constraints, one for each
node of such a quiver gauge theory. This in turn means that the ``links''
between gauge nodes behave as a generalization of matter, as suggested by the
structure of these quivers. We also show that condensing these strings leads
to a sequence of brane recombinations, relying on a parallel with Hanany-Witten moves
\cite{Hanany:1996ie} seen in the type IIA framework to derive the type IIB recombination
moves. We present a complete characterization of quiver-like theories with classical algebras,
and briefly discuss what would be needed to extend this analysis to quiver-like theories with
exceptional gauge group factors.

The explicit characterization of nilpotent orbits in terms of string junctions
also allows us to study Higgs branch flows in which the number of gauge groups
is small. This case is especially interesting because
there are non-trivial correlations on the symmetry breaking patterns,
one emanating from the left flavor symmetry $G_{0}$ and the subsequent D-term constraints on
its gauged neighbors and one emanating from
the right flavor symmetry $G_{k+1}$ and its gauged neighbors in the quiver of line (\ref{quivo}).
This sort of phenomenon occurs whenever the size of the nilpotent orbit
of the flavor algebras is sufficiently
large, and the number of gauge groups $k$ is sufficiently small. We study these ``overlapping T-branes'' in detail in the case of the classical
algebras. In particular, we show how to extract the resulting IR SCFT using
our picture in terms of brane recombination. We leave the case of short
quivers with exceptional gauge groups / flavor symmetries to future work.

The rest of this chapter is organized as follows. First, in section \ref{sec:QUIVER}, we
review in general terms the structure of 6D SCFTs as quiver-like gauge
theories, and we explain how the worldvolume theory on $7$-branes leads to a
direct link between Higgs branch flows and nilpotent orbits of flavor
symmetries. In section \ref{sec:NILPJUNC}, we show how to reconstruct the nilpotent cone of a
flavor symmetry algebra in terms of the combinatorial data of strings
stretched between stacks of $[p,q]$ $7$-branes. Section \ref{sec:RECOMBO} uses
this combinatorial data to provide a systematic method for analyzing Higgs
branch flows in quiver-like theories with classical gauge groups, including cases with
6D conformal matter. In section \ref{sec:GETSHORTY}, we study Higgs branch flows
from overlapping nilpotent orbits in short quivers, and in section \ref{sec:CONC2} we
present our conclusions. A number of additional detailed computations are
included in the Appendices.

\section{6D SCFTs as Quiver-Like Gauge Theories \label{sec:QUIVER}}

In this section, we briefly review the relevant aspects of 6D\ SCFTs which we
shall be studying in the remainder of this chapter. The main item of interest
for us will be the quiver-like structure of all such theories, and the
corresponding Higgs branch flows associated with nilpotent orbits of the flavor
symmetry algebra.

To begin, we recall that the F-theory realization of 6D\ SCFTs involves
specifying a non-compact elliptically-fibered Calabi-Yau threefold
$X\rightarrow B$, where the base $B$ of the elliptic fibration is
a non-compact K\"{a}hler surface. In
minimal Weierstrass form, these elliptic threefolds can be viewed as a
hypersurface:%
\begin{equation}
y^{2}=x^{3}+fx+g.
\end{equation}
The order of vanishing for the coefficients $f$, $g$ and the discriminant
$\Delta=4f^{3}+27g^{2}$ dictate the structure of possible gauge groups, flavor
symmetries and matter content in the 6D\ effective field theory. We are
particularly interested in the construction of 6D\ SCFTs, which requires us to simultaneously collapse a collection of curves in the base to zero size at
finite distance in the Calabi-Yau metric moduli space. This can occur for
curves with negative self-intersection, and compatibility with the condition
that we maintain an elliptic fibration over generic points of each curve
imposes further restrictions \cite{Heckman:2013pva}. Each such configuration can be viewed as being
built up from intersections of non-Higgsable clusters (NHCs) \cite{Morrison:2012np} and possible
enhancements in the singularity type over each such curve. The tensor branch
of the 6D\ SCFT corresponds to resolving the collapsing curves in the base to
finite size, and the Higgs branch of the 6D\ SCFT corresponds to blow-downs and smoothing
deformations of the Weierstrass model such as \cite{Heckman:2015ola}:
\begin{equation}
y^{2}=x^{3}+(f+\delta f)x+(g+\delta g).
\end{equation}

In references \cite{Heckman:2013pva, Heckman:2015bfa}, the full list of possible F-theory geometries
which could support a 6D\ SCFT was determined. Quite remarkably, all of these
theories have the structure of a quiver-like gauge theory with a single spine
of gauge group nodes and only small amounts of decoration by (generalized) matter on the left and right of each quiver. In this description,
$7$-branes with ADE\ gauge groups intersect at points where additional
curves have collapsed. These points are often referred to as \textquotedblleft
conformal matter\textquotedblright\ since they localize at points just as in the case of ordinary matter
in F-theory \cite{DelZotto:2014hpa, Heckman:2014qba}. These configurations indicate
the presence of additional operators in the 6D\ SCFT and, like ordinary matter, can have non-trivial
vevs, leading to a deformation onto the Higgs branch. A
streamlined approach to understanding the vast majority of 6D\ SCFTs was
obtained in \cite{Heckman:2018pqx} where it was found that any 6D\ SCFT can be viewed as
\textquotedblleft fission products,\textquotedblright\ namely as deformations
of a quiver-like theory with partial tensor branch such as:%
\begin{equation}
\lbrack E_{8}]\overset{\mathfrak{g}_{ADE}}{1}\overset{\mathfrak{g}_{ADE}%
}{2}...\overset{\mathfrak{g}_{ADE}}{2}[G_{ADE}] \label{hetinst}%
\end{equation}
or:%
\begin{equation}
\lbrack G_{ADE}]\overset{\mathfrak{g}_{ADE}}{2}...\overset{\mathfrak{g}%
_{ADE}}{2}[G_{ADE}], \label{m5probe}%
\end{equation}
where the few SCFTs which cannot be understood in this way can be obtained by
adding a tensor multiplet and weakly gauging a common flavor symmetry of these fission products through a
process known as fusion. In the above, each compact curve of self-intersection
$-n$ with a $7$-brane gauge group of ADE\ type is denoted as
$\overset{\mathfrak{g}_{ADE}}{n}$. The full tensor branch of these theories is
obtained by performing further blowups at the collision points between the
compact curves (in the D- and E-type cases). To emphasize this quiver-like
structure, we shall often write:%
\begin{equation}
\lbrack G_{0}]-G_{1}-...G_{k}-[G_{k+1}],
\end{equation}
to emphasize that there are two flavor symmetry factors (indicated by square
brackets), and the rest are gauge symmetries.

The 6D SCFTs given by lines (\ref{hetinst}) and (\ref{m5probe}) can also be
realized in M-theory. The theories of line (\ref{hetinst}) arise from an
M5-brane probing an ADE\ singularity which is wrapped by an $E_{8}$
nine-brane. The theories of line (\ref{m5probe}) arise from M5-branes probing
an ADE\ singularity. In what follows, we shall primarily be interested in
understanding Higgs branch flows associated with the theories of line (\ref{m5probe}).

For $G_{ADE}$ of A or D type, the IR SCFTs of these Higgs branch flows can also be realized
in type IIA. $SU$ gauge algebras are obtained from the worldvolume of D6-branes suspended between
spacetime-filling NS5-branes, while $SO$ algebras and $Sp$ gauge algebras also require $O6^-$ and
$O6^+$ branes, respectively, stretched between $\frac{1}{2}$ NS5-branes. These constructions
will prove especially useful in section \ref{sec:RECOMBO}, where we discuss Hanany-Witten moves of
the branes of the type IIA construction.

One of the main ways to cross-check the structure of proposed RG flows is through anomaly
matching constraints. The anomaly polynomial of a 6D SCFT is calculable because the tensor branch description of each such theory is available
from the F-theory description, and the anomaly polynomial obtained on this branch of moduli space can be
matched to that of the conformal fixed point \cite{Ohmori:2014pca, Ohmori:2014kda,
Heckman:2015ola, Cordova:2015fha, Cordova:2018cvg}. To fix conventions, we often write this as a formal
eight-form with conventions (as in reference \cite{Heckman:2018jxk}):
\begin{align}
I_{8}  &  = \alpha c_{2}(R)^{2} + \beta c_{2}(R) p_{1}(T) + \gamma
p_{1}(T)^{2} + \delta p_{2}(T)\nonumber\\
&  + \sum_{i} \left[  \mu_{i} \, \mathrm{Tr} F_{i}^{4}
+ \, \mathrm{Tr} F_{i}^{2} \left(  \rho_{i}
p_{1}(T) + \sigma_{i} c_{2}(R) + \sum_{j} \eta_{ij} \, \mathrm{Tr} F_{j}^{2}
\right)  \right] , \label{anomalypoly}%
\end{align}
where in the above, $c_{2}(R)$ is the second Chern class of the $SU(2)_{R}$ symmetry,
$p_{1}(T)$ is the first Pontryagin class of the tangent bundle, $p_{2}(T)$ is
the second Pontryagin class of the tangent bundle, and $F_{i}$ is the field
strength of the $i^{th}$ symmetry, where $i$ and $j$ run over the
flavor symmetries of the theory. See the review article
\cite{Heckman:2018jxk} as well as the Appendices for
additional details on how to calculate the anomaly polynomial in specific
6D SCFTs.

Returning to the F-theory realization of the 6D\ SCFTs of line (\ref{m5probe}),
there is a large class of Higgs branch deformations associated with
nilpotent orbits of the flavor symmetry algebras.\footnote{We note that
although a T-brane deformation has vanishing Casimirs and may thus appear to
be \textquotedblleft invisible\textquotedblright\ to the geometry, we can
consider a small perturbation away from a T-brane which then would register as
a complex structure deformation. Since we are dealing with the limiting case
of an SCFT, all associated mass scales (as well as fluxes localized on
$7$-branes)\ will necessarily scale away. This also means that each nilpotent
element can be associated with an elliptic threefold \cite{DelZotto:2014hpa}.}
Moreover, nilpotent elements admit a partial ordering which also dictates a
partial ordering of 6D\ fixed points. We say that a nilpotent element
$\mu\preceq\nu$ when there is an inclusion of the orbits under the adjoint
action: Orbit$(\mu)\subseteq$ $\overline{\text{Orbit}(\nu)}$.

In the 6D\ SCFT, there is a triplet of adjoint valued moment maps $D_{\text{adj}}^{1},$
$D_{\text{adj}}^{2},$ $D_{\text{adj}}^{3}$ which couple to the flavor symmetry
current supermultiplet. The nilpotent element can be identified with the
complexified combination $D_{\text{adj}}^{\mathbb{C}}=D_{\text{adj}}%
^{1}+iD_{\text{adj}}^{2}$. Closely related to this triplet of moment maps are
the triplet of D-term constraints for each gauge group factor $G_{j}$ for
$j=1,...,k$. Labeling these as a three-component vector taking values in the
adjoint of each such group $\overrightarrow{D}_{j}$, supersymmetric vacua are
specified in part by the conditions:%
\begin{equation}
\overrightarrow{D}_{j}=0\text{ for all }j,
\end{equation}
modulo unitary gauge transformations. We note that in the weakly coupled
context, the D-term constraints for each gauge group factor are in fact
correlated with one another. In particular, if we specify a choice of moment
map $\overrightarrow{D}_{0}\neq0$ and $\overrightarrow{D}_{k+1}\neq0$ on the
left and right of the quiver, respectively, this propagates to a non-trivial
breaking pattern in the interior of the quiver.

That being said, the actual description of this breaking pattern using
6D\ conformal matter is poorly understood because there is no weakly
coupled description available for these degrees of freedom. So, while we
expect there to be a correlated breaking pattern for gauge groups in the
interior of a quiver, the precise structure of these terms is unclear due to
the unknown structure of the microscopic degrees of freedom in the field theory.

In spite of this, it is often possible to extract the resulting IR fixed point after such a deformation,
even in the absence of a Lagrangian description. The main reason this
is possible is because in the context of an F-theory compactification, we
already have a classification of all possible outcomes which could have
resulted from a Higgs branch flow (since we have a classification of
6D\ SCFTs). In many cases, this leads to a unique candidate
theory after Higgsing, and this has been used to directly determine the
Higgsed theory. Even so, this derivation of the
theory obtained after Higgsing involves a number of steps which are not
entirely systematic, thus leading to potential ambiguities in cases where the number of gauge group
factors in the quiver is sufficiently small that there is a non-trivial correlation in the
symmetry breaking pattern obtained from a pair of nilpotent orbits (one on the left and one on the
right of the quiver). We refer to such quivers as being ``short,'' and the case where there is no correlation between
breaking patterns from different nilpotent orbits as ``long.''

One of our aims in the present
chapter will be to determine the condensation of strings stretched between
different stacks of branes. Our general strategy for analyzing
Higgs branch flows will therefore split into two parts:

\begin{itemize}
\item First, we determine the particular configuration of multi-pronged strings
associated with each nilpotent orbit.

\item Second, we determine how to consistently condense these multi-pronged
string states to trigger brane recombination in the quiver-like gauge theory.
\end{itemize}

\section{Nilpotent Orbits from String Junctions \label{sec:NILPJUNC}}

One of our aims in this chapter is to better understand the combinatorial
structure associated with symmetry breaking patterns for 6D\ SCFTs. In
this section we show how to construct all of the nilpotent
orbits of a semi-simple Lie algebra of type ABCDEFG from the structure of
multi-pronged string junctions. The general idea follows earlier
work on the construction of such algebras, as in \cite{Gaberdiel:1997ud, DeWolfe:1998zf, Bonora:2010bu} (see also \cite{Grassi:2013kha, Grassi:2014ffa, Grassi:2018wfy}). We refer the interested reader to Appendix \ref{sec:Nilpotents}
for additional details and terminology on nilpotent orbits which we shall
reference throughout this chapter.

Recall that in type IIB, we engineer such algebras using $[p,q]$ $7$-branes, namely
a bound state of $p$ D7-branes and $q$ S-dual
$7$-branes. Labeling the monodromy of the axio-dilaton around a source of
$7$-branes by a general element of $SL(2,\mathbb{Z})$:
\begin{equation}
\tau\mapsto\frac{a\tau+b}{c\tau+d}\text{ \ \ for \ \ }\left[
\begin{array}
[c]{cc}%
a & b\\
c & d
\end{array}
\right]  \in SL(2,\mathbb{Z})\text{,}%
\end{equation}
a $[p,q]$ $7$-brane determines a conjugacy class in $SL(2,\mathbb{Z})$ as specified
by the orbit of:\footnote{A note on conventions: One can either consider this matrix or its inverse depending on
whether we pass a branch cut counterclockwise or clockwise. This will not affect our discussion in any material way.}
\begin{equation}
M_{[p,q]} = \left[
\begin{array}
[c]{cc}%
1 + pq & - p^2 \\
q^2 & 1 - pq
\end{array}
\right].
\end{equation}
\\
The relevant structure for realizing the different ADE\ algebras are the monodromies:%
\begin{align}
A= M_{[1,0]} = \left[
\begin{array}
[c]{cc}%
1 & -1\\
0 & 1
\end{array}
\right]  \text{, \ \ }B= M_{[1,-1]} = \left[
\begin{array}
[c]{cc}%
0 & -1\\
1 & 2
\end{array}
\right]  \text{, \ \ } \nonumber \\
C= M_{[1,1]} = \left[
\begin{array}
[c]{cc}%
2 & -1\\
1 & 0
\end{array}
\right]  \text{, \ \ }
X= M_{[2,-1]} = \left[
\begin{array}
[c]{cc}%
-1 & -4\\
1 & 3
\end{array}
\right]  .
\end{align}
The $7$-branes necessary to engineer various
Lie algebras follow directly from the Kodaira classification of possible
singular elliptic fibers at real codimension two in the base of an F-theory
model \cite{Vafa:1996xn, Morrison:1996na, Morrison:1996pp}. They can also
be directly related to a set of basic building blocks in the string junction picture
worked out in \cite{Gaberdiel:1997ud} which we label as in reference \cite{DeWolfe:1998pr}:
\begin{align}
A_{N}  &  :A^{N+1}\\
H_{N}  &  :A^{N+1}C\text{ \ \ (for }N=0,1,2\text{)}\\
D_{N}  &  :A^{N}BC\\
E_{N}  &  :A^{N - 1}BC^{2}\text{ \ \ (for }N=6,7,8\text{)}\\
\widetilde{E}_{N}  &  :A^{N}XC\text{ \ \ (for }N=6,7,8\text{)}.
\end{align}
The $H_{N}$ series in the second line represents an alternative way to realize low
rank $SU$ type algebras. We also note that in the case of the A- and D-
series, it is possible to remain at weak string coupling, while the H- and
E-series require order one values for the string coupling. Here, we have indicated two alternate presentations
of the $E$-type algebras (see reference \cite{DeWolfe:1998pr}). It will prove convenient in what follows to use the
$\widetilde{E}_N$ realization with an $X$-brane. The non-simply laced algebras have the
same $SL(2,\mathbb{Z})$ monodromy type. In the string junction description, this involves further
identifications of some of the generators of the algebra by a suitable outer automorphism. Some aspects of this
case are discussed in \cite{Bonora:2010bu}.

We would like to understand the specific way that nilpotent generators of the
Lie algebra are encoded in this physical description. In all these cases, the
main idea is to first separate the $7$-branes so that we have a physical
realization of the Cartan subalgebra. Then, a string which stretches from one
brane to another will correspond to an 8D vector boson with mass dictated by
the length of the path taken to go from one stack to the other:%
\begin{equation}
\text{mass}\sim\frac{\text{length}}{\ell_{\ast}^{2}},
\end{equation}
with $\ell_{\ast}$ a short distance cutoff. In the limit where
all the $7$-branes are coincident, we get a massless state.

With this in mind, let us recall how we engineer the gauge algebra $\mathfrak{su}(N)$
using D7-branes. All we are required to do in this case is introduce $N$
D7-branes, which are $[p,q]$ $7$-branes with $p=1$ and $q=0$. Labeling the
$7$-branes as $A_{1},...,A_{N}$, we can consider an open string which
stretches from brane $A_{i}$ to brane $A_{j}$. Since this string comes with an
orientation, we can write:%
\begin{equation}
A_{i}\rightarrow A_{j}\text{,}%
\end{equation}
and introduce a corresponding nilpotent $N\times N$ matrix with a single entry
in the $i^{th}$ row and $j^{th}$ column. We denote by $E_{i,j}$ the matrix
with a one in this single entry so that the corresponding nilpotent element is
written as $v_{i,j} E_{i,j}$ with no summation on indices. Conjugation by an $SL(n,\mathbb{C})$
element reveals that the actual entry does not affect the orbit. We will, however, be
interested in RG flows generated by adding perturbations away from a single
entry, so we will often view $v_{i,j}$ as indicating a vev / energy scale. In this manner, we can represent an RG flow triggered by moving onto the Higgs branch of the theory, which is labeled by a nilpotent orbit of a Lie algebra, in terms of a collection of strings stretched between the $7$-branes.

\begin{sidewaystable}
\centering
\resizebox{\textheight}{!}{
\begin{tabular}{|>{\centering}m{.25cm} |>{\centering}m{2.2cm} |>{\centering}m{5.8cm} |>{\centering}m{4cm} |>{\centering}m{5cm}|m{3.8cm}|}
\hline
& Dynkin diagram & IIB with mirror plane & Physical picture from \cite{DeWolfe:1998zf} & Branching rule to $\mathfrak{su}(4) \times \mathfrak{u}(1)$ & \begin{center}Positive roots \end{center} \\
\hline
$A_4$
&\includegraphics[scale=.8]{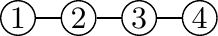}
&\includegraphics[scale=.8]{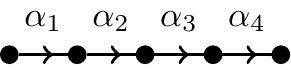}
&\includegraphics[scale=.8]{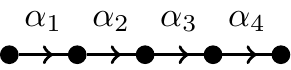}
& $24 \rightarrow 15_0 + 4_1 + \overline{4}_{\scalebox{0.75}[1.0]{-}1} + 1_0$ & \makecell{10 one-pronged strings: \\ $a_i-a_j$} \\ \hline
$B_4$
&\includegraphics[scale=.8]{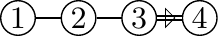}
&\includegraphics[scale=.8]{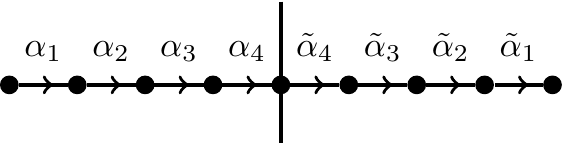}
&\includegraphics[scale=.8]{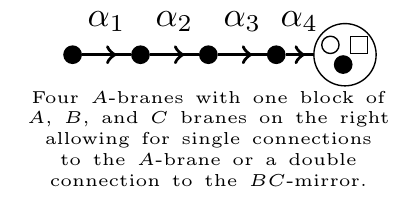}
& $36 \rightarrow 15_0 + 6_2 + \overline{6}_{\scalebox{0.75}[1.0]{-}2}+4_1+\overline{4}_{\scalebox{0.75}[1.0]{-}1} + 1_0$
& \makecell{10 one-pronged strings: \\ $a_i-a_j, \ \tilde{a}_j-\tilde{a}_i$ \\
  6 two-pronged strings: \\ $a_i-\tilde{a}_j, \ a_j-\tilde{a}_i$} \\
  \hline
$C_4$
&\includegraphics[scale=.8]{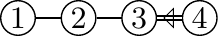}
&\includegraphics[scale=.8]{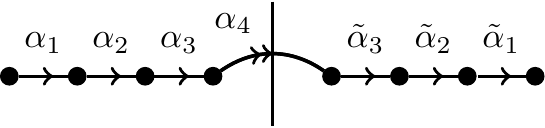}
&\includegraphics[scale=.8]{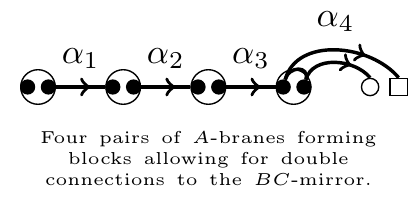}
  & $36 \rightarrow 10_2 + \overline{10}_{\scalebox{0.75}[1.0]{-}2} + 15_0 +1_0$ & \makecell{6 one-pronged strings: \\ $a_i-a_j, \ \tilde{a}_j-\tilde{a}_i$ \\
  4 double strings: \\ $a_i-\tilde{a}_i$ \\
6 two-pronged strings: \\ $a_i-\tilde{a}_j, \ a_j-\tilde{a}_i$} \\ \hline
$D_4$
&\includegraphics{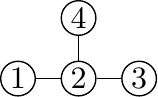}
&\includegraphics{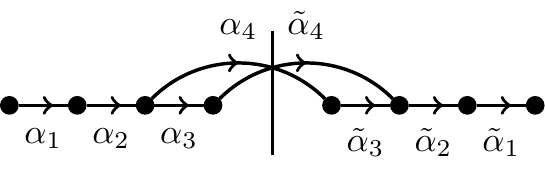}
&\includegraphics{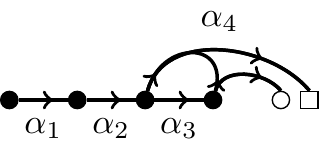}
& $28 \rightarrow 6_2 + \overline{6}_{\scalebox{0.75}[1.0]{-}2}+15_0 + 1_0$
& \makecell{6 one-pronged strings: \\ $a_i-a_j, \ \tilde{a}_j-\tilde{a}_i$ \\
6 two-pronged strings: \\ $a_i-\tilde{a}_j, \ a_j-\tilde{a}_i$} \\ \hline
\end{tabular}}
\caption{Summary of basic properties for the string junction realization of the classical Lie algebras $A_4$, $B_4$, $C_4$, $D_4$. The columns from left to right are: Dynkin diagrams, IIB brane picture, string junction picture from \cite{DeWolfe:1998zf}, branching rule of adjoint decomposition in $\mathfrak{su}(4) \times \mathfrak{u}(1)$, explicit expression of groups of positive roots based on the adjoint decomposition. Here the indices $i$, $j$ run from $1$ to the number of nodes on the left-hand side of the mirror ($BC$). The tilde nodes are the reflected branes and the indices continue running as $\tilde{i} = N-i$ where $N$ is the total number of nodes in the diagram.}
\label{tab:classicalRootTable}
\end{sidewaystable}

\begin{figure}[t!]
\centering
\includegraphics{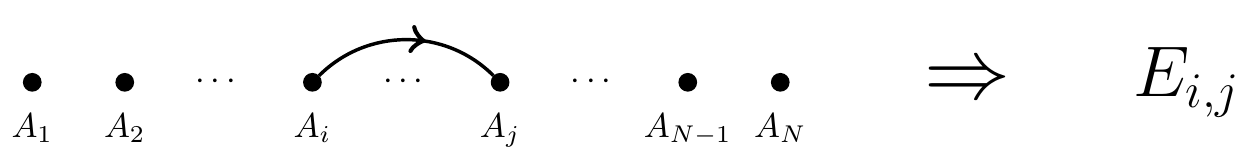}
\caption{Separating a collection of A-type branes leads to a deformation of $\mathfrak{su}(N)$ to the Cartan subalgebra. Open strings
stretched between distinct branes are associated with specific generators in the complexified Lie algebra. In the figure, this is shown for a string stretched from brane $A_i$ to brane $A_j$.}%
\label{fig:hoppadoo}%
\end{figure}

Ordering the branes $A_{1},..,A_{N}$ from left to right in the plane
transverse to the stack of $7$-branes, we see that we can now populate the
strictly upper triangular portion of a matrix in terms of strings
$A_{i}\rightarrow A_{j}$ for $i<j$ (see figure \ref{fig:hoppadoo}). So in other words, we can populate all
possible nilpotent orbits (in this particular basis). Similar considerations
hold for the other algebras, but clearly, this depends on a number of
additional features such as unoriented open strings (in the case of the
classical SO / Sp algebras) and multi-pronged string junctions (in the case of
the exceptional algebras). A related comment is that we are just constructing
a representative nilpotent element in the orbit of the Lie algebra. What we
will show is that for any deformation onto the Cartan, there is a ``minimal
length'' choice, and all the other elements of the orbit are obtained through
the adjoint action of the Lie algebra.

Our plan in the rest of this section will be to establish in detail how to
construct the corresponding nilpotent orbits for each configuration of
strings. Additionally, we show that not only can we generate all orbits, but
that the combinatorial method of \textquotedblleft adding extra
strings\textquotedblright\ automatically generates a partial ordering on the
space of nilpotent orbits, which reproduces the standard partial ordering of the
nilpotent cone. The essential information for the classical Lie algebras, and
in particular the list of simple and positive roots, is illustrated in table
\ref{tab:classicalRootTable}. We elaborate on the content of this table (as
well the exceptional analogs) in the following subsections.

\subsection{$SU(N)$: Partition by Grouping Branes with Strings}

In the case of an $SU(N)$ flavor we simply have $N$ perturbative $A$-branes
with $[p,q]=[1,0]$ charges. The $N-1$ simple roots of $SU(N)$ can be
represented by strings joining two adjacent $A$-branes as shown in figure
\ref{fig:SUNroots}. We refer to these as ``simple strings'' due to their
correspondence to the simple roots. The remaining (non-simple) roots are then
described by strings connecting any two $A$-branes. The positive roots are
represented by strings stretching from left to right while the negative ones
would go in the opposite direction (as indicated by the arrows). That is we
choose a basis for the generators of the $\mathfrak{su}_{N}$ algebra to be
given by:

\begin{itemize}
\item $N(N-1)/2$ nilpositive elements $E_{i,j}$ with $1 \leq i < j
\leq N$ corresponding to strings stretching from the $i^{th}$ to the $j^{th}$
$A$-brane (with the arrow pointing from left to right).

\item $N(N-1)/2$ nilnegative elements $E_{j,i} = X_{k}^{T}$ with $1
\leq i < j \leq N$ corresponding to strings stretching from the $j^{th}$ to
the $i^{th}$ $A$-brane (with the arrow now pointing from right to left).

\item $(N-1)$ Cartans $[E_{i,i+1}, E_{i+1,i}]$ for $1 \leq i \leq N-1$.
\end{itemize}

Through out this chapter we denote $E_{i,j}$ to be matrix with value $+1$ in the
entry $(i,j)$ but zeros everywhere else. The positive simple roots are given
by $\alpha_{i}$ $(1 \leq i \leq rank(G))$, with the corresponding matrix
representation labeled $E_{\alpha_{i}}$. Any non-simple root can then be labeled
explicitly in terms of its simple roots constituents: $\alpha_{i,j,k,\dots
,p,q} = \alpha_{i} + \alpha_{j} + \alpha_{k} + \dots+ \alpha_{p} + \alpha_{q}$
and the corresponding matrix representation is obtained from nested commutators.

In this basis, the simple positive roots are $E_{i,i+1}$ for $1
\leq i \leq N-1$, as illustrated by their corresponding directed strings in
figure \ref{fig:SUNroots}. Furthermore, we use the convention of
\cite{DeWolfe:1998zf} to keep track of the different monodromies. Namely, we only display the directions
transverse to the $7$-brane, thus representing each $7$-brane as a point. In this picture
the associated branch cut for $SL(2,\mathbb{Z})$ monodromy stretches vertically downward to infinity.
This will not enter our analysis in any material way so in order not to overcrowd the
figures, we will mostly not draw the branch cuts.

\begin{figure}[t!]
\centering
\includegraphics{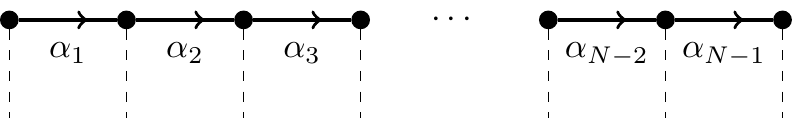}
\caption{Brane diagram of strings/roots stretching between the $A$-branes
yielding an $SU(N)$ flavor symmetry (see \cite{DeWolfe:1998zf}). The dashed
lines represent the position of branch cuts. Since they do not contribute to
our analysis, they are not drawn in subsequent pictures.}%
\label{fig:SUNroots}%
\end{figure}

We have already seen that nilpotent orbits of $SU(N)$ are parametrized by
partitions of $N$ (with no restriction whatsoever). Thus it becomes natural to
classify nilpotent orbits by how branes are grouped together. Namely, we can
group any set of $A$-branes by stretching strings between them, giving rise to
a particular partition of the $N$ branes. This partition is then in one-to-one
correspondence with its corresponding nilpotent orbit. As an equivalence
class, we have many different string configurations belonging to the same
orbit (just like many different matrices have the same Jordan block
decomposition). For instance, the three string junctions of figure
\ref{fig:equivSU} all represent the same $[3,2,1]$ partition:

\begin{itemize}
\item The first string junction picture has a matrix representation $M_{1} =
E_{1,2}+E_{2,3}+E_{4,5}$.

\item The second configuration has matrix representation $M_{2} = E_{1,3}+E_{3,6}+E_{4,5}$.

\item And finally, the third one has matrix representation $M_{3} =E_{1,3}+E_{4,5}+E_{5,6}$.
\end{itemize}

\begin{figure}[t!]
\centering
  \includegraphics{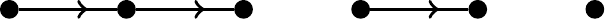}
  \par
  \includegraphics{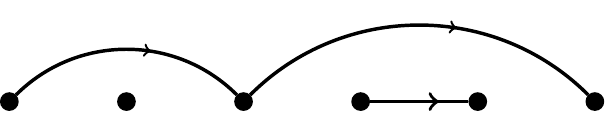}
  \par
  \includegraphics{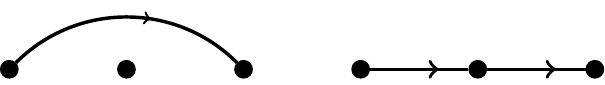}
\caption{Three equivalent ways of describing the partition $[3,2,1]$ in the
set of nilpotent orbits of $SU(6)$. To each picture is associated a different
matrix, but they all have the same Jordan block decomposition and thus belong
to the same equivalence class.}%
\label{fig:equivSU}%
\end{figure}

To each nilpotent orbit of $SU(N)$ we can then associate one of many possible string junction pictures. To keep the picture as simple as
possible, we choose to use only ``simple'' positive strings, that is strings stretching
from left to right between two adjacent $A$-branes. This ensures that we
only make use of simple roots. This typically does not completely fix a string
junction representative, so we are free to make a convenient choice of the remaining possibilities.

By starting with a configuration with no string attached (a $[1^{N}]$
partition) we can add more and more strings to go from the $[2,1^{N-2}]$ orbit
all the way to the $[N]$ partition. This generates a whole Hasse diagram of
nilpotent orbits which exactly matches that which is mathematically predicted.
Figure \ref{fig:SU6hasse} illustrates this diagram for the case of $SU(6)$
where we associate a ``standard'' string junction picture to each
nilpotent orbit according to how the branes are partitioned as we add more and
more strings.

\begin{figure}[t!]
  \centering
  \includegraphics{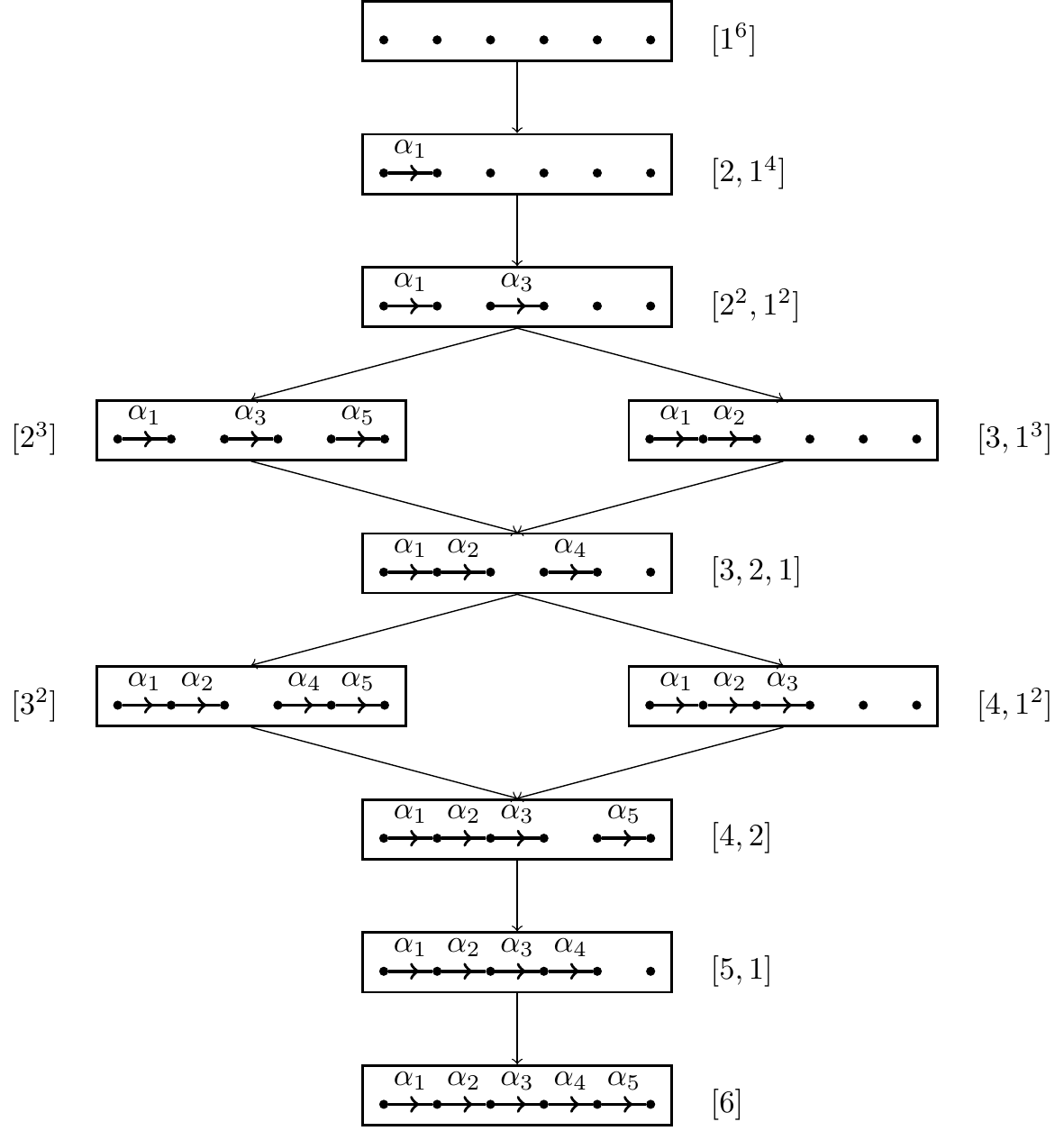}
  \caption{Hasse diagram of $SU(6)$ nilpotent deformations going from top (UV) to bottom (IR) where all simple roots are turned on and all corresponding ``simple strings'' connect the $A$-branes.}
  \label{fig:SU6hasse}
\end{figure}

More precisely, in order to flow from one point of the Hasse diagram to the
next, one simply needs to add a small perturbation, that is, an oriented
string (moving from left to right) corresponding to a positive root. By the definition
of the partial ordering of nilpotent orbits, this guarantees that the RG flow
indeed always takes us deeper into the IR. Weyl transformations / brane permutations can then be used to reduce
the obtained diagram back to one of the standard ones which only relies on the
simple roots.

The flows involving only the addition of a simple root (corresponding to
linking two more branes together) are fairly clear. The only cases where that
is not so obvious are the ones corresponding to flows that are similar to the
one described in figure \ref{fig:SU6hasse} by going from $[2^{2},1^{2}]$ to
$[3,1^{3}]$. For this we can add the string $\alpha_{2}+\alpha_{3}=a_{2}%
-a_{4}$, corresponding to a small deformation $\epsilon\cdot E_{2,4}$. This
particular flow is illustrated in figure \ref{fig:SUspecialFlow}. Generalizing
this procedure to arbitrary $SU(N)$ shows that the intermediate RG flows are
guaranteed to be physically realizable in the same fashion.

\begin{figure}[t!]
  \centering
  \includegraphics{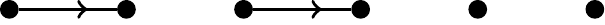}
  \par
{\Huge \rotatebox{-45}{$\Downarrow$} }
\par
{\Huge \vspace{-0.5cm}
  \includegraphics{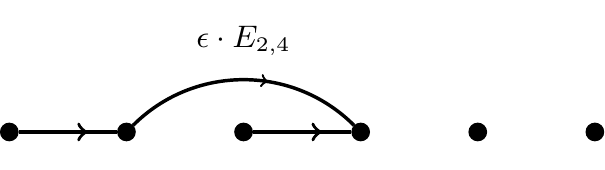}
  \hspace{.25cm} $\Leftrightarrow$ \hspace{.25cm}
  \includegraphics{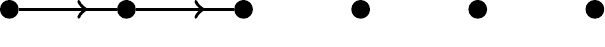}
}\caption{One way of flowing from the $[2^{2},1^{2}]$ nilpotent orbit (top) to
the $[3,1^{3}]$ orbit (bottom). In the top figure we have the matrix
representation $M_{1}=E_{1,2}+E_{3,4}$. The flow is then induced by adding an
extra string stretching between the $2^{nd}$ and $3^{rd}$ branes, as
illustrated in the bottom left figure. This corresponds to the matrix
$M_{2}=E_{1,2}+E_{3,4}+\epsilon\cdot E_{2,4}$. This matrix is similar to
$M_{2}^{\prime}=E_{1,2}+E_{2,3}$ corresponding to the bottom right diagram.
Thus, both bottom string junctions belong to the same nilpotent orbit
$[3,1^{3}]$.}%
\label{fig:SUspecialFlow}%
\end{figure}

\subsection{$SO(2N)$ and $SO(2N-1)$ \label{subsubsec:SO2Nflav}}

In F-theory, the $SO(2N)$ and $SO(2N-1)$ geometries
are realized by the presence of $A^{N}BC$-branes. In type IIB however, the
$BC$-branes turn into an $O7^{-}$ orientifold plane (as discussed in
\cite{Sen:1996vd}) which we refer here as the ``BC-mirror''. This mirror
reflects the $N$ $A$-branes across, yielding a total of $2N$ branes (half of
which are physical, half of which are ``image'' branes). We thus
represent $SO(2N)$ by $2N$ dots separated by a vertical line representing the
$BC$-mirror, and $SO(2N-1)$ by merging one $A$-brane with its mirror image
onto the orientifold so that we have $N-1$ $A$-branes on the left, $N-1$
mirror $A$-branes on the right, and a single $A$-brane squeezed onto the
vertical line representing the mirror.

Furthermore, \cite{DeWolfe:1998zf} provides us with a set of string junctions
to represent the simple roots of $SO(2N)$, as illustrated in figure
\ref{fig:SO2NrootsFtheory}. We can then obtain the corresponding roots for
$SO(2N-1)$ via the standard projection (or branching) of $SO(2N) \rightarrow
SO(2N-1)$. We see that much like $SU(N)$, we can have strings stretching
between any pair of $A$-branes, and the simple strings correspond to those
stretching between adjacent pairs. However, the presence of the $B$ and $C$
branes allows for a new kind of string: a two-pronged string which takes two
$A$-branes and connects them to the $B$ and $C$-branes. All these
configurations are regulated by charge conservation: the $A$-branes
all have charges $[1,0]$ so that a fundamental string can stretch between any pair
of them, but the $B$-brane has charge $[1,-1]$, and the $C$-brane has charge
$[1,1]$. Thus, no string can stretch directly between a $B$ and a $C$-brane.
However, these two branes together have an overall charge of $[2,0]$, which is exactly
twice that of an $A$-brane. Therefore, by combining two $A$-branes with the
$B$ and $C$-branes, charge can be conserved. This combination is achieved
through the introduction of a two-pronged string denoted $\alpha_{N}$ in
figure \ref{fig:SO2NrootsFtheory}.

We then visualize this $SO(N)$ geometry by introducing the orientifold, which reflects the strings as well as the $A$-branes. This is
illustrated in figures \ref{fig:SO2Nroots} and \ref{fig:SO2Nm1roots} for
$SO(2N)$ and $SO(2N-1)$ respectively.

\begin{figure}[t!]
\centering
\includegraphics[scale=.9]{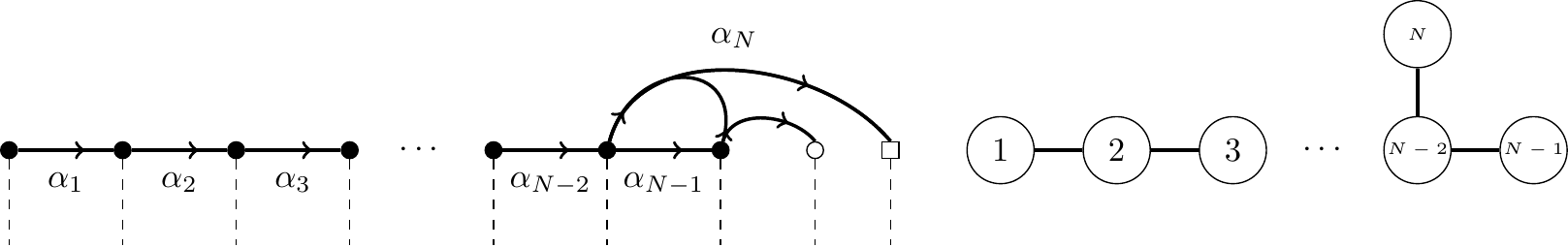}
\caption{Brane diagram of strings/roots stretching between the $A$, $B$, and
$C$-branes, making up the $SO(2N)$ symmetry (see \cite{DeWolfe:1998zf}). The
$A$-branes are denoted by black circles, the $B$-brane by an empty circle, and
the $C$-brane by an empty square. The dashed lines represent the position of
branch cuts, which (once again) are not drawn in subsequent pictures. To the right
we give the corresponding Dynkin diagram with simple roots numbered.}%
\label{fig:SO2NrootsFtheory}%
\end{figure}

\begin{figure}[t!]
\centering
\includegraphics{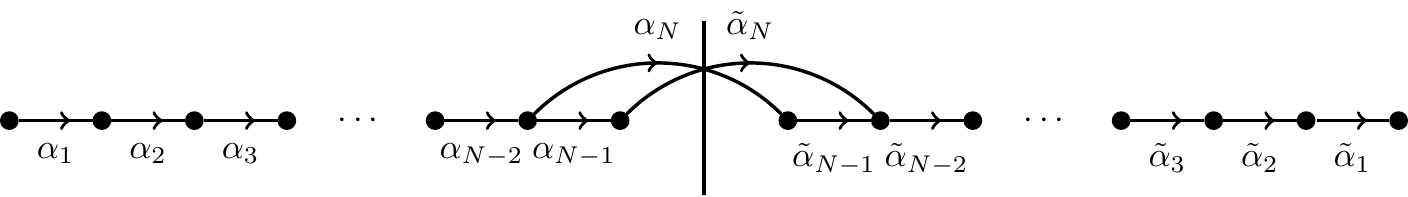}
\caption{Brane diagram of strings/roots stretching, for $SO(2N)$. The $B$ and
$C$-branes are turned into an orientifold, which is denoted by a mirror (vertical line).
The strings corresponding to simple roots are illustrated by arrows stretching
between the branes and reflected across the mirror. We note that the
distinguished root $\alpha_{N}$ corresponds to the two-pronged string and
indeed it is made of two legs moving across the $BC$-mirror in order to
respect the difference in charges between the $A$, $B$, and $C$ branes.}%
\label{fig:SO2Nroots}%
\end{figure}

\begin{figure}[t!]
\centering
\includegraphics{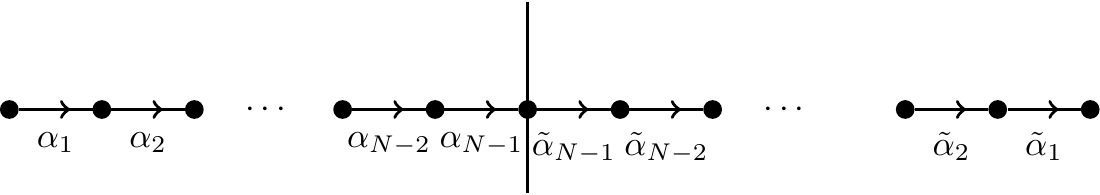}
\caption{Brane diagram of strings/roots stretching, for $SO(2N-1)$. The $B$
and $C$-branes are turned into an orientifold denoted by a mirror (vertical
line) and one of the $A$-branes is squeezed onto it. The strings corresponding
to simple roots are illustrated by arrows stretching between the branes and
reflected across the mirror.}%
\label{fig:SO2Nm1roots}%
\end{figure}

As we can see, the presence of the mirror guarantees that even parts (in the
partition of $2N$ or $2N-1$) appear an even number of times whenever we use
any of the regular one-pronged simple strings. Thus, using the same rules as
with $SU(N)$, we can generate most allowed partitions corresponding to $SO$
groups. We note that unlike $SU(N)$, we also have the presence of a two-pronged
string coming as a result of the distinguished root $\alpha_{N}$ of $SO(2N)$.
This can result in configurations where the partitions are not so obvious from
the string junction picture. We can thus turn to the equivalent matrix
representation and read off the corresponding partition from the equivalence
class it belongs to. To do that, we once again need to specify what basis we
are using. Generalizing the rules from $\mathfrak{su}_{N}$ listed in the
previous section to $\mathfrak{so}_{2N}$, we have the following $N(N-1)$
nilpositive elements:

\begin{itemize}
\item Half of them are: $E_{\mathrm{1-pronged}} = E_{i,j}-(-1)^{j-i}
E_{2N-j+1,2N-i+1}$ with $1 \leq i < j \leq N$ corresponding to one-pronged
strings stretching from the $i^{th}$ to the $j^{th}$ $A$-brane, as well as
their reflections--namely, the strings stretching between the $(2N-j+1)^{th}$
and the $(2N-i+1)^{th}$ nodes, which are on the right-hand side of the mirror.
These correspond to the $\mathfrak{su}_{N} \subset\mathfrak{so}_{2N}$
nilpositive generators.

\item The other half are: $E_{\mathrm{2-pronged}} = E_{i,2N-j+1}- (-1)^{j-i}
E_{j,2N-i+1}$ with $1 \leq i < j \leq N$ corresponding to two-pronged strings
stretching between the $i^{th}$ and $(2N-j+1)^{th}$ nodes as well as the
$j^{th}$ and $(2N-i+1)^{th}$ nodes.
\end{itemize}

The associated $N(N-1)$ nilnegative elements are simply $E^{T}_{\mathrm{1-pronged}}$
and $E^{T}_{\mathrm{2-pronged}}$. These correspond to the same one- and two-pronged
strings but with their directions reversed. Finally, we have $N$ Cartans:
The first $(N-1)$ come from one-pronged strings: $H_{i} = [E_{i,i+1}%
+E_{2N-i,2N-i+1},E_{i+1,i}+E_{2N-i+1,2N-i}]$ for $1 \leq i \leq N-1$. These
correspond to the $\mathfrak{su}_{N} \subset\mathfrak{so}_{2N}$ Cartan
generators. The last generator is then given by $H_{N} = [E_{N-1,N+1}%
+E_{N,N+2},E_{N+1,N-1}+E_{N+2,N}]$

Note the presence of negative values introduced by the reflection across
the $BC$-mirror. We choose our convention such that simple roots only contain
positive entries. The minus signs are then imposed to some non-simple roots
simply because they are given by commutators of simple root. For instance the
non-simple string $\alpha_{1}+\alpha_{2}$ inside $SO(8)$ is represented by the
matrix $[E_{1,2}+E_{7,8}, E_{2,3}+E_{6,7}] = E_{1,2} \cdot E_{2,3} - E_{6,7} \cdot E_{7,8} = E_{1,3}-E_{6,8}$.

As a result of the above equations, the simple positive roots (corresponding
to the simple strings of figure \ref{fig:SO2Nroots}) are then given by the
matrices $E_{i,i+1}+E_{2N-i,2N-i+1}$ for $1 \leq i \leq N-1$
and $X^{SO(2N)}_{N} = E_{N-1,N+1}+E_{N,N+2}$. The positive simple roots for
$SO(2N-1)$ are identical, except for the last one. Indeed, we have:
$E_{i,i+1}+E_{2N-i,2N-i+1}$ for $1 \leq i \leq N-2$ (as
before) but the shorter simple root is $\sqrt{2} \left(E_{N-1,N} + E_{N,N+1}\right)  $.
The remaining non-simple roots are simply
obtained by taking the appropriate commutators.

As an example of a partition which is not immediately obvious from the string
junction picture, we can stretch the two strings $\alpha_{N}$ and $\alpha
_{N-1}$ from figure \ref{fig:SO2Nroots}. The associated matrix makes it
obvious what orbit such configuration belongs to: in particular, it corresponds to
the $2N \times2N$ matrix $M=E_{N-1,N}+E_{N+1,N+2}+E_{N-1,N+1}+E_{N,N+2}$ which
belongs to the nilpotent orbit of $[3,1^{2N-3}]$.

With this set of strings and corresponding matrices we can now associate to
each partition a string junction picture. Just like for $SU(N)$ we have many
choices. For instance, the three diagrams of figure \ref{fig:equivSO} all
represent the same $[3^{2},1^{2}]$ partition:

\begin{itemize}
\item The first string junction picture has a matrix representation $M_{1} =
E_{1,2}+E_{7,8}+E_{2,3}+E_{6,7}$.

\item The second configuration has matrix representation $M_{2} =
E_{2,3}+E_{6,7}+E_{3,4}+E_{5,6}+E_{2,5}-E_{4,7}$.

\item The third has matrix representation $M_{3}%
=E_{1,2}+E_{7,8}+E_{2,5}-E_{4,7}$.
\end{itemize}

\begin{figure}[t!]
\centering
\includegraphics{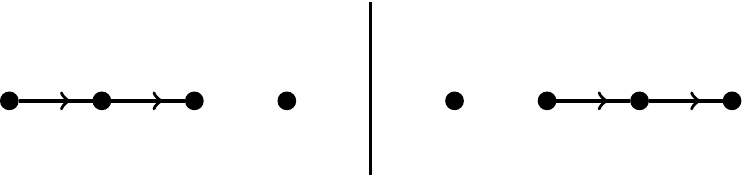}
\par
\includegraphics{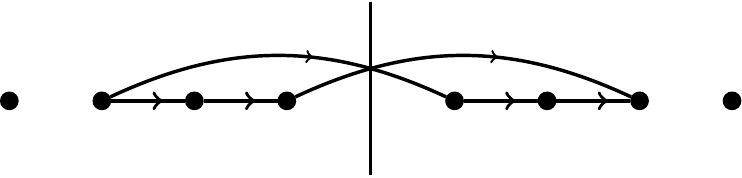}
\par
\includegraphics{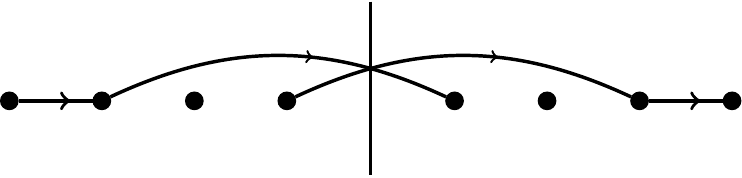}
\caption{Three equivalent ways of describing the partition $[3^{2},1^{2}]$ in
the set of nilpotent orbits of $SO(8)$. To each picture is associated a
different matrix, but there exists an inner automorphism that can bring them
all to the same Jordan block decomposition. Therefore, they belong to the same
equivalence class.}%
\label{fig:equivSO}%
\end{figure}

In order to keep our diagrams as simple as possible, we chose representatives
which only make use of the simple strings from figure \ref{fig:SO2Nroots},
whenever possible. However, unlike $SU(N)$, the $SO(2N)$ and $SO(2N-1)$ algebras
also contain distinguished orbits. These orbits cannot be described
with only simple roots and must therefore involve one or more non-simple
strings. We observe such a special case in the distinguished orbit $[5,3]$ of
$SO(8)$ (see figure \ref{fig:SO8hasse}). Our string junction diagrams then
allow us to recognize distinguished orbits as those requiring the presence of
one or more non-simple strings.

The groups $SO(4N)$ contain ``very even'' orbits. These are orbits
with corresponding partition given by only even parts. Such partitions split into two
separate orbits, such as $[2^{4}]^{I}$ and $[2^{4}]^{II}$ or $[4^{2}]^{I}$ and
$[4^{2}]^{II}$ in $SO(8)$. That is, the matrix representation of a
$[\lambda^{\mu}]^{I}$ and a $[\lambda^{\mu}]^{II}$ configuration have the same
Jordan block decomposition and are therefore related by an \textit{outer}
automorphism. However, they are not related by any \textit{inner} automorphism
and thus do not actually belong to the same nilpotent orbit. This splitting to
two orbits for the very even partitions simply comes from the symmetry of the
Dynkin diagram for $D_{n}$: namely, the exchange of the last two roots
$\alpha_{N-1}$ and $\alpha_{N}$. This means that a very even partition
involving $\alpha_{N-1}$ (a one-pronged string) will be labeled $[\lambda
^{\mu}]^{I}$ while its companion very even partition involving $\alpha_{N}$
instead (a two-pronged string) will be labeled $[\lambda^{\mu}]^{II}$. This is
illustrated in figure \ref{fig:veryEven}.

\begin{figure}[t!]
\centering
\includegraphics[scale=.95]{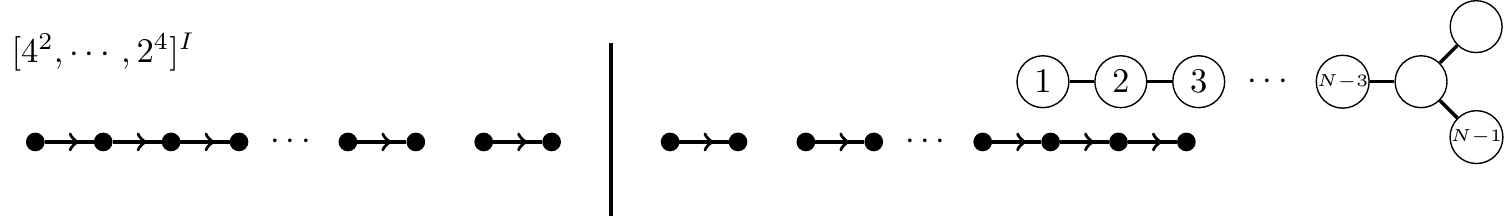}
\par
\includegraphics[scale=.95]{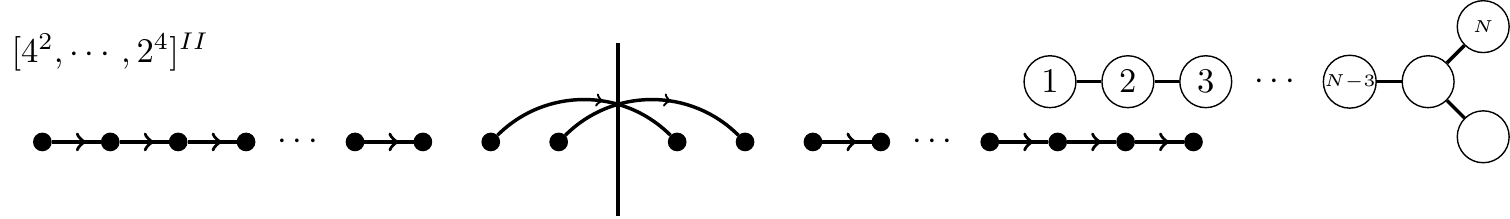}
\caption{Two very even partitions that yield the same partition but do not
belong to the same nilpotent orbit. The first one only involves one-pronged
strings and is labeled $[4^{2},\cdots,2^{4}]^{I}$ while the second one
replaces $\alpha_{N-1}$ with the two-pronged string $\alpha_{N}$ and is
labeled $[4^{2},\cdots,2^{4}]^{II}$. To the right we give the Dynkin diagrams
with the corresponding strings turned on.}%
\label{fig:veryEven}%
\end{figure}

We briefly mention the triality automorphism of $SO(8)$ in figure
\ref{fig:triality}. Namely, we know that the nilpotent orbits with partitions
$[3,1^{5}]$, $[2^{4}]^{I}$, and $[2^{4}]^{II}$ are all related by the triality
outer automorphism. Indeed, they are represented by the following set of
roots: $\{\alpha_{3},\alpha_{4}\}$, $\{\alpha_{1},\alpha_{3}\}$, and
$\{\alpha_{1},\alpha_{4}\}$ respectively. Similarly the partitions $[5,1^{3}%
]$, $[4^{2}]^{I}$, and $[4^{2}]^{II}$ also form a trio. There is no
inner automorphism that exists between these representations, which implies that
they do indeed belong to different nilpotent orbits.

\begin{figure}[t!]
\centering
\includegraphics{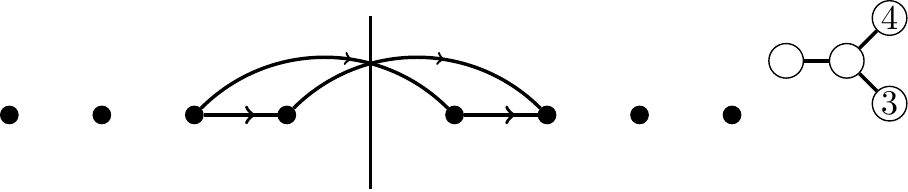}
\par
\includegraphics{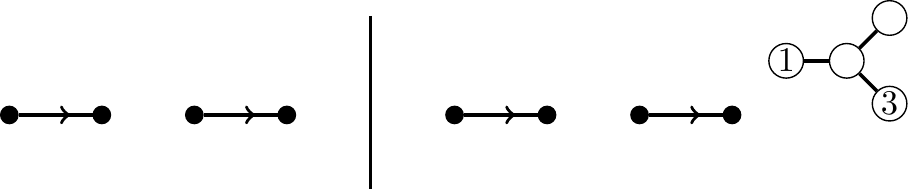}
\par
\includegraphics{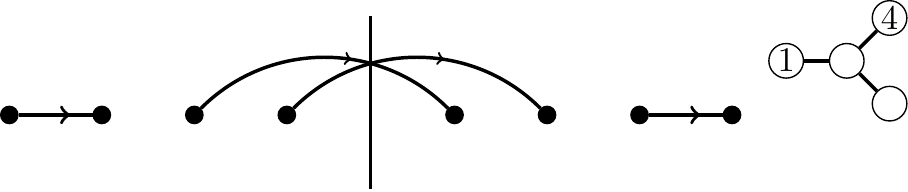}
\caption{Triality of $SO(8)$ illustrated by the three different
representations corresponding to partitions $[3,1^{5}]$ (top), $[2^{4}]^{I}$
(middle), and $[2^{4}]^{II}$ (bottom). The corresponding simple roots used are
illustrated in the adjacent Dynkin diagrams. The first has a matrix
representation $M_{1} = E_{3,4}+E_{5,6}+E_{3,5}+E_{4,6}$, the second is given
by $M_{2} = E_{1,2}+E_{7,8} + E_{3,4}+E_{5,6}$, and the last by $M_{3} =
E_{1,2}+E_{7,8}+E_{3,5}+E_{4,6}$. These all correspond to different nilpotent
orbits because there exists no inner automorphism between these three
matrices.}%
\label{fig:triality}%
\end{figure}

By starting with a configuration with no string attached
($[1^{2N-1}]$ partition for $SO(2N-1)$ or $[1^{2N}]$ for $SO(2N)$) we can add
more and more strings to go from the $[2^2,1^{2N-5}]$ or $[2^{2},1^{2N-4}]$
orbit all the way to the $[2N-1]$ or $[2N]$ partitions. We summarize all of
the nilpotent orbits of $SO(7)$ and $SO(8)$ in figures \ref{fig:SO7hasse} and
\ref{fig:SO8hasse} respectively.

\begin{figure}[t!]
\centering
\includegraphics{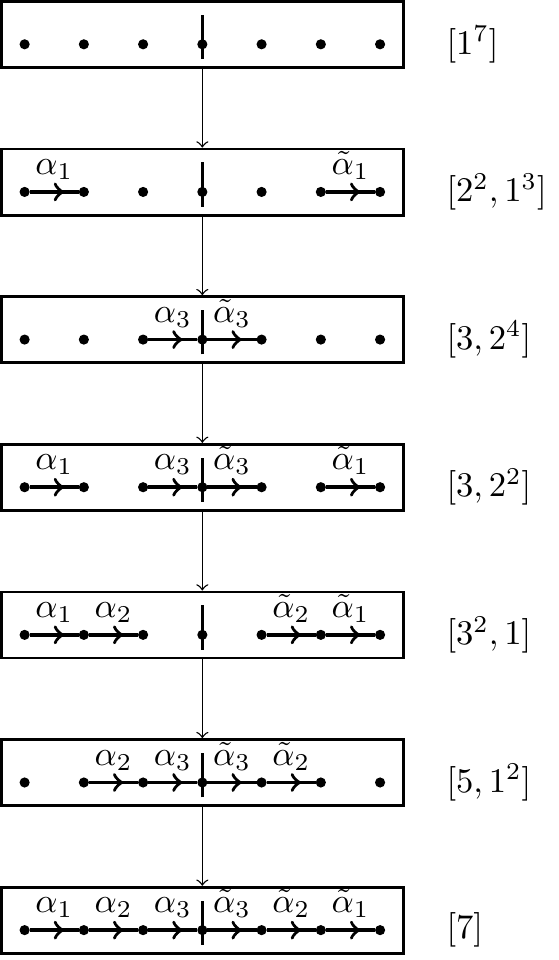}
\caption{Hasse diagram of $SO(7)$ nilpotent deformations going from the
smallest orbits at the top to largest orbits at the bottom. All simple roots
are present and every corresponding simple string is connecting the
$A$-branes. In the case of the last simple root, one $A$-brane is connecting
to the middle $A$-brane located on the $BC$-mirror.}%
\label{fig:SO7hasse}%
\end{figure}

\begin{figure}[!htb]
  \centering
  \includegraphics{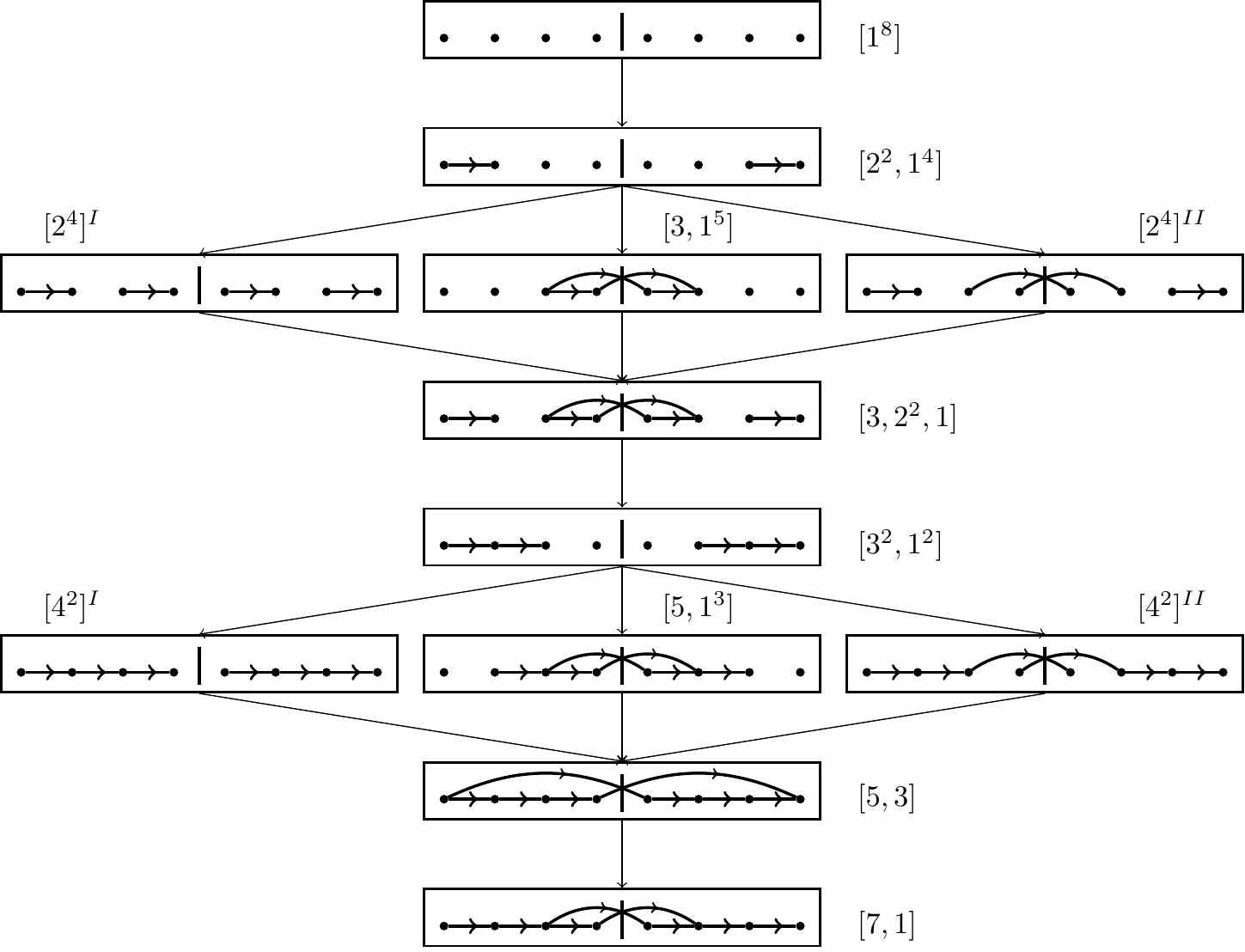}
  \caption{Hasse diagram of $SO(8)$ nilpotent deformations going from top (UV) to bottom (IR) where all simple roots are present and every corresponding simple string is connecting adjacent $A$-branes or in the case of the last simple root, two $A$-branes are connected to the $BC$-mirror.}
    \label{fig:SO8hasse}
\end{figure}

Finally, much like what we have seen in $SU(N)$, most flows include the simple
addition of a root/string and therefore are obvious. However, there are a few
cases that are not so immediately clear. We work them out here in the case of
$SO(8)$ and note that the methods below extend to the higher rank $SO$ groups.

\begin{itemize}
\item {$[2^{2},1^{4}]\rightarrow[3,1^{5}]$:} We can add to $\alpha_{1}$ the
highest positive root $\alpha_{2,1,3,2,4}=\alpha_{1}+2 \alpha_{2}+\alpha
_{3}+\alpha_{4}$ (identified with the matrix $E_{1,7}+E_{2,8}$). This setup is
represented by the matrix $E_{1,2}+E_{7,8}+\epsilon\left(  E_{1,7}+E_{2,8}
\right)  $, which belongs to the same orbit as $E_{3,4}+E_{5,6}+E_{3,5}%
+E_{4,6}$ and corresponds to the diagram involving the set of simple strings
$\{\alpha_{3}, \alpha_{4}\}$.

\item {$[3,2^{2},1]\rightarrow[3^{2},1^{2}]$:} We can add the non-simple
string $\alpha_{2}+\alpha_{3}+\alpha_{4}$ to the initial set $\{\alpha
_{1},\alpha_{3},\alpha_{4}\}$. This gives the matrix $E_{1,2}+E_{7,8}%
+E_{3,4}+E_{5,6}+E_{3,5}+E_{4,6}+\epsilon\left(  E_{2,6}+E_{3,7} \right)  $
which is similar to the matrix $E_{1,2}+E_{7,8}+E_{2,3}+E_{6,7}$.

\item {$[3^{2},1^{2}] \rightarrow[5,1^{3}]$:} We can add the non-simple string
$\alpha_{2}+\alpha_{3}+\alpha_{4}$ to the set of simple roots $\{\alpha_{1},
\alpha_{2}\}$ to obtain the matrix $E_{1,2}+E_{7,8}+E_{2,3}+E_{6,7}%
+\epsilon\left(  E_{2,6}+E_{3,7} \right)  $. This matrix is similar to the one
corresponding to the set of strings $\{\alpha_{2}, \alpha_{3}, \alpha_{4}\}$.

\item {$[5,1^{3}], [4^{2}]^{II} \rightarrow[5,3]$} Starting from the set of simple
roots $\{\alpha_{2}, \alpha_{3}, \alpha_{4}\}$ of $[5,1^{3}]$ we can add the
positive root $\alpha_{1}+\alpha_{2}+\alpha_{3}$ to obtain the equivalent set
$\{\alpha_{1}, \alpha_{2}, \alpha_{3}, \alpha_{2}+\alpha_{3}+\alpha_{4}\}$.

Similarly, starting from the set of simple roots $\{\alpha_{1}, \alpha_{2},
\alpha_{4}\}$ of $[4^{2}]^{II}$ we can add the positive non-simple root
$\{\alpha_{2}, \alpha_{3}, \alpha_{4}\}$ again to obtain the same Weyl
equivalent set $\{\alpha_{1}, \alpha_{2}, \alpha_{3}, \alpha_{2}+\alpha
_{3}+\alpha_{4}\}$.
\end{itemize}

\subsection{$Sp(N)$}

Recall that in F-theory, we realize the $Sp(N)$-type gauge theories by a non-split $I_N$ fiber.
In terms of 7-branes, this involves the transverse intersection of a stack of D7-branes with an $O7^{-}$-plane along a common
6D subspace. In the IIA realization of this algebra, we can also consider a stack of D6-branes on top of an $O6^{+}$-plane.

For our present purposes, we can merge the $A$-branes pairwise on each side of the mirror. This then yields $N$ nodes on each side of the mirror but with the particularity that a two-pronged string can stretch from a single composite node, as seen in table \ref{tab:classicalRootTable}. Zooming out, the two-pronged string -- which corresponds to the long simple root of $Sp(N)$ -- gets squished into a double arrow coming out of the same node and connecting to its mirror-image across the $BC$-branes. This means that, unlike with $SO(2N)$ algebras, we can now draw a double string stretching from the same node and crossing the $BC$-mirror. The simple root $\alpha_N$ of figure  \ref{fig:SpNroots} is one example of the $N$ double string connections that can be stretched that way. In terms of the IIA description, the change in orientation
of the mirror means we can now draw all of the same string junctions as for
$SO(2N)$, but we also have an additional $2N$ possible roots which correspond to
double connections coming out of the same node (something that was not allowed
in $SO(2N)$). The set of simple roots/strings for $Sp(N)$ is given in figure
\ref{fig:SpNroots}.

\begin{figure}[t!]
\centering
\includegraphics{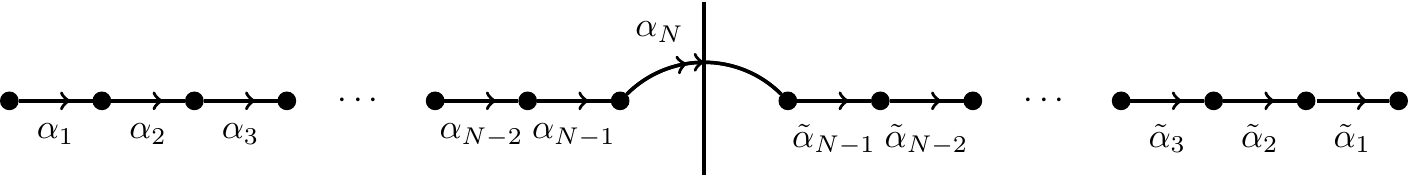}
\caption{Brane diagram of strings/roots stretching, for $Sp(N)$. The
orientifold is once again represented by a mirror (vertical line). The strings
corresponding to simple roots are illustrated by arrows stretching between the
branes and reflected across the mirror. We note that the longer root
$\alpha_{N}$ corresponds to the two-pronged string being squeezed into a
single double arrow crossing the mirror, ensuring that the charge differences
are still respected.}%
\label{fig:SpNroots}%
\end{figure}

The set of simple strings (as illustrated in figure \ref{fig:SpNroots}) along
with the reflecting mirror ensures that odd parts in the partition of
$2N$ must appear with even multiplicity. This exactly matches the constraint
that, in the partitions used to parametrize the nilpotent orbits of $Sp(N)$,
the multiplicity of odd parts must be even. Furthermore, $Sp(N)$ also contains
distinguished orbits, which involve the presence of one or more non-simple root.

Following the same conventions as before, we use the following matrices as the
nilpositive part of the basis for $\mathfrak{sp}_{N}$:

\begin{itemize}
\item $N(N-1)/2$ one-pronged strings $E_{\mathrm{1-pronged}} = E_{i,j}%
-(-1)^{j-i} E_{2N-j+1,2N-i+1}$ with $1 \leq i < j \leq N$ corresponding to
one-pronged strings stretching from the $i^{th}$ to the $j^{th}$ $A$-brane as
well as their reflections. That is the strings stretching between the
$(2N-j+1)^{th}$ and the $(2N-i+1)^{th}$ nodes which are on the right-hand side
of the mirror. These correspond to the $\mathfrak{su}_{N} \subset
\mathfrak{sp}_{N}$ nilpositive generators.

\item $N(N-1)/2$ two-pronged strings $E_{\mathrm{2-pronged}} = E_{i,2N-j+1} +
(-1)^{j-i} E_{j,2N-i+1}$ with $1 \leq i < j \leq N$ corresponding to
two-pronged strings stretching between the $i^{th}$ and $(2N-j+1)^{th}$ nodes
as well as the $j^{th}$ and $(2N-i+1)^{th}$ nodes.

\item $N$ double strings $X_{\mathrm{doubled}} =2 E_{i,2N-i+1}$ with $1 \leq i
\leq N-1$ and the long simple string $X_{N} = E_{N,N+1}$. These correspond to
double-pronged strings merged together into single double connections. They
stretched from the $i^{th}$ to the $(2N-i+1)^{th}$ node.
\end{itemize}

The $N$ doubled strings coming out of the same node are the only new roots
which were not present in $\mathfrak{so}_{2N}$.

We give the Hasse diagram of nilpotent orbits for $Sp(3)$ in figure
\ref{fig:Sp3hasse} to illustrate the possible string junctions. Flows between
each level in the Hasse diagrams follow the same rules as for the $SO$ groups.

\begin{figure}[t!]
\centering
\includegraphics{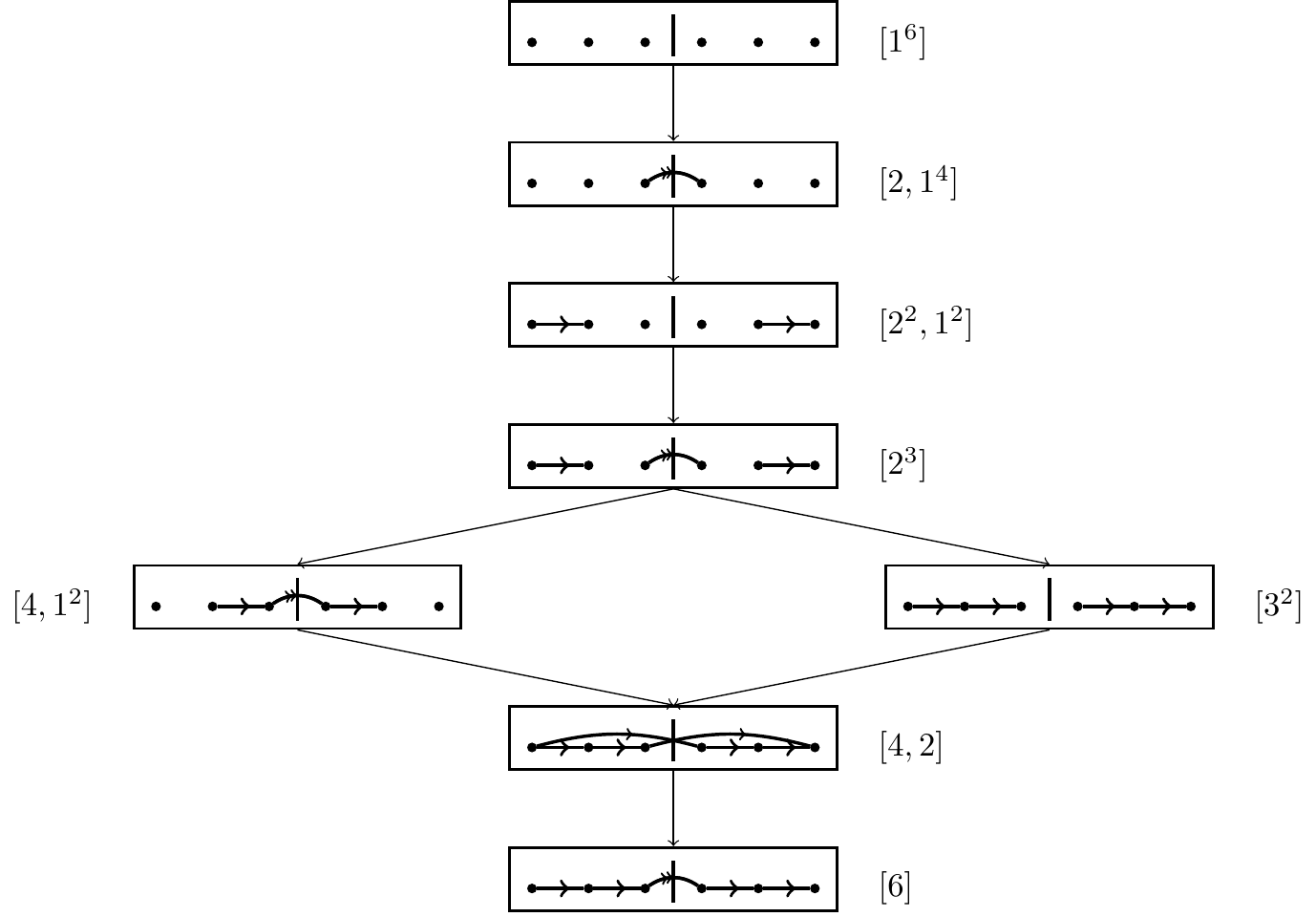}
\caption{Hasse diagram of $Sp(3)$ nilpotent deformations going from top (UV)
to bottom (IR) where all simple roots are turned on and every corresponding
simple strings are connecting the $A$-branes. In the case of the last simple
root, a double connection stretches from the last node and connects across the
mirror, ensuring charge conservation.}%
\label{fig:Sp3hasse}%
\end{figure}

\subsection{An Almost Classical Algebra: $G_{2}$}

\label{subsubsec:AlmostClassical} We next consider the exceptional Lie
group $G_{2}$. Even though the Lie algebra of $G_{2}$ is technically an exceptional
Lie group, the fact that it can easily be embedded inside the Lie algebra of $SO(7)$
makes it behave almost identically. Furthermore, as we are going to encounter
this algebra even when dealing only with classical quivers it is useful to
have a closer look at exactly how one might want to describe it.

First, we note that the monodromy of $G_{2}$ is the
same as for $SO(7)$ and $SO(8)$ that is, there are a
total of four $A$-branes and a $B$ with a $C$ brane. Thus, we can start from
the $SO(7)$ configuration which has four $A$-branes with one stuck on the
$BC$-mirror (see figure \ref{fig:SO7hasse}). Then, we note
that for $G_{2}$, the roots $\alpha_{1}$ and $\alpha_{3}$ are identified while
$\alpha_{2}$ is left untouched. Namely, the branching $SO(7) \rightarrow G_{2}$
takes $\alpha_{1} + \alpha_{3} \rightarrow\alpha_{1}$ and $\alpha_{2}
\rightarrow\alpha_{2}$. Therefore, we obtain the positive roots listed in
figure \ref{fig:G2roots}.

\begin{figure}[t!]
\centering
\includegraphics{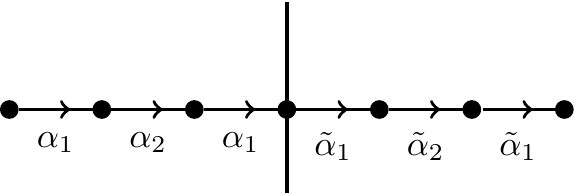}
\caption{Brane diagram of strings/roots stretching, for $G_{2}$. The $B$ and
$C$-branes are turned into an orientifold denoted by a mirror (vertical line)
and one of the $A$-branes is squeezed onto it. Furthermore, the first
$A$-brane is ``linked" to the middle one (as if it were also merged onto the
mirror), so that the first and third root of $SO(7)$ join together as the
first root of $G_{2}$ (as dictated by the quotient which takes $SO(7) \rightarrow
G_{2}$). The strings corresponding to simple roots are illustrated by arrows
stretching between the branes and reflected across the mirror.}%
\label{fig:G2roots}%
\end{figure}The matrix representation is taken directly from $SO(7)$. For the
positive simple roots we have:
\begin{align}
X_{1}  &  \equiv E_{1,2}+E_{6,7}+\sqrt{2} \left(  E_{3,4}+E_{4,5}\right)  ,\\
X_{2}  &  \equiv E_{2,3}+E_{5,6}.
\end{align}
The other four positive roots are given by:
\begin{align}
\left[  X_{1},X_{2}\right]   &  = E_{1,3} - E_{5,7} - \sqrt{2} \left(
E_{2,4}-E_{4,6}\right)  ,\\
\left[  \left[  X_{1},X_{2}\right]  ,X_{1}\right]   &  = 2 \sqrt{2} \left(
E_{1,4}+E_{4,7}\right)  -2\left(  E_{2,5}+E_{3,6}\right)  ,\\
\left[  \left[  \left[  X_{1},X_{2}\right]  ,X_{1}\right]  ,X_{1}\right]   &
= 6 \left(  E_{1,5}-E_{3,7}\right)  ,\\
\left[  \left[  \left[  \left[  X_{1},X_{2}\right]  ,X_{1}\right]
,X_{1}\right]  ,X_{2}\right]   &  = 6 \left(  E_{1,6} + E_{2,7}\right)  .
\end{align}

As a result, we can now give the four non-trivial nilpotent orbits of $G_{2}$
in terms of strings (see figure \ref{fig:G2hasse}). We note that, once again, we
have a simple correspondence with partitions of $7$, illustrated by the
groupings allowed from the associated string junctions. The ordering is a total ordering rather than a mere partial ordering (unlike
for most larger groups), and the flows from one orbit to the other follow from the fact that they are projections of the previously studied $SO(7)$ symmetry.

\begin{figure}[t!]
\centering
\includegraphics{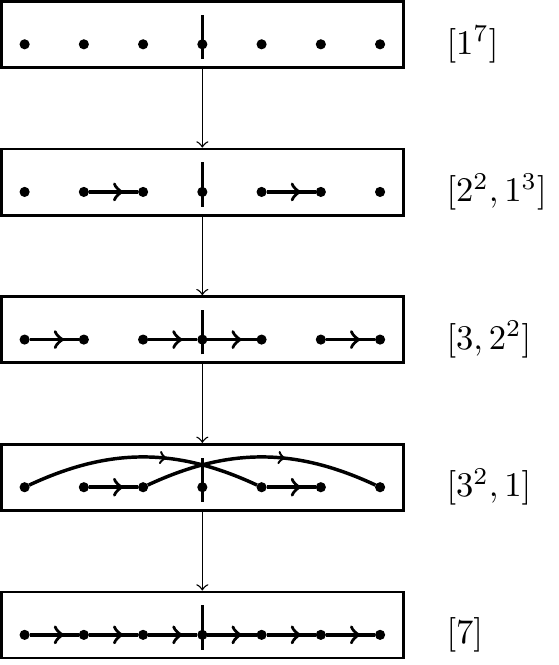}
\caption{Hasse diagram of $G_{2}$ nilpotent deformations going from top (UV)
to bottom (IR) where both simple roots are present so that both corresponding
simple strings are there to connect all $7$-branes and mirror image branes}%
\label{fig:G2hasse}%
\end{figure}

\subsection{Nilpotent Orbits for Exceptional Algebras}

\label{sec:NilpotentExceptionals}
We now turn our attention to the exceptional Lie algebras $E_{6,7,8}$. These
distinguish themselves from the classical algebras in several ways. First,
their nilpotent orbits are not simply described by partitions but rather by
Bala-Carter labels. These labels are in one-to-one correspondence with a
weighted Dynkin diagram and a set of roots. Interestingly, when the matrix
representations of these roots are added together, their Jordan block
decomposition still yields a unique partition. Thus, we can still parametrize
the nilpotent orbits of $E_{6,7,8}$ by partitions of $27$, $56$, and $248$
(corresponding to the dimension of their respective fundamental
representations). These partitions arise from the branching of the
fundamental of $E_{N}$ to the $SU(2)$ associated to the nilpotent orbit. However,
there does not exist a simple set of rules or restriction on these partitions like we
have seen for the classical Lie algebras. Thus this classification is very limited.

By making use of string junctions and the brane configuration describing these
algebras, it is however possible to gain a little more insight into the
structure of nilpotent orbits for these exceptional groups. Physically, we know that the $E_{N}$ symmetries are given by $A^{N-1}BC^{2}$ or
equivalently $A^{N}XC$ brane configurations. The advantage of using the description with
an $X$-brane is that we can now branch $E_{N}$ to
$SU(N) \times U(1)$, where the $SU(N)$ piece is represented by $N$ $A$-branes
and $N-1$ ordinary open strings (i.e. one beginning and one end) stretching between them.
States charged under the $U(1)$ factor necessarily involve multi-prong strings which attach to
this stack of $A$-branes and also involve the $XC$ stack. This procedure matches identically the
initial setup used for describing $SO(2N)$ symmetries. The only difference is
that we now have a generalized mirror made out of an $X$ and a $C$ brane
instead of simply a $B$ and $C$ branes. This means that it now takes a
three-pronged string stretching from three $A$-branes to attach to the
$XC$-mirror in order to conserve the charges. Indeed, the charges from an $X$
and a $C$ brane now sum to $[3,0]$ which is exactly three times that of an $A$
brane. As a result we obtain the brane and string configurations given in
figure \ref{fig:ENrootsFtheory}.
\begin{figure}[t!]
\centering
\includegraphics[scale=.9]{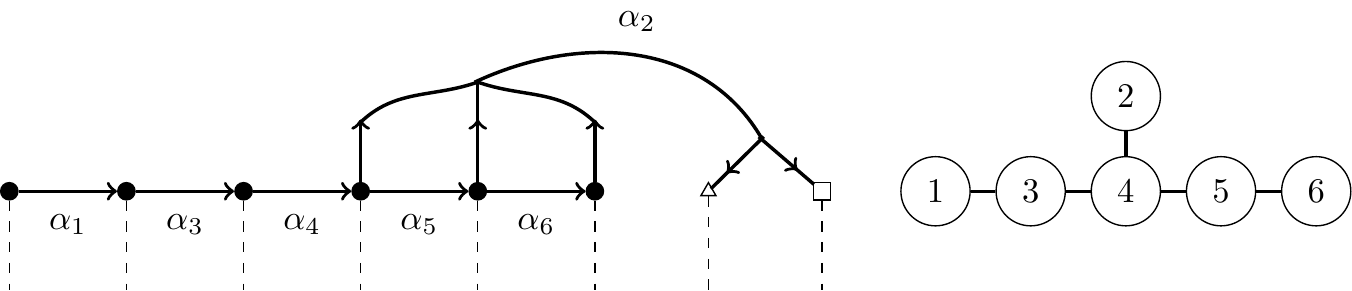}
\vspace{.5cm}
\par
\includegraphics[scale=.9]{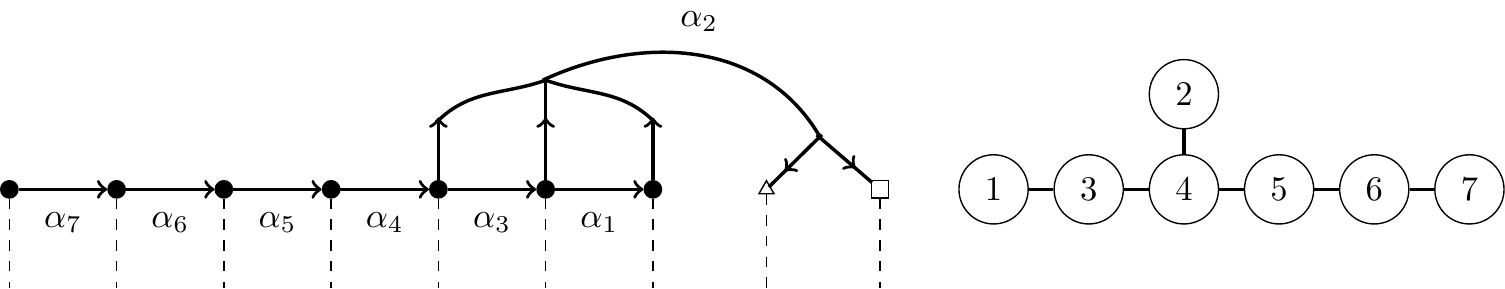}
\vspace{.5cm}
\par
\includegraphics[scale=.9]{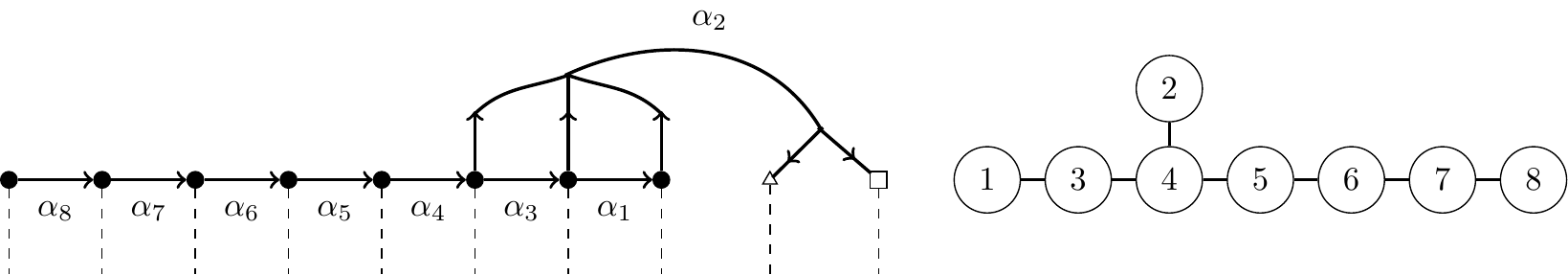}
\caption{Brane diagram of strings/roots stretching between the $A$, $X$, and
$C$-branes making up the $E_{6,7,8}$ symmetry (see \cite{DeWolfe:1998eu}). The
$A$-branes are denoted by black circles, the $X$-brane by an empty triangle
and the $C$-brane by an empty square. The dashed lines represent the position
of branch cuts. Again, these branch cuts are not drawn in subsequent pictures. To the
right we give the corresponding Dynkin diagram with simple roots numbered.}%
\label{fig:ENrootsFtheory}%
\end{figure}

We then treat the $X$ and $C$ branes together as a generalized mirror and use
the short-hand picture of figure \ref{fig:ENroots} where the $XC$-mirror is
represented by an $\times$ inside a circle to differentiate it from the
vertical line that represented the $BC$-mirror for the orientifold seen in
the $SO(N)$ symmetries.

\begin{figure}[t!]
\centering
\includegraphics{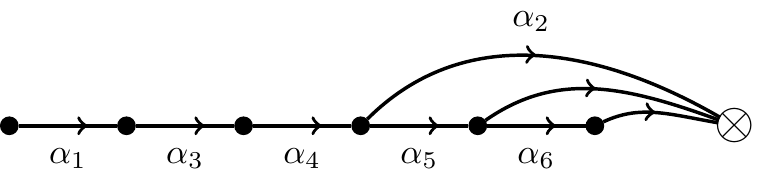}
\vspace{.5cm}
\par
\includegraphics{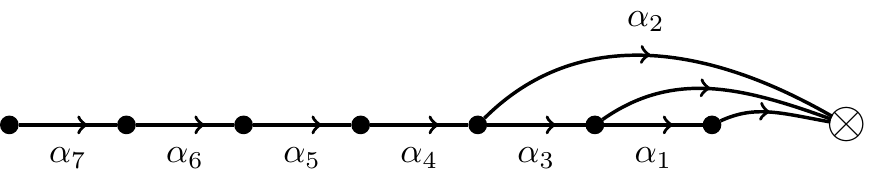}
\vspace{.5cm}
\par
\includegraphics{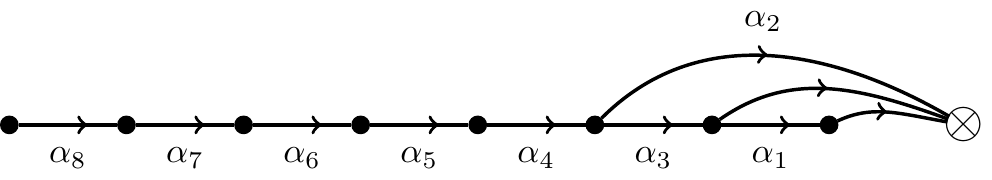}
\caption{Brane diagram of strings/roots stretching, for $E_{6,7,8}$. The $X$
and $C$-branes are turned into a generalized mirror denoted by a crossed
circle. The strings corresponding to simple roots are illustrated by arrows
stretching between the branes. We note that the distinguished root $\alpha
_{2}$ corresponds to the three-pronged string and indeed is made of three-legs
attaching to the $XC$-mirror in order to respect the difference in charges
between the $A$, $X$, and $C$ branes.}%
\label{fig:ENroots}%
\end{figure}

This $XC$-mirror is more complicated than the simply reflecting mirror for
the classical algebras. Indeed, we can think of this mirror as fragmenting the
partitions of $27$, $56$, and $248$ according to their branching rules.
The fundamental representation of $E_{N}$ branches to irreducible representations of
$SU(N) \times U(1)$ as:
\begin{align}
&  \mathbf{27} \rightarrow\mathbf{\overline{15}}_{0} + \mathbf{6}_{1} +
\mathbf{6}_{\scalebox{0.75}[1.0]{-}1}, & \text{for } E_{6} \rightarrow SU(6)
\times U(1),\\
&  \mathbf{56} \rightarrow\mathbf{\overline{21}}_{\scalebox{0.75}[1.0]{-}2} +
\mathbf{21}_{2} +\mathbf{\overline{7}}_{6} + \mathbf{7}%
_{\scalebox{0.75}[1.0]{-}6}, & \text{for } E_{7} \rightarrow SU(7) \times
U(1),\\
&  \mathbf{248} \rightarrow\mathbf{63}_{0} + \mathbf{56}_{3} +
\mathbf{\overline{56}}_{\scalebox{0.75}[1.0]{-}3} + \mathbf{28}%
_{\scalebox{0.75}[1.0]{-}6} + \mathbf{\overline{28}}_{6} + \mathbf{\overline
{8}}_{\scalebox{0.75}[1.0]{-}9} +\mathbf{8}_{9} + \mathbf{1}_{0}, & \text{for
} E_{8} \rightarrow SU(8) \times U(1).
\end{align}
Here, $\mathbf{15}$ is the two-index anti-symmetric representation of $SU(6)$
and $\mathbf{21}$ is the two-index anti-symmetric representation of $SU(7)$.
For the $E_{8}$ case, $\mathbf{63}$ is the adjoint, $\mathbf{28}$ is the
two-index anti-symmetric, $\mathbf{56}$ is the three-index anti-symmetric and
$\mathbf{8}$ is the fundamental representation of $SU(8)$. For the adjoint
representations of $E_{6}$ and $E_{7}$ we also have:
\begin{align}
&  \mathbf{78} \rightarrow + \mathbf{35}_{0} + \mathbf{20}_{1}
+ \mathbf{20}_{\scalebox{0.75}[1.0]{-}1}
+  \mathbf{1}_{2}+ \mathbf{1}_{\scalebox{0.75}[1.0]{-}2}
+  \mathbf{1}_{0}, & \text{for } E_{6} \rightarrow SU(6) \times U(1),\\
&  \mathbf{133} \rightarrow \mathbf{45}_{0} + \mathbf{35}_{\scalebox{0.75}[1.0]{-}4} + \mathbf{\overline{35}}_{4} +\mathbf{7}_{8} +\mathbf{\overline{7}}_{\scalebox{0.75}[1.0]{-}8} +   \mathbf{1}_{0} , & \text{for } E_{7}
\rightarrow SU(7) \times U(1).
\end{align}
By embedding $SU(N)$ inside $E_N$ in this manner, we see that positive strings can be described by any
set of one-pronged strings between the $N$ $A$-branes or any three-pronged string
attaching to three $A$-branes and stretching to the $XC$-mirror.
Furthermore, $E_{6}$ also allows a six-pronged string attaching all of its
$A$-branes to the $XC$-mirror, as illustrated by the trivial representation
$\mathbf{1}_{2}$ in its branching. This string corresponds to the highest root
of $E_{6}$. $E_{7}$ also allows six-pronged strings, as seen by the presence of
$\mathbf{\overline{7}}_{\scalebox{0.75}[1.0]{-}8}$ in its branching (this is
indeed the six index anti-symmetric representation of $SU(7)$). Finally,
$E_{8}$ not only allows six-pronged strings (as seen by the six index
anti-symmetric $\mathbf{\overline{28}}_{6}$ representation), but it also
allows for eight different nine-pronged strings, which connect all eight
$A$-branes to the $XC$-mirror with a double connection stretching from one of
the eight $A$-branes. These rules follow directly from the structure of the
exceptional algebras, as shown in \cite{DeWolfe:1998zf,DeWolfe:1998eu}.
To illustrate these situations, we depict the highest roots of $E_{6}$, $E_{7}$
and $E_{8}$ in figures \ref{fig:highestE6}, \ref{fig:highestE7}, and
\ref{fig:highestE8}.
\begin{figure}[t!]
\centering
\includegraphics{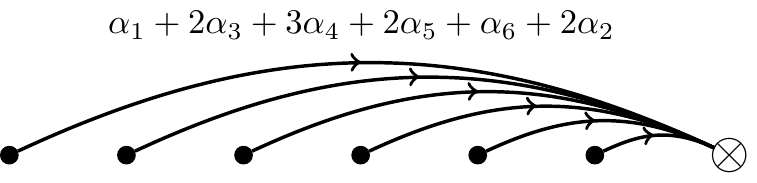}
\caption{Highest roots of $E_{6}$ represented by its corresponding six-pronged
string. It stretches from all six $A$-branes and attaches to the $X$ and $C$
branes represented by the crossed circle.}%
\label{fig:highestE6}%
\end{figure}

\begin{figure}[t!]
\centering
\includegraphics{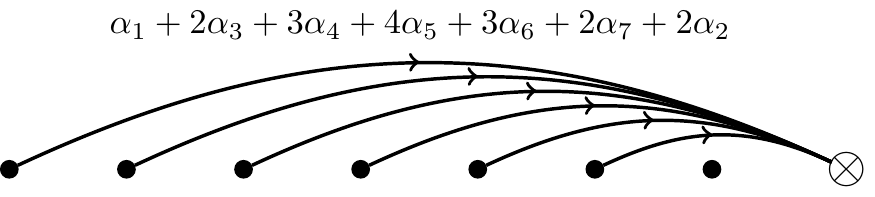}
\caption{Highest roots of $E_{7}$ represented by its corresponding six-pronged
string. It stretches from the six left-most $A$-branes and attaches to the $X$
and $C$ branes represented by the crossed circle.}%
\label{fig:highestE7}%
\end{figure}

\begin{figure}[t!]
\centering
\includegraphics{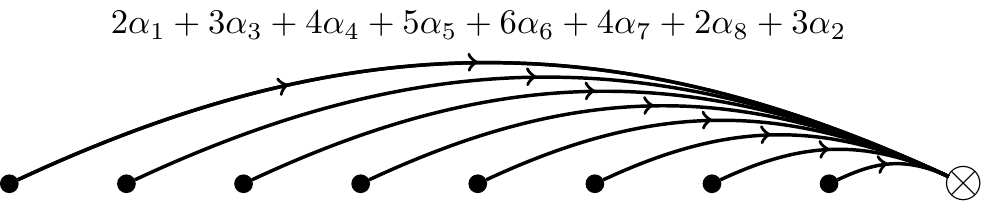}
\caption{Highest roots of $E_{8}$ represented by its corresponding
nine-pronged string. It stretches from all eight $A$-branes (attaching twice
onto the first one) to the $X$ and $C$ branes represented by the crossed
circle.}%
\label{fig:highestE8}%
\end{figure}

In order to describe each nilpotent orbit, we now need to rely more heavily on
the matrix representation. As a result, we associate to each simple string of
figure \ref{fig:ENrootsFtheory} a matrix in the fundamental representation of
$E_{N}$. Any choice of basis will yield the same results, but for reference we
give the simple roots in Appendix \ref{app:ExpMat} and use the method of
\cite{howlett2001matrix} to obtain the remaining non-simple roots.

Next, we proceed just as with the classical algebras. Namely, we start with $N$
$A$-branes next to an $XC$-mirror and start attaching more and more small
string deformations until we reach the deepest nilpotent orbit. To every
string junction diagram we associate a matrix representation which belongs to
some nilpotent orbit. We can differentiate between nilpotent orbits based on
the Bala-Carter label or the partition associated to the matrix (by Jordan
block decomposition). For instance, the diagram involving the first two simple
roots of $E_{6}$ is represented by the matrix $X_{1}+X_{3}$ where
\begin{align*}
X_{1}  &  = E_{1,2} + E_{12,13} + E_{15,16} + E_{17,18} + E_{19,20} +
E_{21,22},\\
X_{3}  &  = E_{2,3} + E_{10,12} + E_{11,15} + E_{14,17} + E_{20,23} +
E_{22,24}.
\end{align*}
This matrix $X_{1}+X_{3}$ has Jordan block decomposition $[3^{6},1^{9}]$ and
is associated to the Bala-Carter label $A_{2}$.

Much as in the case of the classical algebras, multiple diagrams belong to the same
equivalence class. Thus, in order to keep our diagrams as simple as possible,
we choose representative string junction diagrams that only make use of the
simple strings from figure \ref{fig:ENrootsFtheory} whenever possible.
Indeed, once again we identify some distinguished orbits as those which cannot
be described solely by a set of simple roots and necessarily involve non-simple
roots. Furthermore, while any string junction yielding the proper partition
is valid, for simplicity we select configurations with the minimum number of strings required (with as
few non-simple strings as possible) so that the addition of only a single
positive root $\epsilon\cdot X_{k}$ is required to flow to the nearest
nilpotent orbit. We illustrate the nilpotent orbits of $E_{6}$, $E_{7}$, and
$E_{8}$ in figures \ref{fig:E6hasse}, \ref{fig:E7hasse}, \ref{fig:E8hasse}.
The Hasse diagrams labeled by just their Bala-Carter labels can be found in e.g. the Appendix of \cite{Chacaltana:2012zy}, which summarizes several aspects regarding nilpotent orbits of exceptional algebras.

We see that we can move from one nilpotent orbit to the next by small deformations, just like we did for the classical groups.
Furthermore, we can describe every orbit using only simple strings except for
the distinguished ones. These distinguished orbits once again require the
presence of one (or two, for $E_{8}(a_{7})$) non-simple roots.

\subsubsection{The Non-Simply Laced $F_{4} \subset E_{6}$}

\label{subsubsect:F4} Finally, we note that $F_{4} \subset E_{6}$ is obtained
from $E_{6}$ by a very simple identification of simple roots:
\begin{align}
\alpha_{2}^{E_{6}}  &  = \alpha_{1}^{F_{4}},\nonumber\\
\alpha_{4}^{E_{6}}  &  = \alpha_{2}^{F_{4}},\nonumber\\
\alpha_{3}^{E_{6}}+\alpha_{5}^{E_{6}}  &  = \alpha_{3}^{F_{4}},\nonumber\\
\alpha_{1}^{E_{6}}+\alpha_{6}^{E_{6}}  &  = \alpha_{4}^{F_{4}},
\end{align}
where $\alpha_{1}^{F_{4}}$ and $\alpha_{2}^{F_{4}}$ denote the first two short
roots of $F_{4}$ while $\alpha_{3}^{F_{4}}$ and $\alpha_{4}^{F_{4}}$ denote
the longer ones. As a result, we can also simply give the Hasse diagram of
$F_{4}$ as a subset of the one from $E_{6}$.

\begin{figure}[ptb]
\centering
\includegraphics[scale=.95]{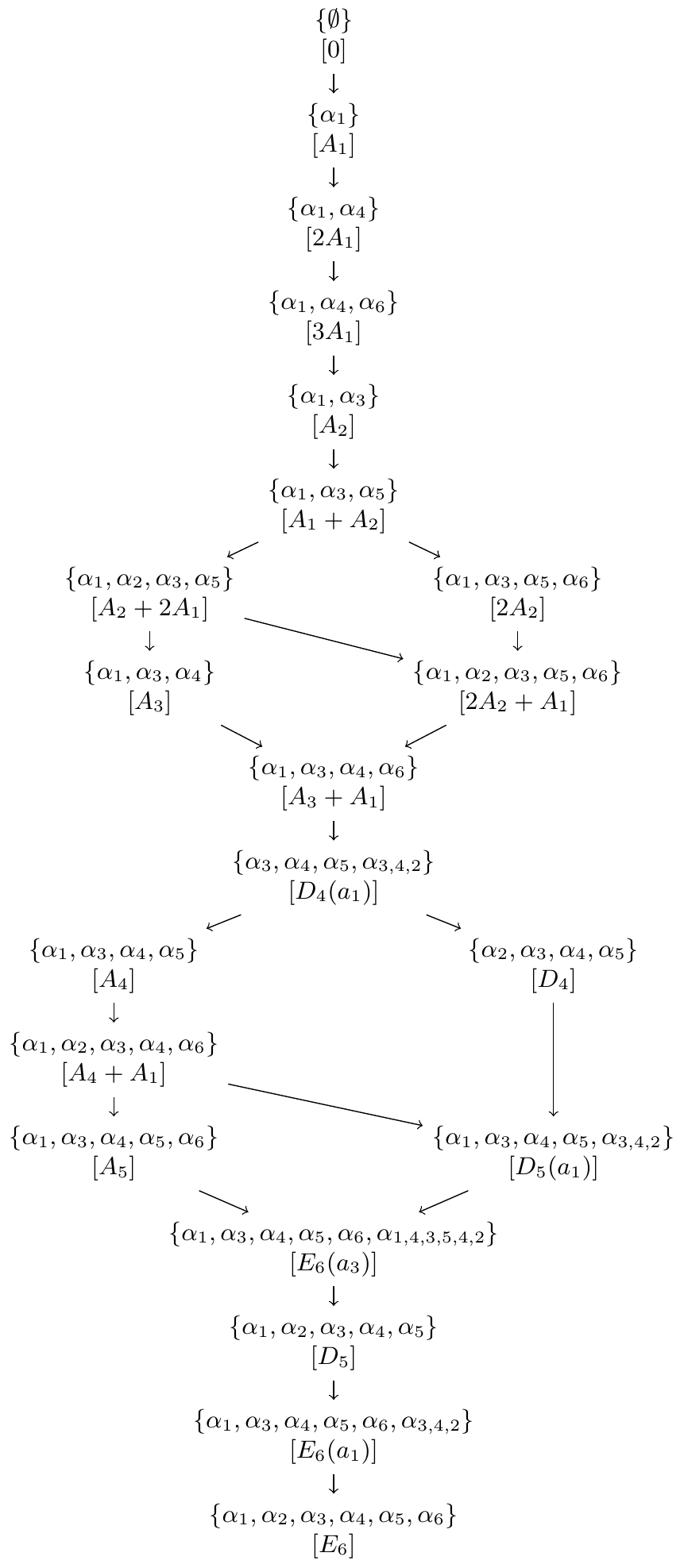}
\caption{Hasse diagram of $E_{6}$ nilpotent deformations going from top (UV)
to bottom (IR) where all simple roots are present, and every corresponding
simple string connects adjacent $A$-branes, or in the case of the second
simple root, three $A$-branes are connected to the $XC$-mirror. For ease of
exposition we only list the set of strings rather than the complete string
junction diagrams for each case.}%
\label{fig:E6hasse}%
\end{figure}

\begin{figure}[ptb]
\centering
\vspace{-1cm}
\includegraphics[scale=.9]{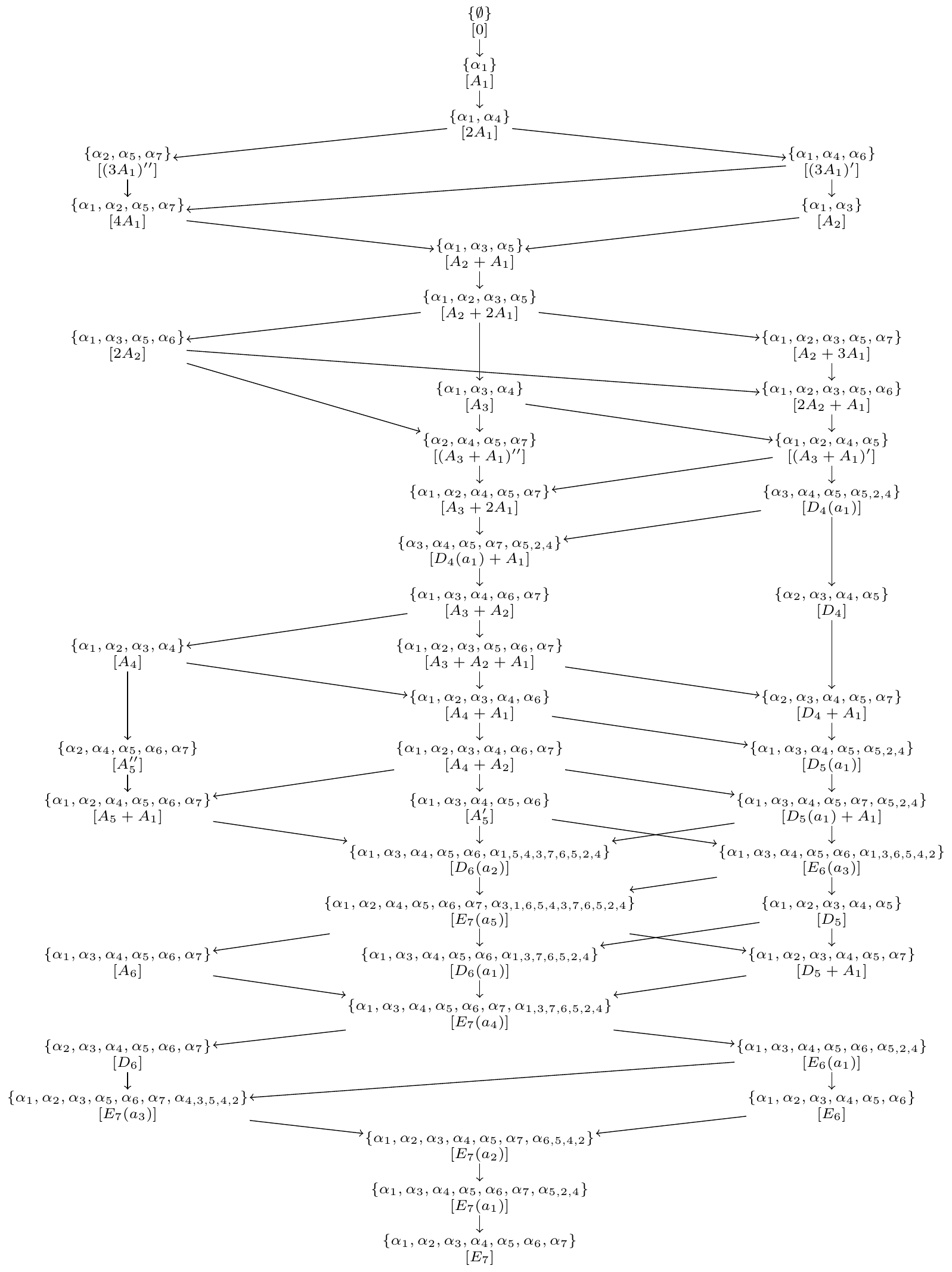}
\caption{Hasse diagram of $E_{7}$ nilpotent deformations going from top (UV)
to bottom (IR) where all simple roots are present, and every corresponding
simple string connects adjacent $A$-branes, or in the case of the second
simple root, three $A$-branes connect to the $XC$-mirror.}%
\label{fig:E7hasse}%
\end{figure}

\begin{figure}[ptb]
\centering
  \vspace{-1cm}
  \includegraphics[scale=.85]{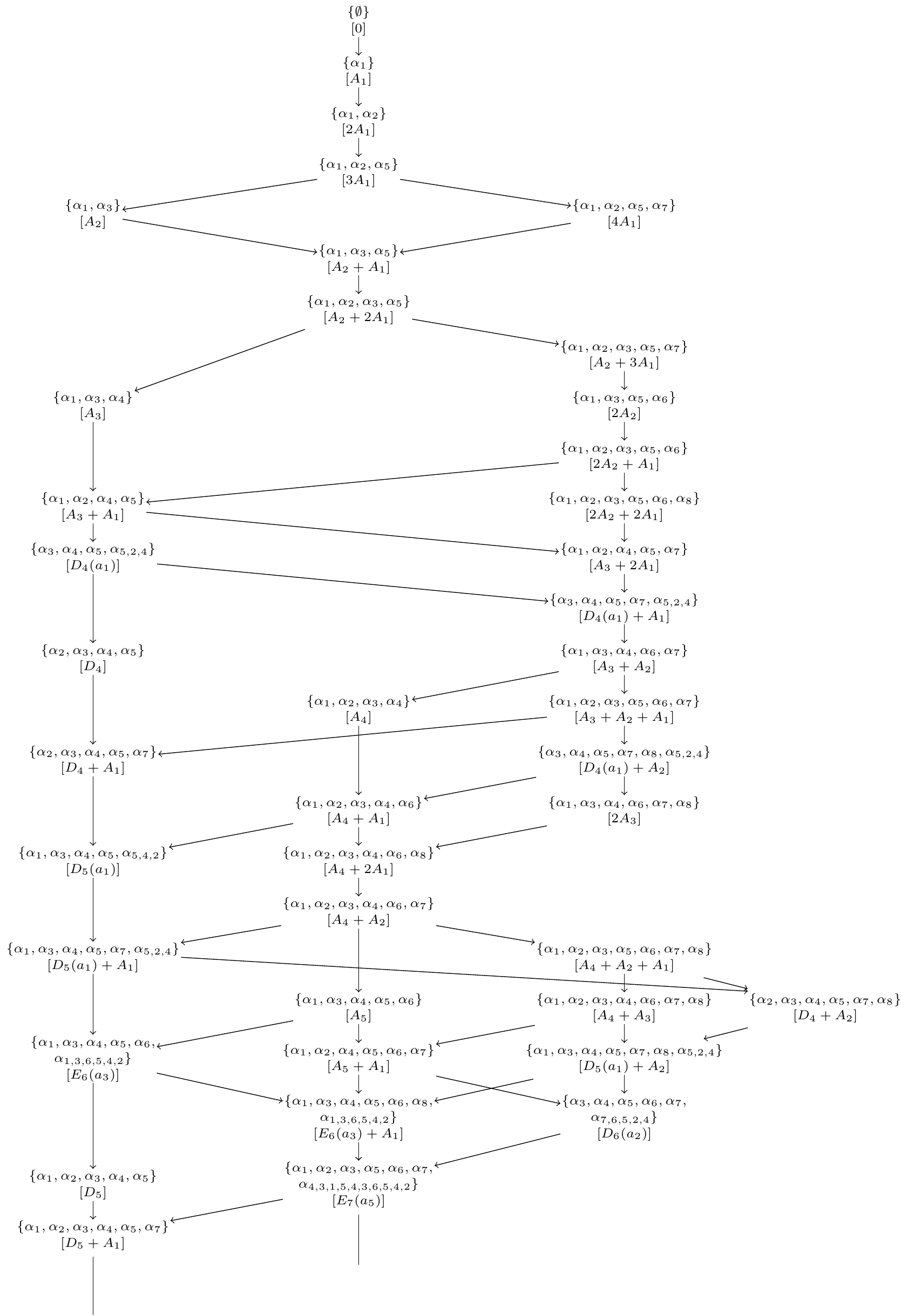}
\end{figure}

\begin{figure}[ptb]
  \includegraphics[scale=.85]{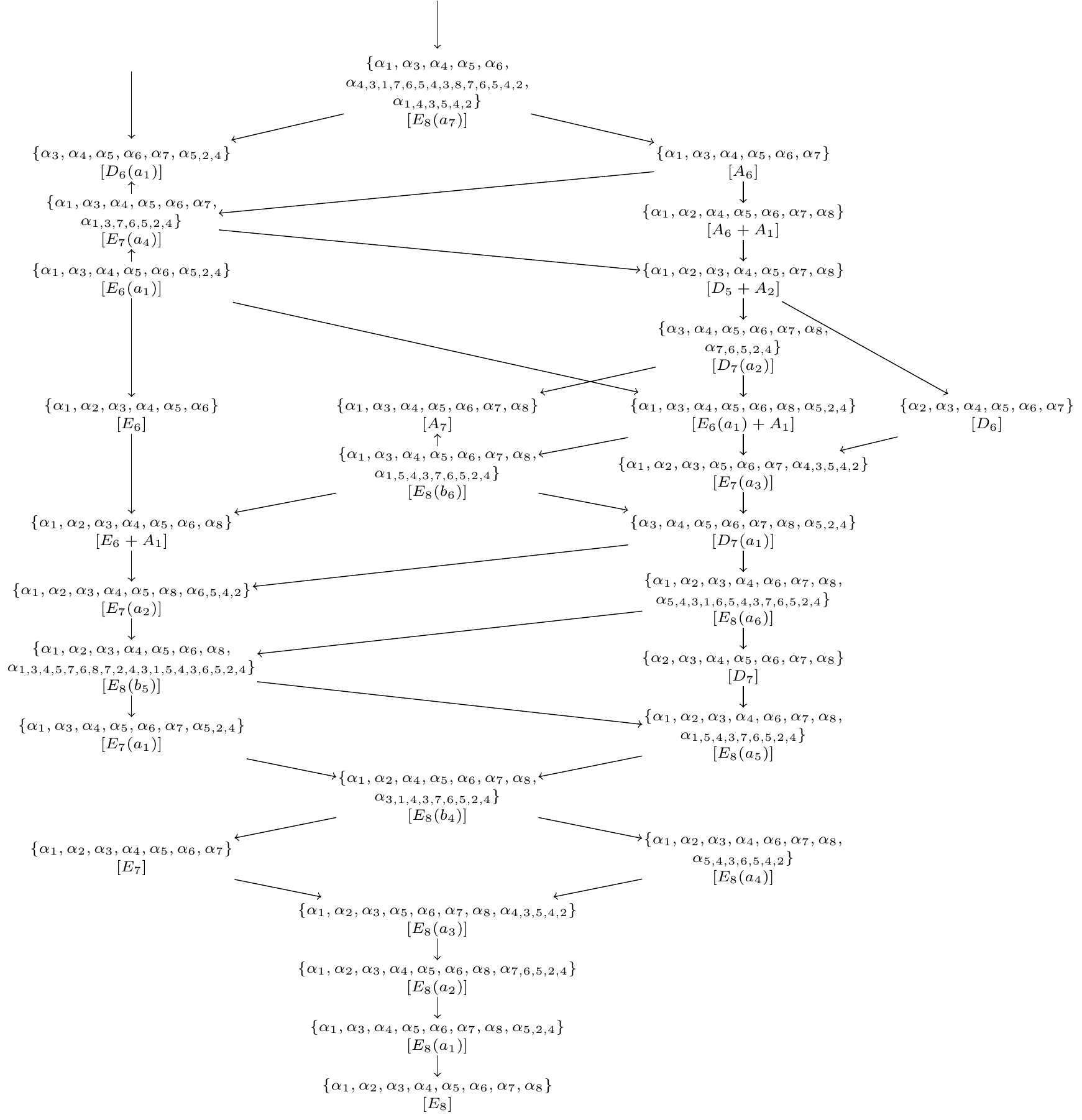}
\caption{Hasse diagram of $E_8$ nilpotent deformations going from top (UV) to bottom (IR) where all simple roots are present, and every corresponding simple string connects adjacent $A$-branes, or in the case of the second simple root, three $A$-branes connect to the $XC$-mirror.}%
\label{fig:E8hasse}%
\end{figure}

\section{Higgsing and Brane Recombination \label{sec:RECOMBO}}

In the previous section, we showed how to generate the entire nilpotent cone of
a semi-simple algebra using the combinatorics of string junctions. In
particular, the operation of \textquotedblleft adding a
string\textquotedblright\ reproduces the expected partial ordering based on
orbit inclusion. We now use this analysis to study Higgs branch flows for
6D\ SCFTs. Our main task here will be to study the effects of brane
recombination triggered by vevs for 6D conformal matter.

We first remark that the picture in terms of string junctions leads to a simple
description of Higgsing with semi-simple deformations. Recall that a semi-simple element is one that is
diagonalizable (in particular, not nilpotent). Since all the quiver-like gauge theories consist of stacks of $A^N$ branes
with either a $BC$ or $XC$ plane, we may join an open string from one stack of $A$-branes to the next, continuing
from left to right across the entire quiver. This leads to a ``peeling off'' of the corresponding $7$-brane, and has the effect of reducing the rank of each of the gauge algebras by one in both the classical case and the exceptional case.

Much more subtle is the case of T-brane deformations.
For the most part, we confine our analysis to the case of quiver-like theories in which all the gauge groups are classical (see figures \ref{fig:SUNUV}, \ref{fig:SO2NUV}, \ref{fig:SO2Nm1UV}, \ref{fig:SpNUV}).
Even in these cases, the matter content of the partial tensor branch can still be strongly coupled, as evidenced by
$SO-SO$ 6D conformal matter. Nonetheless, we will still be able to develop systematic sets of rules to extract the IR fixed point obtained from a given T-brane deformation in such cases.

To some extent, the necessary data is encoded by judiciously applying Hanany-Witten moves involving suspended D6-branes. Such moves were used in \cite{Gaiotto:2014lca}, for instance, to extract different presentations of a given 6D SCFT. To apply the Hanany-Witten analysis of that work to the case at hand, we will need to extend it in two respects. First of all, to cover the case of quiver-like theories with $SO$ gauge algebras, such brane maneuvers sometimes
result in a formally negative number of D6-branes \cite{Heckman:2016ssk, Mekareeya:2016yal}. Additionally, in the case of short quivers, the data specified by
pairs of nilpotent orbits will produce correlated effects in the resulting IR fixed points. To address both points, we will need to extend the available results in the literature.

As we have already mentioned, our main focus will be on tracking brane recombinations as
triggered by the condensation of open strings. In the
context of 6D\ SCFTs, all of this occurs in a small localized region of the base of the non-compact elliptic threefold.
Macroscopic data such as the surviving flavor symmetries corresponds to the asymptotic behavior of non-compact $7$-branes that pass
through this singular region, but which also extend out to the boundary of the
non-compact base. This also means that, provided we hold fixed the total
asymptotic $7$-brane charge present in the configuration, we can consider
any number of \textquotedblleft microscopic processes\textquotedblright\ which
could appear in the physics of brane recombination.

One such process which we shall often use is the
creation of brane / anti-brane pairs localized in the region
near the 6D\ SCFT. We denote such an anti-brane
by $\overline{A}$ and use it in annihilation processes such as:
\begin{equation}\label{AAbar}
A+\overline{A}\rightarrow\text{no branes.}%
\end{equation}
Strictly speaking, such a physical process would generate radiation. The only sense in which
we are really using these objects is to count the overall Ramond--Ramond charge asymptotically far away from
the configuration. In this sense, there will be little distinction between an anti-brane and a ``negative / ghost-brane.''
Since we are primarily interested in determining the end outcome of Higgsing, we use these $\overline{A}$-branes as a
combinatorial tool which must disappear at the final stages of our analysis through processes such as line (\ref{AAbar}).
We refer to this as having a \textquotedblleft Dirac sea\textquotedblright\ of $A/\overline{A}$ pairs of
$7$-branes.

Much as in the case of a general configuration of plus and minus charges in electrodynamics, a lowest energy configuration is obtained by allowing charges to freely move through a material. In much the same way, we shall allow the branes and anti-branes to redistribute. Our main physical condition is that the net $7$-brane charge is unchanged by such processes, and also, that no anti-brane charge remains uncanceled in any
final configuration obtained after Higgsing.

We also remark that from the standpoint of renormalization group flow, these sorts of microscopic details are expected to be irrelevant at long 
distances. Said differently, while there could, a priori, be different UV completions in the full framework of quantum gravity, 
such details will not matter in determining possible fixed points obtained after a Higgs branch deformation. The brane maneuvers indicated 
here are of this sort and are used as a tool to analyze possible fixed points.

Including these formal structures is useful in that it allows us to make sense of the resulting 6D\ SCFT, even when the ranks of the intermediate gauge groups are negative numbers of small magnitude. This procedure has been used
in \cite{Heckman:2016ssk, Mekareeya:2016yal, Apruzzi:2017iqe, Heckman:2018pqx, Frey:2018vpw}
as a way to track the effects of Higgs branch flows
in certain 6D\ SCFTs. We will return to this point in section \ref{sec:GETSHORTY}.

Our main focus in this section will be on determining the Higgs branch flows
associated with the classical algebras, since in these cases there is also a
gauge theory description available for some Higgs branch flows in terms of
vevs of conventional hypermultiplets. Any nilpotent orbit is then described by stretching the appropriate strings as
described in section \ref{sec:NILPJUNC}. We then need to propagate
the deformation by removing some strings as we move deeper into the quiver, which allows us
to read off the resulting gauge symmetries that are left over in the IR. We explain these propagation rules in the following
section.

\begin{figure}[t!]
\centering
\includegraphics{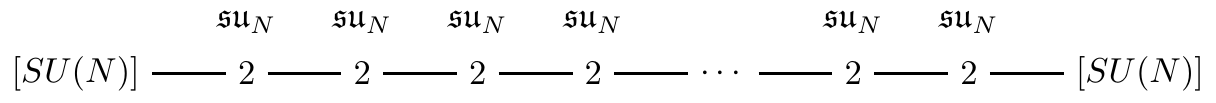}
\caption{Tensor branch of the UV quiver-like theory with just $SU(N)$ gauge algebras.}%
\label{fig:SUNUV}%
\end{figure}

\begin{figure}[t!]
\centering
\includegraphics{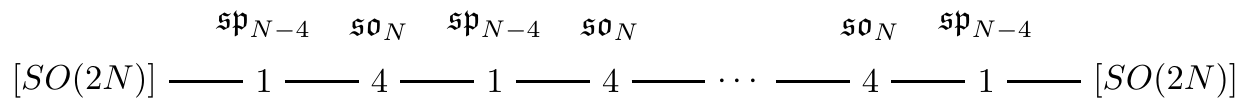}
\caption{Tensor branch of the UV quiver-like theory with just $SO(2N)$ gauge algebras. The full tensor branch also
includes additional $Sp(N-4)$ gauge algebras coming from blowing up the conformal matter between D-type collisions.}%
\label{fig:SO2NUV}%
\end{figure}

\begin{figure}[t!]
\centering
\includegraphics{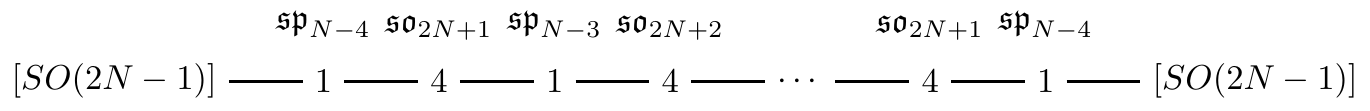}
\caption{Tensor branch of the UV theory with just $SO(2N-1)$ gauge algebras. The full tensor branch also includes
additional $Sp$ gauge algebras coming from blowing up the conformal matter between D-type collisions.
Any deformation with partition $\mu = [\{\mu_i\}]$ in $SO(2N-1)$ is equivalent to the partition $\nu = [\{\mu_i\},3]$ in $SO(2N+2)$.}%
\label{fig:SO2Nm1UV}%
\end{figure}

\begin{figure}[t!]
\centering
\includegraphics{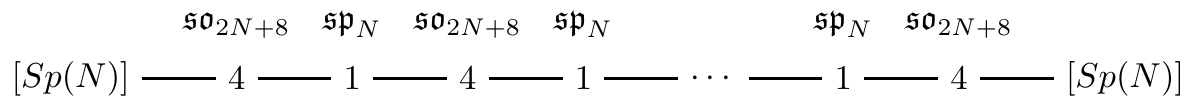}
\caption{UV theory for $Sp(N)$.}%
\label{fig:SpNUV}%
\end{figure}

Before that, however, we need to introduce the possibility of
anti-branes. Indeed, while the nodes in the $SU(N)$ quivers all have the same
number of branes on each level (namely $N$ $A$-branes), the other
classical algebras do not. For instance, a quiver with $SO(2N)$ flavor in the
UV will alternate between $N$ and $N-4$ $A$-branes on the $\mathfrak{so}_{2N}$
and $\mathfrak{sp}_{N}$ levels respectively. This introduces an additional complication in that we may end up with configurations that have
more strings stretching between branes (as dictated by the nilpotent orbit
configuration of section \ref{sec:NILPJUNC}) than are
available according to the gauge group on the quiver node. We remedy this
situation by extracting as many extra $A$ branes as necessary out of the brane
/ anti-brane \textquotedblleft Dirac sea\textquotedblright\ to draw the proper
number of string junctions. These extra branes are then immediately
canceled with the same number of anti-branes.

For example, the theory with $SO(8)$ flavor symmetry
has gauge symmetries alternating between $\mathfrak{sp}_{0}$ (i.e. a trivial gauge group associated with an ``unpaired tensor'' \cite{Morrison:2016djb}) and
$\mathfrak{so}_{8}$, and the nilpotent orbit $[4^{2}]^I$ uses strings
stretching between every brane (i.e. all four $A$-branes and their images have at least one
string attached). However, $\mathfrak{sp}_{0}$ only has the $BC$-mirror
and no $A$-brane. So, in order to describe the $[4^{2}]^I$ nilpotent orbit, we
must introduce four $A$-branes through which we can stretch strings (on each
side of the mirror) and then add them with four anti-branes. This also applies to
the non-simply laced classical algebras, since they can be
obtained from Higgs branch flows of $SO($even$)$ quiver-like theories \cite{Heckman:2015bfa}.

Notably, there are a few cases, even for SO- and Sp-type quivers,
which require non-perturbative ingredients such as E-string / small instanton
deformations. In these cases, the number of tensor multiplets in the
6D\ SCFT\ also decreases. Our method using brane / anti-brane pairs
carries over to these situations and allows us to obtain a complete
picture of Higgs branch flows in these cases as well. We use this
feature in section \ref{sec:GETSHORTY} to determine IR fixed points in the case of short quivers.

Our plan in the rest of this section is as follows: first, we discuss a IIA realization of quiver-like theories
with classical gauge groups, and especially the treatment of Hanany-Witten moves in such setups. After this, we
state our rules for how a T-brane propagates into the interior of a quiver with classical gauge algebras. We then illustrate with several examples the general procedure for Higgsing such theories. This provides a uniform account of brane recombination and also
agrees in all cases with the result expected from related F-theory methods (when available). We also
comment on some of the subtleties associated with extending this to the case of quiver-like theories with exceptional
algebras.

\subsection{IIA Realizations of Quivers with Classical Gauge Groups}\label{subsec:HananyWitten}

To aid in our investigation of Higgs branch flows for 6D SCFTs, it will also prove convenient to
use the type IIA realizations of the quiver-like theories with classical algebras, as used previously in references
\cite{Hanany:1997gh, Brunner:1997gk, Brunner:1997gf, Gaiotto:2014lca}. In the case of quivers with $SU$ gauge group factors,
each classical gauge group factor is obtained from a collection of D6-branes suspended between spacetime filling NS5-branes, with non-compact ``flavor'' D8-branes emanating ``out to infinity.'' The case of $SO$ algebras on the partial tensor branch is obtained by also including $O6^{-}$-planes coincident with each stack of D6-branes. In this case, the NS5-branes can fractionate to $\frac{1}{2}$ NS5-branes. Working in terms of these fractional branes, there is an alternating sequence of $O6^+$ and $O6^-$ planes, and correspondingly an alternating sequence of $SO$ and $Sp$ gauge group factors. This all matches up with the F-theory realization of these theories, where each
$SO$ factor originates from an $I_n^{\ast}$ fiber and each $Sp$ factor from a non-split $I_m$ fiber.

The utility of this suspended brane description is that we can write several equivalent brane configurations which realize the same IR fixed point via ``Hanany-Witten moves,'' much as in the original reference \cite{Hanany:1996ie} and its application to 6D SCFTs in reference \cite{Gaiotto:2014lca}. This provides a convenient way to uniformly organize the data of Higgs branch deformations generated by nilpotent orbits. In fact, we will shortly demonstrate that using these brane moves along with some additional data (such as the appearance of brane / anti-brane pairs) provides an intuitive method for determining the resulting fixed points in both long and short quivers.

Since we will be making heavy use of the IIA realization in our analysis of Higgs branch flows, we now discuss such constructions in greater detail. In our analysis, we will also consider formal versions of Hanany-Witten moves which would seem to involve a negative number of branes. These cases are closely connected with strong coupling phenomena (such as the appearance of small instanton transitions and spinor representations) and can be fully justified in the corresponding F-theory realization of the same SCFT. Indeed, the description in terms of Hanany-Witten moves extends to the F-theory description, so we will interchangeably use the two conventions when the context is clear.

\subsubsection{SU($N$)}
We begin with a quiver-like theory with $L-1$ tensor multiplets and for each one a paired $SU(N)$ gauge group factor. The UV theory has a tensor branch given by the quiver
\begin{center}
\includegraphics{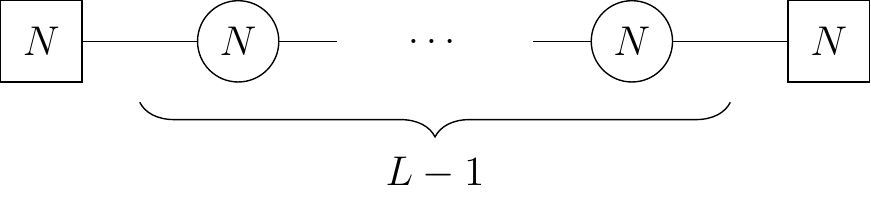},
\end{center}
which is realized in terms of the IIA brane setup:
\begin{center}
\includegraphics{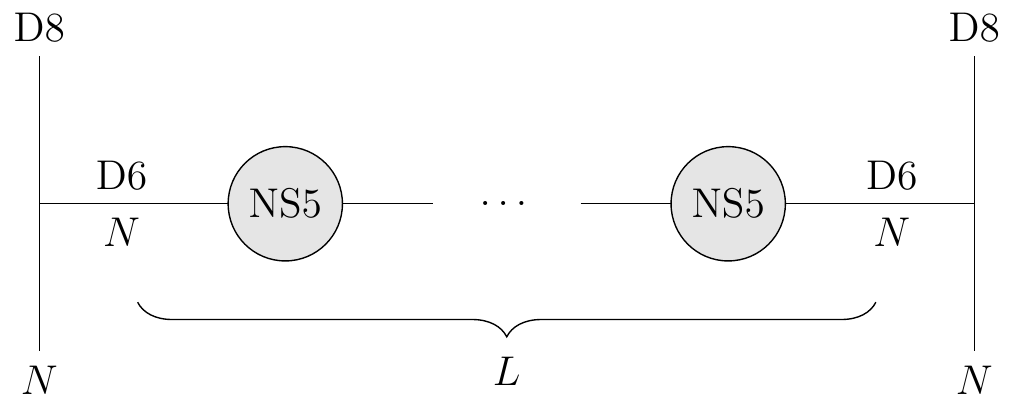}\,.
\end{center}
From the point of view of the D6-branes, the D8-branes specify boundary conditions, which are controlled by the Nahm equations \cite{Diaconescu:1996rk}. These pick three ($X^i$, $i=1,2,3$) out of the $N^2-1$ scalars controlling the Higgs branch and relates them to the distance $t$ of the intersection point by
\begin{equation}
  X^i \sim \frac{T^i}{t}.
\end{equation}
The generators $T^i$ describe an $SU(2)$ subgroup of the flavor symmetry SU($N$), whose embedding is captured by a partition of $N$. This happens on both sides of the quiver. Thus all the data we need in order to study Higgs branch flows of the UV theory are two partitions $\mu_L$ and $\mu_R$ of $N$ and the length $L$.

A partition $\mu$ of $N$ is given in terms of $l\le N$ integers $\mu^i$ with $\mu^1 \ge \mu^2 \ge \dots \ge \mu^l$ and $\mu^1 + \mu^2 + \dots \mu^l = N$. In the corresponding brane realization, the two partitions describe the separation of the stack of $N$ D8-branes on each side into smaller stacks
\begin{center}
\includegraphics{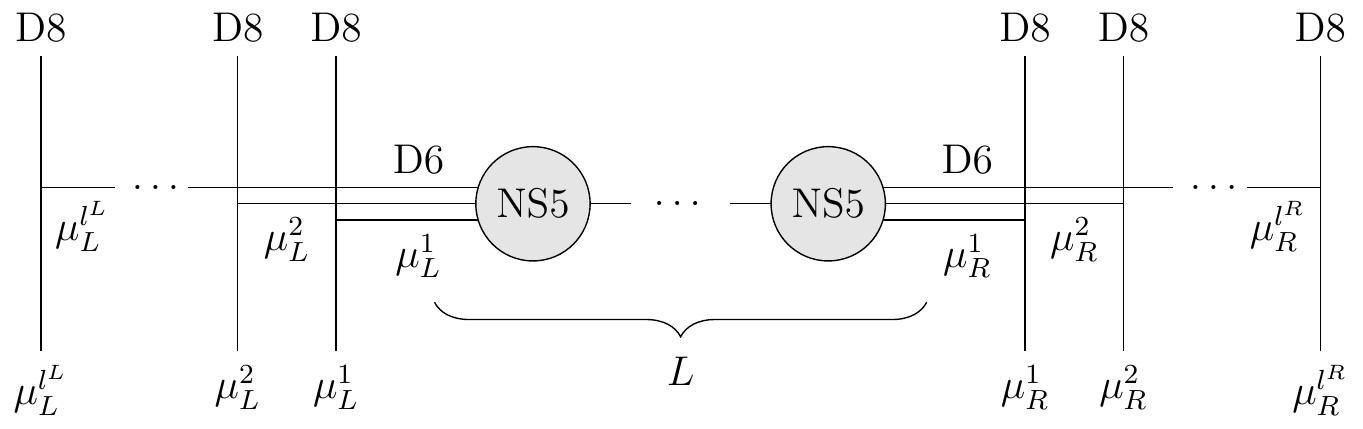}\,.
\end{center}
The brane picture is particularly useful because we can easily read off the IR theory from it. This works by applying Hanany-Witten moves, which swap a D8-brane and an NS5-brane, until all of the D8-branes are balanced. Looking at the stack of $\mu_L^1$  D8-branes left of the first NS5-branes, we can measure its imbalance by the difference $\Delta n$ of D6-branes departing from the right and arriving on the left. A balanced stack would have $\Delta n=0$, but for the setup depicted above we find $\Delta n = \mu_L^1$ instead. After performing the Hanany-Witten move described in figure~\ref{fig:hwmove}, $\Delta n$ becomes
\begin{equation}
  \Delta n' = \Delta n - 1 \qquad \text{with} \qquad
    \Delta = n_2 - n_1 \quad \text{and} \quad
    \Delta' = n_3 - n_2' \,.
\end{equation}
Hence, we have to perform exactly $\Delta n = \mu_L^1$ Hanany-Witten moves to balance this stack.
\begin{figure}[t!]
\centering
\includegraphics{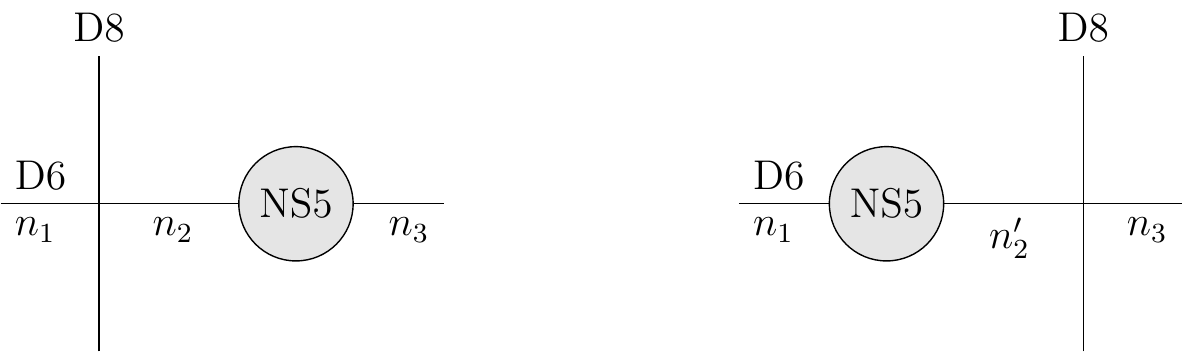}
\caption{The basic operation of swapping a D8- and NS5-branes. The relation between the number of D6-branes stretching between the D8-brane and the NS5-brane before ($n_2$) and after ($n_2'$) the swapping is given by $n_2' = n_1 + n_3 - n_2 + 1$.}\label{fig:hwmove}
\end{figure}
One can always balance all D8-branes provided that the length of the quiver $L$ is large enough. This constraint will become important when we discuss short quivers in section~\ref{sec:GETSHORTY}. Once all D8-branes are balanced, the resulting IR quiver gauge theory can be read off by using the building blocks
\begin{center}
\includegraphics{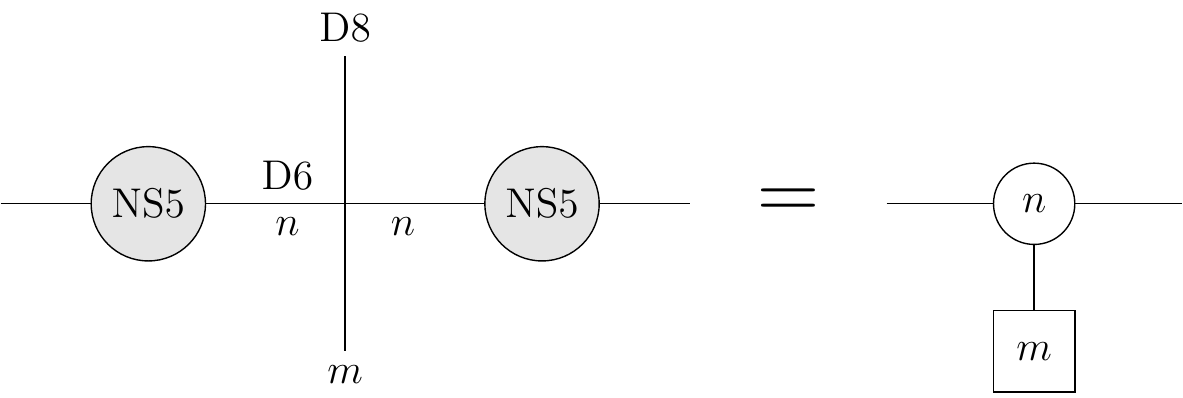}\,.
\end{center}

Applying subsequent Hanany-Witten moves results in a simple, algebraic description of the Higgs branch flows. Let us, for simplicity, consider very long quivers. In this case it is sufficient to just focus on one partition, i.e. $\mu_L$, since the analysis on the right-hand side is perfectly analogous. Using the fact that a stack of $\mu^i_L$ D8-branes moves $\mu^i_L$ NS5-branes to the right until it is balanced, we can read off the flavor symmetries of the IR theory directly from the partition. However, obtaining the number of D6-branes stretched between each pair of adjacent NS5s is slightly more complicated. If we denote this number as $n_i$ between the $i$'s and $i+1$'s NS5s we find the following recursion relation
\begin{equation}
  (n_i)_j = \begin{cases} (n_i)_{j-1} - \mu_L^j + i & \text{for } i < \mu_L^j \\
    (n_i)_{j-1} & \text{otherwise}\,.
  \end{cases}
\end{equation}
Here $(n_i)_j$ denotes the $n_i$ after the $j$'th stack of NS5-branes has been balanced. Hence, the initial condition is $(n_i)_0 = N$, and we are interested in $(n_i)_{l_L}$, which describes the number of D6-branes once all D8-branes have been balanced. An example for $N=6$ is $\mu=[3\,2\,1]$, for which we find
\begin{equation}
  \begin{aligned}
    (n_i)_1 &= \begin{pmatrix} 4 & 5 & 6 & 6 & \dots \end{pmatrix} \\
    (n_i)_2 &= \begin{pmatrix} 3 & 5 & 6 & 6 & \dots \end{pmatrix} \\
    (n_i)_3 &= \begin{pmatrix} 3 & 5 & 6 & 6 & \dots \end{pmatrix}
  \end{aligned}
\end{equation}
with the resulting IR quiver
\begin{center}
  \includegraphics{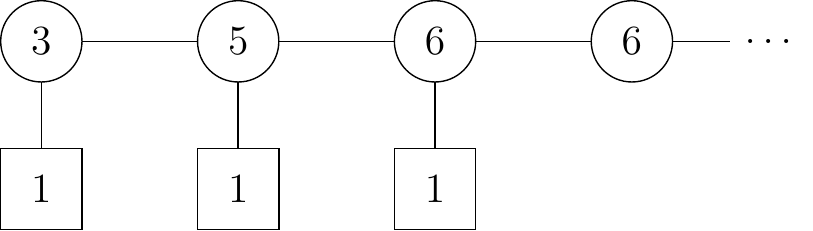}\,.
\end{center}

\subsection{SO($2N$), SO($2N + 1$) and Sp($N$)}
Gauge groups SO($2N$), SO($2N + 1$) and Sp($N$) arise if the setup from the last subsection is extended to include O6 orientifold planes placed on top of the D6-branes. In particular, assume we have $N$ physical D6-branes. Each of these has a mirror image under the $\mathbb{Z}_2$ orientifold action $\Omega$, and thus we have in total $2N$ 1/2 D6-branes. Their Chan-Paton factors transform under $\Omega$ as $\Omega \lambda = M \lambda^T M^{-1}$. Since $\Omega^2 = 1$, we therefore find two different solutions for $M$, which are denoted as $M_\pm = \pm M_\pm^T$. Each of these solutions gives rise to a distinguished orientifold action $\Omega_\pm$. Only massless open string excitations satisfying $\Omega_\pm \lambda = - \lambda^T$ survive the orientifold projection. Depending on whether $\Omega_-$ (O6$^-$) or $\Omega_+$ (O6$^+$) is used, the resulting gauge group is either SO($2N$) or Sp($N$). If a single 1/2 D6-branes is exactly on top of the O6$^-$ plane, it becomes its own mirror and we obtain the gauge group SO($2N+1$). Similar to the D6-branes, a single NS5-branes on the orientifold plane splits into two half NS5-branes:
\begin{center}
\includegraphics[scale=.9]{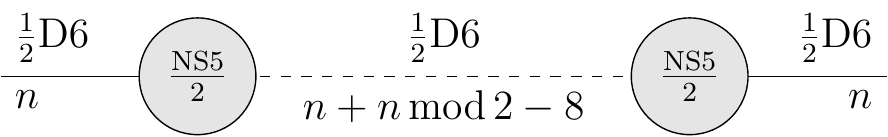}
\end{center}
Here, we depict a stack of 1/2 D6-branes on O6$^-$ with a solid line and a stack of 1/2 D6-branes on O6$^+$ with a dashed line. Because the D6-charge of the O6$^+$ differs by 4 from the one of the O6$^-$ the number of 1/2 D6-branes changes from $n$ to $n + n\,\mathrm{mod}\,2 - 8$ and back.

\begin{figure}[htb]
  \centering
  \begin{tabularx}{.9\textwidth}{l l}
  SO($2N-1$) & \multicolumn{1}{m{2cm}}{\includegraphics[scale=.9]{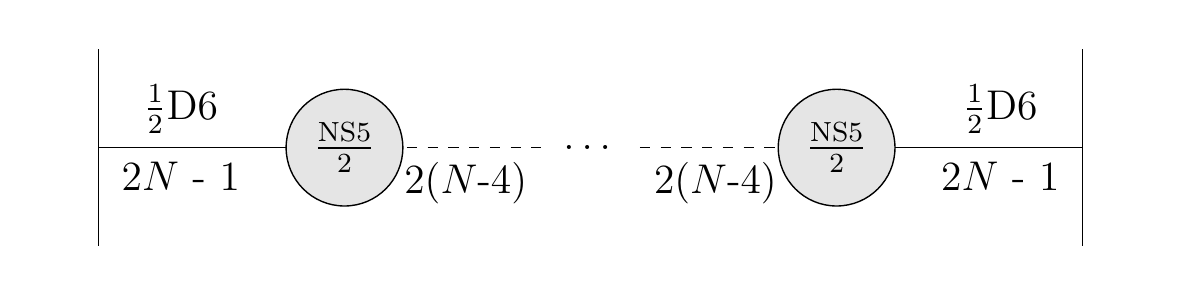}} \\
  SO($2N$)   & \multicolumn{1}{m{2cm}}{\includegraphics[scale=.9]{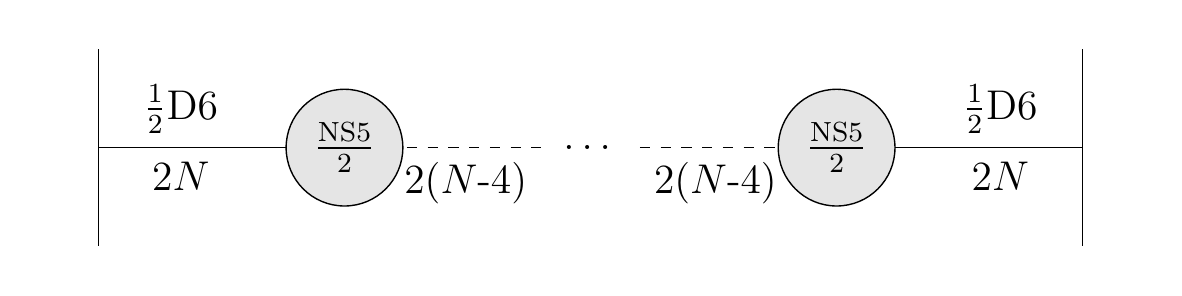}} \\
  Sp($N$)    & \multicolumn{1}{m{2cm}}{\includegraphics[scale=.9]{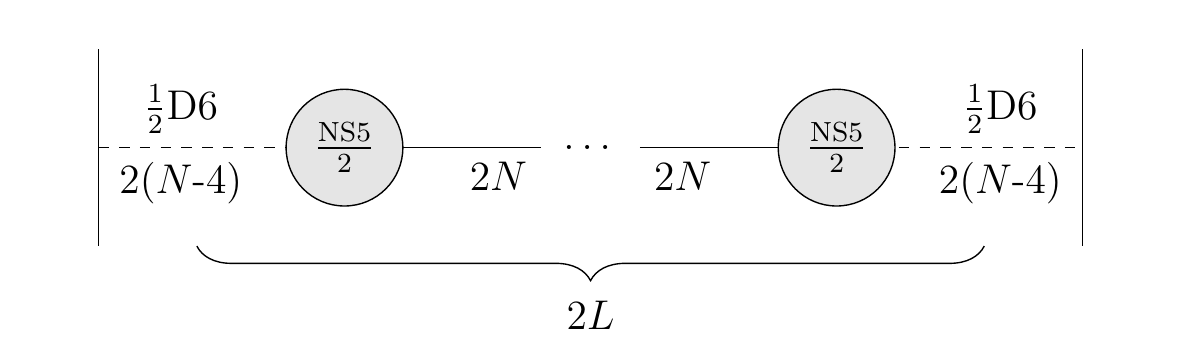}}
  \end{tabularx}
  \caption{Suspended brane realization of UV quivers with SO($2N$-1), SO($2N$), and Sp($N$) flavor symmetries.}
  \label{fig:sospbranes}
\end{figure}
There are three different classes UV SCFTs which we can now realize in terms of suspended branes depicted in figure~\ref{fig:sospbranes}.
To study their Higgs branch flow, we follow the same approach as in the SU($N$) case: first, we choose two partitions, which each describe an embedding of $\mathfrak{su}_2$ into the corresponding flavor symmetry algebra. These control how the stacks of 1/2 D8-branes on the left and right side of the quiver are split into smaller stacks. Finally, we apply Hanany-Witten moves to these stack until they are balanced.

It is convenient to combine the D6-brane charge of the orientifold planes with the contribution from the 1/2 D6-branes. In this case, rules for the Hanany-Witten shown in figure~\ref{fig:hwmove} still apply and we can use the results from the last subsection. The only thing we have to keep in mind is that we are now counting 1/2 D6-branes.

\subsection{Propagation Rules}

In this section, we present a set of rules for working out Higgs branch
deformations in the case of quivers with classical gauge algebras. The main idea is
to consider each stack of $7$-branes wrapped over a curve and
strings that stretch from one stack to the next. To visualize the possible
locations where such strings can begin and end, we will use the same
diagrammatic analysis developed in section \ref{sec:NILPJUNC} to track these breaking
patterns. When such a string is present, it signals the presence of a brane
recombination move, and the corresponding brane becomes non-dynamical (having
become attached to a non-compact $7$-brane on the boundary of the quiver).
On each layer, we introduce a directed graph, as dictated by a choice of nilpotent orbit. This
tells us how to connect the branes into ``blobs'' after recombination. We want to see how these
blobs recombine, both with the non-compact branes at the end of quiver and the compact branes further in the interior.

On each consecutive level of the quiver (i.e. for each gauge algebra in the quiver), we draw the same string configuration with a few modifications according to the following rules for propagating
Higgs branch flows into the interior of a quiver:

\begin{itemize}
\item First, we consider blobs made only of $A$-branes. That is, only
one-pronged strings are involved and there is no crossing or touching the
mirror. These configurations cover all possible orbits of $SU(N)$. In such
cases, the one-pronged strings get removed one at a time (per blob) so that
one $A$-brane is added back (to each blob) at each step. These steps can be
visualized in the example of $SU(6)$ nilpotent orbits given in figure
\ref{fig:SU6quiver}.

\item Next, we consider cases with a two-pronged string, but in which both legs are
disjoint (unlike $\alpha_{N}$ for $Sp(N)$) so that no loop is formed. In this
case, the propagation follows the same rule as for one-pronged strings. Indeed
in such configurations each leg becomes independent and they individually
behave like one-pronged strings. This is the case for $SO(2N)$ whenever the
two-pronged string $\alpha_{N}$ is present but not the string $\alpha_{N-1}$
below it. (See for instance the partition $[2^{4}]^{II}$ for $SO(8)$ in figure
\ref{fig:SO8quiver}).

\item Now suppose (without loss of generality) that branes $A_{1}, A_{2},
\cdots, A_{n}$ are connected via simple one-pronged strings and a two-pronged
string attaches the $i^{th}$ and $n^{th}$ brane to the mirror ($1 \leq i <
n$). Then, for the next $n-i$ levels, the right-most leg moves one step to the
left (attaching to the brane $A_{n-1}, A_{n-2}, \cdots,A_{n-i}$) and the
right-most one-pronged string below it is removed, namely $\alpha_{n}$
followed by $\alpha_{n-1}, \cdots, \alpha_{n-i}$. After these $n-i$ steps, both
legs overlap and the right-most leg cannot move any further. Instead, we then
move the second leg one step to the left so that one leg stretches from
$\alpha_{n-i-1}$ and the other stretches from $\alpha_{n-i}$. We can now
repeat the previous steps once by moving the right-most leg one brane to the
left (and removing $\alpha_{n-i-1}$) so that it overlaps with the left-most
leg. This process ends whenever the two-pronged string with both legs
overlapping is the last one of the group and it is then simply removed for the
next node in the quiver. (See for instance the partitions $[5,3]$ or $[7,1]$
for $SO(8)$ in figure \ref{fig:SO8quiver}).

\item Finally we can have groups of $K$ branes involving the short root
$\alpha_{N-1}$ of $SO(2N-1)$, which connects the $N^{th}$ $A$-brane to the one
merged onto the mirror. In this case, the first step consists of lifting the
short string above the middle brane so that it becomes a doubled-arrow string
crossing the mirror and connecting $K-1$ branes. The next steps in the
propagation are then identical to the ones described in the previous point.
(See for instance the partitions $[7,1^{2}]$ or $[9]$ for $SO(9)$ in figure
\ref{fig:SO9quiver}).
\end{itemize}

We note that in terms of partitions, these steps simply translate into every
part being reduced by $1$, so that the partition $[\mu_{1}, \mu_{2}, \cdots,
\mu_{i}, 1^{k}]$ goes to $[\mu_{1}-1, \mu_{2}-1, \cdots, \mu_{i}-1, 1^{k+i}]$
after each step until there are no more parts with $\mu_{i} > 1$, and we are left with
the trivial partition (corresponding to the total absence of strings).

\subsection{Higgsing and Brane Recombination}\label{ssec:Higgsing}

Once we have propagated the strings according to the above rules, we are ready
to read off the residual gauge symmetry on each node. To do so, we note that
the strings force connected branes on each side of the mirror to coalesce so that a blob of $K$ $A$-branes behaves like a single
$A$-brane. We can then directly read off the gauge symmetry that is described
by the resulting collapsed brane configuration.

For $SU(N)$ quivers, which do not involve a mirror, strings group $A$-branes
without any ambiguity, as no $B$ or $C$ brane is present. Thus, the residual
gauge symmetry is given by the number of groups formed at each level. For
instance, if only one simple string stretches between two $A$-branes, these branes
coalesce, and we are left with $N-1$ separate groups of strings on
this level. This yields the residual gauge symmetry $\mathfrak{su}_{N-1}$ as
illustrated in the first orbit of $SU(6)$ (see figure \ref{fig:SU6quiver}).

Similarly, a blob with $K$ branes connected by strings on each side of
a mirror turns an $\mathfrak{so}_{2N}$ algebra into $\mathfrak{so}_{2(N-K+1)}$,
$\mathfrak{so}_{2N-1}$ into $\mathfrak{so}_{2(N-K+1)-1}$, and $\mathfrak{sp}_{N}$
into $\mathfrak{sp}_{N-K+1}$. The same is true if the blob consists of branes on both sides of the mirror connected by double-pronged strings. However, if the blob consists of branes connected by a double-arrowed string, then the blob of connected branes gets merged onto the mirror.
As a result, an $\mathfrak{so}_{2K}$ algebra will turn into $\mathfrak{so}_{2K-1}$, and $\mathfrak{so}_{2K-1}$ into $\mathfrak{so}_{2K-2}$. (See for instance the [7,1] diagrams at the bottom of figure \ref{fig:SO8quiver}.) We note that
the propagation rules listed above prevent such a configuration from ever
appearing on a level with $\mathfrak{sp}_{N}$ gauge symmetry.

In some cases, the $\mathfrak{so}$ quivers require the
introduction of ``anti-branes.'' In our figures, we denote a brane by a filled in circle (black dot)
and an anti-brane by an open circle. At the final step, all such anti-branes must disappear by
pairing up with other coalesced branes, deleting such blobs from the resulting configuration.
This further reduces the number of leftover blobs which generate the residual gauge symmetry.

Note that there are also situations where the number of anti-branes is larger than the number of available blobs of branes on a given layer. This occurs whenever the number of D6-branes in the type IIA suspended brane realization formally becomes negative, signaling that the perturbative type IIA description has broken down, and F-theory is required to construct the theory in question. Nevertheless, it is still useful to write down a ``formal IIA quiver,'' which includes negative numbers of D6-branes and hence negative gauge group ranks. Additionally, as we will now show with examples, our picture of brane / anti-brane nucleation can be adapted to these situations if we allow extra anti-branes at a given layer to move to other layers and annihilate other blobs of branes.

Consider, for instance, the partition $[5,3]$ of $SO(8)$ requires the presence of four $A$-branes on the first
quiver node, which only has $\mathfrak{sp}_{0}$ symmetry. Thus, we also need to
introduce four anti-branes to compensate. Only one blob of branes is formed
on each side of the mirror, so only one of the four anti-branes is used to
cancel it, and we are left with three anti-branes. The first anti-brane is used
to collapse the $-1$ curve it is on. The second anti-brane is distributed to
the next $\mathfrak{so}$ quiver node and the third anti-brane is distributed
to the next $\mathfrak{sp}$ quiver node, where it is used to either reduce the
gauge symmetry from $\mathfrak{sp}_{K}$ to $\mathfrak{sp}_{K-1}$ or, if $K=0$, to blow down
this next $-1$ curve. The
anti-brane that lands on a quiver node with an $\mathfrak{so}$ algebra also reduces the residual symmetry according to the following rules:
\begin{align}
  \mathfrak{so}_{N} &\overset{\overline{A}}{\rightarrow} \mathfrak{so}_{N-1} \text{ for } N \geq 8, \nonumber\\
  \mathfrak{so}_{7} &\overset{\overline{A}}{\rightarrow} \mathfrak{g}_{2},\nonumber\\
  \mathfrak{g}_{2}  &\overset{\overline{A}}{\rightarrow} \mathfrak{su}_{3},\nonumber\\
  \mathfrak{so}_{6} \simeq \mathfrak{su}_{4} &\overset{\overline{A}}{\rightarrow} \mathfrak{su}_{3},\nonumber\\
  \mathfrak{su}_{3} &\overset{\overline{A}}{\rightarrow} \mathfrak{su}_{2},\nonumber\\
  \mathfrak{so}_{5} \simeq \mathfrak{sp}_{2} &\overset{\overline{A}}{\rightarrow} \mathfrak{sp}_{1} \simeq \mathfrak{su}_{2},\nonumber\\
  \mathfrak{so}_{4} &\overset{\overline{A}}{\rightarrow} \mathfrak{so}_{3} \simeq \mathfrak{su}_{2},\nonumber\\
  \mathfrak{so}_{3} \simeq \mathfrak{su}_{2} &\overset{\overline{A}}{\rightarrow} \mathfrak{su}_{1} \simeq \emptyset.
  \label{eq:rules}
\end{align}
Note that for classical quiver theories, there can never be more than four anti-branes, since the
quiver nodes with $\mathfrak{sp}$ gauge symmetry only have four fewer branes
than their neighboring $\mathfrak{so}$ nodes.

We illustrate all of these steps through the examples of $SU(6)$, $SO(8)$,
$SO(10)$, $SO(9)$, and $Sp(3)$ in figures \ref{fig:SU6quiver},
\ref{fig:SO8quiver}, \ref{fig:SO10quiver}, \ref{fig:SO9quiver}, and
\ref{fig:Sp3quiver} respectively. Explicit examples of $\mathfrak{g}_{2} \overset{\overline{A}}{\rightarrow} \mathfrak{su}_{3}$ and $\mathfrak{su}_{3} \overset{\overline{A}}{\rightarrow} \mathfrak{su}_{2}$ can only be found when dealing with ``short quivers,'' which we discuss in section \ref{sec:GETSHORTY}.
\newcommand{\SUscale}{0.8}
\newcommand{\SUvspace}{0.5cm}

\begin{figure}[H]
\includegraphics[trim={3cm 6cm 3cm 2.6cm},clip,scale=.9]{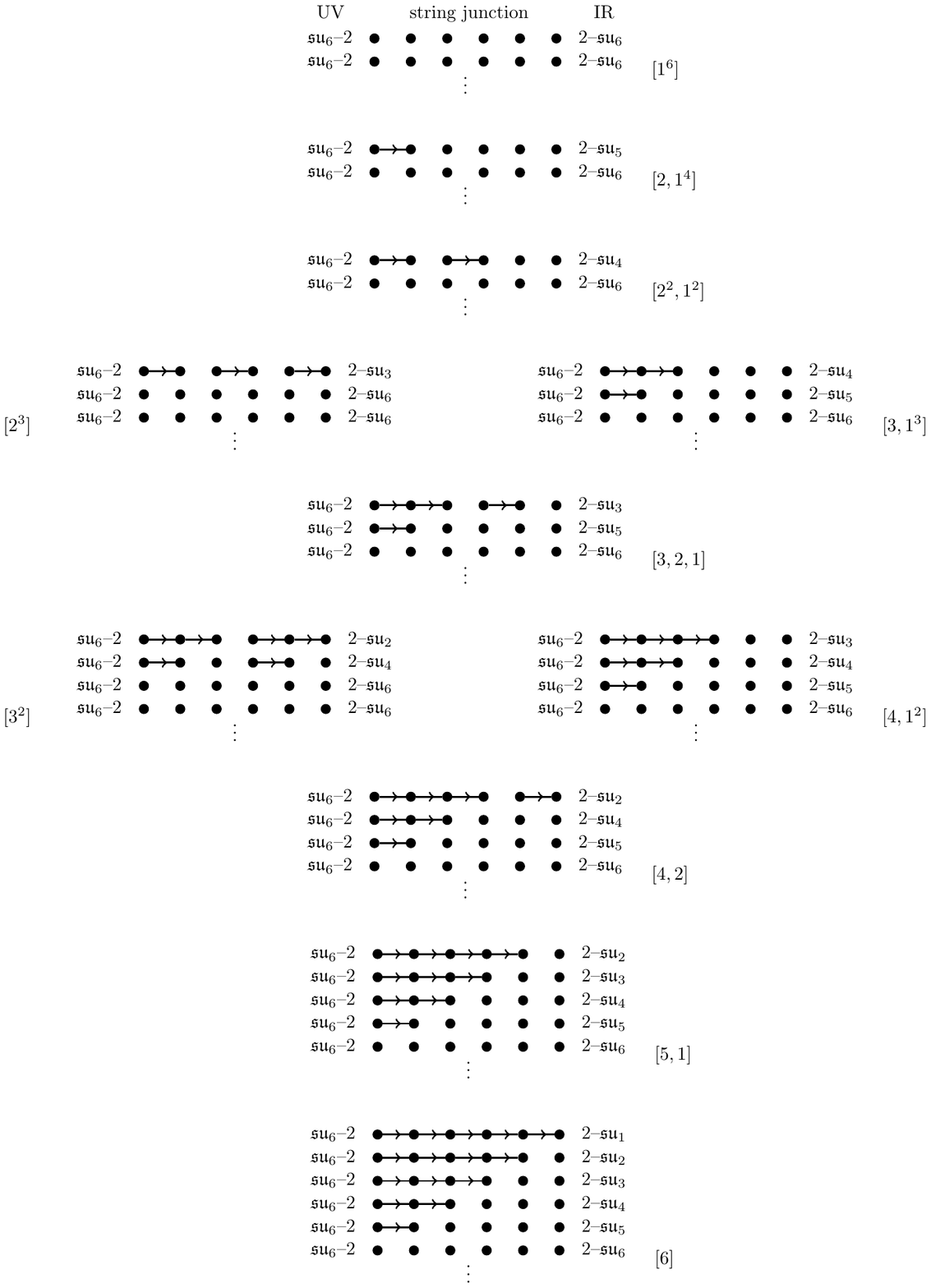}
\caption{Nilpotent deformations of the $SU(6)$ quiver from the UV configuration of figure \ref{fig:SUNUV}. Each subfigure corresponds to the quiver diagram of a nilpotent orbit with strings propagating through. The quivers have been rotated to go from top to bottom (rather than left to right) to fit on the page. On the left-hand side of each subfigure we have the setting in the UV with each $-2$ curve containing an $\mathfrak{su}_6$ gauge algebra, while on the right-hand side we give the IR theory induced by the strings stretched in the middle diagram. The theories are ordered from top to bottom according to their partial ordedring of RG flows, which matches their mathematical ordering. The corresponding partitions are given on the side.}
  \label{fig:SU6quiver}
\end{figure}

\begin{figure}[H]
\includegraphics[trim={1.1cm 7.1cm 0cm 2.6cm},clip,scale=.73]{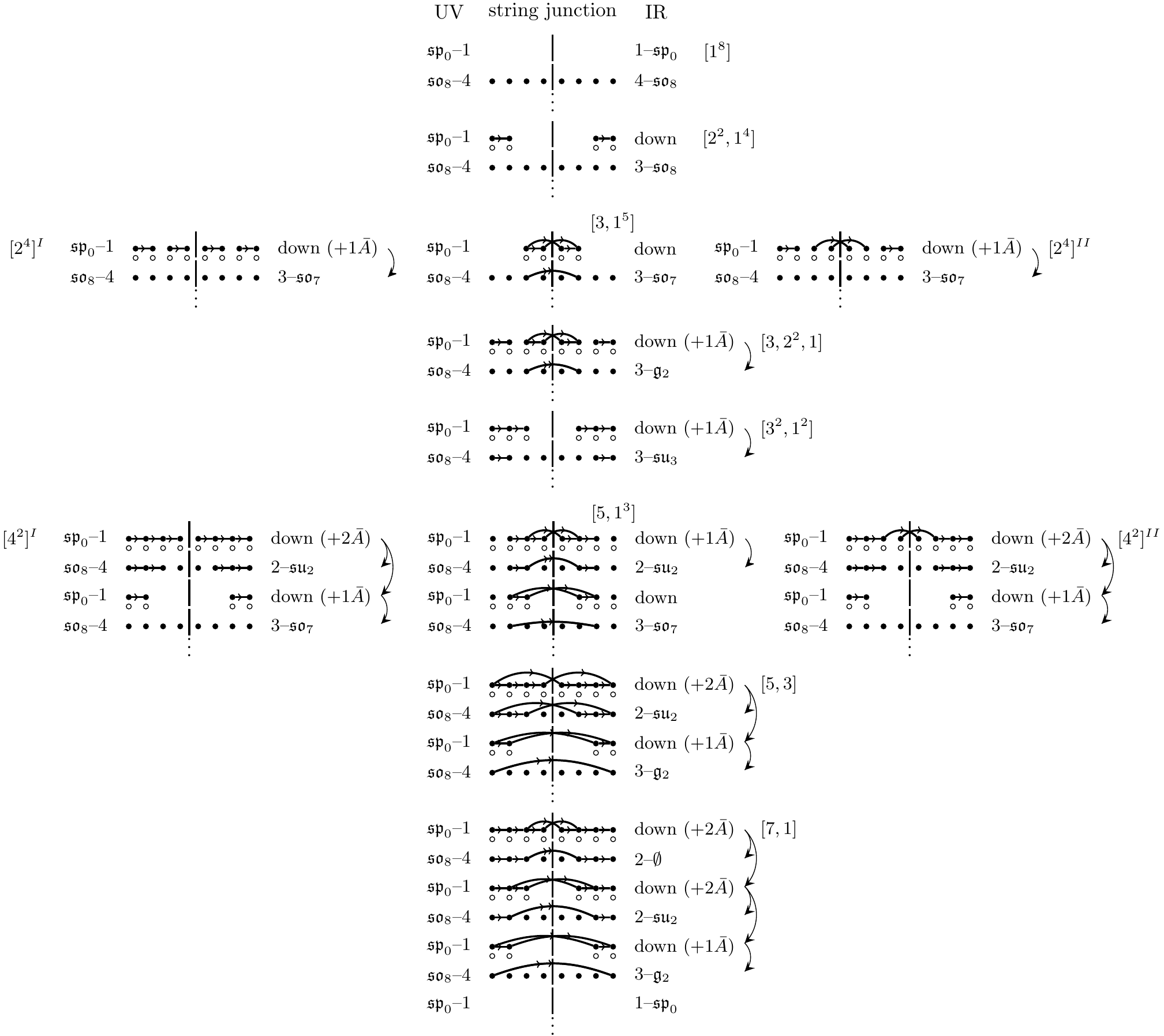}
  \caption{Nilpotent deformations of the $SO(8)$ quiver from the UV configuration of figure \ref{fig:SO2NUV}. Each subfigure corresponds to the quiver diagram of a nilpotent obit with strings propagating into the interior of the quiver. The quivers have been rotated to go from top to bottom (rather than left to right) to fit on the page. On the left-hand side of each subfigure, we have the initial UV theory with alternating $-1$ and $-4$ curves containing $\mathfrak{sp}_0$ and $\mathfrak{so}_8$ respectively. On the right-hand side, we give the IR theory induced by the strings stretched in the middle diagram. The vertical line denotes the $BC$-mirror. Whenever anti-branes are required, they are denoted by white circle below their $A$-brane counterparts. In some cases, there are extra anti-branes indicated in parentheses on the right (which occur when there are more groups of $A$-branes than anti-branes). The first one is used to blow-down the $-1$ curve it is on (indicated by the word ``down''), while the others get distributed on the following quiver nodes as indicated by the side arrows on the right. The theories are ordered from top to bottom according to their partial ordering of RG flows. The corresponding partitions are given on the side.}
  \label{fig:SO8quiver}
\end{figure}

\begin{figure}[H]
\includegraphics[trim={2.6cm 6cm 2.6cm 2.6cm},clip,scale=.93]{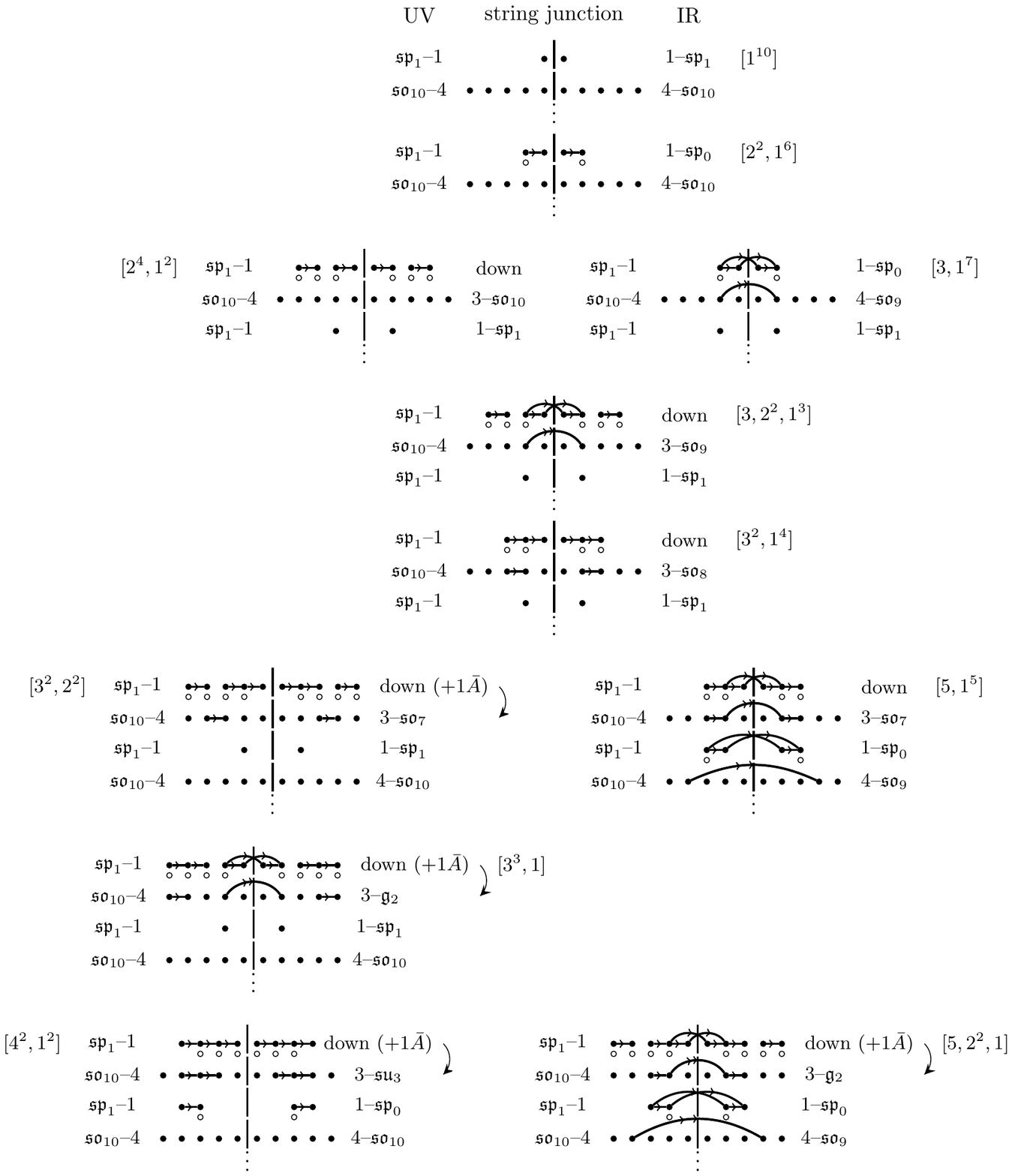}
 \caption{Nilpotent deformations of the $SO(10)$ quiver from the UV configuration of figure \ref{fig:SO2NUV}. See figure \ref{fig:SO8quiver} for additional details on the notation and conventions.}
\end{figure}
\begin{figure}\ContinuedFloat
\includegraphics[trim={3.5cm 10.5cm 2.7cm 2.6cm},clip,scale=.93]{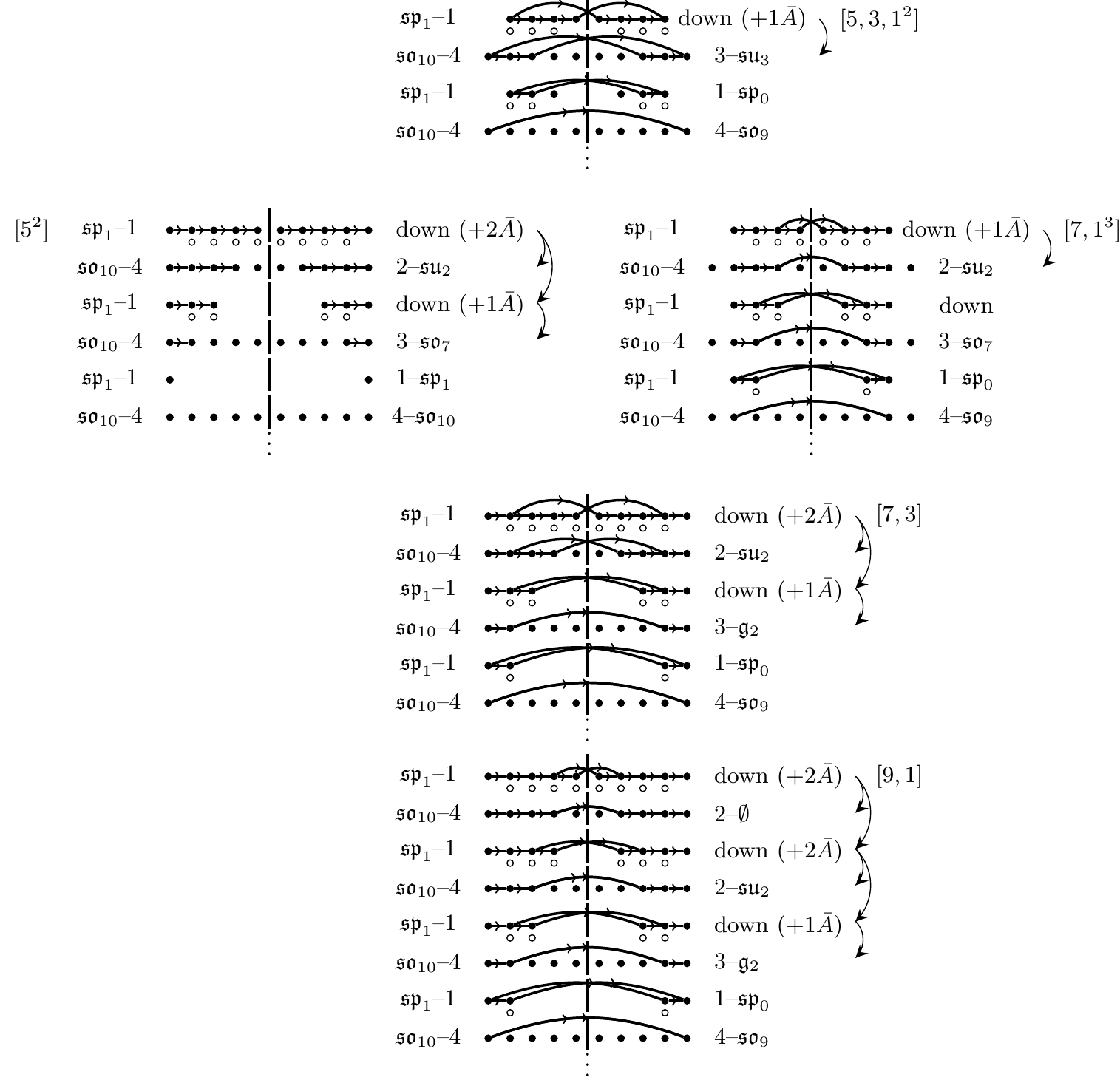}
\caption{(continued) Nilpotent deformations of the $SO(10)$ quiver from the UV configuration of figure \ref{fig:SO2NUV}. See figure \ref{fig:SO8quiver} for additional details on the notation and conventions.}
  \label{fig:SO10quiver}
\end{figure}

\begin{figure}[H]
\includegraphics[trim={3cm 6cm 2.25cm 2.6cm},clip,scale=.93]{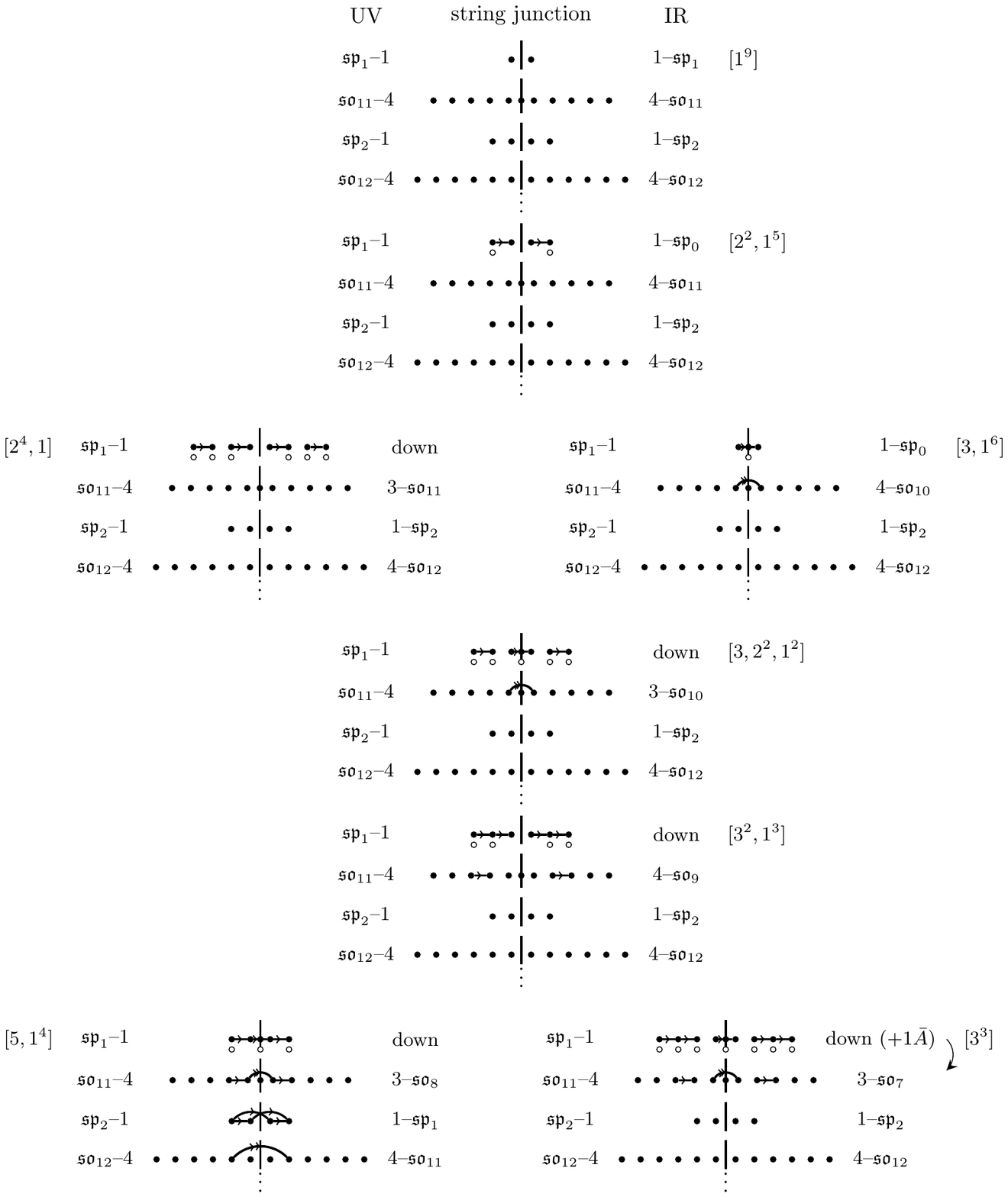}
  \caption{Nilpotent deformations of the $SO(9)$ quiver from the UV configuration of figure \ref{fig:SO2Nm1UV}. See figure \ref{fig:SO8quiver} for additional details on the notation and conventions.}
\end{figure}
\begin{figure}\ContinuedFloat
\includegraphics[trim={3.2cm 9.5cm 2.1cm 2.6cm},clip,scale=.93]{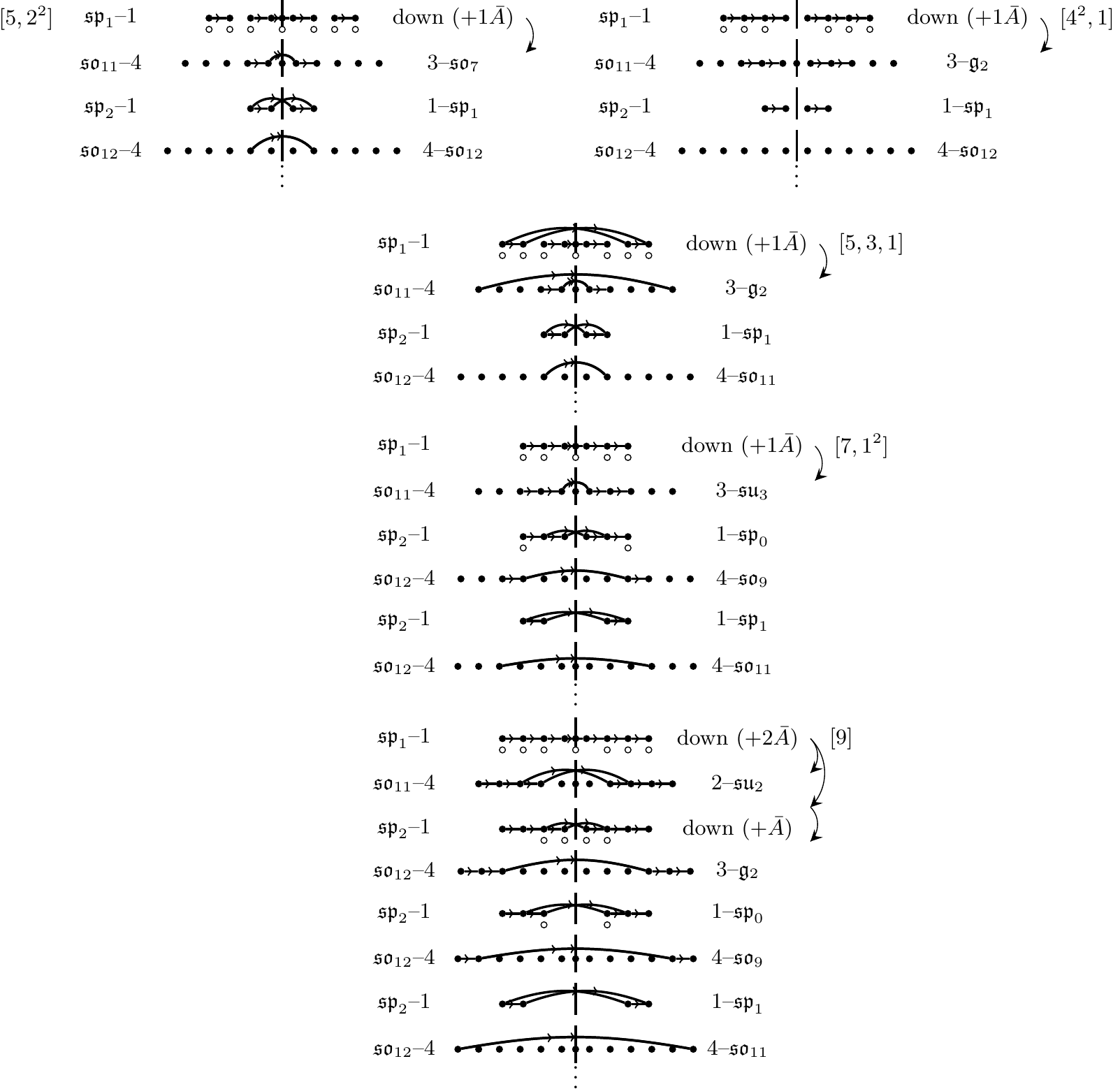}
\caption{(continued) Nilpotent deformations of the $SO(9)$ quiver from the UV configuration of figure \ref{fig:SO2Nm1UV}. See figure \ref{fig:SO8quiver} for additional details on the notation and conventions.}
  \label{fig:SO9quiver}
\end{figure}

\begin{figure}[H]
\includegraphics[trim={3cm 7cm 2.cm 2.6cm},clip,scale=.93]{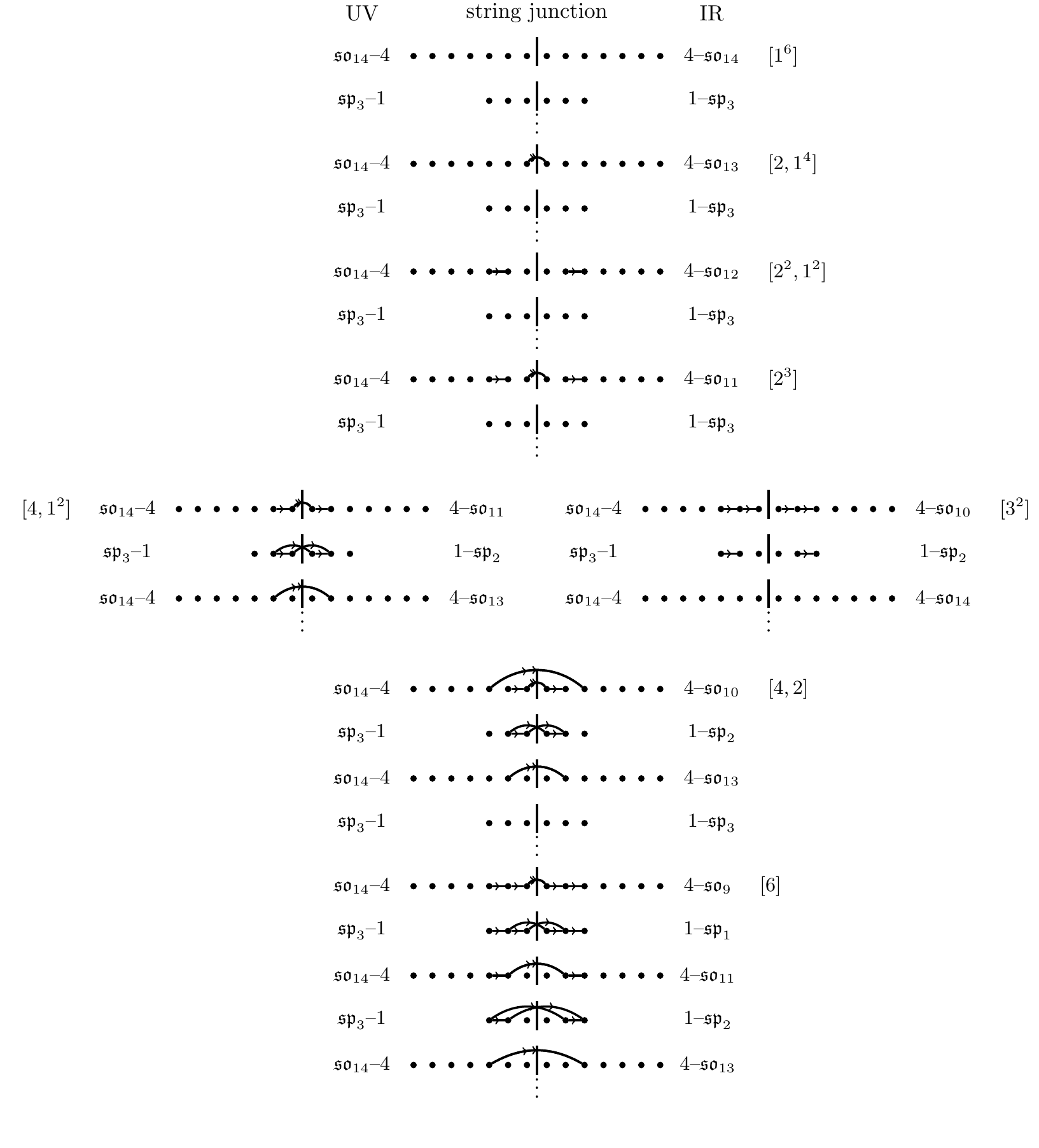}
  \caption{Nilpotent deformations of the $Sp(3)$ quiver from the UV configuration of figure \ref{fig:SpNUV}. See figure \ref{fig:SO8quiver} for additional details on the notation and conventions.}
  \label{fig:Sp3quiver}
\end{figure}

\subsection{Comments on Quiver-like Theories with Exceptional Algebras}

It is natural to ask whether the propagation
rules given for quivers with classical algebras also
extend to theories with exceptional algebras. In principle, we expect this to follow from our description of
the nilpotent cone in terms of multi-pronged string junctions. Indeed, we have already explained that at least for
semi-simple deformations, there is no material distinction between the quivers of classical and exceptional type.

That being said, we expect our analysis of nilpotent deformations to be more subtle in this case.
Part of the issue is that even in the case of the D-type algebras, to really describe the physics of brane recombination, we had to go onto the
full tensor branch so that both $SO$ and $Sp$ gauge algebras could be manipulated (via brane recombination). From this perspective, we need to
understand brane recombination in 6D conformal matter for the following configurations
of $(E_{N},E_{N})$ conformal matter:
\begin{align}
& \lbrack E_{6}],1,3,1,[E_{6}]\\
& \lbrack E_{7}],1,2,3,2,1,[E_{7}]\\
& \lbrack E_{8}],1,2,2,3,1,5,1,3,2,2,1,[E_{8}].
\end{align}
Said differently, a breaking pattern which connects two E-type algebras will necessarily
involve a number of tensor multiplets. For the most part, one can work out a set of ``phenomenological''
rules which cover nearly all cases involving quivers with $E_6$ gauge algebras, but its generalization to $E_7$
and $E_8$ appears to involve some new ingredients beyond the ones introduced already in this chapter. For all these
reasons, we defer a full analysis of these cases to future work.

\section{Short Quivers \label{sec:GETSHORTY}}

In the previous section, we demonstrated that the physics of brane
recombination accurately recovers the expected Higgs branch flows for
6D\ SCFTs. It is reassuring to see that these methods reproduce -- but also
extend -- the structure of Higgs branch flows obtained through other methods.
The main picture we have elaborated on is the
propagation of T-brane data into the interior of a quiver-like gauge theory.

The main assumption made in these earlier sections is the presence of a
sufficient number of gauge group factors in the interior of the quiver so that this
propagation is independent of other T-brane data associated with other
flavor symmetry factors. In this section
we relax this assumption by considering \textquotedblleft short quivers\textquotedblright\ in which
the number of gauge group factors is too low
to prevent such an overlap. There has been very little analysis in the 6D SCFT
literature on this class of RG flows.

Using the brane recombination picture developed in the previous section, we
show how to determine the corresponding 6D\ SCFTs generated by such
deformations. We mainly focus on quivers with classical algebras, since this
is the case we presently understand most clearly. Even here, there is a rather
rich structure of possible RG\ flows.

There are two crucial combinatorial aspects to our analysis. First of all,
we use open strings to collect recombined branes into ``blobs.'' Additionally, to determine the scope of possible
deformations, we introduce brane / anti-brane pairs, as prescribed by the rules of section \ref{sec:RECOMBO}. To track the effects of
having a short quiver, we gradually reduce the number of gauge group factors until the brane moves on either side of the quiver become correlated. As a result, we sometimes reach configurations in which the anti-branes cannot be eliminated. We take this to mean
that we have not actually satisfied the D-term constraints in the quiver-like gauge theory.

The procedure we outline also has some overlap with the formal proposal of reference \cite{Mekareeya:2016yal} (see also \cite{Apruzzi:2017iqe}), which analyzed Higgs branch flows by analytically continuing the rank of gauge groups to negative values. Using our description in terms of anti-branes, we show that in many cases, the theory we obtain has an anomaly polynomial which matches to these proposed theories. We also find,
however, that in short quivers (which were not analyzed in \cite{Mekareeya:2016yal})
this analytic continuation method sometimes does not produce a sensible IR fixed point.
This illustrates the utility of the methods developed in this chapter.

In the case of sufficiently long quiver-like theories, there is a natural partial ordering set by the nilpotent
orbits in the two flavor symmetry algebras. In the case of shorter quivers, the partial ordering becomes more complicated because
there is (by definition) some overlap in the symmetry breaking patterns on the two sides of a quiver. In many cases, different pairs of nilpotent orbit wind up generating the same IR fixed point simply because most or all of the gauge symmetry in the quiver has already been Higgsed.
We show in explicit examples how to obtain the corresponding partially ordered set of theories labeled by pairs of overlapping nilpotent orbits. We refer to these as ``double Hasse diagrams'' since they merge two Hasse diagrams of a given flavor symmetry algebra.

To illustrate the main points of this analysis, we primarily focus on illustrative examples in which the number of gauge group factors in the interior of a quiver is sufficiently small and / or in which the size of the nilpotent orbits is sufficiently large so that there is non-trivial overlap between the breaking patterns on the left and right. For this reason, we often work with low rank gauge algebras such as $\mathfrak{su}(4)$ and $\mathfrak{so}(8)$ and a small number of interior gauge group factors, though we stress that our analysis works in the same way for all short quivers.

The rest of this section is organized as follows. First, we show how to obtain short quivers as a limiting case in which we gradually reduce the number of gauge group factors in a long quiver. We then turn to a study of nilpotent hierarchies in these models, and we conclude this section with a brief discussion of the residual global symmetries after Higgsing in a short quiver.

\subsection{From Long to Short Quivers \label{subsec:introShortQuiver}}

In this subsection, we determine how T-brane data propagating from the two sides of a quiver becomes intertwined as we decrease the number of gauge groups / tensor multiplets. It is helpful to split up this analysis according to the choice of gauge group appearing, so we present examples for each different choices of gauge algebras.

\subsubsection{$SU(N)$ Short Quivers}

We begin with quiver-like theories with $\mathfrak{su}$ gauge algebras.
Applying the Hanany-Witten rules from section \ref{subsec:HananyWitten} to
the type IIA realization of the $SU(N)$ theories, we have that:
\begin{align}
  &k_{\textrm{NS5}} \geq \textrm{Max}\{\mu_{L}^{1}, \mu_{R}^{1}\} + 1
\end{align}
for left and right partitions $\mu_{L} = [\mu^{i}]$, $\mu_{R}=[\mu^{j}]$
respectively. Here, $k_{\textrm{NS5}}$ denotes the number of NS5-branes in the
corresponding type IIA picture. When this condition is violated, it is impossible to balance the D8-branes. Note that $k_{\textrm{NS5}}$ is also equal to one plus the number of $-2$ curves $N_{-2} = N_{T}$ the number of tensor multiplets in the UV quiver, so we may equivalently write this condition as
\begin{equation}
  \mathrm{Max}\{\mu_L^1,\mu_R^1\} \leq N_{-2},
  \label{eq:HWconstraintSU}
\end{equation}
where $N_{-2}$ denotes the number of $-2$ curves in the UV quiver.
This is equivalent to saying that, when only one nilpotent deformation (either $\mu_L$ or $\mu_R$) is implemented over the UV quiver (either the left or right partition), there has to be at least one $-2$ curve whose fiber remains untouched by the deformation.

Assuming this restriction is obeyed, we can straightforwardly produce any short $SU(N)$ quiver given a UV quiver and a pair of nilpotent orbits. Before giving the general formula, however, let us look at a concrete example: consider a UV theory of $SU(5)$ over five $-2$ curves, and apply the
nilpotent deformations of $[3, 2]$ -- $[2^{2}, 1]$, where no interaction
between the orbits take place. This theory can be written as:
\begin{equation}
[3, 2]:\,\, \overset{\mathfrak{su}(2)}{2} \,\,
\underset{[N_f = 1]}{\overset{\mathfrak{su}(4)}{2}} \,\,
\underset{[N_f = 1]}{\overset{\mathfrak{su}(5)}{2}}
\,\,\underset{[SU(2)]}{\overset{\mathfrak{su}(5)}{2}} \,\,
\underset{[N_f = 1]}{\overset{\mathfrak{su}(3)}{2}}\,\,: [2^{2}, 1]
\end{equation}
where the notation $[N_f = 1]$ refers to having one additional flavor on each corresponding gauge algebra.

We now decrease the length of the quiver and gradually turn it into a short quiver. We decrease the number of $-2$ curves one at a time, and when the nilpotent deformation from the left and right overlaps, we simply add the rank reduction
effect together linearly. After each step we get:
\begin{equation}
[3, 2]:\,\, \overset{\mathfrak{su}(2)}{2} \,\,
\underset{[N_f = 1]}{\overset{\mathfrak{su}(4)}{2}} \,\, \underset{[
SU(3)]}{\overset{\mathfrak{su}(5)}{2}} \,\,
\underset{[N_f = 1]}{\overset{\mathfrak{su}(3)}{2}}\,\,: [2^{2}, 1]
\end{equation}
\begin{equation}
[3, 2]:\,\, \overset{\mathfrak{su}(2)}{2} \,\,
\underset{[SU(3)]}{\overset{\mathfrak{su}(4)}{2}} \,\,
\underset{[SU(2)]}{\overset{\mathfrak{su}(3)}{2}}\,\,: [2^{2}, 1]
\end{equation}
At this stage we are unable to decrease the length of the quiver any further without violating the constraint of (\ref{eq:HWconstraintSU}).

We note that each step changes the global symmetry, the
gauge symmetry, or both. In particular, after the second step we no longer
see a node with the UV gauge group $SU(5)$. The global symmetries also change at each step, which will be discussed further in \ref{subsec:globalSymmetryShortQuiver}.

Let us consider another example of a short quiver with $SU(N)$ gauge groups. If we take the UV quiver theory to be:
\begin{equation}
[SU(6)] \,\, \overset{\mathfrak{su}(6)}{2} \,\, \overset{\mathfrak{su}(6)}{2}
\,\, \overset{\mathfrak{su}(6)}{2}\,\, \overset{\mathfrak{su}(6)}{2}\,\,
\overset{\mathfrak{su}(6)}{2} \,\, [SU(6)]\label{eqn:SUshortexample1:1}%
\end{equation}
and apply the following pair of nilpotent deformations denoted by partitions
$\mu_{L,R}$:
\begin{equation}
\mu_{L} = [5, 1],\ \ \mu_{R} = [2^{3}]
\end{equation}
we obtain the resulting IR theory:
\begin{equation}
\underset{[N_f = 1]}{\overset{\mathfrak{su}(2)}{2}} \,\, \overset{\mathfrak{su}%
(3)}{2} \,\, \overset{\mathfrak{su}(4)}{2}\,\,
\underset{[SU(3)]}{\overset{\mathfrak{su}(5)}{2}}\,\,
\underset{[N_f = 1]}{\overset{\mathfrak{su}(3)}{2}} \,.\label{eqn:SUshortexample1:3}%
\end{equation}

We illustrate another example with $SU(5)$ UV gauge group and partitions
$\mu_L=[5]$, $\mu_R = [4,1]$ in figure \ref{fig:SUshortBrane}, making the
brane recombination explicit.

\begin{figure}[ptb]
  \centering
  \includegraphics{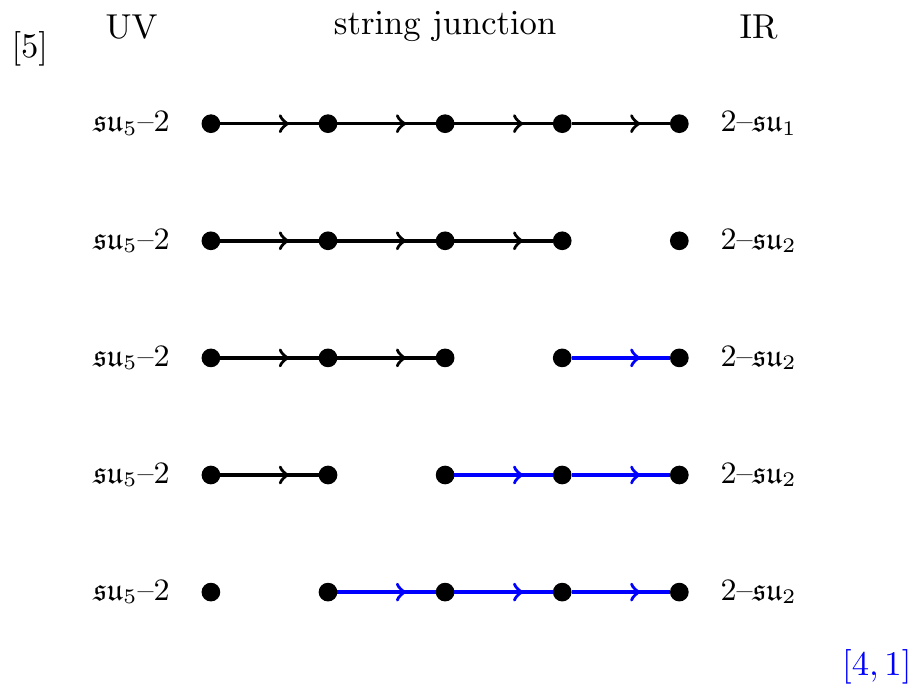}
\caption{An $SU(N)$ short quiver brane picture, the pair of nilpotent
deformation being $\mu_{L}=[5]$, $\mu_{R}=[4, 1]$ on $SU(5)$ UV theory and
four $-2$ curves. The figure is arranged so that the left deformation starts
from the top and propagates downwards (in black) while the right deformation
starts on the bottom and propagates upwards (blue).}%
\label{fig:SUshortBrane}%
\end{figure}

In general, let us define the conjugate partitions of the left and right nilpotent
orbits to be $\rho_{L}:= \mu_L^T$ and $\rho_R := \mu_R^T$ and denote their number of elements as
$N_{L}^{\prime}$ and $N_{R}^{\prime}$, with the index counting from each of their
starting point, respectively. Then, the gauge group rank at the $m^{\text{th}}$ node is
given by %
\begin{equation}
r_{m}=N-\sum_{i=m+1}^{N_{L}^{\prime}} \rho_{i}^{L}-\sum_{j=(N_{-2})-m+1}%
^{N_{R}^{\prime}} \rho_{j}^{R},
\label{eq:SUshort}
\end{equation}
with the UV gauge group equal to $SU(N)$.

\subsubsection{Interlude: $SO$ and $Sp$ Short Quivers}

In the case of quivers with $SU$ gauge groups, the Higgsing of
the corresponding quiver-like gauge theories is controlled by vevs for
weakly coupled hypermultiplets. In this case, the physics of brane recombination primarily serves to simplify the combinatorics associated with
correlated breaking patterns in the quiver. Now, an important feature of the other quiver-like theories with flavor groups $SO$ or
$Sp$ is the more general class of possible Higgs branch flows as generated by 6D conformal matter. Recall
that on the full tensor branch of such a theory, we have a gauge group consisting of alternating classical gauge groups.
These gauge groups typically have bifundamental matter (in half-hypermultiplets of $SO \times Sp$ representations), which in turn
leads to Higgs flows generated by ``classical matter,'' much as in the case of the $SU$ quivers. There are, however,
more general Higgs branch flows connected with vevs for conformal matter. Recall that these are associated with a smoothing deformation for a collapsed $-1$ curve, namely the analog of a small instanton transition as in the case of the E-string theory. The combinatorics associated with this class of Higgs branch flows is more subtle, but as we have already remarked, the brane / anti-brane description correctly computes the resulting IR fixed points in this case as well.

By definition, in the case of a short quiver, the effects of Higgsing on the two sides of the quiver become correlated. It is therefore helpful to distinguish a few specific cases of interest as the size of the nilpotent orbit / breaking pattern continues to grow. As the size of the nilpotent orbit grows, the appearance of a small instanton deformation becomes inevitable. The distinguishing feature is the extent to which small instanton transitions become necessary to realize the corresponding Higgs branch flow. When there is at least one $-1$ curve remaining in the tensor branch description of the Higgsed theory, we refer to this as a case where the nilpotent orbits are ``touching.'' The end result is that so many small instanton deformations are generated that the tensor branch of the resulting IR theory has no $-1$ curves at all. We refer to this as a ``kissing case'' since the partitions are now more closely overlapping. Increasing the size of a nilpotent orbit beyond a kissing case leads to a problematic configuration: There are no more small instanton transitions available (as the $-1$ curves have all been used up). We refer to these as ``crumpled cases.'' In terms of our brane / anti-brane analysis, this leads to configurations with $\overline{A}$ branes which cannot be canceled off. Such crumpled configurations are inconsistent, and must be discarded. Summarizing, we refer to the different sorts of overlapping nilpotent orbit configurations as:
\begin{itemize}

\item A ``touching'' configuration is one in which all gauge groups of the quiver-like theory are at least partially broken, but at least one $-1$ curve remains in the tensor branch of the Higgsed theory.

\item A ``kissing'' configuration is defined as one in which all groups of the quiver-like theory are at least partially broken, and there are no $-1$ curves remaining in the Higgsed theory.

\item A ``crumpled'' configuration is defined as one in which the orbits have become so large that there are left over $\overline{A}$ branes which cannot be canceled off, and therefore such configurations are to be discarded.

\end{itemize}

Of course, there are also nilpotent orbits which are uncorrelated, as will occur whenever the quiver is sufficiently long or the nilpotent orbits are sufficiently small, which we can view as ``independent cases.''
Such ``independent / touching cases'' fall within the scope of the long quiver analysis that we have discussed previously -- the latter just marginally so. We illustrate all four configurations in figure \ref{fig:ToShort} for $SO(10)$ with partitions $\mu_L=\mu_R=[9,1]$ going from an ``independent'' (long) quiver configuration all the way down to a forbidden ``crumpled'' configuration.

Following the IIA realization from section \ref{subsec:HananyWitten}, we
can formally perform Hanany-Witten moves even when small instanton transitions
occur by allowing for a negative number of D6-branes, or in the
string-junction picture by allowing brane / anti-brane
pairs as intermediate steps in our analysis.
 The formula (\ref{eq:HWconstraintSU}) generalizes to the
other quiver-like theories with classical algebras:
\begin{align}
  & k_{\frac{1}{2} \textrm{NS5}} \geq \textrm{Max}\{\mu_{L}^{1}, \mu_{R}^{1}\} + 1, \text{ rounded up to the nearest even number.} \\
  \iff & N_{T} \geq \textrm{Max}\{\mu_{L}^{1}, \mu_{R}^{1}\}. \label{eq:HWconstraint}
\end{align}
Here $k_{\frac{1}{2} \textrm{NS5}}$ is the number of half NS5-branes in the
corresponding type IIA picture, and equals one plus the number of tensor multiplets in the UV quiver ($N_{T}=2N_{-4}+1$) in the UV.
One might worry that this becomes meaningless whenever small instanton transitions occur. Indeed, the quivers described after such transitions all have matter with spinor representations and therefore no perturbative type IIA representation. While we can formally draw suspended brane diagrams with gauge groups of negative ranks, physically there is no corresponding suspended brane diagram. However, by analytically continuing the anomaly polynomials of these quivers to the case of negative ranks, we find perfect agreement with the anomaly polynomials of the actual, physical theory constructed via F-theory. This gives us strong reason to believe that the rules for Hanany-Witten moves should likewise carry over to the formal IIA brane diagrams, which implies that the formal quiver must be of length at least $\textrm{Max}\{\mu_{L}^{1}, \mu_{R}^{1}\}$.

Finally, from the brane / anti-brane analysis, we note that there should not be any residual $\overline{A}$'s in the IR theories. Any configuration yielding extra $\overline{A}$'s that cannot be canceled are said to ``crumple'' and are therefore forbidden. This further restricts the above constraints from Hanany-Witten moves.

As an example, an $SO(2N)$ quivers with partitions
\begin{equation}
  \mu_L=\mu_R=[2 N - 1,1]
\end{equation}
requires that
\begin{equation}
  k_{\frac{1}{2} \textrm{NS5}} \geq 2N + 4,
\end{equation}
which is a strictly stronger lower bound than the one imposed by equation \ref{eq:HWconstraint}. This particular example is illustrated for $SO(10)$ with partitions $\mu_L=\mu_R=[9,1]$ in the ``crumpling'' example of subfigure \ref{subfig:crumple}.

\subsubsection{$SO(2N)$ Short Quivers} \label{subsec:braneShortQuiver}

\begin{figure}[ptb]
  \centering
  \begin{subfigure}{0.38\textwidth}
    \includegraphics[width=0.9\textwidth]{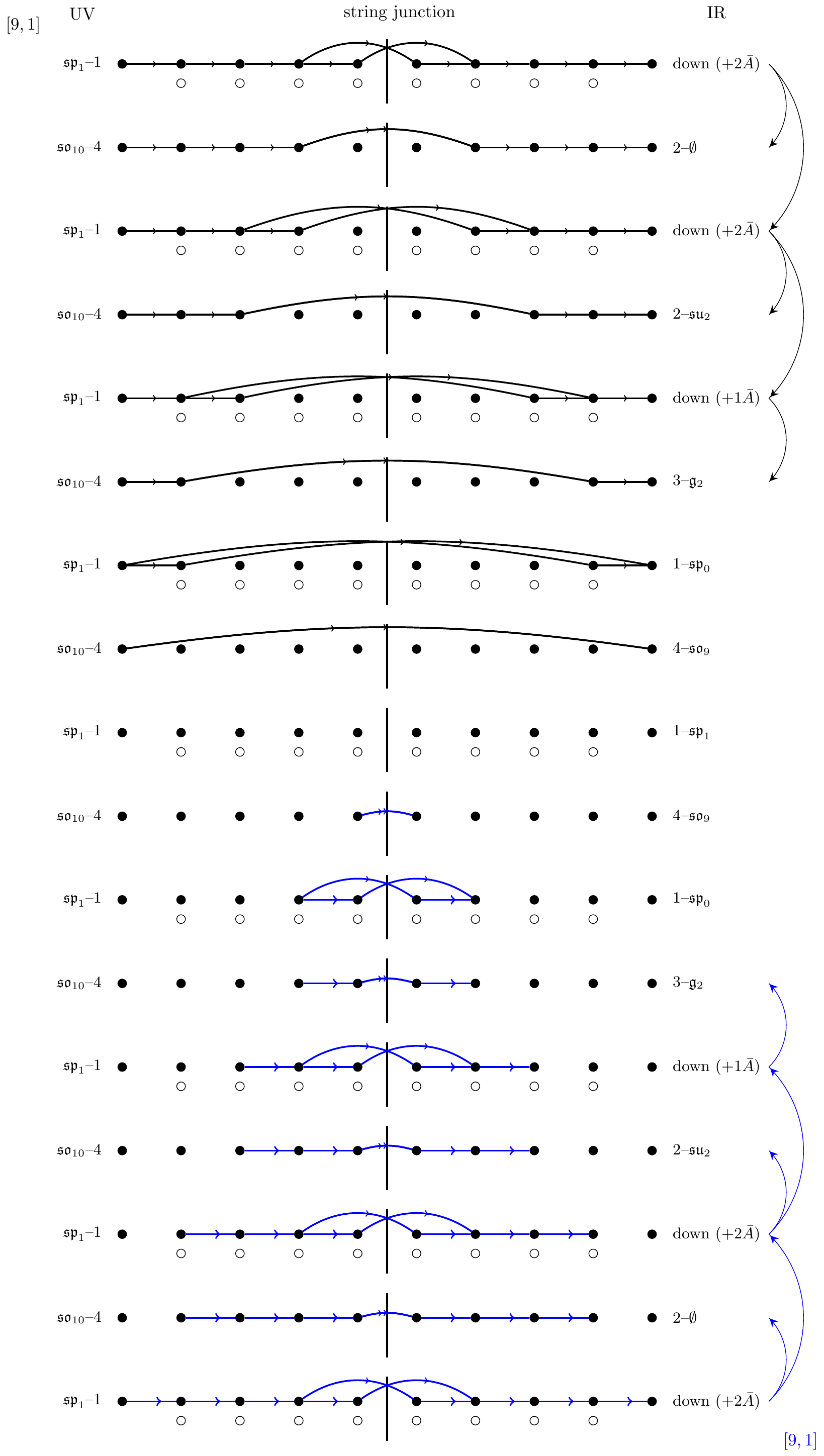}
    \captionsetup{font=footnotesize,labelfont=footnotesize}
    \caption{Independent example: Partitions $\mu_L=\mu_R=[9,1]$ on $17$ curves.} \label{subfig:independent}
  \end{subfigure} \hfill
  \begin{subfigure}{0.38\textwidth}
    \includegraphics[width=0.9\textwidth]{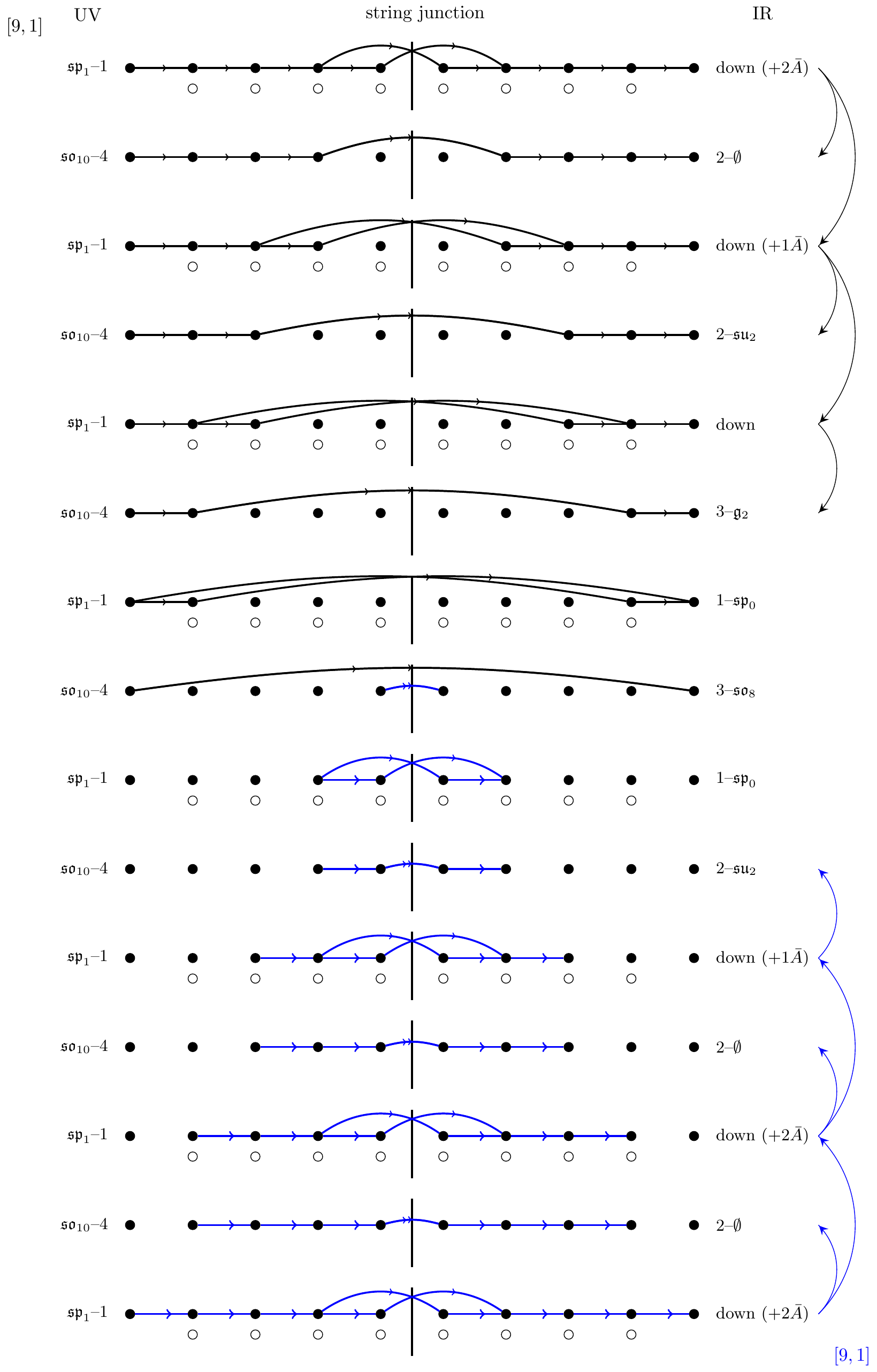}
    \captionsetup{font=footnotesize,labelfont=footnotesize}
    \caption{Touching example: Partitions $\mu_L=\mu_R=[9,1]$ on $15$ curves. Some but not all $-1$ curves participate in small instanton deformations.} \label{subfig:touch}
  \end{subfigure} 

  \vspace{0.5cm}
  \begin{subfigure}{0.38\textwidth}
    \includegraphics[width=0.9\textwidth]{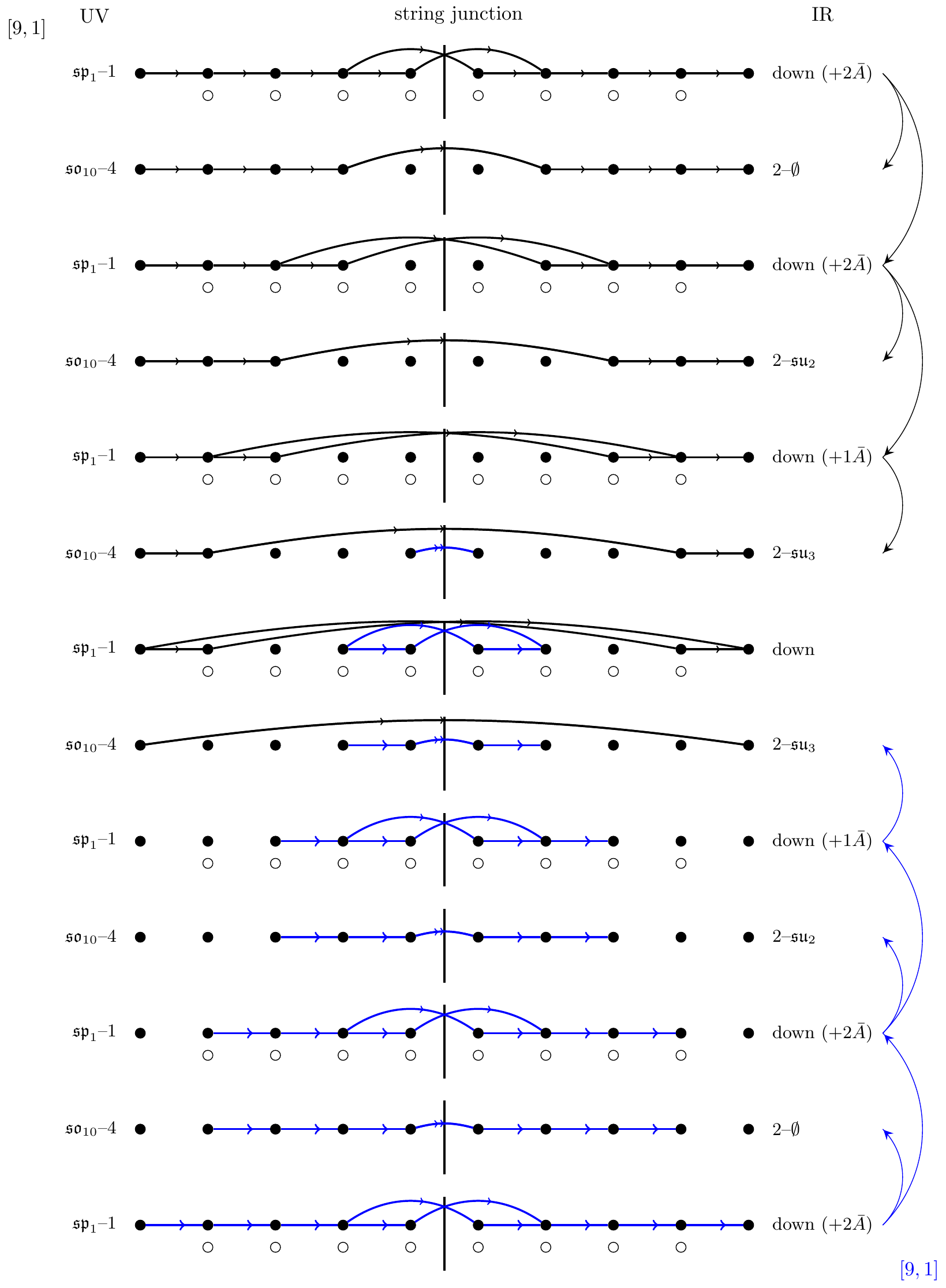}
    \captionsetup{font=footnotesize,labelfont=footnotesize}
    \caption{Kissing configuration: Partitions $\mu_L=\mu_R=[9,1]$ on $13$ curves. Every $-1$ curve participates in a small instanton / smoothing deformation.} \label{subfig:kiss}
  \end{subfigure}  \hfill
  \begin{subfigure}{0.38\textwidth}
    \includegraphics[width=0.9\textwidth]{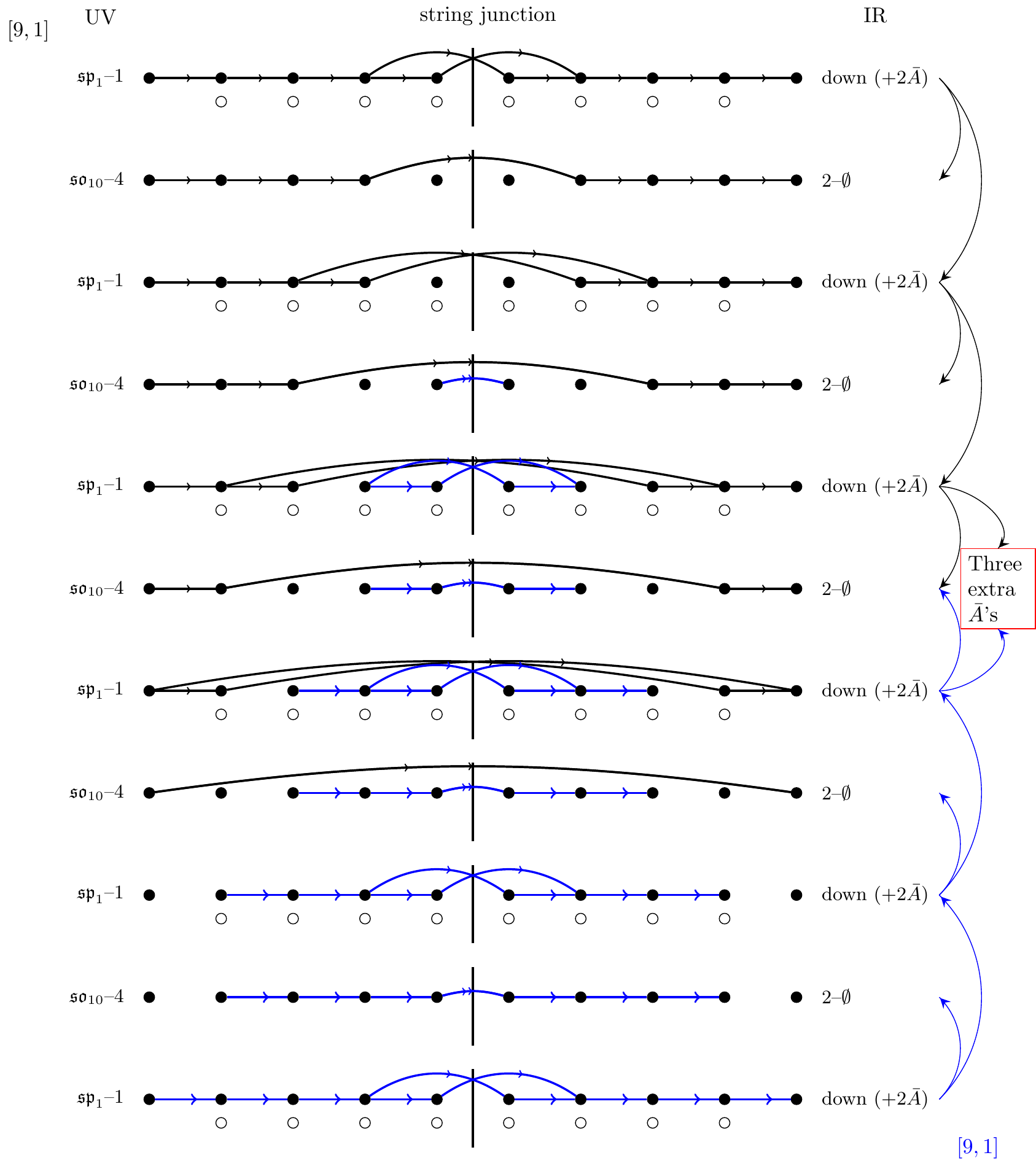}
    \captionsetup{font=footnotesize,labelfont=footnotesize}
    \caption{Crumpled configuration: Partitions $\mu_L=\mu_R=[9,1]$ on only $11$ curves. Too many $\overline{A}$'s are generated.} \label{subfig:crumple}
  \end{subfigure} 
  \caption{Holding fixed the partitions $\mu_L=\mu_R=[9,1]$ we can decrease the number of curves to go from a long quiver (where the deformations are independent) all the way to a forbidden crumpled configuration.}\label{fig:ToShort}
\end{figure}

As we did in the $SU(N)$ case, we now show how to produce short $SO(2N)$ quivers beginning from long ones. For our first example, we consider the following formal $SO(8)$ quiver:
\begin{equation}
[5, 3]: \,\, \overset{\mathfrak{sp}(-3)}{1} \,\, \overset{\mathfrak{so}(4)}{4}
\,\, \overset{\mathfrak{sp}(-1)}{1} \,\, \overset{\mathfrak{so}(7)}{4} \,\, 1
\,\, \overset{\mathfrak{so}(8)}{4} \,\, \overset{\mathfrak{sp}(-1)}{1} \,\,
\overset{\mathfrak{so}(4)}{4}\,\,\overset{\mathfrak{sp}(-3)}{1}\,\, :[4^{2}],
\end{equation}
which is converted into the following F-theory quiver:
\begin{equation}
[5, 3]: \,\, \overset{\mathfrak{su}(2)}{2} \,\, \overset{\mathfrak{g}_{2}}{3}
\,\, 1 \,\, \underset{[SU(2)]}{\overset{\mathfrak{so}(7)}{3}} \,\, {\overset{\mathfrak{su}(2)}{2}}
\,\, [4^{2}].
\end{equation}
If we reduce the length by one, we would get a kissing theory (that is, every $-1$ curve has been blown-down):
\begin{equation}
[5, 3]: \,\, \underset{[N_f = 1]}{\overset{\mathfrak{su}(2)}{2}} \,\, \underset{[SU(2)]}{\overset{\mathfrak{su}%
(3)}{2}}\,\, \underset{[N_f = 1]}{\overset{\mathfrak{su}(2)}{2}} \,\, [4^{2}] \,.
\end{equation}
However, if we try to further reduce the length, we will reach a case that ``crumples'' due to an excess of $\overline{A}$'s that cannot be canceled, and therefore is invalid.

We can also keep the length of the quiver fixed and follow the RG flows along the nilpotent orbits (we will discuss this part in more detail in section
\ref{subsec:DoubleNilpotentHierarchy}). Consider the same example, but now increase the right nilpotent orbit from $[4^2]$ to $[5, 3]$. We still get an ``independent'' theory:
\begin{equation}
[5, 3]: \,\, \overset{\mathfrak{su}(2)}{2} \,\, \overset{\mathfrak{g}_{2}}{3}
\,\, \underset{[SU(2)]}{1} \,\, \overset{\mathfrak{g}_{2}}{3} \,\, \overset{\mathfrak{su}(2)}{2}
\,\, [5, 3] \,.
\end{equation}
If we further increase the right nilpotent orbit to $[7, 1]$, we will instead get
a kissing theory:
\begin{equation}
[5, 3]: [SU(2) \times SU(2)] \,\, {\overset{\mathfrak{su}(2)}{2}} \,\,
\overset{\mathfrak{su}(2)}{2} \,\, \underset{[N_f = 3/2]}{\overset{\mathfrak{su}(2)}{2}} \,\, 2 \,\,
[7, 1] \,.
\end{equation}
At this step, increasing the left orbit also up to $[7, 1]$ would give a crumpled configuration, which is not allowed.

We can describe all of this in general using the string junction picture previously developed.
Following our previous proposal for long quiver brane pictures, we start
from the outermost curves of the quiver, where we initialize our nilpotent deformation in terms of the string junction picture. Then, following the $SO/Sp$
propagation rule, we propagate the clusters from both sides towards the middle
simultaneously. In the case of short quivers, strings from both sides might
end up touching, sharing different intermediate layers, in which case the
gauge group reduction effects from both sides add together. For example,
figure \ref{fig:SOshortBrane} illustrates the action of
$\mu_L=[9,1]$, $\mu_R=[5^2]$ for $SO(10)$ in a theory with $11$ curves.

\begin{figure}[ptb]
  \centering
  \includegraphics[scale=.9]{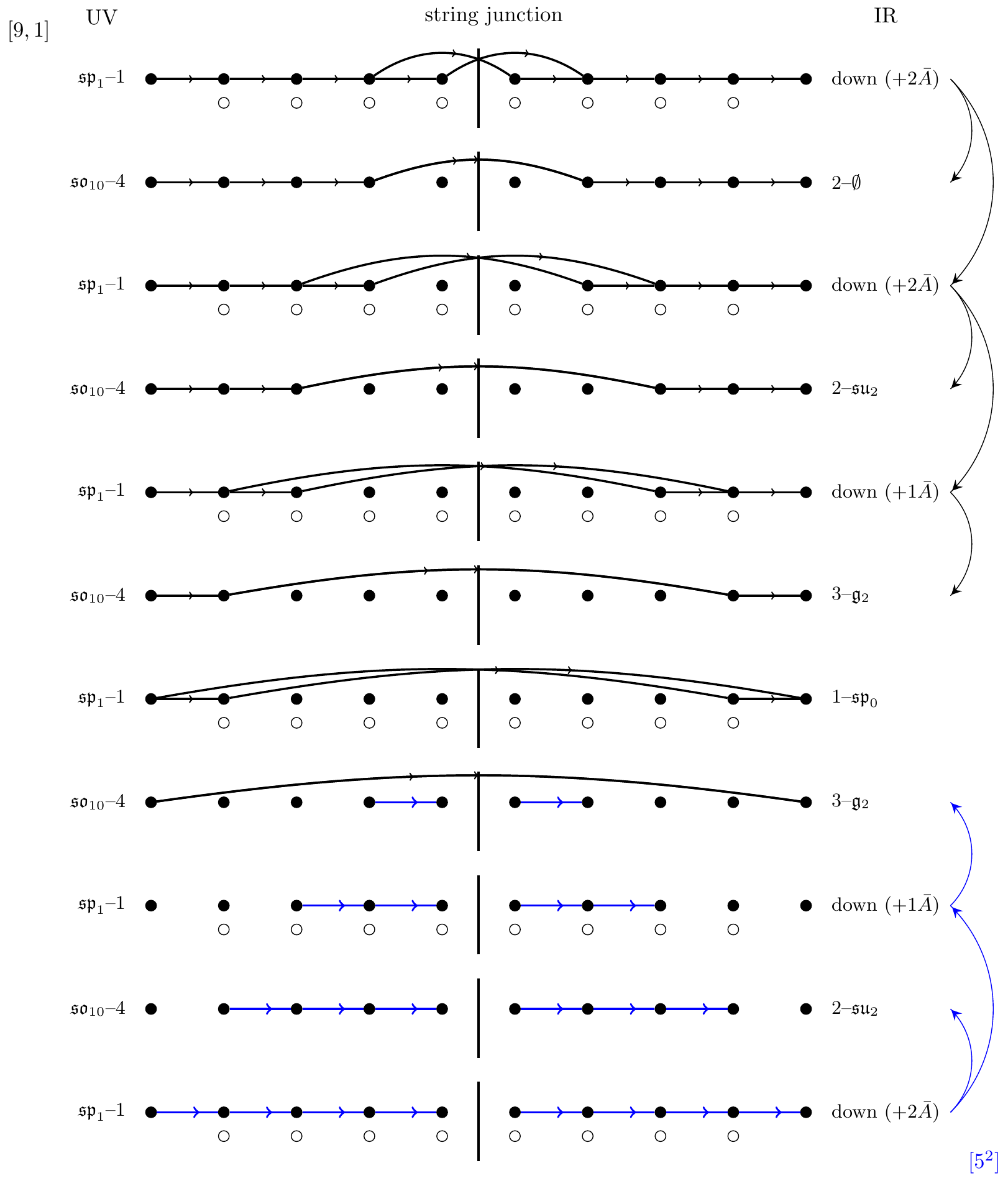}
\caption{An $SO(10)$ short quiver brane picture for nilpotent deformations $\mu_{L} = [9, 1]$, $\mu_{R}=[5^{2}]$. Additional branes are needed in order to construct the associated string diagrams, which in turn introduces anti-branes (depicted by white circles). The figure is arranged so that the left deformation
starts from the top and propagates downwards (in black) while the right
deformation starts on the bottom and propagates upwards (in blue). After the
blowdown and  Higgsing procedures, all
but one of the $-1$ curves are blown down, and the remaining curves now have self-intersection $-2$ or $-3$.}%
\label{fig:SOshortBrane}%
\end{figure}

We note that we can have new situations that could not previously occur in long quivers. The first novelty comes from the fact that levels with $\mathfrak{so}$ gauge algebra can now be Higgsed by two $\overline{A}$'s: one from the left nilpotent deformation and one from the right. As a result, we get configurations where two anti-branes accumulate on the same $-4$ curve and reduce it to a $-2$ curve. The resulting gauge algebra is then given by two applications of the rules for anti-brane reductions given in section \ref{ssec:Higgsing}. Figure \ref{fig:newShort} illustrates this phenomenon for a pair of theories, which respectively involve the reductions:
\begin{align}
  \mathfrak{so}_7 \overset{\overline{A}}{\rightarrow} \mathfrak{g}_2 \overset{\overline{A}}{\rightarrow} \mathfrak{su}_3 \\
  \mathfrak{so}_6 \simeq \mathfrak{su}_4 \overset{\overline{A}}{\rightarrow} \mathfrak{su}_3 \overset{\overline{A}}{\rightarrow} \mathfrak{su}_2.
\end{align}

\begin{figure}[ptb]
  \centering
  \hspace{-1cm}
  \begin{subfigure}{0.5\textwidth}
    \includegraphics[width=\textwidth]{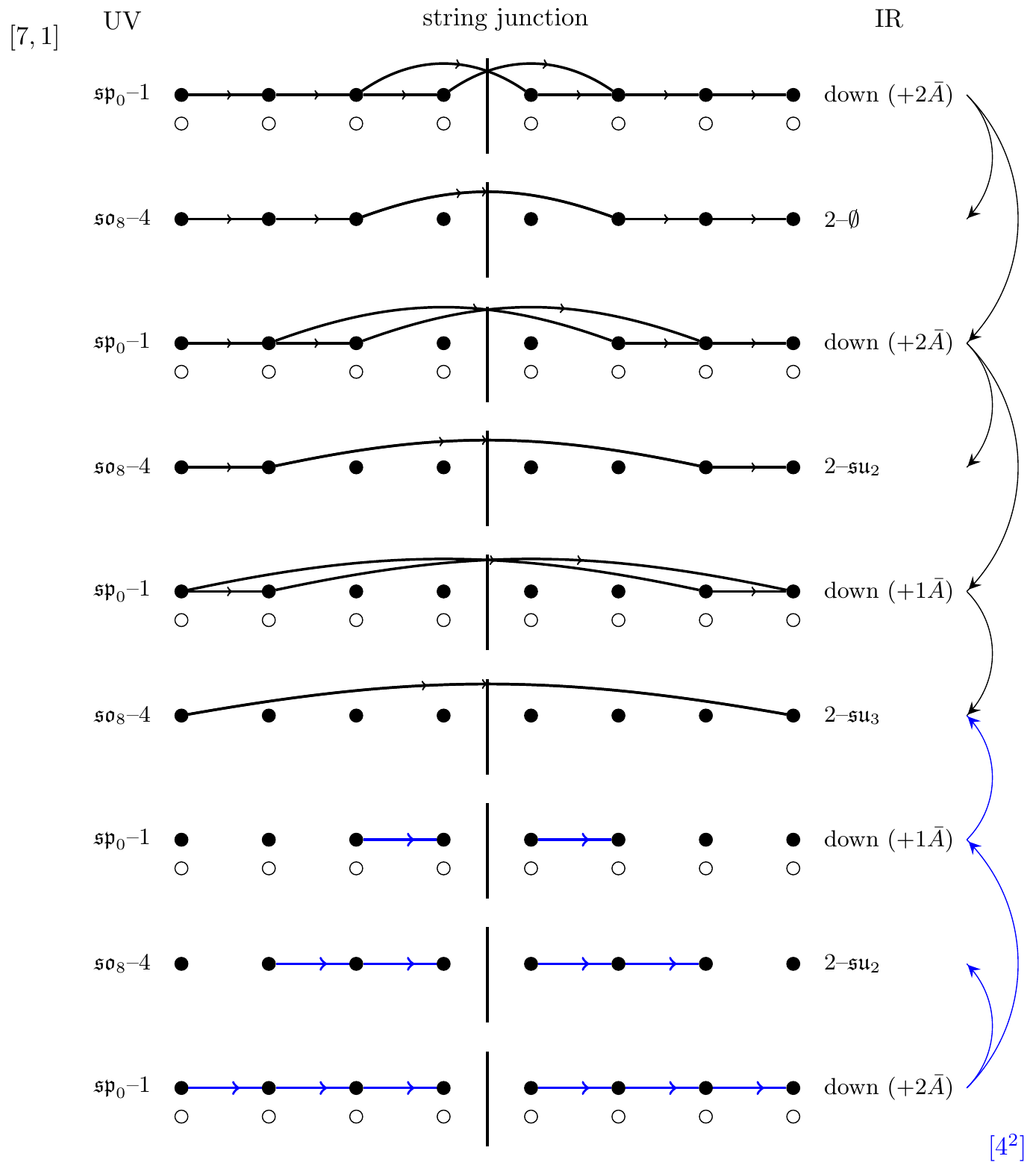}
    \captionsetup{font=footnotesize,labelfont=footnotesize}
    \caption{An example of a configuration that was not found for long quivers: partitions $\mu_L=[7,1]$, $\mu_R=[4^2]$ for a short quiver with $9$ curves. Note that two $\overline{A}$'s land on the third $-4$ curve, one from the top (left partition) and one from the bottom (right partition). There, the gauge group is reduced according to $\mathfrak{so}_7 \overset{\overline{A}}{\rightarrow} \mathfrak{g}_2 \overset{\overline{A}}{\rightarrow} \mathfrak{su}_3$.} \label{subfig:newShort1}
  \end{subfigure} \hspace{0.25cm}
  \begin{subfigure}{0.5\textwidth}
    \includegraphics[width=\textwidth]{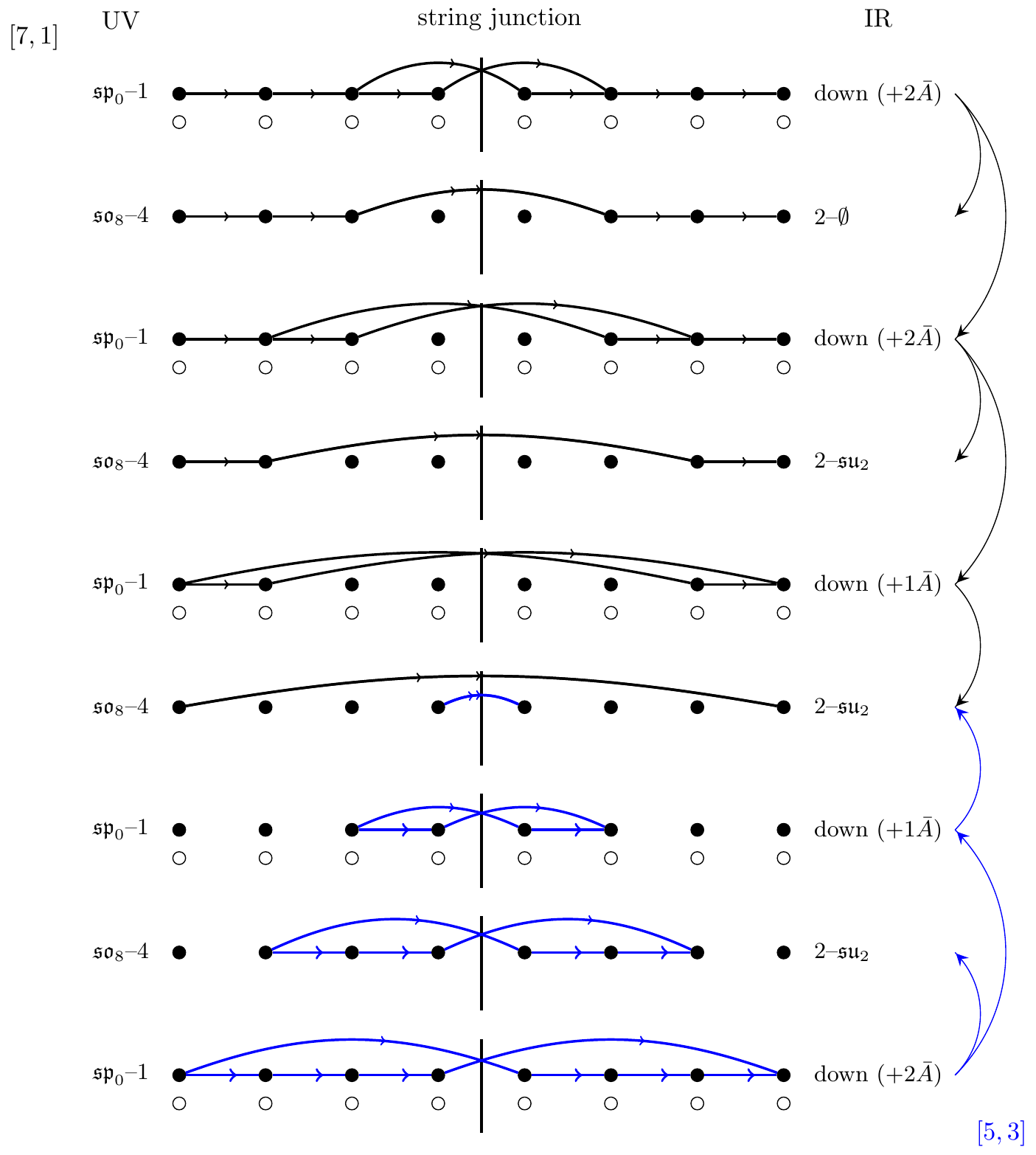}
    \captionsetup{font=footnotesize,labelfont=footnotesize}
    \caption{A second example of a configuration that was not found for long quivers: partitions $\mu_L=[7,1]$, $\mu_R=[5,3]$ for a short quiver with $9$ curves. Note that two $\overline{A}$'s land on the third $-4$ curve, one from the top (left partition) and one from the bottom (right partition). There, the gauge group is reduced according to $\mathfrak{so}_6 \simeq \mathfrak{su}_4 \overset{\overline{A}}{\rightarrow} \mathfrak{su}_3 \overset{\overline{A}}{\rightarrow} \mathfrak{su}_2$.} \label{subfig:newShort2}
  \end{subfigure} \hspace{-1cm}
  \caption{Two interesting examples where two $\overline{A}$'s land on the same $-4$ curve resulting in a chain of Higgsings that was not previously observed for long quivers.}\label{fig:newShort}
\end{figure}

The second novelty is that, in the $SO(8)$ case, partitions related by the triality outer automorphism do not necessarily yield the same IR theory! We saw previously that the long quivers for $\mu=[2^4]^{I,II}$ and $\mu=[3,1^5]$ are identical, as well as long quivers with deformations $\mu=[4^{2}]^{I,II}$ and $\mu=[5,1^{3}]$.
In the case of a long quiver, both of the $[4^{2}]$ and $[5, 1^{3}]$ deformations reduces the UV theory to the following IR theory \cite{Heckman:2016ssk}:
\begin{equation}
\overset{\mathfrak{su}(2)}{2} \,\, \underset{[SU(2)]}{\overset{\mathfrak{so}%
(7)}{3}} \,\, 1 \,\, \overset{\mathfrak{so}(8)}{4} \dots[SO(8)]\,.
\end{equation}
However, if we go to the short quiver cases from a UV theory of three $-4$
curves, we see that the pairs of $[4^{2}]$ -- $[4^{2}]$ and $[4^{2}]$ -- $[5,1^{3}]$ both yield the following quiver theory:
\begin{equation}
\underset{[N_f = 1/2]}{\overset{\mathfrak{su}(2)}{2}} \,\,
\underset{[Sp(2)]}{{\overset{\mathfrak{g_{2}}}{2}}} \,\,
\underset{[N_f = 1/2]}{\overset{\mathfrak{su}(2)}{2}}\,.%
\end{equation}
However, the pair of deformation $[5, 1^{3}]$ -- $[5, 1^{3}]$ gives a different short quiver theory:
\begin{equation}
{\overset{\mathfrak{su}(2)}{2}} \,\,
\underset{[SU(4)]}{{\overset{\mathfrak{su}(4)}{2}}} \,\,
{\overset{\mathfrak{su}(2)}{2}}\,.%
\end{equation}
This is a new effect regarding the outer automorphism of $SO(8)$, which is specific to having a short quiver.
The main point is that is that both $[4^{2}]$ -- $[4^{2}]$ and $[4^{2}]$ -- $[5,1^{3}]$ have one or two $\overline{A}$ branes involved, making it possible to reduce the gauge symmetry to $\mathfrak{g}_{2}$, while the $[5, 1^{3}]$ -- $[5, 1^{3}]$ does not involve $\overline{A}$ branes. Instead, the strings break the UV gauge group down to $\mathfrak{so}(6) \simeq \mathfrak{su}(4)$.

These phenomena are recorded in figures \ref{fig:so8ShortFlowsOne}, \ref{fig:so8ShortFlowsTwo}, and \ref{fig:so8ShortFlowsThree}, but we show explicitly the string junction pictures in figure \ref{fig:compare} for the partitions $\mu_L=\mu_R=[4^{2}]$ vs. the partitions $\mu_L=\mu_R=[5,1^{3}]$. In section \ref{subsubsec:ShortAnomalyRules}, we will justify this surprising conclusion by an analysis of the anomaly polynomials for these respective theories.

\begin{figure}[ptb]
  \centering
  \hspace{-1cm}
  \begin{subfigure}{0.5\textwidth}\vspace{0.78cm}
    \includegraphics[width=\textwidth]{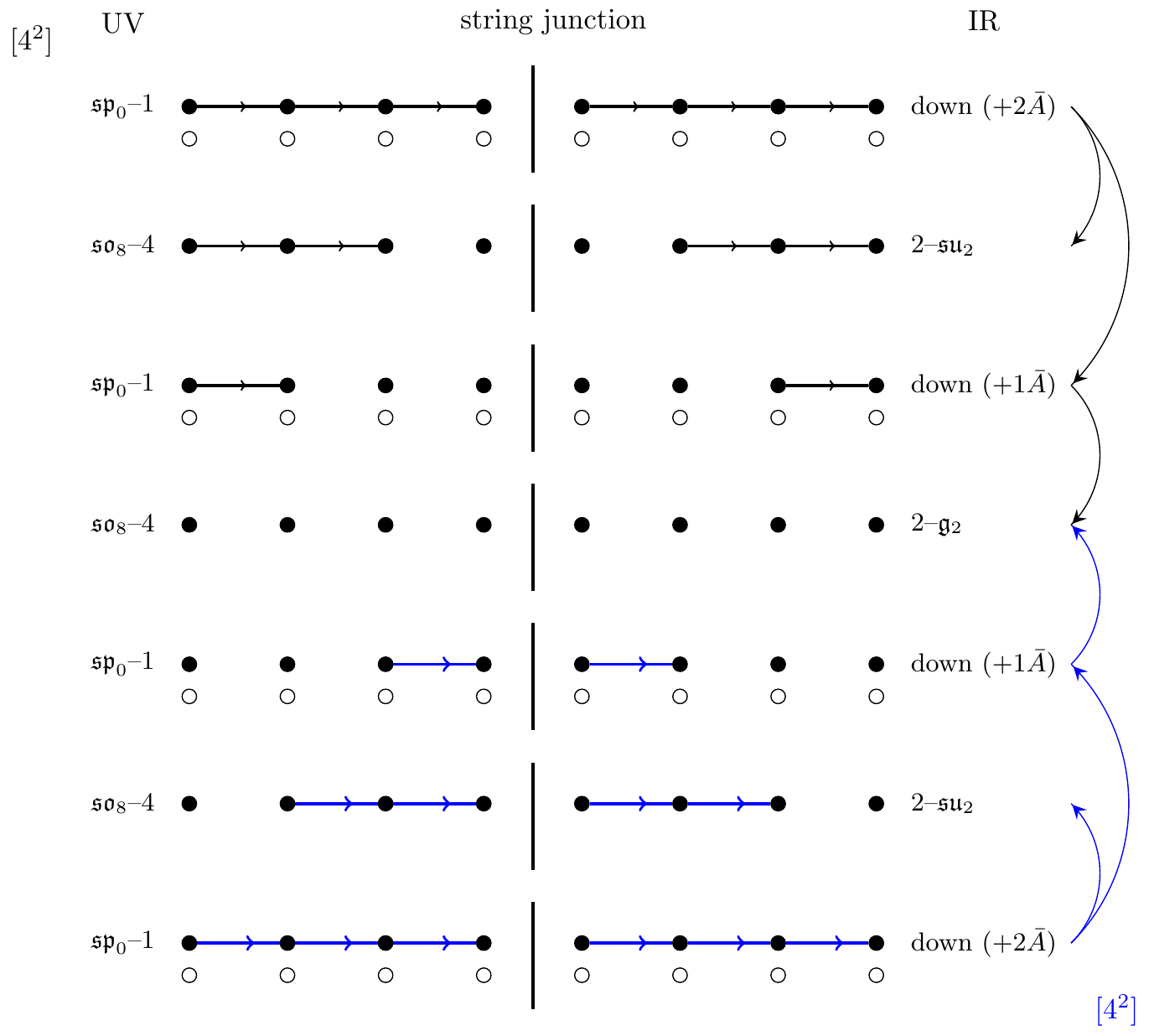}
    \captionsetup{font=footnotesize,labelfont=footnotesize}
    \caption{Partitions $\mu_L=\mu_R=[4^2]$ for a short quiver with $7$ curves. We note that in contrast to long quivers, we obtain a different IR theory than for the partitions $\mu_L=\mu_R=[5,1^3]$. Two $\overline{A}$'s land on the middle $-4$ curve, one from the top (left partition) and one from the bottom (right partition). There, the gauge group is reduced according to $\mathfrak{so}_8 \overset{\overline{A}}{\rightarrow} \mathfrak{so}_7 \overset{\overline{A}}{\rightarrow} \mathfrak{g}_2$.} \label{subfig:compare1}
  \end{subfigure} \hspace{0.25cm}
  \begin{subfigure}{0.5\textwidth}
    \includegraphics[width=\textwidth]{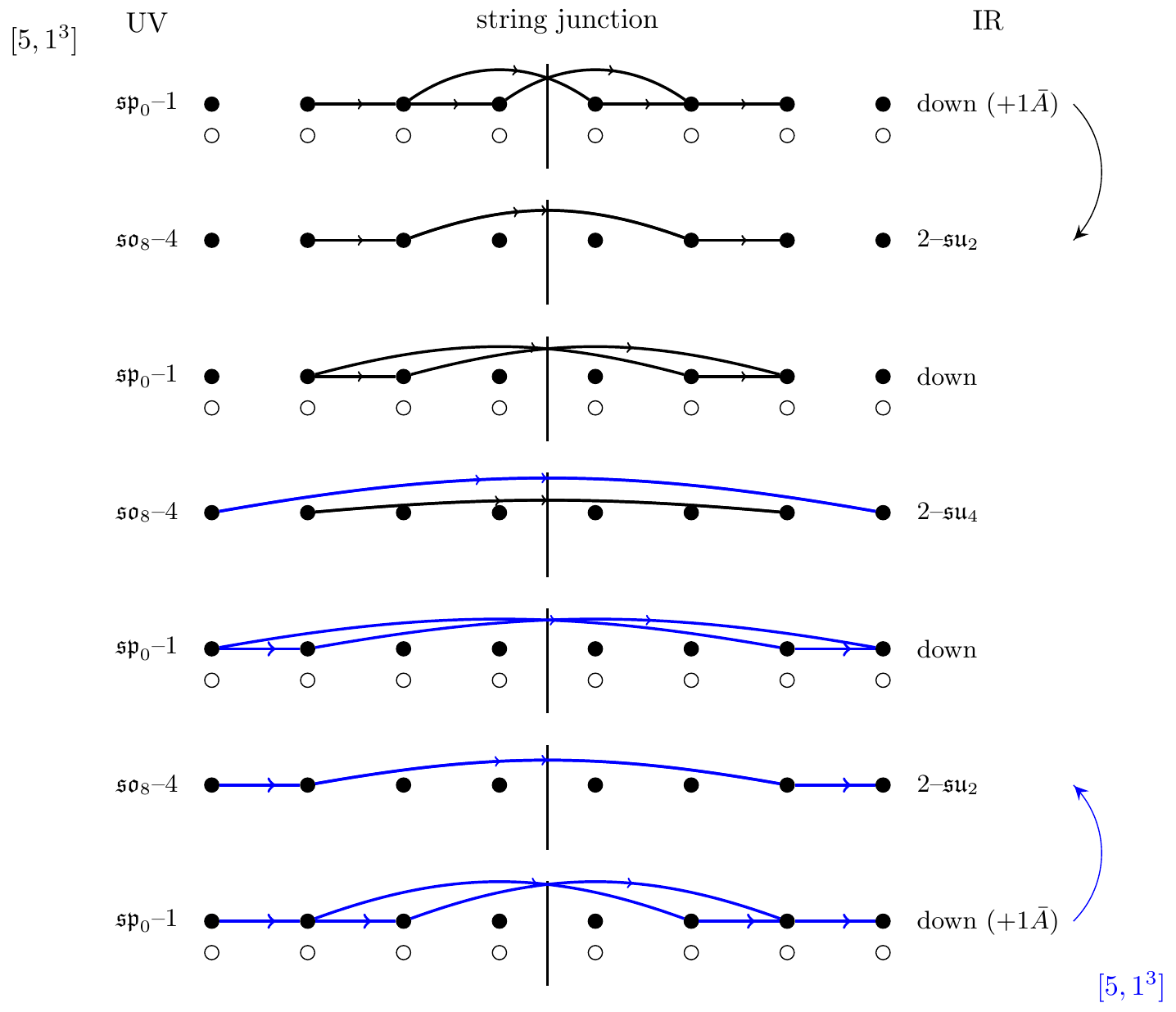}
    \captionsetup{font=footnotesize,labelfont=footnotesize}
    \caption{Partitions $\mu_L=\mu_R=[5,1^3]$ for a short quiver with $7$ curves. We note that in contrast to long quivers we obtain a different IR theory than for the partitions $\mu_L=\mu_R=[4^2]$. On the middle $-4$ curve we now have $\mathfrak{so}_6 \simeq \mathfrak{su}_4$ gauge algebra.} \label{subfig:compare2}
  \end{subfigure} \hspace{-1cm}
  \caption{Nilpotent orbits with $\mu=[5,1^3]$ or $\mu=[4^2]$ yield the same IR theories for long quivers (see figure \ref{fig:SO8quiver} for instance). However, here we see a clear difference for short quivers.}\label{fig:compare}
\end{figure}

\subsubsection{$SO($odd$)$ Case} \label{subsubsec:SOodd}

In general, $SO(2N-1)$ short quivers can be reinterpreted as $SO(2N+2)$ short quivers deformed by a pair of nilpotent orbits.
For example, suppose we start from an $SO(7)$ short quiver UV theory, written as:%
\begin{equation}
[SO(7)] \ \ 1 \ \ \overset{\mathfrak{so}(9)}{4}
\ \ \underset{[N_f = 1]}{\overset{\mathfrak{sp}(1)}{1}}
\ \ \overset{\mathfrak{so}(9)}{4} \ \ 1 \ \ [SO(7)].
\end{equation}
This can be reinterpreted as starting from the following $SO(10)$ UV theory:
\begin{equation}
[SO(10)] \ \ \overset{\mathfrak{sp}(1)}{1}
\ \ \overset{\mathfrak{so}(10)}{4} \ \ \overset{\mathfrak{sp}(1)}{1} \ \ \overset{\mathfrak{so}(10)}{4}
\ \ \overset{\mathfrak{sp}(1)}{1} \ \ [SO(10)],
\end{equation}
and applying the pair of nilpotent deformations $[3, 1^{7}]$ -- $[3,1^{7}]$.

In general, any $SO(2N-p)$ quiver with deformations parametrized by the partitions $\mu_L^{\mathrm{odd}}$, $\mu_R^{\mathrm{odd}}$ of $2N-p$ can be reinterpreted as an $SO(2N)$ quiver with associated partitions $\mu_L^{\mathrm{even}}$, $\mu_R^{\mathrm{even}}$ obtained by simply adding a ``$p$'' to the partitions $\mu_L^{\mathrm{odd}}$ and $\mu_R^{\mathrm{odd}}$, respectively. For instance, for the minimal choice $p=3$ with $\mu_L^{\mathrm{odd}} = [1^9]$, $\mu_R^{\mathrm{odd}} = [7, 1^2]$, we can equivalently express the theory as an $SO(12)$ quiver with $\mu_L^{\mathrm{even}} = [3,1^9]$, $\mu_R^{\mathrm{even}}=[7,3,1^2]$. In this way, the rules we developed for $SO(2N)$ quivers above carry over straightforwardly to $SO(2n-p)$ quivers for $p$ odd.

\subsubsection{$Sp$ Case}
We now turn to quiver-like theories in which the flavor symmetries are a pair of $Sp$-type. The first thing we should note is that no blow-downs can happen.
As a result, there are no ``kissing'' or ``crumpled'' configurations. The only constraint that needs to be imposed comes from the Hanany-Witten moves:
\begin{equation}
  N_{T} \geq \textrm{Max}\{\mu_L^1,\mu_R^1\},
\end{equation}
with $N_{T}$ the number of tensor multiplets in the UV theory.

The behavior of the $Sp$ short quivers is then the same as for $SO(2N)$, where the contributions from each side can overlap, but without any of the complications found due to small instanton transitions or anti-branes. Indeed, no anti-branes are necessary for $Sp$ -- $Sp$ quivers.

\subsubsection{Mixed $[G]$--$[G^\prime]$ Case}
It is interesting to consider mixed quivers where the left and right flavors are not equal. The advantage of our analysis is that it straightforwardly generalizes to these cases. Indeed, without loss of generality let $M \leq N$, then
\begin{itemize}
  \item Quivers with $SU(M)$ -- $SU(N)$, $M<N$, flavor symmetries are obtained from partitions of $N$ with $\mu_L=[\nu_L^i,N-M]$ and $\mu_R=[\mu_R^i]$, where $[\nu_L^i]$ is a partition of $M$.
  \item Quivers with $SO(2M)$ -- $SO(2N)$, $M<N$, flavor symmetries are similarly obtained from partitions of $2N$ with $\mu_L=[\nu_L^i,(N-M)^2]$ and $\mu_R=[\mu_R^i]$, where $[\nu_L^i]$ is a partition of $2M$.
  \item Quivers with $SO(\mathrm{even})$ -- $Sp$ flavors can be viewed as two $SO(\mathrm{even})$ flavor symmetries with the right most $-1$ curve decompactified. Small instanton transitions of the interior $-1$ curves on the right-hand side of this quiver are allowed only if the resulting base is given by 223 or 23.
  \item Any quiver involving $SO($odd$)$ flavor symmetries can be embedded inside an $SO($even$)$ quiver, as seen in subsection \ref{subsubsec:SOodd}. Thus, these reduce to the cases above.
\end{itemize}

\subsection{Anomaly Matching for Short Quivers}\label{subsec:matching}

In this subsection, we propose a method for computing the anomalies of short quivers with classical algebras. We begin by introducing the notion of a ``formal $SO$ quiver.'' We then show how these can be useful in determining the true F-theory quiver of a 6D SCFT via anomaly polynomial matching. In some cases of short quivers, there is a mismatch between the anomaly polynomial computed via the formal $SO$ quiver and the quiver obtained through the string junction picture described previously. However, this mismatch seems to take a universal form, indicating that the string junction approach may nonetheless give the correct answer, even when there is a disagreement with the formal quiver approach. We conclude the subsection with illustrative examples.

\subsubsection{Formal $SO$ theories}

``Formal'' $SO$ quivers involve
analytically continuing the gauge algebra $SO(8+m)$ or $Sp(n)$ so that $m, n
\leq0$. This is only an intermediate step, and the motivation for introducing
such formal quiver is to help determine the actual F-theory quiver via anomaly
polynomial matching (see \cite{Mekareeya:2016yal} for a detailed construction
of such formal quivers). Here, we present a brief review of how this is done.

We start from the long quiver case, where we make a comparison between a long
$SO(8)$ quiver theory and its formal quiver theory and show that the the anomaly polynomials between the two agree. The actual F-theory quiver is obtained by a $[5, 3]$
deformation to the left:
\begin{equation}
[5, 3]: \,\, \overset{\mathfrak{su}(2)}{2} \,\, \overset{\mathfrak{g}_{2}}{3}
\,\, 1 \,\, \overset{\mathfrak{so}(8)}{4} \,\, \cdots\,\, 1 \,\, [SO(8)] \,\,
:[1^{8}]\,.
\label{eq:53}
\end{equation}
On the other hand, we can also express this in terms of a formal quiver by allowing for gauge groups with negative rank:
\begin{equation}
[5, 3]: \,\, \overset{\mathfrak{sp}(-3)}{1} \,\, \overset{\mathfrak{so}(4)}{4}
\,\, \overset{\mathfrak{sp}(-1)}{1} \,\, \overset{\mathfrak{so}(7)}{4} \,\, 1
\,\, \overset{\mathfrak{so}(8)}{4}\,\, \cdots\,\, 1 \,\, [SO(8)] \,\,
:[1^{8}]\,.
\end{equation}
If we truncate both of these theories, keeping only the part of the quiver to the left of the ``$\cdots
$", then their anomaly polynomials are both given by
\begin{equation}
I_8=\frac{6337}{168}c_{2}(R)^{2} + \frac{25}{336}c_{2}(R)p_{1}(T) + \frac
{631}{40320}p_{1}(T)^{2} -  \frac{79}{1440} p_{2}(T).
\label{eq:53I}
\end{equation}
In the case of the formal quiver, this anomaly polynomial computation is performed by analytically continuing the formula for an $Sp-SO$ quiver to negative gauge group rank (see \cite{Mekareeya:2016yal}).

This example illustrates the utility of the formal quiver for anomaly matching. In
our short quiver theories, the actual F-theory quivers can be difficult to read off, whereas these formal $SO$ quivers are easy to determine. As a result, we can use them together with their associated anomaly polynomials relation to check our proposal for the F-theory quiver, as described below.

The general formula for formal quivers--both long and short--is similar to the formula (\ref{eq:SUshort}) for the $SU$ case.
Define the partition of the left and right nilpotent orbits of $SO(2N)$ to be $\mu_{L}^{j},
\mu_{R}^{j}$ and define their conjugate partitions $\rho_{L}^{j}, \rho_{R}^{j}$. We have an alternating
sequence of $SO$ and $Sp$ gauge algebras on the full tensor branch. Indexing the gauge algebras by a parameter $m$
which starts with $Sp(q_1)$ on the left and continues to $SO(p_2)$, ... and terminating with an $Sp$ factor, we have the
assignments:
\begin{equation}
SO(p_m),~~~p_{m}=2N-\sum_{i=m+1}^{N_{L}^{\prime}} \rho_{i}^{L}-\sum_{j=N_T-m+2}%
^{N_{R}^{\prime}} \rho_{j}^{R} \quad(m\,\, \mathrm{even})
\end{equation}
\begin{equation}
Sp(q_m),~~~q_{m}=\frac{1}{2}(2N-\sum_{i=m+1}^{N_{L}^{\prime}} \rho_{i}^{L}-\sum
_{j=N_T-m+2}^{N_{R}^{\prime}} \rho_{j}^{R}) - 4 \quad(m\,\, \mathrm{odd})\,.
\end{equation}
Here, $N_T$ is the number of tensor multiplets in the UV
F-theory description and $N_L^{'}, N_R^{'}$ are the lengths of left and right conjugate partitions, respectively.

Let us illustrate the construction of short quiver formal $SO$ theories by starting with a sufficiently long formal theory and then reducing the length. Consider the $SO(8)$ theory with $[5, 3]$ and $[3^{2}, 1^{2}]$ nilpotent
deformations and four $-4$ curves, so that the
pair of deformations does not overlap:
\begin{equation}
[5, 3]: \,\, \overset{\mathfrak{sp}(-3)}{1} \,\, \overset{\mathfrak{so}(4)}{4}
\,\, \overset{\mathfrak{sp}(-1)}{1} \,\, \overset{\mathfrak{so}(7)}{4} \,\, 1
\,\, \overset{\mathfrak{so}(8)}{4} \,\, 1 \,\, \overset{\mathfrak{so}%
(4)}{4}\,\,\overset{\mathfrak{sp}(-1)}{1}\,\, :[3^{2}, 1^{2}]\,.
\end{equation}
Now we decrease the length of the quiver. In each step, we start from a
shorter UV theory by removing one group of $(-1, -4)$ curves. We get the
following set of theories after each step:%
\begin{equation}
[5, 3]: \,\, \overset{\mathfrak{sp}(-3)}{1} \,\, \overset{\mathfrak{so}(4)}{4}
\,\, \overset{\mathfrak{sp}(-1)}{1} \,\, \overset{\mathfrak{so}(7)}{4} \,\, 1
\,\, \overset{\mathfrak{so}(4)}{4}\,\,\overset{\mathfrak{sp}(-1)}{1}\,\,
:[3^{2}, 1^{2}]
\end{equation}
\begin{equation}
[5, 3]: \,\, \overset{\mathfrak{sp}(-3)}{1} \,\, \overset{\mathfrak{so}(4)}{4}
\,\, {\overset{\mathfrak{sp}(-1)}{1}} \,\,
\overset{\mathfrak{so}(5)}{4}\,\,{\overset{\mathfrak{sp}%
(-2)}{1}}\,\, :[3^{2}, 1^{2}]\,.
\end{equation}

We stop at this point, following the constraints from the Hanany-Witten moves.
We see that the formal gauge algebra goes down to the unphysical values of $\mathfrak{sp}%
(-3)$ and $\mathfrak{so}(2)$.

However, from such a quiver we may still extract its anomaly polynomial by analytically continuing the formulae developed in the physical regime, $\mathfrak{sp}(m), m > 0$ and $\mathfrak{so}(n),
n \geq8$. In the long quiver case, the anomaly polynomial of the formal quiver exactly matches that of the actual quiver \cite{Mekareeya:2016yal}, as in the example in (\ref{eq:53})-(\ref{eq:53I}). This serves as a strong motivation for us to test the relationship between $SO$ short quivers and their formal counterparts via anomaly matching.

\subsubsection{Anomaly Polynomial Matching and Correction Terms}
    \label{subsubsec:ShortAnomalyRules}

    For theories with long quivers, there is a well-defined prescription in the literature for producing the F-theory quiver of a given formal type IIA quiver (see \cite{Mekareeya:2016yal}). For short quiver theories, however, the situation becomes much more complicated, and there is at present no well-defined proposal in the literature. Nonetheless, the rules we have introduced in section \ref{sec:RECOMBO} carry over to the case of short quivers, so we may check that these rules give the correct answer by comparing the anomaly polynomials of the proposed short quiver theories to those obtained from the formal quiver. This check has been done explicitly for all cases in the catalogs \ref{table:SO8tangentialShortQuiver} and \ref{table:SO10TangentshortQuiver} in Appendix \ref{apdx:shortQuiverCatalogs}.

In general, we find that there is frequently a mismatch in the $p_1(T)^2$ and $p_2(T)$ coefficients of the anomaly polynomials computed via the formal quiver vs. the actual F-theory quiver. However, this is not very concerning, as the mismatch can always be canceled by adding an appropriate number of neutral hypermultiplets, each of which contributes $(4p_1(T)^2-7p_2(T))/5760$ to the anomaly polynomial. Indeed, such a mismatch in short quiver theories was previously noted in \cite{Heckman:2018pqx}.

More concerning are the mismatches in the coefficients of the $c_2(R)^2 $ coefficient and the $c_2(R) p_1(T)$ coefficient (denoted $\alpha$ and $\beta$, respectively). These mismatches are relatively rare, arising only in a smaller number of kissing cases (see tables \ref{table:SO8tangentialShortQuiver} and \ref{table:SO10TangentshortQuiver} in Appendix \ref{apdx:shortQuiverCatalogs}). This could be an indication that these theories are sick and should be discarded. However, we note that these mismatches seem to follow a universal set of rules, which indicates that our proposed F-theory quiver may nonetheless represent an accurate translation of the formal quiver.

Theories with mismatches always involve two anti-branes acting on a curve carrying an $\mf{so}$ gauge algebra according to the rules in (\ref{eq:rules}), and it depends on the size of the gauge group. In particular, denoting the mismatch in the anomaly polynomial coefficients $\alpha$ and $\beta$ by $\Delta \alpha$, $\Delta \beta$, respectively, we have:
\begin{enumerate}[label*=\arabic*)]
    \item
    \begin{equation}
    \mathfrak{so}(8) \overset{2\overline{A}}{\rightarrow} \mathfrak{g}_{2}: (\Delta \alpha, \Delta \beta) = (0,0)
    \end{equation}
    (see figure \ref{subfig:compare1} for an example)

    \item
    \begin{equation}
    \mathfrak{so}(7) \overset{2\overline{A}}{\rightarrow} \mathfrak{su}(3):  (\Delta \alpha, \Delta \beta) = (\frac{1}{24},\frac{1}{48})
    \label{eq:rule2}
    \end{equation}
    (see figure \ref{subfig:newShort1} for an example)

    \item
    \begin{equation}
    \mathfrak{so}(6) \simeq \mathfrak{su}(4) \overset{2\overline{A}}{\rightarrow} \mathfrak{su}(2):  (\Delta \alpha, \Delta \beta) = (\frac{1}{12},\frac{1}{24})
    \label{eq:rule3}
    \end{equation}
    (see figure \ref{subfig:newShort2} for an example)

    \item
    \begin{equation}
    \mathfrak{so}(5)  \overset{2\overline{A}}{\rightarrow} \mathfrak{su}(1):  (\Delta \alpha, \Delta \beta) = (\frac{1}{6},\frac{1}{12})
    \end{equation}

    \item
    \begin{equation}
    \textrm{All remaining cases}:  (\Delta \alpha, \Delta \beta) = (0,0).
    \end{equation}
    \end{enumerate}
Note that the kissing condition and Hanany-Witten constraints only allow one $-4$ curve to have 2 $\overline{A}$'s simultaneously attach to the curve. There is one borderline case involving $\mathfrak{so}(4)$ gauge symmetry and a pair of $\overline{A}$'s. In both long and short quivers, we have a consistent rule $\mathfrak{so}(4) \overset{\overline{A}}{\rightarrow} \mathfrak{su}(2)$, but adding an additional $\overline{A}$ brane appears to be problematic in general. Including this case would generate a curve without any gauge symmetry, which in many examples leads to a quiver where the ``convexity condition'' required of gauge group ranks is violated. This is best illustrated with an example.
Consider the UV quiver:
\begin{equation*}
  [1^{16}] \,\,
  \overset{\mathfrak{sp}(4)}{1}\,\,\overset{\mathfrak{so}(16)}{4}\,\,
  \overset{\mathfrak{sp}(4)}{1} \,\,\overset{\mathfrak{so}(16)}{4}\,\,
  \overset{\mathfrak{sp}(4)}{1} \,\,\overset{\mathfrak{so}(16)}{4}\,\,
  \overset{\mathfrak{sp}(4)}{1}\,\,
          [1^{16}]
\end{equation*}
If we were to na\"ively assume that $\mathfrak{so}(4)  \overset{2\overline{A}}{\rightarrow} \emptyset$ without crumpling, then the deformation $\mu_L=\mu_R = [7^2,1^2]$ would yield the following sick IR theory:
\begin{equation*}
  [7^2,1^2] \,\,
  \overset{\mathfrak{su}(2)}{2}\,\,
  \overset{\emptyset}{2}\,\,
  \overset{\mathfrak{su}(2)}{2}\,\,
          [7^2,1^2]
\end{equation*}
From this, we conclude that whenever $\mathfrak{so}(4)$ is hit by two $\overline{A}$'s simultaneously, it must crumple, so we forbid these configurations.

In summary, in cases without a double $\overline{A}$ Higgsing chain (``All remaining cases") we never have such a mismatch, and in many cases with a double $\overline{A}$ Higgsing chain, there is also no mismatch. There are a few cases where there is a mismatch, which always involve two $\overline{A}$'s in the Higgsing chain. The above proposal has been explicitly verified in the $SO(8)$ and $SO(10)$ catalogs of
Appendix \ref{apdx:shortQuiverCatalogs}.

What is the physical interpretation of these mismatches? We note that in case (1), where there is no mismatch, the gauge group is reduced from $\mathfrak{so}(8)\overset{2\overline{A}}{\rightarrow} \mathfrak{g}_2$, and the brane picture and the string junction root system make perfect sense. However, when there is a mismatch (as in cases (2)-(5)), we always start from an SO brane picture with an orientifold and somehow end up with a SU brane without an orientifold. We leave further explanation of this issue for future work.

\subsubsection{Examples}\label{ssec:examples}

In this section, we present a number of examples to demonstrate our procedure of anomaly matching explicitly and to reveal some of the subtleties of our procedure regarding different quiver lengths, different UV gauge groups, and different types of Higgsing.

\begin{itemize}

\item{\textbf{Example 1}}

We start with the pair of orbits $[5, 1^{3}], [5, 1^{3}]$ on an $SO(8)$ UV
theory with tensor branch given by three $-4$ curves. The resulting description in F-theory
is:
\begin{equation}
\overset{\mathfrak{su}(2)}{2}
\,\,\underset{[SU(4)]}{\overset{\mathfrak{su}(4)}{2}}
\,\,\overset{\mathfrak{su}(2)}{2}
\end{equation}
This theory gives the same anomaly polynomial as the corresponding formal $SO$ quiver:
\begin{equation}
[5, 1^3]: \,\, \overset{\mathfrak{sp}(-2)}{1} \,\, \overset{\mathfrak{so}(5)}{4}
\,\, \overset{\mathfrak{sp}(-1)}{1} \,\, \overset{\mathfrak{so}(6)}{4}\,\, \overset{\mathfrak{sp}(-1)}{1} \,\,
\overset{\mathfrak{so}(5)}{4}\,\,\overset{\mathfrak{sp}(-2)}{1}\,\, :[5,1^3]\,.
\label{eq:SOexample}
\end{equation}
The anomaly polynomial reads:%
\begin{equation}
I_8=\frac{77}{4}c_{2}(R)^{2} - \frac{3}{8}c_{2}(R)p_{1}(T) + \frac{73}{2880}%
p_{1}(T)^{2} - \frac{49}{720}p_{2}(T).
\end{equation}

\item{\textbf{Example 2}}

For a second example, we deform the UV theory of three $-4$ curves by the pair
of orbits of $[4^{2}], [4^{2}]$ (our analysis does not distinguish between the two nilpotent orbits
associated with this partition). The formal theory:
\begin{equation}
[4^{2}]: \,\, \overset{\mathfrak{sp}(-3)}{1} \,\, \overset{\mathfrak{so}%
(4)}{4} \,\, \overset{\mathfrak{sp}(-1)}{1} \,\,
{\overset{\mathfrak{so}(8)}{4}} \,\, \overset{\mathfrak{sp}%
(-1)}{1} \,\, \overset{\mathfrak{so}(4)}{4}\,\,\overset{\mathfrak{sp}%
(-3)}{1}\,\, :[4^{2}]
\end{equation}
gives the following anomaly polynomial:%
\begin{equation}
\frac{463}{24}c_{2}(R)^{2} - \frac{17}{48}c_{2}(R)p_{1}(T) + \frac{73}%
{2880}p_{1}(T)^{2} - \frac{101}{1440}p_{2}(T).
\end{equation}
If we subtract off the contribution of one neutral hypermultiplet $I_{\textrm{neutral}} =
\frac{7p_{1}(T)^{2} - 4p_{2}(T)}{5760}$, we get the F-theory quiver anomaly polynomial:
\begin{equation}
I_{F} = I_{\textrm{formal}} - I_{\textrm{neutral}} = \frac{463}{24}c_{2}(R)^{2} - \frac{17}{48}%
c_{2}(R)p_{1}(T) + \frac{139}{5760}p_{1}(T)^{2} - \frac{97}{1440}p_{2}(T)
\end{equation}
which can be obtained from the F-theory quiver:
\begin{equation}
[4^{2}]: \,\, \underset{[N_f = 1/2]}{\overset{\mathfrak{su}(2)}{2}} \,\, \underset{[Sp(2)]}{\overset{\mathfrak{g}_{2}%
}{2}}\,\, \underset{[N_f = 1/2]}{\overset{\mathfrak{su}(2)}{2}} \,\,: [4^{2}].
\end{equation}

This result is actually quite surprising: the nilpotent deformations considered in these past two examples are related by triality of $SO(8)$. Indeed, their long F-theory quivers are identical, and they have identical anomaly polynomials, even though their formal quivers differ. However, we have just seen that their kissing cases actually differ! We have confirmed this surprising result via anomaly polynomial matching.

\item{\textbf{Example 3}}

Next, we consider a pair of cases with an anomaly polynomial mismatch.

\begin{itemize}

\item{3a}

Consider the theory with $\mu_L=[7, 1]$, $\mu_R = [4^2]$ on an SO(8) UV quiver with four $-4$ curves. The brane pictures for this example are depicted in figure \ref{subfig:newShort1}.
The theory has the following IR quiver:

\begin{equation}
  [7, 1]: \,\,2   \,\, \underset{[N_f = 1/2]}{\overset{\mathfrak{su}(2)}{2}} \,\, \underset{[SU(2)]}{\overset{\mathfrak{su}(3)}{2}}\,\, \underset{[N_f = 1]}{\overset{\mathfrak{su}(2)}{2}} \,\, :[4^2] \,.
\end{equation}

The curve carrying $SU(3)$ na\"ively has $\mf{so}(7)$ gauge algebra, but it is hit by two $\overline{A}$'s, one from the right and one from the left. As a result, the gauge algebra is reduced according to $\mathfrak{so}(7)\overset{2\overline{A}}{\rightarrow}\mathfrak{su}(3)$. This puts us in the situation of rule 2, shown in (\ref{eq:rule2}), so we expect an anomaly correction term of the form $(\Delta \alpha,\Delta\beta) =(1/24, 1/48)$.

Indeed, the formal quiver in this case is given by
\begin{equation}
[7,1]: \,\, \overset{\mathfrak{sp}(-3)}{1} \,\, \overset{\mathfrak{so}%
(3)}{4} \,\, \overset{\mathfrak{sp}(-2)}{1}\,\, \overset{\mathfrak{so}%
(5)}{4} \,\, \overset{\mathfrak{sp}(-1)}{1} \,\,
{\overset{\mathfrak{so}(7)}{4}} \,\, \overset{\mathfrak{sp}%
(-1)}{1} \,\, \overset{\mathfrak{so}(4)}{4}\,\,\overset{\mathfrak{sp}%
(-3)}{1}\,\, :[4^{2}] \,.
\end{equation}
The anomaly polynomial of the F-theory quiver is given by
\begin{equation}
I_{F} = \frac{1331}{60}c_{2}(R)^{2} - \frac{5}{24}%
c_{2}(R)p_{1}(T) + \frac{37}{1440}p_{1}(T)^{2} - \frac{31}{360}p_{2}(T),
\end{equation}
which is indeed the same as $I_{\textrm{formal}}-c_2(R)^2/24-c_2(R) p_1(T)/48- 2 I_{\textrm{neutral}}$.

\item{3b}

Consider the $SO(8)$ theory with nilpotent deformations $[3, 2^2, 1]$ and $[2^4]$ on a UV quiver with a single $-4$ curve. The F-theory quiver is given by:
\begin{equation}
  [3, 2^2, 1]: \,\, \underset{[SU(6)]}{\overset{\mathfrak{su}(3)}{2}} \,\,  :[2^4] \,.
\end{equation}
Here, we again have one anti-brane from both the left and the right, which collide on the $-4$ curve and reduce it as $\mathfrak{so}(7)\overset{2\overline{A}}{\rightarrow}\mathfrak{su}(3)$. The formal quiver is given by
\begin{equation}
[3,2^2,1]: \,\, \overset{\mathfrak{sp}(-2)}{1} \,\, \overset{\mathfrak{so}%
(7)}{4} \,\, \overset{\mathfrak{sp}(-2)}{1}\,\, :[2^{4}]\,.
\end{equation}
The anomaly polynomial of the F-theory quiver is given by
\begin{equation}
I_{F} = \frac{47}{24}c_{2}(R)^{2} - \frac{7}{48}%
c_{2}(R)p_{1}(T) + \frac{31}{1920}p_{1}(T)^{2} - \frac{13}{480}p_{2}(T)\,,
\end{equation}
which is equal to $I_{\textrm{formal}}-c_2(R)^2/24-c_2(R) p_1(T)/48- 4 I_{\textrm{neutral}}$, as expected from (\ref{eq:rule2}).

\end{itemize}
Note that the rule from (\ref{eq:rule2}) has worked correctly for both examples, despite the difference in size of their respective quivers.

\item{\textbf{Example 4}}

As a final example, let us consider a pair of theories with a similar mismatch in the anomaly polynomial but  different UV gauge groups.

\begin{itemize}

\item{4a}

 First, we consider the theory with $SO(8)$ UV gauge groups, nilpotent deformations $[7, 1]$ and $[5, 3]$, and a theory with four $-4$ curves, whose brane diagrams are depicted in figure \ref{subfig:newShort2}. The IR quiver takes the form:
\begin{equation}
  [7, 1]: \,\, 2 \,\, \underset{[N_f = 3/2]}{\overset{\mathfrak{su}(2)}{2}} \,\, {\overset{\mathfrak{su}(2)}{2}}\,\, {\overset{\mathfrak{su}(2)}{2}} \,\, [SU(2)\times SU(2)] \,\, :[5, 3]\,.
\end{equation}
Here, the middle $\mf{su}(2)$ gauge algebra comes from two anti-branes acting on an $\mf{so}(6)$. Per rule 3 of (\ref{eq:rule3}), we expect a mismatch of the form $(\Delta\alpha,\Delta\beta) = (1/12,1/24)$. Indeed, the formal quiver is given by
\begin{equation}
[7,1]: \,\, \overset{\mathfrak{sp}(-3)}{1} \,\, \overset{\mathfrak{so}%
(3)}{4} \,\, \overset{\mathfrak{sp}(-2)}{1}\,\, \overset{\mathfrak{so}%
(5)}{4} \,\, \overset{\mathfrak{sp}(-1)}{1} \,\,
{\overset{\mathfrak{so}(6)}{4}} \,\, \overset{\mathfrak{sp}%
(-1)}{1} \,\, \overset{\mathfrak{so}(4)}{4}\,\,\overset{\mathfrak{sp}%
(-3)}{1}\,\, :[5,3]\,.
\end{equation}
The anomaly polynomial of the F-theory quiver is given by
\begin{equation}
I_{F} = \frac{1943}{120}c_{2}(R)^{2} - \frac{5}{48}%
c_{2}(R)p_{1}(T) + \frac{47}{1920}p_{1}(T)^{2} - \frac{41}{480}p_{2}(T),
\end{equation}
which is indeed the same as $I_{\textrm{formal}}-c_2(R)^2/12-c_2(R) p_1(T)/24- 2 I_{\textrm{neutral}}$.

\item{4b}

 Finally, consider the $SO(10)$ theory with nilpotent deformations $[5^2]$, $[3^2, 2^2]$ on a quiver with two $-4$ curves. This gives:
\begin{equation}
  [5^2]: \,\, [SU(2)] \,\, {\overset{\mathfrak{su}(2)}{2}}\,\, {\overset{\mathfrak{su}(2)}{2}} \,\, [SU(2)\times SU(2)] \,\, :[3^2, 2^2] \,.
\end{equation}
The $\mf{su}(2)$ gauge algebra on the right-hand side again comes from two anti-branes acting on $\mf{so}(6)$. The formal quiver is given by
\begin{equation}
[5^2]: \,\, \overset{\mathfrak{sp}(-3)}{1} \,\, \overset{\mathfrak{so}%
(4)}{4} \,\, \overset{\mathfrak{sp}(-1)}{1}\,\, \overset{\mathfrak{so}%
(6)}{4} \,\, \overset{\mathfrak{sp}(-2)}{1}\,\, :[3^2,2^2] \,.
\end{equation}
The anomaly polynomial of the F-theory quiver is given by
\begin{equation}
I_{F} = \frac{23}{6}c_{2}(R)^{2} - \frac{1}{12}%
c_{2}(R)p_{1}(T) + \frac{11}{720}p_{1}(T)^{2} - \frac{2}{45}p_{2}(T),
\end{equation}
which is indeed the same as $I_{\textrm{formal}}-c_2(R)^2/12-c_2(R) p_1(T)/24- 4 I_{\textrm{neutral}}$, as expected from (\ref{eq:rule3}).

\end{itemize}
Note that the rule from (\ref{eq:rule3}) has worked correctly for both examples, despite the difference in size of their respective quivers as well as their UV gauge groups.

\end{itemize}

Further examples of anomaly polynomial matching can be found in the catalogs in Appendix \ref{apdx:shortQuiverCatalogs}.

\subsection{Nilpotent Hierarchy of Short Quivers \label{subsec:DoubleNilpotentHierarchy}}

Using our analysis above, we now determine a partial ordering for 6D SCFTs based on
pairs of nilpotent orbits, which works in both long and short quivers. We refer to this as a ``double Hasse diagram,'' since it generalizes the independent Hasse diagrams realized by nilpotent orbits on each side of a long quiver (see \cite{Heckman:2016ssk, Bourget:2019aer}) to the case of a short quiver, where the nilpotent deformations overlap. We will see that as we reduce the length of the quiver, several nilpotent orbits will end up generating the same IR fixed point. Said another way, different pairs of nilpotent orbits actually give rise to the same IR theory.

Constructing the double Hasse diagrams proceeds in two steps. First we apply the product order to the tuple of left and right partitions $\mu_L$ and $\mu_R$. It is defined by $(\mu_L, \mu_R) \preceq (\nu_L, \nu_R)$ which holds if and only if $\mu_L \preceq \nu_L$ and $\mu_R \preceq \nu_R$. However, because several deformations in the UV can flow to the same IR theory, we refine this partial ordering in the second step by merging all partitions which result in the same IR quiver. We obtain the same result from a microscopic perspective by appropriately adding strings to the left and right sides of the string junction picture, exactly as we did for the long quivers.

\subsubsection{Example: $SU(4)$}

As a first example, we consider an $SU(4)$ double Hasse diagram. We begin with the UV theory:%
\begin{equation}
[1^{4}]: \underset{[SU(4)]}{\overset{\mathfrak{su}(4)}{2}} \,
\overset{\mathfrak{su}(4)}{2 }\, \underset{[SU(4)]}{\overset{\mathfrak{su}%
(4)}{2}} \, \,:[1^{4}]\,.
\end{equation}

Then we turn on nilpotent deformations on both sides, as in the single-sided versions that were plotted in \cite{Heckman:2016ssk}. Note that $SU(4)$ only has
five nilpotent orbits - $[1^{4}], [2, 1^{2}], [2^{2}], [3, 1], [4]$, but the
$[4]$ orbit is prohibited on $N_{-2} = N_{T} = 3$ curves by the Hanany-Witten moves constraint of equation (\ref{eq:HWconstraintSU}).
We are then left with the double Hasse diagram of figure \ref{fig:su4ShortFlowsThree}. This generalizes straightforwardly to all $SU(N)$  quivers.

\begin{figure}
\includegraphics{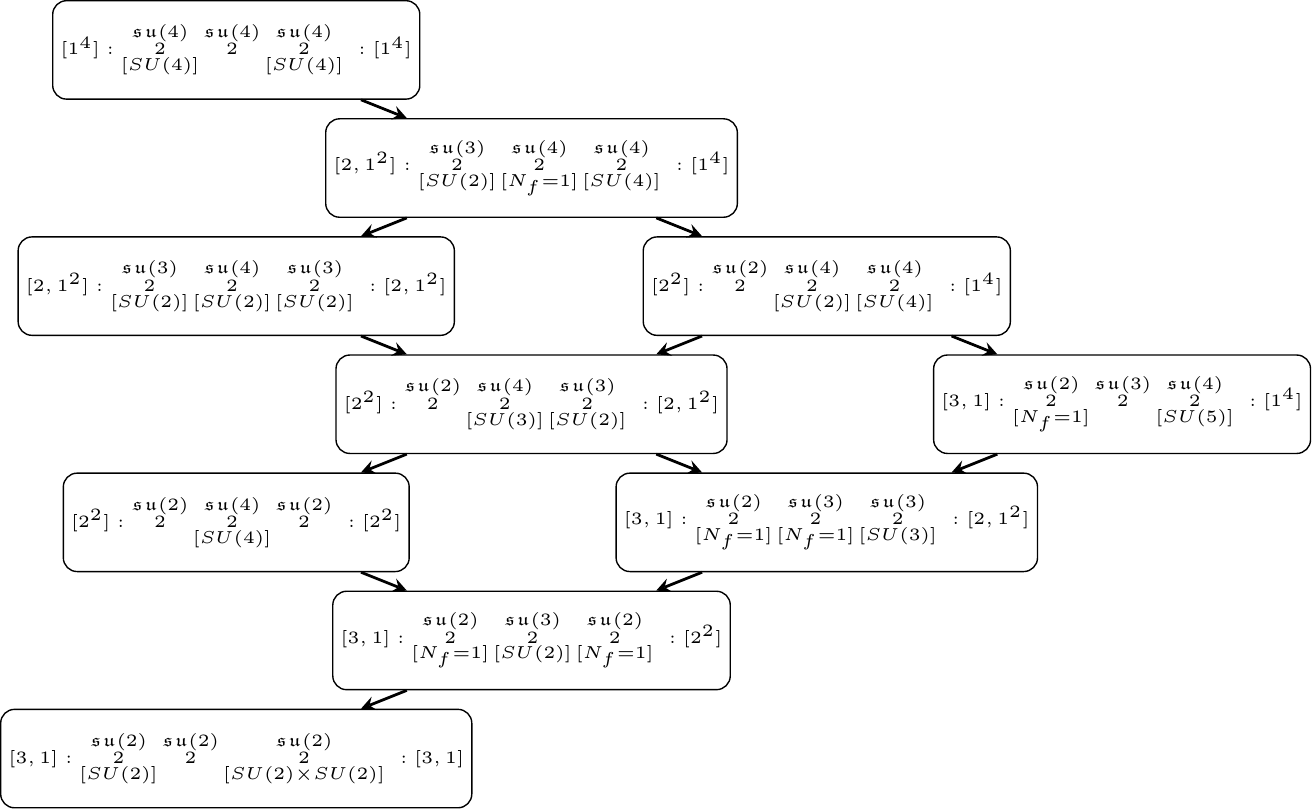}
\caption{Half of the double Hasse diagram of $SU(4)$ short quivers. The full diagram is obtained by reflection across the left-most nodes, as the quivers can always be flipped under the reflection $\mu_L \leftrightarrow \mu_R$.}
\label{fig:su4ShortFlowsThree}%
\end{figure}

\subsubsection{Example: $SO(8)$}
\begin{figure}
\includegraphics{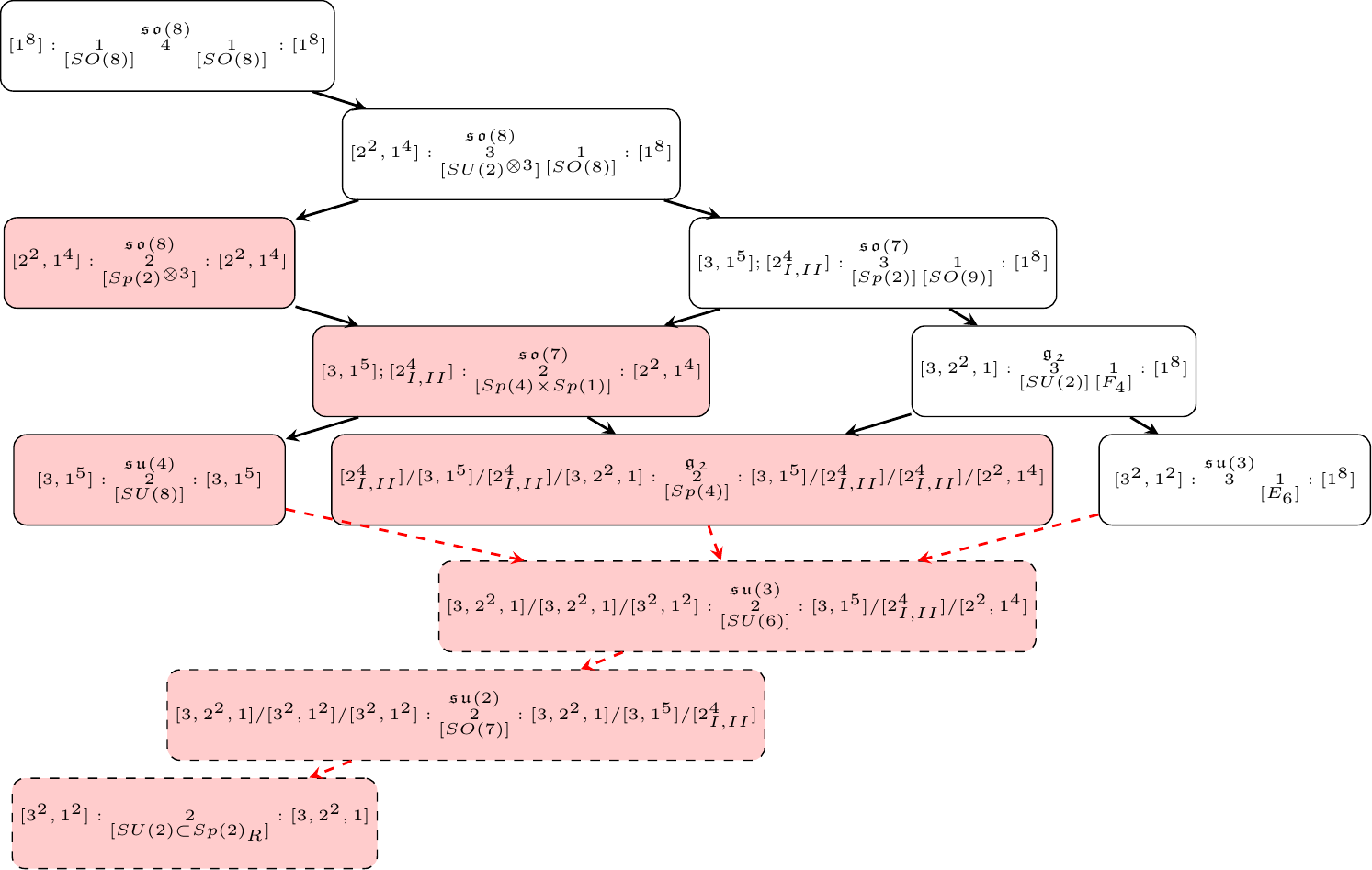}
\caption{Double Hasse diagram for $SO(8)$ short quiver theories with one $-4$ curve in the UV theory. This diagram is again half of a full figure, following the same convention as in figure \ref{fig:su4ShortFlowsThree}. ``Kissing'' configurations are highlighted in red. For concision, several pairs of nilpotent deformations that yield the same IR theory are written in the same box. We separate partitions with semicolons $\mu_L$; $\nu_L$ -- $\mu_R$; $\nu_R$ to denote all possible combinations $\mu_L$ -- $\mu_R$, $\mu_L$ -- $\nu_R$, $\nu_L$ -- $\mu_R$, and $\nu_L$ -- $\nu_R$. On the other hand, slashes denote one-to-one pairings, so $\mu_L/\nu_L$ -- $\mu_R/\nu_R$ means $\mu_L$ -- $\mu_R$ and $\nu_L$ -- $\nu_R$ only. We also mark theories with $(\Delta \alpha, \Delta \beta)$ anomaly mismatches with dashed frames and draw the RG flows towards these cases using red dashed arrows. Note that, whenever there is a dashed frame with more than one possible pair of nilpotent orbits, at least one pair of nilpotent orbits out of them has $(\Delta \alpha, \Delta \beta)$ anomaly mismatch, and in some cases not all of them have such mismatches. See table \ref{table:SO8tangentialShortQuiver} for more details of anomaly mismatches in $SO(8)$ short quiver theories.}%
\label{fig:so8ShortFlowsOne}%
\end{figure}
\begin{figure}
\includegraphics[scale=.8]{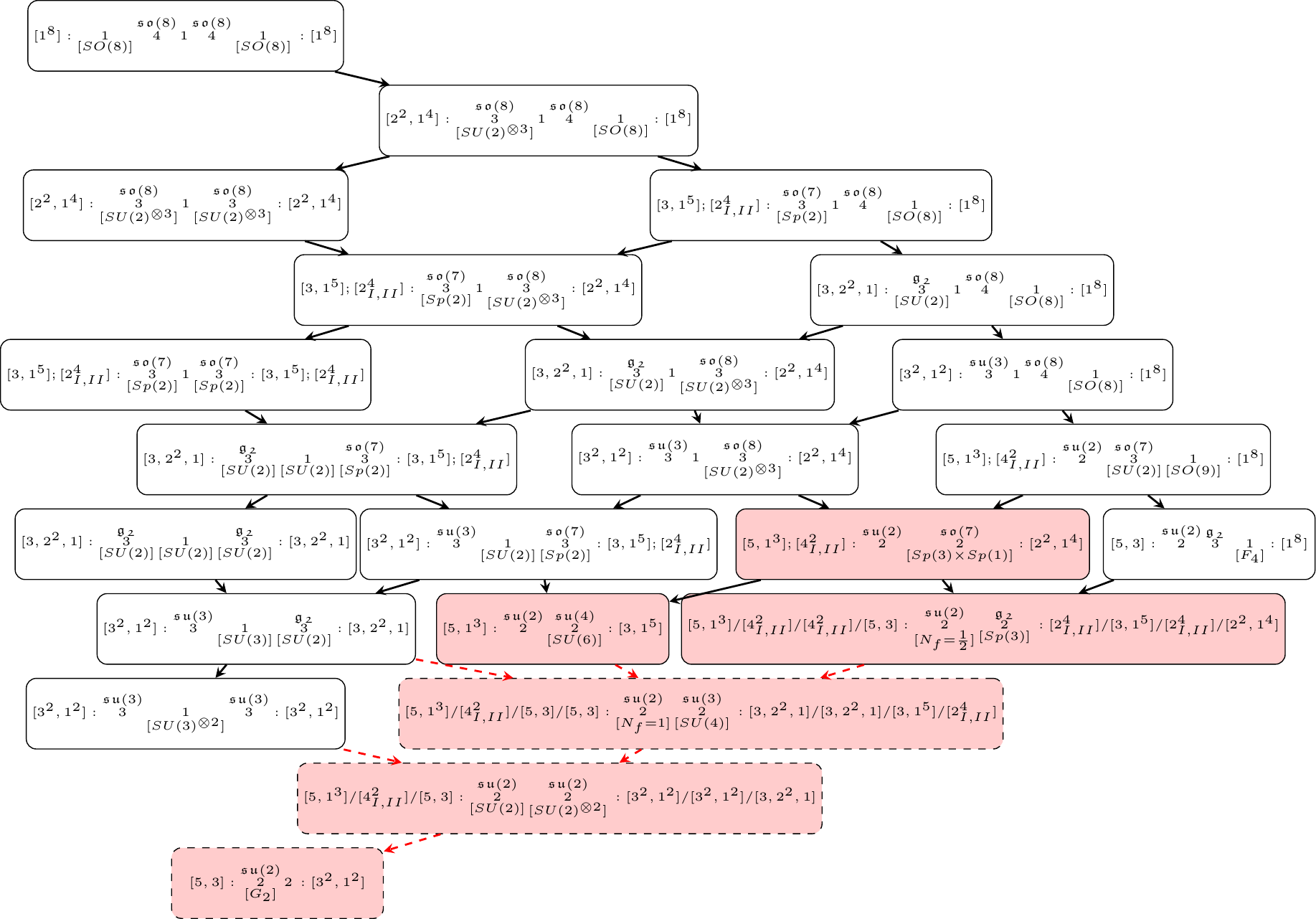}
\caption{Double Hasse diagram of $SO(8)$ short quiver theories over two $-4$ curves in the UV theory. The notation is the same as in figure \ref{fig:so8ShortFlowsOne}.}%
\label{fig:so8ShortFlowsTwo}%
\end{figure}
\begin{figure}[ptb]
\includegraphics[width = \textwidth]{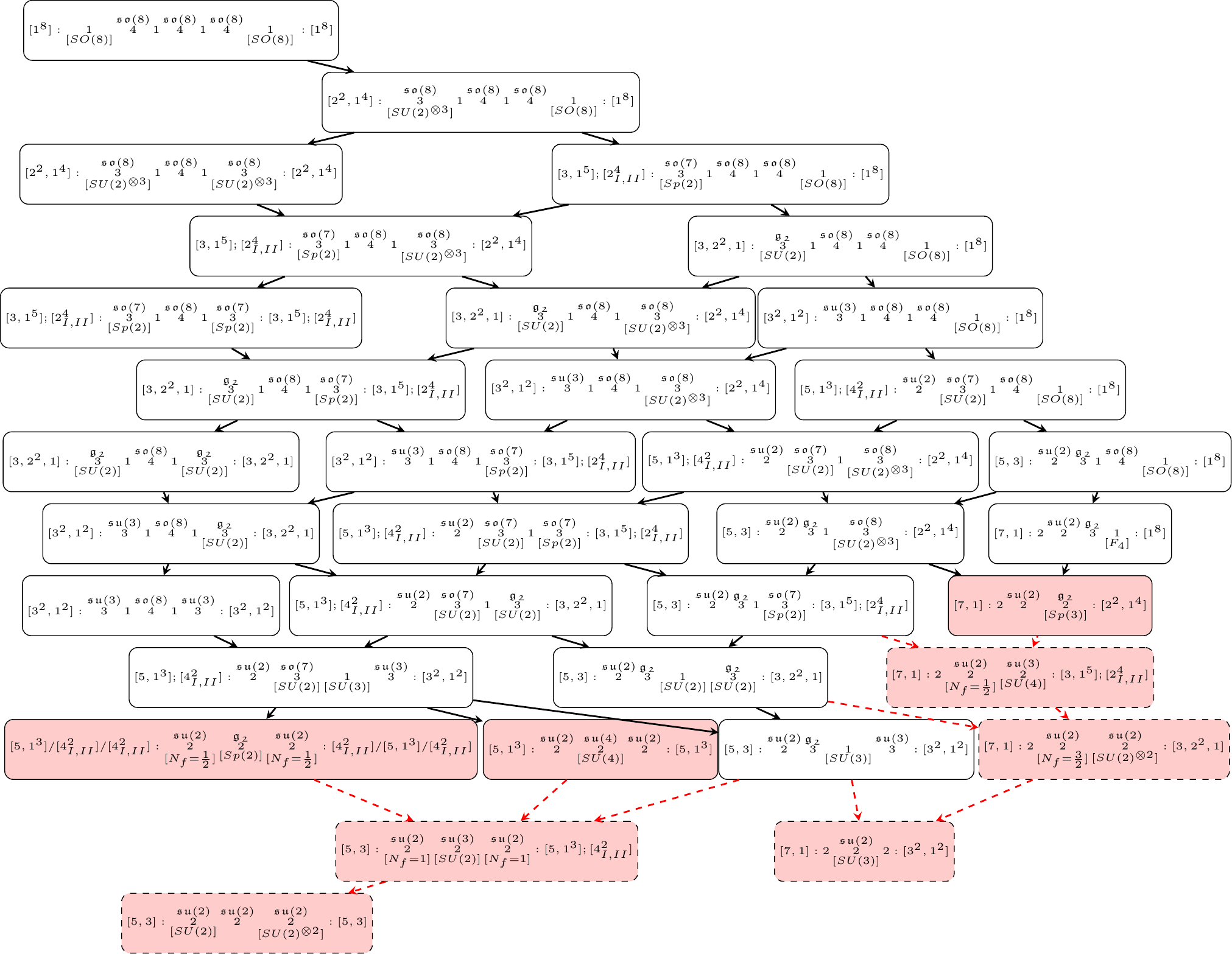}
\caption{Double Hasse diagram of $SO(8)$ short quiver theories over three $-4$ curves in the UV theory. The notation is the same as in figure \ref{fig:so8ShortFlowsOne}.}%
\label{fig:so8ShortFlowsThree}
\end{figure}

Next we look at the double Hasse diagrams for the $SO(8)$ UV theories. For $SO(2N), N>4$ the story is similar, but we choose to illustrate with $SO(8)$ for simplicity. We look at UV quivers with one, two and three $-4$ curves respectively:%
\begin{align}
[1^{8}]: \underset{[SO(8)]}{1} \,  &  {\overset{\mathfrak{so}(8)}{4}} \,
\underset{[SO(8)]}{1} \,:[1^{8}]\\
[1^{8}]: \underset{[SO(8)]}{1} \, {\overset{\mathfrak{so}(8)}{4}} \  &  1 \,
{\overset{\mathfrak{so}(8)}{4}} \, \underset{[SO(8)]}{1} \,:[1^{8}]\\
[1^{8}]: \underset{[SO(8)]}{1} \, {\overset{\mathfrak{so}(8)}{4}} \, 1 \,  &
{\overset{\mathfrak{so}(8)}{4}} \, 1 \, {\overset{\mathfrak{so}(8)}{4}} \,
\underset{[SO(8)]}{1} \,:[1^{8}] \,.
\end{align}
The associated double Hasse diagrams are shown in figures \ref{fig:so8ShortFlowsOne}, \ref{fig:so8ShortFlowsTwo}, and \ref{fig:so8ShortFlowsThree}. We see that as the number of curves decreases, the Hanany-Witten constraints
forbid more and more deformations that were allowed in the long quiver. In each diagram, we highlight in red the ``kissing'' configurations which have all of their $-1$ curves blown-down. We also use dashed lines to indicate theories with an anomaly polynomial mismatch with their associated formal quiver, and we denote flows to these theories with dashed lines.

It is worth pausing here to elaborate on a surprising point noted in example 2 of section \ref{ssec:examples} above: $SO(8)$ nilpotent orbits related by triality always give the same long quiver theory, but they do not not always generate the same short quiver theory. When they do yield the same quiver they are drawn in the same box, but when they give rise to distinct theories, we use separate boxes to denote them.

As an example in which the two disagree, consider the short quivers $[3,1^5]$ -- $[3,1^5]$ and $[2^4]$ -- $[3,1^5]$ on a UV quiver with a single $-4$ curve.
These yield respectively,
\begin{align}
[3, 1^5]:\,\, \overset{\mathfrak{su}_4}{2} \,\, [SU(8)] \,\, : [3, 1^5], \\
[2^4]:\,\, \overset{\mathfrak{g}_2}{2} \,\, [Sp(4)] \,\, : [3, 1^5]\,.
\end{align}
For the first case, with $[3,1^5]$ -- $[3,1^5]$, there are two double strings stretching on the middle curve, so the original $\mathfrak{so}_8$  is Higgsed to $\mathfrak{so}_6 \simeq \mathfrak{su}_4$.
On the other hand the quiver with $[2^4]$ -- $[3,1^5]$ has a single double string stretching on the middle curve (coming from the right deformation) and one extra $\overline{A}$ coming from the left, so the original $\mathfrak{so}_8$ is Higgsed to $\mathfrak{so}_7 \overset{\overline{A}}{\rightarrow} \mathfrak{g}_2$.

The rules that lead us to these quivers can be verified in other examples as well. For instance, consider an $SO(10)$ theory with three $-4$ curves in the UV quiver, deformed by $\mu_L = [7, 1^3]$, $\mu_R = [5, 3, 1^2]$. The resulting theory is given by
\begin{equation}
  [7,1^3]: {\overset{\mathfrak{su}(2)}2}  \,\,   \underset{[SU(4)]}{{\overset{\mathfrak{su}(4)}2}}  \,\,{\overset{\mathfrak{su}(2)}2}:[5,3,1^2] \,.
\end{equation}
In the brane picture, the $\mf{su}(4)$ on the middle $-2$ curve comes from two double strings, one each from the left and right, exactly parallel to the $[3,1^5], [3,1^5]$ case above.

Similarly,
for $\mu_L = [7, 3]$, $\mu_R = [5, 3, 1^2]$, the kissing theory is given by
\begin{equation}
  \underset{[N_f = 1/2]}{\overset{\mathfrak{su}(2)}2}  \,\,   \underset{[Sp(2)]}{{\overset{\mathfrak{g_{2}}}2}}  \,\,\underset{[N_f = 1/2]}{\overset{\mathfrak{su}(2)}2} \,.
\end{equation}
The second $-2$ curve now has a $\mathfrak{g}_2$ gauge algebra, which in the brane picture comes from a single double string coming from one side and an extra $\overline{A}$ coming from the other, just as in the case of the $[2^4]$, $[3,1^5]$ theory above.

 This example nicely illustrates the utility of the string junction approach for determining the nilpotent hierarchy of short quivers, as the short quivers in two cases (which are different) cannot be determined unambiguously from their associated long quivers alone (which are identical).

 Finally, it is also worth noting that additional RG flows have opened up in these short quivers that were not available in the case of long quivers. For instance, in an $SO(8)$ long quiver of fixed size, there is no RG flow from the theory with $\mu_L=[3,2^2,1], \mu_R=[1^8]$ to the theory with $\mu'_L = \mu'_R = [2^4]$, because although $\mu_R \preceq \mu_R'$, we also have $\mu_L \npreceq \mu_L'$.

 However, for a sufficiently-short quiver with these nilpotent orbits, there is a flow from the former to the latter. In particular, there is a flow from
 \begin{equation}
[3,2^2,1]:    \underset{[Sp(1)]}{{\overset{\mathfrak{g}_2}3}}  \,\, \underset{[F_4]}1:[1^8]
\end{equation}
to the theory
 \begin{equation}
[2^4]:    \underset{[Sp(4)]}{{\overset{\mathfrak{g}_2}2}}  :[2^4] .
\label{eq:ex}
\end{equation}
This is related to the fact that short quivers are often degenerate: in particular, the theory of (\ref{eq:ex}) can also be realized by the nilpotent orbits $\mu_L' = [3,2^2,1]$, $\mu_R' = [2^2,1^4]$, which \emph{do} satisfy $\mu_R \preceq \mu_R'$, $\mu_L \preceq \mu_L'$.

\subsection{Flavor Symmetries \label{subsec:globalSymmetryShortQuiver}}

The structure of nilpotent orbits also provides a helpful guide to the analysis of flavor symmetries in 6D SCFTs \cite{Heckman:2016ssk}.
Given a nilpotent orbit, the commutant subalgebra specifies an unbroken symmetry inherited from the UV. For the classical groups, the resulting flavor symmetry algebra associated with a given nilpotent orbit is given simply in terms of the data of partition (see e.g. \cite{Chacaltana:2012zy}):
\begin{equation}
\begin{array}{l@{\quad\text{when}\ }l}
\mathfrak{s}[ \oplus_{i} \mathfrak{u}(r_i)  ] & \mathfrak{g}=\su(N) ,\\
 \op{i\, \text{odd}}\mathfrak{so}(r_i) \oplus \op{i\, \text{even}} \mathfrak{sp}(r_i/2)  & \mathfrak{g}=\mathfrak{so}(2N+1)\, \text{or}\, \mathfrak{so}(2N), \\
\op{i\, \text{odd}}\mathfrak{sp}(r_i/2) \oplus \op{i\, \text{even}} \mathfrak{so}(r_i)  & \mathfrak{g}=\mathfrak{sp}(N). \\
\end{array}\label{classicalflavour}
\end{equation}

In a long quiver, the flavor symmetry inherited from the
parent UV theory is thus given by the products of these flavor symmetries.
For short quivers, on the other hand, we typically observe enhancements of the flavor symmetry whenever flavors coming from the left and from the right end up sharing the same node. As usual, this is easiest to see in theories with $\mf{su}$ gauge symmetries. Here, if flavor symmetries $[SU(m)]_L$ and $[SU(n)]_R$ share the same node, the symmetry enhances from $[SU(m)] \times [SU(n)]$ to $[SU(m+n)]$. For $SO/Sp$ quivers without any small instanton transitions, flavor symmetries of $[SO(m)]_L$ and $[SO(n)]_R$ get enhanced to $[SO(m+n)]$, and similarly for the $Sp$ cases. To illustrate this fact, we start with the theory
\begin{equation}
  [3, 2]:\,\, \overset{\mathfrak{su}(2)}{2} \,\,
\underset{[N_f = 1]}{\overset{\mathfrak{su}(4)}{2}} \,\,
\underset{[N_f = 1]}{\overset{\mathfrak{su}(5)}{2}}
\,\,\underset{[SU(2)]}{\overset{\mathfrak{su}(5)}{2}} \,\,
\underset{[N_f = 1]}{\overset{\mathfrak{su}(3)}{2}}\,\,: [2^{2}, 1] \,.
\end{equation}
We can then shorten the quiver to have only $4$ curves:
\begin{equation}
[3, 2]:\,\, \overset{\mathfrak{su}(2)}{2} \,\,
\underset{[N_f = 1]}{\overset{\mathfrak{su}(4)}{2}} \,\, \underset{[
SU(3)]}{\overset{\mathfrak{su}(5)}{2}} \,\,
\underset{[N_f = 1]}{\overset{\mathfrak{su}(3)}{2}}\,\,: [2^{2}, 1].
\end{equation}
After this first step, we already see an enhancement: the $[SU(3)]$ factor comes from two components: $SU(2)$ from the left and $U(1)$ from the right. Removing yet another curve, we have:
\begin{equation}
[3, 2]:\,\, \overset{\mathfrak{su}(2)}{2} \,\,
\underset{[SU(3)]}{\overset{\mathfrak{su}(4)}{2}} \,\,
\underset{[SU(2)]}{\overset{\mathfrak{su}(3)}{2}}\,\,: [2^{2}, 1].
\end{equation}
Here the enhancement is even greater. Indeed, both of the $[SU(3)]$ and $[SU(2)]$ flavors come from similar enhancements.

Ignoring Abelian factors, enhancements occur in the following two cases:
\begin{itemize}
\item  When flavor symmetries coming from the left and from the right end up sharing the same node.

\item When a $-1$ curve has its surrounding gauge symmetry lowered by short quiver effects (as detailed below). This can happen either for a $-1$ at the edge of the quiver or in the interior.

\end{itemize}
As a first example of the former, consider the theory with nilpotent orbits $[3, 1^5]$ and $[2^4]$ on an $SO(8)$ UV quiver with two $-4$ curves:
  \begin{equation}
    \underset{[Sp(4)]}{\overset{\mathfrak{g_{2}}}2}.
  \end{equation}
  We see that the flavor symmetry $Sp(2) \times Sp(2)$ present in the case of a long quiver has been enhanced to $Sp(4)$.

As another example of the former case, consider the theory with nilpotent orbits $\mu_L=\mu_R=[3,1^{2N-3}]$ on an $SO(2N)$ quiver with one $-4$ curve, which can equivalently be regarded as an $SO(2N-3)$ quiver with $\mu_L=\mu_R=[1^{2N-3}]$:
\begin{equation}
  [SO(2N-2)] \ \ \overset{\mathfrak{sp}(N-5)}{1} \ \ \overset{\mathfrak{so}(2N-2)}{4} \ \ \overset{\mathfrak{sp}(N-5)}{1} \ \ [SO(2N-2)] \,.
\end{equation}
We see that the flavor symmetries of the left and right have been enhanced from $SO(2N-3)$ to $SO(2N-2)$.

Finally, as an example of the latter case, consider the theory of nilpotent orbits $[7, 1]$ and $[1^8]$ on an $SO(8)$ UV quiver with three $-4$ curves:
    \begin{equation}
      2 \, \overset{\ksu(2)}2 \, {\overset{\mathfrak{g_{2}}}3}  \, {1} \,\, [F_4].
    \end{equation}
    The flavor symmetry on the right has been enhanced from $SO(8)$ to $F_4$.

In all cases, we find that the flavor symmetry of a short quiver is enhanced relative to the flavor symmetry of a long quiver associated with the same nilpotent deformations.

\section{Conclusions \label{sec:CONC2}}

In this chapter we have developed general methods for determining
the structure of Higgs branch RG\ flows in 6D\ SCFTs. In particular, we have analyzed several
aspects of vevs for \textquotedblleft conformal
matter.\textquotedblright\ We have seen that the entire nilpotent cone of a simple Lie algebra,
including its structure as a partially ordered set can be obtained from simple
combinatorial data connected with string junctions stretched between bound
states of $7$-branes. Recombination moves involving
intersecting branes as well as brane / anti-brane pairs fully determine the Higgs branch of
quiver-like 6D\ SCFTs with classical gauge algebras. An added benefit of
this approach is that it also extends to short quiver-like theories where
Higgsing from different nilpotent orbits leads to correlated symmetry breaking constraints.
In the remainder of this section we discuss some other potential areas for future investigation.

In this chapter we have primarily focused on Higgsing in quiver-like theories with classical algebras.
We have also seen that we can understand the nilpotent cone of the E-type algebras using multi-pronged
string junctions. This suggests that by including additional $7$-brane
recombination effects, it should be possible to cover
these cases as well. This would provide a nearly complete picture of Higgs
branch flows for 6D\ SCFTs engineered via F-theory.

This work has primarily focused on the case of 6D\ SCFTs in which Higgs
branch deformations can be understood in terms of localized T-brane
deformations of a non-compact $7$-brane. We have already noted how
\textquotedblleft semi-simple\textquotedblright\ deformations fit into
this picture. The other class of Higgs branch deformations which appear quite
frequently involve discrete group homomorphisms from finite subgroups of
$SU(2)$ into $E_{8}$ \cite{Heckman:2015bfa, Mekareeya:2017jgc, Frey:2018vpw}. Obtaining an analogous correspondence in this
case would cover another broad class of Higgs branch deformations in 6D SCFTs.

The main emphasis of this work has centered on combinatorial data connected
with Higgs branch flows and $7$-brane recombination. That being said, it is
also clear that explicit complex structure deformations of the associated
F-theory models should describe some of these deformations as well, a point which
deserves to be clarified.

Moreover, the overarching aim in this work has been to better understand the
structure of all possible 6D\ RG flows obtained from deformations of different
conformal fixed points. The fact that we now have a fairly systematic way to
also understand deformations of short quivers suggests that the time may be
ripe to obtain a full classification of such RG\ flows.

Lastly, this chapter has focused on theories in six dimensions. It would also be interesting to see how similar methods can be applied to systems in four dimensions. Seeing how powerful and intuitive string junctions can be, we now would like to return to 4D theories. Specifically, we will look at a class of $\N=2$ SCFTs engineered with a generalization of orientifolds known as S-folds.

\chapter{S-folds, String Junctions and 4D $\N = 2$ SCFTs}\label{chapter3}

\section{Introduction}\label{sec:INTRO3}

Building upon the last chapter we will now look to introduce additional geometrical structure to connect four dimensional SCFTs. One of the important ingredients in many string theory realizations of quantum
field theories is the use of singular geometries in the presence of various
configurations of branes. For example, in perturbative type II string theory,
all of the classical gauge groups can be realized by open strings ending on
D-branes, possibly in the presence of orientifold planes. It is also possible
to realize exceptional groups via the heterotic string, and with singular
geometries in type II / M- / F-theory compactifications.  This point of view
has led to the prediction of entirely new sorts of quantum field theories in
diverse dimensions.

As a striking example, stringy considerations led to the discovery of 4D
$\mathcal{N} = 3$ SCFTs \cite{Garcia-Etxebarria:2015wns}. These
$\mathcal{N} = 3$ theories are inherently strongly coupled, and many of them
have a realization in string theory as a stack of D3-branes on top of an
S-fold plane.\footnote{There are $\mathcal{N} = 3$ theories that come from
$\mathcal{N}=4$ super Yang--Mills with an exceptional gauge algebra which do not
have a D3-brane realization\cite{Garcia-Etxebarria:2016erx}.} The S-fold is a
generalization of the usual orientifold plane where the $\mathbb{Z}_2$
reflection symmetry is replaced by a $\mathbb{Z}_k$ symmetry, however this
only leads to a consistent supersymmetric field theory when the axio-dilaton
of Type IIB string theory is locally fixed to specific $k$-dependent values.
For additional work on $\mathcal{N} = 3$ SCFTs, see, for example, references
\cite{Garcia-Etxebarria:2015wns, Garcia-Etxebarria:2016erx,
Aharony:2015oyb,Nishinaka:2016hbw,Aharony:2016kai,Imamura:2016abe,Imamura:2016udl,Agarwal:2016rvx,
Cordova:2016emh,Lemos:2016xke,Arras:2016evy,Lawrie:2016axq,vanMuiden:2017qsh,
Amariti:2017cyd,Bourton:2018jwb,Assel:2018vtq,Tachikawa:2018njr,Ferrara:2018iko,Bonetti:2018fqz,
Garozzo:2018kra,Arai:2018utu,Garozzo:2019hbf,Arai:2019xmp,Garozzo:2019ejm,Amariti:2020lua,Zafrir:2020epd}.

Of course, rather than resorting to the full machinery of string theory one
might instead ask whether general principles of self-consistency can be used
to chart the landscape of possible quantum field theories. A notable example
of this sort of reasoning was carried out in a series of papers
\cite{Argyres:2015ffa,Argyres:2015gha,Argyres:2016xua,Argyres:2016xmc,Argyres:2016yzz,Martone:2020nsy,Argyres:2020wmq}
which established a complete classification of possible 4D $\mathcal{N} = 2$
SCFTs with a one-dimensional Coulomb branch.  A particularly interesting
feature of these results is that, at the time they were found, only some of
these theories had known string theory realizations. A key feature of this
analysis is the appearance of specific flavor symmetry algebras, as dictated
by how the Casimir invariants of the flavor symmetry translate to deformations
of the associated Seiberg--Witten curve.

Some of these 4D $\mathcal{N} = 2$ SCFTs now have known stringy realizations,
both in terms of compactifications of 6D SCFTs \cite{Ohmori:2018ona,
Giacomelli:2020jel}, as well as in terms of D3-brane probes of S-folded
7-branes \cite{Apruzzi:2020pmv}.  That being said, there are still some
theories predicted in references
\cite{Argyres:2015ffa,Argyres:2015gha,Argyres:2016xua,Argyres:2016xmc,Argyres:2016yzz,Martone:2020nsy,Argyres:2020wmq}
which have yet to be constructed.

Our aim in this chapter will be to develop a general framework for understanding
the impact of S-folds on the flavor symmetries experienced by probe D3-branes
in the presence of an ambient stack of 7-branes. To this end, we develop a
prescription which generalizes the standard orientifold projection
construction for open strings, but now for more general S-folds acting on
string junctions. Doing so, we show that the structure of the resulting flavor
symmetry algebra is closely tied to the appearance of discrete torsion in the
S-fold. This is quite analogous to what happens for O3-planes, where there are
four distinct choices depending on whether a $\mathbb{Z}_2$ discrete torsion
has been activated in either the RR or NS sector. We show that the presence of
discrete torsion, in tandem with the geometric $\mathbb{Z}_k$ action on the
local geometry, leads to a well-defined set of rules which act on the
endpoints of the string junction states. This in turn leads to a general
quotienting procedure for the resulting flavor symmetry algebras. In fact, the
string junction provides more, since we can also deduce which representations
of a given flavor symmetry algebra are actually present.  For earlier work on
the use of string junctions and its relation to symmetries realized on a
7-brane, see e.g. references \cite{Gaberdiel:1997ud, DeWolfe:1998zf,
Bonora:2010bu, Grassi:2013kha, Hassler:2019eso}.  For earlier work on string
junctions in $\mathcal{N} = 3$ SCFTs, see reference \cite{Imamura:2016udl}.

The 4D $\mathcal{N} = 2$ theories that we consider will be the following. We
will start with the rank $N$ generalizations of the  Argyres--Douglas $H_0$,
$H_1$, and $H_2$ theories \cite{Argyres:1995jj}, the theory of $SU(2)$ with four fundamentals, and
the Minahan--Nemeschansky $E_6$, $E_7$, and $E_8$ theories \cite{Minahan:1996fg, Minahan:1996cj}. These theories
will be labeled as the ``parent'' theories and they are related to each other
via mass deformations from the $E_8$ Minahan--Nemeschansky theory. Furthermore
each of these parent theories has a realization as a worldvolume theory on a
stack of D3-branes in a 7-brane background (see e.g. \cite{Banks:1996nj, Noguchi:1999xq}).
We will consider the ``S-fold descendant theories'', or simply ``descendants'', as the theories obtained by
further inclusion of an S-fold plane on top of the D3-brane stack, either with or
without discrete torsion.

One of the main results of our analysis is that the resulting flavor symmetry
depends on the discrete torsion of the S-fold. In particular, we find that
when no torsion is switched on, there is a simple geometric picture available
which matches to a quotient of the associated F-theory geometry for the
7-branes. When a discrete torsion is present on the S-fold, we find that the
resulting flavor symmetry of a probe D3-brane is also different. In these
cases, the standard F-theory geometry is not valid, but we can instead deduce
its structure from the corresponding Seiberg--Witten curve of the 4D
$\mathcal{N} = 2$ SCFT.

Indeed, using this procedure, we show how to match each possible S-fold
quotient of 7-branes to a corresponding theory appearing in the list of rank
one 4D $\mathcal{N} = 2$ SCFTs appearing in references
\cite{Argyres:2015ffa,Argyres:2015gha,Argyres:2016xua,Argyres:2016xmc,Argyres:2016yzz,Martone:2020nsy,Argyres:2020wmq},
where the rank one theories are classified by the associated Kodaira fiber
type obtained from the Seiberg--Witten curve. In matching to our 7-brane
realization, we can visualize this process in terms of an overall quotienting
/ smoothing deformation. See table \ref{tab:AMSFOLD} for a summary of this
correspondence, and figure \ref{fig:N2rankone} for a summary of how these
different theories are related by mass deformations and discrete quotients.
Implicit in our considerations is that if we remove all the 7-branes, then we
realize $\mathcal{N} = 3$ theories, and discrete quotients thereof. An
additional comment here is that there are a few theories from
\cite{Argyres:2015ffa,Argyres:2015gha,Argyres:2016xua,Argyres:2016xmc,Argyres:2016yzz,Martone:2020nsy,Argyres:2020wmq}
which do not appear to have a simple 7-brane realization. We take this to mean
that the resulting quotients used to construct these additional theories may
not arise from purely geometric ingredients present in the ultraviolet, but
may instead involve structures which only emerge in the infrared.

\begin{table}
  \centering
  \begin{tabular}{c|c}
    Quotient & Rank One 4D $\mathcal{N} = 2$ SCFTs \\ \hline
    $IV^*/\mathbb{Z}_2$ & $[II^*, F_4]$ \\
    $I_0^*/\mathbb{Z}_2$ & $[III^*, B_3]$ \\
    $IV/\mathbb{Z}_2$ & $[IV^*, A_2]$ \\
    $I_0^*/\mathbb{Z}_3$ & $[II^*, G_2]$ \\
    $III/\mathbb{Z}_3$ & $[III^*, A_1]$ \\
    $IV/\mathbb{Z}_4$ & $[II^*, B_1]$ \\
    $IV^*/\widehat{\mathbb{Z}}_2$ & $[II^*, C_5]$ \\
    $I_0^*/\widehat{\mathbb{Z}}_2$ & $[III^*, C_3C_1]$ \\
    $IV/\widehat{\mathbb{Z}}_2$ & $[IV^*, C_2U_1]$ \\
    $I_0^*/\widehat{\mathbb{Z}}_3$ & $[II^*, A_3 \rtimes \mathbb{Z}_2]$ \\
    $III/\widehat{\mathbb{Z}}_3$ & $[III^*, A_1U_1 \rtimes \mathbb{Z}_2]$ \\
    $IV/\widehat{\mathbb{Z}}_4$ & $[II^*, A_2 \rtimes \mathbb{Z}_2]$
  \end{tabular}
  \caption{For each possible discrete quotient of an F-theory
  Kodaira fiber as associated with a probe D3-brane in the presence of a 7-brane and an S-fold with or without discrete torsion,
  we find a corresponding interacting rank one theory as given in table 1 of \cite{Argyres:2016yzz}.}
  \label{tab:AMSFOLD}
\end{table}

\begin{figure}
  \centering
  \resizebox{.98\textwidth}{!}{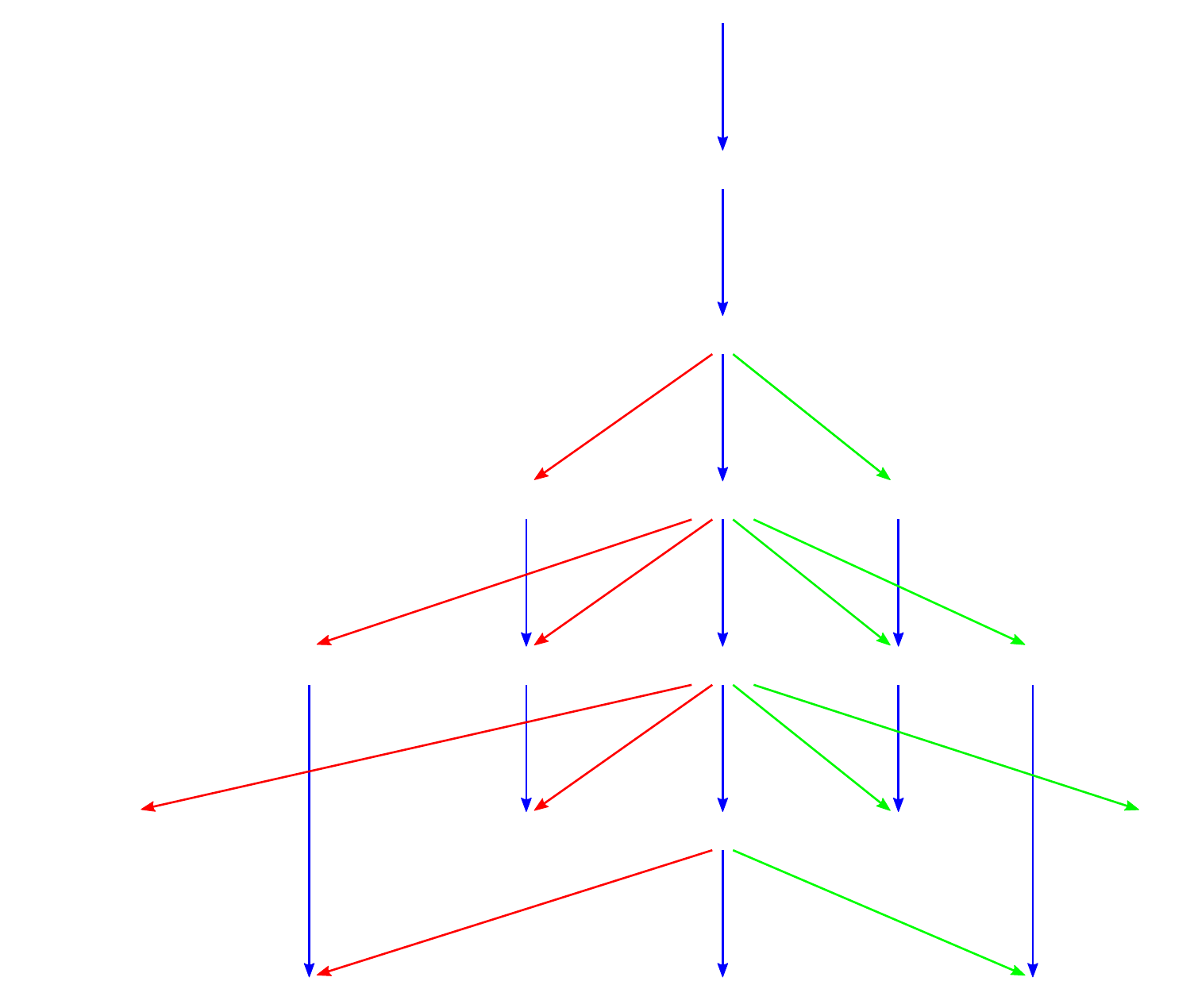}
  \vspace{1em}
  \caption{Realization of the different rank one 4D $\mathcal{N} = 2$ SCFTs
  starting from the $E_8$ Minahan--Nemeschansky theory, written as $[II^{\ast}
  , E_8]$.  We can perform mass deformations (as indicated by downward blue arrows), or
  we can act by a discrete twist by an outer automorphism of an algebra,
  possibly composed with an inner automorphism.  All of the different choices
  can be realized by a suitable choice of S-fold projection with (diagonal red arrows and $\widehat{\mathbb{Z}}_{k}$)
  or without (diagonal green arrows and $\mathbb{Z}_{k}$) discrete torsion. Here, we use the conventions of
  references
  \cite{Argyres:2015ffa,Argyres:2015gha,Argyres:2016xua,Argyres:2016xmc,Argyres:2016yzz,Martone:2020nsy,Argyres:2020wmq},
  which labels a given theory by its Kodaira fiber type, as well as the
  associated flavor symmetry algebra. We note that while this notation does
  not necessarily uniquely specify a particular 4D SCFT, it does so for the
  theories listed here. The notation $\chi_a$ refers to the fact that the
  theory has a chiral deformation parameter which has scaling dimension
  $a$. The theories connected to the $[II^*, E_8]$ theory by blue arrows will be referred to as ``parent''
  theories, and the theories determined via the red/green arrows from a given
  parent will be referred to as the ``descendants'' of that parent.}\label{fig:N2rankone}
\end{figure}

The theories we construct include some notably subtle cases such as theories with $F_4$ flavor
symmetry. Indeed, an important point in this case is that there are some
putative 4D $\mathcal{N} = 2$ SCFTs with $F_4$ global symmetry which are now
known to be inconsistent \cite{Beem:2013sza, Shimizu:2017kzs}. These
inconsistent cases are those in which the Higgs branch of the 4D theory would
have coincided with the instanton moduli space of $F_4$ gauge theory.  Our
brane realization makes clear that we are dealing with a different theory
since in our case, we have a bulk $E_6$ 7-brane in the presence of a
codimension four S-fold with no discrete torsion. A D3-brane sitting on top of
the S-fold sees an $F_4$ flavor symmetry, while moving it inside the 7-brane
but off the S-fold results in an $E_6$ flavor symmetry.  This is also in line
with the fact that the anomalies of reference \cite{Argyres:2016yzz} are
different from the ones of the putative (and sick) $F_4$ theory ruled out in
\cite{Shimizu:2017kzs}. As an additional comment, in F-theory there are no
7-branes with 8D gauge group $F_4$, in line with the feature that such an
object does not exist either from the standpoint of F-theory, or generalized
Green--Schwarz anomalies \cite{Garcia-Etxebarria:2017crf}.

Turning the discussion around, we can also see how the emergent
Seiberg--Witten geometry for these $\mathcal{N} = 2$ theories provides an
operational definition of F-theory in S-fold backgrounds with discrete
torsion.  As a point of clarification, we note that in the single D3-brane
case there can be additional enhancements in the flavor symmetry.  The
F-theory geometry is then obtained by performing a mass deformation to the
generic flavor symmetry, and performing a further rescaling in the local
coordinates.

The rest of this chapter is organized as follows. First, in section
\ref{sec:SFOLD} we present a brief review of S-folds.  In section
\ref{sec:7BRANE} we discuss the specific case of S-folds without discrete
torsion and their realization in F-theory compactifications. In section
\ref{sec:JUNCTIONS} we present a general prescription for reading off the
flavor symmetry of D3-branes probing an S-folded 7-brane.  We then use this to
provide a geometric proposal for F-theory geometry in the presence of discrete
torsion in section \ref{sec:DISCRETE}.  As a further check on our proposal, we
also compute the leading order contributions to the conformal anomalies $a$
and $c$ in the limit of a large number of probe D3-branes in section
\ref{sec:ANOMO}. Section \ref{sec:CONC3} presents our conclusions. Some
additional details on brane motions in the presence of S-folds are presented
in Appendix \ref{app:BRANE}, and an explicit example of string junction projections is worked out in Appendix \ref{app:E6SJ}.

\section{S-folds} \label{sec:SFOLD}

In this section we present a brief review of S-folds. In particular, we emphasize that these objects can sometimes carry a discrete torsion.
S-fold planes are a generalization of orientifold planes introduced in
\cite{Garcia-Etxebarria:2015wns} and further studied in
\cite{Aharony:2016kai}. Initially they were used to build four dimensional
$\mathcal N=3$ supersymmetric field theories on the worldvolume of D3-branes
in the proximity of an S-fold. This was generalized in \cite{Apruzzi:2020pmv}
by adding 7-branes on top of the S-fold thus producing $\mathcal N=2$
theories. In this section we will review the construction of
\cite{Garcia-Etxebarria:2015wns} and discuss various properties of S-folds
that we shall need in the following. We will discuss the inclusion of 7-branes in section \ref{sec:7BRANE}.

\subsection{S-fold Quotients}

S-folds arise from particular terminal singularities in F-theory backgrounds
\cite{Garcia-Etxebarria:2015wns}. The singularity is produced by an orbifold
action that acts simultaneously on the base and elliptic fiber. This implies
that the geometric quotient on the base is accompanied by an $SL(2,\mathbb Z)$
action on the elliptic curve, thus explaining the name of these objects.
More concretely we consider an F-theory solution on $\mathbb
C_{(z_1,z_2,z_3)}^3 \times \mathbf T_w^2$ quotiented by a $\mathbb Z_k$ action
with generator $\sigma_k$ acting on the coordinates as
\al{ \label{eq:quot}\sigma_k : (z_1,z_2,z_3,w) \quad \rightarrow \quad (\zeta_k z_1, \zeta_k^{-1} z_2,\zeta_k z_3,\zeta_k^{-1} w)\,.
}
Here $\zeta_k$ is a $k$-th primitive root of unity. The singularity produced
is terminal as it does not admit any crepant resolution \cite{TERMINATORone,
TERMINATORtwo}.  One important observation is that in order to have a well
defined action on the torus the only allowed values of $k$ are $k=2,3,4,6$.
Compatibility with the quotient fixes the value of the complex structure
$\tau$ of the torus when $k>2$, while leaving it a free parameter for $k=2$.
The allowed values of $\tau$ as well as the $SL(2,\mathbb Z)$ action $\rho$ on
the elliptic fiber are collected in Table \ref{tab:SFOLD}. This background
preserves 12 supercharges for all values of $k>2$ and adding D3-branes probing
the singularity does not further break any additional supersymmetry (see e.g.
\cite{Becker:1996gj}). The $k=2$ case preserves 16 supercharges and therefore
produces an $\mathcal N=4$ supersymmetric theory, and the S-fold in this case
simply corresponds to the usual O3${}^{-}$-plane. Let us note that for $k = 3$ we have
chosen to use the value $\tau = \exp (2 \pi i / 3)$ which is, under a
$T$-transformation of $SL(2,\mathbb{Z})$, equivalent to taking $\exp(2 \pi i /
6)$, the ``standard'' value in the fundamental domain. This has no material
effect on any statements we make about the flavor symmetry algebra since we
can always conjugate all $SL(2,\mathbb{Z})$ generators by this
$T$-transformation anyway. The reason for this choice is to make the
$\mathbb{Z}_{k}$ action of the S-fold more manifest.

\begin{table}[!t]
\begin{center}
\begin{tabular}{c| c| c}
 & $\tau$ & $\rho$ \\[2mm]
 \hline
\rule{0pt}{1.5\normalbaselineskip} $k=2$ & free & $\left(\begin{array}{cc} -1&0\\0&-1\end{array}\right)$\\[4mm]
\hline
\rule{0pt}{1.5\normalbaselineskip}  $k=3$ & $e^{\frac{2 \pi i}{3}}$& $\left(\begin{array}{cc} -1&-1\\1&0\end{array}\right)$\\[4mm]
\hline
\rule{0pt}{1.5\normalbaselineskip}  $k=4$ & $i$& $\left(\begin{array}{cc} 0&-1\\1&0\end{array}\right)$\\[4mm]
\hline
\rule{0pt}{1.5\normalbaselineskip}  $k=6$ & $e^{\frac{2 \pi i }{6}}$& $\left(\begin{array}{cc} 0&-1\\1&1\end{array}\right)$
 \end{tabular}
 \caption{Allowed values of the Type IIB axio-dilaton $\tau$ and $SL(2,\mathbb Z)$ monodromies for various S-folds when
 no 7-branes are present.}
 \label{tab:SFOLD}
 \end{center}
 \end{table}

\subsection{Discrete Torsion}\label{sec:tor}

As in the case of orientifold 3-planes, it is possible to construct different
variants of S-folds by considering trapped three-form fluxes at the
singularity, i.e. discrete torsion. To understand the different allowed
possibilities for discrete torsion, it is helpful to consider the
asymptotic profile of the spacetime far from the singularity, as captured by a
quotient of $S^5$. As in \cite{Witten:1998xy, Aharony:2016kai}, it suffices to
consider $N$ D3-branes probing a $\mathbb Z_k$  S-fold plane. The holographic
dual in the large $N$ limit is given by Type IIB string theory on
$\text{AdS}_5 \times S^5/\mathbb Z_k$. To understand which fluxes can be
introduced it is necessary to study the cohomology of $S^5/\mathbb Z_k$, in
particular the third cohomology group which corresponds to the introduction of
three-form fluxes. In Type IIB we have two possible choices of three-form
fluxes and in the following the first component will be the NSNS flux and the
second one will be the RR flux. Usually we would simply need to compute the
cohomology with coefficients in $\mathbb Z\oplus \mathbb Z$, however due to
the fact that the S-fold action is non-trivial on the fluxes it is necessary
to take cohomology with coefficients in $(\mathbb Z \oplus \mathbb Z)_\rho$
where $\rho$ is the $SL(2,\mathbb Z)$ element listed for every S-fold in table
\ref{tab:SFOLD}.  This computation was done in \cite{Aharony:2016kai} where it
was shown that $H^3(S^5/\mathbb Z_k,(\mathbb Z \oplus \mathbb Z)_\rho)$ is the
cokernel of the map $(\mathbf{id}-\rho) : \mathbb Z^2 \rightarrow \mathbb Z^2$. The
resulting cohomology groups are
\al{ H^3(S^5/\mathbb Z_2,(\mathbb Z \oplus \mathbb Z)_\rho) &= \mathbb Z_2 \oplus \mathbb Z_2\,,\\
H^3(S^5/\mathbb Z_3,(\mathbb Z \oplus \mathbb Z)_\rho) &= \mathbb Z_3\,,\\
H^3(S^5/\mathbb Z_4,(\mathbb Z \oplus \mathbb Z)_\rho) &= \mathbb Z_2\,,\\
H^3(S^5/\mathbb Z_6,(\mathbb Z \oplus \mathbb Z)_\rho) &= \mathbb \mathbb{I}.
}
The $k=2$ case reproduces the well-known example of the four different O3-planes
\cite{Witten:1998xy}. We list here all the inequivalent choices of discrete torsion for the various S-fold planes
\al{ &k =2 \,,  \qquad \left\{ (0,0),(1,0),(0,1),(1,1)\right\}\,,\\
&k =3 \,,  \qquad \left\{ (0,0),(1,0),(2,0)\right\}\,,\\
&k =4 \,,  \qquad \left\{ (0,0),(1,0)\right\}\,,\\
&k =6 \,,  \qquad \left\{ (0,0)\right\}\,.
}
One final piece of information that will be useful in the following is the D3-brane charge carried by the S-fold plane. The charge of the $\mathbb Z_k$ S-fold plane is \cite{Aharony:2016kai}
\al{ \varepsilon_{\text{D3}} = \pm\frac{1-k}{2k}\,,
}
where the plus sign refers to the case without discrete torsion and the minus sign to the case with discrete torsion.

\section{F-theory and S-folds without Torsion}\label{sec:7BRANE}

Having reviewed some basic features of S-folds, we now turn to the structure of local F-theory models
in the presence of an S-fold. Here, we study how this is detected by the worldvolume
theory of a spacetime filling D3-brane. Recall that in F-theory, the appearance of 7-branes is encoded in
the local profile of the Type IIB axio-dilaton. Strictly speaking, this geometric correspondence between the
Coulomb branch of the D3-brane moduli space and the F-theory geometry is only valid in the purely geometric phase of F-theory,
where no discrete torsion is present. Indeed, in section \ref{sec:DISCRETE}
we will later turn the discussion around and argue that
the associated Seiberg--Witten curve provides an operational \textit{definition}
of F-theory in such backgrounds.

The rest of this section is organized as follows. First, we discuss the action
of S-folds on a local Weierstrass model. These local Weierstrass models are
chosen such that they correspond to an F-theory background for the ``parent''
theories, to wit, the rank $N$ generalizations of the Argyres--Douglas,
$SU(2)$ with four flavors, and Minahan--Nemeschansky theories. After this, we turn to an explicit
analysis of the various possible S-fold quotients of such geometries,
organizing our discussion by the corresponding $\mathbb{Z}_2$, $\mathbb{Z}_3$
and $\mathbb{Z}_4$ group action.  In the case of $\mathbb{Z}_{6}$, the
admissible minimal Kodaira fibers are trivial and we get an $\mathcal{N} = 3$
theory from D3-branes probing such a singularity. Following this procedure, we
show how to recover some examples of the Seiberg--Witten geometries, and thus
physical data like the flavor symmetry algebras,
for 4D
$\mathcal{N} = 2$ SCFTs of the sort predicted in references
\cite{Argyres:2015ffa,Argyres:2015gha,Argyres:2016xua,Argyres:2016xmc,Argyres:2016yzz,Martone:2020nsy,Argyres:2020wmq}.
As a point of clarification, the flavor symmetry which is really detected in
this way is the generic one present for multiple D3-branes probing the S-fold.
There is also an $SU(2)$ flavor symmetry as associated with the rotational
group in the worldvolume of the 7-brane (but transverse to the D3-brane), and
in the special case of a single D3-brane, there can be an ``accidental''
enhancement in the infrared. In the worldvolume theory of the D3-brane, $z$
will refer to the Coulomb branch coordinate in the covering space, and $u$
will refer to the Coulomb branch coordinate in the quotient geometry. The
$M_i$ will refer to a degree $i$ Casimir invariant built from the mass
deformations of the 7-brane flavor symmetry algebra.

\subsection{Weierstrass Models}

In order to understand which kinds of 7-brane configurations are allowed in
the presence of an S-fold plane it is convenient to understand the F-theory
Weierstrass model on the orbifolded base. Specifically we consider F-theory on
the base $B=\mathbb C^3_{(z_1,z_2,z_3)}/\mathbb Z_k$ where the generator of
$\mathbb Z_k$ acts on the coordinates of the base as in \eqref{eq:quot}. For
additional details on the procedure see, for example, \cite{DelZotto:2017pti}.
The Weierstrass model on such a base is given as usual by the polynomial
\al{ y^2 = x^3 + f(z_1,z_2,z_3) x + g(z_1,z_2,z_3)\,.
}
However, due to the orbifold action on the base coordinates $f$ and $g$ become
$\mathbb Z_k$-equivariant polynomials. By the condition that the elliptic
fibration be a Calabi--Yau variety the coefficients of the Weierstrass
model, $f$ and $g$, are required to be sections of $\mathcal O(-4K_B)$ and $\mathcal
O(-6K_B)$, respectively. Homogeneity fixes $x$
to be a section of $\mathcal O(-2 K_B)$ and $y$ to be a section of $\mathcal
O(-3K_B)$. For an orbifold a section of $\mathcal O(- l K_B)$ must transform
with a factor $\text{det} (\gamma)^{l}$ where $\gamma$ is the matrix
representation of any orbifold group element acting on the coordinates. To
write down possible Weierstrass models it is convenient to expand $f$ and $g$
as polynomials in the variables $z_i$
\al{ f &= \sum_{a,b,c\geq 0} f_{abc}z_1^a z_2^b z_3^c\,,\\
g &= \sum_{a,b,c\geq 0} g_{abc}z_1^a z_2^b z_3^c\,.
}
Requiring $f$ and $g$ to transform appropriately under the orbifold action puts restrictions on the allowed polynomial coefficients $f_{abc}$ and $g_{abc}$. We list in the following the possible choices for the different S-fold planes.

\begin{itemize}
\item[-]$\underline{k = 2}$. In this case both $f$ and $g$ are invariant under the orbifold action. This fixes $f_{abc} = g_{abc}= 0 $  for $a-b+c \neq 0 \mod 2$. The lowest order terms are the constant ones giving generically a smooth elliptic curve with constant complex structure over $\mathbb C^3$.\footnote{Note that this does not mean that the orbifold action is trivial on the elliptic curve. Indeed the coordinate $y$ changes sign under the action of the generator of $\mathbb Z_2$.}

\item[-]$\underline{k = 3}$. In this case the orbifold action implies that $g$ is invariant and $f \rightarrow e^{2 \pi i/3 } f$. This fixes $f_{abc} = 0$ for $a-b+c \neq 1 \mod 3$ and $g_{abc} = 0 $ for $a-b+c \neq 0 \mod 3$. 
%
%

\item[-]$\underline{k = 4}$. In this case the orbifold action implies that $f$ is invariant and $g \rightarrow - g$. This fixes $f_{abc} = 0$ for $a-b+c \neq 0 \mod 4$ and $g_{abc} = 0 $ for $a-b+c \neq 2 \mod 4$. 
%
%

\item[-]$\underline{k = 6}$. In this case the orbifold action implies that $g$ is invariant and $f \rightarrow e^{4 \pi i/3 } f$. This fixes $f_{abc} = 0$ for $a-b+c \neq 4 \mod 6$ and $g_{abc} = 0 $ for $a-b+c \neq 0 \mod 6$. 
%
%

\end{itemize}

In the following we will be interested in a restricted class of Weierstrass models that preserve $\mathcal N=2$ supersymmetry. This can be achieved by taking all 7-branes to wrap the $(z_1,z_2)$-plane, implying that $f$ and $g$ will only depend on $z_3$. Moreover to simplify the notation we shall denote by $z$ the coordinate $z_3$ in the covering space.

We exclusively focus on Weierstrass models where the axio-dilaton is constant
so that we can realize an SCFT on the worldvolume of the D3-brane. F-theory
constructions with constant coupling were discussed in \cite{Dasgupta:1996ij}.
Additionally, we require that the singularity type remain minimal, which
imposes the further condition that the degrees of $f$ and $g$ as polynomials
in $z$ are $\text{deg}(f)<4 $ and $\text{deg}(g)<6$. For each possible S-fold
quotient, we list the covering space theory prior to the quotient in table
\ref{tab:weierstrass}. Note that the $k=6$ quotient does not allow any
dependence on $z$ in the Weierstrass model without incurring non-minimal
Kodaira fibers, and thus there can be no 7-branes present. This implies that
the theory will have enhanced $\mathcal N\geq 3$ supersymmetry.

A careful comparison of tables \ref{tab:SFOLD} and \ref{tab:weierstrass} also
reveals that the correlation of values of $k$ with $\tau$ are different in the presence or absence of
7-branes. This is to be expected because the presence of 7-branes impacts the profile of the axio-dilaton.

\begin{table}[!t]
\begin{center}
\begin{tabular}{c| c|c|c}
 Quotient & Weierstrass Model & Kodaira Fiber Type & $\tau$ \\
 \hline
\rule{0pt}{1.\normalbaselineskip} $k=2$ & $y^2= x^3 + z^4$ & $IV^*$ & $e^{\frac{2 \pi i}{3}}$ \\[1pt]
\hline
\rule{0pt}{1.\normalbaselineskip} $k=2$ & $y^2= x^3 + z^2 x$ & $I_0^*$ & $ i $ \\[1pt]
\hline
\rule{0pt}{1.\normalbaselineskip} $k=2$ & $y^2= x^3 + z^2 $& $IV$ & $e^{\frac{2 \pi i}{6}}$ \\[1pt]
\hline
\rule{0pt}{1.\normalbaselineskip} $k=3$ & $y^2= x^3 + z^3 $ & $I_0^*$ & $e^{\frac{2 \pi i}{6}}$ \\[1pt]
\hline
\rule{0pt}{1.\normalbaselineskip} $k=3$ & $y^2= x^3 + z x $ & $III$ & $ i $\\[1pt]
\hline
\rule{0pt}{1.\normalbaselineskip} $k=4$ & $y^2= x^3 + z^2 $ & $IV$ & $e^{\frac{2 \pi i}{6}}$ \\[1pt]
\hline
\rule{0pt}{1.\normalbaselineskip} $k=6$ & $y^2= x^3    + g_0  $ & $\varnothing$ & $e^{\frac{2 \pi i}{6}}$ \\[1pt]
\hline
 \end{tabular}
 \caption{Allowed values of S-fold projection compatible with a specified minimal Kodaira fiber type.
 Here we drop all higher order singularities and focus on the specific situation where the axio-dilaton is constant.}
 \label{tab:weierstrass}
 \end{center}
 \end{table}

The relevance of the Weierstrass model is that it will allow us to read off
the Seiberg--Witten curve of the resulting $\mathcal N=2$ theory for the case
of a single D3-brane probe. Indeed in this case the Seiberg--Witten curve can be identified with
the elliptic fiber of the F-theory model and the coordinate $z$ becomes the Coulomb branch parameter of
the theory. In the following we will discuss each possible case leading to a
rank one SCFT writing down the Seiberg--Witten curve and match the results to
the ones known in the literature. We would like to stress that the procedure works only in the case \emph{without} discrete torsion, and in the presence of discrete torsion we do not have a procedure to read off the Seiberg--Witten curve from the geometry. We will confirm the various identifications via a string junction analysis in section \ref{sec:JUNCTIONS} where we will also be able to identify the theories on the probe D3-branes also in the presence of discrete torsion.
Before turning to the discussion of each case separately we would like to point out that in the above
we have been using the covering space coordinates. It is also helpful to work directly in terms of a
local coordinate in the quotient geometry. In general for a $\mathbb Z_k$ quotient we would need to use $u = z^k$ which is invariant under the quotient. To find the appropriate invariant combinations for $x$ and $y$ we can use the fact that under the general rescaling \cite{0907.0298,Argyres:2016yzz,Apruzzi:2020pmv}
 \al{ x \mapsto \lambda^2 x\,, \quad y \mapsto \lambda^3 y\,,
 }
 which modifies $f$ and $g$ as
 \al{f \mapsto \lambda^{-4} f \,, \quad g \mapsto \lambda^{-6} g\,,
 }
the elliptic fibration is left invariant. By choosing $\lambda = z^{1-k}$ the rescaled $x$ and $y$ variables will be invariant under the $\mathbb Z_k$ quotient.

Using this information we will be able to write down the Seiberg--Witten curves for the various rank one theories.

\subsection{$\mathbb Z_2$ Quotients }

In this subsection we turn to $\mathbb{Z}_2$ quotients of an F-theory model. This sort of quotient
can be taken for parent theories with an $E_6$ 7-brane, as realized by a type $IV^{\ast}$ fiber,
a $D_4$ 7-brane, as realized by a type $I_{0}^{\ast}$ fiber, and an $H_2$ 7-brane, as realized by
a type $IV$ fiber.

\subsubsection{Quotient of $E_6$}

The Weierstrass model for an $E_6$ singularity can be written as
\al{ y^2 = x^3 + z^4\,.
}
Homogeneity fixes the scaling dimension of $z$ to be $\Delta(z) = 3$. The
maximal deformation of the singularity compatible with the $\mathbb Z_2$
quotient involves introducing the following $M_i$:
\al{y^2= x^3 +x \left(M_8 + M_2 z^2 \right)+ z^4 +  M_6 z^2 +M_{12}\,.
}
Here we chose the convention to label the mass deformations of the 4D
$\mathcal{N} = 2$ SCFT as degree $i$ Casimir invariants $M_i$ where the
scaling dimension is $\Delta(M_i) = i$. We can now move to the quotient space
by performing the aforementioned rescaling. Let us be explicit in this first
case. The scaling is
\begin{equation}
  x \rightarrow z^{-2} x \,, \quad y \rightarrow z^{-3} y \,,
\end{equation}
which leads to an overall factor on the $y^2$ and $x^3$ terms in the Weierstrass
equation of $z^{-6}$. Removing this denominator is equivalent to the rescaling
\begin{equation}
  f \rightarrow z^4 f \,, \quad g \rightarrow z^6 g \,,
\end{equation}
as described in the general case in \cite{0907.0298}. After this rescaling
we perform the replacement with the quotiented coordinate, $u$, via $u
= z^2$. The resulting model becomes
\al{y^2= x^3 +x \left(M_8 u^2 + M_2 u^3 \right)+ u^5 +  M_6 u^4 +M_{12} u^3\,,
}
where we have used the same notation $x$ and $y$ for before and after the
rescaling. In this case turning
off all mass deformations we obtain a $II^*$ singular fiber at the origin.
Comparing with \cite{Argyres:2015gha} we see that this Weierstrass model
matches the Seiberg--Witten curve of the
$[II^*,F_4]$ theory.

\subsubsection{Quotient of $D_4$}

The $D_4$ singularity admits two different minimal Weierstrass presentations,
one of which is compatible with the $\mathbb{Z}_2$ quotient and the other
which is compatible with the $\mathbb{Z}_3$ quotient.
For the $\mathbb Z_2$ quotient we have the Weierstrass model
\al{ y ^2 = x^3 + x z^2\,.
}
Homogeneity fixes the scaling dimension of $z$ to be $\Delta(z) = 2$, and the
deformation of the singularity compatible with the $\mathbb Z_2$ quotient is
given by the introduction of the Casimirs $M_2$, $M_4$, and $M_6$:
\al{y^2= x^3 +x \left(M_4 + z^2 \right)+  M_2 z^2 +M_{6}\,.
}
Again we move to the quotient space by performing the rescaling,
as described above. After rescaling the model becomes
\al{y^2= x^3 +x \left(M_4 u^2 +  u^3 \right)+   M_2 u^4 +M_{6} u^3\,.
}
In this case turning off all mass deformations we obtain a $III^*$ singular
fiber at the origin, and if we compare with \cite{Argyres:2015gha} we see that this
Weierstrass model matches the Seiberg--Witten curve of the $[III^*,B_3]$
theory listed therein.

\subsubsection{Quotient of $H_2$}

The Weierstrass model for an $H_2$ singularity, also known as a type $IV$
fiber, is
\al{ y ^2 = x^3 + z^2\,.
}
As usual the scaling dimension of $z$ is fixed by homogeneity of the
Weierstrass equation. We have $\Delta(z) =3/2$. The singularity can be
deformed in such a way that is compatible with a $\mathbb{Z}_2$ quotient by
introducing $M_2$ and $M_3$ as follows:
\al{y^2= x^3 +x M_2 +  z^2 +M_{3}\,.
}
The resulting model in the quotient space is obtained by performing the
now-familiar rescaling:
\al{y^2= x^3 +x M_2 u^2 +   u^4 +M_{3} u^3\,.
}
We can see that turning off all mass deformations we obtain a $IV^*$ singular
fiber at the origin. Comparing with \cite{Argyres:2015gha} we see that this
Weierstrass model is precisely giving the Seiberg--Witten curve of the $[IV^*,A_2]$
theory.

\subsection{$\mathbb Z_3$ Quotients }

We next turn to $\mathbb{Z}_{3}$ quotients of a local F-theory geometry. This
can be carried out for a $D_4$ 7-brane, as realized by a type $I_{0}^{\ast}$
fiber, and an $H_1$ 7-brane, as realized by a type $III$ fiber.

\subsubsection{Quotient of $D_4$}

The other Weierstrass model for the $I_0^*$ singularity, the one compatible with the $\mathbb Z_3$ symmetry, is:
\al{ y ^2 = x^3 + z^3\,,
}
and homogeneity fixes the scaling dimension of $z$ to be $\Delta(z) =2$. The
deformation of the singularity compatible with the $\mathbb Z_3$ quotient is
\al{y^2= x^3 +M_2 x  z  + M_6 + z^3\,.
}
We can now move to the quotient space by performing the aforementioned rescaling. The resulting model becomes
\al{y^2= x^3 +x M_2 u^3 +   u^5 +M_{6} u^4\,.
}
In this case turning off all mass deformations we obtain a $II^*$ singular
fiber at the origin. Comparing with \cite{Argyres:2015gha} we see that this
Weierstrass model matches the Seiberg--Witten curve of the $[II^*,G_2]$
theory.

\subsubsection{Quotient of $H_1$}

The Weierstrass model for an $H_1$ singularity, or type $III$ fiber, compatible with the $\mathbb Z_3$ symmetry is
\al{ y ^2 = x^3 + x z\,.
}
Homogeneity fixes the scaling dimension of $z$ to be $\Delta(z) =4/3$. The deformation of the singularity compatible with the $\mathbb Z_3$ quotient is
\al{y^2= x^3 + x  z  + M_2 \,.
}
As usual we can move to the quotient space by performing the rescaling
described above. The resulting model becomes
\al{y^2= x^3 +x  u^3 +   M_{2} u^4\,.
}
In this case turning off all mass deformations we obtain a $III^*$ singular
fiber at the origin, and a comparison with \cite{Argyres:2015gha} shows that
this Weierstrass model reproduces the Seiberg--Witten curve of the
$[III^*,A_1]$ theory.

\subsection{$\mathbb Z_4$ Quotient of $H_2$}

Finally, we turn to the case of $\mathbb{Z}_{4}$ quotients. In this case there is only a single
choice available, as given by an $H_2$ 7-brane, namely a type $IV$ fiber.
Recall that the Weierstrass model for an $H_2$ singularity is
\al{ y ^2 = x^3 + z^2\,,
}
and that homogeneity of the polynomial fixes the scaling dimension of $z$ to
be $\Delta(z) =3/2$. The deformation of the singularity compatible with the
$\mathbb Z_4$ quotient allows the introduction of only a single Casimir invariant $M_2$:
\al{y^2= x^3 + M_{2} x    + z^2 \,.
}
We can pass to the quotient space geometry by performing the aforementioned rescaling.
The resulting Weierstrass model is
\al{y^2= x^3 +M_2 x  u^3 +    u^5\,.
}
Turning off all mass deformations we obtain a $II^*$ singular fiber at the
origin. Comparing with \cite{Argyres:2015gha} we see that this Weierstrass
model matches the Seiberg--Witten curve of the $[II^*,B_1]$ theory.

\section{String Junctions}\label{sec:JUNCTIONS}

In the previous section we presented a general analysis of how to read off the Seiberg--Witten curve for the worldvolume theory
of a probe D3-brane in the presence of a 7-brane and an S-fold without discrete torsion.
Geometrically, this provides a satisfying picture for how to realize a subset of possible 4D $\mathcal{N} = 2$ SCFTs,
but it also leaves open the question as to whether we can also understand quotients with discrete torsion. An additional issue
is that in all cases the information of the flavor symmetry is encoded indirectly in the Seiberg--Witten curve via the unfolding of the singularity.

To provide a systematic analysis of cases with and without discrete torsion, we now analyze the spectrum of string junctions
in the presence of an S-fold. The rules we develop lead to a different quotienting procedure for the flavor symmetry algebra,
and the available options are all contained in the options predicted in references
\cite{Argyres:2015ffa,Argyres:2015gha,Argyres:2016xua,Argyres:2016xmc,Argyres:2016yzz,Martone:2020nsy,Argyres:2020wmq}.
Again, we must add the caveat that our analysis really leads to a derivation of the generic flavor symmetry, namely the one which is present for
multiple probe D3-branes.

To better understand how S-fold projection works, we first review the standard
case of orientifold projection for oriented perturbative strings, we follow
this with the rules for S-fold projection in the case of $\mathbb{Z}_2$,
$\mathbb{Z}_3$, and $\mathbb{Z}_4$ quotients. We then turn to the explicit
S-fold projections for string junctions attached to 7-branes.

In what follows, we will find it useful to arrange the bound states of
$[p,q]$ 7-branes so that the group action amounts to a simple rearrangement
operation on these stacks. We refer to these branes according to the resulting
$SL(2,\mathbb{Z})$ monodromy on the axio-dilaton, writing the monodromy as:
\begin{equation} \label{monomove}
M_{[p,q]} = \left[
\begin{array}
[c]{cc}%
1 + pq & - p^2 \\
q^2 & 1 - pq
\end{array}
\right] \,,
\end{equation}
for a $[p,q]$ 7-brane. We will frequently refer to the branes:
\begin{align}
A & = M_{[1,0]} = \left[
\begin{array}
[c]{cc}%
1 & -1\\
0 & 1
\end{array}
\right]  \text{, \ \ }B= M_{[1,-1]} = \left[
\begin{array}
[c]{cc}%
0 & -1\\
1 & 2
\end{array}
\right]  \text{, \ \ }
C= M_{[1,1]} = \left[
\begin{array}
[c]{cc}%
2 & -1\\
1 & 0
\end{array}
\right]  \text{, \ \ } \nonumber \\
D & = M_{[0,1]} = \left[
\begin{array}
[c]{cc}%
1 & 0\\
1 & 1
\end{array}
\right]  \text{, \ \ }
X= M_{[2,-1]} = \left[
\begin{array}
[c]{cc}%
-1 & -4\\
1 & 3
\end{array}
\right]  \text{, \ \ }
Y= M_{[2,1]} = \left[
\begin{array}
[c]{cc}%
3 & -4\\
1 & -1
\end{array}
\right] \,.
\end{align}
We will also need to rearrange our branes to make the S-fold quotient more manifest.
We accomplish this with different brane arrangements (see Appendix \ref{app:BRANE}). This includes:
\begin{align}
E_{6} &: A^{5} B C^2 \sim A^{6} XC \sim AAA C AAA C \\
D_{4} &: A^{4} BC \sim AA C AA C \sim AABBDD \\
H_{2} &: A^{3} C \sim AC AC \sim AYAY \sim DADA \\
H_{1} &: A^{2} C \sim ABD \,.
\end{align}
These 7-branes correspond to the
F-theory backgrounds that give rise to the parent theories on the probe
D3-branes, when there is no S-fold. We again stress that the symmetry algebra
obtained when we include the S-fold is the one enjoyed by the probe D3-branes.

The utility in introducing these different brane systems is that we can then read off the corresponding root system as well as representations from string junctions stretched between these different constituent branes. As a point of notation, we write $a_{i}$ to denote weights associated with $A$-branes, with similar conventions for the $B$, $C$, and $D$ branes, and where the presence of a minus sign indicates the orientation of
the string. For example, the roots of $SU(N)$ for a stack $A^{N}$ would then be represented as $(a_i - a_j)$ for $i,j = 1,...,N$ and $i \neq j$. A junction with endpoints on different types of branes is represented similarly by an oriented graph with weights. Elements of the Cartan subalgebra correspond to string junctions which begin and end on the same branes.

\subsection{Orientifold Projection}

Before delving into how the S-fold planes act on the string junctions stretching between 7-branes, we first review how the usual orientifold planes that appear in perturbative string theory act on string states. Recall that in the presence of a stack of $2N$ D-branes, open string states containing a vector are labeled by Chan--Paton factors $\lambda_{ij}$ for $i,j = 1,\dots, 2N$. Each $\lambda_{ij}$ state is an open string stretching between the $i$-th and the $j$-th brane. When the stack of D-branes sits on top of an orientifold plane it is necessary to specify the action of worldsheet orientation reversal on these states. The general action is
\al{ \Omega : \lambda \ \mapsto\  - M \lambda^T M^{-1}\,.
}
The minus sign appears because of the effect of worldsheet parity on the open
string oscillators and transposition appears because the endpoints of an open
string are interchanged. $M$ is an additional conjugation on the endpoints and
consistency fixes it to be either symmetric or anti-symmetric. When $M$ is
chosen to be symmetric the resulting Lie algebra on the stack of D-branes will
be $D_N$ and when $M$ is anti-symmetric the Lie algebra will be $C_N$. Given
this we will label the symmetric choice $M_{\text{SO}}$ and the anti-symmetric
one $M_{\text{Sp}}$. In the following we will choose\footnote{Note that it is
customary in the literature to choose $M_{\text{SO}}$ to be the identity
matrix. Our choice will give isomorphic algebras after projection and leads to
a simpler geometric picture in terms of branes probing the orientifold plane.}
\al{ M_{\text{SO}} &= \left(\begin{array}{cc}0 & J_N \\ J_N &0\end{array}\right)\,,\\
 M_{\text{Sp}} &= \left(\begin{array}{cc}0 & iJ_N \\ -iJ_N &0\end{array}\right)\,.
}
Here $J_N = \delta_{i+j,N+1}$ with $i,j = 1,\dots, N$, namely the anti-diagonal matrix in which non-zero entries are all equal to 1.
We can therefore explicitly write the action of $\Omega$ on the various string states which we label as $|ij\rangle$ for a string stretching between the $i$-th and $j$-th brane. We will find it convenient to use the notation  $i' = 2N+1-i$. The map is:
\al{ \Omega |ij\rangle = \gamma_{\Omega}|j' i'\rangle\,.
}
Here, the choice of phase factor is specified via (see figure \ref{fig:oplane}):
\begin{align}
\text{Sp projection} &\rightarrow \gamma_\Omega = 1\\
  \text{SO projection} &\rightarrow \left\{\begin{array}{ll}\gamma_\Omega = -1, & \text{string crosses orientifold} \\\gamma_\Omega=-1,& i=j\\ \gamma_\Omega = 1, &\text{otherwise.} \end{array}\right.
\end{align}

\begin{figure}[t!]
\begin{center}
\includegraphics[scale=.8]{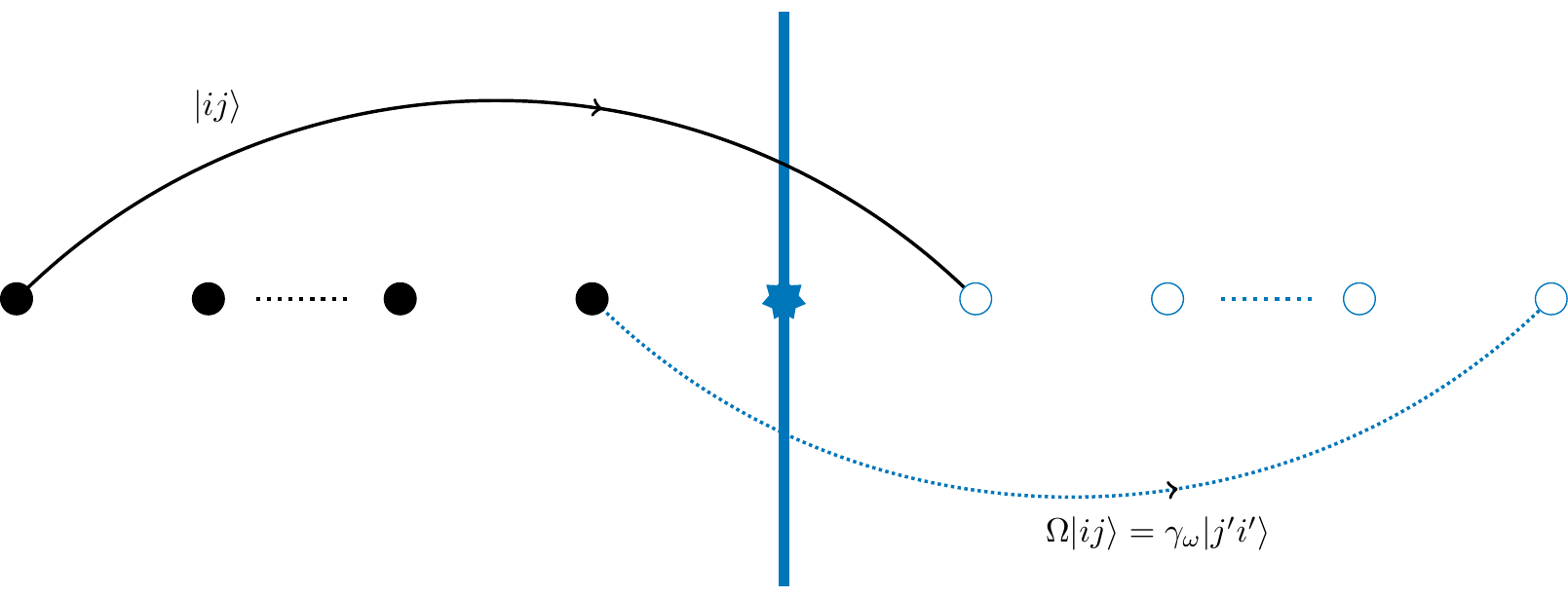}
\end{center}
\caption{Illustration of orientifold projection acting on perturbative open strings. We denote the orientifold image branes by
open shapes, and image strings by dashed blue lines.}
\label{fig:oplane}
\end{figure}

Finally, it is important to understand which projection corresponds to which
orientifold plane. The system that more closely resembles the ones we will
study in the following is a stack of D7-branes on top of an orientifold
3-plane. Recall that there exist four different orientifold planes usually
called $\text{O3}^-$, $\widetilde{\text{O3}}^-$, $\text{O3}^+$,  and
$\widetilde{\text{O3}}^+$. In terms of the discrete torsion introduced in
section \ref{sec:tor} they have torsion $(0,0)$, $(0,1)$, $(1,0)$ and $(1,1)$
respectively. The first two give a $D$-type algebra on a stack of D3-branes
and a $C$-type algebra on a stack of D7-branes, the last two give a $C$-type
algebra on a stack of D3-branes and a $D$-type algebra on a stack of
D7-branes. The action on other kinds of 7-branes can be obtained via
$SL(2,\mathbb Z)$ conjugation knowing that the $\text{O3}^-$ plane is
invariant under $SL(2,\mathbb Z)$ and that the action of $SL(2,\mathbb Z)$ for
the other planes can be inferred by looking at the action on the plane's
discrete torsion. For example an $\text{O3}^+$ plane will give a $C$-type
algebra on a stack of $[0,1]$ 7-branes. With this information we can easily
infer that when a string junction of charge $[p,q]$  crosses an orientifold
3-plane of discrete torsion $(a,b)$ worldsheet parity will produce a sign
$(-1)^{a p -  b q}$ on the string state. In the following we will generalize
this to other S-folds. As a final comment, we note that when mutually
non-local 7-branes are present, we find that all that matters is whether
discrete torsion is switched on or not; this is different from the situation
with all 7-branes mutually local.  In particular, when all 7-branes are
mutually local then the spectrum is ``blind'' to some sector of discrete
torsion; for example, when all 7-branes are mutually local D7s then the
Ramond--Ramond component of the discrete torsion cannot be detected by the
7-branes.

\subsection{S-fold Projection}

In the following we will consider different $\mathbb Z_k$ projections on the
set of string junctions. To get invariant states we will call $\Pi_k$ the generator
of the $\mathbb Z_k$ action on the string state and we will sum over the $\mathbb
Z_k$ images to get the states after projection, meaning that we shall consider
the combination
\al{ \frac{1}{k} \left( \mathbb I + \sum_{l=1}^{k-1} \Pi_k^{l}\right)\,.
}
This action is considered over the generators of the complexified Lie algebra,
not on the root vectors. In particular, Lie algebra generators that are mapped
to themselves may be projected out due to some phases in $\Pi_k$. Indeed since
the only requirement for $\Pi_k$ is that its $k$-th power is the identity it
is possible to twist it by some $\mathbb Z_k$ phases corresponding to
different choices of discrete torsion; these choices were reviewed in section
\ref{sec:tor}. What needs to be fixed is the phase that is acquired by the
various junctions in the presence of discrete torsion. Note that this
information is relevant only for junctions whose root vectors are invariant
under the S-fold projection as the addition of these  phases may project them
out. We will write down the phase for a $[p,q]$-string crossing the S-fold
with torsion $(a,b)$. The phase is fixed by requiring invariance under the
torsion equivalence relations described in section \ref{sec:tor}. The various
cases are
\al{ &k= 2 \,, \qquad (-1)^{a p - b q}\,,\\
&k=3 \,, \qquad e^{\frac{2 \pi i }{3} (a p - b p -a q +b q)}\,,\\
&k=4 \,, \qquad (-1)^{    a p - b p -a q +b q}\,,
}
where we omit the case $k=6$ since no discrete torsion is available for this value.
See figure \ref{fig:Sfold} for a depiction of S-fold projection on string junction states.

In the above discussion, we have made reference to a specific duality frame.
Given that we are working at strong coupling, it is natural to ask about the
behavior of our S-fold projection under $SL(2,\mathbb{Z})$ duality
transformations. Note that while
the expression for the phase is invariant under global $SL(2,\mathbb Z)$
transformations for $k=2$, for $k > 2$ it is necessary to conjugate the
pairing between junction charges and discrete torsion under global
$SL(2,\mathbb Z)$ transformations in order to ensure that the phase is
unchanged\footnote{In practice we will conjugate the pairing for all values of $k$.}. This
should not come as a surprise as for $k>2$ we are implicitly referring to a
specific choice of an $SL(2,\mathbb Z)$ frame when discussing the torsional
fluxes: indeed the equivalence relations among discrete torsion discussed in
section \ref{sec:tor} refers to a matrix $\rho$ that is not invariant under
global $SL(2,\mathbb Z)$  transformations. Given that the product appearing in
the phase is fixed by requiring compatibility with these equivalence relations
it will necessarily be different when going to a new $SL(2,\mathbb Z)$ frame
in order to ensure that the new equivalence relations are respected.

\begin{figure}[t!]
\begin{center}
\includegraphics[scale=.8]{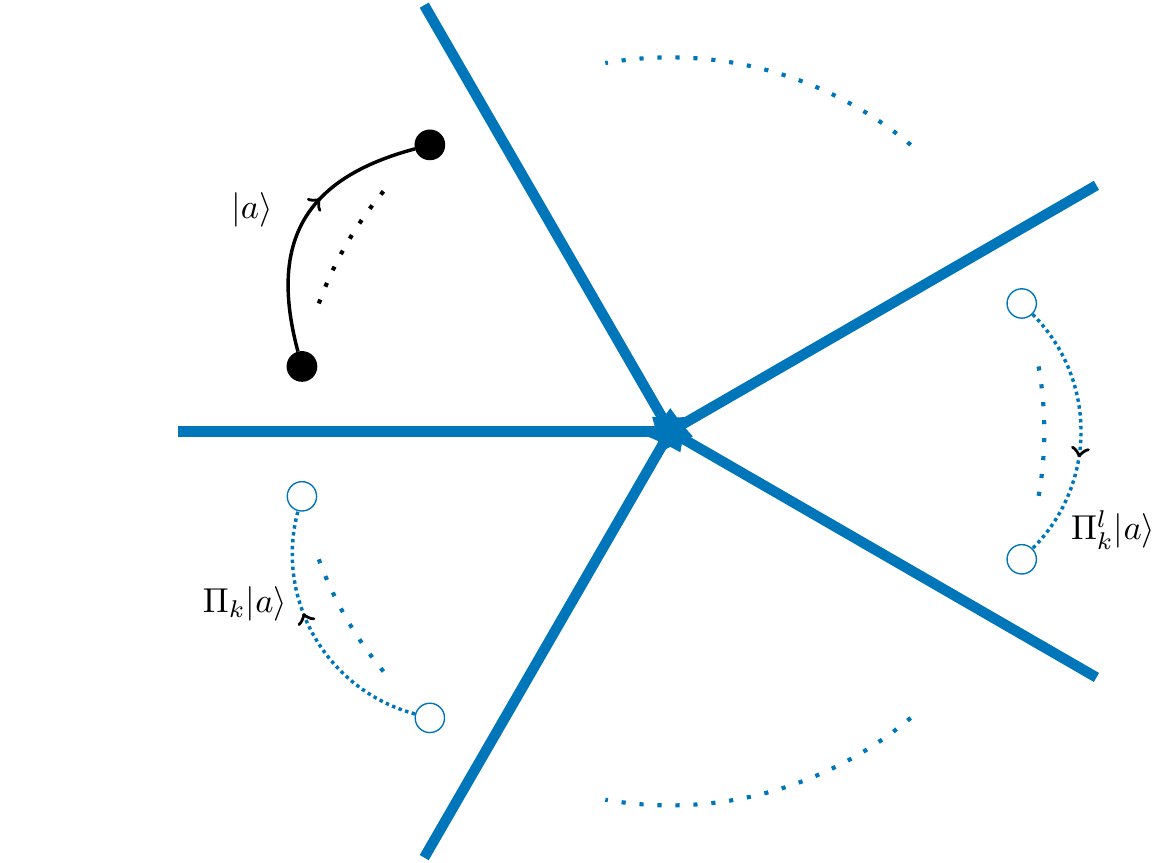}
\end{center}
\caption{Projection rules for S-fold planes acting on string junctions. We denote the orientifold image branes by
open shapes, and image strings by dashed blue lines.}
\label{fig:Sfold}
\end{figure}

To proceed further, we now examine the different choices of S-fold projections on different stacks of 7-branes.

\subsection{$\mathbb Z_2$ Quotients of  $E_6$}

We now turn to an analysis of $\mathbb{Z}_2$ quotients of $E_6$, namely we consider
the action of O3-planes on string junctions attached to an $E_6$ 7-brane.
We start by writing $E_6$ in a $\mathbb Z_2$-symmetric fashion.
The usual brane configuration $A^6 X C$ \cite{DeWolfe:1998zf} can be permuted
to a configuration $ A^3  C  A^3  C$. We discuss the permutations in Appendix \ref{app:BRANE}.
The set of $72$ junctions giving the roots of $E_6$ is
\begin{align} &\pm \left( {a}_i -  a_j\right)\,,\quad  1\leq j < i \leq 6\,,\nonumber\\
&\pm \left(\sum_{i=1}^3   a_i - \sum_{j=4}^6    a_j  -   a_k +  a_l +  c_1 -   c_2\right)\,,\quad  1\leq k \leq 3\,, \, 4 \leq l \leq 6\,,\nonumber\\
& \pm \left(   a_i -   a_j +  c_1 -   c_2\right) \,, \quad 1\leq k \leq 3\,, \, 4 \leq l \leq 6\,,\nonumber\\
&\pm \left(\sum_{i=1}^3   a_i - \sum_{j=4}^6    a_j +  c_1 -    c_2\right)\,,\nonumber \\
&\pm \left(\sum_{i=1}^3   a_i - \sum_{j=4}^6    a_j +2  c_1 - 2   c_2\right)\,,\nonumber \\
&\pm\left(  c_1 -   c_2\right)\,.
\label{eq:E6roots}\end{align}
A set of simple roots is given by
\al{\left\{ a_1-  a_2,  a_2-  a_3,  a_3 -   a_4,  a_4-  a_5, a_5-  a_6,  c_1-  c_2\right\}.
}
We will now turn to studying the effects of the S-fold projection, both
without and with discrete torsion (for all possible choices) turned on.
\begin{figure}[t!]
\begin{center}
\includegraphics[scale=.8]{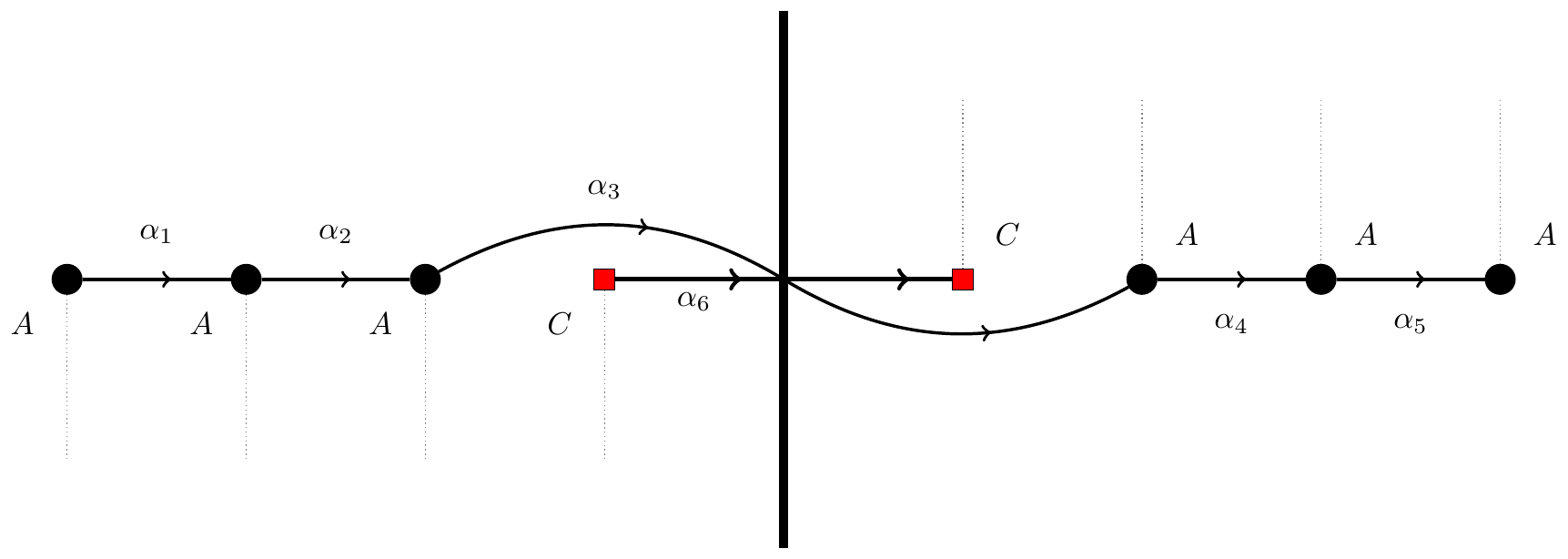}
\end{center}
\caption{$\mathbb Z_2$ symmetric configuration for $E_6$ theory.}
\label{fig:E6Z2}
\end{figure}

\subsubsection{$\mathbb Z_2$ Quotient without Discrete Torsion}

Consider first the case without any discrete torsion. After the projection
$48$ string junctions survive (see Appendix \ref{app:E6SJ} for a fully worked
example of which string junctions are projected out for the quotients of
$E_6$).  Given the symmetry of the system we can write all junctions
specifying only the charges on half the set of branes for sake of convenience.
The remaining junctions after projection are
\al{ \label{eqn:dora} &\pm \frac{1}{2}\left( {a}_i -  a_j\right)\,,\quad  1\leq j < i \leq 3\,,\nonumber\\
&\pm \frac{1}{2}\left( {a}_i +  a_j\right)\,,\quad  1\leq j < i \leq 3\,,\nonumber\\
& \pm   a_i\,, \quad 1\leq i \leq 3\,, \nonumber\\
& \pm \left(  a_i+  c_1\right), \quad 1\leq i \leq 3\,, \nonumber\\
&\pm \left(\sum_{i=1}^3   a_i -    a_j +  c_1 \right)\,, \quad 1\leq j \leq 3\,,\nonumber \\
&\pm\frac{1}{2} \left(\sum_{i=1}^3   a_i -    a_j +2  c_1 \right)\,, \quad 1\leq j \leq 3\,,\nonumber \\
&\pm \frac{1}{2}\left(\sum_{i=1}^3   a_i +    a_j +2  c_1 \right)\,, \quad 1\leq j \leq 3\,,\nonumber \\
& \pm   c_1\,,\nonumber\\
&\pm \left(\sum_{i=1}^3   a_i +   c_1\right)\,,\nonumber\\
&\pm \left(\sum_{i=1}^3   a_i + 2  c_1\right)\,.
}
This gives in total $48$ junctions, as expected for $F_4$. One choice of simple roots is
\begin{align}\left\{\frac{1}{2}\left(a_1 -   a_2\right), \frac{1}{2}\left(a_2 -   a_3\right), a_3,  c_1\right\}\,.\label{eq:F4simpleroots}
\end{align}
It is possible to check that using the intersection matrix of the brane system of $E_6$ one obtains the Cartan matrix of $F_4$,
thus indicating that the resulting algebra is $F_4$.

From the above considerations, we conclude that D3-branes probing this
S-folded 7-brane configuration will enjoy an $F_4$ global symmetry. At
first glance, this would appear to be at odds with reference
\cite{Shimizu:2017kzs} which demonstrated that for 4D $\mathcal{N} = 2$ SCFTs
with Higgs branch given by the single instanton moduli space of $F_4$ gauge
theory, there is a global inconsistency in the anomalies of the associated
theory. An important point to emphasize here, however, is that the same class
of assumptions also allows one to extract the values of various
anomalies including $\kappa_F = 5$, $a = 4/3$ and $c = 5/3$, which is rather
different from the values of references
\cite{Argyres:2015ffa,Argyres:2015gha,Argyres:2016xua,Argyres:2016xmc,Argyres:2016yzz,Martone:2020nsy,Argyres:2020wmq},
which have $\kappa_F = 6$, $a = 41/24$ and $c = 13/6$. Our analysis is compatible
with these considerations and indicate that the structure of the Higgs branch
is more subtle. Indeed, this is in line with the fact that moving the D3-brane
off the S-fold but still inside the $E_6$ 7-brane, the local spectrum of
string junction states is actually $E_6$. The brane picture indicates that it
is more appropriate, then, to view the Higgs branch moduli space for the
D3-brane as an instanton in an $E_6$ gauge theory but in the presence of a
codimension four S-fold defect.

\begin{figure}[t!]
\begin{center}
\includegraphics[scale=.8]{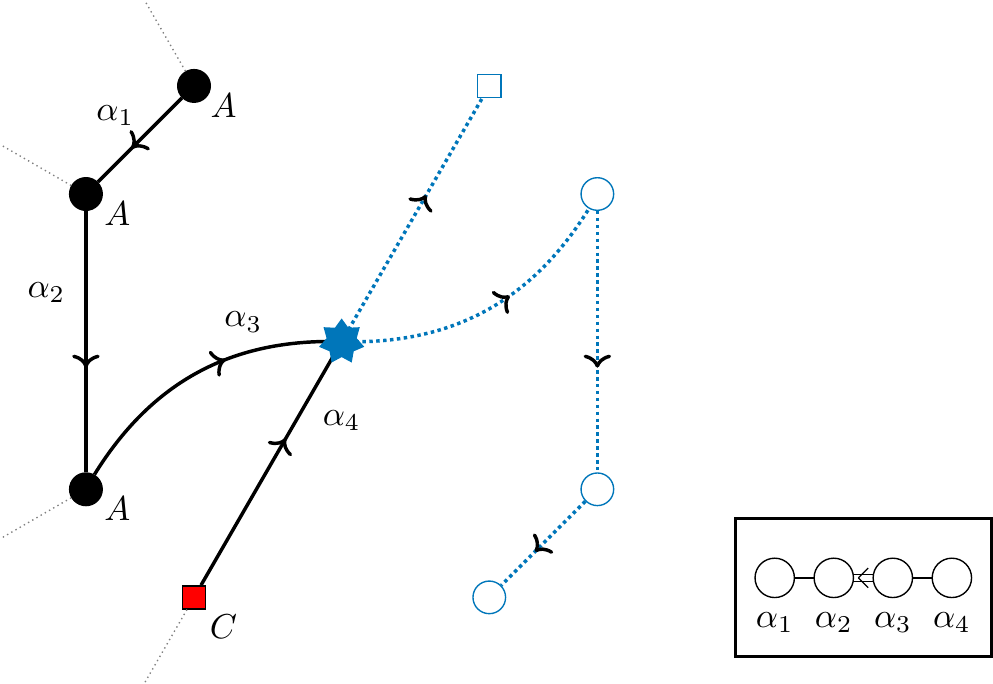}
\end{center}
\caption{String junctions for the S-fold projection of $E_6$ to $F_4$. We denote the orientifold image branes by
open shapes, and image strings by dashed blue lines.}
\label{fig:F4}
\end{figure}

\subsubsection{$\mathbb Z_2$ Quotient with Discrete Torsion}

Consider next the case of an orientifold projection with discrete torsion for
string junctions attached to an $E_6$ 7-brane. We find that in all cases the
junctions that are not invariant under the $\mathbb Z_2$ action are not
affected by the torsion. These are the ones with $1/2$ factors in the formulas
written in (\ref{eqn:dora}). For all choices of non-trivial discrete torsion
we find that $16$ additional junctions are projected out, though which ones in
particular depends on the choice of the discrete torsion. This gives in all
cases a set of $32$ junctions after projection. We illustrate the different
string junction configurations which survive the S-fold projection in figure
\ref{fig:e6sfolds}. This shows that although different string junctions
survive for each choice of discrete torsion, the actual flavor symmetry
algebra realized in all these cases is the same. Moreover, this analysis
establishes that in all cases the root system is the one of $C_4$.

\begin{figure}[t!]
\begin{center}
\includegraphics[width=\textwidth]{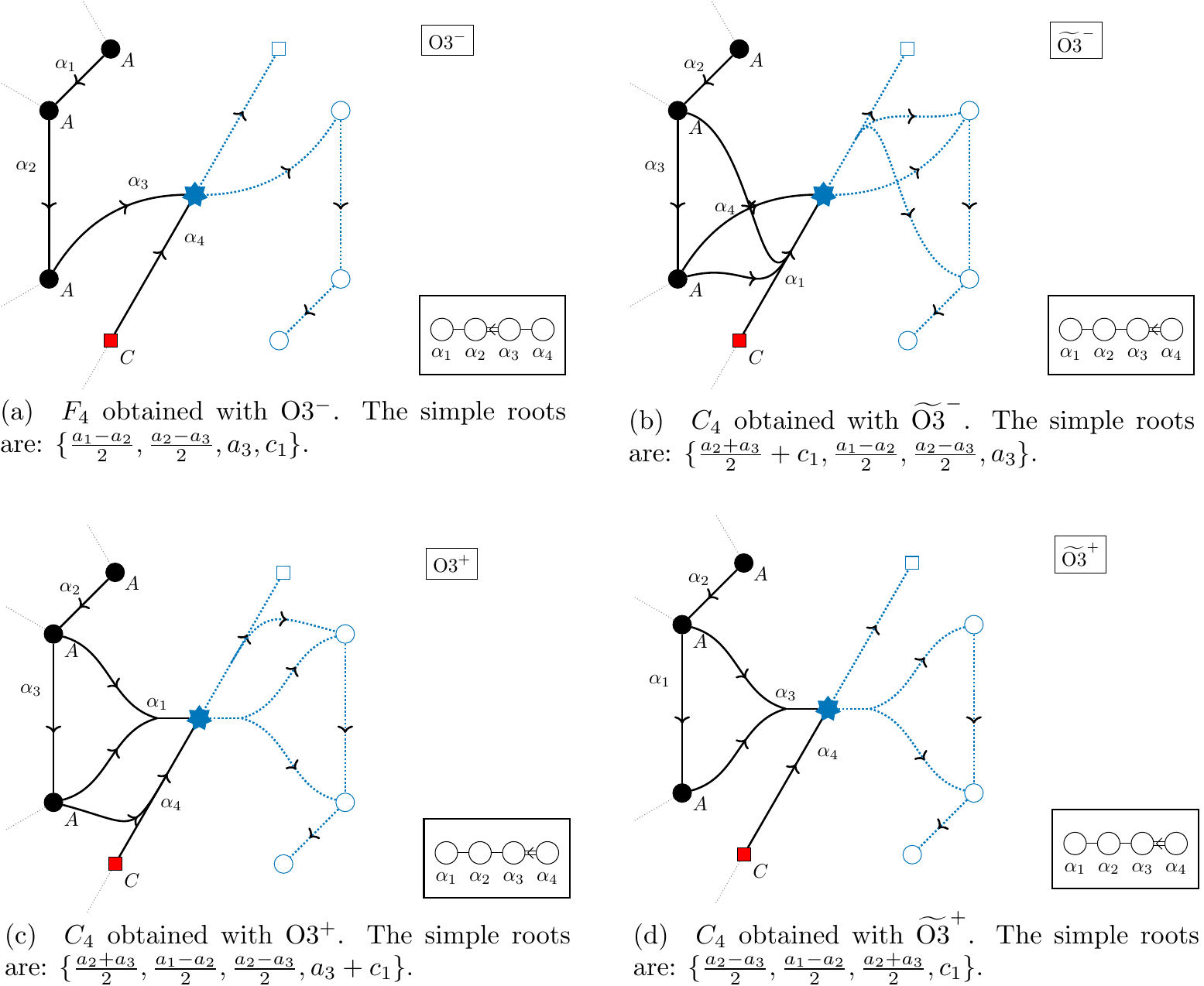}
\end{center}
\caption{Depiction of the different S-fold projections for an $E_6$ stack of 7-branes. Applying this projection results in two
physically distinct configurations, the one without discrete torsion (a), and the ones with discrete torsion (b,c,d).
We denote the orientifold image branes by open shapes, and image strings by dashed blue lines.}
\label{fig:e6sfolds}
\end{figure}

\subsection{$\mathbb Z_2$ Quotients of  $D_4$}

Consider next $\mathbb{Z}_2$ quotients of $D_4$. Recall that in F-theory, this is associated with a type $I_{0}^{\ast}$ fiber.
In this case, it is helpful to use the fact that the $E_6$ stack can be written as $AAA C AAAC$, so removing an A-brane from each grouping, we arrive at $AAC AAC$, the $\mathbb{Z}_2$ symmetric grouping for $D_4$. Therefore, one obtains the roots of $D_4$ by selecting the $E_6$ junctions without $  a_3$ and $  a_6$.  Moreover, for notational simplicity we shall rename $  a_{i+3}$ as $  a_{i+2}$ for $i = 1,2$. There are $24$ remaining junctions (as expected), and we list them here:
\al{ &\pm \left( {a}_i -  a_j\right)\,,\quad  1\leq j < i \leq 4\,,\nonumber\\
&\pm \left(  a_1 +   a_2 -   a_3 -  a_4 +  c_1 -   c_2\right)\,,\nonumber\\
& \pm \left(   a_i -   a_j +  c_1 -   c_2\right) \,, \quad 1\leq i \leq 2\,, \, 3 \leq j \leq 4\,,\nonumber\\
&\pm\left(  c_1 -   c_2\right)\,.
}
A set of simple roots is given by
\al{\left\{a_1-  a_2, a_2 -   a_3,a_3-  a_4,  c_1-  c_2\right\}.
}
Let us now turn to the different S-fold (really orientifold)
projections in this case.

\subsubsection{$\mathbb Z_2$ Quotient without Discrete Torsion}

Consider first the S-fold projection of $D_4$ without discrete torsion. In this case we
find $18$ junctions which survive, and this is the dimension of the root system of both $B_3$ and $C_3$.
As before, after the quotient we can write the junction specifying the charges on only half the set of branes.
The junctions after the projection are
\al{ &\pm \frac{1}{2}\left(  a_1 -   a_2\right)\,,\nonumber\\
&\pm\frac{1}{2} \left( {a}_1 +  a_2\right)\,,\nonumber\\
& \pm   a_i\,, \quad 1\leq i \leq 2\,, \nonumber\\
& \pm \left(  a_i+  c_1\right), \quad 1\leq i \leq 2\,, \nonumber\\
&\pm \left(  a_1 +   a_2 +  c_1 \right)\,,\nonumber \\
&\pm \frac{1}{2}\left(  a_1 +   a_2 +2  c_1 \right)\,,\nonumber \\
& \pm   c_1\,.
}
Computing the Cartan matrix we finds it corresponds to the Lie algebra $B_3$. One choice of simple roots is
\al{\left\{\frac{1}{2}\left(a_1 -   a_2\right), a_2, c_1\right\}\,.
}
As an additional comment, we observe that the above brane construction can be
viewed as specifying a mass deformation from a theory with $F_4$ global
symmetry to one with $B_3$ symmetry. This is indeed precisely the sort of
deformation observed from purely bottom up considerations in reference
\cite{Argyres:2015gha}. One can see this mass deformation as a blue arrow
between the $[II^*, F_4]$ and the $[III^*, B_3]$ theories in figure \ref{fig:N2rankone}.

\subsubsection{$\mathbb Z_2$ Quotient with Discrete Torsion}

Consider next the case of $D_4$ 7-branes in the presence of an orientifold (i.e. $\mathbb{Z}_2$ S-fold)
with discrete torsion. The result is that after the projection, $10$ string junctions survive for all different choices of
discrete torsion other than the trivial one. In all cases the resulting algebra is $C_2 \oplus A_1$.

\subsection{$\mathbb Z_2$ Quotients of  $H_2$}

The final case allowed with the $\mathbb Z_2$ S-fold is an $H_2$ 7-brane, namely a type $IV$ fiber at the origin.
We can obtain it by starting from the $AAC AAC$ realization of the $D_4$ case and dropping an $A$-brane from
both stacks, resulting in the configuration $ACAC$. The junctions can thus be obtained from the $D_4$ ones and this yields:
\al{ &\pm \left( {a}_1 -  a_2\right)\,,\nonumber\\
& \pm \left(   a_1 -   a_2 +  c_1 -   c_2\right) \,,\nonumber\\
&\pm\left(  c_1 -   c_2\right)\,.
}
A set of simple roots is given by:
\al{\{a_1-  a_2, c_1 -   c_2\}\,.
}
So, we get $6$ junctions as expected for $H_2$, giving an $A_2$ algebra.
Let us now turn to S-fold projections of this flavor symmetry.

\subsubsection{$\mathbb Z_2$ Quotient without Discrete Torsion}

Consider first the $\mathbb{Z}_2$ quotient without discrete torsion of an
$H_2$ flavor 7-brane. In this case, it is interesting to note that all string
junctions are invariant under the $\mathbb Z_2$ action when there is no
discrete torsion.  Consequently, we retain the same flavor symmetry algebra.
In the context of 4D $\mathcal{N} = 2$ SCFTs \cite{Argyres:2015gha}, we
observe that we can also consider the associated flow, via mass deformation,
from $[III^*,B_3]$ to $[IV^*,A_2]$, as in figure \ref{fig:N2rankone}, which is
compatible with our brane picture.

\subsubsection{$\mathbb Z_2$ Quotients with Discrete Torsion}

We next consider the $\mathbb Z_2$ projection with discrete torsion of the $H_2$ theory.
The result is that after the projection, $2$ string junctions survive for all different
choices of discrete torsion other than the trivial one. In all cases the resulting algebra
is $A_1 \oplus U(1)$. Here we observe the appearance of a $U(1)$ factor in the symmetry algebra. We see this
since there are string junctions stretched to just the $C$ brane of the configuration $A^{3} C$ realizing $H_2$ and its
subsequent $\mathbb{Z}_2$ quotient. This is also in accord with the quotient group action on the symmetry algebra
of the parent theory.

\subsection{$\mathbb Z_3$ Quotients of $D_4$}

As we already saw in section \ref{sec:7BRANE}, the $D_4$ configuration of 7-branes also admits a $\mathbb{Z}_3$ S-fold quotient. Here,
we study the resulting algebras both in the absence and in the presence of discrete torsion. To proceed, we observe that the $\mathbb Z_3$ symmetric choice of branes is $AABBDD$ where $D$ is a $[0,1]$-brane. In this presentation the junctions giving the root system of $D_4$ are
\al{ &\pm\left(a_1 -a_2\right)\,,\quad \pm\left(b_1 -b_2\right)\,, \quad \pm\left(d_1 -d_2\right)\,,\nonumber\\
&\pm \left(a_i - b_j -d_k\right)\,, 	\quad i=1,2\,, \ j = 1,2\,,\ k = 1,2\,,\quad \nonumber\\
&\pm\left(a_1 + a_2 -b_1-b_2-d_1-d_2\right)\,.
}
One choice of simple roots is
\al{\{-a_1+a_2,a_1-b_1-d_1,d_1-d_2,b_1-b_2\}\,.
}

\subsubsection{$\mathbb Z_3$ Quotient without Discrete Torsion}

With this in place, we are ready to discuss $\mathbb{Z}_3$ S-fold projections of $D_4$ 7-branes. Consider first the case of
S-fold projections without discrete torsion. The $\mathbb Z_3$ action maps the branes as follows
\al{a_i \rightarrow -b_i\,, \quad b_i \rightarrow d_i \,, \quad d_i \rightarrow -a_i\,.
}
After the projection the remaining junctions are
\al{&\pm\left(a_i - b_i -d_i\right)\,, \quad i = 1,2\,,\nonumber\\
& \pm\frac{1}{3}\left(-a_1 + a_2 +b_1-b_2+d_1-d_2\right)\,,\quad  \pm\left(a_1 + a_2 -b_1-b_2-d_1-d_2\right)\,,\nonumber\\
&\pm \frac{1}{3}\left( 2a_1 +a_2 -2 b_1 - b_2 -2 d_1 - d_2\right)\,, \quad \pm\frac{1}{3}\left( a_1 +2a_2 - b_1 -2 b_2 - d_1 -2 d_2\right)\,.
}
The simple roots after projection can be chosen to be
\al{\left\{\frac{1}{3}\left(-a_1 + a_2 +b_1-b_2+d_1-d_2\right),a_1 - b_1 -d_1\right\}\,,
}
whose intersection gives the Cartan matrix of $G_2$, which matches to the $[II^*,G_2]$ theory of reference
\cite{Argyres:2015gha}.

\subsubsection{$\mathbb Z_3$ Quotients with Discrete Torsion}

We next consider the $\mathbb Z_3$ projection with discrete torsion of the $D_4$ theory. In this case the reason why some junctions may be projected out is that after summing over the $\Pi_3$ images they get a factor $1 + \zeta + \zeta^2 = 0$ where $\zeta$ is a primitive third
root of unity. One can check that for both choices of discrete torsion the junctions $\pm (a_i -b_i-d_i)$ for $i=1,2$ and $\pm (a_1+a_2-b_1-b_2-d_1-d_2)$ are projected out. This leaves in total 6 junctions giving the $A_2$ algebra.\footnote{Going from $G_2$ to $A_2$ follows because the root system of $G_2$ is nothing but the root system of $A_2$ with the addition of the weights of the $\mathbf 3$ and $\bar{\mathbf 3}$ representations. Including discrete torsion projects out these vectors leaving only $A_2$ behind.}

\subsection{$\mathbb Z_3$ Quotients of $H_1$}

Let us now turn to $\mathbb{Z}_3$ quotients of the $H_1$ stack of 7-branes. We can use our analysis of the $D_4$ stack of 7-branes to
aid in this analysis. To this end, we begin with the realization of the $D_4$ algebra using the $\mathbb{Z}_3$ symmetric
stack $AABBDD$. We get to the $H_1$ stack by removing one $A$ brane, one $B$ brane and one $D$ brane. The remaining junctions are
\al{\pm (a - b - d)\,,
}
thus giving an $A_1$ algebra.

\subsubsection{$\mathbb Z_3$ Quotient without Discrete Torsion}

Consider first the $\mathbb{Z}_3$ S-fold projection in the absence of discrete
torsion. This junction is already invariant under the $\mathbb Z_3$ quotient
suggesting that the theory can be identified with the $[III^*,A_1]$ of
\cite{Argyres:2015gha}. Indeed there is a flow $[II^*,G_2] \rightarrow [III^*,A_1]$ for
the corresponding 4D $\mathcal{N} = 2$ SCFTs.

\subsubsection{$\mathbb Z_3$ Quotients with Discrete Torsion}

Next consider the $\mathbb{Z}_3$ S-fold projection with discrete torsion. In
both cases of $\mathbb Z_3$ discrete torsion there are no junctions surviving
leaving only one single Cartan generator behind. The flavor symmetry is
therefore simply $U(1)$.

\subsection{$\mathbb Z_4$ Quotients of $H_2$}

We next turn to the $\mathbb{Z}_4$ S-fold projection of the $H_2$ stack of 7-branes.
The brane system can be conjugated to a $DADA$ system  where again $D$ is a $[0,1]$-brane. The junctions giving the roots are
\al{ \pm\left(a_1-a_2\right)\,, \quad \pm\left(d_1-d_2\right)\,, \quad \pm\left(a_1-a_2 + d_1-d_2\right)\,.
}
\subsubsection{$\mathbb Z_4$ Quotient without Discrete Torsion}

Consider first the $\mathbb{Z}_4$ S-fold projection without discrete torsion on the $H_2$ stack of 7-branes.
The $\mathbb Z_4$ projection maps
\al{ a_1 \rightarrow d_1\,, \quad d_1 \rightarrow- a_2 \,, \quad a_2 \rightarrow d_2\,, \quad  d_2 \rightarrow -a_1\,.
}
After projection one finds only the junctions
\al{ \pm \left(a_1 + d_1 -a_2-d_2 \right)\,.
}
The algebra is therefore $A_1$ thus giving the $[II^*,B_1]$
theory.\footnote{Note that at the level of Lie algebras we have $A_1 \simeq
B_1$.}

\subsubsection{$\mathbb Z_4$ Quotient with Discrete Torsion}

In the case of the $\mathbb{Z}_4$ S-fold projection with discrete torsion of
the $H_2$ stack of 7-branes, we find by a similar analysis that the algebra is $A_1$, i.e. there is no distinction
in the flavor symmetry algebras for the cases with and without discrete
torsion.

\subsection{Collection of Flavor Symmetry Algebras}

In this section we collect our results on the resulting flavor symmetry
algebras. First, we remind the reader that the particular non-zero values of the discrete torsion are
irrelevant; the spectrum of physical states, as determined from the string junctions, is
identical for all cases with non-zero discrete torsion. We then
summarize the different algebras and a choice of root system in tables
\ref{tab:Z2}, \ref{tab:Z3}, \ref{tab:Z4}. In table \ref{tab:ENHANCO} we summarize
the relevant patterns, indicating quotients without discrete torsion as $\mathbb{Z}_k$
and those with discrete torsion as $\widehat{\mathbb{Z}}_k$.

\renewcommand{\arraystretch}{1.5}

\begin{table}[]
    \centering
    \begin{tabular}{|c|c|c|c|}\hline
        S-fold & $E_6/\bZ_2$ & $D_4/\bZ_2$ & $H_2/\bZ_2$  \\ \hline\hline
        $\mathrm{O3}^-$  & $F_4: $ & $B_3:$ & $A_2:$\\
         & $\{\frac{a_1-a_2}{2}, \frac{a_2-a_3}{2}, a_3, c_1 \}$ &$ \{\frac{a_1-a_2}{2}, a_2, c_1 \}$  &  $\{a_1 , c_1 \}$
        \\ \hline\hline
        $\mathrm{\widetilde{O3}}^-$  &
        $C_4:$ & $C_2 \oplus A_1:$ & $A_1 \oplus U(1):$\\
        & $\{\frac{a_2+a_3}{2}+c_1, \frac{a_1-a_2}{2}, \frac{a_2-a_3}{2}, a_3 \}$ & $ \{ \frac{a_1 - a_2}{2}, a_2 \}\oplus \{\frac{a_1 + a_2}{2}+c_1 \}$ & $ \{a_1 \}$
        \\ \hline
        $\mathrm{O3}^+$ &
        $C_4:$ & $C_2 \oplus A_1: $ & $A_1 \oplus U(1): $ \\
        & $ \{\frac{a_2+a_3}{2}, \frac{a_1-a_2}{2}, \frac{a_2-a_3}{2}, a_3+c_1 \}$ & $\{ \frac{a_1 - a_2}{2}, c_1 + a_2 \} \oplus \{\frac{a_1 + a_2}{2} \} $ &  $\{a_1 + c_1 \}$
        \\ \hline
        $\mathrm{\widetilde{O3}}^+$  & $C_4:$ & $C_2 \oplus A_1:$ & $A_1 \oplus U(1):$ \\
         & $ \{\frac{a_2-a_3}{2}, \frac{a_1-a_2}{2}, \frac{a_2+a_3}{2}, c_1 \}$ & $ \{ \frac{a_1 - a_2}{2}, -a_1-a_2-c_1 \}\oplus \{\frac{a_1 - a_2}{2} \}$ &$ \{c_1 \}$
        \\ \hline
    \end{tabular}
    \caption{Simple roots of $\mathbb{Z}_2$ S-folds (i.e. orientifold projection) with all possible choices of discrete torsion.}
    \label{tab:Z2}
\end{table}

\begin{table}[]
    \centering
    \begin{tabular}{|c|c|c|}\hline
        S-fold & $D_4/\bZ_3$ & $H_1/\bZ_3$  \\ \hline\hline
    Trivial & $G_2:$ & $A_1:$\\
    torsion& $\{\frac{1}{3}\left(-a_{1}+a_{2}+b_{1}-b_{2}+d_{1}-d_{2}\right), $ &  $\{a - b - d\}$ \\
    & $a_{1}-b_{1}-d_{1}\}$ &
    \\\hline\hline
    Non-trivial & $A_2:$ & $U(1)$ \\
    torsion& $\{\frac{1}{3}\left(-a_{1}+a_{2}+b_{1}-b_{2}+d_{1}-d_{2}\right),$ & \\
    & $\frac{1}{3}\left(2 a_{1}+a_{2}-2 b_{1}-b_{2}-2 d_{1}-d_{2}\right)\}$ &
    \\\hline
    \end{tabular}
    \caption{Simple roots of $\mathbb{Z}_3$ S-folds with all possible choices of discrete torsion.}
    \label{tab:Z3}
\end{table}

\begin{table}[]
    \centering
    \begin{tabular}{|c|c|c|}\hline
        S-fold & $H_2/\bZ_4$\\ \hline\hline
    Trivial torsion & $A_1: \{a_1 + d_1 - a_2 - d_2\}$
    \\\hline
    Non-trivial torsion & $A_1: \{a_1 + d_1 - a_2 - d_2\}$
    \\\hline
    \end{tabular}
    \caption{Simple roots of $\mathbb{Z}_4$ S-folds with all possible choices of discrete torsion. Here having non-trivial torsion does not affect the gauge algebra or the simple root system.}
    \label{tab:Z4}
\end{table}

\renewcommand{\arraystretch}{1}

\begin{table}[]
\centering
\begin{tabular}{c c c c c c c}
\textbf{parent } & $\mathbb Z_2$ & $\widehat{\mathbb Z}_2$ & $\mathbb Z_3$ & $\widehat{\mathbb Z}_3$ & $\mathbb Z_4$ & $\widehat{\mathbb Z}_4$\\
\hline
$E_8$ \\
$E_7$\\
$E_6$ & $F_4$ & $C_4 $\\
$D_4$ & $B_3$ & $ C_2 \oplus A_1 $ & $G_2$ & $A_2$\\
$H_2$ & $A_2$ & $A_1 \oplus U(1) $ & & & $A_1$ & $A_1$\\
$H_1$ & & & $A_1$ & $U_1 $\\
$H_0$
\end{tabular}
    \caption{Summary of symmetry algebras obtained from an S-fold projection of a parent stack of 7-branes. We find that there are two qualitative quotients, based on $\mathbb{Z}_k$ without discrete torsion, and based on $\widehat{Z}_{k}$ with discrete torsion.}
    \label{tab:ENHANCO}
\end{table}

The aforementioned flavor algebras are always realized on the worldvolume of
the 7-branes and for all ranks of the SCFT. However it is expected that in the case of
rank one theories, a quotient with discrete torsion can result in an enhancement of the geometric
flavor symmetry and that realized by a 7-brane. This geometric symmetry is $SU(2)$
for $\widehat{\mathbb{Z}}_2$ quotients and $U(1)$ for the other $\widehat{\mathbb{Z}}_{k}$ quotients.
We can determine that there is likely an enhancement when the level of the
$SU(2)$ and the level of the 7-brane flavor symmetry (both
of which we can calculate) match. The expected enhancements \cite{Argyres:2016yzz} are:
\begin{itemize}
\item[-] For the $\widehat {\mathbb Z}_2$ quotient of $E_6$ the rank one theory is expected to have $C_5$ flavor symmetry;
\item[-] For the $\widehat {\mathbb Z}_2$ quotient of $D_4$ the rank one theory is expected to have $C_3 \oplus A_1$ flavor symmetry;
\item[-] For the $\widehat {\mathbb Z}_2$ quotient of $H_2$ the rank one theory is expected to have $C_2 \oplus U_1$ flavor symmetry;
\item[-] For the $\widehat {\mathbb Z}_3$ quotient of $D_4$ the rank one theory is expected to have $A_3 \rtimes \mathbb Z_2$ flavor symmetry;
\item[-] For the $\widehat {\mathbb Z}_3$ quotient of $H_1$ the rank one theory is expected to have $A_1 \oplus U_1 \rtimes \mathbb Z_2$ flavor symmetry;
\item[-] For the $\widehat {\mathbb Z}_4$ quotient of $H_2$ the rank one theory is expected to have $A_2 \rtimes \mathbb Z_2$ flavor symmetry.
\end{itemize}

\subsection{Admissible Representations}

So far we have focused on the structure of the Lie algebra of the flavor symmetry.
The string junction picture also allows us to access the admissible representations.
We will discuss only the cases where the center of the simply connected group of a given Lie algebra is non-trivial. We begin by first discussing S-fold projections without discrete torsion, and then turn to the case of examples with discrete torsion.
If there happen to be other sources of flavor symmetries, this can lead to additional global structure. For example, $E_8$ has an $E_6 \times SU(3) / \mathbb{Z}_3$ subgroup, but also has representations in the $(\mathbf{27} , \mathbf{3})$. If we ignore the $SU(3)$ factor, then we would loosely refer to this as realizing an $E_6$ group. In the probe D3-brane theories,
we also know that there is an $SU(2)$ flavor symmetry associated with symmetries internal to the 7-brane but transverse to the D3-brane,
so determining the full structure of the 4D flavor symmetry must reference this feature as well. We leave this determination for future work. What we can assert from the string junction picture is whether we see evidence for a given type of representation, and so to indicate this information we will mildly abuse terminology and refer to $G_{\mathrm{rep}}$ as specifying the ``the flavor group'' and its admissible representations.

\subsubsection{S-fold Projections without Discrete Torsion}

We now turn to S-fold projections without discrete torsion in which, for a given Lie algebra, the associated
simply connected Lie group has a non-trivial center. This limits us to the following cases:

\begin{itemize}
\item[-] The $\mathbb Z_2$ quotient of a $D_4$ stack of 7-branes without discrete torsion yields a $B_3$ algebra, which means that the flavor group is either $Spin(7)$ or $Spin(7)/\mathbb Z_2 \simeq SO(7)$. One quick way to check which representations are allowed is to use the fact that the $B_3$ theory descends from the $F_4$ theory. Decomposing the adjoint of $F_4$ one finds
\al{ F_4 &\rightarrow Spin(7) \otimes SO(2)\\
\mathbf {52} & \rightarrow \mathbf 1_0 \oplus \mathbf 7_{2} \oplus \mathbf 7_{-2} \oplus \mathbf {21}_0 \oplus \mathbf 8_1 \oplus \mathbf 8_{-1}\,.
}
Note that the $\mathbf 8$ is the spinor representation of $Spin(7)$ so indeed the flavor group is $Spin(7)$.
\item[-] The $\mathbb Z_2$ quotient of a $H_2$ stack of 7-branes without discrete torsion yields an $A_2$ algebra, which means that the flavor group is either $SU(3)$ or $SU(3)/\mathbb Z_3 \simeq PSU(3)$. Similarly to the previous case we can use the fact that the $A_2$ theory descends from the $B_3$ theory. Decomposing the adjoint of $Spin(7)$ one finds
\al{ Spin(7) &\rightarrow SU(3) \otimes U(1)\\
\mathbf {21} & \rightarrow \mathbf {1}_0 \oplus \mathbf 8_0\oplus \mathbf 3_4 \oplus \mathbf{\bar 3}_{-4}\oplus \mathbf 3_2 \oplus \mathbf{\bar 3}_{-2}\,.
}
Since the $\mathbf 3$ representation of $A_2$ is present this fixes the flavor symmetry group to be $SU(3)$.

\item[-] The $\mathbb Z_3$ quotient of an $H_1$ theory without discrete torsion gives the flavor algebra $A_1$, which means that the flavor group could be either $SU(2)$ or $SU(2)/\mathbb Z_2 \simeq SO(3)$. We can follow the logic outlined before noting that this theory comes from the $G_2$ theory. Decomposing  the adjoint of $G_2$ we find
\al{ G_2 & \rightarrow SU(2) \otimes SU(2)\,,\\
\mathbf{14} & \rightarrow (\mathbf 3,\mathbf 1) \oplus (\mathbf 1, \mathbf 3) \oplus (\mathbf 4,\mathbf 2)\,.
}
It is possible to check by computing the charges of the junctions that after breaking $G_2$ the junctions lie in the $\mathbf 4$ representation of the unbroken group, implying that this group is $SU(2)$ rather than $SO(3)$ given that the $\mathbf 4$ is charged under the center.
\end{itemize}

\subsubsection{S-fold Projections with Discrete Torsion}

Let us now turn to the related case of S-fold projections with discrete torsion. Again, we confine our analysis to
those Lie algebras which have a simply connected Lie group with non-trivial center. The relevant cases are:

\begin{itemize}
\item[-] The $\mathbb Z_2$ quotient of the $E_6$ theory with discrete torsion gives the flavor algebra $C_4$, which means that the flavor group can be either $USp(8)$ or $USp(8)/\mathbb Z_2$. In this case we note that all junctions must descend from junctions of the parent $E_6$ theory and its weight lattice is generated by the junctions giving the $\mathbf {27}$ representation. Decomposing it we find
\al{ E_6 & \rightarrow USp(8)\,,\\
\mathbf{27} & \rightarrow \mathbf{27}\,.
}
The $\mathbf {27}$ of $USp(8)$ is the two-index anti-symmetric representation which is not charged under the center. This implies that no junctions charged under the center can be generated, implying that the flavor group is $USp(8)/\mathbb Z_2$.

\item[-] The $\mathbb Z_2$ quotient of the $D_4$ theory with discrete torsion gives the flavor algebra $C_2 \oplus A_1$. Here there are various possibilities for the global structure of the gauge group. Knowing that this theory descends from the $C_4$ theory we can decompose the adjoint of $C_4$
\al{ USp(8) & \rightarrow USp(4) \otimes SU(2) \otimes U(1)\,,\\
\mathbf {36} & \rightarrow (\mathbf 4,\mathbf 2)_1 \oplus (\mathbf 4 , \mathbf 2)_{-1} \oplus (\mathbf 1,\mathbf 3)_0 \oplus (\mathbf 1,\mathbf 3)_{2} \oplus (\mathbf 1,\mathbf 3)_{-2} \oplus (\mathbf 1, \mathbf 1)_0 \oplus (\mathbf{10},\mathbf 1)_0\nonumber\,.
}
We see that the only representations charged under the center of $USp(4)$ and $SU(2)$ appear together, which suggests that the group is $\left(USp(4) \otimes SU(2)\right)/\mathbb Z_2$. Note that other quotients like for instance $USp(4)/\mathbb Z_2 \otimes SU(2)/\mathbb Z_2$ are not compatible with the representations appearing given that the fundamental representations of $USp(4)$ and $SU(2)$ appear in the previous decomposition. Following a similar logic starting from the $\mathbf {27}$ representation of $USp(8)$ which is the smallest representation available confirms this result.

\item[-] The $\mathbb Z_2$ quotient of the $H_2$ theory with discrete torsion gives the flavor algebra $A_1 \oplus U_1$. In this case we can decompose the adjoint of $C_2 \oplus A_1$ as
\al{ USp(4) \otimes SU(2) &\rightarrow SU(2) \otimes U(1)_a \otimes U(1)_b\,,\\
(\mathbf{10},\mathbf 1)\oplus (\mathbf 1, \mathbf 3) &\rightarrow  \mathbf 1_{(0,0)}\oplus \mathbf 1_{(0,0)} \oplus \mathbf 3_{(0,0)}\oplus \left(\mathbf 2_{(1,1)} \oplus   \mathbf 1_{(2,2)} \oplus \mathbf 1_{(2,-2)} \oplus \text{h.c}\right)\,.
}
The broken generator is $U(1)_b$ leaving $SU(2) \otimes U(1)_a$. Therefore the flavor symmetry group seems to be  $\left(SU(2) \otimes U(1)\right)/\mathbb Z_2$. The conclusion does not change when looking at other representations of $\left(USp(4) \otimes SU(2)\right)/\mathbb Z_2$.

\item[-] The $\mathbb Z_3$ quotient of the $D_4$ theory with discrete torsion gives the flavor algebra $A_2$, which means that the flavor group can be either $SU(3)$ or $SU(3)/\mathbb Z_3 \simeq PSU(3)$. In this case we note that all junctions must descend from junctions of the parent $E_6$ theory and its weight lattice is generated by the junctions giving the $\mathbf {8_s}$, the $\mathbf {8_c}$ and the $\mathbf {8_v}$ representations. Decomposing them we find
\al{ Spin(8) & \rightarrow SU(3)\,,\\
\mathbf{8_s} & \rightarrow \mathbf{8}\,,\\
\mathbf{8_c} & \rightarrow \mathbf{8}\,,\\
\mathbf{8_v} & \rightarrow \mathbf{8}\,.
}

The $\mathbf 8$ representation of $A_2$ is of course the adjoint which is uncharged under the center. This means that no representation charged under the center is present, giving the flavor symmetry $PSU(3)$.

\item[-] The $\mathbb Z_4$ quotient of the $H_2$ theory with discrete torsion gives the flavor algebra $A_1$, which means that the flavor group can be either $SU(2)$ or $SU(2)/\mathbb Z_2 \simeq SO(3)$. In this case we note that all junctions must descend from junctions of the parent $H_2$ theory and its weight lattice is generated by the junctions giving the $\mathbf 3$ representation. Decomposing them we find
\al{ SU(3) & \rightarrow SU(2)\,,\\
\mathbf{3} & \rightarrow \mathbf{3}\,.
}
The $\mathbf 3$ representation of $A_1$ is of course the adjoint which is uncharged under the center. This means that no representation charged under the center is present giving the flavor symmetry $SO(3)$.

\end{itemize}

\section{F-theory and S-folds with Discrete Torsion}\label{sec:DISCRETE}

One useful application of the F-theory construction is that it allows one to read
off the Seiberg--Witten curve from the geometry for the rank one
theories. However, as we stressed before, this procedure works only in the
absence of discrete torsion. Given this identification between geometry and
the low-energy field theory data it is tempting to push this identification
beyond the case without discrete torsion. We propose that the F-theory
geometry in the presence of discrete torsion is the Seiberg--Witten curve of
the theory on a single probe D3-brane. In this section we will list all the
maximally mass deformed Seiberg--Witten curves from \cite{Argyres:2015gha} for
the various theories we obtained in the presence of discrete torsion. One
subtle point is that in the case of a single D3-brane, there can be additional
enhancements in the flavor symmetry relative to the case of multiple
D3-branes. In these cases, we interpret the F-theory geometry as the one
obtained by taking a mass deformation of the enhanced symmetry algebra which
takes us to the generic flavor symmetry, and then taking a further scaling
limit so that the terms with the mass deformation are scaled out. In all
cases, this is associated with the degree two Casimir invariants of the flavor
symmetry algebra. In what follows, we leave this operation implicit in our
discussion. With notation as earlier, we use the Coulomb branch parameter $u$
to indicate the directions transverse to the 7-brane in the quotiented
geometry.

\begin{itemize}
\item[-] The Seiberg--Witten curve for the $\mathbb Z_2$ quotient with discrete torsion of the $E_6$ theory is
\al{ y^2 = x^3 &+ 3 x \left[2 u^3 M_2 +u^2 \left(M_4^2 -2 M_8\right)+2 u M_4 M_{10}-M_{10}^2\right]\nonumber\\
&+ 2 \left[u^5 + u^4 M_6+u^3 \left(2 M_4^3-3 M_4 M_8 -3 M_2 M_{10}\right)\right.\\
& \left.+3u^2 M_8 M_{10} -3 u M_4 M_{10}^2 +M_{10}^3 \right]\nonumber\,.
}

\item[-] The Seiberg--Witten curve for the $\mathbb Z_2$ quotient with
  discrete torsion of the $D_4$ theory is
\al{ y^2 = x^3 &+  x \left[ 12 u^3 -u^2\left(M_4 + 4 M_2^2\right)+12 u M_2 M_6-3 M_6^2\right]\nonumber\\
    &-12 u^4 \left(2 M_2 + 3 \widetilde{M}_2 \right)+2 u^3 \left(M_2 M_4 + 6 M_6\right)\\
& -u^2 \left(16 M_2^2 + M_4\right)M_6+12u M_2 M_6^2 -2 M_6^3\nonumber\,.
}
Note the presence of two independent degree two Casimirs, $M_2$ and
    $\widetilde{M}_2$. This occurs whenever the flavor symmetry is
    semi-simple, in this case it is $C_3 \oplus A_1$.

\item[-] The Seiberg--Witten curve for the $\mathbb Z_2$ quotient with discrete torsion of the $H_2$ theory is
\al{ y^2 = x^3 &-  x \left[ 3 u^2 \left( M_2 + M_1^2 \right)+12u M_1 M_4+3 M_4^2\right]\nonumber\\
&-864 u^4 + 2 u^3 M_1 \left(M_1^2 -3 M_2\right)-3 u^2 (5M_1^2 + M_2)M_4\\
&-12 u M_1 M_4^2 -2 M_4^3 \nonumber\,.
}

\item[-] The Seiberg--Witten curve for the $\mathbb Z_3$ quotient with discrete torsion of the $D_4$ theory is
\al{ y^2 = x^3 +3 x u^2 \left(2 u M_2 - M_4^2 \right)+2u^3\left(u^2 +M_4^3+ u M_6\right)\,.
}
This was identified in \cite{Argyres:2016xua} and reproduces the curve already found in \cite{Chacaltana:2016shw}.

\item[-] The Seiberg--Witten curve for the $\mathbb Z_3$ quotient with discrete torsion of the $H_1$ theory is
  \al{y^2 = x^3 +3 x \left(u^3 -u^2 \widetilde{M}_2^2 \right)+ 2 \left(u^4 M_2 +u^3
    \widetilde{M}_2^3\right)\,.
}
\item[-] The Seiberg--Witten curve for the $\mathbb Z_4$ quotient with discrete torsion of the $H_2$ theory is
\al{ y^2 = x^3 -\frac{1}{8} x \left(2 u -M_6\right)^3 M_2 -\frac{1}{8} (2 u -M_6)^4 (u + 2 M_6)\,.
}

\end{itemize}

As an additional comment, we note that here, we have mainly focused on the situation where we treat the $M_i$ as
mass parameters. Of course, since the S-fold introduces a codimension four defect in the worldvolume of the 7-brane,
we can also include additional position dependence in these mass parameters. Doing so would produce F-theory backgrounds
which we can characterize as elliptically fibered Calabi--Yau threefolds in the presence of discrete torsion.

\section{Anomalies}\label{sec:ANOMO}

As a further check on our proposal, in this section we study the scaling of
the conformal anomalies $a$ and $c$ in the limit of large $N$, that is, when
we have a large number of probe D3-branes. We shall also determine the flavor symmetry
anomaly $\kappa_G$ associated with two flavor currents and an R-symmetry current,
namely $\mathrm{Tr}(\mathcal{R} GG)$, where $\mathcal{R}$ denotes the current
for the $U(1)_{\mathcal{R}}$ factor of the R-symmetry $SU(2) \times
U(1)_{\mathcal{R}}$ of a 4D $\mathcal{N} = 2$ SCFT and $G$ refers to a flavor
symmetry current associated with a 7-brane. Since we are dealing with
topological features of the theory,
we will extrapolate our results back to small values of $N$, much as in
reference  \cite{Aharony:2007dj}. From our analysis, we can read off both the
order $N^2$ and order $N$ contributions to the conformal anomalies, however we
will not be able to access the $\mathcal{O}(N^0)$ contributions via these
methods.
This will allow us to compare with the results of reference \cite{Giacomelli:2020jel}, which studies
certain 4D SCFTs from $T^2$ compactifications of 6D $\mathcal{N} = (1,0)$ SCFTs, as well as
with reference \cite{Apruzzi:2020pmv}, which studies some examples of D3-brane probes of
S-folds with discrete torsion. In the rank one case, $N = 1$, we will find
consistency with the rank one theories of \cite{Argyres:2016yzz}, though in those cases we
will have to subtract a free hypermultiplet to match with the interacting
SCFT.

The computation is done using  holography as in \cite{Aharony:2007dj}. The large $N$ dual of the background we are considering is Type IIB on $\text{AdS}_5 \times S^5/\mathbb Z_k$ with 7-branes. We will separate the various terms appearing in the central charges according to their $N$ scaling, with leading order being $N^2$.

\begin{itemize}
\item[-] $\mathcal O(N^2)$: this term comes from the total D3-brane charge induced by the background. The general formula is
\al{ a|_{\mathcal O(N^2)}= c|_{\mathcal O(N^2)} = \frac{M^2 \pi^3}{4 V_5}\,,
}
where $M$ is the D3-brane charge and $V_5$ is the volume of the internal
    five-manifold. In our case $M = N + \varepsilon$ where $\varepsilon =
    \pm(1-k)/2k$ is the charge of the S-fold plane\footnote{Recall that the
    plus sign corresponds to the case without discrete torsion, and the minus
    sign to that with discrete torsion, regardless of the particular choice of
    the discrete torsion.} and $V_5 = \pi^3/k \Delta$.
    The reason for the last identification is that the volume of the
    five-sphere is reduced by a factor of $k$ by the S-fold quotient
    \cite{Aharony:2016kai,Apruzzi:2020pmv} and by a factor of $\Delta$ due to
    the deficit angle of the 7-branes \cite{Aharony:2007dj}.\footnote{$\Delta$
    is both the deficit angle and the dimension of the Coulomb branch
    operator. The values of $\Delta$ are: $\Delta = 6$ for the $E_8$ theory,
    $\Delta = 4$ for the $E_7$ theory, $\Delta = 3$ for the $E_6$ theory,
    $\Delta = 2$ for the $D_4$ theory, $\Delta = 3/2 $ for the $H_2$ theory,
    $\Delta = 4/3$ for the $H_1$ theory and $\Delta = 6/5$ for the $H_0$
    theory.}

\item[-] $\mathcal O(N)$: this term comes from the Chern--Simons terms on the 7-branes. The general formula is
\al{ a|_{\mathcal O(N)} &= \frac{M (\Delta-1)}{2}\,,\\
c|_{\mathcal O(N)} &= \frac{3M (\Delta-1)}{4}\,.
}
As before $M = N + \varepsilon$. Notice that there is no dependence on $k$.
    This is because both the volume wrapped by the 7-branes and the volume of
    the sphere are both affected in the same way by the quotient (the Chern--Simons action is proportional to the ratio of these volumes). Moreover these terms disappear whenever $\Delta = 1$, that is in the case when there are no 7-branes.\footnote{The number of 7-branes is $n_7 = 12 (\Delta-1)/\Delta$.}
%
%
\end{itemize}

While we have, in principle, been determining the terms at quadratic and
linear orders in $N$, we in fact have determined contributions at
$\mathcal{O}(1)$ from the $\varepsilon$ terms in $M$. We will disregard these
terms, as we cannot determine the $\mathcal{O}(1)$ terms anyway, and we are in
fact required to subtract these terms if the central charges are to match those
occurring for the $\mathcal{N} \geq 3$ theories \cite{Aharony:2016kai}.
Adding the quadratic and linear terms together we get
\al{\label{eqn:ac} a &= \frac{k \Delta}{4} N^2 +  \frac{\left(k \Delta \varepsilon + \Delta-1\right)}{2}  N \,,\\
c &=\frac{k \Delta}{4} N^2 +  \frac{\left( 2 k \Delta \varepsilon + 3 \Delta -3\right)}{4}  N \,.
}
Recall that $\varepsilon = \pm(1-k)/2k$. We can use these formulas and can
check that they agree with the known results for rank one 4D SCFTs
\cite{Argyres:2016yzz}, although in these cases we need to subtract a center
of mass hypermultiplet. In addition, we are able to compute $\kappa_G$, the anomaly
associated with $\mathrm{Tr}(\mathcal{R} GG)$, with $G$ the flavor symmetry generated by the
7-branes in the presence of the S-fold. The results for the cases with
discrete torsion are in \cite{Giacomelli:2020jel}, and here we focus on the
cases without discrete torsion. In general, following \cite{Aharony:2007dj}, one
finds that the central charge for the flavour symmetry $G$ on the 7-branes and
the geometric $SU(2)$ flavour symmetry are
\al{ \kappa_G = 2 N \Delta\,, \quad \kappa_{SU(2)} = kN^2 \Delta -N (\Delta-1-2 k \Delta \varepsilon)\,.
}
Let us note that in the special case where $N = 1$, we always find that either
$\kappa_{SU(2)} = 0$, or that
there is an accidental enhancement in the infrared where the $SU(2)$ merges with the 7-brane flavor symmetry.
We tabulate the values that we get for all cases without discrete torsion writing both the rank $N$ and rank one values,
indicating as well the Kodaira fiber type prior to the quotient. As expected, these are the same values displayed in
reference \cite{Argyres:2016yzz} (for the rank $N$ case the results here match
with \cite{Giacomelli:2020jel}, worked out from compactifications of a 6D
SCFT):

\begin{center}
\begin{tabular}[]{ |c |c |c|c|c|}
\hline
 & $24 a$ & $12 c$& $\kappa_G$& $(24 a,12c,\kappa_G)|_{N=1}$\\
 \hline
$ IV^* / \mathbb Z_2 $
&  $36N^2 +6 N$ & $18N^2 +9N$ & $6N$&(42,27,6)\\
\hline
$ I_0^* / \mathbb Z_2 $
& $24N^2$ & $12 N^2+3N$& $4N$& (24,15,4) \\
\hline
$ IV / \mathbb Z_2 $
& $18N^2-3N$ & $9N^2$ & $3N$ & (15,9,3) \\
\hline
$ I_0^* / \mathbb Z_3 $
&$36 N^2-12 N$  & $18 N^2 -3N$&$4N$& (24,15,4) \\
\hline
$ III / \mathbb Z_3 $
& $24N^2 -12 N$ & $12N^2-5N$&$8N/3$&(12,7,8/3) \\
\hline
$ IV / \mathbb Z_4 $
&  $36N^2-21N$ & $18N^2-9N$&$3N$ &(15,9,3)\\
\hline
\end{tabular}
\end{center}
Here we denoted the theories using the fiber type before taking the quotient
and the type of quotient applied. All the values obtained match with
\cite{Argyres:2016yzz}. Note that the formulas for $a$ and $c$ match the
$\mathcal N=3$ case (obtained when $\Delta = 1$) provided that the $\mathcal
O(1)$ term coming from the center of mass of the system of D3-branes is added
back. For completeness, we can also list the same information in the cases
with discrete torsion, again focusing on the rank one case. As expected, these
are the same values displayed in reference \cite{Argyres:2016yzz} (see also
\cite{Apruzzi:2020pmv,Giacomelli:2020jel}). We can determine these values in
the following manner. We use the formulae in (\ref{eqn:ac}) to
determine the leading and subleading contributions in $N$. The
$\mathcal{O}(1)$ terms were determined in \cite{Apruzzi:2020pmv}, where it was
argued that the parent theory should include $k(\Delta - 1)$ additional free
hypermultiplets before the quotient, and we include them here verbatim.

\begin{center}
\begin{tabular}[]{ |c |c|c|c|c|}
\hline
 & $24 a$ & $12 c$& $\kappa_G$ & $(24 a,12c,\kappa_G)|_{N=1}$ \\
 \hline
$IV^* / \widehat{\mathbb Z}_2 $
& $36N^2+42N+4$ & $18 N^2+27 N+4$ & $6N+1$ & (82,49,7) \\
\hline
$ I_0^* / \widehat{\mathbb Z}_2 $
& $24N^2+24N+2$ &$12N^2 + 15N+2$ & $(4N+1,8N)$ &(50,29,(5,8)) \\
\hline
$ IV / \widehat{\mathbb Z}_2 $
&$18N^2+15+1$ & $9N^2+9N+1$ &$3N+1$ &(34,19,(4,-)) \\
\hline
$ I_0^* / \widehat{\mathbb Z}_3 $
& $36N^2+36N+3$ &$18N^2+21N+3$&$12N+2$&(75,42,14) \\
\hline
$ III / \widehat{\mathbb Z}_3 $
& $24N^2+20N+1$ & $12 N^2 + 11N +1$&-&(45,24,-) \\
\hline
$ IV / \widehat{\mathbb Z}_4 $
&  $36N^2 + 33N+2$ & $18N^2+18N +2$&$12N+2$&(71,38,14) \\
\hline
\end{tabular}
\end{center}
In the above, we have included a ``$-$'' in some entries to reflect
the fact that our present methods do not fix the level of the $U(1)$ flavor current.

\section{Conclusions}\label{sec:CONC3}

S-folds are a non-perturbative generalization of O3-planes which figure in the
stringy construction of novel 4D quantum field theories. In this chapter we have
proposed a procedure for how S-fold projection acts on the spectrum of string junctions
attached to a stack of 7-branes and probe D3-branes. We have developed a
general prescription for reading off the resulting flavor symmetry algebra
under S-fold projection. This procedure leads to new realizations of many of
the rank one 4D $\mathcal{N} = 2$ SCFTs which arise from mass deformations
and/or discrete gaugings of the rank one $E_8$ Minahan--Nemeschansky theory.
We have also argued that the Seiberg--Witten curves associated with some of
these theories provide an operational definition of F-theory in the presence
of an S-fold background with discrete torsion.  In the remainder of this
section we discuss some avenues for future investigation.

An interesting feature of our analysis is that there is a close correspondence between possible S-fold quotients
of 7-branes, and admissible rank one 4D $\mathcal{N} = 2$ SCFTs. That being said, there are a few examples
which appear in reference \cite{Argyres:2016yzz} which seem to involve some additional ingredients.
The Kodaira fiber types and flavor symmetries
for these cases are $[II^{\ast} , C_2]$, $[III^{\ast} , C_1]$, $[IV^{\ast}_{1}, \varnothing]$, $[II^{\ast} , C_1]$.
In some cases, we can understand the origin of these theories as arising from a mass deformation of another theory, followed by an additional
discrete quotient. That being said, it remains to be understood whether these operations can be fully realized purely in
geometric terms.

There are in principle other ways to generate the same class of rank one 4D $\mathcal{N} = 2$ SCFTs.
In particular, compactifications of 6D SCFTs with suitable discrete twists provide
an alternative way to realize many such examples (see e.g.
\cite{Giacomelli:2020jel}). Since there is now a classification of possible F-theory backgrounds which
can generate 6D SCFTs (see e.g. \cite{Heckman:2013pva, Heckman:2015bfa} and \cite{Heckman:2018jxk} for a review),
it would be interesting to systematically classify all possible ways of incorporating such
discrete effects, thus providing a complementary viewpoint on many of the same questions.

In this chapter we have mainly focused on structures associated with 4D $\mathcal{N} = 2$ SCFTs.
It would be quite natural to investigate the structure of related systems with only 4D $\mathcal{N} = 1$ supersymmetry.
For example starting from a 4D $\mathcal{N} = 2$ SCFT, deformations by nilpotent mass deformations often trigger
flows to such theories \cite{Heckman:2010qv, Maruyoshi:2016tqk, Apruzzi:2018xkw}.

O3-planes often play an important role in the construction of consistent Type IIB string vacua.
Having analyzed the effect of S-fold projection on the flavor symmetries of probe D3-branes in the vicinity of
7-branes, it is also natural to consider possible ways in which such ingredients might be used in compact F-theory models.

In the first part of this thesis we have explored strong coupling effects of four-dimensional and six-dimensional superconformal field theories. In particular we have used many geometrical tools to extract information about fixed points as well as their overall hierarchy. We've also seen the effect of non-perturbative deformations and studied the resulting RG flows. However, there is more to learn about conformal field theories by exploring dualities between them. In fact, it is good at this point to return to a more fundamental duality of string theory, namely T-duality. In its simplest form, abelian T-duality is simply the observation that a theory with strings propagating on a circle of radius $R$ is equivalent to that of strings propagating on a circle of radius $1/R$, where momentum and winding numbers are simply interchanged. This abelian form of T-duality is well known. However its non-abelian generalization has only recently drawn more interest, and the relevant mathematical tools are only just being unearthed from the mathematical literature. In the second part of this thesis we will explicitly explore Poisson-Lie T-duality and give evidence in favor of it being a full duality between integrable $\sigma$-models. In particular, we will begin by showing that the Poisson-Lie T-duality transformation rules map conformal field theories to conformal field theories, before moving on to its effects on RG flows.

\part{String Theory and Poisson-Lie T-Duality}

\chapter{$\alpha$'-Corrected Poisson-Lie T-Duality}\label{chapter4}

\section{Introduction}
Abelian T-duality is an important cornerstone in the framework of string theory. It is applicable to target space geometries that possess abelian isometries and a natural question is if it is possible to extend T-duality to more general situations. Non-abelian T-duality \cite{delaOssa:1992vci} arose from this idea and is based on the observation that the Buscher procedure \cite{Buscher:1987sk}, which describes abelian T-duality in the closed string $\sigma$-model, can be applied to non-abelian isometries, too. However, there are two major obstacles compared to the abelian case. First, the dual background has a smaller isometry group than the original one. Hence, it seems in general impossible to invert the transformation which is crucial to have a duality. Second, it is problematic to extract global properties of the dual target space. They are for example required to construct an operator mapping on higher genus Riemann surfaces \cite{Alvarez:1993qi}. Poisson-Lie (PL) T-duality arises from an elegant solution to the first problem. It is based on the seminal observation \cite{Klimcik:1995ux} that both $\sigma$-models, which describe either the target space or its dual, originate from the same structure, a Drinfeld double. It governs the Hamiltonian dynamics of the models, and their equivalence is guaranteed by a canonical transformation. Remarkably, non-abelian T-duality constraints the Drinfeld double significantly. But the idea implemented in \cite{Klimcik:1995ux} works as well without this restriction. Hence, PL T-duality provides a more general notion of T-duality whose name originates from the fact that it relates target spaces that are PL groups. Like abelian and non-abelian T-duality are only applicable to target space geometries with isometries, a related notion exists for PL T-duality. It is based on non-commutative conserved currents on the worldsheet \cite{Klimcik:1995ux} which generate PL symmetry. Despite their intriguing mathematical structure and physical properties, research activity in $\sigma$-models with PL symmetric target spaces was moderate for almost two decades. Most arguably because they inherit the problems on global properties that non-abelian T-duality already faces. Just six years ago, when their relation to integrable string worldsheet theories was fully appreciated \cite{Delduc:2013qra}, significant new interest arose. Due to the astonishing success with which integrability was applied in the AdS/CFT correspondence to explore 4D maximally supersymmetric Yang-Mills theory in the large $N$ limit beyond the perturbative regime \cite{Beisert:2010jr}, the demand for new integrable $\sigma$-models is high and a vast new field of applications opens up for PL symmetry and T-duality. 

In this context, a particularly important question is how PL T-duality is affected by quantum corrections. They are controlled in string theory by two parameters: $\alpha'$ and $g_\mathrm{S}$. The former captures the extended nature of the string and the latter its ability to split. Abelian T-duality is a genuine symmetry of string theory and therefore applies to all orders in $\alpha'$ and $g_\mathrm{S}$ \cite{Rocek:1991ps}. For PL T-duality the situation is more subtle. Because of the notorious problem with higher genus worldsheets, there is currently not much to say about the fate of $g_\mathrm{S}$-corrections. However, this does not rule out the possibility of extending the validity of PL T-duality beyond the leading order in $\alpha'$. On the contrary, recently computed $\alpha'$-corrections of integrable deformations point very clearly in this direction \cite{Hoare:2019ark,Hoare:2019mcc,Borsato:2020bqo}. Hence, the objective of this chapter is to construct leading order $\alpha'$-corrections to the PL T-duality transformation rules in a bosonic $\sigma$-model and to argue that they preserve conformal invariance. Key to this endeavour are three techniques: The formulation of PL symmetric target space geometries in the framework of Double Field Theory (DFT) \cite{Hassler:2017yza}, the $\alpha'$-corrected DFT flux formulation introduced by Marqu\'es and Nu$\tilde{\mathrm{n}}$ez \cite{Marques:2015vua}, and finite generalized Green-Schwarz (gGS) transformations recently presented by Borsato, L{\'o}pez, and Wulff \cite{Borsato:2020bqo}.

\section{PL T-Duality and DFT:} Directly at the level of the metric, $B$-field, and dilaton, PL symmetric target spaces might look very complicated. But fortunately, their underlying structure becomes much simpler in the framework of DFT \cite{Siegel:1993th,Hull:2009mi,Hohm:2010pp}, where they are expressed in the language of generalized geometry. More precisely, the metric and the $B$-field can be unified in a generalized frame field \cite{Hohm:2010xe} $E_A{}^I$ on the generalized tangent space. It is governed by the frame algebra \cite{Hassler:2017yza}
\begin{equation}\label{eqn:framealg}
  \mathcal{L}_{E_A} E_B{}^I = F_{AB}{}^C E_C{}^I
\end{equation}
where $\mathcal{L}$ denotes the generalized Lie derivative
\begin{equation}
  \mathcal{L}_{E_A} E_B{}^I = E_A{}^J \partial_J E_B{}^I + \big( \partial^I E_{AJ} - \partial_J E_A{}^I \big) E_B{}^J
\end{equation}
and $F_{AB}{}^C$ are the structure constants of a Lie algebra $\mathfrak{g}$, generating the corresponding Lie group $G$. Uppercase, Latin characters denote doubled indices, running from $1,\dots,2 D$. They come in two different kinds: flat indices ranging from $A$ to $H$ and curved indices starting with $I$. Both are related by the generalized frame field. They are raised and lowered with
\begin{equation}
  \eta^{IJ} = \begin{pmatrix}
    0 & \delta_i{}^j \\
    \delta^i{}_j & 0
  \end{pmatrix}\,, \quad
  \eta^{AB} = \begin{pmatrix}
    \eta^{ab} & 0 \\
    0 & -\eta^{\bar a\bar b}
  \end{pmatrix}\,,
\end{equation}
and their respective inverses, where $\eta^{ab}=\eta^{\bar a\bar b}$ has either Lorentzian or Euclidean signature. Furthermore, we always deal with the canonical solution to the section condition $\partial_I = ( 0 \,\, \partial_i )$.

Frame fields $E_A{}^I$ that satisfy~\eqref{eqn:framealg}, can be constructed systematically on the coset $H \backslash G$, if $H$ is a maximally isotropic subgroup of $G$ \cite{Hassler:2019wvn,Demulder:2018lmj}. Isotropy is defined in terms of an O($D$,$D$) invariant pairing $\langle \cdot \,,\, \cdot \rangle$ on $\mathfrak{g}$. It is equivalent to $\eta_{AB}$, once an appropriate set of $2D$ linearly independent generators $t_A \in \mathfrak{g}$ is chosen. In this case, we identify $\langle t_A, t_B \rangle = \eta_{AB}$ and define a maximally isotropic subgroup $H$ as a subgroup of $G$ which has the maximal number of linearly independent generators that are pairwise annihilated by the pairing. Taking into account the signature of $\eta_{AB}$, it follows that $\dim H = D$. Depending on $G$ and the pairing, different subgroups (labeled $H_1$, $H_2$, \dots) might have this property. This observation is directly related to PL T-duality because each of them results in a generalized frame field describing a different, but still physically equivalent, target space geometry. At this point, the term duality is slightly misleading because it implies that there are at most $H_1$ and $H_2$ which is not true. Thus in general, one might prefer to refer to PL plurality. There are two major ingredients that enter the construction of the generalized frame field. The right-action of $G$ on the coset $H\backslash G$ gives rise to $2 D$ vector fields $k_A{}^i \partial_i$. They furnish the frame algebra
\begin{equation}
  L_{k_A} k_B{}^i = F_{AB}{}^C k_C{}^i
\end{equation}
under the standard Lie derivative $L$. It matches the vector part of \eqref{eqn:framealg} and therefore it is natural to identify $E_A{}^i = k_A{}^i$. To complete the construction, we also need the corresponding one form part \cite{Hassler:2019wvn}
\begin{equation}
  E_{Ai} \dd x^i = \langle t_A, l \rangle - \frac12\langle \iota_{k_A} l, l\rangle - \iota_{k_A} B_\mathrm{WZW}
\end{equation}
with $l=m^{-1} \dd m$, $m \in H \backslash G$. In general, it contains a locally defined $B$-field which captures the WZW-term of the underlying $\sigma$-model
\begin{equation}
  \dd B_\mathrm{WZW} = \frac1{3!} \langle l \overset{\wedge}{,} l \wedge l \rangle \,.
\end{equation}

For latter convenience, we parameterize the result in terms of three quantities: the frame field $e_a{}^i$ whose inverse transpose is denoted by $e^a{}_i$, the $B$-field $B_{ij}$ and a Lorentz transformation $\Lambda_{\bar a}{}^{\bar b}$ with the defining property $\Lambda_{\bar a}{}^{\bar c} \Lambda_{\bar b}{}^{\bar d} \eta_{\bar c\bar d} = \eta_{\bar a\bar b}$,
\begin{equation}\label{eqn:decompframeA}
  E_A{}^I = \frac{1}{\sqrt{2}} \begin{pmatrix}
    \delta_a{}^b & 0 \\
    0 & \Lambda_{\bar a}{}^{\bar b}
  \end{pmatrix}\begin{pmatrix}
    e_{bi} + e_b{}^j B_{ji} &  e_b{}^i \\
    - e_{\bar bi} + e_{\bar b}{}^j B_{ji} & e_{\bar b}{}^i
  \end{pmatrix}\,.
\end{equation}
While $e^a{}_i$=$e^{\bar a}{}_i$, which gives rise to the metric $g_{ij}$=$e^a{}_i e^b{}_j \eta_{ab}$, and $B_{ij}$ shape the target space directly, the role of $\Lambda_{\bar a}{}^{\bar b}$ is more subtle. In a bosonic $\sigma$-model at the classical level, it is irrelevant. Still, it is crucial for \eqref{eqn:framealg} to hold and we will see that it plays a significant role for $\alpha'$-corrections to PL T-duality. Remarkably, the same is true for the R/R sector of type II superstrings where $\Lambda_{\bar a}{}^{\bar b}$ already affects the transformation rules to leading order $\alpha'$ \cite{Hassler:2017yza}. Similar to abelian T-duality, the PL T-duality transformation rules for the metric and $B$-field can be elegantly written in terms of the generalized metric
\begin{equation}\label{eqn:genmetric}
  \mathcal{H}^{IJ} = \begin{pmatrix} g_{ij} - B_{ik} g^{kl} B_{lj} & -B_{ik} g^{kj} \\
    g^{ik} B_{kj} & g^{ij}
    \end{pmatrix}
\end{equation}
as a coordinate dependent O($D$,$D$) transformation \cite{Hassler:2017yza}. We construct the latter by assuming that $G$ has at least two different maximally isotropic subgroups $H$ and $\widetilde H$ because only then PL T-duality is applicable. For both, we construct the generalized frame fields, $E_A{}^I$ and $\widetilde{E}_A{}^I$, to eventually extract
\begin{equation}\label{eqn:defO}
  O_I{}^J = E^A{}_I \widetilde{E}_A{}^J\,.
\end{equation}
It mediates the O($D$,$D$) transformation which relates both PL T-dual backgrounds,
\begin{equation}\label{eqn:trgenmetric0}
  \widetilde{\mathcal{H}}^{IJ} = O_K{}^I O_L{}^J \mathcal{H}^{KL}\,.
\end{equation}

A huge advantage of this approach is that it emphasizes the invariance of $F_{AB}{}^C$ in \eqref{eqn:framealg} under PL T-duality. Furthermore, the flux formulation of DFT \cite{Geissbuhler:2013uka} allows us to rewrite the low-energy effective action
\begin{equation}\label{eqn:Seff}
  S = \int\dd^{D}x\,
    \sqrt{g} e^{-2\phi} \big(R + 4 (\partial \phi)^2 
    - \frac{1}{12} H^2 \big)
\end{equation}
and its field equations exclusively in terms of $F_{ABC}$ and
\begin{equation}
  F_A = 2 E_A{}^I \partial_I d + E^{BI} \partial_I E_B{}^J E_{AJ}
\end{equation}
with the generalized dilaton $d = \phi - \frac12 \log \sqrt{g}$. It is natural to assume that since $F_{ABC}$ is invariant under PL T-duality, $F_A$ should be, too. Imposing this additional constraint fixes the transformation of the generalized dilaton
\begin{equation}\label{eqn:trgend}
  \partial_I \widetilde{d} = O^J{}_I \partial_J d + \frac12 O_I
\end{equation}
with
\begin{equation}
  O_I = \partial_J \left( \widetilde{E}_A{}^J - E_A{}^J \right) E^A{}_I\,.
\end{equation}
Because $O_I{}^J$ and $O_I$ depend simultaneously on the coordinates of $H \backslash G$ and of its dual $\widetilde{H} \backslash G$, one might be worried to end up with target space fields that depend on unphysical coordinates after the transformation. Fortunately, for PL symmetric target spaces this situation is ruled out. It is common lore that it can be very hard to spot this symmetry directly at the level of the target space fields. Hence, it is more common to start from the doubled description in terms of the Lie group $G$ combined with the constants $F_{ABC}$, $F_A$ and then extract both PL T-dual target spaces according to the diagram of figure \ref{fig:alphaLetterFigdiag}

\begin{figure}
\centering
\includegraphics[scale=1]{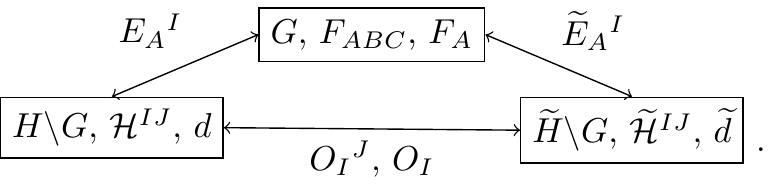}
\caption{By starting from the doubled description (top) in terms of the Lie group $G$ combined with the constants $F_{ABC}$, $F_A$, one can extract both PL T-dual target spaces, thus making the duality between the two theories (bottom) more evident.}%
\label{fig:alphaLetterFigdiag}%
\end{figure}

Particularly interesting are target space geometries whose metric, $B$-field and dilaton solve the field equations of the effective action \eqref{eqn:Seff} because they give rise to conformal field theories (CFTs) on the worldsheet (at least at the one-loop level). Hence, we conclude that because the field equations do not change under PL T-duality, solutions are mapped to solutions and therefore conformal invariance is preserved. At one loop this statement can be further refined. A CFT can be perturbed by a relevant deformation which triggers an RG flow from the UV to either another CFT or a gapped phase in the IR. PL symmetric $\sigma$-models are one-loop renormalizable \cite{Valent:2009nv} and again, their $\beta$-functions can be expressed exclusively in terms of  $F_{ABC}$ and $F_A$ \cite{Sfetsos:2009vt}. Hence, PL T-duality does not only preserve fixed points but rather the complete RG flow.

\section{$\alpha'$-Corrected DFT:} We will now show how this argumentation extends beyond the leading order of $\alpha'$. A major challenge is that beyond one loop, all relevant quantities like the effective action or $\beta$-functions become renormalization scheme dependent. Different schemes are related by field redefinitions. Eventually, this dependence drops out for physical observables but during all intermediate steps, it is essential to keep track of it. Consequentially, there is no universal expression for the four-derivative effective action comparable to ~\eqref{eqn:framealg}, but rather one for every scheme. Popular schemes are the Metsaev-Tseytlin (MT) \cite{Metsaev:1987zx}, Hull-Townsend (HT) \cite{Hull:1987yi} and the generalized Bergshoeff-de Roo (gBR) \cite{Bergshoeff:1989de,Marques:2015vua} scheme. Choosing an appropriate scheme can simplify calculations significantly. In particular, it affects how symmetries of the theory are realized. An example is that while in the MT or HT scheme the action of diffeomorphisms and Lorentz transformations is the same as at one loop, the $B$-field Lorentz transformations in the gBR scheme receive a correction. Intriguingly, this correction is required to facilitate the Green-Schwarz (GS) anomaly cancellation mechanism for the heterotic superstring.

Because~\eqref{eqn:framealg} has proven to be a fundamental identity for all PL symmetric backgrounds, we prefer a scheme where it still holds unchanged. Furthermore, the effective action in this scheme should be exclusively captured by $F_{AB}{}^C$ and $F_A$ like before. Fortunately, a scheme with exactly these properties exists \cite{Marques:2015vua,Baron:2017dvb} and we will refer to it as Marqu\'es-Nu\~{n}ez (MN) scheme. While not affecting generalized diffeomorphisms, which is essential to keeping the construction of generalized frame fields from above applicable, it modifies double Lorentz transformations. At leading order, the latter leave by definition the generalized metric invariant. This is the reason why we could safely ignore $\Lambda_{\bar a}{}^{\bar b}$ in \eqref{eqn:trgenmetric0}. Beyond that order, it has to be included and results in $\alpha'$ corrected transformation rules. More precisely, except for $\eta_{IJ}$, all quantities will receive $\alpha'$-corrections. They are labeled by $\mathcal{H}_{IJ} = \mathcal{H}^{(0)}_{IJ} + \alpha' \mathcal{H}^{(1)}_{IJ} + \mathcal{O}(\alpha'^2)$. Finite double Lorentz transformations, also called gGS transformations, are denoted by $\mathcal{H}_{IJ} \rightarrow \mathcal{H}_{IJ} + \Delta_\Lambda \mathcal{H}_{IJ}$ where 
\begin{equation}
  \Lambda_A{}^B = \begin{pmatrix}
    \Lambda_a{}^b & 0 \\
    0 & \Lambda_{\bar a}{}^{\bar b}
  \end{pmatrix}
\end{equation}
is the parameter of the transformation. For our purpose, it is sufficient to restrict it to the form of the first matrix in the generalized frame field~\eqref{eqn:decompframeA} and thus set $\Lambda_a{}^b = \delta_a{}^b$. A major challenge is that \cite{Marques:2015vua} does not present finite gGS transformations $\Delta_\Lambda$, but only the infinitesimal version $\delta_\lambda$, with $\Lambda = \exp(\lambda)$. While it should be possible to formally integrate $\delta_\lambda$, we find it more convenient to make an educated guess of how a finite counterpart might look like and then show that it is compatible with the infinitesimal transformations in \cite{Marques:2015vua}. A similar approach allowed \cite{Borsato:2020bqo} to present finite gGS transformations for the metric and $B$-field. However, at this level, the elegant structure of PL T-duality is not manifest. Hence, we prefer to discuss doubled quantities, like the generalized metric or dilaton. Remarkably, their transformation cannot be written exclusively in terms of $\eta_{IJ}$, $\mathcal{H}_{IJ}$, $E_A{}^J$ and $F_{ABC}$. It additionally depends on an involution $K_I{}^J$, with the leading contribution
\begin{equation}
  K^{(0)}_I{}^J = \begin{pmatrix}
    - \delta^i{}_j & 0 \\
    2 B^{(0)}_{ij} & \delta_i{}^j
  \end{pmatrix}\,,
\end{equation}
which equips the target space with an almost Born structure \cite{Freidel:2013zga} (we do not require its integrability).

\section{Finite gGS Transformations and PL T-Duality:} It is this structure which eventually facilitates to write down a proposal for the finite gGS transformation of the generalized metric
\begin{equation}\label{eqn:doubleLgenmetric}
  \Delta_\Lambda^{(1)} \mathcal{H}_{IJ} = K^{(0)}_{(I|}{}^K \left( \mathcal{H}^{(0)}_{KL} \Delta^{(1)}_\Lambda K_{|J)}{}^L + 2 \Delta^{(0)}_\Lambda \mathcal{G}_{|J)K} \right)
\end{equation}
with
\begin{equation}
  \mathcal{G}^{(0)}_{IJ} = - \frac12 F^{(0)}_{\pb I\p A}{}^{\p B} F^{(0)}_{\pb J\p B}{}^{\p A}\,.
\end{equation}
Here, we adopt the notation of \cite{Marques:2015vua} to indicate indices that are projected by either $P_{IJ}= \frac12 ( \eta_{IJ} - \mathcal{H}_{IJ} )$ as $V_{\p I} = P_I{}^J V_J$ or $\overline{P}_{IJ} = \frac12 ( \eta_{IJ} + \mathcal{H}_{IJ} )$ as $V_{\pb I} = \overline{P}_I{}^J V_J$. Taking into account that the generalized frame field transforms to leading order as $E_A{}^I \rightarrow \Lambda_A{}^B E_B{}^I$, it is straightforward to obtain
\begin{equation}\label{eqn:trafoG}
  \Delta^{(0)}_\Lambda \mathcal{G}_{IJ} = F^{(0)}_{(\pb I|\p A}{}^{\p B} \Theta_{|\pb J)\p B}{}^{\p A} - \frac12 \Theta_{\pb I\p A}{}^{\p B} \Theta_{\pb J\p B}{}^{\p A}
\end{equation}
where $\Theta_{IA}{}^B$ captures the left-invariant Maurer-Cartan form (the invariant left-action is $\Lambda_A{}^B \rightarrow \Lambda'_A{}^C \Lambda_C{}^B$ where $\Lambda'_A{}^B$ is constant)
\begin{equation}
  \Theta_{IA}{}^B = \partial_I \Lambda^C{}_A \Lambda_C{}^B
\end{equation}
with the corresponding Maurer-Cartan equation $2 \partial_{[I} \Theta_{J]A}{}^B = [ \Theta_I, \Theta_J ]_A{}^B$. Note that the proposed transformation \eqref{eqn:doubleLgenmetric} guarantees that the algebraic relations of the Born structure are preserved at order $\alpha'$.  To explicitly evaluate it, we additionally impose
\begin{equation}\label{eqn:doubleLK}
  \Delta^{(1)}_\Lambda K_{IJ} = \Delta^{(0)}_\Lambda \mathcal{B}_{IJ} - K^{(0)}_I{}^K \Delta^{(0)} \mathcal{B}_{KL} K^{(0)}_J{}^L
\end{equation}
with
\begin{equation}\label{eqn:trafoB}
  \Delta^{(0)}_\Lambda \mathcal{B}_{IJ} = F^{(0)}_{[\pb I|\p A}{}^{\p B} \Theta_{|\pb J]\p B}{}^{\p A} + \mathcal{B}_{\pb I\pb J}^{\mathrm{WZW}}
\end{equation}
and
\begin{equation}
  3 \partial_{[I}\mathcal{B}_{JK]}^{\mathrm{WZW}} = \Theta_{[I|A}{}^B \Theta_{|J|B}{}^C \Theta_{|K]C}{}^A \,.
\end{equation}

Eventually, we have to show that our proposal for finite gGS transformations is compatible with the known infinitesimal results mediated by $\delta_\lambda$ with the antisymmetric parameter $\lambda_{AB}$. In order to extract the latter from the former, the finite transformations are perturbed by the right action $\Lambda\rightarrow  \Lambda + \Lambda\lambda$, which only affects
\begin{equation}
  \tilde{\delta}_\lambda \Theta_{IA}{}^B = - \partial_I \lambda_{A}{}^{B} - \lambda_A{}^C \Theta_{IC}{}^B - \Theta_{IA}{}^C \lambda^B{}_C\,.
\end{equation}
The generalized frame field and the projected structure coefficients $F_{\pb I\p A}{}^{\p B}$ are invariant under this transformation. They are rather governed by $\delta^{(0)}_\lambda E_A{}^I = \lambda_A{}^B E_B{}^I$ which implies, due to the frame algebra~\eqref{eqn:framealg},
\begin{equation}
  \delta^{(0)}_\lambda F_{\pb I\p A}{}^{\p B} = \partial_{\pb I} \lambda_{\p A}{}^{\p B} + \lambda_{\p A}{}^{\p C} F^{(0)}_{\pb I \p C}{}^{\p B} + F^{(0)}_{\pb I\p A}{}^{\p C} \lambda^{\p B}{}_{\p C}\,.
\end{equation}
It is important to keep in mind that this transformation does not affect $\Theta_{IA}{}^B$. Hence $\tilde{\delta}_\lambda$ should be understood as an auxiliary transformation whose main purpose is to write the Taylor expansion of $\Delta_\Lambda$ around the identity transformation in the compact form
\begin{equation}
  \Delta_\Lambda = \sum_{n=1}^\infty \frac1{n!}(\left. \tilde{\delta}_\lambda)^n \Delta_\Lambda \right|_{\Theta = 0} = \sum_{n=1}^\infty \frac1{n!}(\delta_\lambda)^n\,.
\end{equation}
By taking into account the definition of a finite transformation as the exponential map of its infinitesimal version, we are able to read off $\delta_\lambda$ directly from the leading contribution of this expansion. Additionally, one has to verify that all subleading contributions match as well. Otherwise, the proposal for $\Delta_\Lambda$ would be inconsistent and should be discarded. Fortunately, both \eqref{eqn:trafoG} and \eqref{eqn:trafoB} satisfy the relation
\begin{equation}
  \tilde \delta_\lambda \Delta^{(0)}_\Lambda - \delta^{(0)}_\lambda \Delta^{(0)}_\Lambda = \delta^{(0)}_\lambda
\end{equation}
that implies $(\tilde{\delta}_\lambda)^n \Delta^{(0)}_\Lambda |_{\Theta=0} = (\delta^{(0)}_\lambda)^n$ and therefore guarantees the correctness of the proposed transformations.

To make contact with the known expressions for $\delta_\lambda$ in the literature, we first calculate
\begin{equation}\label{eqn:deltaRBWZW}
  \tilde\delta_\lambda \mathcal{B}^{\mathrm{WZW}}_{IJ} = \partial_{[I|} \lambda_A{}^B \Theta_{|J]B}{}^A + \partial_{[I} \xi_{J]}
\end{equation}
which is only defined up to a shift by a closed two-form. By the Poincar\'e lemma, this two-form is in local patches exact where it can be parameterized by $\xi_I = ( 0 \,\, \xi_i )$. Equation~\eqref{eqn:deltaRBWZW} implies
\begin{equation}
  \delta^{(0)}_\lambda \mathcal{B}_{IJ} = \partial_{[\pb I|} \lambda_{\p A}{}^{\p B} F^{(0)}_{|\pb J]\p B}{}^{\p A} + (\partial\xi)_{[\pb I\pb J]}
\end{equation}
and ultimately $\delta^{(1)}_\lambda K_{IJ}$. Note that in the last term projections are applied after taking the derivative. This is important because both operations do not commute. Contact with \cite{Marques:2015vua} is made through the infinitesimal transformation of the generalized frame field
\begin{equation}
  \delta^{(1)}_\lambda E^A{}_I E^{(0)}_{AJ} = \partial_{[\p I|} \lambda_{\p A}{}^{\p B} F^{(0)}_{|\pb J]\p B}{}^{\p A} + (\partial \xi)_{[\p I \pb J]}
\end{equation}
where we extended (3.24) by a compensation $B$-field transformation. The Born structure gives rise to $K_I{}^J P_J{}^K = \overline{P}_I{}^J K_J{}^K$, $K_I{}^J \partial_J = \partial_I$ and eventually allows us to establish
\begin{equation}
  \begin{aligned}
    \delta^{(1)}_\lambda K_{IJ} &= 2 \delta^{(1)}_\lambda E^A{}_{[I|} E^{(0)}_A{}^K K^{(0)}_{K|J]} \\
    &= \delta^{(0)}_\lambda \mathcal{B}_{IJ} - K^{(0)}_I{}^K \delta^{(0)}_\lambda \mathcal{B}_{KL} K^{(0)}_J{}^L\,.
  \end{aligned}
\end{equation}
Hence, the finite transformation~\eqref{eqn:doubleLK} is indeed compatible with the known infinitesimal version. The same applies to~\eqref{eqn:doubleLgenmetric}, which we rewrite as
\begin{equation}
  \begin{aligned}
    \Delta^{(1)}_\Lambda \mathcal{H}_{IJ} = 2 K^{(0)}_{(I|}{}^K \Bigl(& \Delta^{(0)}_\Lambda \mathcal{B}_{K|J)} + \Delta^{(0)}_\Lambda \mathcal{G}_{K|J)} \Bigr)
  \end{aligned}
\end{equation}
to find
\begin{equation}
  \delta^{(1)}_\Lambda \mathcal{H}_{IJ} = 2 \partial_{(\p I|} \lambda_{\p A}{}^{\p B} F^{(0)}_{|\pb J)\p B}{}^{\p A} + 2 (\partial \xi)_{(\p I\pb J)}\,.
\end{equation}
We obtain a match with (3.27) of \cite{Marques:2015vua} and conclude our discussion of finite gGS transformations.

\section{Conclusions}

The $\alpha'$-corrected PL T-duality transformation rules in the MN scheme arise after~\eqref{eqn:trgenmetric0} is adapted to take into account the non-trivial action of double Lorentz transformations on the generalized metric beyond the leading order in $\alpha'$, namely
\begin{equation}\label{eqn:trgenmetric1}
  \widetilde{\mathcal{H}}^{IJ} = O_K{}^I O_L{}^J \left( \mathcal{H}^{KL} + \Delta^{(1)}_{\widetilde{\Lambda}\Lambda^{-1}} \mathcal{H}^{KL} \right)\,.
\end{equation}
Like $O_I{}^J$ in \eqref{eqn:defO}, the transformation parameter of the gGS transformation, $(\widetilde{\Lambda} \Lambda^{-1})_A{}^B = \widetilde{\Lambda}_A{}^C \Lambda^B{}_C$, is directly extracted from the corresponding generalized frame fields. The generalized dilaton is invariant under this transformation and thus \eqref{eqn:trgend} still applies. This however does not imply that the dilaton $\phi$ is resistant to $\alpha'$-corrections. It depends on both, the generalized dilaton and the determinant of the target space metric, and the latter receives corrections. For completeness, let us note that the four derivative effective action in the MN scheme is given in (3.38) of \cite{Marques:2015vua}. The field redefinitions which are required to go to the gBR and the MT scheme can also be found in this paper (in equations (3.67) and (B.7), respectively). In the presented DFT formulation, it is manifest that~\eqref{eqn:trgenmetric1} will not change the action nor the corresponding field equations, since both can be exclusively written in terms of the structure coefficients $F_{AB}{}^C$ \cite{Baron:2017dvb}. Hence, it is guaranteed that two-loop conformal invariance of a PL symmetric $\sigma$-model is preserved. An important but more subtle question is if this result can be extended to RG flows between CFTs, like it is possible at one loop. Here a significant challenge is that the relation between $\beta$-functions and field equations of the effective action becomes more and more complicated with increasing loop order. However, recently presented $\alpha'$-corrected RG flows for integrable and PL symmetric $\eta$- and $\lambda$-deformations \cite{Hoare:2019ark,Hoare:2019mcc} suggest that it is possible to overcome this problem in the future.

Another aspect that deserved further investigation is the, at least for us initially surprising, connection to Born geometry. In contrast to Riemannian geometry where the Levi-Civita connection is unique, DFT does not possess a completely determined, torsion-free covariant derivative which is compatible with both, $\eta_{IJ}$ and the generalized metric. Consequentially, the generalized Riemann tensor contains undetermined contributions \cite{Hohm:2011si}. They drop out in all physically relevant quantities at the two derivative level, like the generalized Ricci scalar and tensor. However, it is not possible to construct the Riemann tensor squared term that captures $\alpha'$-corrections of the effective action directly from the generalized Riemann tensor. Born geometry already was argued to help to obtain a unique connection by additionally requiring compatibility with $K$ \cite{Freidel:2018tkj}. Considering our results, one might hope that it also gives valuable insights into the generalized geometry of $\alpha'$-corrections.

Following up this short chapter, the next natural question to ask is whether or not this duality holds for RG flows between CFTs. The next chapter addresses this question explicitly by looking at the two-loop $\beta$-functions of integrable $\sigma$-models, and how they transform under Poisson-Lie T-duality.

\chapter{$O(D,D)$-Covariant Two-Loop $\beta$-Functions and Poisson-Lie T-Duality}\label{chapter5}

\section{Introduction}
Two seemingly completely different theories, for example, one strongly coupled and the other one weakly coupled, may still exhibit the same physics. This remarkable phenomenon is governed by dualities and even if it is not generic, it can provide deep insights into the theories involved. A genuine duality is not restricted to the classical level but still applies after quantization. Unfortunately, the dualities that are understood best, only apply to a very limited class of theories. A prominent example is abelian T-duality in string theory. It is restricted to target spaces with abelian isometries which are of course by no means generic. Yet, it provides many crucial insights into string theory. Therefore, it is unarguably an important challenge to advance our knowledge about dualities and their properties. In this process, one encounters the problem that the notion of duality outlined above is very strong. But often only certain properties of a theory are relevant to solve a problem. In this case, it is sufficient to ask: Is it possible to find two different theories that share at least these properties? This approach has the considerable advantage that it is much less constraining. A remarkable example along these lines is Poisson-Lie (PL) T-duality \cite{Klimcik:1995ux}. 

In fact the term PL T-duality is slightly ambiguous because it is sometimes used as a synonym for a whole family of different dualities. All started with non-abelian T-duality \cite{delaOssa:1992vci}. It is based on the observation that the Buscher procedure \cite{Buscher:1987sk}, which mediates abelian T-duality on the closed string $\sigma$-model, can be extended to non-abelian isometries. There are however two major problems one encounters in this generalization \cite{Giveon:1993ai,Alvarez:1993qi}:
\begin{enumerate}[leftmargin=1.5em,itemindent=1em,itemsep=0cm,label*=\arabic*)]
\item\label{item:prob1} The Buscher procedure employs a Lagrange multiplier that enforces a flat connection on the worldsheet. However, the connection might still have non-trivial monodromies around non-contractible cycles on the worldsheet. In the abelian case, this problem is resolved by using a periodic Lagrange multiplier \cite{Rocek:1991ps}. Physically this choice leads to the celebrated momentum winding exchange under abelian T-duality and allows for the identification of the topology of the dual target space. Unfortunately, this idea does not work for non-abelian isometries. Therefore, the global properties of non-abelian T-duality are not fully understood and a topic of active research. 
  \item\label{item:prob2} A second problem is that the resulting, dual target space geometry has in general a smaller isometry group which seemingly prohibits the duality to be inverted. This is particularly severe because by definition a duality has to be invertible.
\end{enumerate}
PL T-duality solves problem~\ref{item:prob2} by the seminal observation \cite{Klimcik:1995ux} that both, the original and the dual, $\sigma$-models originate from the same underlying structure, a Drinfeld double. Drinfeld doubles are in one-to-one correspondence with PL groups, which actually form the corresponding target spaces and give the duality its name. Remarkably, non-abelian T-duality is based on a further refined class of Drinfeld doubles with an abelian, maximally isotropic subgroup. Therefore, PL T-duality, which works for arbitrary Drinfeld doubles, not just shows that non-abelian T-duality is invertible but additionally gives rise to a broader family of dualities that do not need isometries at all. Intriguingly, this already rich notion of duality can be even pushed beyond Drinfeld doubles by relaxing the Poisson structure of the PL group to a quasi-Poisson structure \cite{Klimcik:2001vg,Severa:2001qm}. Physically, this leads to a Wess-Zumino-Witten (WZW) term and describes $H$-flux in a non-trivial cohomology class. Eventually, the duality was extended from groups to cosets by the dressing coset construction \cite{Klimcik:1996np}. Thus, the term ``PL T-duality'' may refer to any member of the family
\begin{center}
  dressing cosets $\supset$ PL with WZW term $\supset$ PL $\supset$ non-abelian T-duality\,.
\end{center}
In this chapter, we use it for all of them except for dressing cosets, which we hope to address in the future based on \cite{Demulder:2019vvh}.

Problem \ref{item:prob1} is still an issue since it prohibits discussions of PL T-duality on higher genus Riemann surfaces that appear in the $g_\mathrm{s}$ expansion of the string path integral. Moreover, beyond non-abelian T-duality we do not know a gauging procedure comparable to Buscher's original approach which could be used to check if the path integrals of dual theories match. Hence, PL T-duality is deemed to not be a genuine symmetry of string theory but at most a map between different conformal field theories (CFTs). However, quantum corrections to the classical string are not exclusively controlled by $g_\mathrm{s}$. Additionally, the $\alpha'$-expansion incorporates quantum effects for fixed worldsheet topologies. Fortunately, $\alpha'$-corrections are accessible even without solving problem \ref{item:prob1} and at the leading, one-loop order in this expansion, it is known that \cite{\refonelooprenorm}
\begin{enumerate}[leftmargin=1.5em,itemindent=1em,itemsep=0cm,label*=\arabic*)]
\item PL symmetric\footnote{PL symmetric refers to the properties a target space geometry must have to permit PL T-duality. We give an exact definition in section~\ref{sec:genframes}.} $\sigma$-models are renormalizable.
  \item The RG flows of two PL T-dual $\sigma$-models are identical because they share the same $\beta$-functions.
\end{enumerate}
These two points are important hints that PL T-duality is not just a classical phenomenon but captures quantum effects as well. An immediate question is if they continue to hold at higher loop orders. We will answer it in the affirmative at two loops in this chapter by explicitly computing the one- and two-loop $\beta$-functions of the bosonic string. For string theory, most relevant are points in the moduli space where these functions vanish, and CFTs at fixed points of the RG flow emerge. In this case, it is instructive to expand the $\beta$-functions in the couplings $\lambda^a$. As we discuss in much more detail below, the resulting expansion is scheme dependent. However, there exists a particular scheme in which it reads \cite{Gaberdiel:2008fn,Codello:2017hhh}
\begin{equation}
  \beta^a = \mu \frac{\dd \lambda^a}{\dd \log\mu} = - ( 2 - \Delta_a ) \lambda^a + \sum\limits_{b,c} C^a{}_{bc} \lambda^b \lambda^c
  + \dots\,,
\end{equation}
where $\Delta_a$ and $C^c{}_{ab}$ denote the anomalous dimensions and coupling constants which appear in the OPE
\begin{equation}
  \left\langle \mathcal{O}_a(x) \mathcal{O}_b(y) \dots \right\rangle = 
    \sum_c \frac1{|x-y|^{\Delta_a + \Delta_b - \Delta_c}} C^c{}_{ab} \left\langle \mathcal{O}_c(x) \dots \right\rangle
\end{equation}
of the classically marginal operators $\mathcal{O}_a$ that correspond to the couplings $\lambda^a$. There are other primary fields in the CFT, too. Hence, the $\beta$-functions do not capture the CFT data completely. Still, as PL T-duality does not affect $\beta$-functions (at least up to two loops), the two CFTs it connects are clearly not unrelated and share at least a common subsector formed by the operators $\mathcal{O}_a$.

Hence, we conclude: A quantum version of PL T-duality is not out of reach and definitely worth studying. Especially, since this duality is tightly linked to integrable deformations of two dimensional $\sigma$-models\footnote{Recently, \cite{Lacroix:2020flf} constructed $\mathcal{E}$-models \cite{Klimcik:1995dy} for a large class of integrable $\sigma$-models and thereby makes their PL symmetry manifest.}. Prominent examples include Yang-Baxter deformations \cite{Klimcik:2002zj}, which are either governed by the homogeneous or inhomogeneous, classical Yang-Baxter equation, and $\lambda$-deformations \cite{Sfetsos:2013wia}. While all of them were discovered independently, they are actually linked by a web of PL T-dualities (and analytic continuations) \cite{\reflambdaetaPL}. Because the S-matrix of integrable models is strongly constrained it only depends on a small number of free parameters. Ultimately, these parameters originate from couplings in the underlying $\sigma$-model. Of course, this relation is extremely complicated but it suggests that if integrability is not broken by quantum effects, only these couplings are affected by RG flows. Motivated by this observation, it was possible to show that $\eta$- and $\lambda$-deformation are indeed two-loop renormalizable \cite{Hoare:2019ark,Hoare:2019mcc,Georgiou:2019nbz}. This is an important clue that PL symmetric $\sigma$-models, might be renormalizable beyond one-loop. Moreover, insights from double field theory (DFT) \cite{Siegel:1993th,Siegel:1993xq,Hull:2009mi} were used to show that PL T-duality with adapted transformation rules maps CFTs to CFTs \cite{\refPLalphaprime}. Motivated by these findings we will use DFT techniques to compute $\beta$-functions for PL $\sigma$-models and show that they are renormalizable. Because the framework we are using is independent of the chosen duality frame, our results automatically imply that all $\beta$-functions are preserved under PL T-duality.

Because the computations which we present are technically challenging, we split their presentation into two parts. In section~\ref{sec:results}, we summarize our results and demonstrate them for the $\lambda$- and $\eta$-deformation. All required tools are reviewed, but no derivations are given. For readers who are mainly interested in computing the $\beta$-functions of particular PL $\sigma$-models, for example integrable deformations, reading this section should be sufficient. Detailed derivations are discussed in section~\ref{sec:derivation}. In particular, we exploit that the $\beta$-functions we are dealing with are governed by a gradient flow \cite{Metsaev:1987zx,Hull:1987yi,Friedan:2009ik}. We show how this flow arises in the conventional $\sigma$-model and then rewrite all its constituents in an O($D$,$D$)-covariant form. After capturing the target space geometry of a PL $\sigma$-model in terms of a generalized frame field and the corresponding generalized fluxes \cite{\reflongDFTPLlist}, this manifestly covariant form permits us to directly read off the results presented in section~\ref{sec:results}. However, the O($D$,$D$)-covariant $\beta$-functions, which we derive, are completely general and hold for arbitrary target space geometries. Section~\ref{sec:conclusion} concludes the chapter with several still open questions and ideas for future research.

\section{One and Two-Loop \texorpdfstring{$\beta$}{beta}-Functions}\label{sec:results}
In the following, we present a summary of the main result of this chapter, the two-loop $\beta$-functions for a bosonic, PL symmetric $\sigma$-model of the form \cite{Fradkin:1984pq}
\begin{equation}\label{eqn:sigmamodel}
  S = \frac1{4\pi\alpha'} \int_\Sigma \dd z \dd \overline{z} ( \sqrt{h} h^{ab} g_{ij} \partial_a X^i \partial_b X^j + \ci \epsilon^{ab} B_{ij} \partial_a X^i \partial_b X^j + \alpha' \sqrt{h} R^{(2)} \phi )\,.
\end{equation}
The couplings of this model are the target space metric $g_{ij}$, $B$-field $B_{ij}$, and dilaton $\phi$. As we explain in section~\ref{sec:genframes}, PL symmetry constraints them significantly. After imposing it, only a finite number of couplings survive. We discuss their $\beta$-functions first at one-loop and eventually at two loops in sections~\ref{sec:oneloop} and \ref{sec:twoloops}, respectively. Along the way we introduce all required DFT techniques and apply them to the $\lambda$- and $\eta$-deformation on a Lie group $G$ \cite{Sfetsos:2013wia} to have an explicit example. Poisson-Lie T-duality is completely manifest in our framework and preserves the $\beta$-functions. This will allow us to deduce the RG flow of the $\eta$-deformation \cite{Klimcik:2002zj} directly from the results of the $\lambda$-deformation since both are related by PL T-duality and analytic continuation \cite{\reflambdaetaPL}.

\subsection{PL Symmetry and Generalized Frame Fields}\label{sec:genframes}
A very powerful way to describe PL symmetric target space geometries is in terms of a generalized frame field $E_A{}^I$ on the generalized tangent bundle $T M \oplus T^* M$ of the target space manifold $M$. Each element of this bundle has a vector and a one-form component. A generalized frame $E_A = E_A{}^i \partial_i + E_{A i} d x^i$ consists of $A=1, \dots, 2D$  such elements, where $D$ denotes the dimension of the target space. They are linearly independent and defined on every point $M$. We distinguish two different sets of indices: $A$ to $H$ are called flat and from $I$ on they are called curved. While the latter are naturally associated to the generalized tangent space, the former are valued in a doubled Lie algebra $\mfd$ with generators $T_A$ and the commutator relations
\begin{equation}
  [T_A, T_B] = F_{AB}{}^C t_C\,.
\end{equation}
Additionally, $\mfd$ is equipped with a ($D$,$D$)-signature pairing 
\begin{equation}
  \langle T_A, T_B \rangle = \eta_{AB}\,,
\end{equation}
which is invariant under the adjoint action of $\mfd$. As a direct consequence $F_{ABC} = F_{AB}{}^D \eta_{DC}$ is totally anti-symmetric. We follow the standard convention in DFT and lower/raise indices with $\eta_{AB}$/its inverse $\eta^{AB}$. Without loss of generality, they can always be brought into the form
\begin{equation}
  \eta_{AB} = \begin{pmatrix} \eta_{ab} & 0 \\
    0 & - \eta_{\bar a\bar b} \end{pmatrix} \qquad 
  \text{and} \qquad
  \eta^{AB} = \begin{pmatrix} \eta^{ab} & 0 \\
    0 & - \eta^{\bar a\bar b} \end{pmatrix}\,,
\end{equation}
where lowercase indices run only from $1$ to $D$ and $\eta_{ab}=\eta_{\bar a\bar b}=\eta^{ab}=\eta^{\bar a\bar b}$ is the invariant metric of the target space's Lorentz group. The generalized frame field translates between the structure on $\mfd$ and the generalized tangent space. More specifically, it relates $\eta^{AB}$ to the canonical pairing
\begin{equation}
  \eta^{IJ} = E_A{}^I E_B{}^J \eta^{AB} = \begin{pmatrix} 0 & \delta_i^j \\ \delta^i_j & 0 \end{pmatrix}
\end{equation}
on $T M \oplus T^*M$.

In this framework, PL symmetry is encoded by the partial differential equation \cite{Hassler:2017yza}
\begin{equation}\label{eqn:framealgebra}
  \mathcal{L}_{E_A} E_B{}^I = F_{AB}{}^C E_C{}^I\,,
\end{equation}
where $\mathcal{L}$ denotes the generalized Lie derivative
\begin{equation}
  \mathcal{L}_{E_A} E_B{}^I = E_A{}^J \partial_J E_B{}^I + \big( \partial^I E_{AJ} - \partial_J E_A{}^I \big) E_B{}^J
  \,.
\end{equation}
As its name suggests, it serves the same purpose as the Lie derivative in conventional geometry. But due to the structure of the generalized tangent space, it not only captures diffeomorphisms on $M$ but also $B$-field transformations. There is a slight subtlety concerning the partial derivatives $\partial_I$ in this expression. In DFT, they in general incorporate not only the $D$ coordinates of the target space but also $D$ additional coordinates on an auxiliary space. But in this setup, the generalized Lie derivative does not close into an algebra automatically. It only does if additional constraints are satisfied. The most restrictive one is the section condition, or strong constraint. It requires that arbitrary combinations of fields, denoted by $\cdot$, are annihilated by $\partial_I \cdot \partial^I \cdot = 0$. A trivial solution to this constraint is given by $\partial_I = \begin{pmatrix} 0 & \partial_i \end{pmatrix}$. It renders DFT equivalent to generalized geometry and we will use it for the rest of the chapter. It is interesting to note that in the framework of generalized geometry, PL symmetric backgrounds mimic the structure of group manifolds in conventional geometry. More precisely, $E_A{}^I$ corresponds to $D$ vector fields that are dual to the left-invariant Maurer-Cartan form while the generalized Lie derivative is replaced by the standard Lie derivative.

Each generalized frame field, even when it is not a solution of \eqref{eqn:framealgebra}, can be brought into the form
\begin{equation}\label{eqn:decompframe}
  E_A{}^I = \frac{1}{\sqrt{2}} \begin{pmatrix}
    \delta_a{}^b & 0 \\
    0 & \Lambda_{\bar a}{}^{\bar b}
  \end{pmatrix}\begin{pmatrix}
    e_{bi} + e_b{}^j B_{ji} &  e_b{}^i \\
    - e_{\bar bi} + e_{\bar b}{}^j B_{ji} & e_{\bar b}{}^i
  \end{pmatrix} := \Lambda_A{}^B \Eh_B{}^I \,,
\end{equation}
where $B_{ij}$ is the $B$-field on the target space and $e_a{}^i = e_{\bar a}{}^i$ denotes a conventional frame field. The latter encodes the metric $g^{ij} = e_a{}^i \eta^{ab} e_b{}^j$ and a Lorentz frame. Note that we use the standard convention that lowercase, curved indices, like $i$, are lowered and raised by this metric and its inverse. Additionally, the generalized frame field incorporates a double Lorentz transformation $\Lambda_{\bar a}{}^{\bar b}$ with the defining property $\Lambda_{\bar a}{}^{\bar c} \Lambda_{\bar b}{}^{\bar d} \eta_{\bar c\bar d} = \eta_{\bar a\bar b}$. At a first glance, it seems irrelevant because it does not affect the target space geometry encoded by the metric and the $B$-field. However, except for a few special cases, it is crucial to solving the constraint \eqref{eqn:framealgebra} for PL symmetry. Moreover, we will see later that it plays a central role beyond one-loop. If the doubled Lie group $\DD$ associated to $\mfd$ has a maximally isotropic subgroup $H$, it is always possible to explicitly construct $E_A{}^I$ on the coset $H \backslash \DD$ \cite{\reflongDFTPLlist}. This construction has become standard and we will not repeat it here. Frequently, the explicit target space geometry is convoluted and while it can always be constructed, it is more elegant to extract as much information as possible directly from the doubled formalism. We will do exactly this for the one- and two-loop $\beta$-functions in the next subsections. A considerable advantage of this approach is that PL T-duality only affects the generalized frame field but not the structure coefficients $F_{ABC}$ and the pairing $\eta_{AB}$. Hence all quantities which can be exclusively written in terms of the latter are manifestly invariant under PL T-duality. Different dual target space geometries arise if $\DD$ has different maximally isotropic subgroups $H_i$. For each of them a different frame field on a different target space $M_i = H_i \backslash \DD$ can be constructed. 

The dilaton $\phi$ is encoded in the generalized dilaton
\begin{equation}\label{eqn:defgend}
  d = \phi - \frac14 \log \det g\,.
\end{equation}
Its condition for PL symmetry can be written in full analogy with the generalized frame field as
\begin{equation}\label{eqn:defFA}
  \mathcal{L}_{E_A} e^{-2 d} = - F_A e^{-2 d}\,, \qquad F_A = \text{const.}\,,
\end{equation}
where $e^{-2 d}$ transforms as a weight $+1$ density under the generalized Lie derivative, namely
\begin{equation}
  \mathcal{L}_{E_A} e^{-2 d} = E_A{}^I \partial_I e^{-2d} + e^{-2d} \partial_I E_A{}^I\,.
\end{equation}
$F_A$ is in one-to-one correspondence with the Lie algebra element $F^A T_A = F \in \mfd$. This element has to be in the center of $\mfd$, meaning that it is constrained by $[ T_A , F ] = 0$ for all generators $T_A$. Moreover, it has to be isotropic and therefore satisfy $\langle F, F \rangle = 0$. We find these two conditions directly from the closure of the generalized Lie derivatives \cite{Geissbuhler:2013uka}. 

\subsubsection{\texorpdfstring{$\lambda$}{Lambda}- and \texorpdfstring{$\eta$}{Eta}-Deformation}\label{sec:genframelambda}
The $\lambda$-deformation on a semisimple group manifold $G$ \cite{Sfetsos:2013wia} is a good example to demonstrate this structure explicitly. It is governed by the doubled group $\DD = G \times G$ with the maximally isotropic subgroup $H=G_{\mathrm{diag}}$ \cite{Klimcik:2015gba} that is used to construct the generalized frame field $E_A{}^I$. The frame field $e_a{}^i$ in \eqref{eqn:decompframe} is written in terms of the inverse transpose of the left- and right invariant Maurer-Cartan forms
\begin{equation}\label{eqn:leftrightinv}
  t_a l^a{}_i d x^i = \sqrt{\frac{k}{2}} g^{-1} \dd g\,, \qquad
  t_a r^a{}_i d x^i = \sqrt{\frac{k}{2}} \dd g g^{-1}\,, \qquad
  [t_a, t_b] = f_{ab}{}^c t_c\,,
\end{equation}
($l_a{}^i l^b{}_i = \delta_a^b$, $r_a{}^i r^b{}_i = \delta_a^b$) and reads
\begin{equation}\label{eqn:eailambda}
  e_a{}^i = \frac{\kappa + 1}{2\sqrt{\kappa}} l_a{}^i + \frac{\kappa - 1}{2\sqrt{\kappa}} r_a{}^i\,,
\end{equation}
where $k$ and $\kappa$ are free parameters. To construct the $B$-field, a locally defined two-form, $B_0$, whose exterior derivative results in the three-form
\begin{equation}
  H_0 = -\frac1{3 \sqrt{2 k}} f_{abc} l^a \wedge l^b \wedge l^c = \dd B_0
\end{equation}
is required. It gives rise to
\begin{equation}
  B = B_0 - \frac{\kappa + 1}{2\sqrt{\kappa}} l^a{}_i e_{aj} d x^i \wedge d x^j
\end{equation}
and completes, together with
\begin{equation}\label{eqn:lambdalambdadef}
  \Lambda_{\bar a}{}^{\bar b} = \frac{(\kappa + 1) \delta_{\bar a}{}^{\bar b} - 2 \sqrt{\kappa} e_{\bar a}{}^i l^{\bar b}{}_i}{\kappa - 1}\,,
\end{equation}
the constituents of the generalized frame field \eqref{eqn:decompframe}. Apparently, these expressions look rather complicated and they turn out to become even more involved once a parameterisation for the group element $g$ is fixed. This is because the standard target fields obscure the underlying structure of the $\lambda$-deformation. The structure coefficients of $\mfd$ encode the same information but in a much more streamlined form. They arise from \eqref{eqn:framealgebra} and read
\begin{equation}\label{eqn:genfluxlambda}
  F_{abc} = \frac{\kappa^2 + 3}{4 \sqrt{\kappa k}} f_{abc} \,, \quad
  F_{ab\bar c} = \frac{\kappa^2 - 1}{4 \sqrt{\kappa k}} f_{ab\bar c} \,, \quad
  F_{a\bar b\bar c} = \frac{\kappa^2 - 1}{4 \sqrt{\kappa k}} f_{a\bar b\bar c} \,, \quad
  F_{\bar a\bar b\bar c} = \frac{\kappa^2 + 3}{4 \sqrt{\kappa k}} f_{\bar a\bar b\bar c}\,.
\end{equation}
Note that the remaining components are fixed by the total antisymmetry of $F_{ABC}$. Furthermore, in this form the symmetry $\kappa \rightarrow - \kappa$ and $k \rightarrow -k$ \cite{Itsios:2014lca} of the $\lambda$-deformation is immediately manifest. The semisimple doubled Lie algebra $\mfd=\mathfrak{g}\times\mathfrak{g}$ has no center and thus $F_A=0$. Starting from \eqref{eqn:defgend} and \eqref{eqn:defFA}, one can use this fact to extract the derivative of the dilaton
\begin{equation}
  \partial_i \phi = \frac12 \omega^a{}_{ai}\,,
\end{equation}
where $\omega_{ia}{}^b$ denotes the spin connection corresponding to the frame field \eqref{eqn:eailambda}.

PL T-duality relates the $\lambda$-deformation to the $\eta$-deformation up to an analytic continuation \cite{\reflambdaetaPL}. A generalized frame field for the latter can be easily constructed \cite{Demulder:2018lmj}. The detailed expressions for the metric and $B$-field are not relevant for our discussion. All information we rely on is contained in the structure coefficients $F_{ABC}$ and thus it is not surprising that the $\lambda$- and $\eta$-deformation are both captured by \eqref{eqn:genfluxlambda} after the identification
\begin{equation}\label{eqn:paramlambdaeta}
  \begin{aligned}
    \text{$\lambda$-deformation:} & &
      \kappa&=\frac{1 - \lambda}{1 + \lambda} &\qquad
      k&=k \\
    \text{$\eta$-deformation:} & &
      \kappa&= - \ci \eta &
      k&= \frac{\ci}{4 \eta t}\,.
  \end{aligned}
\end{equation}
Both form two different branches on the space of structure coefficients $F_{ABC}$, representing $\DD=G\times G$ and $\DD=G^{\mathbb{C}}$, respectively. There is a one-dimensional subspace where both meet. It is defined by the limit $\kappa \to 0$ and $k \to \infty$. In this case, we have $\lambda=1$ and $\eta=0$, whereas $t$ remains a free parameter with $\kappa = h/(4k)$. The corresponding model is the principal chiral model (PCM) on the group manifold $G$, and $\DD$ is contracted to $T^* G$.

\subsection{One-Loop}\label{sec:oneloop}
A $\sigma$-model has an infinite number of couplings that are encoded in the metric $g_{ij}$, the $B$-field $B_{ij}$ and the dilaton $\phi$. As some of them are redundant, we first note that infinitesimal diffeomorphisms and gauge transformations,
\begin{equation}
  \dg_{ij} = 2 \nabla_{(i} \xi_{j)}\,, \qquad
  \dB_{ij} = H_{ijk} \xi^k + 2 \nabla_{[i} \chi_{j]}\,, \qquad \text{and} \qquad
  \dphi = \xi^i \nabla_i \phi\,,
\end{equation}
that are generated by the vector $\xi^i$ and the one-form $\chi_i$, do not affect any local observables. Thus, it is useful to define equivalence classes of $\beta$-functions which only differ by those transformations. Each class has a canonical representative for which the $\beta$-functions do not generate any diffeomorphisms or gauge transformations. We denote it with a bar and define an arbitrary member of its equivalence class by
\begin{equation}\label{eqn:defbh}
  \bh^E_{ij} = \bb^E_{ij} + 2 \nabla_{(i} \xi_{j)} + H_{ijk} \xi^k + 2 \nabla_{[i} \chi_{j]} \,,
    \qquad
  \bh^\phi = \bb^\phi + \xi^i \nabla_i \phi\,,
\end{equation}
where $E_{ij} = g_{ij} + B_{ij}$ unifies the metric and $B$-field into a single object. Furthermore we use the standard convention where the RG flow is governed by
\begin{equation}\label{eqn:RGflownormal}
  \mu \frac{\dd E_{ij}}{\dd\log\mu} = \beta^E_{ij} \qquad \text{and} \qquad
  \mu \frac{\dd \phi}{\dd\log\mu} = \beta^\phi\,.
\end{equation}

At one-loop $\bb^E_{ij}$ reads \cite{Friedan:1980jf,Curtright:1984dz}
\begin{equation}\label{eqn:bb1E}
  \bb^{(1)E}_{ij} = R_{ij} - \frac14 H^2_{ij} - \frac12 \nabla_k H^k{}_{ij}
    \qquad \text{with} \qquad 
    H^2_{ij} = H_{imn} H_j{}^{mn}\,.
\end{equation}
It is the first non-vanishing term in the expansion
\begin{equation}
  \bb^{E}_{ij} = \alpha' \bb^{(1)E}_{ij}  + \alpha'^2 \bb^{(2)E}_{ij}  + \dots\,.
\end{equation}
We adopt the same notation for all other quantities that admit an $\alpha$-expansion, too. Saying that a quantity$^{(n)}$ comes with a factor of $\alpha'^n$. Because derivatives contribute a factor of $\sqrt{\alpha}$, we can alternatively conclude that quantities at the level $n$ normally contain $2n$ derivatives. The notable exceptions are the vector $\xi^i$ and the one form $\chi_i$ in \eqref{eqn:defbh}. They contain $2n - 1$ derivatives. Computing \eqref{eqn:bb1E} directly is cumbersome and one might ask if there is an easier way to obtain the RG flow. At this point working with doubled quantities, as they naturally appear in DFT, is very convenient. As already demonstrated in the last section, they are particularly powerful to describe PL symmetric target space whose flows we ultimately want to address. The doubled version of the first equation in \eqref{eqn:RGflownormal} becomes
\begin{equation}\label{eqn:flowdoubled}
  \mu \frac{\dd \Eh_A{}^I}{\dd \log\mu} \Eh_{BI} = \Bh^{(1)E}_{AB}
\end{equation}
in the framework of DFT. In this equation we prefer the partially double Lorentz fixed generalized frame field $\Eh_A{}^I$ over $E_A{}^I$ because its remaining, unfixed symmetries coincide with the diffeomorphisms, $B$-field and Lorentz transformations that are manifest symmetries of \eqref{eqn:bb1E}. The doubled $\beta$-function on the right-hand side is based on $\bh^E_{ij}$ that arises from \eqref{eqn:defbh} with 
\begin{equation}\label{eqn:diffgauge1}
  \xi^{(1)i} = \nabla^i \phi\,, \qquad \chi^{(1)}_i = 0\,.
\end{equation}
More specifically, its off-diagonal contributions
\begin{equation}\label{eqn:betaEdoubled}
  \Bh^E_{AB} = \frac12 \begin{pmatrix} 
    0 & \bh^E_{a\bar b} \\
    - \bh^E_{b\bar a} & 0
  \end{pmatrix}
\end{equation}
are formed by $\bh^E_{ij}$ in flat indices. This embedding is motivated by the observation that all physical information is contained in the off-diagonal blocks while the diagonal blocks only generate double Lorentz transformations. Therefore, we set them to zero. In order to extract the physically relevant blocks, the projectors
\begin{equation}
  P_A{}^B = \begin{pmatrix} \delta_a{}^b & 0 \\
    0 & 0 \end{pmatrix}
  \qquad \text{and} \qquad
  \Pb_A{}^B = \begin{pmatrix} 0 & 0 \\
    0 & \delta_{\bar a}{}^{\bar b} \end{pmatrix}
\end{equation}
are required. Note the factor of $1/2$ in the definition \eqref{eqn:betaEdoubled}. It appears because $\bh^E_{ij}$ governs the flow of the metric and $B$-field directly, whereas $\Bh^E_{AB}$ captures the flow of a (generalized) frame field. The former is the square of the latter and of course the derivative of a square always introduces a factor of $2$. Due to this factor, we have to carefully distinguish between $\bh^E_{a\bar b}$ and $\Bh^E_{a\bar b}$.

Our primary objective is to find an expression for $\Bh^{(1)E}_{AB}$ that reproduces \eqref{eqn:bb1E} and can be written exclusively in terms of the doubled quantities we encountered so far, namely $\Fh_{ABC}$, $\Fh_A$, $\Dh_A = \Eh_A{}^I \partial_I$, $P^{AB}$, $\Pb^{AB}$. Hats over the $F$'s indicate that they are still computed by \eqref{eqn:framealgebra} and \eqref{eqn:defFA} but for $\Eh_A{}^I$ instead of $E_A{}^I$. Therefore, neither $\Fh_{ABC}$ nor $\Fh_A$ is constant. Using the parameterisation given in \eqref{eqn:decompframe}, we obtain the generalized fluxes
\begin{equation}
  \begin{aligned}
    \Fh_{abc} &= \frac1{2\sqrt{2}} ( H_{abc} - 6 \omega_{[abc]} ) & \qquad
    \Fh_{\bar a b c} &= \frac1{2\sqrt{2}} ( H_{\bar abc} - 2 \omega_{\bar a bc} ) \\
    \Fh_{a\bar b\bar c} &= \frac1{2\sqrt{2}} ( H_{a\bar b\bar c} + 2 \omega_{a\bar b\bar c} ) &
    \Fh_{\bar a\bar b\bar c} &= \frac1{2\sqrt{2}} ( H_{\bar a\bar b\bar c} + 6 \omega_{\bar a\bar b\bar c} )
  \end{aligned}
\end{equation}
and
\begin{equation}\label{eqn:Fha}
  \Fh_a = \Fh_{\bar a} = \sqrt{2} \Dh_a \phi - \frac1{\sqrt{2}} \omega^b{}_{ba}
    \qquad \text{with} \qquad
    \Dh_a = e_a{}^i \partial_i
\end{equation}
written in terms of the spin connection $\omega_{abc}$ and the $H$-flux. Eventually, one is able to come up with the one-loop, doubled $\beta$-function 
\begin{equation}\label{eqn:beta(1)Edoubled}
  \Bh^{(1)E}_{AB} = 2 P_{[A}{}^C \Pb_{B]}{}^D \left( \Fh_{CEG} \Fh_{DFH} P^{EF} \Pb^{GH} + \Fh_{CDE} \Fh_F P^{EF} + \Dh_D \Fh_C - \Dh_E \Fh_{CDF} P^{EF} \right)\,,
\end{equation}
which agrees with the starting point \eqref{eqn:bb1E}. A detailed derivation of this equation is given in section~\ref{sec:oneloopdetail}.

We will encounter more equations like this one. To see their structure more clearly, one might represent them in diagrammatic form. To this end, we identify the projectors $P_{AB}$ and $\Pb_{AB}$ with two different propagators
\begin{align}
  P_{AB} &= \tikz[baseline=(A.base)]{\draw (0,0) node[anchor=east] (A) {$A$} -- (1,0) node[anchor=west] {$B$};} &      
    \text{and} & &
    \Pb_{AB} &= \tikz[baseline=(A.base)]{\draw[dashed] (0,0) node[anchor=east] (A) {$A$} -- (1,0) node[anchor=west] {$B$};}   
  \,,
\intertext{while the fluxes become the vertices}
  \Fh_{ABC} &= \tikz[baseline=(B.base)] {
    \draw (0:0) node[A] {} -- (0:0.4) node[anchor=west] (B) {$B$};
    \draw (0:0) -- (120:0.4) node[anchor=east] {$A$};
    \draw (0:0) -- (240:0.4) node[anchor=east] {$C$};
  } & \text{and} & &
  \Fh_A &= \tikz[baseline=(A.base)] {
    \draw (0:0) node[FA] {} -- (0:0.4) node[anchor=west] (A) {$A$};
  }\,.
\end{align}
Finally, we denote a derivative with an arrow, for example
\begin{equation}
  \Dh_A \Fh_B = \tikz[baseline=(A.base)] { 
    \draw[der] (0:0) node[anchor=east] (A) {$A$} -- (0:0.4) node[FA,anchor=west] {};
    \draw (0:0.4) -- (0:0.8) node [anchor=west] {$B$};}\,.
\end{equation}
Dummy indices are suppressed in these diagrams and, if unambiguous, also external indices can be dropped. Making use of these conventions, \eqref{eqn:beta(1)Edoubled} can be written as
\begin{equation}\label{eqn:diagbeta1}
  \bh^{(1)E}_{a\bar b} = - 2 \dFsquare01 - 2\dFFA0 + 2\dDAFA + 2\dFADA0\,.
\end{equation}
For the $\beta$-function of the generalized dilaton, the same argument applies and one can check that (again all the details are given in section~\ref{sec:oneloopdetail})
\begin{equation}\label{eqn:diagbeta1d}
  \begin{aligned}
    \bh^{(1)d} &= \mu\frac{\dd d}{\dd\log\mu} = \bh^{(1)\phi} - \frac14 g^{ij} \bh^{(1)E}_{ij} = 
      -\frac14 R + \frac1{48} H^2 + (\nabla \phi)^2 - \nabla^2 \phi \\
      &= \frac{1}{12} \Fsquare000 + \frac14\Fsquare010 + \frac12\FAFA{0} - \FADA{0}
  \end{aligned}
\end{equation}
holds.

\subsubsection{Double Lorentz Transformation}
Instead of $\Fh_{ABC}$ and $\Fh_A$, we would rather use $F_{ABC}$ and $F_A$ as they are the natural objects for a PL symmetric $\sigma$-model. They are connected to each other by the double Lorentz rotation $\Lambda_A{}^B$ defined in \eqref{eqn:decompframe}. Although the generalized fluxes and their derivatives transform anomalous under this rotation, the particular combination in which they enter the $\beta$-function cancels all anomalous contributions. This is a standard result in the flux formulation of DFT \cite{Hohm:2010xe,Geissbuhler:2013uka}, but since double Lorentz rotations become much more subtle beyond one-loop, we want to review how it arises: The finite transformation $\Lambda_A{}^B$ is a composition of infinitesimal transformations, namely $\Lambda_A{}^B = \exp ( \lambda_A{}^B )$ with $\lambda_{AB} = - \lambda_{BA}$. Covariant quantities, like $\B^{(1)E}_{AB}$, transform as
\begin{equation}\label{eqn:deltalambdabeta1}
  \delta_\lambda^{(0)} \Bh^{(1)E}_{AB} = \lambda_A{}^C \Bh^{(1)E}_{CB} + \lambda_B{}^C \Bh^{(1)E}_{AC}
\end{equation}
under the infinitesimal action $\delta_\lambda^{(0)}$. Note that this relation only holds to leading order in $\alpha'$, indicated by the superscript ${}^{(0)}$ on the action. As already mentioned, there are also  non-covariant quantities, like the generalized fluxed $\Fh_{ABC}$. To treat them in a methodical way, we introduce the ``anomalous'' contribution to  the transformation
\begin{equation}
  \Al = \delta_\lambda - \lambda \cdot\,. 
\end{equation}
With $\lambda \cdot$, we denote the standard action of $\lambda$ on every flat index. For example, the generalized fluxes have the leading order anomalous transformation
\begin{equation}\label{eqn:anFABC}
  \Al^{(0)} \Fh_{ABC} = \delta_\lambda^{(0)} \Fh_{ABC} - 3 \lambda_{[A}{}^D \Fh_{BC]D} = 3 \Dh_{[A} \lambda_{BC]}\,.
\end{equation}
Let us see in more detail how the right-hand side of this equation arises. Because $\Al$ is a linear operator ($\Al (a + b) = \Al a + \Al b$) that acts as a derivative ($\Al ( a b ) = \Al a b + a \Al b$), all we need to evaluate \eqref{eqn:anFABC} from the definition $\Fh_{ABC} = 3 \Dh_{[A} \Eh_B{}^I \Eh_{C]I}$ is $\Al^{(0)} \Eh_A{}^I = 0$ and the commutator of $\Al^{(0)}$ and $\Dh_A$. The later is given by
\begin{equation}
  [ \Al^{(0)}, \Dh_A ] \Eh_B{}^I = \Dh_A \lambda_B{}^C \Eh_C{}^I\,.
\end{equation}
In the same vein one obtains $\Al^{(0)} \Fh_A = \Dh_B \lambda^B{}_A$ (after taking into account $\Al d=0$) and $\Al P^{AB} = - \Al \Pb^{AB} = 0$. Eventually, we can directly evaluate $\Al^{(0)} \Bh^{(1)E}_{AB}$ from \eqref{eqn:diagbeta1} and find that it vanishes. Hence, we come full circle and arrive again at \eqref{eqn:deltalambdabeta1}. 

A finite double Lorentz transformation arises from the exponential map
\begin{equation}\label{eqn:expdeltalambda}
  e^{\delta_\lambda} = e^{\Al + \lambda\cdot} = e^\lambda e^{\Al} = \Lambda \cdot e^{\Al} \,.
\end{equation}
$\Lambda \cdot$ denotes the group action of $\Lambda_A{}^B$ on every free index. Applying this relation to $\Bh^{(1)E}$, we eventually obtain 
\begin{equation}\label{eqn:BtoBh}
  \B^{(1)E}_{AB} = \Lambda_A{}^C \Lambda_B{}^D \Bh^{(1)E}_{CD}
\end{equation}
and prove that it is valid to drop all the hats in \eqref{eqn:beta(1)Edoubled} and use the rotated $\beta$-function $\beta^{(1)E}$ instead of $\Bh^{(1)E}$. It is important to stress that both only are written in different double Lorentz frames, but still describe exactly the same physics. However, the latter is much better adapted to PL symmetric target space geometries because all quantities are just constant. Hence all terms that contain derivatives $D_A$ drop out. Double Lorentz transformations do not affect the $\beta$-function of the generalized dilaton in \eqref{eqn:diagbeta1d} and we thus identify
\begin{equation}
  \bh^{(1)d} = \beta^{(1)d}\,.
\end{equation}

\subsubsection{Renormalizable \texorpdfstring{$\sigma$}{sigma}-Models}\label{sec:renormalizable}
All information about the $\sigma$-model of the bosonic string \eqref{eqn:sigmamodel} is condensed in $F_{ABC}$ and $F_A$. We might take these two objects as being parameterized by $N$ coupling constants $c^\mu$ where $\mu=1, \dots, N$. The $\beta$-functions for these couplings arise directly from $\B^E_{AB}$ and $\beta^d$ through the relations
\begin{equation}\label{eqn:betamu}
  \begin{aligned}
    \beta^\mu \partial_\mu F_{ABC} &= 3 D_{[A} \B^E_{BC]} + 3 \B^E_{[A}{}^D F_{BC]D} \\
    \beta^\mu \partial_\mu F_A &= D^B \B^E_{BA} + \B^E_A{}^B F_B + 2 D_A \beta^d\,.
  \end{aligned}
\end{equation}
For general target space geometries, neither $F_{ABC}$ nor $F_A$ is constant. They rather have different values on every point of the target space manifold $M$. Hence, one needs infinitely many coupling constants $c^\mu$ to accommodate this information. In contrast, PL $\sigma$-models have by definition constant $F_{ABC}$'s and $F_A$'s. Therefore, PL symmetry just permits a finite number of couplings. If this property is preserved under RG flow, it renders the PL $\sigma$-model renormalizable. From \eqref{eqn:betamu} it follows that this is the case if
\begin{equation}\label{eqn:renormalizable}
  D_A \B^E_{BC} = 0 \qquad \text{and} \qquad D_A \beta^d = 0
\end{equation}
holds, which is clearly the case for the one-loop $\beta$-functions presented in \eqref{eqn:diagbeta1} and \eqref{eqn:diagbeta1d}. Hence, we conclude that PL $\sigma$-models are one-loop renormalizable. This observation is by now well established \cite{\refonelooprenorm}. However, all previous works we are aware of only incorporate the metric and the $B$-field but not the dilaton.

Another advantage of encoding all $\sigma$-model couplings in terms of $F_{ABC}$ and $F_A$ is that their transformation under infinitesimal generalized diffeomorphisms, which unify diffeomorphisms and $B$-field transformations, is very simple, namely
\begin{equation}
  \delta F_{ABC} = \Xi^I \partial_I F_{ABC} \qquad \text{and} \qquad
  \delta F_A = \Xi^I \partial_I F_A\,.
\end{equation}
Here $\Xi^I = \begin{pmatrix} \chi_i & \xi^i \end{pmatrix}$ contains the parameters introduced in \eqref{eqn:defbh}. PL symmetric backgrounds are invariant under such transformations because $F_A$ and $F_{ABC}$ are constant.

Remarkably, PL $\sigma$-models are just a particular example of a more general scheme: At one-loop, all target space geometries which admit a consistent truncation result in renormalizable $\sigma$-models. Both notions are related because the one-loop $\beta$-functions \eqref{eqn:diagbeta1} and \eqref{eqn:diagbeta1d} are equivalent to the field equations of the bosonic string's two-derivative target space effective action. One might understand field equations of a classical field theory as describing an infinite number of coupled degrees of freedom. Consistent truncations are based on the observation that it is sometimes possible to decouple a finite number of them from the rest, which then can be safely truncated. This technology is extremely useful to simplify the hard task of finding solutions to the field equations. Here, we see that it also has a natural interpretation in terms of two-dimensional, renormalizable field theories.

\subsubsection{\texorpdfstring{$\lambda$}{Lambda}- and \texorpdfstring{$\eta$}{Eta}-Deformation}\label{sec:lambda1}
We have now all we need to compute the one-loop $\beta$-function of the coupling $\kappa$ and $k$ in the $\lambda$- and $\eta$-deformation. Only the first diagram in \eqref{eqn:diagbeta1} contributes to
\begin{equation}
  \beta^{(1)E}_{a\bar b} = - 2 \dFsquare01 = 2 F_{\bar c d a} F^{d \bar c}{}_{\bar b} = - \frac{(\kappa^2 - 1)^2}{8 k \kappa} c_G \eta_{a\bar b}\,.
\end{equation}
Here we use the normalisation $f_{ac}{}^d f_{bd}{}^c = - c_G \eta_{ab}$ for the structure coefficients of $G$'s Lie algebra with the dual Coxeter number $c_G$. From \eqref{eqn:betamu}, we extract
\begin{equation}\label{eqn:beta1kappa}
  \beta^k = 0 \qquad \text{and} \qquad \beta^\kappa = \frac{c_G}{8 k} (\kappa^2 - 1)^2\,.
\end{equation}
$\kappa$ is related to $\lambda$ and $\eta$ by \eqref{eqn:paramlambdaeta}, which eventually gives rise to
\begin{equation}
  \beta^\lambda = - \frac{\lambda^2 c_G}{k(\lambda + 1)^2}
    \qquad \text{and} \qquad
  \beta^\eta = \frac{\eta t c_G}2 ( 1 + \eta^2 )^2\,.
\end{equation}
These results match with the ones provided in the literature \cite{Itsios:2014lca,Sfetsos:2015nya}. We also compute the $\beta$-function for the generalized dilaton
\begin{equation}\label{eqn:lambdadefbeta1d}
  \beta^{(1)d} = \frac{1}{12} \Fsquare000 + \frac14\Fsquare010 = \frac{\kappa^4 - 6 \kappa^2 - 3}{96 k \kappa} c_G \dim G \,.
\end{equation}
At $\lambda=0$ the RG-flow has a fixed point, the WZW-model on the group $G$.

\subsection{Two Loops}\label{sec:twoloops}
Beyond one-loop the $\beta$-functions become scheme dependent. Therefore, we first have to fix a particular scheme in which we present our results. As we will see, making a good choice is crucial because only in a distinguished scheme PL symmetry becomes manifest and the computations manageable. Different schemes arise from an ambiguity in choosing counter terms during the renormalization of the $\sigma$-model. An alternative perspective is that different schemes are related by field redefinitions, which are diffeomorphisms on the space of couplings. Naively, choosing a scheme is the same as committing to a particular set of coordinates in general relativity. Obviously when dealing with a problem with rotational symmetry, it is a good idea to choose spherical coordinates instead of Cartesian coordinates. We know that the final, physical observables do not depend on this choice. But it is much easier to extract them in adapted coordinates.

\subsubsection{Scheme Transformation}
There is one aspect of scheme transformations for $\sigma$-models which makes them slightly more complicated than the standard diffeomorphisms that we are used to from general relativity. Because a $\sigma$-model has an infinite number of coupling constants, one has to deal with diffeomorphisms on an infinite dimensional manifold. The tangent space of this manifold is spanned by the vectors $\delta_\Psi$ with $\Psi = \begin{pmatrix} \delta E_{AB} & \delta d \end{pmatrix}$. In working with them, it is very helpful to remember what happens after a projection onto a finite dimensional submanifold (this is exactly what PL symmetry will allow us to do later). In this case, $\Psi$ reduces to a column vector $\Psi^\mu$ and $\delta_\Psi$ becomes $\Psi^\mu \partial_\mu$. The derivative $\delta_\Psi$ is defined by its action on
\begin{equation}\label{eqn:variations}
  \begin{aligned}
   \delta_\Psi F_{ABC} & = 3 D_{[A} \delta E_{BC]} + 3 \delta E_{[A}{}^D F_{BC]D} & \qquad \delta_\Psi P_{AB} &= 0\\
   \delta_\Psi F_A &= D^B \delta E_{BA} + \delta E_A{}^B F_B + 2 D_A \delta d & \delta_\Psi \Pb_{AB} &= 0 \\
   \delta_\Psi D_A &= \delta E_A{}^B D_B + D_A \delta_\Psi\,.
  \end{aligned}
\end{equation}
Note that these relations allow us to rewrite \eqref{eqn:betamu} in the cleaner form
\begin{equation}
  \beta^\mu \partial_\mu = \delta_{\begin{pmatrix} \B^E & \beta^d \end{pmatrix}}
\end{equation}
and we see that above we actually restricted the infinite dimensional coupling space to the finite dimensional space of couplings which are compatible with PL symmetry.

An infinitesimal scheme transformation of the $\beta$-functions $\B = \begin{pmatrix} \B^E & \beta^d \end{pmatrix}$ with the parameter $\Psi$ is mediated by the Lie derivative
\begin{equation}
  L_\Psi \B = \delta_\Psi \B - \delta_\B \Psi - T(\Psi, \B) \,.
\end{equation}
The last term takes into account that the derivative $\delta_\Psi$ in general has torsion, which is defined by
\begin{equation}
  \delta_\Psi \delta_{\Psi'} - \delta_{\Psi'} \delta_{\Psi} = \delta_{T(\Psi,\Psi')}\,.
\end{equation}
From the definition \eqref{eqn:variations}, one indeed obtains the non-vanishing torsion
\begin{equation}
  T(\Psi, \Psi') = \begin{pmatrix} 2 \delta E^C{}_{[A} \delta{E}'_{B]C} & 0 \end{pmatrix}\,.
\end{equation}
Since $\delta E_A{}^B$ generates an O($D$,$D$) transformation, the non-trivial part of the torsion tensor may be written as $[\delta E, \delta E']$. This rewriting shows that the torsion we encounter originates from the O($D$,$D$) structure of the generalized tangent space. At a first glance, our choice of derivative might seem peculiar because it clearly differs from the canonical, torsion-free variation with respect to $g_{ij}$, $B_{ij}$, and $\phi$. In the end, one can check that both give rise to the same results. However, using this $\delta_\Psi$ simplifies the computations considerably and therefore we prefer it.

Infinitesimal scheme transformations are sufficient for our purpose because we are just concerned with contributions to the $\beta$-functions up to the order $\alpha'^2$, and for all $\Psi$ which we consider, $\Psi^{(0)}$ always vanishes. Consequentially $\B^{(1)}$ is not affected and $\B^{(2)}$ is corrected by
\begin{equation}\label{eqn:schemetrbeta}
  \B^{(2)} \rightarrow \B^{(2)} + L_{\Psi^{(1)}} \B^{(1)}\,.
\end{equation}
In principal, one could apply more general transformations with a non-trivial $\Psi^{(0)}$. But they would spoil the manifest symmetries of the one-loop results obtained in the last subsection. Hence, we are restricted to transformations that start with $\Psi^{(1)}$ and \eqref{eqn:schemetrbeta} applies.

\subsubsection{\texorpdfstring{$\beta$}{Beta}-Functions}
Like in the last subsection, we again start with the known result for the two-loop $\beta$-functions of the metric, $B$-field, and dilaton in the Metsaev-Tseytlin (MT) scheme \cite{Metsaev:1987zx}. The reason why we preferred this scheme over other popular options, like the Hull-Townsend (HT) scheme, is purely technical and will be explained in section~\ref{sec:twoloopsdetails}. Compared to the discussion at one-loop, the most striking difference is that the two-loop $\beta$-functions, which are given in \eqref{eqn:betag2}-\eqref{eqn:betaphi2}, are considerably more complicated. However, we can still relate them a member in their equivalence class, which is suited to be written exclusively in terms of $\Fh_{ABC}$, $\Fh_A$, $\Dh_A$, $P^{AB}$ and $\Pb^{AB}$, by the infinitesimal diffeomorphism and gauge transformation
\begin{equation}\label{eqn:requireddiff+gauge}
  \xi^{(2)i} = - \frac1{48} \nabla^i H^2 + \frac14 H^{ijk} \bh^{(1)E}_{jk} \,, \qquad 
  \chi^{(2)}_i = \omega_i{}^{ab} \bh^{(1)E}_{ab}\,.
\end{equation}
Still, this is not sufficient and we furthermore have to change the scheme by \eqref{eqn:schemetrbeta} with $\Psi^{(1)} = \begin{pmatrix} \Delta^{(1)} E_{ij} & 0 \end{pmatrix}$ with $\Delta E_{ij} = \Delta g_{ij} + \Delta B_{ij}$ and
\begin{equation}\label{eqn:schemetr}
  \Delta^{(1)} g_{ij} = - \frac12 \omega_{ia}{}^b \omega_{jb}{}^a + \frac38 H^2_{ij}\,, \qquad 
  \Delta^{(1)} B_{ij} = - \frac12 H_{[i|a}{}^b \omega_{|j]b}{}^a \,.
\end{equation}

After a cumbersome computation, that we detail in the next section, one finds that $\bh^{(2)E}$ has in total 342 terms(=diagrams). They are invariant under the $\mathbb{Z}_2$ action $Z$ that swaps the projectors $P$ and $\Pb$. To illustrate how $Z$ acts on the level of diagrams, take for example
\begin{equation}\label{eqn:Z2}
  Z( \HCNa00001 ) = - \HCNa01111\,.
\end{equation}
Here, we first swap solid and dashed lines ($P \leftrightarrow \Pb$) and then bring the external solid line to the left and the dashed one to the right. This swapping of the external lines corresponds to $a \rightarrow \bar b$ and $\bar b \rightarrow a$, or equivalently $A\leftrightarrow B$, of the antisymmetric $\Bh^{(2)E}_{AB}$ and therefore introduces a minus sign. Because the two-loop $\beta$-functions of the bosonic string satisfies
\begin{equation}
  Z(\bh^{(2)E}_{a\bar b}) = \bh^{(2)E}_{a\bar b}\,,
\end{equation}
we actually only have to cope with 172 different diagrams for $\bh^{(2)E}$ while the others are fixed by the $\mathbb{Z}_2$ symmetry. It is not very illuminated to present this bulky result here. Fortunately, for PL symmetric target spaces it can be simplified considerably. But to benefit from the structure introduced in section~\ref{sec:genframes}, we again have to switch to unhatted quantities by applying the double Lorentz transformation $\Lambda_A{}^B$.

\subsubsection{Double Lorentz Transformation}
At this point we encounter another important subtlety that we need to handle beyond one-loop: Double Lorentz transformations of the generalized frame field pick up the anomalous contribution
\begin{equation}\label{eqn:genGStr}
  \Al^{(1)} \Eh_A{}^I \Eh_{BI} = - P_{[A}{}^C \Pb_{B]}{}^D \Dh_C \lambda_{EF} \Fh_{DGH} P^{EG} P^{FH} - P \leftrightarrow \Pb\,.
\end{equation}
It originates from the non-Lorentz-covariant scheme transformation \eqref{eqn:schemetr} and was dubbed generalized Green-Schwarz transformation (gGS) \cite{Marques:2015vua}. The name is motivated by the observation that the $B$-field of the heterotic string receives a non-Lorentz-covariant contribution to its transformation at the subleading order of $\alpha'$. This correction is captured by the first term on the left hand side of \eqref{eqn:genGStr} and gives rise to the celebrated Green-Schwarz anomaly cancellation mechanism \cite{Green:1984sg}. Moreover, gGS transformations play a central role in constructing $\alpha'$-corrections in DFT, where the one-loop and two-loop effective target space actions, $\Sh^{(1)}$ and $\Sh^{(2)}$, are related by \cite{Marques:2015vua}
\begin{equation}\label{eqn:DLS2}
  \Al^{(0)} \Sh^{(2)} = \delta_{\AlE} \Sh^{(1)} = L_{\AlE} \Sh^{(1)}\,.
\end{equation}
Actually, this relation is so strong that it fixes $S^{(2)}$ completely. Note that the generalized dilaton is not affected and $\Al d = 0$ still holds. Following the steps that we demonstrated at one-loop, one obtains the anomalous transformation of the two-loop $\beta$-functions,
\begin{equation}\label{eqn:DLbeta2}
  \Al^{(0)} \Bh^{(2)} = L_{\AlE} \Bh^{(1)}\,,
\end{equation}
which is of the same form as \eqref{eqn:DLS2}. We present the derivation of this important relation in section~\ref{sec:twoloopsdetails}. For the moment, we are rather interested in a finite version of the left hand side. Because $\Al$ acts as a derivative on the Lie derivative\footnote{One can show that for two arbitrary vectors $X$ and $Y$,
\begin{equation}
  \Al (L_X Y) = L_{\Al X} Y + L_X (\Al Y)
\end{equation}
holds.} one obtains
\begin{equation}\label{eqn:beta2DL}
  \B^{(2)} = \Lambda \cdot \Bigl( \Bh^{(2)} + L_{\begin{pmatrix} \DL^{(1)} \Eh & 0 \end{pmatrix}} \bh^{(1)} \Bigr)\,.
\end{equation}
According to our convention, $\DL^{(1)} \Eh$ denotes the term in the finite gGS transformation
\begin{equation}
  \DL \Eh = \left( e^{\Al} - 1 \right) \Eh
\end{equation}
of the generalized frame field $\Eh_A{}^I$ at the leading order in $\alpha'$.

Hence, we conclude that to go from hatted to unhatted quantities at the two-loop level, not only a rotation by $\Lambda_A{}^B$, but also a scheme transformation is required. Fortunately, neither affects any observables of the theory.  Consequentially, we can drop the hats in the expression for $\bh^{(2)E}$ as we did already at one-loop. PL symmetry removes all terms with flat derivatives $D_A$ and we are left with 20 diagrams contributing to
\begin{equation}\label{eqn:diagbeta2}
  \begin{split}
    \beta^{(2)E}_{a\bar b} = 
      &\HCNa00001 \co{+}  \HCNa00101 \co{+2} \HCNb00011 \co{+2} \HCNb00100 \co{+4} \HCNb01001 \\
            \co{-4} &\HCNb01010 \co{+2} \HCNb11000 \co{-2} \HCNc00010 \co{-4} \HCNc00011 \co{+} \HCNc00110 \\
            \co{+4} &\HCNc01010 \co{+} \HCNc10001 \co{-2} \HCNc10010 \co{+} \dFAFFFa1000 \co{-2} \dFAFFFc0000 \\
            \co{-2} &\dFAFFFc0010 \co{+} \dFAFFFbprime0000 \co{+2} \dFAFFFbprime0010 \co{+} \dFAFFFbprime0100 {-2} \dFAFFFb0100 
      \\
       + & P \leftrightarrow \Pb\,.
  \end{split}
\end{equation}
To avoid problems with the translation from diagrams to a tensor expression, we give here the explicit result
\newcommand{\refpoint}[1]{\tikz[remember picture,overlay]{\node (#1) {};}}
\begin{align}
  \B^{(2)E}_{AB} =&
 -F_{IJM}F_{CDF}F_{EGH}F_{KLN}P_{[A}{}^{C}P^{DE}P^{FG}P^{HI}P^{JK}\Pb_{B]}{}^{L}\Pb^{MN} \refpoint{HCNa1} \quad\refpoint{HRZ}\qquad\qquad\qquad \nonumber \\&+
 F_{ILN}F_{CDF}F_{EGK}F_{HJM}P_{[A}{}^{C}P^{DE}P^{FG}P^{HI}\Pb_{B]}{}^{J}\Pb^{KL}\Pb^{MN} \refpoint{HCNa2} \nonumber \\&+
 2F_{CDF}F_{EHK}F_{GIM}F_{JLN}P_{[A}{}^{C}P^{DE}P^{FG}P^{HI}\Pb_{B]}{}^{J}\Pb^{KL}\Pb^{MN} \refpoint{HCNb1} \nonumber \\&+
 2F_{IKL}F_{CDF}F_{EHM}F_{GJN}P_{[A}{}^{C}P^{DE}P^{FG}P^{HI}P^{JK}\Pb_{B]}{}^{L}\Pb^{MN} \nonumber \\&- 
 4F_{ILN}F_{CDK}F_{EFH}F_{GJM}P_{[A}{}^{C}P^{DE}P^{FG}P^{HI}\Pb_{B]}{}^{J}\Pb^{KL}\Pb^{MN} \nonumber \\&-
 4F_{IJN}F_{CDK}F_{EFM}F_{GHL}P_{[A}{}^{C}P^{DE}P^{FG}P^{HI}\Pb_{B]}{}^{J}\Pb^{KL}\Pb^{MN}   \nonumber \\&+ 
 2F_{CKM}F_{DFJ}F_{EHL}F_{GIN}P_{[A}{}^{C}P^{DE}P^{FG}P^{HI}\Pb_{B]}{}^{J}\Pb^{KL}\Pb^{MN} \refpoint{HCNb2} \nonumber
  \tikz[remember picture,overlay,br/.style={decorate,decoration={brace}}] {
    \draw[br] ($(HRZ|-HCNa1)+(0,1em)$) -- ($(HRZ|-HCNa2)-(0,.5em)$) node[midway, anchor=west] {\HCNaT};
    \draw[br] ($(HRZ|-HCNb1)+(0,1em)$) -- ($(HRZ|-HCNb2)-(0,.5em)$) node[midway, anchor=west] {\HCNbT}; } 
  \displaybreak \\&-
 2F_{IKN}F_{CDF}F_{EHL}F_{GJM}P_{[A}{}^{C}P^{DE}P^{FG}P^{HI}P^{JK}\Pb_{B]}{}^{L}\Pb^{MN}  \refpoint{HCNc1} \nonumber \\&-
 4F_{ILN}F_{CDF}F_{EHK}F_{GJM}P_{[A}{}^{C}P^{DE}P^{FG}P^{HI}\Pb_{B]}{}^{J}\Pb^{KL}\Pb^{MN}  \nonumber \\&+
 F_{ILN}F_{CDF}F_{EHJ}F_{GKM}P_{[A}{}^{C}P^{DE}P^{FG}P^{HI}\Pb_{B]}{}^{J}\Pb^{KL}\Pb^{MN}   \nonumber \\&+
 4F_{IJL}F_{CDK}F_{EFM}F_{GHN}P_{[A}{}^{C}P^{DE}P^{FG}P^{HI}\Pb_{B]}{}^{J}\Pb^{KL}\Pb^{MN} \refpoint{HCNc2} \nonumber \\&-
 F_{CDK}F_{EJM}F_{FHL}F_{GIN}P_{[A}{}^{C}P^{DE}P^{FG}P^{HI}\Pb_{B]}{}^{J}\Pb^{KL}\Pb^{MN}   \nonumber \\&+
 2F_{ILN}F_{CDK}F_{EFJ}F_{GHM}P_{[A}{}^{C}P^{DE}P^{FG}P^{HI}\Pb_{B]}{}^{J}\Pb^{KL}\Pb^{MN}  \refpoint{HCNc2} \nonumber \\&+
 F_{DJK}F_{EFH}F_{GIL}P_{[A}{}^{D}P^{CE}P^{FG}P^{HI}\Pb_{B]}{}^{J}\Pb^{KL}F_{C}   \refpoint{FAa}  \nonumber \\&+ 
 2F_{IKL}F_{DFH}F_{EGJ}P_{[A}{}^{D}P^{CE}P^{FG}P^{HI}P^{JK}\Pb_{B]}{}^{L}F_{C}  \refpoint{FAb1} \nonumber \\&+
 2F_{DFK}F_{EHL}F_{GIJ}P_{[A}{}^{D}P^{CE}P^{FG}P^{HI}\Pb_{B]}{}^{J}\Pb^{KL}F_{C}  \refpoint{FAb2} \nonumber \\&+ 
 F_{DFH}F_{EJL}F_{GIK}P_{[A}{}^{D}P^{CE}P^{FG}P^{HI}P^{JK}\Pb_{B]}{}^{L}F_{C}  \refpoint{FAc1} \nonumber \\&-
 2F_{DFK}F_{EHJ}F_{GIL}P_{[A}{}^{D}P^{CE}P^{FG}P^{HI}\Pb_{B]}{}^{J}\Pb^{KL}F_{C}  \nonumber \\&- 
 F_{DFH}F_{EJK}F_{GIL}P_{[A}{}^{D}P^{CE}P^{FG}P^{HI}\Pb_{B]}{}^{J}\Pb^{KL}F_{C}  \nonumber \\&-
 2F_{IJL}F_{DEF}F_{GHK}P_{[A}{}^{D}P^{CE}P^{FG}P^{HI}\Pb_{B]}{}^{J}\Pb^{KL}F_{C} \refpoint{FAc2} \nonumber \\&
+ P \leftrightarrow \Pb\,.
  \tikz[remember picture,overlay,br/.style={decorate,decoration={brace}}] {
    \draw[br] ($(HRZ|-HCNc1)+(0,1em)$) -- ($(HRZ|-HCNc2)-(0,.5em)$) node[midway, anchor=west] {\HCNcT};
    \draw[br] ($(HRZ|-FAa) +(0,1em)$) --  ($(HRZ|-FAa)-(0,.5em)$) node[pos=-.2, , anchor=west] {\dFAFFFaT};
    \draw[br] ($(HRZ|-FAb1)+(0,1em)$) --  ($(HRZ|-FAb2)-(0,.5em)$) node[midway, anchor=west] {\dFAFFFcT};
    \draw[br] ($(HRZ|-FAc1)+(0,1em)$) --  ($(HRZ|-FAc2)-(0,.5em)$) node[midway, anchor=west] {\dFAFFFbT};}
\end{align}

The finite gGS transformation $\Delta_\Lambda \widehat E$ in \eqref{eqn:beta2DL} is a pivotal ingredient the $\alpha'$-corrected PL T-duality transformation rules \cite{\refPLalphaprime}. Thus, it is not surprising that it appears here. In the next subsection, we explain how it is used to extract the metric, $B$-field, and dilaton in the MT scheme. Besides this technical point, it is important to remember that our discussion started from known results for ${\Bb}$ and eventually identified them with the PL duality invariant $\B$ by following the steps
\begin{equation}
  \tikz[baseline=(betabar.center)]{
    \node (betabar) {$\Bb$};
    \node[at=(betabar.east),anchor=west,xshift=6cm] (betahat) {$\Bh$};
    \node[at=(betahat.east),anchor=west,xshift=6cm]  (beta) {$\B$\,.};
    \draw[<->] (betabar.east) -- (betahat.west) node[above,midway] {gauge \& scheme transformation};
    \draw[<->] (betahat.east) -- (beta.west) node[above,midway] {finite gGS scheme transformation};
  }
\end{equation}
Let us stress again that all $\beta$-functions in this diagram capture the same physics. Hence, for practical purposes one might start directly with $\B$ and, if required, reconstruct the much more complicated $\Bb$ by inverting the transformations we found. We will do exactly this for the $\lambda$-deformation below. For completeness, let us just state the results for the two-loop $\beta$-function of the generalized dilaton, either in terms of diagrams
\begin{equation}\label{eqn:diagbeta2d} 
  \b^{(2)d} = \frac14 \FFFFa001011 - \frac14 \FFFFa010010 + \frac14 \FFFFb001100 + 
  \frac13 \FFFFb101001 + \frac14 \FFFAFA0000 + P \leftrightarrow \Pb\,
\end{equation}
or as the tensor expression
\begin{align}
  \b^{(2)d} &= \frac14  F_{ACG} F_{BEH} F_{DIK} F_{FJL} P^{AB} P^{CD} P^{EF} \Pb^{GH} \Pb^{IJ} \Pb^{KL} \nonumber \\
  &-\frac14 F_{ACI} F_{BEJ} F_{DGK} F_{FHL} P^{AB} P^{CD} P^{EF} P^{GH} \Pb^{IJ} \Pb^{KL} \nonumber \\
  &+\frac14  F_{ACI} F_{BEK} F_{DGL} F_{FHJ} P^{AB} P^{CD} P^{EF} P^{GH} \Pb^{IJ} \Pb^{KL} \nonumber \\
  &+\frac13  F_{ACE} F_{BGI} F_{DHK} F_{FJL} P^{AB} P^{CD} P^{EF} \Pb^{GH} \Pb^{IJ} \Pb^{KL} \nonumber\\
  &-\frac14 F_{CEG} F_{DFH} P^{CD} P^{EF} P^{AG} P^{BH} F_{A} F_{B}
  + P \leftrightarrow \Pb \,.
\end{align}

\subsubsection{Renormalizable PL \texorpdfstring{$\sigma$}{sigma}-Models}
Again, the argument from section~\ref{sec:renormalizable} applies: Because both $\B^{(2)E}$ and $\beta^{(2)d}$ satisfy \eqref{eqn:renormalizable}, PL $\sigma$-models are two-loop renormalizable and PL T-duality leaves RG-flows invariant. It would be interesting to see if this result can be extended to more general backgrounds by extending the currently available tools for consistent truncations to include $\alpha'$-corrections. We comment more on this point in the conclusion in section~\ref{sec:conclusion}.

\subsubsection{\texorpdfstring{$\lambda$}{Lambda}- and \texorpdfstring{$\eta$}{Eta}-Deformation}\label{sec:lambda2}
Using \eqref{eqn:diagbeta2} and \eqref{eqn:diagbeta2d}, it is straightforward to calculate the two-loop $\beta$-functions of the $\lambda$- and $\eta$-deformation. A considerable simplification arises because all four components of the generalized fluxes $F_{ABC}$ in \eqref{eqn:genfluxlambda} just differ by a prefactor and $F_A=0$. Therefore, the remaining diagrams in $\beta^{(2)E}_{a\bar b}$ decompose into two contributions: A topological piece, which is independent of the particular projectors involved, and a coefficient capturing the projector structure. We encounter three different diagram topologies. They are denoted by 
\begin{equation}
  A \sim \HCNaT \,, \quad
  B \sim \HCNbT \,, \quad \text{and} \quad
  C \sim \HCNcT \,.
\end{equation}
From this structure it follows that the two-loop $\beta$-function has to have the form
\begin{equation}
  \beta^{(2)E}_{a\bar b} = \frac{2A + B + 2C}{2 ( 16 \kappa k )^2} c_G^2 \eta_{a\bar b}\,,
    \quad \text{or} \quad
  \beta^{(2)k} = 0 \quad \text{and} \quad 
  \beta^{(2)\kappa} = - \frac{2 A + B + 2 C}{2 (16 k)^2 \kappa} c_G^2
\end{equation}
after taking into account \eqref{eqn:betamu}. The coefficients
\begin{equation}\label{eqn:ABC}
  \begin{aligned}
    A &= 2 \left[ ( -1 ) x^2 y^2 + ( +1 ) x y^3 \right]\\
    B &= 2 \left[ ( +2 ) x^2 y^2 + ( -2 +4 ) x y^3 + ( -4 + 2 ) y^4 \right]\\
    C &= 2 \left[ ( +2 -4 +1 )xy^3 + ( +4 +1 -2 )y^4 \right]
  \end{aligned}
\end{equation}
follow directly from the rules: For each vertex in a diagram of \eqref{eqn:diagbeta2} with no dashed propagators (no $\Pb$s) or all dashed propagators (three $P$'s) put a $x=\kappa^2 + 3$, for all other vertices put a $y=\kappa^2 - 1$. Furthermore, every internal $\Pb$ contributes with a minus sign. Note that swapping $P\leftrightarrow \Pb$ neither changes the topology nor the contributing powers of $x$ and $y$ for a diagram. Thus, we just can introduce an overall factor of two on the left hand side of each line in \eqref{eqn:ABC} and restrict the discussion to the 13 diagrams printed in \eqref{eqn:diagbeta2}. Remarkably, this is sufficient to obtain the two-loop $\beta$-function
\begin{align}
  \beta^{(2)k} &= 0 & &\text{and} &
  \beta^{(2)\kappa} &= - \frac{(3\kappa^2 + 1)(\kappa^2 - 1)^3 c_G^2}{128 k^2 \kappa}\,,
\intertext{or equivalently}
  \beta^{(2)\lambda} &= - \frac{\lambda^3 ( 1 - \lambda + \lambda^2 ) c_G^2}{(1 - \lambda)(1 + \lambda)^5 k^2} 
    & &\text{or} &
  \beta^{(2)\eta} &= - \frac{(1-3\eta^2)(1+\eta^2)}{16\eta} (\eta t c_G)^2
  \,.
\end{align}
Our result matches with the one presented in equation (3.9) of \cite{Georgiou:2019nbz} for the $\lambda$-deformation.

For the dilaton the two relevant topologies are
\begin{equation}
  A \sim \FFFFaT \,, \quad \text{and} \quad B \sim \FFFFbT
  \quad \text{with} \quad \beta^{(2)d} = \frac{2 A + B}{2 (16 \kappa k)^2} c_G^2 \dim G\,.
\end{equation}
By applying the same rules as for $\beta^{(2)E}_{a\bar b}$, we obtain
\begin{equation}\label{eqn:lambdadefbeta2d}
  A = - y^4\,, \quad B = \frac12 y^4 - \frac23 x y^3\,, \quad \text{and} \quad
  \beta^{(2)d} = \frac{(1 - \kappa^2)^3 (3 +13 \kappa^2)}{3072 (\kappa k)^2} c_G^2 \dim G\,.
\end{equation}
Fixed points of the RG-flow give rise to CFTs. Their central charge is related to the value of $\beta^d$ as \cite{Metsaev:1987zx}
\begin{equation}\label{eqn:cfrombetad}
  c = 6 \beta^d \,.
\end{equation}
Taking into account that $\beta^{(0)d} = D/6$, where $D$ denotes the dimension of the target space, we extract for the fixed point at $\lambda = 0$ the central charge
\begin{equation}
  c = \dim G \left( 1 - (2 k)^{-1} c_G \right) + \mathcal{O}(k^{-3})
\end{equation}
by combining \eqref{eqn:lambdadefbeta1d} and \eqref{eqn:lambdadefbeta2d}. Matching it with the central charge of the level $\kh$ WZW-model on the Lie group $G$ \cite{Knizhnik:1984nr},
\begin{equation}
  c = \frac{2 \kh \dim G}{2 \kh + c_G}\,,
\end{equation}
we see that $k = \kh + 1/2 c_G$. Again this observation is in agreement with equation (3.8) of \cite{Georgiou:2019nbz}.

\subsection{Finite Generalized Green-Schwarz Transformation}\label{sec:finitegGS}
From a conceptual point of view, finite gGS transformations are straightforward. They just exponentiate the infinitesimal version $\delta_\lambda$. Formally, this was already done in \eqref{eqn:expdeltalambda} but at the end of the day, one needs an explicit prescription how this transformation acts on the metric, $B$-field, and dilaton. This is more complicated than one might initially think because it requires an infinite tower of $\alpha'$-corrections. As we restrict our discussion to $\beta$-functions up to two loops, we can fortunately circumvent this problem and just need to compute the first contribution. More precisely, we consider
\begin{equation}
  e^{\delta_\lambda} \Eh_A{}^I \Eh_{BI} = \Lambda_A{}^C \Lambda_B{}^D ( \Lambda_{CD} + \DL \Eh_{CD} )
\end{equation}
with
\begin{equation}
  \DL \Eh_{AB} = \frac12 \begin{pmatrix} 0 & \DL g_{a\bar b} + \DL B_{a\bar b} \\
    - \DL g_{b \bar a} + \DL B_{b \bar a} & 0
  \end{pmatrix}\,.
\end{equation}
To evaluate the scheme transformation \eqref{eqn:beta2DL} that links $\Bh$ with $\B$, one has to compute $\DL^{(1)} g_{a\bar b}$ and $\DL^{(1)} B_{a\bar b}$, respectively. There are slightly different ways how one can do this \cite{Borsato:2020bqo,Hassler:2020tvz}. Of course all of them lead to the same result \cite{Borsato:2020bqo}
\begin{equation}\label{eqn:finitegGStr}
  \begin{aligned}
    \DL^{(1)} g_{ij} &= - \frac12 \Theta_{(i| \bar a}{}^{\bar b} \omega^{(-)}_{|j)\bar b}{}^{\bar a} + 
      \frac14 \Theta_{i\bar a}{}^{\bar b} \Theta_{j\bar b}{}^{\bar a}\\
    \DL^{(1)} B_{ij} &= - \frac12 \Theta_{[i| \bar a}{}^{\bar b} \omega^{(-)}_{|j]\bar b}{}^{\bar a} + B^\Theta_{ij}
  \end{aligned}
\end{equation}
with
\begin{equation}
  \Theta_{i\bar a}{}^{\bar b} = \partial_i \Lambda^{\bar c}{}_{\bar a} \Lambda_{\bar c}{}^{\bar b}\,, \qquad
  \omega^{(-)}_{i\bar a}{}^{\bar b} = \omega_{i\bar a}{}^{\bar b} - \frac12 H_{i\bar a}{}^{\bar b}\,, 
\end{equation}
and
\begin{equation}
  \dd B^\Theta = -\frac1{12} \Theta_{i\bar a}{}^{\bar b} \Theta_{j\bar b}{}^{\bar c} \Theta_{k\bar c}{}^{\bar a} \dd x^i \wedge \dd x^j \wedge \dd x^k\,.
\end{equation}
In order to keep these equations as simple as possible, we frequently switch between flat and curved indices by contracting with the frame $e_a{}^i = e_{\bar a}{}^i$ from \eqref{eqn:decompframe} or its inverse transpose $e^a{}_i = e^{\bar a}{}_i$. Not surprisingly, the resulting field redefinitions are still quite complicated and cumbersome even for simple, low dimensional examples. Hence, one should rather perform all calculations in the adopted scheme of $\B$. Let us revisit the simplest $\lambda$-deformation on SU(2) \cite{Sfetsos:2013wia} to emphasize this claim.

\subsubsection{SU(2) \texorpdfstring{$\lambda$}{lambda}-Deformation}
Generators in the fundamental representation of the Lie algebra $\mathfrak{su}(2)$ can be conveniently written in terms of the three Pauli matrices $\sigma_a$ as 
\begin{equation}\label{eqn:su(2)gen}
  t_a = - \frac{i}{\sqrt{2}} \sigma_a\,.
\end{equation}
Note that we use an exotic normalisation that results in $c_{\mathrm{SU(2)}} = 4$ rather than the standard value of $2$. It will become obvious shortly that this choice is required to match with the results in \cite{Hoare:2019mcc}. With the group element
\begin{equation}
  g = \begin{pmatrix}
    \sqrt{1-\alpha^2} - i \alpha \cos\gamma\sin\beta & -\alpha ( \cos\beta - i \sin\beta \sin\gamma) \\
    \alpha (\cos\beta + i \sin\beta\sin\gamma) & \sqrt{1-\alpha^2} + i \alpha \cos\gamma\sin\beta
  \end{pmatrix}\,,
\end{equation}
we obtain the leading order metric, $H$-flux and dilaton,
\begin{equation}
  \begin{aligned}
    \dd s^{2(0)} &= \frac{k}{\kappa (1 - \alpha^2)}\dd \alpha^2 + \frac{\alpha^2 \kappa k}{\Delta} 
      \dd s^2(S^2) \\
    H^{(0)} &= \frac{k \alpha^2 \left[ 2 \kappa^2 + (1-\kappa^2) \Delta \right]}{%
      \sqrt{1-\alpha^2}\Delta^2} \dd \alpha \wedge \mathrm{vol}(S^2)\\
    \phi^{(0)} &= - \frac12 \log \Delta & 
    \quad \text{with} \quad \Delta &=\kappa^2 + \alpha^2 (1-\kappa^2)\,,
  \end{aligned}
\end{equation}
after implementing the discussion in section~\ref{sec:genframelambda}. They match equation (3.10) in \cite{Hoare:2019mcc} and fix the normalisation \eqref{eqn:su(2)gen} we use for the generators $t_a$. In order to make the expression more readable, we use the round-two sphere $S^2$ with the metric $\dd s^2(S^2) = \dd \beta^2 + \sin\beta \dd \gamma^2$ and the volume form $\mathrm{vol}(S^2) = \sin\beta \dd \beta \wedge \dd \gamma$ as a reference. Evaluating \eqref{eqn:defbh} for \eqref{eqn:bb1E} and \eqref{eqn:diffgauge1}, one obtains the one-loop $\beta$-function for the metric and $B$-field, which can be written as
\begin{equation}
  \bh^{(1)E}_{ij} = - c_G \frac{(\kappa^2 - 1)^2}{8 k \kappa} \Lambda_{ji}\,,
\end{equation}
where $\Lambda_{ij}$ is the curved version of $\Lambda_{\bar a\bar b}$ in \eqref{eqn:lambdalambdadef}. As expected, this equation agrees with \eqref{eqn:BtoBh}. Another remarkable property, which is not directly obvious at the level of the target space fields, is
\begin{equation}
  \frac{\dd\left(g_{ij}^{(0)} + B_{ij}^{(0)}\right)}{\dd \kappa} = - \frac1{\kappa} \Lambda_{ji}\,.
\end{equation}
It can be used to verify the $\beta$-function for $\kappa$ in \eqref{eqn:beta1kappa} and emphasize that already the one-loop computations involving $\bh$ are more opaque than the ones for $\beta$. In the same vein, one checks the $\beta$-function of the generalized dilaton.

Using \eqref{eqn:finitegGStr}, we evaluate the corrections to the metric and $B$-field,
\begin{equation}
  \begin{aligned}
    \DL^{(1)} \dd s^2 &= \frac{(1-\kappa^2)\Delta + 2 \kappa^2}{\Delta^2} \left[
      (1-\alpha^2)^{-1} \dd \alpha^2 + \alpha^2 \dd s^2 ( S^2 ) \right]\\
    \DL^{(1)} B &= \frac{\alpha^2 \kappa \sin\beta}{\Delta^2 \sqrt{1 - \alpha^2}} \left[
      4 \gamma \dd\alpha\wedge\dd\beta + 2 \alpha (\alpha^2 - 1) (\kappa^2 - 1) 
      \dd\beta\wedge\dd\gamma \right]\,,
  \end{aligned}
\end{equation}
which originate from the finite gGS transformation with the parameter $\Lambda_{\bar a\bar b}$. Combining them with the scheme transformation \eqref{eqn:schemetr}, we obtain the $\alpha'$-corrections
\begin{equation}
  \begin{aligned}
    \dd s^{2(1)} &=  - \frac{8 \kappa^4 - 8 \kappa^2(1 - \kappa^2)\Delta - 3 (1 - \kappa^2)^2 \Delta^2}{
        4 \kappa k \Delta^2} d s^{2 (0)}\\
    H^{(1)} &= \frac{\kappa^4 \left[12 + \Delta(3\Delta-14)\right] + 2 \kappa^2(3-2\Delta)\Delta +
          \Delta^2}{k \left[ \kappa^2 (2 - \Delta) + \Delta \right] \Delta^2} \kappa H^{(0)}\\
    d^{(1)} &= 0\,.
  \end{aligned}
\end{equation}
As a check, one can evaluate the two-loop $\beta$-function \eqref{eqn:bh2dpure}\footnote{Equation~\eqref{eqn:bh2dpure} does not include the infinitesimal diffeomorphism $\xi^{(1) i}(\bh^{(1)B})$ from \eqref{eqn:difftodoubleK} which generates the second term in \eqref{eqn:requireddiff+gauge}. Thus, we add it to get the $\bh^{(2)}$ in \eqref{eqn:bh2dexample}.} for the generalized dilaton for this corrected target space geometry. With the help of the xCoba Mathematica package, we find
\begin{equation}\label{eqn:bh2dexample}
  \bh^{(2)d} = \frac{(1 - \kappa^2)^3 (3 +13 \kappa^2)}{3072 (\kappa k)^2} c_G^2 \dim G \,,
\end{equation}
which matches \eqref{eqn:lambdadefbeta2d}. We could continue to compute the $\beta$-functions of the metric and $B$-field. For them, performing the scheme transformation is more involved. Moreover, one has to account for a further correction from a partial double Lorentz frame fixing, as explained in section~\ref{sec:twoloopsdetails}. Because it will not provided any further insights, we will not present this complicated calculation here.

\section{Doubled Gradient Flow}\label{sec:derivation}
The results in the last section are self-contained and can be used without additional insights into how they were obtained. Still, it is of course interesting to see how we systematically derive expressions like \eqref{eqn:diagbeta1} and \eqref{eqn:diagbeta2}. Thus, we will go step by step through the derivation in the following.

A crucial observation is that it is in general highly complicated to compute O($D$,$D$)-covariant $\beta$-functions directly. To avoid this problem, we exploit the fact that they alternatively arise from a gradient flow,
\begin{equation}\label{eqn:gradientflow}
  \delta_\Psi S = \intvol \Psi \cdot K ( \B )\,,
\end{equation}
where $K(\B)$ is an invertible rank two tensor on the coupling space. In order to obtain $\B$, it is sufficient to know $S$ and $K$. Splitting \eqref{eqn:gradientflow} order by order in $\alpha'$, one finds
\begin{equation}\label{eqn:SandK}
  \begin{aligned}
    \delta_\Psi S^{(1)} &= \intvol \Psi \cdot K^{(0)} (\B^{(1)}) \\
    \delta_\Psi S^{(2)} &= \intvol \Psi \cdot \left[ K^{(0)}(\B^{(2)}) + K^{(1)}(\B^{(1)}) \right]\,.
  \end{aligned}
\end{equation}
Because $K^{(0)}$ does not contain any derivatives, it is just a matrix and can be inverted easily. With the inverse, which is fixed completely by a one-loop computation, it is straightforward to extract $\B^{(1)}$ and $\B^{(2)}$. At a first glance, this route might seem more complicated than just trying to directly rewrite the known results for one and two-loop $\beta$-functions of the bosonic string in a doubled, O($D$,$D$)-covariant way. However, we will see that it is much easier. Especially, since the covariant expressions for $S^{(1)}$ and $S^{(2)}$ are already known \cite{Marques:2015vua}. Furthermore, $K^{(0)}$ follows nearly immediately from known results in DFT. Hence, the remaining challenge is to find $K^{(1)}$ and bring it in an O($D$,$D$)-covariant form. In doing so, a considerable advantage is that $K^{(1)}$ just contains two derivatives and dealing with Bianchi identities simplifies significantly compared to $S^{(2)}$ or $\B^{(2)}$.

We start with the one-loop computation in the next subsection. It contains all the major ingredients of the gradient flow \eqref{eqn:gradientflow} in a simple setting. After introducing all required quantities, we demonstrate how the $\beta$-functions from section~\ref{sec:oneloop} arise. Subsequently, we address the two-loop $\beta$-functions in section~\ref{sec:twoloopsdetails}. They require to additionally discuss scheme transformations, partial double Lorentz gauge fixing, and gGS transformations. Manifest PL symmetry does not only simplifies the $\beta$-functions considerably but also $K$, which governs the gradient flow. Hence, we explain in section~\ref{sec:cfunction} how one computes the $c$-function and the corresponding gradient flow metric of PL $\sigma$-model. In the spirit of section~\ref{sec:results}, we discuss the $\lambda$-deformation as an explicit example.

\subsection{One-Loop}\label{sec:oneloopdetail}
The starting point of our derivation is the one-loop $\beta$-functions $\bh^{(1)E}_{ij}$ and $\bh^{(1)\phi}$ from section~\ref{sec:oneloop}. For convenience, we decompose the former into its metric and $B$-field contribution, $\bh^{E}_{ij} = \bh^{g}_{ij} + \bh^{B}_{ij}$. Hence, the three $\beta$-functions
\begin{align}
    \bh^{(1)g}_{ij} &= R_{ij} - \frac14 H^2_{ij} + 2 \nabla_i \nabla_j \phi \,, \nonumber \\
    \bh^{(1)B}_{ij} &= -\frac12 \nabla_l H^l{}_{ij} + \nabla_l \phi H^l{}_{ij} \,, \nonumber \\  
    \label{eqn:betah1}
    \bh^{(1)\phi}  &= - \frac12 \nabla^2 \phi - \frac1{24} H^2 + (\nabla \phi)^2
\end{align}
form the basis of our discussion. Each line contains an infinitesimal diffeomorphism with $\xi^{(1)i}=\nabla^i \phi$, which relates the respective $\bh$-function to $\bb^{(1)g}_{ij}$, $\bb^{(1)B}_{ij}$ and $\bb^{(1)\phi}$. 

In order to understand how these $\beta$-functions arise from a gradient flow, we vary the one-loop effective target space action, 
\begin{equation}\label{eqn:action1}
  \Sh^{(1)} = \int \dd^D x \sqrt{g} e^{-2\phi} \left( R - \frac1{12} H^2 - 4 (\nabla \phi)^2 + 4 \nabla^2 \phi \right)\,,
\end{equation}
of the bosonic string. By comparing the result
\begin{equation}\label{eqn:varaction1}
  \delta_\Psi \Sh^{(1)} = \intvol \left( - \delta g^{ij} \bh^{(1)g}_{ij} - \delta B^{ij} \bh^{(1)B}_{ij} + 8 \delta d \bh^{(1)d} \right)
\end{equation}
with the first equation in \eqref{eqn:SandK}, we verify that the $\beta$-functions \eqref{eqn:betah1} indeed arise from a gradient flow. Like before, we use the $\beta$-function for the generalized dilaton \eqref{eqn:diagbeta1d} instead of $\bh^\phi$. Moreover, we can easily read off $K^{(0)}(\Bh)$. For the following discussion it is crucial that the first two terms in the integral come both with a minus sign. For the metric, this sign is subtle as we can either vary with respect to the metric or its inverse (both differ by a sign). All metric variations we perform are with respect to the metric $g_{ij}$. Thus, the natural index position for $\delta g$ is $\delta g_{ij}$ and the corresponding $\beta$-function has both indices raised. Due to the superscripts $\bh^{(1)g}$ is carrying, it is usually more convenient to use exactly the opposite notation, like in \eqref{eqn:varaction1}. This is perfectly fine, as long as we keep in mind that the variation is still with respect to the metric and not its inverse.

Our next objective is to rewrite \eqref{eqn:varaction1} in terms of the O($D$,$D$)-covariant quantities from section~\ref{sec:genframes}. To this end, we first obtain the variation of the generalized frame $\Eh_A{}^I$,
\begin{equation}\label{eqn:deltahatE}
  \dEh_A{}^I \Eh_{BI} = \dEh_{AB} + \dEgf_{AB}
\end{equation}
with
\begin{equation}
  \dEh_{AB} = \frac12 \begin{pmatrix}
    0 & \dg_{a\bar b} + \dB_{a\bar b} \\
    -\dg_{b \bar a} + \dB_{b \bar a} & 0
  \end{pmatrix} \qquad \text{and} \qquad
  \dEgf_{AB} = \frac12 \begin{pmatrix}
    \dB_{ab} & 0 \\
    0 & \dB_{\bar a\bar b}
  \end{pmatrix}\,.
\end{equation}
$\delta g_{ab}$ and $\delta B_{ab}$ denote the flattened variations of the metric and $B$-field ($\delta g_{ab} = e_a{}^i e_b{}^j \delta g_{ij}$ and $\delta B_{ab}=e_a{}^i e_b{}^j \delta B_{ij}$). All fluctuations of the generalized frame field in \eqref{eqn:deltahatE} split into two parts because $\Eh_A{}^I$ is partially gauge fixed to a distinguished double Lorentz frame. If we would only apply $\dEh_{AB}$, whose form is identical to the $\Bh_{AB}$ in \eqref{eqn:betaEdoubled}, we would destroy this gauge fixing. Hence, we have to additionally apply the compensating gauge transformation $\dEgf_{AB}$. The same pattern applies to the $\beta$-function and we readily define
\begin{equation}
  \BhEgf = \frac12 \begin{pmatrix}
    \bh^{B}_{ab} & 0 \\
    0 & \bh^{B}_{\bar a\bar b}
  \end{pmatrix}\,.
\end{equation}
Because the one-loop action $\Sh^{(1)}$ is invariant under double Lorentz transformations, gauge fixing terms drop out from the doubled version,
\begin{equation}\label{eqn:varaction1doubled}
  \delta_\Psi \Sh^{(1)} = \intvol \left( \delta\Eh^{AB} K^{(0)}_{AB}{}^{CD} \Bh^{(1)E}_{AB} + 8 \delta d \bh^{(1)d} \right)\,,
\end{equation}
of \eqref{eqn:varaction1}. But they will become relevant at two loops, as we discuss in section~\ref{sec:PDLgf}. From \eqref{eqn:varaction1doubled}, we read off $K^{(0)}$ for the metric and the $B$-field. It takes the surprisingly simple form
\begin{equation}\label{eqn:K0doubled}
  K^{(0)}_{ABCD} = 2 \eta_{AC} \eta_{BD}\,.
\end{equation} 

In DFT, the action $\Sh^{(1)}$ is expressed in terms of the generalized Ricci scalar,
\begin{equation}
  \Sh^{(1)} = \intvol \scRh^{(1)}\,.
\end{equation}
There are two different ways to write $\scRh^{(1)}$, either in terms of a generalized metric or generalized fluxes. We adopt the latter, the flux formulation \cite{Siegel:1993th,Siegel:1993xq,Hohm:2010xe,Geissbuhler:2013uka}, where it reads
\begin{equation}\label{eqn:genR}
  \widehat{\mathcal{R}}^{(1)} = P^{AB} P^{CD} \left( \overline{P}^{EF} + \frac13 P^{EF} \right) \Fh_{ACE} \Fh_{BDF}  + 2 P^{AB} ( 2 \Dh_A \Fh_B - \Fh_A \Fh_B )\,.
\end{equation}
Finally, we compute the variation\footnote{%
The variations are exactly the ones given in \eqref{eqn:variations}. Furthermore, performing integration by parts with
\begin{equation}
  \intvol \Dh_A \cdot = \intvol \Fh_A
\end{equation}
is required.%
} of this action with respect to $\delta\Eh_{AB}$ and $\delta d$, 
\begin{equation}
  \delta_\Psi \Sh^{(1)} = \intvol \left( \delta \Eh^{AB} \widehat{\mathcal{G}}_{AB}^{(1)} - 2 \delta d \widehat{\mathcal{R}}^{(1)} \right)
\end{equation}
with
\begin{equation}
  \mathcal{G}^{(1)}_{AB} = 4 P_{[A}{}^C \Pb_{B]}{}^D \left( \Fh_{CEG} \Fh_{DFH} P^{EF} \Pb^{GH} + \Fh_{CDE} \Fh_E P^{EF} + \Dh_D \Fh_C - \Dh_E \Fh_{CDF} \Pb^{EF} \right)\,,
\end{equation}
and compare the result with \eqref{eqn:varaction1doubled}. One directly reads off $\bh^{(1)E}_{a\bar b} = \widehat{\mathcal{G}}^{(1)}_{a\bar b}$, $\bh^{(1)d}=-\frac14\widehat{\mathcal{R}}^{(1)}$ and thereby obtains the results discussed in section~\ref{sec:oneloop}.

\subsection{Two Loops}\label{sec:twoloopsdetails}
Beyond one-loop, $\beta$-functions become scheme dependent and we have to choose a scheme to start with. There are two popular options for the bosonic string, the Metsaev-Tseytlin \cite{Metsaev:1987zx} (MT) scheme and the Hull-Townsend (HT) scheme. Both are connected by a scheme transformation which is detailed in appendix~\ref{app:HTtoMT}. We found it a little easier to extract $K^{(1)}$ from the results presented in \cite{Metsaev:1987zx} and therefore we start from the two-loop $\beta$-functions in the MT scheme\footnote{Note that $H^2_{ab} = H_{acd}H_b{}^{cd}$, $H^4_{ab} = H_{acd}H^{cef}H_{eg}{}^d H^{g}{}_{bf}$ and that the signs of the last two terms in the first line of \eqref{eqn:betaB2} are flipped compared to \cite{Metsaev:1987zx}. It seems that there happened a misprint in \cite{Metsaev:1987zx}, because the combination of the signs in \eqref{eqn:betaB2} is the one which arises from the variation of the target space effective action in appendix~\ref{app:var2laction}. It is also required to obtain the $B$-field $\beta$-function in the HT scheme after the appropriate scheme transformation (see appendix~\ref{app:HTtoMT} for details).} \cite{Metsaev:1987zx}
\begin{align}\label{eqn:betag2}
  \bb^{(2)g}_{ab} &= 
    \frac{1}{2} \Big[ R_{acde}R_b{}^{cde} + \frac{1}{8} (H^4)_{ab} + 
    \frac{3}{4} \nabla_c H_{ade} \nabla^c H_b{}^{de} + \frac{1}{8} H^{cd}{}_a H_{db}{}^{e}{}(H^2)_{ce} 
    \\ & \qquad
    - R^{cdef} H_{acd} H_{bef} - \frac{5}{2} R_{(a}{}^{cde} H_{b)cf}{} H_{dbe}{}^{f} - 
    \frac{1}{2} R^c{}_{ab}{}^d (H^2)_{cd} + \frac{1}{12} \nabla_a H_{cde} \nabla_b H^{cde} \Big] \,, 
    \nonumber \\ \label{eqn:betaB2}
  \bb^{(2)B}_{ab} &= 
    \frac{1}{4} \Big[ 2R_{[a|cde} \nabla^c H_{|b]}{}^{de} + \nabla^c H^{}_{de[a} H_{b]}{}^{fd} H_{cf}{}^e + 
      2 \nabla_c (H^{2})_{d[a} H_{b]}{}^{dc} - \frac{1}{2} H^2_{cd} \nabla^c H^d{}_{ab} \Big]\,, \\
      \label{eqn:betaphi2}
  \bb^{(2)\phi} &= 
    \frac1{16} \Big[ R_{abcd} R^{abcd} + \frac5{24} H^4 + \frac{4}{3} \nabla_d H_{abc} \nabla^d H^{abc} +
    \frac38 H_{ab}^2(H^2)^{ab} 
    \\ & \qquad\qquad
    - \frac{11}2 R^{abcd} H_{abe} H_{cd}{}^e - 2 H^2_{ab} \nabla^a \nabla^b \phi \Big] \,. \nonumber
\end{align}
Note that we use flat indices because this is more in line with the objects we expect to find in the O($D$,$D$)-covariant rewriting we are looking for. But as the covariant derivative $\nabla_i$ annihilates by construction the frame field $e_a{}^i$, which is used to go from flat to curved indices, switching between the two becomes just a relabeling. Like we have seen in the last subsection, instead of $\Bb$, the gradient flow usually involves a different member in the same equivalence class, $\Bh$, which is obtained by an infinitesimal diffeomorphism and/or a $B$-field gauge transformation. More precisely, we take $\xi^{(2) i} = -\frac1{48}\nabla^i H^2$ and $\chi^{(2)}_i = 0$ in \eqref{eqn:defbh} to fix $\bh^{(2)E}_{ij}$ and $\bh^{(2)\phi}$ respectively.

After a cumbersome computation, which is summarized in appendix~\ref{app:var2laction}, we find that the variation of the two-loop target space effective action
\begin{equation}\label{eq:S2}
  \Sh^{(2)} = \intvol \frac14 \Big[ R_{abcd}R^{abcd} - \frac12 R^{abcd} H_{abe}     H_{cd}{}^{e} + \frac1{24} H^4 - \frac18 (H^2_{ab})^2 \Big]
\end{equation}
gives rise to
\begin{align}\label{eqn:varS2}
  \delta_\Psi \Sh^{(2)} = \intvol \Bigr[& - \delta g^{ab} \bh^{(2)g}_{ab} - \delta B^{ab} \bh^{(2)B}_{ab} + 8 \delta d \b^{(2)d} + \Kh^{(1)d}(\bh^{(1)B})\\
  & + \delta g^{ab} \Kh^{(1)g}_{ab} ( \bh^{(1)g}, \bh^{(1)B} ) + \delta B^{ab} \Kh^{(1)B}_{ab}( \bh^{(1)g}, \bh^{(1)B} )  \Bigl] \nonumber
\end{align}
with
\begin{align}\label{eqn:Kh(1)d}
  \Kh^{(1)d}(\beta^B) &= H^{abc} \nabla_a \beta^B_{bc}\,, \\ \label{eqn:Kh(1)g}
  \Kh^{(1)g}_{ab}(\beta^g,\beta^B) &= 
    -\nabla^2\beta^g_{ab} + \nabla^c \nabla^{\vphantom{g}}_{(a|} \beta^g_{c|b)} - 
    \frac34 H_{acd} H_{be}{}^{d} (\beta^g)^{ec} - \frac14 H^2_{(a|c} (\beta^g)_{|b)}{}^{c} 
    \\ &\quad 
    - 2\left(\nabla_{(a|}\beta^g_{cb)} - \nabla_c \beta^g_{ab}\right)\nabla^c\phi + 
    H^{cd}{}_{(a|} \nabla_c \beta^B_{d|b)} + \frac12 H^{cd}{}_{(a|} \nabla_{|b)} \beta^B_{cd}
    \nonumber \\ &\quad 
    - \frac12\nabla_{(a} H_{b)cd}\beta^{B cd} + \beta^B_{ca}\beta^{B c}{}_{b} \,,
    \nonumber \\ \label{eqn:Kh(1)B}
  \Kh^{(1)B}_{ab}(\beta^g,\beta^B) &=  
    -H_{[a|}{}^{cd} \nabla_c \beta^g_{d|b]} - \frac12 R_{ab}{}^{cd} \beta^B_{cd} - 
    \frac14 H_{abc} H^{dec} \beta^B_{de} + \frac14 H_{acd}H_{be}{}^{c} \beta^{B ed}
    \\ &\quad 
    -\frac12 H_{de[a|}H^{dec}\beta^B_{c|b]} \,. \nonumber
\end{align}
It is actually rather non-trivial to bring $\delta_\Psi \Sh^{(2)}$ into this form. That it is still possible demonstrates the power of the gradient flow equations \eqref{eqn:SandK}.

\subsubsection{Physically Equivalent Choices for \texorpdfstring{$\Kh^{(1)}$}{Khat(1)}}\label{sec:fixKh1}
The expressions \eqref{eqn:Kh(1)d} to \eqref{eqn:Kh(1)B}, we obtained for $\Kh^{(1)}$, cannot be brought into a doubled, O($D$,$D$)-covariant form as given. However, they can be modified to overcome this problem. More specifically there are at least four different ways to change an arbitrary $\Kh$ while keeping the physics it describes unchanged:
\begin{enumerate}[leftmargin=1.5em,itemindent=0cm,itemsep=0cm,label*=\arabic*)]
  \item Assume that the $\beta$-functions are shifted by a combination of an infinitesimal gauge transformation and diffeomorphism which is parameterized by $\Xi^I = \begin{pmatrix} \chi_i & \xi^i \end{pmatrix}$ and mediated by the generalized Lie derivative $\mathcal{L}$, 
    \begin{equation}
      \Bh \rightarrow \Bh + \mathcal{L}_{\Xi} \Bh\,.
    \end{equation}
    Moreover, take $\Xi$ to be a function of the one-loop $\beta$-functions, which contains one additional derivative. If we want to keep the second gradient flow equation in \eqref{eqn:SandK} invariant, we have to adapt $\Kh^{(1)}$ according to
    \begin{equation}
      \Kh^{(1)} \rightarrow \Kh^{(1)} - K^{(0)} \mathcal{L}_{\Xi^{(1)}} \,.
    \end{equation}
    We will do exactly this with
    \begin{equation}\label{eqn:difftodoubleK}
      \xi^{(1) i}(\bh^{(1)B}) = \frac14 H^{ijk} \bh^{(1)B}_{ij} \qquad \text{and} \qquad
      \chi^{(1)}_i(\bh^{(1)B}) =  \frac12 \omega_i{}^{ab} \bh^{(1)B}_{ab}\,.
    \end{equation}
  \item Additionally, the invariance of the action $\Sh$ under generalized diffeomorphisms gives rise to relations between $\beta$-functions. In particular, one can use
    \begin{equation}
      \delta_{\mathcal{L}_\Xi \begin{pmatrix} \Eh & d \end{pmatrix}} \Sh^{(1)} = 0 
    \end{equation}
    to obtain the identities
    \begin{equation}
      \begin{aligned}
        0 &= \bh^{(1)g}{}_{ab} \nabla^b \phi - \frac12 \nabla^b \bh^{(1)g}{}_{ab} +
          \frac14 H_a{}^{bc} \bh^{(1)B}_{bc} - \nabla_a \bh^{(1)d} \\
        0 &= \nabla^b \Bigl( e^{-2\phi} \bh^{(1)B}_{ab} \Bigr)\,.
      \end{aligned}
    \end{equation}
  \item Shifting $\Kh^{(1)B}_{ab}(\beta^g, \beta^B)$ by
    \begin{equation}
      \frac12 \left( \bh^{(0)B}_{c[a} \beta^g{}_{b]}{}^c - 
        \beta^{B c}{}_{[a} \bh^{(1)g}_{b]c} \right)
    \end{equation}
    does not affect \eqref{eqn:varS2}, because for $\beta^g=\bh^{(1)g}$ and $\beta^B=\bh^{(1)B}$ it vanishes.
  \item Eventually, we perform a scheme transformation from the MT scheme to the generalized Bergshoeff-de Roe scheme (gBdR). This transformation is required to bring the action $\Sh^{(2)}$ into an O($D$,$D$)-covariant form \cite{Marques:2015vua}. Thus, it is natural to apply it to $\Kh^{(1)}$, too. In our conventions, this transformation is parameterized by
  \begin{equation}\label{eqn:MTtogBdR}
    \begin{aligned}
      \Delta^{(1)} g_{ij} &= - \frac12 \omega_{ia}{}^b \omega_{jb}{}^a + \frac38 H^2_{ij}\,,\\
      \Delta^{(1)} B_{ij} &= - \bh^{B(1)}_{ij} - \frac12 H_{[ia}{}^b \omega_{j]b}{}^a \,, \qquad &
      \Delta^{(1)} d &= 0
    \end{aligned}
  \end{equation}
  and implemented by the Lie derivative on the coupling space. We already discussed the latter for $\beta$-functions. Here, we extend it in the canonical way to $K^{(1)}$, namely
  \begin{equation}
    K^{(1)}(\Psi', \B) \rightarrow K^{(1)}(\Psi', \B) + L_{\Psi^{(1)}} K^{(0)}(\Psi', \B) 
  \end{equation}
  with
  \begin{equation}\label{eqn:LieK(0)}
    \begin{aligned}
      L_\Psi K^{(0)}(\Psi', \B) = &K^{(0)}(\delta_{\Psi'} \Psi, \B) + 
        K^{(0)}(\Psi', \delta_\B \Psi) + \\
        &K^{(0)}(T(\Psi',\Psi), \B) + K^{(0)}(\Psi',T(\B,\Psi))\,,
    \end{aligned}
  \end{equation}
  where we understand $K$ as a pairing between two vectors,
  \begin{equation}
    K(\Psi, \B) = \intvol \Psi\cdot K(\B)\,,
  \end{equation}
  on the infinite dimensional coupling space. Evaluating \eqref{eqn:LieK(0)} for \eqref{eqn:MTtogBdR} is cumbersome, especially because \eqref{eqn:MTtogBdR} contains Lorentz symmetry violating terms. We approach this challenge by writing the one-loop $\beta$-functions in terms of the spin connection $\omega_{abc}$, the flat derivative $\Dh_a$, $\Fh_a$ from \eqref{eqn:Fha} and the $H$-flux $H_{abc}$ with the following, non-vanishing, variations
  \begin{equation}
    \begin{aligned}
      \delta_\Psi \omega_{abc} &= D_{[c} \delta g_{b]a} + \delta g_{d[b} \omega_{c]a}{}^d + \delta g_{ad} \omega_{[cb]}{}^d - \frac12 \delta g_{ad} \omega^d{}_{bc}\,, \\
      \delta_\Psi H_{abc} &= - \frac32 \delta g_{[a}{}^d H_{bc]d} + 3 \nabla_{[a} \delta B_{bc]}\,, \\
      \delta_\Psi \Fh_{a} &= \sqrt{2} \Dh_a \delta d + \frac1{2\sqrt{2}} \Dh_b \delta g_a{}^b - \frac12 \Fh_b \delta g_a{}^b \qquad \text{and} \qquad & [ \delta_\Psi, \Dh_a ] = - \frac12 \delta g_a{}^b \Dh_b \,.
    \end{aligned}
  \end{equation}
\end{enumerate}

To keep the following discussion more tractable, we split $K$ into a symmetric and an antisymmetric part,
\begin{equation}
  K_\pm(\Psi, \Psi') = \frac12 \left[ K(\Psi, \Psi') \pm K(\Psi', \Psi) \right]\,.
\end{equation}
Most of $\Kh^{(1)}_+$ can be re-expressed in terms of O($D$,$D$)-covariant quantities. Unfortunately, the situation for the asymmetric part is much worse. Hence, one might hope that there is a way to get rid of $\Kh^{(1)}_-$ and, while doing so, also to obtain the missing terms that are required to complete the doubling of $\Kh^{(1)}_+$. Remarkably, this is indeed possible by applying a scheme transformation, which is linear in the one-loop $\beta$-functions, namely
\begin{equation}\label{eqn:2ndschemetr}
    \Delta^{(1)} g_{ij} = 0 \,, \qquad
    \Delta^{(1)} B_{ij} = \bh^{B(1)}_{ij}\,,\qquad 
    \Delta^{(1)} d = 0\,.
\end{equation}
But instead of applying it to all quantities in \eqref{eqn:SandK}, we only transform the $\beta$-functions
\begin{equation}
  \Bh'^{(2)} = \Bh^{(2)} + L_{\Psi'} \Bh^{(1)}\,.
\end{equation}
While the first equation of \eqref{eqn:SandK} is not affected by this transformation, the second one becomes
\begin{equation}
  \delta_\Psi \Sh^{(2)} = \intvol \Psi \left[ K^{(0)}( \Bh'^{(2)} ) + \Kh'^{(1)} ( \Bh^{(1)} ) \right]
\end{equation}
with
\begin{equation}
  \Kh'^{(1)}(\B^{(1)}) = \Kh^{(1)} (\B^{(1)}) - K^{(0)}(L_{\Psi'} \B^{(1)})\,.
\end{equation}
Here $\Psi'$ generates the scheme transformation \eqref{eqn:2ndschemetr} and, as intended, all terms of $K'^{(1)}_-$ vanish. For the sake of brevity, we drop the prime from now on.

\subsubsection{\texorpdfstring{O($D$,$D$)}{O(D,D)}-Covariant Rewriting of \texorpdfstring{$\Kh^{(1)}$}{Khat(1)}}\label{sec:matchingdiagrams}%
\begin{table}[!t]
  \centering
  {\tabulinesep=1.2mm
\begin{tabu}{ |c|c|c|c|c|c|c|c|c|c| } 
  \hline
  & & & & \multicolumn{3}{|c|}{\# of diagrams} &  \multicolumn{3}{|c|}{terms in $\Kh^{(1)}$ to match}\\
  \hline
  type & $\Fh_{ABC}$ & $\Fh_A$ & $\Dh_A$ & class $A$ & class $B$ & class $C$ &
    class $A$ & class $B$ & class $C$ \\ \hline
  I   & 0 & 0 & 2 & 0& 1& 1 & 0 & 2 & 2\\
  II  & 1 & 0 & 1 & 4& 4& 0 & 9 & 16 & 0\\
  III & 0 & 1 & 1 & 0& 4& 2 & 0 & 2 & 2\\
  IV  & 2 & 0 & 0 & 3& 3& 2 & 16 & 16 & 0\\
  V   & 1 & 1 & 0 & 2& 2& 0 & 3 & 8 & 0\\
  VI  & 0 & 2 & 0 & 0& 1& 1 & 0 & 0 & 0\\
 \hline
\end{tabu}}
\caption{Different combinations of the three tensors $\Fh_{ABC}$, $\Fh_{A}$, and $\Dh_A$ with the number of different possible diagrams obtained by combining two of them. We reference each combination with a Roman numeral from I to VI and further specify one of three classes, $A$, $B$, or $C$.\label{tab:diagclasses}}
\end{table}%
Written in terms of the spin connection $\omega_{abc}$, the flat derivative $\Dh_a$, the $H$-flux $H_{abc}$, and $\Fh_a$, $\Kh^{(1)}(\B^{(1)})$ consists of 76 different terms which can be recast using exclusively $P^{AB}$, $\Pb^{AB}$, $\Fh_{ABC}$, $\Fh_A$ and $\Dh_A$. For this job, the diagrams, which we have introduced in section~\ref{sec:oneloop}, are a convenient tool because they make keeping track of all different terms which could possibly appear much easier. Hence, we first have to  determine all diagrams with
\begin{enumerate}[leftmargin=1.5em,itemindent=1em,itemsep=0cm,label*=\arabic*)]
  \item two external legs, one with a $P$ and the other one with a $\Pb$
  \item internally $\beta^{E}_{a\bar b} = \tikz[baseline={($(beta.west)+(0,-4pt)$)}]{
      \node[draw=gray!60!black,shape=circle,inner sep=7pt,{label={center:{\Large $\beta$}}}] (beta) {};
      \draw[dashed] (beta.east) -- +(1,0);
      \draw (beta.west) -- +(-1,0);
    }$, representing the argument of $\Kh^{(1)}(\B)$
  \item two derivatives\,.
\end{enumerate}
$\Fh_{ABC}$, $\Fh_A$ and $\Dh_A$ contribute one derivative each. Thus, only two of them can be present in a diagram at a time. This results in six different combinations that we call types and number by roman numerals. Moreover, there are three different classes of diagrams where $\beta^{(1)E}_{a\bar b}$ is connected to
\begin{enumerate}[leftmargin=1.5em,itemindent=1em,itemsep=0cm,label*=\Alph*)]
  \item no external leg
  \item one external leg
  \item both external legs\,.
\end{enumerate}
Finally, note that all diagrams have to come in pairs because $\Kh^{(1)}$ is even under the $\mathbb{Z}_2$ symmetry defined in \eqref{eqn:Z2}. Hence, we only draw one diagram of each pair and understand that it has to be complemented by its partner, which arises under the swapping $P \leftrightarrow \Pb$. The resulting number of admissible diagrams for all types and the corresponding classes is summarized in table \ref{tab:diagclasses}. Going through this list, we find the factors listed in table~\ref{tab:diagcoeffs} in front of the relevant diagrams by equating coefficients. At this point, we have to refine our prescription to construct the diagrams slightly because diagrams of type I have two derivatives acting on $\beta^{E}_{a\bar b}$. But these two derivatives do not commute and thus we have to decide which one comes first. Our convention is that we go from top to bottom and left to right. The order how we encounter derivatives is the order we write them down in the tensorial expression.
\begin{table}[!t]
  \centering
  {\tabulinesep=1.2mm
  \begin{tabu}{|l|l|l|l|l|l|}
    \hline
    \multicolumn{2}{|c|}{type} & \multicolumn{4}{|c|}{diagrams $-$ $P\leftrightarrow \Pb$}  \\ \hline  
    I   & B & \multicolumn{4}{|l|}{$\co{+ 2} \DDl0 $}\\   \hline 
    I   & C & \multicolumn{4}{|l|}{$\co{- 2} \DADA0$} \\  \hline 
    II  & A & \multicolumn{4}{|l|}{$\co{- 2} \FDla10 \co{+ 0} \DFa01  \co{+ 2} \DFb01 \co{+ 2} \DFc01 $} \\  \hline 
    II  & B & \multicolumn{4}{|l|}{$\co{- 2} \FDlb00 \co{+ 2} \FDlb01 \co{+ 4} \DFd11 \co{+ 4} \DFd01 $} \\  \hline 
    III & B & \multicolumn{4}{|l|}{$\co{+ 0} \FADla0 \co{- 2} \DFAa0  \co{+ 0} \DFAb0 \co{+ 0} \DFAc0 $} \\  \hline 
    III & C & \multicolumn{4}{|l|}{$\co{+ 2} \DFAd0  \co{+ 0} \FADlb0 $} \\  \hline 
    IV  & A & \multicolumn{4}{|l|}{$\co{+ 0} \FFa010 \co{+ 2} \FFc001 \co{+ 4} \FFc010 $} \\  \hline 
    IV  & B & \multicolumn{4}{|l|}{$\co{- 1} \FFb000 \co{+ 2} \FFb010 \co{- 1} \FFb110$} \\   \hline 
    V   & A & \multicolumn{4}{|l|}{$\co{- 2} \FFAb10 \co{+ 0} \FFAa01$} \\   \hline 
    V   & B & \multicolumn{4}{|l|}{$\co{+ 2} \FFAc00 \co{- 2} \FFAc10$} \\   \hline
  \end{tabu}}
  \caption{Diagrams which might contribute to the doubling of $\Kh^{(1)}$ and their respective coefficients. The equivalent tensor expression is given in \eqref{eqn:K1doubled}.}
  \label{tab:diagcoeffs}
\end{table}%
To avoid any confusion and since it is a main result of our work, the explicit tensor expression corresponding to the diagrams in table~\ref{tab:diagcoeffs} reads
\begin{align*}\label{eqn:K1doubled}
    \Kh^{(1)}_{AB}&(\B) = -2P_{[A}{}^{C}P^{DE}\Pb_{B]}{}^{F}D_{E}D_{C}\B_{DF}
 +2P_{[A}{}^{C}P^{DE}\Pb_{B]}{}^{F}D_{E}D_{D}\B_{CF} \\&
 +2F_{CFH}P_{[A}{}^{C}P^{DE}\Pb_{B]}{}^{F}\Pb^{GH}D_{E}\B_{DG}
 +2P_{[A}{}^{C}P^{DE}\Pb_{B]}{}^{F}\Pb^{GH}\B_{DG}D_{E}F_{CFH} \\&
 +2F_{EFH}P_{[A}{}^{C}P^{DE}\Pb_{B]}{}^{F}\Pb^{GH}D_{C}\B_{DG}
 -2P_{[A}{}^{C}P^{DE}P^{FG}\Pb_{B]}{}^{H}\B_{DH}D_{G}F_{CEF} \\&
 -2P_{[A}{}^{C}P^{DE}\Pb_{B]}{}^{F}\Pb^{GH}\B_{CG}D_{E}F_{DFH}
 -4F_{EFH}P_{[A}{}^{C}P^{DE}\Pb_{B]}{}^{F}\Pb^{GH}D_{D}\B_{CG} \\&
 -4F_{CEG}P_{[A}{}^{C}P^{DE}P^{FG}\Pb_{B]}{}^{H}D_{F}\B_{DH}
 +2P_{[A}{}^{D}P^{CE}\Pb_{B]}{}^{F}F_{C}D_{D}\B_{EF} \\&
 -2P_{[A}{}^{D}P^{CE}\Pb_{B]}{}^{F}F_{C}D_{E}\B_{DF}
 +2F_{CHI}F_{EFJ}P_{[A}{}^{C}P^{DE}\Pb^{IJ}\Pb_{B]}{}^{F}\Pb^{GH}\B_{DG} \\&
 -4F_{CEI}F_{FHJ}P_{[A}{}^{C}P^{DE}\Pb^{IJ}\Pb_{B]}{}^{F}\Pb^{GH}\B_{DG}
 +F_{DGI}F_{FHJ}P_{[A}{}^{C}\Pb^{IJ}\Pb_{B]}{}^{D}\Pb^{EF}\Pb^{GH}\B_{CE} \\&
 -2F_{DFI}F_{EHJ}P_{[A}{}^{C}P^{DE}\Pb^{IJ}\Pb_{B]}{}^{F}\Pb^{GH}\B_{CG}
 +F_{DFH}F_{EGJ}P_{[A}{}^{C}P^{DE}P^{FG}\Pb^{IJ}\Pb_{B]}{}^{H}\B_{CI} \\&
 -2F_{DFH}P_{[A}{}^{D}P^{CE}\Pb_{B]}{}^{F}\Pb^{GH}F_{C}\B_{EG}
 +4F_{EFH}P_{[A}{}^{D}P^{CE}\Pb_{B]}{}^{F}\Pb^{GH}F_{C}\B_{DG} \\&
 - P \leftrightarrow \Pb \,. \stepcounter{equation}\tag{\theequation}
\end{align*}

\subsubsection{Partial Double Lorentz Gauge Fixing}\label{sec:PDLgf}
There are still a few terms in $\Kh^{(1)}(\B)$ which cannot be matched by the procedure above. However, we will now show that they are just an artifact of the partially double Lorentz fixed generalized frame field $\Eh_A{}^I$ used in the calculation. Its variation \eqref{eqn:deltahatE} contains the compensating double Lorentz transformation $\dEgf_{AB}$ and by restricting \eqref{eqn:SandK} to it, we find
\begin{equation}\label{eqn:S(2)DL}
  \delta_{\Psi^{\mathrm{gf}}} \Sh^{(2)} = \intvol \dEgf \cdot \Kh^{(1)}( \Bh^{(1)} )\,.
\end{equation}
Since $\Sh^{(2)}$ is not invariant under double Lorentz transformations, $\Kh^{(1)}$ has to have contributions which relate physical degrees of freedom with gauge transformations. We fix them by remembering that $\Sh^{(2)}$ has been constructed such that the relation \cite{Marques:2015vua}
\begin{equation}
  \delta_{\Psi^{\mathrm{gf}}} \Sh^{(2)} = \intvol ( \A^{(1)}_{\Psi^{\mathrm{gf}}} \Eh ) \bh^{(1)E}
\end{equation}
holds. An analogous mechanism governs gauge fixed, two-loop $\beta$-functions, too. More precisely, they split into the two contributions
\begin{equation}\label{eqn:GFdecompbetaE}
  \Bh^{E(2)} = \Bh'^{E(2)} + \A^{(1)}_{\Bh^{(1)E\mathrm{gf}}} \Eh\,,
\end{equation}
where $\Bh'^{(1)E}$ is not gauge fixed. In the final result, we neither want to include \eqref{eqn:S(2)DL} nor the second term on the left hand side of \eqref{eqn:GFdecompbetaE}. The reason is that we are looking for two-loop $\beta$-functions which do not depend on a particular gauge fixing. All terms that we therefore drop can be neatly combined in the symmetric, double Lorentz gauge fixing term
\begin{equation}
  \widehat{K}^{(1)\mathrm{gf}}(\delta\Eh,\Bh) = \intvol \left[ -\frac14 \left( \delta g^a{}_b H^{bcd} - 2 \delta B^a{}_b\omega^{bcd} \right) D_a \bh^B_{cd} + \left( \bh^E \leftrightarrow \delta E \right) \right]
\end{equation}
and we eventually find that $\Kh^{(1)}$ can be written as
\begin{equation}
  \Kh^{(1)}(\delta\Eh, \Bh) = \intvol \delta \Eh^{AB} \Kh^{(1)}_{AB}(\bh) + \Kh^{(1)\mathrm{gf}}( \delta\Eh, \bh )\,.
\end{equation}
Note that we have dropped the prime on the $\Bh'^E$ to avoid cluttering our notation. From now on all doubled $\beta$-functions are free of any gauge fixing.

\subsubsection{Extracting the \texorpdfstring{$\beta$}{beta}-Functions}
Since, we have been successful in writing $\Kh^{(1)}_{AB}(\B)$ in the O($D$,$D$)-covariant form \eqref{eqn:K1doubled}, it is straightforward to compute the two-loop $\beta$-functions. The procedure goes along the same line as at one-loop in section~\ref{sec:oneloopdetail}: First, we rewrite the gradient flow \eqref{eqn:varS2} in terms of doubled quantities. More specifically, we take the components of
\begin{equation}
  \Kh^{AB}(\B) = \begin{pmatrix}
    0 & \Kh^{a\bar b}(\B) \\
    -\Kh^{b\bar a}(\B) & 0 
  \end{pmatrix}\,,
    \qquad \text{with} \qquad
    \Kh^{a\bar b}(\B) = \Kh^{g\, a\bar b}(\B) + \Kh^{B\, a\bar b}(\B)\,,
\end{equation}
to rewrite \eqref{eqn:varS2} as
\begin{equation}
  \delta_\Psi \Sh^{(2)} = \intvol \left[ \delta E^{AB} K^{(0)}_{AB}{}^{CD} \left( \bh^{(2)E}_{CD} - \frac12 \Kh^{(1)}_{CD}(\Bh) \right) + 8 \delta d  \bh^{(2)d} \right]\,.
\end{equation}
For the discussion in section~\ref{sec:fixKh1}, we know that the action $\Sh^{(2)}$ has to be in the gBdR scheme to be compatible with our $\Kh^{(1)}_{AB}$ from section~\ref{sec:matchingdiagrams}. In this scheme, it can be written in the O($D$,$D$)-covariant form \cite{Marques:2015vua,Baron:2017dvb}
\begin{equation}
  \Sh^{(2)} = \int d^D x e^{-2d} \scRh^{(2)} \qquad \text{with} 
  \qquad \scRh^{(2)} = - \scRh^+ - \scRh^- \,,
\end{equation}
where the explicit expression for $\scRh^\pm$ is given in (2.33) of \cite{Baron:2017dvb}. In terms of diagrams $\scR^{(2)}$ reads
\begin{equation}
  \begin{array}{lll}
    \scRh^{(2)} =
    &\co{-}  \FFFFa100110 \co{+} \FFFFa010010 &\co{-\frac43} \FFFFb101001 \co{-} \FFFFb001100 \co{-}\FFFAFA1001 \\[0.8cm]
      &\co{+4} \FFFD10010   \co{-} \FFFD01100 &\co{+}  \FFFD11100 \co{+2} \FFFADl1001 \co{+2}\FFFADb1010 \\[0.8cm]
      &\co{+2} \FFFADa1100  \co{+\frac12} \DDFF0010 &\co{-\frac12} \DDFF0011 \co{+} \FFDDl1010 \co{+} \FFDDlprime1010 \\[0.8cm]
      &\co{-} \FFDlDl1001 + P \leftrightarrow \Pb \,.
  \end{array}
\end{equation}
All that is left to be done is compute the variation of this action. It has the form
\begin{equation}
  \delta_\Psi \Sh^{(2)} = \intvol \left( \delta E^{AB} \scGh^{(2)}_{AB} - 2 \delta d \scRh^{(2)} \right)
\end{equation}
and immediately allows for the identification
\begin{equation}
  \bh^{(2)E}_{a\bar b} = \scGh^{(2)}_{a\bar b} + \widehat{K}^{(1)}_{a\bar b}( \Bh^{(1)} )\,,
    \qquad
  \bh^{(2)d} = - \frac14 \scRh^{(2)}\,.
\end{equation}
We already computed $\Kh^{(1)}_{a\bar b}$ and therefore we only need $\scGh^{(2)}_{a\bar b}$ to obtain the final result. We compute it with the xTensor package of the xAct suite and get the results presented in section~\ref{sec:twoloops}.

\subsubsection{Generalized Green-Schwarz Transformation}
The last thing we have to do to make full contact with section~\ref{sec:twoloops} is to prove that \eqref{eqn:DLbeta2} holds. To this end, we take a closer look at the identity 
\begin{equation}\label{eqn:deltaPsiLchiS2}
    \delta_\Psi L_{\chi^{(1)}} \Sh^{(1)} = L_{\chi^{(1)}} ( \widehat{K}^{(0)} ) ( \Psi, \Bh^{(1)} ) +
      \widehat{K}^{(0)}( \Psi, L_{\chi^{(1)}} \Bh^{(1)} )\,,
\end{equation}
which arises if we apply $L_\chi$ to both sides of the first equation in \eqref{eqn:SandK}. We now identify $\chi^{(1)} = \Al^{(1)} \begin{pmatrix} \Eh & d \end{pmatrix}$ to further simply this relation by using
\begin{equation}
  L_{\chi^{(1)}} \Sh^{(1)} = \Al^{(0)} \Sh^{(2)} \qquad \text{and} \quad
  L_{\chi^{(1)}} \Bh^{(1)} = \Al^{(0)} \Bh^{(2)}\,,
\end{equation}
which are equivalent to \eqref{eqn:DLS2} and \eqref{eqn:DLbeta2}, respectively. Together with \eqref{eqn:deltaPsiLchiS2} they imply
\begin{equation}
  \delta_\Psi \Al^{(0)} \Sh^{(2)} = L_{\chi^{(1)}} (\widehat{K}^{(0)}) (\Psi, \Bh^{(1)} ) +
    \widehat{K}^{(0)} (\Psi, \Al^{(0)} \Bh^{(2)}) = \Al^{(0)} \delta_\Psi \Sh^{(2)}\,.
\end{equation}
Note that we are able to swap $\delta_\Psi$ and $\Al$ because the variation parameter $\Psi$ does not transform anomalously under double Lorentz transformations and therefore $\Al \Psi = 0$ holds. This equation can be alternatively obtained by applying $\Al^{(0)}$ to the left and right side of the second equation of \eqref{eqn:SandK}, if we further impose
\begin{equation}\label{eqn:DLK0andK1}
  L_{\Al^{(1)} \begin{pmatrix} \Eh & d \end{pmatrix}} \widehat{K}^{(0)} = \Al^{(0)} \widehat{K}^{(1)} \,.
\end{equation}
Equally, one might conclude that if this identity holds for $\widehat{K}^{(0)}$ and $\widehat{K}^{(1)}$, it implies \eqref{eqn:DLbeta2}. This result is not very surprising, because we expect $\Kh$, like the action and the $\beta$-functions, to transform covariantly under gGS transformation. Indeed one can check that the expressions we have presented in \eqref{eqn:K0doubled} and \eqref{eqn:K1doubled} satisfy \eqref{eqn:DLK0andK1}. This result provides an important consistency check. Moreover, it would be interesting to see if, similar to the action $\Sh^{(2)}$, $\widehat{K}^{(1)}$ can be completely fixed by just imposing its covariance under gGS transformations.

\subsection{\texorpdfstring{$c$}{c}-Function and Gradient Flow Metric}\label{sec:cfunction}
We argue in section~\ref{sec:renormalizable} that PL symmetry restricts the $\sigma$-model $\beta$-functions to a finite dimensional subspace of the coupling space. The same is true for $K_{AB}(\B)$, which looses all derivatives on a PL symmetric background and thus can be written as
\begin{equation}
  K(\Psi, \B)= \delta E_{AB} \beta^{E}_{CD} K^{ABCD} V\,,
    \qquad \text{with} \qquad
  V = \intvol\,.
\end{equation}
Here, $K^{ABCD}$ only depends on the couplings that enter through $F_{ABC}$ and $F_A$. In the same vein, we rewrite the low-energy effective target space action,
\begin{equation}\label{eqn:Scfunction}
  S = V \scR = -4 V \beta^d = -\frac23 V c\,,
\end{equation}
where the last identity originates from \eqref{eqn:cfrombetad}. Now, the gradient flow \eqref{eqn:gradientflow} takes a form that matches (14) in Zamolodchikov's famous paper \cite{Zamolodchikov:1986gt}, namely
\begin{equation}
  \partial_\nu c  = 12 G_{\mu\nu} \beta^\nu
\end{equation}
with the gradient flow metric
\begin{equation}\label{eqn:Zmetric}
  G_{\mu\nu} = - \frac18 J_\mu{}^{AB} J_\nu{}^{CD} K_{ABCD}\,,
\end{equation}
and the Jacobian
\begin{equation}
  J_\mu{}^{AB} = \partial_\mu E^{AI} E^B{}_I\,.
\end{equation}

Because $K^{(n)}_{ABCD}$ is symmetric under the exchange of the indices $AB\leftrightarrow CD$, $G^{(n)}_{\mu\nu}$ is a symmetric tensor, at least for $n=0\,, 1$. Hence, one might conclude that the latter is the Zamolodchikov metric \cite{Zamolodchikov:1986gt}. But the gradient flow away from the conformal point has a more general form and incorporates corrections \cite{Friedan:2009ik}. Therefore, we prefer the term gradient flow metric for $G_{\mu\nu}$. On the other hand, the action $S$ in \eqref{eqn:Scfunction} has the ``central charge'' form of \cite{Tseytlin:1987bz} and thus, what we call $c$ should match Zamolodchikov's definition. A thorough comparison between the quantities, we identified here, and results from the fixed point CFT and its conformal perturbation theory is required to settle these points completely. This analysis is beyond the scope of this chapter. But as a first step, we discuss the $\lambda$-deformation in the following, which was extensively studied from a CFT perspective \cite{Itsios:2014lca,Georgiou:2016iom,Georgiou:2016zyo}.

\subsubsection{\texorpdfstring{$\lambda$}{Lambda}-Deformation}\label{sec:lambdagradient}
We already have computed $c$ of the $\lambda$- and $\eta$-deformation for one and two loops in the sections~\ref{sec:lambda1} and \ref{sec:lambda2}, respectively. For convenience, we repeat it here,
\begin{equation}
  c = D - \frac{1 + 2 \lambda + 2 \lambda^3 + \lambda^4}{2 k (1 - \lambda) (1 + \lambda)^3} c_G D + 
    \frac{\lambda^3 (4 - 5\lambda + 4\lambda^2 )}{2 k^2 (1-\lambda)^2 (1+\lambda)^6} c_G^2 D\,,
\end{equation}
in terms of $\lambda$ instead of $\kappa$. While the first two terms match (3.30) of \cite{Georgiou:2019nbz} perfectly, the last term deviates. A possible explanation is that our $c$ and theirs actually capture different quantities. The derivation of $c$ in \cite{Georgiou:2019nbz} starts from the Zamolodchikov metric, obtained by conformal perturbation theory. Combining the Zamolodchikov metric and the $\beta$-functions, $\partial_\mu c$ is calculated and then integrated to obtain $c$. As explained above, our $G_{\mu\nu}$ is expected to differ from the Zamolodchikov metric away from the conformal point.

Because there is only one coupling that flows, $G_{\mu\nu}$ is solely formed by $G_{\lambda\lambda}$. Evaluating \eqref{eqn:Zmetric} with
\begin{equation}
  J_\lambda{}^{AB} = \frac1{1-\lambda^2}\begin{pmatrix} 0 & \eta_{a\bar b} \\
    - \eta_{b\bar a} & 0 \end{pmatrix}
\end{equation}
works along the same line as for the $\beta$-functions. 
The result 
\begin{equation}\label{eqn:Gll}
  G_{\lambda\lambda} = \frac{D}{2(1 - \lambda^2 )^2} \left( 1 + \frac{Q(\lambda)}{k (1-\lambda) (1+\lambda)^3} c_G \right)\,.
\end{equation}
matches (3.16) of \cite{Georgiou:2019nbz}. There it is argued that the function $Q$ have to have the form
\begin{equation}
  Q(\lambda) = c_0 + c_1 \lambda + c_2 \lambda^2 + c_1 \lambda^3 + c_0 \lambda^4
\end{equation}
to be compatible with the symmetry $\lambda \leftrightarrow \lambda^{-1}$, $k \leftrightarrow -k$. We find a $Q(\lambda)$ of this form, but instead of $c_0=c_1=c_2=0$ \cite{Georgiou:2019nbz}, we obtain
\begin{equation}
  c_0 = -1\,, \qquad c_1 = 2\,, \quad \text{and} \quad c_2 = -4\,.
\end{equation}
This is not very surprising because already our $c$-function is different from theirs.

It should be possible to better understand this discrepancy by using alternative techniques to obtain the values of these coefficients. In particular, $c_0$ is accessible from the level $\kh$ WZW-model on the group manifold $G$, which arises at the fixed point $\lambda=0$ of the RG flow. At this distinguished point, the marginal operator that triggers the flow is
\begin{equation}\label{eqn:Olambda}
  \mathcal{O}_\lambda(z,\zb) = \frac{\gamma}{k} \eta_{a\bar b} \, j^a(z) \jb^{\bar b}(\zb) \,,
\end{equation}
where $\gamma$ is a numerical factor. Most important is that $\mathcal{O}_\lambda$ is proportional to $k^{-1}$ and not $\kh$. This dependence enters through the left and right invariant forms \eqref{eqn:leftrightinv}. The Ka\v{c}-Moody currents, which $\mathcal{O}_\lambda$ is formed of, are governed by the OPE
\begin{equation}
  j^a(z) j^b(w) = \frac{\kh \eta^{ab}}{(z-w)^2} + \frac{f^{ab}{}_c}{z-w} j^c(w) + \dots \,.
\end{equation}
The anti-chiral currents $\jb^a(\zb)$ are governed by an analogous version. Moreover, they commute with all chiral currents $j^b(z)$. We now know everything we need to compute the Zamolodchikov metric \begin{equation}
  G_{\lambda\lambda}(0) = \lim_{z\rightarrow w} | z - w |^4 \left\langle \mathcal{O}_\lambda(z,\zb) \mathcal{O}_\lambda(w,\wb) \right\rangle = \gamma^2 D \frac{\kh^2}{k^2} = \gamma^2 D \left ( 1 - \frac{c_G}{k} + \frac{c_G^2}{4 k^2} \right)
\end{equation}
from (6c) in \cite{Zamolodchikov:1986gt}. Matching this result with \eqref{eqn:Gll}, we recover $c_0 = -1$ and furthermore fix $\gamma^2 = 1/2$. This is consistent with the observation that, at least at the fixed point, additional corrections \cite{Friedan:2009ik} vanish and therefore $G_{\lambda\lambda}(\lambda=0)$ becomes the Zamolodchikov metric. The difference to $c_0 = 0$ in \cite{Georgiou:2019nbz} originates from a different normalisation of $\mathcal{O}_\lambda$, since they use $\hat{k}$ instead of $k$ in \eqref{eqn:Olambda}
. Clearly, more work is required to understand this difference and to try to reproduces the remaining two coefficients, $c_1$ and $c_2$, from conformal perturbation theory.

\section{Conclusions}\label{sec:conclusion}
In this chapter, we have established three main results for the bosonic string:
\begin{enumerate}[leftmargin=1.5em,itemindent=1em,itemsep=0cm,label*=\arabic*)]
  \item In an appropriate scheme, the two-loop $\beta$-functions for the metric, $B$-field, and dilaton can be written in a manifestly O($D$,$D$)-covariant form.
  \item PL $\sigma$-models are one and two-loop renormalizable.
  \item The respective RG flows are invariant under PL T-duality.
\end{enumerate}
One might expect that the best way to obtain them is to start from a worldsheet theory with manifest, classical PL symmetry and apply the background field method like in \cite{Sfetsos:2009vt,Severa:2016lwc,Pulmann:2020omk}. However, this idea has not been implemented successfully yet. Therefore, we chose a different approach which heavily relies on previous insights in DFT and on the option to obtain the one and two-loop $\beta$-functions from a gradient flow. An important lesson learned is that it is crucial to work in the right scheme. The latter is tightly linked to the deformation of double Lorentz symmetry on the target space and the corresponding gGS transformations. So it might be promising to revisit the worldsheet approach with this knowledge.

The one-loop RG flow has a natural interpretation in terms of a generalized Ricci flow (see \cite{Garcia-Fernandez:2020ope} for a recent review), the generalized geometry version of the celebrated Ricci flow \cite{hamilton1982} used in Perelman's resolution of the Poincar\'e and Thurston geometrisation conjecture \cite{Perelman:2006un,Perelman:2006up,Perelman:2003uq}. Therefore, all involved quantities possess a (generalized) geometric origin. It is tempting to speculate that something similar might be true for the two-loop flow. Since fundamental symmetries of generalized geometry (like double Lorentz transformations) are deformed in its derivation, it is likely that also the underlying notion of geometry has to be adapted. PL symmetric target space geometries provide intriguing clues on the required modifications: Remember that a significant class of such target spaces is formed by PL groups. But PL groups are just the classical limit of a quantum group (see for example \cite{Chaichian:1996ah} for an introduction). Quantum groups can be approached from different angles. Most significant for us is that they give rise to non-commutative geometries. Hence, we conjecture that $\beta$-functions beyond one-loop might be governed by non-commutative geometry where the deformation parameter is related to the string length $\sim \sqrt{\alpha}'$. A related clue in this direction is that integrable deformations, like the $\lambda$- and $\eta$-deformation, which we discuss in section~\ref{sec:results}, possess a hidden quantum group symmetry \cite{Hollowood:2015dpa,Delduc:2016ihq}. The respective deformations parameters, $q=\exp(i \pi / k)$ and $q=\exp(4 \pi \eta t)$, are RG invariants at one and two loops. It is instructive to restore $\alpha'$ in these expressions. We know that $F_{ABC}$ comes with one derivative and therefore a factor of $\sqrt{\alpha'}$. Hence, we are actually dealing with $\qh=q^{\alpha'}$. In the semiclassical limit, $\alpha'\rightarrow 0$, a $\qh$ deformed quantum group transitions into a Poisson-Hopf algebra with the deformation parameter $q$. It is the latter which partially captures the global symmetries of the classical $\eta$-deformation \cite{Delduc:2016ihq}. Consequentially, we might understand $\alpha'$-corrections as the driving force from the classical Poisson-Hopf algebra to the associated quantum group. Of course, these speculations have to be supplemented with further quantitative evidence. But if we assume that they are justified, it would imply that we could extract all order $\beta$-functions and their generating low-energy, effective target space actions. Another reason to be optimistic that our results can be extended beyond two loops is that gGS transformations and the corresponding O($D$,$D$)-covariant action are in principle (even though they become extremely complicated) available to arbitrary order in $\alpha'$ \cite{Baron:2020xel}.

Two immediate applications for our results are integrable deformations and consistent truncations with higher derivative corrections. The former are motivated by the observation that nearly all currently known integrable $\sigma$-models possess PL symmetry. Already at one-loop, they have interesting RG flows (examples include \cite{\refoneLRGintegr}) with generic features like multiple fixed points \cite{Georgiou:2020eoo}. Recently, first efforts were made to push this analysis to two-loop \cite{Hoare:2019ark,Hoare:2019mcc,Georgiou:2019nbz}. At the level of the target space fields this is challenging, as we have demonstrated in section~\ref{sec:finitegGS} for the $\lambda$-deformation. But with the formalism we develop in this chapter, it becomes a much simpler task. Moreover, PL T-dualities between different integrable deformations are manifest. Due to this fact, we could obtain the flows of the $\lambda$- and $\eta$-deformation from a single calculation. We furthermore noticed that at one-loop, renormalizable $\sigma$-models are in one-to-one correspondence with consistent truncations of the low energy effective theory in the target space. Due to their potential to produce new, sophisticated solutions in (gauged) supergravity they have been intensively studied (for an early work see for example \cite{Duff:1985jd}). But only recently, systematic constructions of such truncations have been discussed and the framework of generalized geometry/double/exceptional field theory is predestined for them \cite{Cassani:2019vcl}. All of the work in this direction, that we are aware of, is based on a two-derivative action and its field equations. Since PL $\sigma$-models are two-loop renormalizable, they result in a large class of consistent truncations involving up to four derivatives. Hence, one might use them as guiding examples to construct a higher derivative version of the current constructions. Another important step that is required to make contact with $\alpha'$-corrected half-maximal gauged supergravities, is to extend our results to the heterotic string.

To conclude, in the second part of this thesis, we have seen that Poisson-Lie T-duality could very well be a full duality of string theory, much like its abelian counterpart. While we focused on the bosonic string there is strong evidence that the above results can be extended to other string theories.
However we leave this to future work and choose to move on to the final part of this thesis. There, we will turn our interest towards other dualities, looking specifically for supersymmetric interfaces between four-dimensional theories. The resulting three-dimensional profiles can be used to unify configurations of phenomenological interest, and some of the resulting three-dimensional interfaces could have concrete condensed matter applications. We begin by studying a duality between 4D $\N=1$ vacua resulting from compactifications of either M-theory on singular $G_2$ spaces or F-theory compactified on elliptically fibered Calabi-Yau fourfolds. This will take us to a supersymmetric three-dimensional theory defined by M-theory on a local $Spin(7)$. We will then end by exploring geometric approaches to probing a larger class of 3D interfaces, with some of them being at strong coupling.

\part{Special holonomies and 3D interfaces}
\chapter{Geometric Unification of Higgs Bundle Vacua}\label{chapter6}
\section{Introduction} \label{sec:INTRO6}

One of the very promising features of string theory is that it contains all of the qualitative
ingredients necessary to couple the Standard Model of particle physics to quantum gravity. That being said,
there could in principle be more than one way that our 4D world -- or some close approximation thereof --
might emerge from this fundamental framework.

Much like we have seen in the previous chapters, one of the lessons of string dualities is that seemingly
different string compactifications may nevertheless describe aspects of the \textit{same} physical system,
just in different (and possibly overlapping) regimes of validity.
With this in mind, it is therefore natural to ask whether there is a common feature present in
different approaches to realizing the Standard Model in string theory. This would in turn provide a
more unified approach to constructing and studying string vacua of phenomenological relevance.

Canonical approaches to realizing 4D $\mathcal{N} = 1$ vacua from strings
include compactification of heterotic strings on Calabi-Yau threefolds \cite{Candelas:1985en},
M-theory on singular $G_2$ spaces \cite{Acharya:2000gb, Acharya:2001gy},
and F-theory on elliptically fibered Calabi-Yau fourfolds \cite{Beasley:2008dc, Donagi:2008ca}.
At first glance, the actual methods used in studying the resulting low energy effective field theories
appear quite different, in tension with expectations from string dualities.

There are, however, some striking similarities between these different approaches,
especially in the particle physics / ``open string sector.''
At a practical level, the actual method for constructing many string vacua begins with the gauge theory of a
spacetime filling brane wrapped on a compact manifold in the extra dimensions.
For example, in the large volume approximation, heterotic strings
are captured by a Ho\v{r}ava--Witten nine-brane wrapped on a Calabi-Yau threefold equipped with a stable holomorphic vector bundle,
in M-theory it is intersecting six-branes wrapped on three-manifolds,
and in F-theory it is intersecting seven-branes wrapped on K\"ahler surfaces.
There are localized versions of dualities which connect these different constructions. For example, heterotic strings on a
$T^2$ is dual to F-theory on an elliptically fibered K3 surface, and this can be used to provide a physical
justification for the spectral cover construction of holomorphic vector bundles used in heterotic models \cite{Donagi:1998xe}.
In local M- and F-theory constructions, these different approaches are captured by Higgs bundles.
This suggests a close connection between these different approaches to realizing 4D physics.

In the resulting 4D effective field theory generated by such a compactification, the general
expectation is that specific details of a given compactification will be encoded in
the Wilson coefficients of higher dimension operators. At a formal level, one can
consider slowly varying these coefficients as a function of position in a 4D $\mathcal{N} = 1$
supersymmetric effective field theory.
Such interpolating profiles would then provide a way to directly connect the corresponding 4D string vacua obtained
from different compactifications. On general grounds, such interpolating profiles could at best preserve 3D Lorentz invariance and
3D $\mathcal{N} = 1$ supersymmetry. Let us emphasize here that in the 4D effective field theory, these interfaces need not be associated with a domain wall, since the interpolating mode may not be a light state. Instead, it can appear as an interpolating profile of Kaluza-Klein modes.

In this chapter we place these general expectations on firm footing by generating such interpolating solutions for the Higgs bundles
used in the construction of 4D $\mathcal{N} = 1$ models based on local M- and F-theory constructions. To accomplish this, we make use of the fact that M-theory on a $Spin(7)$ space results in a 3D $\mathcal{N} = 1$ effective field theory on the spacetime $\mathbb{R}^{2,1}$. The internal gauge theory in question arises from a local four-manifold of ADE singularities, as captured by a spacetime filling six-brane wrapped on this four-manifold.\footnote{The corresponding Higgs bundle for this system was studied recently in reference \cite{Heckman:2018mxl} (see also \cite{Heckman:2019dsj}) in the context of 4D ``$\mathcal{N} = 1/2$'' F-theory backgrounds.}

Here, we consider some further specializations in the structure of this four-manifold so that it is locally a product of a three-manifold and an interval. Reduction on the interval leads to the three-dimensional gauge theory system for local M-theory models \cite{Pantev:2009de} which we shall refer to as the ``PW system.'' We also show that if the four-manifold has an asymptotic region in which it is well-approximated by a K\"ahler surface, then the four-dimensional gauge theory reduces to that used in the study of 4D F-theory models \cite{Beasley:2008dc, Beasley:2008kw,Donagi:2008ca,Donagi:2008kj} which we will refer to as the ``BHV system.'' In each of these specializations, some of the fields of the local $Spin(7)$ system asymptotically approach zero. In this way, the local $Spin(7)$ Higgs bundle configuration serves as a way to glue together Higgs bundles used in the construction of 4D vacua!

This also provides a complementary perspective on geometric approaches to constructing special holonomy spaces from lower-dimensional
spaces. For example, the twisted connected sums construction of $G_2$ manifolds given in reference \cite{kovalevTCS} (see also \cite{Corti:2012kd}) makes use of asymptotically cylindrical Calabi-Yau threefolds which are glued together. In the generalized connected sums proposal for $Spin(7)$ manifolds given in reference \cite{Braun:2018joh}, the building blocks include asymptotically cylindrical spaces $X_{CY_4}$ and $Y_{G_2} \times S^1$, with $X_{CY_4}$ a Calabi-Yau fourfold and $Y_{G_2}$ a $G_2$ space.

A local version of the twisted connected sum construction enters our analysis of interpolating Higgs bundles. In the case of local M-theory constructions specified by a six-brane on a three-manifold $Q$, the ambient space is the non-compact Calabi-Yau threefold $T^{\ast} Q$. In the case of local F-theory constructions, with seven-branes wrapped on a K\"ahler surface $S$, it is the non-compact Calabi-Yau threefold given by the canonical bundle $\mathcal{O}(K_S) \rightarrow S$, and in the local $Spin(7)$ models on a four-manifold $M$, it is instead the non-compact $G_2$ space defined by the bundle of self-dual two-forms $\Omega^{2}_{+} \rightarrow M $. From the perspective of a 4D effective field theory,
we can parameterize these different choices in terms of a non-compact coordinate $\mathbb{R}_{t}$ with local coordinate $t$ such that in the asymptotic region $t \rightarrow - \infty$, we approach a local BHV system, while in the asymptotic region $t \rightarrow + \infty$, we approach a local PW system. In this fibration, the F-theory region of the compactification is specified by a local spacetime coordinate on a line $\mathbb{R}_{\text{F-th}}$ which becomes part of the internal compactification geometry in the local PW system. Conversely, in the M-theory region of the compactification, there is a local spacetime coordinate on a line $\mathbb{R}_{\text{M-th}}$ which becomes part of the internal compactification geometry in the local BHV system. Viewed in this way, the gluing region specified by the ambient $G_2$ space for the local $Spin(7)$ Higgs bundle amounts to a gauge theoretic generalization of the twisted connected sum construction, in which various $S^1$ factors have been decompactified. See figure \ref{fig:BHVPW} for a depiction of this local interpolating profile.

\begin{figure}[t!]
\begin{center}
\includegraphics[scale = 0.5, trim = {2cm 2.0cm 2cm 4.0cm}]{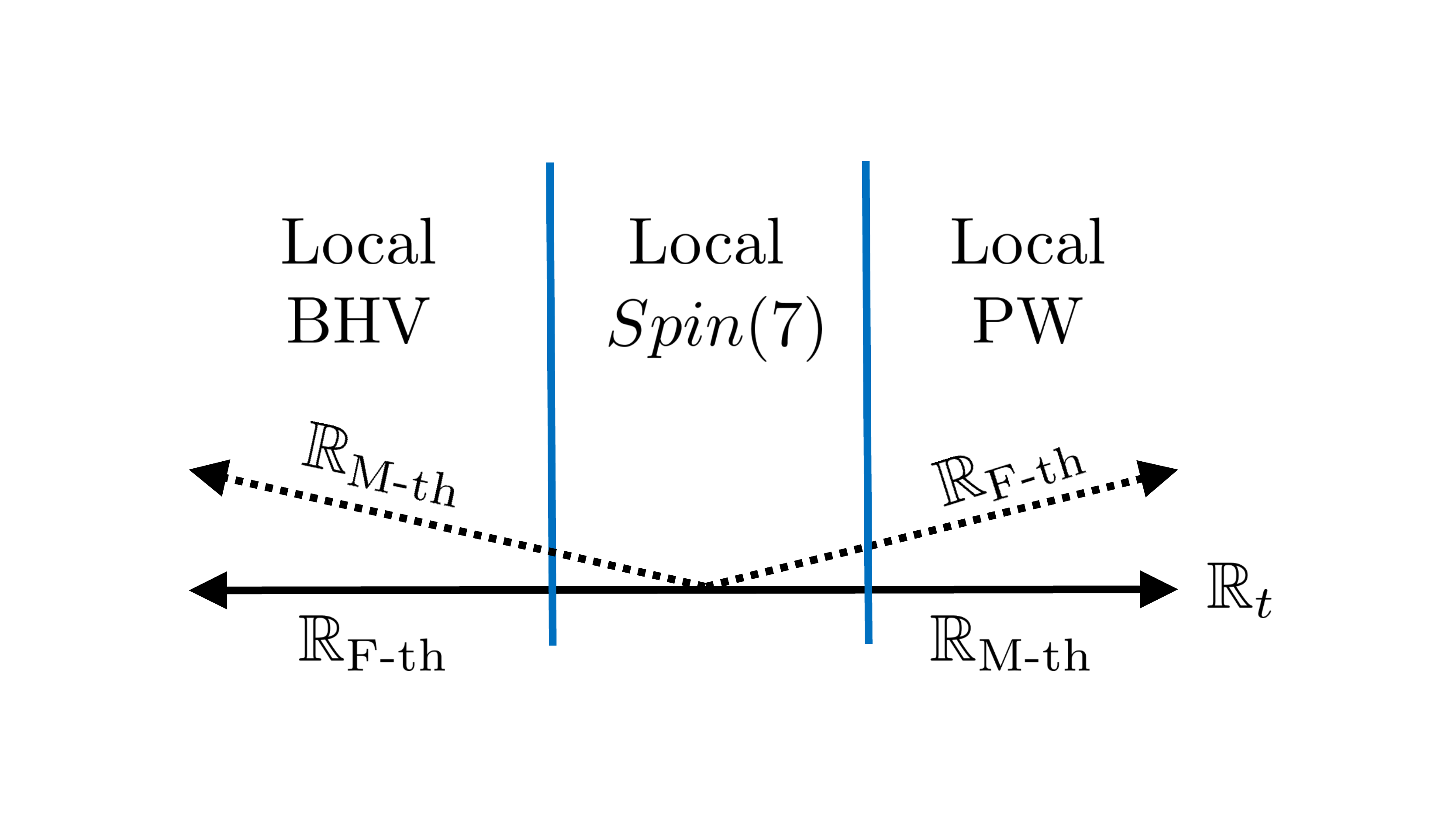}
\caption{Depiction of an interpolating profile between F-theory on a non-compact elliptically
fibered Calabi-Yau fourfold (left) and M-theory on a non-compact $G_2$ space (right). In the 4D effective field theory,
this involves an interpolating profile in a direction $\mathbb{R}_{t}$. In the transition between the F-theory and
M-theory vacua, the local coordinate of the 4D spacetime becomes part of the internal geometry on the opposite side
of the interpolating region. These interpolating profiles are captured by a local BHV system (see \cite{Beasley:2008dc})
in the F-theory region and a local PW system (see \cite{Pantev:2009de}) in the M-theory region.
The interpolating profile between these two 4D vacua is captured by M-theory on a local $Spin(7)$ geometry.}
\label{fig:BHVPW}
\end{center}
\end{figure}

One of the benefits of this local gauge theory analysis is that it also provides a systematic tool for extracting the physical content
from singular spaces of special holonomy. This is especially helpful in the context of local $G_2$ and $Spin(7)$ spaces since holomorphic techniques used in the study of Calabi-Yau spaces are unavailable. Indeed, our gauge theory analysis allows us to make further predictions for the sorts of singularities one should expect to encounter in local $Spin(7)$ spaces. We find that matter fields of the 3D effective field theory can localize on real two-cycles as well as real one-cycles of a compact four-manifold. Interactions between these matter fields can receive various quantum corrections controlled by expansion in large volume parameters of the four-manifold. This is in accord with the fact that the superpotential of a 3D $\mathcal{N} = 1 $ theory is not protected by holomorphy. Interpreting our 3D theories as specifying interpolating profiles between 4D vacua, the resulting matter fields correspond to localized degrees of freedom trapped at the interface between different 4D vacua.

The rest of this chapter is organized as follows. In section \ref{sec:HIGGS} we introduce the Higgs bundles
associated with 5D, 4D and 3D vacua, and then turn in section \ref{sec:EFT} to the interpretation in effective
field theory. In section \ref{sec:CAINANDABEL} we specialize to a class of ``abelian'' solutions in which the Higgs field is diagonalizable,
analyzing the geometry of intersecting branes and localized matter in these systems. We then turn in section \ref{sec:AMIMYBROTHERSKEEPER}
to some examples of interfaces in 5D and 4D vacua associated with the PW system, and
in section \ref{sec:YESIAMNEWJACKCITY} we construct interpolating solutions between BHV and PW systems. Section
\ref{sec:CONC6} contains our conclusions. Some additional technical details on the analysis of solutions to the local $Spin(7)$
equations are presented in an Appendix.

\section{Higgs Bundle Vacua} \label{sec:HIGGS}

In this section we introduce the different Higgs bundles associated with local M- and F-theory models.
We refer to the corresponding effective field theories
generate by these compactifications as ``Higgs bundle vacua.''
As a warmup, we first discuss the case of 5D $\mathcal{N} = 1$ vacua as generated by
M-theory on a curve of ADE singularities. We then turn to local models for M- and F-theory which
result in 4D vacua, and then turn to 3D vacua.

\subsection{5D $\mathcal{N} = 1$ Vacua}

As a warmup, we first discuss the case of M-theory on a non-compact Calabi-Yau threefold
given by a curve of ADE singularities. This is by far the most well studied class of examples, and will
also be used here as an underlying building block in our more general considerations.

With this in mind, consider a Calabi-Yau threefold given by
$C$ a complex curve of ADE singularities. The singularity type of this fibration can degenerate
at points of the curve, and this is associated with localized hypermultiplets. The corresponding Higgs bundle data is in this
case captured by the Hitchin system with gauge algebra of ADE type coupled to point localized defects. We remark that more general
non-simply laced gauge algebras are possible when the fibration has non-trivial monodromy which would interchange some of the
divisors in the resolved fiber. We will not dwell on this possibility here, but it is always available.

Physically, we can view this configuration as defining a six-brane wrapped on the curve $C$ which intersects other six-branes
at points of the curve. Indeed, this analysis generalizes what one expects from a IIA background with
D6-branes wrapped on the non-compact Calabi-Yau twofold $T^{\ast} C$. In a holomorphic presentation,
we can also write this Calabi-Yau as the total space of the canonical bundle,
namely $\mathcal{O}(K_{C}) \rightarrow C$.

Returning to the Higgs bundle formulation of this system, we have a gauge field as well as
an adjoint-valued $(1,0)$ form $\phi_{\text{Hit}}$. The BPS equations of motion governing the six-brane are:
\begin{align}
\label{eq:Hitch1}\overline{\partial}_{A} \phi_{\text{Hit}} & = 0\\
\label{eq:Hitch2}F_{A} + \frac{i}{2}[\phi_{\text{Hit}}^{\dag} , \phi_{\text{Hit}}] & = 0,
\end{align}
and 5D vacua are specified as solutions to the BPS equations of motion modulo gauge transformations.\footnote{A note on convention. Here and in the following we choose a unitary frame, meaning that the dagger operation is simply the hermitian conjugate. Moreover throughout the chapter we will take the generators of the Lie algebra to be anti-hermitian.}
Contributions from localized matter can also be included as source terms on the right-hand side of these equations.

The eigenvalues of $\phi_{\text{Hit}}$ are $(1,0)$ forms, and define sections (possibly meromorphic) of $K_{C}$. This in turn means that the ambient space in which the six-brane ``moves'' is $\mathcal{O}(K_{C}) \rightarrow C$.
One can also work in terms of a symplectic, rather than holomorphic presentation, in which case the Higgs field is an adjoint-valued
one-form. Then, the ambient space would be presented as  $T^{\ast} C$ in a presentation as a symplectic space.

As a final remark, we note that the same structure also appears in 6D vacua of F-theory models. In that case, we have an elliptically fibered Calabi-Yau threefold, and a component of the discriminant locus will correspond to a seven-brane wrapping a curve. Supersymmetric
vacua of the 6D theory are then governed by the same Hitchin equations. We also note that upon circle reduction of the 6D system,
we reduce to the 5D configuration, as captured by a local M-theory model.

\subsection{4D $\mathcal{N} = 1$ Vacua}

We now turn to some of the different possible routes to realizing 4D $\mathcal{N} = 1$ vacua
using Higgs bundles. One of our goals will be to use the analogous Higgs bundle constructions for 3D $\mathcal{N} = 1$ vacua
to generate interpolating profiles between these 4D vacua.

Recall that in type IIA and IIB vacua, the ``open string sector'' arises from intersecting branes, possibly in the
presence of non-trivial gauge field fluxes. D6-branes in Calabi-Yau threefolds which wrap special Lagrangian three-cycles can intersect at
points. At such points, chiral matter is localized. D7-branes in Calabi-Yau threefolds which wrap holomorphic surfaces
intersect along curves, and in the presence of suitable gauge field fluxes also give rise to 4D chiral matter.

These constructions have a natural lift to M- and F-theory, where the
structure of intersecting branes is instead encoded in geometry.
In M-theory on a $G_2$ space, the gauge theory sector arises from a three-manifold of ADE singularities, and further degenerations in the singularity type at real one-cycles produce 5D hypermultiplets compactified on the cycle, while enhancements at points of the three-manifold give rise to 4D chiral matter. There is clearly a close connection between the geometric enhancements of singularity types and the physics of 4D spacetime filling six-branes in the analogous IIA vacua. That being said, the M-theory approach provides a more flexible framework since additional non-perturbative effects can be captured. This includes, for example, the appearance of E-type gauge groups.

In F-theory on an elliptically fibered Calabi-Yau fourfold, the gauge theory sector can be modeled as
a K\"ahler surface of ADE singularities, and further degenerations along curves of the surface
produce 6D hypermultiplets. Switching on background gauge field fluxes through such curves
then leads to chiral matter in the 4D effective field theory. Again, based on the dimensionality of various enhancements, it is appropriate to
refer to these gauge theories as specified by 4D spacetime filling seven-branes, in analogy with IIB vacua.

Higgs bundles provide a general way to model the vacua generated by such intersecting brane configurations.
The essential point is that the existence of $\mathcal{N} = 1$ supersymmetry in the uncompactified 4D spacetime
dictates a unique topological twist for the brane in the internal directions. In the case of M-theory with intersecting six-branes wrapped on a three-manifold $Q$, the field content of the Higgs bundle includes a gauge connection and an adjoint valued one-form $\phi_{\mathrm{PW}}$,
as discussed by Pantev and Wijnholt (PW) in reference \cite{Pantev:2009de}. There is a close connection to IIA strings on the non-compact
Calabi-Yau threefold $T^{\ast} Q$. Indeed, the eigenvalues of the Higgs field of the local M-theory model take values in the cotangent bundle,
and parameterize local motion of the branes in the ambient geometry. Similarly, in the case of F-theory with intersecting seven-branes, the field content of the Higgs bundle includes a gauge connection and an adjoint valued $(2,0)$ form $\phi_{\text{BHV}}$, as discussed in \cite{Beasley:2008dc, Donagi:2008ca}. In this case, there is a close connection to type IIB strings on the non-compact Calabi-Yau threefold given by the total space of the canonical bundle, namely $\mathcal{O}(K_S) \rightarrow S$; the eigenvalues of the $(2,0)$ form parameterize the motion of branes wrapped on holomorphic surfaces in this non-compact threefold.

The ``bulk'' degrees of freedom of these gauge theories can also be coupled to various lower-dimensional defects localized on subspaces of a compactification. These appear as additional source terms in the BPS equations of motion, a point we shall return to soon. In fact, the appearance of these localized sources can \textit{also} be modeled in terms of a corresponding Higgs bundle construction, being associated to the spectrum of localized perturbations about a given background solution.

To illustrate these general considerations and since we will need to make use of them in more detail later, we now turn to the specific bulk BPS equations of motion for local M- and F-theory models. We refer to these as the ``PW'' and ``BHV'' systems, respectively.

\subsubsection{PW System}

Consider first local M-theory models. The system of equations appearing in \cite{Pantev:2009de} describes supersymmetric solutions for six-branes compactified on a three-cycle $Q$ inside a $G_2$ space. This again gives a 4D $\mathcal{N} =1$ supersymmetric theory In this case the fields appearing are a gauge field $A$ and an adjoint valued one-form $\phi_{\text{PW}}$. The supersymmetric equations of motion are
\al{D_A \phi_{\text{PW}} =0\,,\\
D_A * \phi_{\text{PW}} =0\,,\\
F = \left[\phi_{\text{PW}},\phi_{\text{PW}}\right]\,.
}
Including matter fields amounts to adding in additional source terms to the right-hand side of these equations. Vacua are given by solutions to the supersymmetric equations of motion modulo gauge transformations. These vacua are also captured by the critical points of a complexified Chern-Simons functional:
\begin{equation}
W_{\mathrm{PW}}= \int_{Q} \mathrm{Tr}\left( \mathcal{A} \wedge d \mathcal{A} + \frac{2}{3} \mathcal{A} \wedge \mathcal{A} \wedge \mathcal{A} \right)
\end{equation}
modulo complexified gauge transformations. In the above, we have introduced a complexified connection $\mathcal{A} = A + i \phi_{\text{PW}}$.

Though we shall often leave it implicit, the field content of this gauge theory also provides important \textit{geometric}
information on the local structure of M-theory compactified on a $G_2$ space with singularities. To see this, observe that for a three-manifold of ADE singularities, we can perform a resolution of the singular fibers. This results in a basis of compactly supported harmonic two-forms $\omega_{\alpha}$ which are in correspondence with the generators of the Cartan for the given gauge group. A variation in the associated three-form $\Phi_{(3)}$ of the local $G_2$ space results in a decomposition:
\begin{equation}
\delta \Phi_{(3)} = \sum_{\alpha} \phi_{\mathrm{PW}}^{\alpha} \wedge \omega_{\alpha},
\end{equation}
namely, the eigenvalues of our adjoint-valued one-form $\phi_{\mathrm{PW}}$ directly translate to metric data of the local $G_2$ space. Off-diagonal elements are encoded in additional physical degrees of freedom such as M2-branes wrapped on collapsing two-cycles.

\subsubsection{BHV System}

Turning next to local F-theory models, the system of BPS equations derived in \cite{Beasley:2008dc} controls supersymmetric configurations of seven-branes wrapped on a K\"ahler surface $S$. The field content of the Higgs bundle is specified by fixing a gauge group $G$, and consists of a gauge field $A$, and an adjoint valued $(2,0)$ form $\phi_{\mathrm{BHV}}$. The BPS equations for this system are
\al{ \overline{\partial}_A \phi_{\mathrm{BHV}} =0\,, & \\
F_{(0,2)} =0\,, & \\
J_{S} \wedge F + \frac{i}{2} \left[\phi_{\mathrm{BHV}}^\dag,\phi_{\mathrm{BHV}}\right]=0\,.
}
Here we introduced $J_{S}$ which is the K\"ahler form on the four-cycle wrapped by the seven-branes.
The last equation is the equivalent for the BHV system of the usual equation controlling stability of holomorphic vector bundles in Calabi--Yau threefolds \cite{Uhlenbeck1986}. Matter fields localized on complex curves, as well as cubic interactions between these matter fields can all be included by introducing appropriate source terms on the right-hand side of these equations of motion \cite{Beasley:2008dc}. One can also characterize 4D supersymmetric, Lorentz invariant vacua as critical points of a superpotential:
\begin{equation}
W_{\mathrm{BHV}} = \int_{S} \mathrm{Tr} (\phi_{\mathrm{BHV}} \wedge F_{(0,2)})
\end{equation}
modulo complexified gauge transformations.

Much as in the case of the local $G_2$ construction, the field content of this gauge theory also provides important geometric
information on the local structure of F-theory compactified on a singular elliptically fibered
Calabi-Yau fourfold. To see this, observe that for a K\"ahler surface of ADE singularities, we can perform a resolution of the singular fibers. This results in a basis of compactly supported harmonic two-forms $\omega_{\alpha}$ which are in correspondence with the generators of the Cartan for the given gauge group. A variation in the associated holomorphic four-form $\Omega_{(4,0)}$ of the Calabi-Yau fourfold
results in a decomposition:
\begin{equation}
\delta \Omega_{(3,1)} = \sum_{\alpha} \phi_{(2,0)}^{\alpha} \wedge \omega_{\alpha},
\end{equation}
namely, the eigenvalues of our adjoint-valued $(2,0)$ form directly translate to metric data.

\subsection{3D $\mathcal{N} = 1$ Vacua}

Let us now turn to the related case of M- and F-theory compactifications which generate 3D $\mathcal{N} = 1$ vacua, namely
systems with at least two real supercharges. One simple way to generate examples with 3D $\mathcal{N} = 2$ supersymmetry
(four real supercharges) is to take a 4D $\mathcal{N} = 1$ theory and compactify further on a circle.
From the standpoint of compactification, we can then consider M-theory on
$Y_{G_2} \times S^1$ or F-theory on $X_{CY_4} \times S^1$ (in the obvious notation). Using the standard duality between circle reductions of F-theory and M-theory vacua, note that we can alternatively consider M-theory compactified on the Calabi-Yau fourfold $X_{CY_4}$, in which the volume modulus of the elliptic fiber is now a physical parameter (in a local model it is non-dynamical).
This already provides us with two possible Higgs bundles, one associated with the PW system (via compactification on a $G_2$ space) and the other associated with the BHV system (via compactification on a Calabi-Yau fourfold).

We can also consider more general compactifications which only preserve 3D $\mathcal{N} = 1$ supersymmetry
by taking M-theory on a $Spin(7)$ space (see e.g. \cite{Becker:2000jc, Cvetic:2001pga, Cvetic:2001ye, Cvetic:2001zx,
Gukov:2001hf, Gukov:2002es, Gukov:2002zg}).
The analog of local models in this context involves a four-manifold $M$ of ADE singularities.
There can also be local enhancements in the singularity type along subspaces. Indeed, comparing the 3D $\mathcal{N} = 2$ vacua obtained from
$X_{G_2} \times S^1$ and $X_{CY_4}$, we anticipate that enhancements in the singularity type could occur over real one-cycles as well as over
two-dimensional Riemann surfaces. In M-theory, this will be captured by a configuration of intersecting six-branes, possibly with gauge field fluxes switched on. In this case, the appropriate Higgs bundle involves a gauge field and an adjoint-valued self-dual two-form $\phi_{\mathrm{SD}}$ (see e.g. \cite{Heckman:2018mxl}).

Again, there is a close connection between the resulting vacua and those obtained from IIA on a local $G_2$ space. To see this, observe that
the eigenvalues of $\phi_{\mathrm{SD}}$ take values in $\Omega^{2}_{+} \rightarrow M$. The bundle of self-dual two-forms leads to a non-compact $G_2$ space in the sense that there is a distinguished three-form $\Phi_{(3)}$. Indeed, in the special case where $M$ is $S^4$ or $\mathbb{CP}^2$ there is a corresponding complete metric on this space \cite{bryant1989}. More generally, however, the condition of completeness can be relaxed, at the expense of introducing some singularities. This is additional physical data of the system associated with the appearance of light degrees of freedom as one approaches a UV cutoff. For this reason, we also view this more general class of seven-manifolds as local $G_2$ spaces.

We obtain 3D $\mathcal{N} = 1$ vacua from the corresponding BPS equations of motion for this
system \cite{Vafa:1994tf, Heckman:2018mxl} (for an analytic perspective, see also \cite{mares}):
\al{ D_A \phi_{\text{SD}} &=0 \,,\\
F_{\text{SD}} + \phi_{\text{SD}} \times \phi_{\text{SD}} &=0\,,}
where we can include the contributions from localized matter by adding source terms to
the right-hand sides of these equations. Here, $F_{\text{SD}} =\tfrac{1}{2} (F+*F)$ is
the self-dual part of the field strength. We have also introduced a cross product which
in local indices can be written as \cite{Vafa:1994tf}:
\al{ (\phi_{\text{SD}}\times \phi _{\text{SD}})_{ij} = \frac{1}{4} \left[\phi_{\text{SD}\, ik },\phi_{\text{SD}\, jl } \right] g^{kl}\,,}
where $g_{ij}$ refers to the metric on $M$.
Using the distinguished three-form $\varepsilon$ on $\Omega^{2}_{+} (M)$, we can also write \cite{Heckman:2018mxl}:
\begin{equation}
(\phi_{\text{SD}}\times \phi _{\text{SD}})_{a} = \varepsilon_{abc} \phi^{b}_{\text{SD}} \phi^{c}_{\text{SD}},
\end{equation}
where here, we are treating $\phi^{a}_{\mathrm{SD}}$ as a three-component vector in the vector space $\Omega^{2}_{+}$.

Much as in the case of the related 4D vacua, these vacua are labeled by critical points of a 3D $\mathcal{N} = 1$
superpotential:
\al{ W_{\mathrm{Spin(7)}} = \int_{M} \text{Tr} \left( \phi_{\text{SD}} \wedge \left[F_{\mathrm{SD}} + \frac{1}{3} \phi_{\text{SD}} \times \phi_{\text{SD}} \right]\right)\,.}
modulo gauge transformations. In this case, we note that this object is a real function associated with a
D-term (integrated over the full superspace).

The field content of this gauge theory also provides important geometric
information on the local structure of M-theory compactified on a singular $Spin(7)$ space.
For a four-manifold of ADE singularities, we can perform a resolution of the singular fibers. This results in a basis of compactly supported harmonic two-forms $\omega_{\alpha}$ which are in correspondence with the generators of the Cartan for the given gauge group. A variation in the associated Cayley four-form $\Psi_{(4)}$ of the $Spin(7)$ space
results in a decomposition:
\begin{equation}
\delta \Psi_{(4)} = \sum_{\alpha} \phi_{\text{SD}}^{\alpha} \wedge \omega_{\alpha},
\end{equation}
namely, the eigenvalues of the adjoint-valued self-dual two-form directly translate to metric data. Observe also that self-duality of the
Higgs field directly descends from the corresponding condition on the Cayley four-form.

Given a background solution to the local $Spin(7)$ equations, we can also study the spectrum of light degrees of freedom. These
are the ``zero modes'' of a given background. To write down the differential equations that govern the profile
of zero modes we take the BPS equations and expand them at linear order in the fields:
\al{ A&= \langle A \rangle + a\,,\\
 \phi_{\text{SD}}&= \langle \phi_{\text{SD}} \rangle + \varphi\,,
}
and keep only terms linear in $(a,\varphi)$ in the equations. Note that due to the topological twist, $a$ and $\varphi$ are each the real scalar component of a 3D $\mathcal{N}=1$ scalar multiplet and thus specify the matter of the engineered effective field theory.  In the following, for the sake of notational simplicity we shall drop the $\langle \cdot \rangle$ notation when we refer to background values of the fields. The resulting zero mode equations are
\begin{align}\label{zmeqs}
 D^+_A a + \phi_{\text{SD}} \times \varphi = 0, && D_A \varphi - [\phi_{\text{SD}}, a ] =0.
\end{align}
Here $D^+_A = D_A + \ast_4 D_A$. As we will discuss in detail later, \eqref{zmeqs} has both bulk solutions when the commutators with $\phi_{\text{SD}}$ vanish, or localized modes centered around the zero-loci of the adjoint action of
$\phi_{\text{SD}}$. Solutions should be considered equivalent when related to one another
via an infinitesimal gauge transformation
\al{\left\{ \begin{array}{l} a \sim a+ D_A \xi\\ \varphi \sim \varphi + [\phi_{\text{SD}},\xi] \end{array}\right. \,,
}
with $\xi$ an adjoint valued zero-form. Another way to phrase this is to associate to the local $Spin(7)$ system the following complex
\be
\begin{tikzcd}
 0 \arrow[r]& \Omega^0 (\text{ad} E) \arrow[r,"\delta_0"]& \Omega^1(\text{ad} E) \oplus \Omega^2_+ (\text{ad} E) \arrow[r,"\delta_1"]& \Omega^2_+(\text{ad} E) \oplus \Omega^3(\text{ad} E) \arrow[r]& 0 \,,
 \end{tikzcd}
\ee
where $\text{ad} E$ denotes forms in the adjoint representation of the Lie algebra. Moreover $\Omega_+^2$ denotes the bundle of self-dual two forms. The two differentials act as:
\al{ \delta_0 (\xi ) &= \left(\begin{array}{ l}D_A \xi \\ \, [\phi_{\text{SD}},\xi]  \end{array}\right)\,,\\
\delta_1 (\alpha,\beta) &= \left(\begin{array}{l} D^+_A \alpha + \phi_{\text{SD}} \times \beta \\ D_A \beta - [\phi_{\text{SD}},\alpha]  \end{array}\right)\,.
}
The space of infinitesimal deformations of the local $Spin(7)$ system (namely, the tangent bundle to the moduli space) is given by:
\al{ T\mathcal M_{\text{Spin(7)}} = \frac{\text{ker } \delta_1}{\text{im }\delta_0}\,.}
Note also that this complex naturally includes the 3D $\mathcal{N}=1$ vector multiplets as $\text{ker } \delta_0$.
This is so because the vector multiplets are scalars on $C$ and the
associated gauge group is the commutant which is not broken
by a Higgs mechanism.

\subsubsection{Specialization to 3D $\mathcal{N} = 2$ Vacua}

Having stated the general system of equations (as well as linearized fluctuations) for local $Spin(7)$ spaces,
we can also see how further specialization can result in a 3D $\mathcal{N} = 2$ vacuum solution,
as captured by M-theory on $Y_{G_2} \times S^1$ or $X_{CY_4}$. We begin with the PW system,
and then turn to the BHV system.

\paragraph{-- Reduction to PW System}

To relate the field content of the local $Spin(7)$ equations to those of the PW system, consider the special case where the four-manifold $M$ of the local $Spin(7)$ equations takes the form $M = Q \times S^1$ with $Q$ a three-manifold. Denote by $t$ the local coordinate on this $S^1$ factor.\footnote{In our interpretation of interpolating vacua, we will soon decompactify this direction.}
In this case, an adjoint-valued self-dual two-form $\phi_{\text{SD}}$ on $M$ descends to a decomposition of the form:
$\phi_{\text{SD}} =  \phi\wedge d t + \ast_3 \phi$, with $\phi$ an adjoint-valued one-form on $Q$. Observe also that the gauge field on $Q \times S^1$ has the degrees of freedom associated with $Q$, as well as the additional direction $A_{t}$. In terms of this decomposition, the local $Spin(7)$ equations can be written as:
\al{ F-[\phi,\phi] + *\left(D_t A - d_3 A_t \right) &= 0\,,\\
D_A \phi + * D_t \phi &= 0\,,\\
D_A * \phi &=0\,.
}
Here, the Hodge star is always taken in the three directions transverse to $t$ and $d_3$ denotes the exterior derivative in the directions transverse to $t$. We see that we recover the PW system upon setting $A_t = 0$ and $\partial_t A = \partial_t \phi = 0$, meaning that the PW system is the truncation of the $Spin(7)$ system to solutions that are invariant under translations in the $t$ direction and with $A_t=0$ which is compatible with the expectations from dimensional reduction.

\paragraph{-- Reduction to BHV System}

We now show that a different truncation reproduces the BHV system of equations. Along these lines, suppose the local four-manifold
$M$ is actually a K\"ahler surface $S$. In this case, self-dual two-forms decompose into $(2,0)$ forms and a $(1,1)$-form proportional to the
K\"ahler form:
\begin{equation}
\phi_{\text{SD}} \rightarrow \phi_{(2,0)} \oplus \phi_{(1,1)} \oplus \phi_{(0,2)}^{\dag}.
\end{equation}
We recognize the $(2,0)$ form as the same Higgs field appearing in the BHV system. Here, $\phi_{(1,1)} = \phi_{\gamma } \cdot J_S$ with $\phi_{\gamma}$ an adjoint valued function and $J_S$ is the K\"ahler form of $S$. In this decomposition, the local $Spin(7)$ equations become:
\al{ \overline{\partial}_A \phi_{(2,0)} - \frac{i}{2} \partial_A \phi_{(1,1)} & = 0\,,\\
F_{(0,2)} - \frac{i}{2} \phi_{(1,1)} \times \phi_{(0,2)}^\dag & = 0 \,,\\
J_S \wedge F + \frac{i}{2} \left[\phi_{(0,2)}^\dag,\phi_{(2,0)}\right] & =0\,.
}
Upon taking configurations for which $\phi_\gamma = 0$, we recover the BHV equations of motion.

\subsection{Deformations of the Hitchin System}

As the above examples illustrate, the structure of the local $Spin(7)$ equations reduces, upon further specialization,
to the Higgs bundles of the PW and BHV systems for 4D $\mathcal{N} = 1$ vacua.  Similar considerations hold for reduction of
the PW system on a three-manifold $Q$ given by a fibration of a Riemann surface over an interval \cite{Barbosa:2019bgh}.

We now show that starting from a solution to these more specialized
solutions, perturbations will in general produce a trajectory in the moduli space of the $Spin(7)$ equations.
The related analysis for PW systems viewed as perturbations of the Hitchin system was carried out in \cite{Barbosa:2019bgh},
and we refer the interested reader there for further discussion of this case. Specializing to the case of four-manifolds which
can be written as a Riemann surface $C$ fibered over a cylinder $\mathbb{C}^{\ast} \simeq \mathbb{R} \times S^1$, we show that the
BHV system of equations can also be viewed as perturbations of the Hitchin system. We then show that similar considerations
hold for deformations of the Hitchin system to the $Spin(7)$ equations.

To proceed with this analysis, it will be helpful to introduce an explicit coordinate system. Let $w = t + i \theta$ denote the
coordinates of the cylinder, and $x,y$ real coordinates on $C$. We can then express
the self-dual two-form $\phi_{\text{SD}}$ on $M$ as the triplet:
\begin{equation}
  \phi_{\text{SD}} = \phi_\alpha(dx\wedge d\theta - dt\wedge dy)+\phi_\beta(dt\wedge dx + dy\wedge d\theta)+\phi_\gamma(dt\wedge d\theta+dx\wedge dy)\,.
\end{equation}\label{eq:SDtriplet}
We will assume that we have a flat metric, and expand along the $t$ direction as follows:
\begin{equation}
  A_i(x,y,\theta,t) = \sum_{k=0}^{\infty} A_i^{(k)}(x,y,\theta)t^k, \qquad
  \phi_i(x,y,\theta,t) = \sum_{k=0}^{\infty} \phi_i^{(k)}(x,y,\theta)t^k.
  \label{eq:powerseries}
\end{equation}
In what follows, we shall also work in a ``temporal gauge'' where $A_{t}(x,y,\theta,t) = 0$.

\subsubsection{Generating BHV Solutions}\label{sssec:PowerBHV}

As a warmup, we first show how to generate BHV solutions from perturbations of the Hitchin system.
The expanded BHV equations lead to non-trivial differential equations on the coefficients,
\begin{align}
  \begin{split}
  & \mathcal{G}_{ab}^{(j)} \equiv \partial_x \phi_\beta^{(j)}-\partial_y \phi_\alpha^{(j)}
  +\sum_{n=0}^j\left(\left[A_x^{(j-n)},\phi_\beta^{(n)}\right]-\left[A_y^{(j-n)},\phi_\alpha^{(n)}\right]\right) = 0,\\
  & \mathcal{H}_{ab}^{(j)} \equiv \partial_x \phi_\alpha^{(j)}+\partial_y \phi_\beta^{(j)}
  +\sum_{n=0}^{j-1}\left(\left[A_x^{(j-n)},\phi_\alpha^{(n)}\right]+\left[A_y^{(j-n)},\phi_\beta^{(n)}\right]\right) = 0,
  \end{split}\label{eq:difBHV}
\end{align}
together with five equations which fix the higher order coefficients in terms of the preceding one,
\begin{align}
  \begin{split}
  (j+1)A_\theta^{(j+1)} &= -F_{xy}^{(j)}+\left[\phi_\alpha, \phi_\beta\right]^{(j)},\\
  (j+1)A_x^{(j+1)} &= -F_{y\theta}^{(j)},\\
  (j+1)A_y^{(j+1)} &= F_{x\theta}^{(j)},\\
  (j+1)\phi_\alpha^{(j+1)} &= -D_\theta^{(j)}\phi_\beta^{(j)},\\
  (j+1)\phi_\beta^{(j+1)}&=D_\theta^{(j)}\phi_\alpha^{(j)}.
  \end{split}\label{eq:recBHV}
\end{align}
We will assume that $A_{x,y}^{(0)}$ and $\phi_{\alpha,\beta}^{(0)}$ are such that the zeroth order differential equations from \eqref{eq:difBHV} are solved, and the higher order coefficients are fixed by the linear equations \eqref{eq:recBHV}. The one remaining free parameter is $A_\theta^{(1)}$, which sets the ``trajectory'' of the solution. Once we have this initial data, we can show that the BHV equations are automatically solved to all orders in $t$ (see Appendix \ref{app:Power} for further details).

Indeed, it is  sufficient to solve the zeroth order differential equations
\begin{align}
  \begin{split}
  &D_x^{(0)} \phi_\beta^{(0)}  - D_y^{(0)} \phi_\alpha^{(0)} = 0,\\
  &D_x^{(0)} \phi_\alpha^{(0)} + D_y^{(0)} \phi_\beta^{(0)}  = 0,
  \end{split}\label{eq:zerothBHV}
\end{align}
and then one can simply propagate through equations \eqref{eq:recBHV} to build up the higher order terms. Note that this pair of differential equations are part of the Hitchin system on the Riemann surface spanned by $x$ and $y$ as they are the real and imaginary parts of equation \eqref{eq:Hitch1}. The last equation of the Hitchin system, that is equation \eqref{eq:Hitch2}, is deformed to the zeroth order of the first equation of \eqref{eq:recBHV}: this equation implies that an exact solution of the Hitchin system is obtained only for $A_\theta^{(0)} = 0$, meaning that the free parameter $A_\theta^{(0)}$ controls the deformation of the Hitchin system.

\subsubsection{Generating Local $Spin(7)$ Solutions}\label{sssec:PowerSpin7}
Similarly, it is possible to build a local $Spin(7)$ system that is neither just BHV or PW, via this power series expansion. Making use of the power series expansion \eqref{eq:powerseries}, we can expand the $Spin(7)$ equations to yield a single set of differential equations:
\begin{equation}
  \partial_x \phi_\beta^{(j)}-\partial_y \phi_\alpha^{(j)} + \partial_\theta \phi_\gamma^{(j)}
  +\sum_{n=0}^j\left(\left[A_x^{(j-n)},\phi_\beta^{(n)}\right]-\left[A_y^{(j-n)},\phi_\alpha^{(n)}\right]+\left[A_\theta^{(j-n)},\phi_\gamma^{(n)}\right]\right)
  = 0,\label{eq:difSpin7}
\end{equation}
together with six recursion relations,
\begin{align}
  \begin{split}
  jA_\theta^{(j)} &= -\partial_x A_y^{(j-1)}+\partial_y A_x^{(j-1)}
  -\sum_{n=0}^{j-1}\left(\left[A_x^{(j-1-n)},A_y^{(n)}\right]-\left[\phi_\alpha^{(j-1-n)},\phi_\beta^{(n)}\right]\right),\\
  jA_x^{(j)} &= -\partial_y A_\theta^{(j-1)}+\partial_\theta A_y^{(j-1)}
  -\sum_{n=0}^{j-1}\left(\left[A_y^{(j-1-n)},A_\theta^{(n)}\right]-\left[\phi_\gamma^{(j-1-n)},\phi_\alpha^{(n)}\right]\right),\\
  jA_y^{(j)} &= \partial_x A_\theta^{(j-1)}-\partial_\theta A_x^{(j-1)}
  +\sum_{n=0}^{j-1}\left(\left[A_x^{(j-1-n)},A_\theta^{(n)}\right]+\left[\phi_\gamma^{(j-1-n)},\phi_\beta^{(n)}\right]\right),\\
  j\phi_\gamma^{(j)} &= -\partial_x \phi_\alpha^{(j-1)}-\partial_y \phi_\beta^{(j-1)}
  -\sum_{n=0}^{j-1}\left(\left[A_x^{(j-1-n)},\phi_\alpha^{(n)}\right]+\left[A_y^{(j-1-n)},\phi_\beta^{(n)}\right]\right),\\
  j\phi_\alpha^{(j)} &= -\partial_\theta \phi_\beta^{(j-1)}+\partial_x \phi_\gamma^{(j-1)}
  -\sum_{n=0}^{j-1}\left(\left[A_\theta^{(j-1-n)},\phi_\beta^{(n)}\right]-\left[A_x^{(j-1-n)},\phi_\gamma^{(n)}\right]\right),\\
  j\phi_\beta^{(j)}&=\partial_\theta \phi_\alpha^{(j-1)}+\partial_y \phi_\gamma^{(j-1)}
  +\sum_{n=0}^{j-1}\left(\left[A_\theta^{(j-1-n)},\phi_\alpha^{(n)}\right]+\left[A_y^{(j-1-n)},\phi_\gamma^{(n)}\right]\right).
  \end{split}\label{eq:recSpin7}
\end{align}
Once again, it is possible to show that it is sufficient to solve the zeroth order differential equation
\begin{equation}
  D_x^{(0)} \phi_\beta^{(0)}-D_y^{(0)} \phi_\alpha^{(0)} + D_\theta^{(0)} \phi_\gamma^{(0)} = 0,
  \label{eq:zerothSpin7}
\end{equation}
and then one can simply propagate through equations \eqref{eq:recSpin7} to build up the higher order terms (see Appendix \ref{app:Power} for more details). Thus, if we are given $A_{x,y,\theta}^{(0)}$ and $\phi_{\alpha,\beta,\gamma}^{(0)}$ such that the zeroth order equations in \eqref{eq:zerothSpin7} are solved, then we can construct a full solution of the local $Spin(7)$ equations by specifying all the higher order coefficients as in \eqref{eq:recSpin7}.

\subsubsection{Abelian Case}
It is instructive to further specialize to the case where all gauge fields vanish. We refer to this as
an abelian solution because now the Higgs field has trivial cross product with itself.
Taking $A_i = 0$ gives some major simplifications. The local $Spin(7)$ recursion relations \eqref{eq:recSpin7} now become:
\begin{align}
  \begin{split}
  \phi_\alpha^{(j)} &= \frac{1}{j}
  \begin{cases}
    (-1)^{\sfrac{j}{2}} \left(\partial_x^2+\partial_y^2+\partial_\theta^2\right)^{\sfrac{j}{2}}\phi_\alpha^{(0)}\,, \quad \text{if $j$ is even}\\
    (-1)^{\sfrac{(j-1)}{2}} \left(\partial_x^2+\partial_y^2+\partial_\theta^2\right)^{\sfrac{(j-1)}{2}} \left(\partial_x\phi_\gamma^{(0)}-\partial_\theta\phi_\beta^{(0)}\right) \,, \quad \text{if $j$ is odd}
  \end{cases} \\
  \phi_\beta^{(j)} &= \frac{1}{j}
  \begin{cases}
    (-1)^{\sfrac{j}{2}} \left(\partial_x^2+\partial_y^2+\partial_\theta^2\right)^{\sfrac{j}{2}}\phi_\beta^{(0)}\,, \quad \text{if $j$ is even}\\
    (-1)^{\sfrac{(j-1)}{2}} \left(\partial_x^2+\partial_y^2+\partial_\theta^2\right)^{\sfrac{(j-1)}{2}} \left(\partial_\theta\phi_\alpha^{(0)}+\partial_y\phi_\gamma^{(0)}\right) \,, \quad \text{if $j$ is odd}
  \end{cases} \\
  \phi_\gamma^{(j)} &= \frac{1}{j}
  \begin{cases}
    (-1)^{\sfrac{j}{2}} \left(\partial_x^2+\partial_y^2+\partial_\theta^2\right)^{\sfrac{j}{2}}\phi_\gamma^{(0)}\,, \quad \text{if $j$ is even}\\
    (-1)^{\sfrac{(j-1)}{2}} \left(\partial_x^2+\partial_y^2+\partial_\theta^2\right)^{\sfrac{(j-1)}{2}} \left(-\partial_x\phi_\alpha^{(0)}-\partial_y\phi_\beta^{(0)}\right) \,, \quad \text{if $j$ is odd.}
  \end{cases}
  \end{split}\label{eq:abel}
\end{align}

\section{Effective Field Theory of Interpolating Solutions \label{sec:EFT}}

In the previous section we introduced Higgs bundles for minimally
supersymmetric 5D, 4D, and 3D vacua. In particular, we saw that many of
these Higgs bundles admit an interpretation as interpolating
between perturbations of a lower-dimensional Higgs bundle.

In this section we turn to the effective field theory associated with these
interpolating solutions. As a first comment, we note that although we are
clearly considering a change in the vacuum of the higher-dimensional field theory,
this need not be directly associated with a domain wall solution. The general
reason for this is that the fields participating in this interpolating profile
could, a priori, be quite heavy, and actually higher than the Kaluza-Klein
scale for the EFT. From this perspective, the appropriate description will
instead be given by integrating out these modes from the start. In the
resulting theory, this will instead leave its imprint in a profile of possibly
position dependent Wilson coefficients of the effective field theory.

To show how this comes about, we begin by studying interpolating solutions for
5D\ vacua from the standpoint of the 4D effective field theory generated by
the PW\ system. We then turn to interpolating solutions for 4D\ vacua from the
standpoint of the 3D\ effective field theory generated by the local $Spin(7)$
system. To set notation, in what follows we shall consider a D-dimensional
theory \textquotedblleft compactified\textquotedblright\ on either the
non-compact line $\mathbb{R}$ with local coordinate $t$, or a cylinder
$\mathbb{C}^{\ast} \simeq \mathbb{R\times}S^{1}$ with local coordinate $w=t+i\theta$.
Our general strategy will be to package all of the fields of the
higher-dimensional theory in terms of lower-dimensional fields labeled by
points of this extra-dimensional geometry. Writing down all possible
interaction terms of the lower-dimensional theory will then provide a general
way to track possible interpolating profiles between higher-dimensional vacua
obtained in the asymptotic limits as $t\rightarrow$ $-\infty$ and
$t\rightarrow+\infty$.

\subsection{Interpolating 5D\ Vacua}

To begin, we return to the case of interpolating 5D\ vacua, as captured by
M-theory on a non-compact Calabi-Yau threefold specified by a curve of
ADE\ singularities. As we have already mentioned, the Higgs bundle in this
case is the Hitchin system coupled to defects. We take the interpolating gauge
theory for this model to be a Pantev--Wijnholt system on a three-manifold $Q$,
given as a fibration of a Riemann surface over a non-compact line. For
simplicity, we focus on the case where the metric is a product of that on
the Riemann surface and the interval.

Let us begin by packaging the field content of the Higgs bundle fields of the
six-brane gauge theory wrapped on a curve $C$. Recall that the bosonic
field content of the six-brane gauge theory consists of a gauge field $A_{7D}$
as well as a triplet of scalars. After compactifying on a Riemann surface, we
can sort all of these fields into 5D supermultiplets. Owing to the topological twist, all
fields in the same supermultiplet must have the same differential form content
in the internal space. In the 5D\ $\mathcal{N}=1$ effective field theory, we
have a 5D\ vector multiplet with a real adjoint valued scalar,
which we label as $\phi_{t}$, in accord with its interpretation
in the associated PW system defined on $Q=%
\mathbb{R}_t
\times C$. In the 5D\ effective field theory, we also get hypermultiplets
indexed by points of $C$, coming from the gauge field and Higgs field of
the Hitchin system.

In terms of 4D $\mathcal{N}=1$ fields, the 5D vector
multiplet descends to a 4D\ $\mathcal{N}=2$ vector multiplet. The complex
adjoint valued scalar of this system is given by a complexified gauge
connection which we write as:%
\begin{equation}
\mathbb{D}_{t}=d_{t}+A_{t}+i\phi_{t}=d_{t}+\mathcal{A}_{t},
\end{equation}
where in the last equality we have used the complexified connection introduced
earlier in our discussion of the PW\ system. There are also the degrees of
freedom of the Hitchin system. These can also be packaged in terms of a
complexified connection which we write as:%
\begin{equation}
\mathbb{D}_{C}=d_{C}+A_{C}+i\phi_{C}=d_{C}+\mathcal{A}_{C}.
\end{equation}
Observe that on a Riemann surface, there are an equal number of A- and
B-cycles; these canonically pair to form the degrees of freedom of a
hypermultiplet. To emphasize this, we write the pair as $\mathbb{D}_{A}\oplus\mathbb{D}_{B}$.
Summarizing, we have found three adjoint
valued chiral multiplets.

In terms of 4D $\mathcal{N}=1$ fields, the interaction terms of the 5D\ field
theory are constrained by 4D $\mathcal{N}=2$ supersymmetry. In 4D
$\mathcal{N}=1$ language, the superpotential for the bulk fields of the
Hitchin system then takes the form (see e.g. \cite{Marcus:1983wb, ArkaniHamed:2001tb, Beasley:2008dc, Apruzzi:2016iac}):%
\begin{equation}
W_{\text{bulk}}=\underset{%
\mathbb{R}
\times C}{\int}\sqrt{2} \, \text{Tr}\left(  \mathbb{D}_{A}\cdot\mathbb{D}%
_{t}\cdot\mathbb{D}_{B}\right)  ,
\end{equation}
where the \textquotedblleft$\cdot$\textquotedblright\ indicates a wedge
product operation as well as multiplication of matrices in the adjoint
representation of the gauge group (i.e. by commutators in the Lie algebra). We
can also couple this system to additional 5D hypermultiplets (in some
representation of the gauge group)\ localized at points of $C$.
This proceeds through the generalization:%
\begin{equation}
W=\underset{%
\mathbb{R}
\times C}{\int}\sqrt{2}\left(  \text{Tr}(\mathbb{D}_{A}\cdot\mathbb{D}%
_{t}\cdot\mathbb{D}_{B})+\underset{p}{\sum} \, \delta_{p} \, \mathbb{H}_{p}^{c}%
\cdot\mathbb{D}_{t}\cdot\mathbb{H}_{p}\right)  , \label{PresidentWya}%
\end{equation}
in the obvious notation.

Supersymmetric vacua of the 5D\ system are recovered from the F-term equations
of motion coming from varying $W_{\text{eff}}$ with respect to the different
chiral superfields. Doing so, we obtain the F-term equations of motion:%
\begin{align}
\lbrack\mathbb{D}_{A},\mathbb{D}_{B}]  & = \underset{p}{\sum} \, \delta
_{p} \,  \mathbb{H}_{p}^{c} \cdot \mathbb{H}_{p} \\
\lbrack\mathbb{D}_{t},\mathbb{D}_{A}]  &  =0\\
\lbrack\mathbb{D}_{t},\mathbb{D}_{B}]  &  =0.
\end{align}
We recognize the first equation as that of the Hitchin system coupled to
defects. The remaining two equations are simply those associated with the
PW\ system on $Q=%
\mathbb{R}_t
\times C$.

At first, this might suggest that the resulting solutions will generically
preserve 4D\ $\mathcal{N}=2$ supersymmetry rather than just $\mathcal{N}=1$
supersymmetry. We can see that this is not the case based on the
structure of possible solutions. In $\mathcal{N}=2$ terms, the Coulomb branch
of the field theory amounts to setting hypermultiplet vevs to zero, namely
$\mathbb{D}_{A}=\mathbb{D}_{B}=\mathbb{H}_{p}^{c}=\mathbb{H}_{p}=0$ with
$\mathbb{D}_{t}$ non-zero. The Higgs branch is specified by setting
$\mathbb{D}_{t}=0$. There are mixed Coulomb / Higgs branch directions in the
moduli space, but these do not involve the same directions in the gauge
algebra. In the PW\ system, we can have more general solutions since only $\mathcal{N} = 1$
supersymmetry needs to be retained. Of course, if we treat
the above equations as simply specifying the field content
of a 4D\ effective field theory, we could only obtain $\mathcal{N}=2$ vacua.
However, by allowing all modes of the higher-dimensional theory to
participate, there is no need to work exclusively in terms of purely massless
4D\ fields. From this perspective, the interpolating solutions we
have introduced are, by necessity, associated with massive modes of the
higher-dimensional theory.

Another way to state the same conclusion is to return to the 5D\ effective
field theory, but to allow position dependent higher dimension
operators in the 5D\ effective Lagrangian:%
\begin{equation}\label{Wilson}
\mathcal{L}_{\text{eff}}\supset\underset{i}{\sum}c_{i}(t)\frac{O_{i}\left(
x_{4D},t\right)  }{\Lambda^{\Delta_{i}-5}},
\end{equation}
where $\Delta_{i}$ labels the dimension of some operator $O_{i}$. In
principle, we can write down all possible higher order terms compatible with
4D $\mathcal{N}=1$ supersymmetry. To illustrate how this works in practice,
let us return again to the superpotential of equation (\ref{PresidentWya}),
but now expanded around a zero mode of the 4D theory:%
\begin{align}
\mathbb{D}_{A}  &  = \delta \mathbb{D}_{A} + \mathbb{D}_{A}^{(KK)}\\
\mathbb{D}_{B}  &  = \delta \mathbb{D}_{B} + \mathbb{D}_{B}^{(KK)}\\
\mathbb{D}_{t}  &  = \delta \mathbb{D}_{t} + \mathbb{D}_{t}^{(KK)}\\
\mathbb{H}_{p}  &  = \delta \mathbb{H}_{p} + \mathbb{H}_{p}^{(KK)}\\
\mathbb{H}_{p}^{c}  &  = \delta \mathbb{H}_{p}^{c} + \mathbb{H}_{p}^{c(KK)}%
\end{align}
In the above, we note that there could of course be multiple zero modes and
KK\ modes. All of this has been condensed in the present notation.
Substituting these expressions into the superpotential and integrating out all
massive modes, we obtain interaction terms such as:%
\begin{align}
W = & \underset{\mathbb{R} \times C^{(1)} \times C^{(2)}}{\int}\sqrt{2}\left(
\delta \mathbb{D}_{A}\cdot \delta \mathbb{D}_{t}\cdot \delta \mathbb{D}_{B}+
\underset{p}{\sum} \, \delta_p \, \delta \mathbb{H}_{p}^{c}\cdot \delta \mathbb{D}_{t}\cdot \delta \mathbb{H}_{p} \right)\\
+ & \underset{\mathbb{R} \times C^{(1)} \times C^{(2)}}{\int}\sqrt{2}\left(\delta \mathbb{D}_{t} \cdot \delta \mathbb{D}_{B} \cdot \frac{1}{D_{A}^{\prime}}
\cdot \delta \mathbb{D}_{t} \cdot \delta \mathbb{D}_{B}+ \delta \mathbb{D}_{t} \cdot \delta \mathbb{D}_{A} \cdot \frac{1}{D_{B}^{\prime}} \cdot \delta \mathbb{D}_{t} \cdot \delta \mathbb{D}_{A} \right)\\
+ & \underset{\mathbb{R} \times C^{(1)} \times C^{(2)}}{\int}\sqrt{2} \left(\delta \mathbb{D}_{A} \cdot \delta \mathbb{D}_{B}+\underset{p}{\sum} \, \delta_{p} \,%
\delta \mathbb{H}_{p}^{c} \cdot \delta \mathbb{H}_{p}\right) \cdot  \frac{1}{D_{t}^{\prime}} \cdot \left(  \delta \mathbb{D}_{A} \cdot \delta \mathbb{D}_{B}+\underset{p}{\sum}\, \delta_{p} \, \delta \mathbb{H}_{p}^{c} \cdot \delta \mathbb{H}_{p}\right)
\end{align}
where the expressions $1/D^{\prime}$ denote Green's functions on $\mathbb{R} \times C^{(1)} \times C^{(2)}$
with the zero modes omitted. In this expression, we have also
absorbed the different notions of ``trace.'' Let us note that we have confined our answer to dimension
six operators because in the above, we have only presented the F-terms. For the D-terms, there is no
such restriction, and it is also more difficult to perform the corresponding effective field theory analysis.

The derivation of this expression for the effective superpotential follows from using the F-term equations of motion,
and then plugging these solutions back in. Such a result is therefore exact in the F-terms, but it also
implicitly depends on unprotected (non-holomorphic) D-term data. To illustrate how this works in practice,
consider for example the interaction term $\mathbb{D}_A \cdot \mathbb{D}_t \cdot \mathbb{D}_{B}$. Substituting in, we get
terms such as $\delta \mathbb{D}_A \cdot \mathbb{D}_{t}^{(KK)} \cdot \delta \mathbb{D}_{B} + M \mathbb{D}_{t}^{KK} \cdot \mathbb{D}_{t}^{KK}$. In this case, the equation of motion for $\mathbb{D}_{t}^{(KK})$ is of the form:
\begin{equation}
\mathbb{D}_{t}^{(KK)} \sim \delta \mathbb{D}_{A} \frac{1}{M}  \delta \mathbb{D}_{B} + ...,
\end{equation}
where the ``...'' refers to other terms obtained by varying the superpotential with respect to $\mathbb{D}_{t}^{(KK)}$. Here, the factor of ``$1/M$'' refers to the masses of the KK states. Now, feeding this back into the terms $\delta \mathbb{D}_A \cdot \mathbb{D}_{t}^{(KK)} \cdot \delta \mathbb{D}_{B} + M \mathbb{D}_{t}^{KK} \cdot \mathbb{D}_{t}^{KK}$, we arrive at one of the claimed interaction terms. Scanning over all couplings between two zero modes and one KK mode, we obtain the interaction terms indicated above. Similar considerations hold when we integrate out the KK modes associated with the other bulk degrees of freedom, as well as the modes such as $\mathbb{H} \oplus \mathbb{H}^{c}$ which are localized on a curve.

The key feature of these expressions is that these propagators clearly involve a non-trivial
dependence on all three coordinates of the three-manifold $Q$. As such, we
should expect the 5D effective field theory to have position dependent Wilson
coefficients, thus demonstrating the general claim. 
The global form of these expressions involves 
integrating expressions for the zero mode profiles such as $f_{1}(t,x_1,y_1)$ and $f_{2}(t,x_2,y_2)$ against these Green's functions through schematic expressions such as:
\begin{equation}
\underset{\mathbb{R} \times C^{(1)} \times C^{(2)}}{\int} f_{1}(t,x_1,y_1) \left[ \frac{1}{D^{\prime}} \right] (t \vert x_1,y_1; x_2,y_2) f_{2}(t,x_2,y_2),
\end{equation}
and the associated Wilson coefficients for the superpotential are then given via:
\begin{equation}
c_{\mathrm{quartic}}(t) = \underset{C^{(1)} \times C^{(2)}}{\int} f_{1}(t,x_1,y_1) \left[ \frac{1}{D^{\prime}} \right] (t \vert x_1,y_1; x_2,y_2) f_{2}(t,x_2,y_2),
\end{equation}
in the obvious notation.

On general grounds, we also expect that the appearance of localized matter may
also generate singularities in the form of a given interpolating
solution. As a first example, observe that a background value for a localized
hypermultiplet produces a delta function localized source term in the Hitchin
system coupled to defects. With this in mind, the appearance of a singularity
somewhere in the $t$ direction can also be interpreted -- in the PW\ system --
as a background expectation value for matter localized on some
lower-dimensional cycle in $Q$. The appearance of such singularities is of
course well known in other contexts, and determines a defect operator.
We will return to the effect of these defect operators on the background equations later in section \ref{ssec:DEFECTS}.
Near these singularities, the profiles of the
higher-dimensional fields will also exhibit higher order singularities. There
is then some additional data associated with the boundary conditions for
fields.

\subsection{Interpolating 4D\ Vacua}

In the previous subsection we showed that interpolating profiles for Higgs
bundles on a Riemann surface have a natural interpretation in terms of
5D\ vacua with position dependent Wilson coefficients for higher dimension
operators in the effective field theory. We now perform a similar analysis in
the case of Higgs bundles used to define 4D\ vacua, and the corresponding
interpolating profiles. In this case, there is already an important subtlety
because we have already mentioned two distinct ways to generate 4D\ vacua,
namely from M-theory on local $G_{2}$ spaces, or from F-theory on local
Calabi-Yau fourfolds.

Our general expectation is that we can use the 3D\ effective field theory
defined by M-theory on a local $Spin(7)$ space as the \textquotedblleft
glue\textquotedblright\ which can interpolate between these different
profiles. In the case of the PW\ system, this interpretation is
straightforward, since it is defined on a three-manifold, and further fibering
this over an interval will result in a non-compact four-manifold. In the case
of the BHV\ system, however, additional care is required because both the
BHV\ and local $Spin(7)$ systems make reference to a four-manifold!

Keeping these subtleties in mind, we shall therefore reverse the order of
analysis. We begin with the 3D $\mathcal{N}=1$ effective field theory
generated by M-theory compactified on a $Spin(7)$ manifold. We will then use
this starting point to give an interpretation in terms of a compactification of a 4D\ $\mathcal{N}=1$ theory.

We start with the local $Spin(7)$ system and summarize
the field content of the six-brane gauge theory
wrapped on a four-manifold $M$. Owing to the topological twist, fields in the
same supermultiplet will again sort by their differential form content.
From the bulk of the six-brane gauge theory, we have a 3D\ $\mathcal{N}=1$ vector multiplet. Additionally, we have a 3D $\mathcal{N}=1$ scalar multiplet given by an adjoint-valued self-dual two-form $\Phi_{\text{SD}}$, and another 3D
$\mathcal{N}=1$ scalar multiplet $\mathbb{D}$ given by dimensional reduction
of the internal components of the gauge connection on $M$. There can also
be matter fields localized on Riemann surfaces and one-cycles, but in the
interest of brevity we suppress these contributions for now. Focusing on the
scalar multiplets, the superpotential of the 3D $\mathcal{N}=1$ system is:%
\begin{equation}
W_{\text{bulk}}=\underset{M}{\int}\text{Tr}\left(  \Phi_{\text{SD}}\wedge\left(
\mathbb{F}_{\text{SD}}+\frac{1}{3}\Phi_{\text{SD}}\times\Phi_{\text{SD}}\right)  \right)  ,
\end{equation}
in the obvious notation. Here, we have not distinguished between the zero
modes of a particular solution and all of the Kaluza-Klein modes.

We now assume that our four-manifold $M$ can be written as a product of a
Riemann surface $C$ and a cylinder, i.e. $M= C \times%
\mathbb{R}
\times S^{1}$. The connection to a PW\ system is straightforward; We take the
three-manifold of the PW\ system to be $Q= C \times S^{1}$, fibered over the
real line factor. As we have already noted, the local $Spin(7)$ equations
specialize to those of the PW\ system. Including the contributions in the $%
\mathbb{R}
$ direction, we also clearly see that there is a whole tower of KK\ modes
which participate in this process. This is quite analogous to what we already
saw in the context of 5D interpolating\ vacua for Hitchin systems as specified
by the PW\ system. Again, the interpretation is in terms of a 4D\ effective
field theory but with position dependent coefficients for higher-dimension operators.
By using the local $Spin(7)$ system, we see that it is possible to interpolate between
different perturbations of PW systems. Geometrically, this provides a way to glue together two non-compact $G_2$ spaces
to produce a non-compact $Spin(7)$ space. We refer to this as a ``PW--PW'' gluing. We will discuss some examples of these interpolations in section \ref{sec:4Dinter}.

Consider next the other specialization in the local $Spin(7)$ equations, as
captured by the BHV\ system. We would like to understand the effective field theory
interpretation for gluing two BHV solutions via a local $Spin(7)$ system,
as well as possible ways to glue a BHV solution to a
PW solution. Since we have already discussed how to glue together
PW solutions, it suffices to consider the gluing of a PW and BHV system.
The physical interpretation of this situation is clearly more subtle because the
$%
\mathbb{R}_t
$ factor in the BHV system remains \textit{inside} the four-manifold! In what sense, then, can
we claim that there is an asymptotic limit captured by a 4D\ $\mathcal{N}=1$
effective field theory?

The important clue here is that the 4D\ interpretation
of the BHV system takes place in F-theory rather than M-theory. Recall that in
the standard match between M- and F-theory, M-theory compactified on an
elliptically fibered Calabi-Yau $X$ is dual to F-theory on $X \times S^1$. In
this correspondence, the volume of the elliptic curve on the M-theory side of
the correspondence is inversely related to the size of the $S^{1}$ on the
F-theory side. In particular, the component of the seven-brane gauge field along this
$S^1$ direction becomes ``T-dual'' in the local M-theory picture to one of the components of the
one-form Higgs field in the PW system. Said differently, a direction in the cotangent
bundle $T^{\ast} Q$ of the local PW system is actually part of the 4D spacetime on the F-theory side.

With this in mind, we shall denote the spacetime direction used for the interpolating
profile by writing $\mathbb{R}_{\text{M-th}}$ when referring to 4D M-theory vacua obtained from compactification on a $G_2$ space,
and $\mathbb{R}_{\text{F-th}}$ when referring to 4D F-theory vacua obtained from compactification on an elliptically
fibered Calabi-Yau fourfold. As we have already remarked,
on the F-theory side $\mathbb{R}_{\text{F-th}}$ is a spacetime direction, while $\mathbb{R}_{\text{M-th}}$
should be treated as an internal direction. Conversely, on the M-theory side $\mathbb{R}_{\text{M-th}}$ is a spacetime direction,
while $\mathbb{R}_{\text{F-th}}$ should
be treated as an internal direction.

In terms of the field content of the two local models, there is a corresponding interchange in the gauge field and scalar degrees of freedom.
On the PW side, we have a 7D gauge field which we split up as $A_{7D} = A_{3D} \oplus A_{\text{M-th}} \oplus A_{Q}$ and a triplet of real scalars $\phi_{1},\phi_{2},\phi_{3}$. On the BHV side, we have an 8D gauge field which we split up as $A_{8D} = A_{3D} \oplus A_{\text{F-th}} \oplus A_{Q} \oplus A_{4}$, and a pair of real scalars $\phi_{1}, \phi_{2}$. The non-trivial interchange is then:
\begin{align}
\text{PW} \,\,\, & \leftrightarrow \,\,\, \text{BHV} \\
A_{\text{M-th}} \,\,\, & \leftrightarrow \,\,\, A_{4} \\
\phi_{3} \,\,\, & \leftrightarrow \,\,\, A_{\text{F-th}}.
\end{align}

This is in accord with the twisted connected sums \cite{kovalevTCS} and generalized connected sums
\cite{Braun:2018joh} constructions in which an $S^1$ in the base is interchanged with one in the fiber.
The main difference with these cases is that here, we have decompactified these two $S^1$ factors. Additionally,
we have given a 4D spacetime interpretation, in accord with the fact that it is actually connecting M- and F-theory vacua.

In all of these cases, we see that a quite similar analysis of the effective field theory allows us to package the 4D theory in terms of 3D fields, parameterized by an additional spatial direction. In the effective Lagrangian, we therefore have position dependent Wilson coefficients
of the form:
\begin{equation}
L_{\text{eff}}\supset\underset{i}{\sum}c_{i}(t)\frac{O_{i}\left(
x_{3D},t\right)  }{\Lambda^{\Delta_{i}-4}},
\end{equation}
where $\Delta_{i}$ labels the dimension of some operator $O_{i}$ in the 4D theory.

\subsection{Domain Walls for 4D Vacua}

A general point we have emphasized in the above considerations is that the interpolating geometry of $Spin(7)$ solutions will appear in
the 4D effective field theory as varying the profile of Wilson coefficients for higher dimension operators
in the effective field theory. Since these coefficients are not directly associated with light degrees of freedom
of the 4D theory, it is appropriate to view these interpolating profiles as specifying ``interfaces.'' In subsequent sections we will
construct some explicit examples of such interpolating profiles.

Domain walls are also important and constitute a qualitatively different sort of interpolating profile. In this case,
we have two distinct critical points for a 4D $\mathcal{N} = 1$ superpotential, indicating distinct vacua which cannot be connected
through any sort of adiabatic variation. Our aim in this section will be to illustrate some general properties of such domain wall
solutions. Compared with interpolating profiles for parameters, establishing the existence of such domain wall solutions is
considerably more involved. For this reason, we limit our discussion to general remarks, leaving a more detailed analysis for future
work.

Our starting point is a 4D $\mathcal{N} = 1$ theory with chiral superfields $\Phi^i = \phi^i + ...$, a superpotential $W[\phi^i]$, and a K\"ahler potential $K(\phi^i,\overline{\phi^i})$.
A half-BPS domain wall in the direction $t$ is characterized by the flow equation:
\al{\label{eq:BPSDW} D_t \phi^i = e^{i \eta} G^{i \bar \jmath} \partial_{\bar \jmath} \,\overline { W}\,,
}
where $G^{i \bar \jmath}$ is the inverse K\"ahler metric on the target space of the chiral multiplets of the theory. Here, $\eta$ is a constant that determines which linear combination of supercharges is preserved by the domain wall. It is a well known result \cite{Cvetic:1991vp} that the tension of the domain wall is proportional to the difference between the values of the superpotential in the two vacua. In order to make contact with the 4D $\mathcal{N} =1$ vacua defined by the PW and BHV systems, it is necessary to know the superpotential in each case. We begin with the PW system and then turn to the BHV system.

In the PW system on a three-manifold $Q$, the chiral multiplets of the theory are given by the
combination $\mathcal A = A + i \phi$ and the superpotential is \cite{Pantev:2009de}:
\al{ W_{\text{PW}} = \int_{Q} \text{Tr}\left(\mathcal A \wedge d \mathcal A + \frac{2}{3} \mathcal A \wedge \mathcal A \wedge \mathcal A\right)\,,
}
that is, the superpotential is nothing but the Chern--Simons functional for the complexified connection $\mathcal A$ on the internal three-manifold. Taking a flat K\"ahler metric this gives the domain wall equations:
\begin{equation}\label{KWflow}
 D_t \mathcal{A} = e^{i \eta}*_3  \; \overline{\mathcal{F}} \; ,
\end{equation}
where the Hodge star is in the internal manifold and $\mathcal F$ is the curvature of the connection $\mathcal A$. This has to be combined with the D-flatness condition $D_A * \phi = 0$. In the case when $\eta = 0$, one can exactly recover (\ref{KWflow}) from the local $Spin(7)$ system after choosing an isomorphism $\Omega_{SD}^2(Q \times \mathbb{R}_t)\simeq \Omega^1 (Q)$ and fixing a gauge $A_t=0$. The appearance of the $\eta$-phase in the domain wall BPS equations can be explained as follows: the four manifold $Q \times \mathbb R_t$  has a reduced holonomy group and therefore there is a $U(1)$-freedom in the choice of which supersymmetry generator is preserved in 3D.
These more general equations can be put into the form of the Kapustin--Witten (KW) equations \cite{Kapustin:2006}:
\begin{align}
D_{A} \ast \phi & = 0 \, ,\\
(F-\phi \wedge \phi)_{\mathrm{SD}} & =  +u (D_A \phi)_{\mathrm{SD}}\,,\\
(F-\phi \wedge \phi)_{\mathrm{ASD}} & = -u^{-1} (D_A \phi)_{\mathrm{ASD}} \,,
\end{align}
where the subscripts ``SD'' and ``ASD'' refer to self-dual and anti-self-dual two-forms,
$\phi$ is an adjoint valued one-form, and $u= \tfrac{1+\cos \eta}{\sin \eta}$, and $\phi_t=0$. This last condition is necessary to recover equation (\ref{KWflow}), in addition to the fact that there is no local $Spin(7)$ interpretation of $\phi_t$.\footnote{Imposing this condition on $\phi_t$  is actually much weaker than what one might think because as shown in the original paper \cite{Kapustin:2006}, $\phi_t$ is covariantly constant and commutes with the other spacial components $\phi_\mu$. Moreover, by a vanishing theorem, $\phi_t=0$ follows from the boundary condition $\phi_t|_{\pm \infty}=0$.} Note that these equations are also known as complexified instantons for a complex gauge group $G_{\mathbb{C}}$, since they can be rewritten as $e^{-i\eta / 2}\mathcal{F}= *e^{i\eta / 2}\bar{\mathcal{F}}$, while imposing the moment map $\mu=D_A* \phi=0$ for $G$-gauge transformations. As noted in \cite{Witten:2010}, the flow equations (\ref{KWflow}) are believed to give rise to a sort of complexification of Instanton Floer Homology, whose gradient flows between critical points would exactly correspond to half-BPS domain walls for these 4D $\mathcal{N}=1$ theories. In other words, given two complex flat connections on $Q$ at each infinity, $\mathcal{A}_-$ and $\mathcal{A}_+$, such that $\Delta W(\mathcal{A})\neq 0$ (implying that they belong to two different components of the character variety of $Q$) counting the solutions to such flows enumerates domain walls with tension $\Delta W$.
	
Solutions are quite difficult to establish, and few examples are known. Nevertheless, we can make some general statements. The fact that $\text{Im}(e^{-i\eta}W)$ is constant along the flow indicates that the existence of a solution is heavily reliant on our choice of $\eta$. In fact, an index theory calculation \cite{Witten:2010} implies that finitely many solutions are generically expected, provided that we are allowed to vary $\eta$ and that for some $\eta_0$, $\text{Im}(e^{-i\eta_0}W(\mathcal{A}_+))=\text{Im}(e^{-i\eta_0}W(\mathcal{A}_-))$. A detailed example is presented in \cite{Witten:2010}, in the case of $Q = S^3\backslash K$ where $K$ is the trefoil knot and $G_{\mathbb{C}}=SL(2,\mathbb{C})$. The knot arises from a Wilson operator and sources the complex curvature as $\frac{e^{-i\eta}\mathcal{F}}{2\pi}=\delta_K \mu_R$, leading to the following singularities in $A$ and $\phi$ (up to a gauge transformation on $S^3\backslash K$ that removes a $\frac{dr}{r}$-singularity in $\phi$)
\begin{equation}\label{wilson sing}
	 A=\alpha d\theta+\dots, \; \; \; \; \phi=-\gamma d\theta+\dots
\end{equation}	
where $\alpha-i\gamma=\mu_R$. Note that the singularities of the fields are translationally invariant along $\mathbb{R}_t$ , so a flow between minima\footnote{Actually in this example, one must consider flows between minima of $W(\mathcal{A})+I_R(\mathcal{A})$ where the shift $I_R(\mathcal{A})$ captures the Wilson operator insertion into the path integral. The M-theory interpretation of the Wilson operator is a flavor brane, where after a suitable unhiggsing of $G$ to some larger group, one could derive this coupling by giving a zero-mode localized along $K$  (in the representation $R$ of $G$) a vev.} of $W(\mathcal{A})$ is an honest domain wall, and not a codimension-one disorder operator that will occupy more of this chapter. The details in deriving such a flow and properly treating the gauge ambiguity of $W(\mathcal{A})$ is quite involved, even in this ``simple'' example, so we refer the reader to section (5.2) of \cite{Witten:2010} for details. Defining a complexified Floer theory is of deep mathematical interest and it would be intriguing to explore the recent work of \cite{mano1} and \cite{mano2} to derive more examples of these half-BPS domain walls in 4D $\mathcal{N}=1$ systems (see also \cite{Witten:2010}).

We can follow the same logic for the BHV system: now the chiral multiplets are $\Phi_{(2,0)}$ and $\mathbb{D}_{(0,1)} = \overline{\partial} + \mathbb{A}$ and the superpotential is
\al{ W_{\text{BHV}} = \int \text{Tr}\left(\Phi_{(2,0)} \wedge \mathbb{F}_{(0,2)}\right)\,.
}
In this case the interpretation of the local $Spin(7)$ equations as domain wall equations are a bit more subtle as both the BHV and local $Spin(7)$ systems are on a four-manifold. As we have already mentioned in our analysis of the 4D and 3D effective field theory, an additional direction emerges from also including the volume modulus of the elliptic fiber present in a local F-theory model. More concretely to obtain the $Spin(7)$ equations from the BHV domain wall equations one has to choose all fields to be independent of the domain wall direction using only the connection in this direction to break the 4d Lorentz group. This implies that the covariant derivative becomes simply a commutator with the component of the gauge field along the domain wall direction, and as discussed before this component is identified with the additional self-dual two form $\phi_3$ appearing in the $Spin(7)$ system. This does not fully capture the $Spin(7)$ equations as gradients of $\phi_3$ in the internal direction are not visible, however they will appear upon including in the EFT massive modes of the gauge field coming from dimensional reduction. Along these lines, we also see that we can even expect domain walls which separate vacua specified in different duality frames, as is the case in the PW system (defined via IIA / M-theory) and the BHV system (defined via IIB / F-theory).

\section{Abelian Solutions} \label{sec:CAINANDABEL}

Having presented some general observations on
Higgs bundle vacua and interpolating profiles,
in this section we turn to an analysis of ``abelian solutions'' which solve the local $Spin(7)$ equations,
namely the special case where we assume the Higgs field is diagonal.

Geometrically, this class of diagonalizable configurations are those for which the classical
geometry of a $Spin(7)$ space is expected to match to the local gauge theory description. In more general solutions
as captured by T-brane configurations (see e.g. \cite{Aspinwall:1998he, Donagi:2003hh,Cecotti:2009zf,Cecotti:2010bp,Donagi:2011jy,Anderson:2013rka,
Collinucci:2014qfa,Cicoli:2015ylx,Heckman:2016ssk,Collinucci:2016hpz,Bena:2016oqr,
Marchesano:2016cqg,Anderson:2017rpr,Collinucci:2017bwv,Cicoli:2017shd,Marchesano:2017kke,
Heckman:2018pqx, Apruzzi:2018xkw, Cvetic:2018xaq, Collinucci:2018aho, Carta:2018qke, Marchesano:2019azf, Bena:2019rth, Barbosa:2019bgh, Hassler:2019eso}), some of the gauge theory degrees of freedom come
from M2-branes wrapped on collapsing two-cycles. At a practical level, another
reason to focus on abelian solutions is that they are easier to analyze. Moreover,
perturbations in such configurations, as obtained from switching on localized matter field
vevs lead to more general solutions. We leave the latter point implicit in much of what follows, but we expect
the analysis to be quite similar to what occurs in the case of T-brane vacua, as in references \cite{Cecotti:2010bp, Donagi:2011jy,Anderson:2013rka,Anderson:2017rpr}.

We refer to an ``abelian configuration'' as one in which the data of the vector bundle and the Higgs field
are independent of one another. More precisely, in terms of the gauge group $G$, we pick a subgroup $H \times K \subset G$
such that the Higgs field takes non-trivial values in the Lie algebra of $H$, with $\phi_{\mathrm{SD}} \times \phi_{\mathrm{SD}} = 0$.
In this case, the  local $Spin(7)$ equations reduce to:
\begin{equation}
F_{\mathrm{SD}} = 0 \,\,\,\text{and}\,\,\, d \phi_{SD} = 0.
\end{equation}

This system of equations has the great advantage of being linear and therefore it is much simpler to build solutions. Moreover the gauge field configuration and the profile of the self-dual two form are independent. Therefore our low energy effective field theory will consist of two decoupled sectors: self-dual instantons and the profile of a harmonic self-dual two-form. Viewed as an M-theory background, we can relate the former with the presence of M2-brane charge.\footnote{The intuition comes from weakly coupled type IIA string theory: in the D6-brane action there is a term of the form $ \int_{\text{D6}} C_3 \wedge \text{tr}(F \wedge F)$ (here we omitted some proportionality factors), meaning that a stack of D6-branes with an instanton configuration on it will source D2-brane charge.} The moduli space of instantons is a well-studied object, and so in what follows we primarily focus on the profile of the Higgs field.

Turning next to the profile of the Higgs field, we see that since we are dealing with a triplet of commuting matrices, we can speak of $\mathrm{rk}(H)$ independent eigenvalues, each of which is a self-dual two-form on $M$. In what follows, we shall actually entertain two-forms which are singular along a submanifold in $M$. Our reason for doing so is that such solutions have a natural interpretation in terms of sources
in the local $Spin(7)$ equations.

Focusing on a linear combination of such eigenvalues, which by abuse of notation we also refer to as $\phi_{\mathrm{SD}}$, we see that at least locally, we can introduce an ansatz which solves the equation $d \phi_{\mathrm{SD}} = 0$ by writing $\phi_{\mathrm{SD}} = d \beta + \ast d \beta$ where $\beta$ is a one-form gauge potential for the non-compact gauge group $\mathbb{R}^{\ast}$, i.e. the real non-compact form of $U(1)$. Letting $F_{\mathrm{ncpct}}$ denote the field strength for this gauge potential, we see that the condition $d \phi_{\mathrm{SD}} = 0$ is tantamount to solving the Maxwell field equations for this gauge theory, i.e.:
\begin{equation}
d F_{\mathrm{ncpct}} = 0 \,\,\,\text{and}\,\,\, d \ast F_{\mathrm{ncpct}} = 0.
\end{equation}
The analogy to the Maxwell equations also suggests possible ways in which the right-hand side of this equation may be modified in the presence of sources. In other local gauge theory systems, such sources indicate the presence of background matter fields which have non-zero vev. For example, in the PW system, we can have source terms localized at points of the three-manifold. Extending these to one-cycles in a four-manifold, such sources are the analog of ``electrons'' with a worldline in Euclidean space. By a similar token, the source terms of the BHV system localized along a two-cycle are analogous to wires carrying a current in Euclidean space. One might also ask whether it is possible to introduce sources on codimension one subspaces. We find that this does not solve the differential equations associated with the local triplet of self-dual two-forms. As a final comment, we note that solutions to the self-duality equations on a four-manifold $M$ have a close connection to the twistor space of $M$. This is not an accident; In subsection \ref{ssec:SPECTRAL} we develop the related geometry of spectral covers based on four-manifolds embedded in $\Omega^{2}_{+}(M)$. Note that the unit norm self-dual two-forms determine an $S^2$, and this total space is just the twistor space of $M$.

Our plan in the rest of this section will be to further explore this special class of abelian configurations, focusing almost exclusively on the behavior of the Higgs field (since in this case it decouples from the gauge bundle). We begin with an analysis of zero modes in such backgrounds, and also present some examples of localized matter in such configurations. After this, we turn to the spectral cover for these local $Spin(7)$
geometries. We also show how perturbations away from a purely abelian configuration produce more general spectral covers.

\subsection{Spectral Covers \label{ssec:SPECTRAL}}

In this section we discuss some spectral methods for analyzing the profile of intersecting brane configurations generated from
a non-zero Higgs field. In related contexts such as intersecting seven-branes in F-theory \cite{Beasley:2008dc, Beasley:2008kw, Donagi:2008ca, Donagi:2008kj, Donagi:2009ra} and intersecting six-branes in M-theory \cite{Pantev:2009de, Braun:2018vhk}, spectral cover methods
provide a helpful tool in analyzing the resulting geometries.

Recall that for the local $Spin(7)$ system, the ambient geometry experienced by a stack of six-branes is given by the total space of the bundle of self-dual two-forms over $M$. We pick a section $v$ of $\Omega^2_+(M)$ such that $(v=0) = M$ specifies the location of the original
brane system. For ease of exposition, we fix our gauge group to be $G = SU(N)$, and work with respect to the fundamental representation. We will indicate some generalizations of these considerations later.

In the fundamental representation of $SU(N)$, the Higgs field is an $N \times N$ matrix.
Introducing the $N \times N$ identity matrix, the spectral equation is:
\begin{equation} \label{speceq}
\mathrm{det} \left( v \mathbb{I}_{N} - \phi_{N \times N} \right) = 0.
\end{equation}
It describes a four-dimensional subspace inside $\Omega^{2}_{+}(M)$, as specified by
the spectral cover $\widetilde{M} \rightarrow M$. Observe that as written,
line \eqref{speceq} determines three hypersurface constraints.

For representations other than the fundamental of $SU(N)$ one should construct a suitable matrix representation of the action of $\phi_{\text{SD}}$ and construct a similar hypersurface. A similar description also
holds for different Lie algebras replacing the determinant with a suitable polynomial
in $v$ with the coefficients given by the Casimir invariants of $\phi_{\text{SD}}$. One can also
work with the analogs of the parabolic and cameral covers \cite{1995alg.geom.5009D}.

Now, in contrast to the case of the Hitchin system and BHV system, there is no
natural ``holomorphic'' combination of variables available. A similar issue also
arises in the case of the PW system, where there is also a triplet of real constraints.
This packaging in terms of real constraints also complicates the interpretation
in terms of intersecting branes. For all of these reasons, we now
focus on the case of abelian configurations for which $\phi_{\mathrm{SD}} \times \phi_{\mathrm{SD}} = 0$,
in which case many of these issues can be bypassed.

In the case where the profile of $\phi_{\text{SD}}$ is abelian,
we can choose the self-dual Higgs field to be valued in $\Omega^{2}_{+}(M) \otimes \mathfrak{h}$, with
$\mathfrak{h}$ the Cartan subalgebra of $\mathfrak{g}$. Returning to the case of $H = SU(N)$, we pick
$\phi_{\text{SD}} = \text{diag} \left(\lambda_1,\dots,\lambda_N\right)$
where the eigenvalues are self-dual two forms subject to the condition
$\sum_i^N \lambda_i =0$. In this case the spectral cover in the fundamental representation
simplifies significantly, becoming
\al{ \prod_{i=1}^N \left(v - \lambda_i\right) = 0\,.
}
This means that the spectral cover is the union of $N$ sheets (though the positions of only $N-1$ sheets are independent inside $\Omega^2_+(M)$).

One of the useful applications of spectral cover methods is to use the intersection pattern of sheets to glean some information about the presence of localized matter. Indeed, one expects that for generic values of $\phi_{\text{SD}}$ the gauge group is completely Higgsed to its maximal torus. However on the loci where two sheets meet there will be a local enhancement of the gauge group which, following the unfolding procedure of \cite{Katz:1996th}, indicates the presence of localized matter. Geometrically we therefore expect to have localized matter whenever two eigenvalues coincide, and this sheet intersection can occur in different codimension on $M$ depending on the profile of the eigenvalues. It is possible to have matter localized on a codimension two subspace inside $M$, namely matter localized on a two-dimensional cycle inside $M$, when two components of the triplet of the eigenvalues become identical with the third one being zero. Since locally one component of $\phi_{\text{SD}}$ vanishes, this is the kind of localized matter appearing in BHV solutions (matter on curves). The other case is to have matter localized on a codimension three subspace inside $M$, namely matter localized on a one-dimensional cycle inside $M$. This case requires all three components of a pair of eigenvalues to coincide with no component being identically zero, and it is the kind of matter which appears in PW systems.

We can also include ``abelian fluxes'' in the same geometric setting. Indeed, we are free to also consider vector bundles which split up as a direct sum of bundles with $U(1)$ structure group. For a gauge group $SU(N)$, this will appear as a decomposition:
\begin{equation}
V = \mathcal{L}_{1} \oplus ... \oplus \mathcal{L}_{N},
\end{equation}
such that the first Chern class of $V$ vanishes. This can also be used to define a corresponding ``universal line bundle'' on $\widetilde{M}$,
much as in other spectral cover constructions. In the context of 4D BHV models, such fluxes are necessary to realize a chiral matter spectrum, and this will also affect the zero mode spectrum of the 3D model.

Given the presence of localized matter at the intersection of sheets one may wonder how the geometry is modified when the matter fields acquire a non-vanishing vacuum expectation value. This would result in a recombination of different sheets, producing a T-brane configuration. However, in contrast to the BHV system, the absence of a holomorphic structure means the resulting spectral cover may not be as useful in extracting the appearance of localized matter. A similar issue was noted in PW systems with T-brane configurations \cite{Barbosa:2019bgh}. We leave a full analysis of this case for future work.

\subsection{Zero Mode Profiles}

In this section we turn to an analysis of the zero mode profiles generated from working around a fixed Higgs field background.
To have a non-zero abelian configuration in the first place we must assume that there is a suitable set of harmonic
self-dual two-forms on $M$. On a compact four-manifold $M$, we thus require $b_{2}^{+} > 0$. We can also work
more generally by allowing singularities in the profile of the Higgs field. Denoting by $P$ the point set of singularities,
we only  demand the existence of a harmonic self-dual two-form on $M \backslash P$. In the latter case, the condition of compactness is
instead replaced by a notion of suitable falloff for fields near the deleted regions of $M$. In what follows, we do not dwell on this
point, and assume a sufficiently well-behaved compactly supported cohomology theory in all cases considered.

Given a solution to the local $Spin(7)$ equations,
zero modes correspond to linearized fluctuations:
\begin{align}
A & = \langle A \rangle + a \\
\phi_{\mathrm{SD}} & = \langle \phi_{SD} \rangle + \varphi.
\end{align}
Here, we will be interested in the special case where $\phi_{\text{SD}}$ takes
values in the Cartan subalgebra $\mathfrak h \subset \mathfrak{g}$.
To understand the matter content, it is convenient to decompose
the adjoint representation of $G$ into representations of
$H \times K$ where $K$ now refers to the commutant of $H$ inside $G$.
By abuse of notation, we also write $H = U(1)^r$ since now we are dealing with
abelian configurations anyway. The relevant breaking pattern is:
\begin{equation}\label{basic reps}
G\rightarrow K \times U(1)^r    \implies \text{Adj}(G)\rightarrow \text{Adj}(K)_0 \oplus \mathbf{1}^{\otimes k}_0\bigoplus_i\left( \mathbf{R}_{i,\mathbf q_i} \oplus \overline{\mathbf{R}}_{i,-\mathbf{q}_i}\right)\,.
\end{equation}
Here, $\mathbf R_i$ are some representations of $K$ and $\mathbf q_i$ denotes the vector of $U(1)$ charges.
To proceed further, we separate our analysis into modes which have all $U(1)$ charges
zero (bulk modes), and modes with at least one non-zero $U(1)$ charge (localized modes).

\subsubsection{Bulk Modes}

We expect to have bulk modes corresponding to uncharged representations
which are not affected by the background of $\phi_{\text{SD}}$. Their zero mode equations are
\begin{align}
(da)^+=0\,, \qquad   d\varphi =0\,,
\end{align}
which for a generic metric implies $da=0$, therefore we have $b_2^+ + b_1$ bulk
scalar multiplet zero-modes in both the adjoint representation of $K$ and
in the uncharged representation $ \mathbf{1}^{\otimes r}_0$. By standard considerations
we will also generate a 3D $\mathcal{N}=1$ vector multiplet for $K\times U(1)^r$.

\subsubsection{Localized Modes}

Consider next the profile of fluctuations which have non-trivial $U(1)$ charge. As per our
discussion of spectral covers, we expect these to be located at the intersection of two sheets
of the spectral cover (for a choice of some representation $\mathcal{R}$).
Given a Higgs field $\phi_{\mathcal{R}}$ in a representation
$\mathcal{R}$ of $H$, we get a collection of
eigenvalues $\mathrm{Eigen}(\phi_{\mathcal{R}}) = \{\lambda_{1},...,\lambda_{\dim \mathcal{R}} \}$,
each of which is a section of $\Omega^{2}_{+}(M)$. We expect to find localized
matter at the vanishing locus for:
\begin{equation}
\lambda_{ij} \equiv \lambda_{i} - \lambda_{j}.
\end{equation}
Of course, this difference in eigenvalues is again a self-dual two-form.
To avoid overloading the notation, in what follows we shall reference this difference
in eigenvalues as $\lambda_{\mathrm{SD}}$. We will also compare with the related difference in eigenvalues
$\lambda_{\mathrm{BHV}}$ and $\lambda_{\mathrm{PW}}$ for the BHV and PW systems.

Harmonic self-dual two-forms such as $\lambda_{\text{SD}}$ are objects of some interest in the analytic gauge theory community.\footnote{In the case where $M$ is compact and $b_{2}^{+} > 0$. We expect similar considerations to also hold in cases where the self-dual form has non-trivial poles.} This is mainly because $\lambda_{\text{SD}}$ can be treated as a so-called \textit{near}-symplectic form, which means that it is a symplectic form on the complement of the vanishing locus $Z \equiv \{ \lambda_{\text{SD}}=0 \}$ in $M$. As we will confirm below, the locus $Z$ is where the zero-modes are localized so its behavior is crucial for understanding the resulting physics. Since $\lambda_{\text{SD}}$ is locally specified by three real degrees of freedom, $Z$ will generically be codimension-three, although with fine-tuning it may enhance to (co)dimension-two (which is generic from the BHV/holomorphic point-of-view). Because the only compact one-dimensional object is $S^1$, $Z$ is generically a collection of disjoint circles. As shown by Taubes \cite{taubes}, for any class in $H^2_+(M,\mathbb{R})$ and positive integer $n$, there is some $\lambda_{\text{SD}}$ with $n$ circle components in $Z$. Essentially this means that there is no global restriction on $\lambda_{\text{SD}}$ when knowing behavior in a local patch, and in fact an argument in \cite{taubes} says that if we know $\lambda_{\text{SD}}$ and its $Z$-components in some open set $U$ we can perturb it slightly to generate any number of $Z$-components on $M \backslash U$. Interestingly, our calculation of the 3D gauge theory zero modes is very similar to the calculation of Gromov--Witten and Seiberg--Witten invariants on $Q \times S^1$ for $Q$ a three-manifold \cite{gerig}.

We now look at a local patch of a single circle in $Z$, which will be $B \times S^1$, where $B$ is the three-ball/disk. As proved in \cite{honda}, there are exactly two possible forms that $\lambda_{\text{SD}}$ may take, the more obvious one is the so-called ``untwisted form'' and a certain $\mathbb{Z}/2\mathbb{Z}$-quotient yields the ``twisted form.'' The untwisted form can be described with coordinates $(x^1,...,x^4)\in B \times S^{1}$ as
\begin{equation}\label{untwist}
     \lambda_{\text{SD}}=x_1 (dx^{41} + dx^{23}) + x_2(dx^{42}+dx^{31})-2x_3(dx^{43}+dx^{12}),
\end{equation}
where in the above, we have used a condensed notation for wedge products,
writing for example $dx^{ab} = dx^{a} \wedge dx^{b} = dx^{a} dx^{b}$.
By inspection of equation \eqref{untwist}, we observe that this can be recast in terms of the one-form of PW as
\begin{equation}\label{4d3d}
\lambda_{\text{SD}}=*_3 \lambda_{\text{PW}}+dx^4\wedge \lambda_{\text{PW}}  \; \; \; \; \; \lambda_{\text{PW}}=x^1 dx^1+x^2 dx^2 -2x^3 dx^3.
\end{equation}
This means that the untwisted circle generates 3D matter that is a Kaluza-Klein reduction of a 4D chiral multiplet associated to the vanishing locus of $\lambda_{\mathrm{PW}}$ on $B$, so our 3D zero-mode is actually the reduction of a 4D $\mathcal{N} = 1$ chiral multiplet.

In a little more detail, the $S^1$ isometry of the background allows us to reduce the zero-mode equations to that of the
PW system, which thus
yields an explicit solution in the patch. To see how this comes about, let $\omega_i$ ($i$=1,2,3) be the local basis of self-dual two-forms in equation \eqref{untwist}. Then, we may write a candidate zero mode fluctuation in the Higgs field as $\varphi=\sum_i\varphi_i\omega_i=*_3 \varphi+dx^4\wedge \varphi$. By abuse of notation, we shall refer to $\lambda$ and $\varphi$ interchangeably as either self-dual two-forms on $B \times S^1$, or as one-forms on $B$. Consider next the fluctuations of the gauge field $A$. Since we are dealing with small perturbations, we can choose to gauge away the fluctuation along the circle. The field content is then captured by ($\varphi$, $a$), one-forms on $B$.
Normalizing the relevant $U(1)$ charge for the fluctuations to one, the zero mode equations reduce to:
\begin{equation}\label{PWzm}
    d_{3}a-\lambda\wedge\varphi=\partial_4(*_{3}a),
\end{equation}
\begin{equation}\label{PWzm2}
    d_{3}\varphi+\lambda\wedge a=-\partial_4(*_{3}\varphi),
\end{equation}
\begin{equation}\label{PWzm3}
    d^\dag_{3}\varphi+a\cdot \lambda = 0,
\end{equation}
where the subscript ``$4$'' denotes the circle direction. Because the background is
invariant under the $S^1$ rotation, the right-hand side of each equation is
zero for massless 3D modes. We then see that our equations are exactly of the form
of the PW zero-mode equations, allowing us to package the zero-modes as $\psi \equiv a+i\varphi$
\begin{align}
    d_\lambda \psi=0,\,\,\,\text{and}\,\,\,  d^\dagger_\lambda \psi =0,
\end{align}
where $d_\lambda \equiv d+i\lambda$. We observe here that this really describes four real equations
whereas in the previous treatment we only indicated three real equations in lines (\ref{PWzm})-(\ref{PWzm3}).
The first zero mode equation $d_\lambda \psi=0$ directly matches to equations (\ref{PWzm}) and (\ref{PWzm2}), while
the zero modes in the conjugate representation of the 4D theory are captured by equation (\ref{PWzm3}) and an additional Lorentz gauge type condition on $a$ which has no bearing on the spectrum of the physical theory.

As seen in \eqref{4d3d}, we have $\lambda=idf$ in $B$ where $f$ is a harmonic Morse function of index $+1$ but we could have alternatively written down an $f$ with index $-1$. This is relevant because due to the partial topological twist of the PW system on $Q$,  system chiral modes are one-forms on $Q$  localized at the ($+1$)-index critical points of $f$  and anti-chiral modes are two-forms localized at ($-1$)-index critical points. See \cite{Pantev:2009de} and \cite{Braun:2018vhk} for more details. If in the coordinates of \eqref{4d3d}, we have a localized 4D chiral mode, there is, in this coordinate system, a Gaussian falloff proportional to $\exp(-(x_1)^2 - (x_{2})^2 - (x_{3})^2)$ in the zero mode \cite{Pantev:2009de, Braun:2018vhk}. Including all fields in the same supermultiplet and dimensionally
reducing along the one-cycle, we obtain a 3D $\mathcal{N}=2$ chiral multiplet.

The other local possibility for $\phi_{\text{SD}}$ is the twisted form, which gets its name because we can start with the untwisted solution on $B \times [0 , 2 \pi]$ which furthermore wraps a one-cycle in $M$. We then glue the two ends of the interval as
\begin{equation}
x_1 \mapsto x_1, \; \;  x_2 \mapsto -x_2, \; \;  x_3 \mapsto -x_3 \; \; x_4 \mapsto x_4 - 2\pi,
\end{equation}
and we see that this will not lead to any 3D zero modes as the wavefunctions in the previous paragraph are odd under such a transformation and are gapped out in similar spirit to a Scherk-Schwarz compactification. We note that while Taubes proved that the total number of circles can be an arbitrary number, we do have the somewhat weak constraint which is attributed to Gompf in reference \cite{taubes}:
\begin{equation}
\#(\text{untwisted circles})-1+b^1-b^2_+ \equiv 0 \; \text{mod} \; 2.
\end{equation}

\subsection{Defects and Singularities \label{ssec:DEFECTS}}

In the previous subsection we presented a general discussion on the local structure of matter obtained from an
abelian Higgs field configuration. In addition to this zero locus where sheets of the spectral cover meet, there can
also be various singularities present in the profile of the Higgs field. In the BHV system, these singularities have a natural interpretation
as originating from vevs of matter fields localized on a subspace. In this section we develop an analogous treatment for local $Spin(7)$ systems
with matter on a curve $C$ as well as on a line $L$.

To begin, we need to work out the possible couplings between bulk matter fields and defects of the system. Some elements of this analysis
were presented in \cite{Heckman:2018mxl}, but we give a more complete treatment here. Recall that we will have two different kinds of matter fields depending on the localization patterns inside $M$. For the case of matter fields on a two-cycle $C$, these fields will appear as 5D hypermultiplets and it will be convenient to package them as pairs of 4d $\mathcal N=1$ chiral multiplets in conjugate representations calling them $\chi$ and $\chi^c$. The topological twist implies that these fields will transform as sections of $K_{C}^{1/2}$ (tensored with the restriction of vector bundles specified by the six-branes). The presence of these defects introduces new terms in the superpotential, specifically one gets the interaction:
\al{ W_{C} = \int_{C} \langle \chi^c, D_{C} \chi\rangle+\langle \bar{\chi}^c, D_{C} \overline{\chi}\rangle+\int_{C} i^{\ast}_{C}(\phi_{\text{SD}}) \left[\mu\left(\overline{\chi},\chi\right)-\mu\left(\overline{ {\chi}^c},\chi^c\right)\right] \,,
}
where the pairing $\langle \cdot , \cdot \rangle$ contracts the matter field representations to give a gauge singlet and the moment map $\mu$ maps a representation and its conjugate to the adjoint and $i^{\ast}_{C}(\phi_{\mathrm{SD}})$ denotes the pullback of the self-dual two-form onto the curve $C$. Similarly, the notation $D_{C}$ refers to a covariant derivative obtained from the pullback of the bulk gauge connections on the six-branes to the curve $C$. Here and in the following we will put a bar over any 4D $\mathcal N=1$ chiral multiplet to denote its conjugate anti-chiral multiplet.

In addition to this there can be matter fields localized on a one-cycle $L$ inside $M$. In this case the matter fields will appear as 4D $\mathcal N=1$ chiral multiplets dimensionally reduced along the line $L$. We refer to such fields as $\sigma$. In this case the topological twist will be trivial and the matter fields will simply be scalars on $L$. Again, when these fields are present there will be additional superpotential interactions
\al{ W_{L } = \int_{L} \langle \bar \sigma, D_{L} \sigma \rangle\,,
}
where again the pairing $\langle \cdot , \cdot \rangle$ contracts the matter
field representations to give a gauge singlet. See figure \ref{fig:FOURLOCO}
for a depiction of localized matter in a local $Spin(7)$ system.

\begin{figure}[t!]
\begin{center}
\includegraphics[scale = 0.5, trim = {3cm 2.0cm 3cm 6.0cm}]{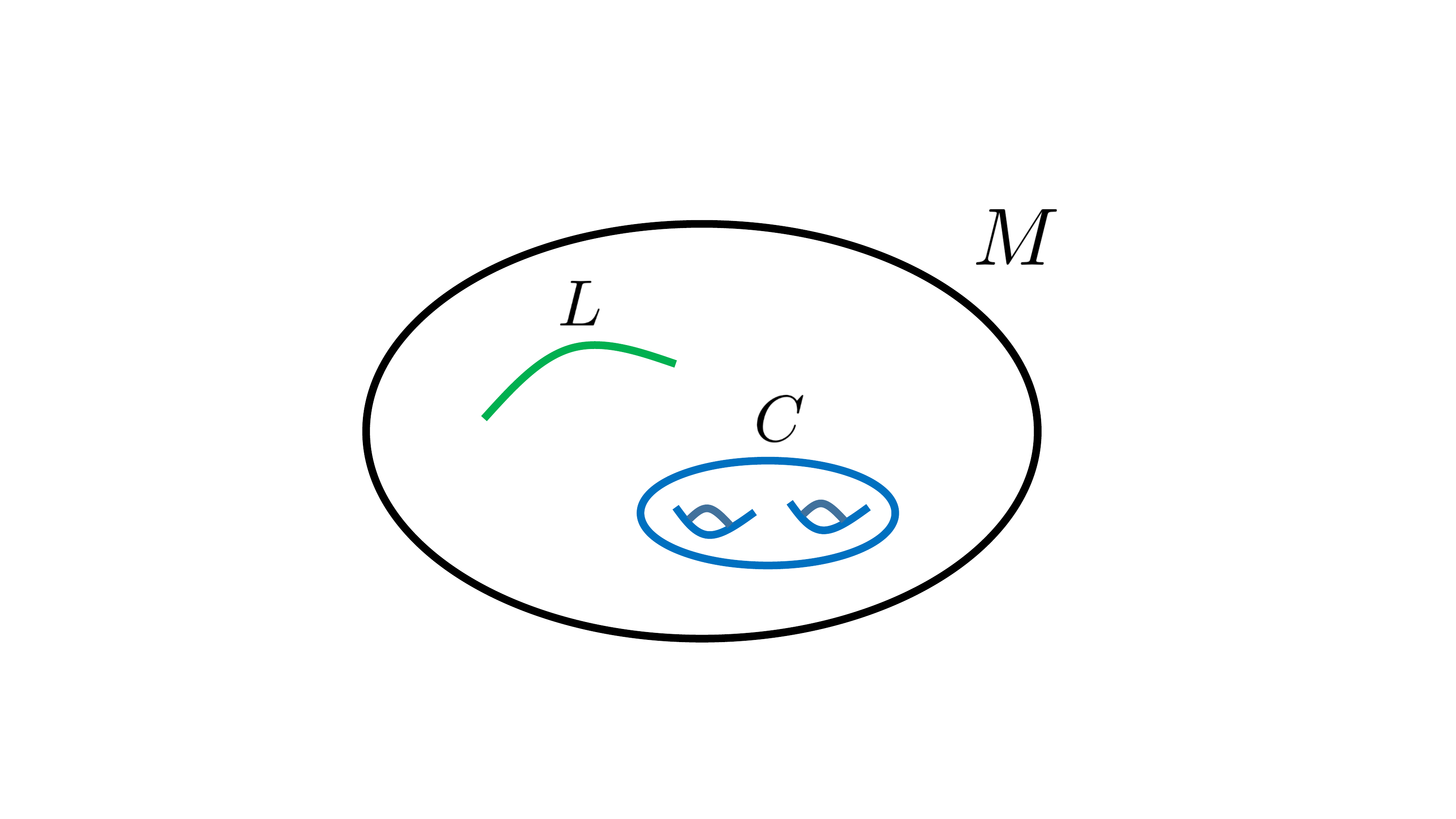}
\caption{Depiction of the four-dimensional gauge theory on a four-manifold $M$
associated with the local $Spin(7)$ system. Matter fields
can be localized on two-cycles $C$, as in the case of the BHV system.
It can also be localized along a real one-cycle $L$,
which amounts to taking matter of the PW system and compactifying further on this line.}
\label{fig:FOURLOCO}
\end{center}
\end{figure}

The presence of localized matter fields
generates a corresponding source term in the local $Spin(7)$ equations. Summing over possible curves and lines, we have
the modified equations of motion:
\al{ F_{\text{SD}} + \phi_{\text{SD}} \times \phi_{\text{SD}} &=\sum_{C} \delta_{C} \left[\mu\left(\overline{\chi},\chi\right)-\mu\left(\overline{ {\chi}^c},\chi^c\right)\right]\,,\\
D_A \phi_{\text{SD}} &= \sum_{C} \delta_{C} \left(\langle \chi^c , \chi\rangle + \langle \overline{{\chi}^c} ,\overline{\chi}\rangle \right) + \sum_{L} \delta_{L} \langle \bar \sigma,\sigma\rangle\,.
}

The presence of these source terms also means that the Higgs field can now acquire possible singularities.
Solutions to the BPS equations in the presence of sources follows
directly appealing to self-dual classical electrodynamics,
albeit with the non-compact gauge group $\mathbb{R}^{\ast}$. Our solution for a singular line with local
coordinate $x^{4}$ is (i.e. ``the worldline of an electron'') has leading behavior:
\begin{equation}
\phi_{4i} \sim \langle \overline{\sigma} , \sigma\rangle \frac{x_i}{2 r^3} \,\,\,\text{and}\,\,\, \phi_{ij} \sim \langle \overline{\sigma}, \sigma \rangle\epsilon^{ijk}\frac{x_k}{2 r^3}\,,
\end{equation}
where we have introduced local coordinates transverse to the line $x^{1}, x^{2}, x^{3}$ with $r^2 = (x^{1})^2 + (x^{2})^2 + (x^{3})^2$.

We can also entertain singularities along a Riemann surface $C$. A singular surface can always be expressed locally in complex coordinates, this is because one can show using the conformal invariance of the BPS equations that $\phi_{\text{SD}}$ specifies an almost complex structure on $M\backslash C$ \cite{honda}, so in a $\mathbb{C}^2$ patch we have the leading behavior:
\begin{equation}
\phi_{\text{SD}} \sim \langle \chi^{c} , \chi \rangle \frac{dz\wedge dw}{z} + h.c.\,,
\end{equation}
where $w$ is a local coordinate along $C$ and $z$ is a coordinate transverse to $C$ such that $C = (z = 0)$.

At the level of gauge theory solutions, one may also consider twisted defects, but since there are no 3D massless states that can have vevs, we ignore this possibility. Also, note that in the presence of defects we should really replace all statements of Betti numbers, cohomology groups, and so on with their relative cohomology analogs with respect to the singular locus of $\phi_{\text{SD}}$.

\section{Interfaces and PW Solutions} \label{sec:AMIMYBROTHERSKEEPER}

In section \ref{sec:EFT} we discussed in general terms how the PW system can be viewed as
defining an interpolating profile between 5D $\mathcal{N} = 1$ vacua,
as captured by the Hitchin system, and that the local $Spin(7)$ system
can be viewed as defining an interpolating profile between 4D $\mathcal{N} = 1$ vacua,
as captured by the PW system. Having given a more general discussion of singularities in
local $Spin(7)$ systems, we now turn to some explicit examples of this sort. As a warmup, we first
present an example of an interface between 5D vacua, and we then turn to an example of an interface between 4D vacua.
In both cases, we find that our abelian Higgs field configuration contains singularities in the
interpolating region of the geometry. We show more generally that abelian interpolating configurations
of this sort always contain such singularities.

\subsection{Codimension-One Defects}

Recall that earlier in section \ref{sec:EFT} we mentioned that our M-theory compactification gives a correspondence between Floer-like solutions to the Kapustin-Witten equations on $Q\times \mathbb{R}_t$ that interpolate between two flat $G_{\mathbb{C}}$-connections on $Q$ and half-BPS domain walls of 4D $\mathcal{N}=1$ systems with tension $T= \vert \Delta W \vert$ set by the difference in the value of the superpotential in the two minima. These domain walls separate different vacua of the theory, and are associated with the interpolation of a light degree of freedom, at least when its mass is below that set by $T^{1/3}$. This begs the question: what is the interpretation of the domain wall solutions we discussed from the perspective of a 4D observer who does not have access to the full higher-dimensional system? When we integrate out to a scale $\Lambda \ll T^{1/3}$, the dynamics the domain wall may be considered fixed and we end up in a situation of studying a field theory in the presence of a codimension-one timelike defect operator. This situation has several different incarnations in the field theory/string theory  literature, and we will fix our nomenclature by calling it an interface. We could have also called this object a disorder operator because, in analogy with the t' Hooft operators of 4D gauge theories, its insertion in the path integral has the effect of changing the space of fields one integrates over to include a certain singularity along the operator, in addition to the fact that they both have an interpretation as an infinitely massive charged excitation. We also see a close relationship between interfaces and boundary conditions, they are essentially synonymous due to what is sometimes called ``flipping'', see for instance \cite{Gaiotto:2008sa}. We call our field theory on the right/left-hand side of the wall with consistent coupling to the interface at $t=0$ as $\mathfrak{T}_L$ and $\mathfrak{T}_R$. This is equivalent to considering a boundary condition for $\mathfrak{T}_R\ominus \mathfrak{T}_L$ that exists just on the right-hand side, where the product $\ominus$ means we take the decoupled sum of the theories but with a $t\rightarrow -t$ action on $\mathfrak{T}_L$.

\subsection{5D Interfaces}
We now turn to interfaces for 5D vacua as obtained from compactifications of M-theory backgrounds.
We primarily focus on M-theory vacua obtained from a local curve of ADE singularities,
with local model given by the Hitchin system.

\begin{figure}[t!]
\begin{center}
\includegraphics[scale = 0.5, trim = {2cm 2.75cm 2cm 5.0cm}]{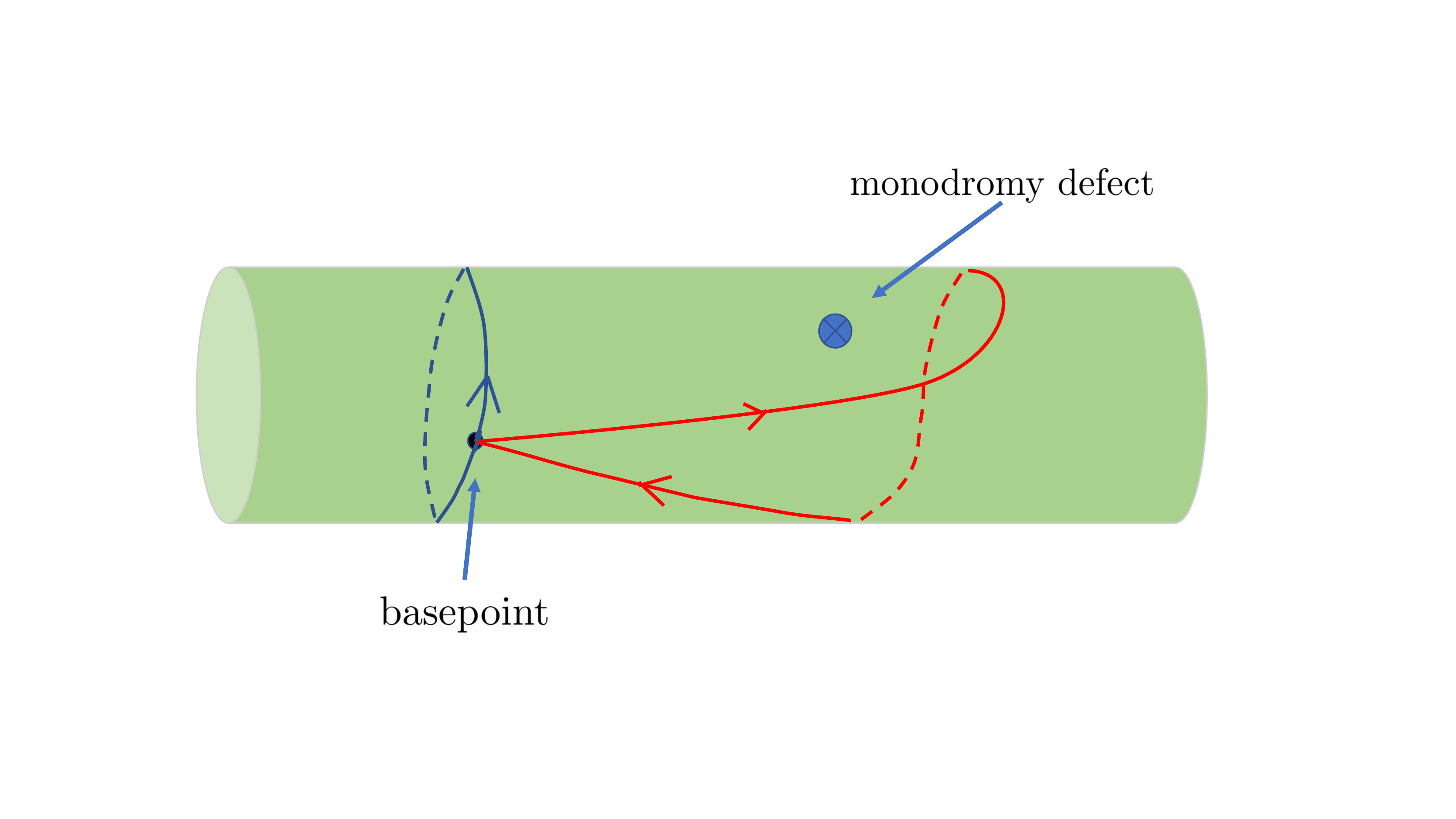}
\caption{Depiction of a monodromy defect operator. This structure occurs along a codimension two subspace.}
\label{fig:MONO}
\end{center}
\end{figure}

We begin with some general considerations. Recall that on $C$ a genus $g$ curve with marked points, solutions to Hitchin's equations are given by complex flat connections with prescribed holonomies around the marked points. This means that the BPS solutions on $C\times \mathbb{R}_t$ with a non-trivial interpolation must have some sort of singularities since flat connections on this three-manifold can always be pulled back to $C$. This agrees with the fact there should not be domain walls interpolating between different vacua of a 5D $\mathcal{N}=1$ theory since $\pi_0(\mathcal{M}_{\textnormal{vac.}})=1$. To study a change in monodromy, we must focus on singularities localized on a one-cycle in $C$, at say $t=0$, because the effect of a point-localized source can be decoupled by shifting counters around the source, while a line-localized source cannot be avoided by all of the 1-cycle counters due to the nondegenerate pairing on $\pi_1(C)$. These defects are known as monodromy defect operators and for the case of 5D interfaces we can build up any representation $\rho:\pi_1(C)\rightarrow G_{\mathbb{C}}$, and thus can interpolate between any two Hitchin solutions given by representations $\rho_L$ and $\rho_R$ by complex conjugation and $t$-reflection.

More specifically, we define a monodromy defect operator much as in \cite{Witten:2011zz} on some manifold $X$ by excising a codimension-two submanifold $U$ and prescribing some monodromy $\mathcal{M}\in G_{\mathbb{C}}$ around it in $X\backslash U$ with the lowest order singularity possible in $\mathcal{A}$. In our case of the three manifold $X= C \times \mathbb{R}_t$, the defect operator is a Wilson loop with the singularity structure of (\ref{wilson sing}). We can then engineer any $\rho_R$ from a trivial representation $\rho_L=\boldsymbol{1}$ by an interpolating representation $\rho_{\textnormal{int}}:\pi_1(X\backslash U,x_0)\rightarrow G_{\mathbb{C}}$ where we chose a basepoint on the left side $(z_0,t_0) \equiv x_0\in C\times (-\infty,0)$. The idea is that $\rho_{\textnormal{int}}$ is trivial when restricting to paths on the lefthand side but paths that only wrap cycles on the right-hand side will necessarily wrap at least one component of $U$ and have nontrivial monodromy. Writing the generators of $\pi_1(C)$ as $A_i$, $B_i$ where $i=1,\dots,g$, the automorphism $A_i \leftrightarrow B_i$ allows us to assign a holonomy to a path that wraps $A_i$ for $t>0$ given by the monodromy $\mathcal{M}_{B_i}$, and similarly $\rho(B_i)=\mathcal{M}_{A_i}$. Because this assignment is at the level of generators we can build any monodromy representation this way. See figure \ref{fig:MONO} for a depiction of a
monodromy defect operator.

\begin{figure}[t!]
\begin{center}
\includegraphics[scale = 0.5, trim = {2cm 1.0cm 2cm 1.0cm}]{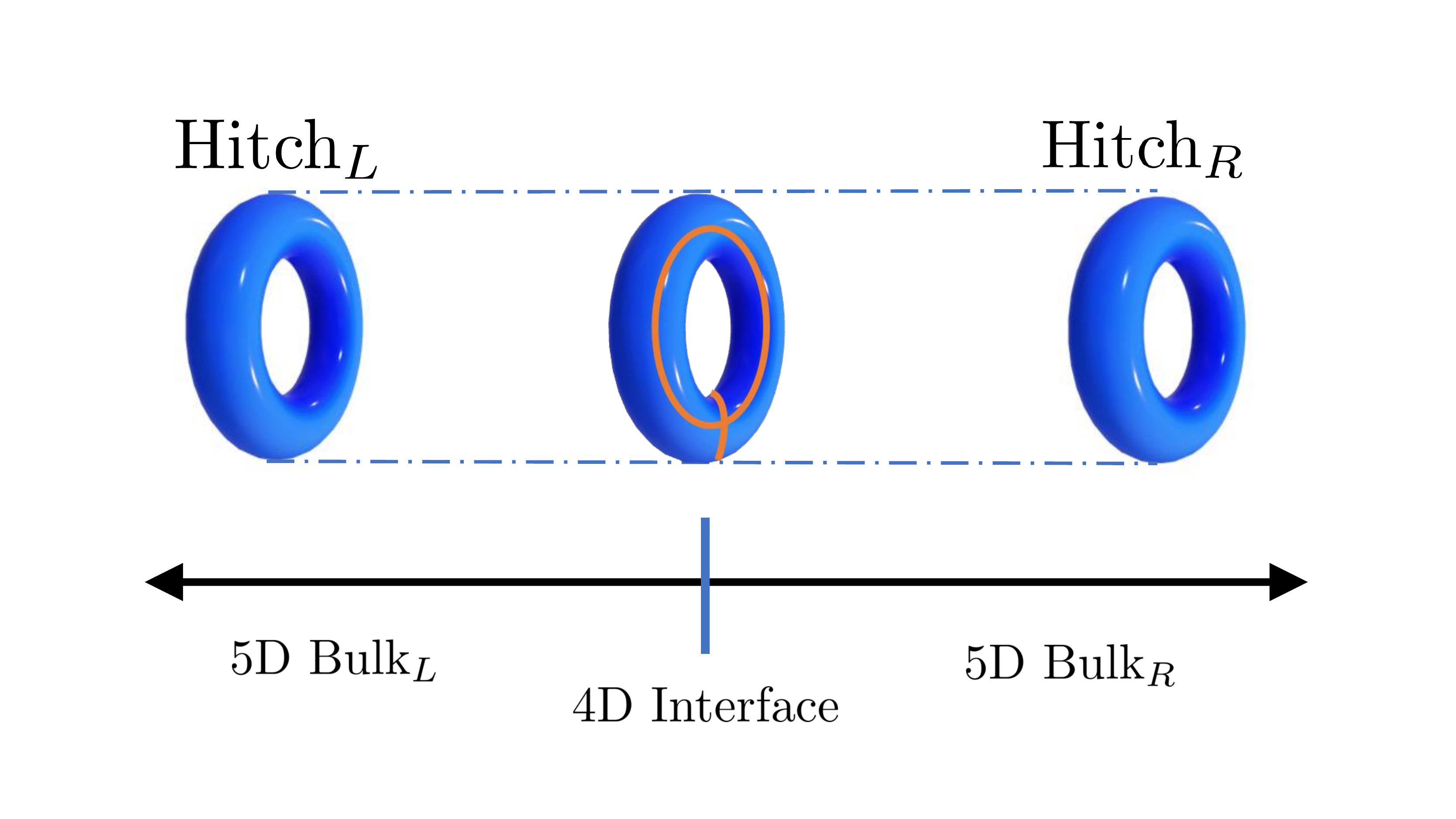}
\caption{Depiction of an interpolating profile between two 5D $\mathcal{N} = 1$
vacua with a 4D interface. The compactification geometry is captured by asymptotically
Calabi-Yau threefold geometries given by a curve of ADE singularities. The interpolating geometry is a non-compact
$G_2$ space. The local gauge theory associated with these cases is a Hitchin system on the left and right, and a PW system
in the interpolating region. We have also indicated the locations of monodromy defect operators of the PW system
by orange lines, namely one-cycles in the non-compact three-manifold.}
\label{fig:5Dvac}
\end{center}
\end{figure}

We now provide an explicit interpolating example for the Hitchin system on a curve $C = T^2$ with marked points.
The presence of marked points will be used to build a position dependent Higgs field since in this case we have $\phi_{\mathrm{Hit}}$ is a meromorphic section of $K_{T^2} \otimes \mathcal{O}(- \sum_{i} p_i)$. We take the three-manifold of the interpolating PW system to be $C \times \mathbb{R}$. In what follows we keep the gauge field $A$ switched off. The BPS equations $d\phi_{PW}=d^{\dagger}\phi_{PW}=0$ are linear so we can simply decompose a solution to the PW system as a linear combination of ``left and right'' pieces, writing:
\begin{equation}
\phi_{\text{PW}} = \phi^{L} + \phi^{R}.
\end{equation}
Introducing coordinates $(x,y)$ for the $T^2$, we can define complex coordinates $u=t+ix$ and $v=t+iy$ to take advantage of the fact that the real or imaginary part of a holomorphic function is harmonic in two dimensions. A simple interpolating solution that behaves as $\phi^{L,R}\rightarrow 0$ for $t\rightarrow \pm \infty$ is
\begin{equation}
   \phi^{L}=\mathrm{Re}\left[f^L_1(u)\frac{-\tanh(u)+1}{2}du+f^L_2(v)\frac{-\coth(v)+1}{2}dv\right],
\end{equation}
\begin{equation}
    \phi^{R}=\mathrm{Re}\left[f^R_1(u)\frac{\tanh(u)+1}{2}du+f^R_2(v)\frac{\coth(v)+1}{2}dv\right],
\end{equation}
which solves the 5D BPS equations of motion because the hyperbolic tangent function has simple poles with residue $+1$, while those of hyperbolic cotangent are $-1$. For example, near $u= i\pi/2$, $\phi^R\sim \textnormal{Re} \left[\frac{f_1^R(i\pi/2)}{2(u-i\pi/2)}du \right]$.
Note also that the periodicity in the $T^2$ directions means that there are an equal number of poles concentrated on the A- and B-cycles of the $T^2$. See figure \ref{fig:5Dvac} for a depiction of the fibered
Hitchin system and the resulting interpretation as an interface for 5D vacua.

\subsection{4D Interfaces}\label{sec:4Dinter}

In the previous section we presented an interpolating profile between two abelian Hitchin systems. The main feature of the solutions previously presented is that we essentially summed up two distinct Hitchin system solutions which only preserved a common 4D $\mathcal{N} = 1$ subalgebra along the interpolating profile coordinate of a non-compact three-manifold.

\begin{figure}[t!]
\begin{center}
\includegraphics[scale = 0.5, trim = {2cm 1.0cm 2cm 1.0cm}]{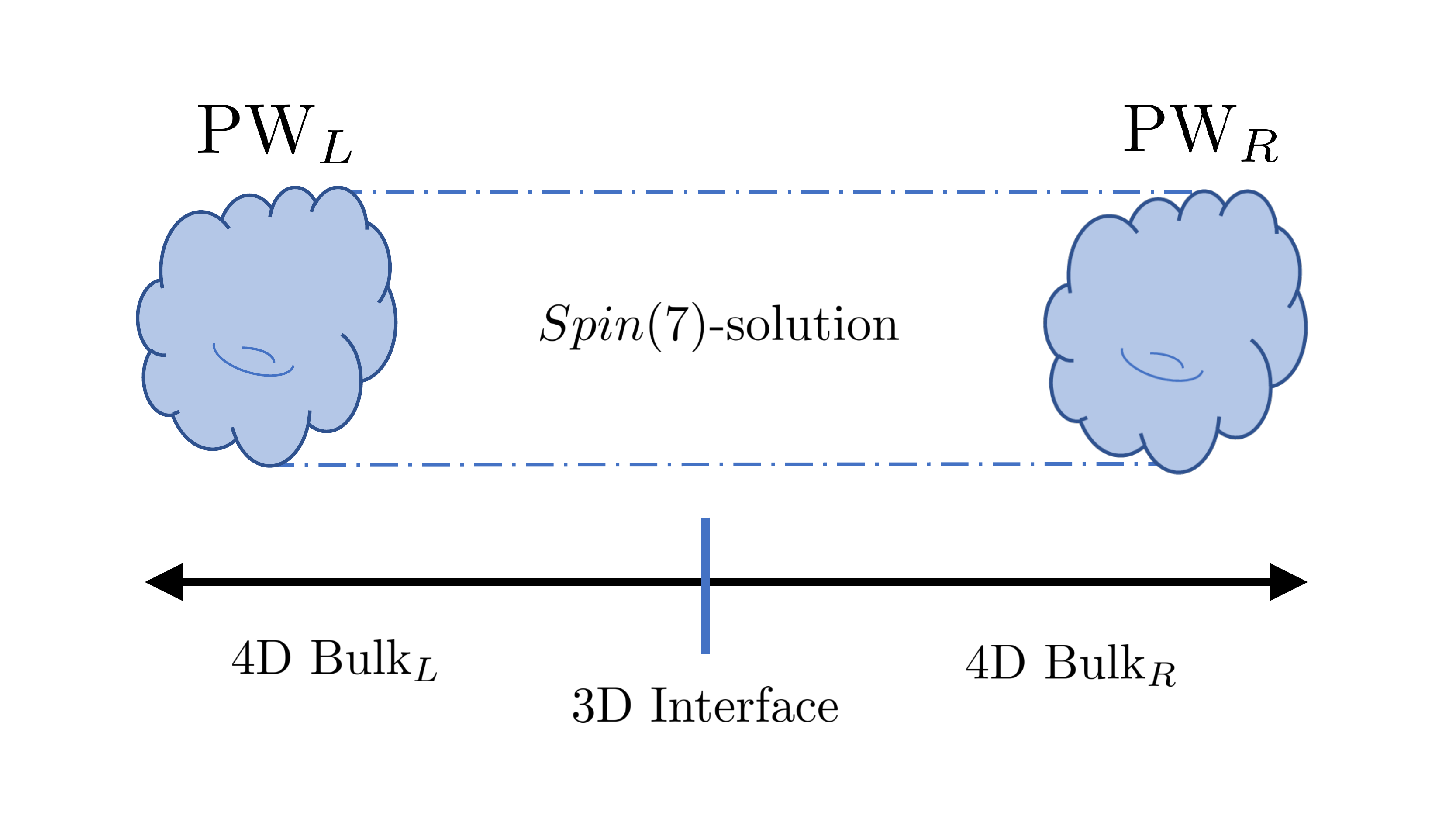}
\caption{Depiction of an interpolating profile between two 4D $\mathcal{N} = 1$
vacua with a 3D interface. The compactification geometry is captured by asymptotically
$G_2$ spaces given by a three-manifold of ADE singularities. The interpolating geometry is a non-compact
$Spin(7)$ space. The local gauge theory associated with these cases is a PW system on the left and right, and a local
$Spin(7)$ system in the interpolating region.}
\label{fig:PWPWvac}
\end{center}
\end{figure}

In this section we present examples of abelian PW systems which are connected by an interpolating profile in a local $Spin(7)$ system. To begin, we observe that the ``summing up Hitchin systems'' construction generalizes to three-manifolds $Q$ with marked one-cycles. The main point is that we can write $T^3$ as a product $S^1 \times S^1 \times S^1$, and so we can pick different pairs of $S^1$ factors to generate curves for a Hitchin system. Letting $(x,y,z)$ denote local real coordinates on these three $S^1$ factors, we can consider three $T^{2}$ factors, namely  $C^{(1)} = T^{2}_{(y,z)}$, $C^{(2)} = T^{2}_{(z,x)}$, $C^{(3)} = T^{2}_{(x,y)}$. For each of these Riemann surfaces, we can also include marked points, which then specify marked one-cycles on the three-manifold $Q$. For each such factor we can specify a corresponding Hitchin system which is trivial along the complementary $S^1$. Each such Hitchin system automatically solves the PW equations, and would, on its own, preserve 4D $\mathcal{N} = 2 $ supersymmetry. The key point we wish to emphasize is that we can switch on more than one Hitchin system, and thus obtain a solution on a compact $Q$ which only retains 4D $\mathcal{N} = 1$ supersymmetry. Adding another solution will not break any further supersymmetry. Summarizing, we get a class of abelian solutions on $Q$ (with marked one-cycles) by writing:
\begin{equation}
\phi_{\mathrm{PW}} = \phi_{\text{Hit}}^{(1)} + \phi_{\text{Hit}}^{(2)} + \phi_{\text{Hit}}^{(3)},
\end{equation}
namely a sum of independent Hitchin system solutions on the curves $C^{(i)}$. See figure \ref{fig:PWPWvac}
for a depiction of a PW--PW gluing.

The advantage of this presentation is that we can now use our previous results on 5D interfaces to generate 4D interfaces. Indeed, for each Hitchin system solution, we can construct an alternative non-compact three-manifold which we can label as $Q^{(i)} = C^{(i)}\times \mathbb{R}$.
For each case, we can also construct an interpolating solution, since the complementary circle is again a ``spectator'' in the analysis. Now, each of these PW solutions can also be repackaged as a self-dual two-form on the four-manifold $Q^{(i)} \times S^{1}_{(i)}$, as per our discussion in section \ref{sec:CAINANDABEL}. Consequently, our solutions can be summed, producing an interpolating $Spin(7)$ solution!

\subsection{Interpolation Singularities}

In the previous examples of interpolating solutions we saw the appearance of a singularity in the $t$ direction, which we interpret
as the presence of a monodromy defect operator in the internal gauge theory, or equivalently as a vev for localized matter. It is natural to ask
whether this is an artifact of these particular solutions or whether the appearance of such singularities is a more generic feature.
In what follows we again focus on abelian configurations.

Along these lines, consider the local $Spin(7)$ equations on the non-compact four-manifold $M = Q \times \mathbb{R}$ with $Q$ a three-manifold.
We show that if there are no singularities in the profile of the Higgs field, we generate a contradiction. To show this, we assume the contrary.
Recall that the self-dual two-form $\phi_{\mathrm{SD}}$ can be repackaged as a one-form $\phi_{\mathrm{PW}}$ of the PW system:
\begin{equation}
 (D_t *_3 \phi_{\mathrm{PW}} + d \phi_{\mathrm{PW}} )\wedge dt=0, \; \; \; \; d^{\dagger} \phi_{\mathrm{PW}}=0\,.
\end{equation}
Integrating the first equation and taking the 3D Hodge dual we have
\begin{equation}\label{abnogo}
\phi_{\mathrm{PW}}(t=\infty)-\phi_{\mathrm{PW}}(t=-\infty)=d^\dagger \left( \int_{-\infty}^{\infty}dt *_3 \phi_{\mathrm{PW}} \right),
\end{equation}
but by assumption, $\phi_{\text{PW}}(t=\pm \infty)$ is harmonic on $Q$ meaning that the right-hand side of (\ref{abnogo}) must vanish by the
Hodge decomposition. We note that this same argument also extends to flat gauge field connections which commute with the Higgs field.
Note that by modifying the argument one can see that the singularities in $\phi_{\mathrm{PW}}$
that are translationally invariant along the $\mathbb{R}$ direction do not affect the conclusion,
but singularities localized in the $t$-direction violate the above assumptions. For example, there are additional contributions to the integral of equation \eqref{abnogo} in this case.

\section{Interpolating BHV--PW Solutions} \label{sec:YESIAMNEWJACKCITY}

In the previous sections we have shown that there is a natural interpretation of the local $Spin(7)$ equations as specifying
an interpolating profile for Higgs bundle vacua obtained from
the PW and BHV systems. This is in accord with the geometric proposal of reference \cite{Braun:2018joh},
which argued that there is a generalized connected sums construction of $Spin(7)$ spaces via $Y_{G_2} \times S^1$ and $X_{CY_4}$
building blocks. The aim of the present section will be to develop the analogous construction in the local setting. One important
feature of these local models is that singularities are necessarily part of the local geometry. One can thus view the present considerations
as a complementary approach to analyzing possible interpolating vacua as generated by GCS-like constructions. Additionally, these local models
also provide some information on data such as the metric through the profile of the interpolating Higgs field. An additional feature of our
considerations is that there is also a close connection between the twisted connected sums construction of $G_2$ spaces and our local systems. Indeed, the ambient geometry of the local $Spin(7)$ system is a non-compact $G_2$ space, and that of the PW and BHV systems are non-compact Calabi-Yau threefolds.

Our strategy for realizing the local model analog of the GCS construction will be to actually start with deformations of the Hitchin system on a curve $C$, and to then fiber this to produce local $Spin(7)$ solutions which asymptotically approach either the PW system or the BHV system. In both cases, we consider a fibration over a cylinder $\mathbb{C}^{\ast} \simeq \mathbb{R} \times S^1$, where in the case of the PW system, we assume that the profile of fields on this additional circle factor is trivial, and in the case of the BHV system we assume that the profile fields is holomorphic in the cylinder coordinate (in a sense we make precise later). The key idea in our construction is that deep in the interpolating region, both the PW and the BHV system approach a Hitchin system on a curve $C$. As we explain, this is close in spirit to what happens in the
GCS construction of reference \cite{Braun:2018joh}.

An important clarifying remark is that there are really two ways in which a PW system will enter our analysis. On the one hand, we have a compact three-manifold $Q = C \times S^1$, and a solution to the Hitchin system, which trivially extends to a solution to the PW system. On the other hand, we have a ``non-trivial'' PW system given by working with the three-manifold $\widetilde{Q} = C \times \mathbb{R}_{t}$. The spacetime interpretation of course depends on whether we view $\mathbb{R}_t$ as part of a 4D spacetime, or an ``internal direction''
which we imagine is eventually compactified (perhaps as in the GCS construction). As we have already discussed in section \ref{sec:EFT}, taking the PW system to be defined on $Q$, we obtain an interpolating profile between 4D vacua. On the other hand, if we take the PW system to be defined on $\widetilde{Q}$, then there is a sense in which we can view our construction as building a particular class of 3D $\mathcal{N} = 1$ theories. Both physical systems are of intrinsic interest, and so in what follows we shall primarily focus on the geometry of the gauge theory solutions.
With this in mind, in this section we shall treat $t$ as an internal coordinate on the four-manifold used to define the local $Spin(7)$ system. It will remain as a local coordinate of the four-manifold used in the local BHV system, but will correspond to a direction normal to the three-manifold appearing in the PW system.

As an additional comment, in the context of local models where we keep the cylinder non-compact, we can of course extend this analysis to start building more general interpolating solutions, alternating between PW and BHV configurations. This provides another way, for example, to realize PW--PW interfaces, simply by constructing a PW--BHV--PW profile. Similarly, we can realize a BHV--PW--BHV profile using the same sort of analysis.

The rest of this section is organized as follows. We begin by reviewing some general features of the generalized connected sums construction,
and then turn to the local model version of this construction. With this in place, we then present an explicit abelian configuration of the local $Spin(7)$ system which asymptotically approaches the BHV and PW systems.

\subsection{Review of Generalized Connected Sums}\label{ssec:REVIEW}

In this section we review the construction of \cite{Braun:2018joh} that builds $Spin(7)$-manifolds by gluing two non-compact eight manifolds. The two building blocks employed in the construction are a non-compact Calabi--Yau fourfold and a product of a non-compact $G_2$-holonomy manifold with a circle. Both building blocks will have a non-compact cylindrical region and the idea behind the construction is that by a suitable gluing of the two blocks happening in this region one can obtain a compact $Spin(7)$-manifold. We first describe the two building blocks and their asymptotic cylindrical regions:
\begin{itemize}
\item[-]  \textbf{Calabi--Yau Block} This building block is a non-compact Calabi--Yau fourfold $X$ which possesses a region $X^{cyl}$ diffeomorphic to the product of a cylinder $\mathbb{C}^{\ast} \simeq \mathbb R \times S^1$ and a compact Calabi--Yau threefold $Z$. The complement of $X^{cyl}$ inside $X$ is compact. One common way to build such manifolds is to excise the anti-canonical class from a Fano K\"ahler manifold \cite{MR1040196,MR1123371,MR1048287}, however in \cite{MR3109862,MR3399097} it was shown that weak-Fano K\"ahler manifolds can also be used as building blocks.
\item[-] \textbf{$G_2$ Block} This building block is the product of a non-compact $G_2$ manifold $Y$ with a circle. The requirement is that outside a compact submanifold $Y$ is diffeomorphic to a Calabi--Yau threefold times an interval.
\end{itemize}

The basic observation is that the two building blocks have the same asymptotic structure, namely, they both asymptote to the product of a cylinder with a Calabi--Yau threefold. By cutting the cylinders at finite distance and gluing the two sides one builds a compact eight dimensional manifold and the proposal of \cite{Braun:2018joh} is that upon taking a sufficiently long tube one can find a suitable deformation of the metric that gives a $Spin(7)$ structure without torsion.

To give some more intuition behind the fact that the resulting compact manifold is a $Spin(7)$-manifold we can take a look at the various calibrating forms of the two building blocks and how they are glued together. Let us start with the Calabi--Yau building block: since a Calabi--Yau fourfold is an eight-manifold of $SU(4)$-holonomy it is a particular case of a $Spin(7)$-manifold. Indeed by using the holomorphic four-form $\Omega_4$ and the K\"ahler form $J$ one can build a four-form
\al{ \Psi_L = \text{Re}\, (\Omega_4) + \frac{1}{2} J \wedge J \,,
}
which is closed and self-dual. In the $G_2$ building block we have a similar situation, that is an eight manifold with a holonomy group that is a subgroup of $Spin(7)$ (in this case $G_2$). In this case we can use the associative three-form $\Phi$ of the $G_2$ manifold to build the four-form
\al{\Psi_R = d \sigma_{1} \wedge \Phi + * \,\Phi\,,
}
where $d\sigma_1 $ is the one-form on the circle and the Hodge star is taken on the $G_2$ manifold. This four-form is again closed and self-dual.

We are interested in what happens in the gluing region, again we start by spelling out the details for the Calabi--Yau building block. In the cylindrical region the holomorphic three-form and the K\"ahler form asymptotically approach respective forms on $Z \times
\mathbb{C}^{\ast}$, that is
\al{ \Omega_4 &\sim (d\sigma_1 + i d\sigma_2) \wedge \Omega_Z\,,\\
J &\sim  d\sigma_1\wedge d\sigma_2 + J_{Z}\,.
}
Here $\sigma_1$ and $\sigma_2$ are coordinates along the circle and interval directions of the cylinder respectively. Moreover by writing $\sim$ we mean equivalence up to terms that are exponentially suppressed in the $\sigma_2$ direction. On the $G_2$ side of the story we need to characterize the asymptotic behavior of the associative three-form in terms of the calibrating forms of the asymptotic Calabi--Yau threefold $Z$
\al{\Phi \sim \text{Re}\, (\Omega_Z)+ d\sigma_2 \wedge J_{Z}\,,
}
where we called $\sigma_2$ the coordinate along the interval and the meaning of $\sim$ is the same as above. Looking at the asymptotic behaviors one can see that the two self-dual four-forms match in the asymptotic region and are the only forms that are preserved after the gluing is performed.

To interpret this geometry as specifying an interface between 4D vacua as in section \ref{sec:EFT}, we would now need to decompactify the $S^1$ direction associated with the $\sigma_1$ coordinate. Additionally, we would have to change the interpretation of $\sigma_2$ as instead being purely in the ``internal'' directions of the compactification geometry. In the associated local model construction, we will again see the appearance of a cylindrical geometry, but this will be purely ``internal.'' To avoid confusion, we have therefore chosen to label the cylindrical coordinates in this subsection differently from the ones which will appear in our local model construction. It would of course be quite interesting to study how explicit decompactification limits connect the global and local pictures. For now, we shall remain agnostic on the precise form of such a procedure.

\subsection{Generalized Connected Sums and Local Models}\label{ssec:INTERPOLATION}

Having reviewed how GCS $Spin(7)$-manifolds are built, we now turn to the local model version of this construction. The expectation is that we have two classes of building blocks in the local model setting as well, each corresponding to 4D $\mathcal{N} = 1$ (and its reduction to 3D $\mathcal{N} = 2$) supersymmetric configurations on the corresponding building block. We first describe the two local model building blocks
\begin{itemize}
\item[-] \textbf{BHV Building Block} This building block corresponds to supersymmetric configurations on a four-cycle inside a Calabi--Yau fourfold. Such configurations are solutions to the BPS equations written in \cite{Beasley:2008dc} and we shall call this a BHV block. In the local $Spin(7)$ BPS equations these configurations are obtained whenever one component of the triplet of self-dual two forms $\phi_{\mathrm{SD}}$ is turned off. In the asymptotic cylindrical region of the Calabi--Yau fourfold the solution has to approach a Hitchin system on a Riemann surface $C$ times a trivial configuration on the cylinder. Note that we can view this as a patch of a compact K\"ahler surface with some locus deleted. An example is $C \times \mathbb{P}^1$ where we mark two points on the $\mathbb{P}^1$.
\item[-]\textbf{PW Building Block} This building block corresponds to supersymmetric configurations on a three-cycle $Q = C \times S^1$ inside a $G_2$ manifold (the additional circle direction plays no r\^ole). Such configurations are solutions to the BPS equations written in \cite{Pantev:2009de} and we shall therefore call this a PW block. Specifically, a PW block is obtained whenever all the fields appearing in the local $Spin(7)$ BPS equations are independent of one direction and the gauge field along that direction is turned off. In the asymptotic region of the $G_2$ manifold the solution has to approach a Hitchin system on a Riemann surface times a trivial configuration along the interval direction.
\end{itemize}
We see that the two building blocks have the same asymptotic behavior and therefore we expect that by cutting the cylinder at a finite distance and gluing the two sides one can build a solution interpolating between the two which would correspond to the local model version of the GCS construction.

One important aspect that we would like to clarify about the GCS construction refers to how quickly one might expect to approach a BHV or PW solution on either side of the glued manifold. We shall focus our attention to the tubular region where the gluing occurs. Here the geometry of the four-manifold simplifies as it is diffeomorphic to $\mathbb{C}^{\ast} \times C$, that is, a cylinder times a Riemann surface. To fix our conventions about the choice of coordinates we take $(t, \theta)$ on the cylinder so that the metric is
\al{ ds^2 = dt^2 + d\theta^2 + ds^2_{C}\,.
}
After gluing the two sides in the tubular region we expect to have a full-blown solution to the local $Spin(7)$ system, that is a solution that is not also a solution to any simpler system of equations. Nevertheless we also expect that the effect of the gluing will be localized in the tubular region and therefore will fade away as we approach the asymptotic regions of the cylinder where we should recover the original building blocks. We start by describing the approach to a BHV solution. Recall that a BHV solution is recovered from a general local $Spin(7)$ solution whenever one of the components of the triplet of self-dual two form $\phi_{\text{SD}}$ vanishes (following the notation used in section \ref{sec:HIGGS} we will call this component $\phi_\gamma$). By inspection of the power series around a point with BHV boundary conditions it is possible to see that $\phi_\gamma$ and its derivatives fall off exponentially, that is there is a coefficient $\lambda >0$\footnote{This can be obtained by using the conformal map between a cylinder and $\mathbb C^*$. If we require that $\phi_\gamma$ vanishes at $\infty$ in $\mathbb C^*$ and require it to be analytic around this point we obtain the exponential behavior when reverting back to the coordinates on the cylinder.}
\al{ \left| \phi_\gamma\right| \sim e^{ \lambda t}\,,
}
and where we took the BHV building block to be located at large negative values of $t$. A similar story occurs when approaching PW solution: recall that a PW solution is recovered from a local $Spin(7)$ one when the component of the gauge field along the circle direction of the cylinder vanishes and all remaining fields do not depend on the circle direction. Again by inspection of the power series around a point with PW boundary conditions one gets the following asymptotic behaviors
\al{ \left|A_\theta\right| \sim e^{-\lambda_1 t}\,,\\
\left| \partial_\theta \psi\right| \sim e^{-\lambda_2 t}\,,
}
for some positive constants $\lambda_{1,2}$. Here we placed the PW boundary at large positive values of $t$ and called $\psi$ all field components other than $A_\theta$. Moreover the asymptotic behavior of $A_\theta$ is defined up to gauge transformations that are bounded in the limit $t \rightarrow \infty$.

We now connect this discussion to a local version of the gluing used by Kovalev \cite{kovalevTCS,Corti:2012kd} in the TCS construction. The idea is that once we consider a four manifold $M$ the total space of the bundle of self-dual two forms is a local $G_2$ space.\footnote{Again, we allow for a metric which is not complete, and for possible singularities in the associative three-form. In the physical setting, possible divergences correspond to the appearance of additional degrees of freedom as the model is ``UV completed''.}
Our aim will be to show how this ambient space splits into non-compact building blocks of the sort appearing in the TCS construction. We will start by setting our notation: our four manifold coordinates will be $x^i$ with $i = 1,\dots,4$, the coordinates on the fibers of the bundle of self-dual two forms will be $y^a $ with $a=1,2,3$. We use a condensed notation for wedge products, writing for example $dx^{ab} = dx^{a} \wedge dx^{b} = dx^{a} dx^{b}$.
The total space of the bundle of self-dual forms is a $G_2$ space and its associative three-form is:
\al{ \Phi_{G_2} = dy^{123}- dy^1\left(dx^{14}+dx^{23}\right)-dy^2 \left(dx^{24}+dx^{31}\right)-dy^3 \left(dx^{34}+dx^{12}\right)\,.
}
Note that our manifold $M$ which is the zero section of the bundle is a co-associative cycle (that is $\Phi_{G_2}|_{M} = 0$) before turning on a profile for $\phi_{\text{SD}}$.

We now look at the two building blocks (BHV and PW) and how they embed as Calabi--Yau threefolds inside the $G_2$ space. Note that given a Calabi--Yau threefold $Z$ with holomorphic three-form $\Omega_Z$ and K\"ahler form $J_Z$ we can build an associative three-form on $Z \times \mathbb R_{\zeta}$ as
\al{ \Phi_{Z \times \mathbb R_{\zeta}} = \text{Re}\left(\Omega_Z\right) + J_Z \wedge d\zeta\,.
}

\paragraph{-- BHV Building Block} In this case we assume $M$ is a K\"ahler surface and we have a non-compact Calabi-Yau threefold given by the total space of the canonical bundle: $\mathcal{O}(K_{M}) \rightarrow M$. Denote by $y_1,y_2$ the two real coordinates in the normal bundle direction. In this case the holomorphic three-form and K\"ahler form are
\al{ \Omega_{\mathrm{BHV}} &= i \left(dx^1 - i dx^2\right)\left(dx^3 - i dx^4\right)\left(dy^1 + i dy^2\right)\,,\\
J_{\mathrm{BHV}} & = -dx^{12}- dx^{34} + dy^{12}\,.
}
One can check that taking $\zeta_{\text{BHV}} = y^3$, we recover the correct associative three-form.

\paragraph{-- PW Building block} In this case we need to take the cotangent bundle $T^* Q$ to a three manifold $Q \subset M$.
We choose the three manifold $Q$ to have local coordinates $x_i$ with $i=1,2,3$. In this case we can take
\al{\Omega_{\mathrm{PW}} &= i \left(dx^1+i dy^1 \right)\left(dx^2+ i dy^2 \right)\left(dx^3 + i dy^3\right)\,,\\
J_{\mathrm{PW}} &= dx^1 dy^1 + dx^2 dy^2 + dx^3 dy^3\,,
}
and with $\zeta_{\text{PW}} = x^4$ we recover the correct associative three-form.

\paragraph{-- Donaldson Gluing} We would now like to consider the Donaldson gluing that is employed in the TCS construction and see if it applies to our case as well. The main difference from the TCS construction is that we work in a decompactified limit, so rather than exchanging $S^1$ directions in the base and fiber, we expect to instead exchange $\mathbb{R}$ factors.

In the  region where the gluing occurs the two Calabi--Yau manifolds become diffeomorphic to the product of a K3 surface with an $\mathbb R^2$ factor. Using coordinates  $t$ and $\tilde t$ in the $\mathbb R^2$  and calling $J_{K3}$ and $\Omega_{K3}$ the K\"ahler form and holomorphic two form on the K3 surface, respectively, we find that the associative three-form on the $G_2$ manifold $K3 \times \mathbb R_t \times  \mathbb R_{\tilde t}\times \mathbb R_\psi$ is
\al{ \Phi = d \psi \wedge dt \wedge d \tilde t + d\psi \wedge J_{K3} + d\tilde t \wedge \text{Re}\left(\Omega_{K3}\right)+ d t \wedge \text{Im}\left(\Omega_{K3}\right)\,.
}
We would like to discuss this in the case of the building blocks we are considering. On the BHV side we have $\psi_{\text{BHV}} = y^3$ and we take $t_{\text{BHV}} = x^4$ as well as $\tilde t_{\textrm{BHV}} =x^3$.\footnote{Strictly speaking, the correct condition to impose is on the differentials of these coordinates.
In the following we will gloss over this distinction.} From this we get:
\al{ \text{Im}\left(\Omega_{\mathrm{K3,BHV}}\right) &= dx^1 dy^1 +dx^2 dy^2\,,\\
\text{Re}\left(\Omega_{\mathrm{K3,BHV}}\right) &= dx^2 dy^1- dx^1 dy^2\,,\\
J_{\mathrm{K3,BHV}} &= -dx^1 dx^2 +dy^1 dy^2\,.
}
On the PW side the identifications are $t_{\text{PW}} = -x^4$, $\psi_{\text{PW}} = x^3$ and $\tilde t_{\text{PW}} = y^3$, so we obtain:
\al{
\text{Im}\left(\Omega_{\mathrm{K3,PW}}\right)&=-  dx^1 dy^1 -dx^2 dy^2\,,\\
\text{Re}\left(\Omega_{\mathrm{K3,PW}}\right) &=-dx^1 dx^2 +dy^1 dy^2\,,\\
 J_{\mathrm{K3,PW}}&=  dx^2 dy^1-dx^1 dy^2\,.
}
The gluing is therefore achieved by the matching conditions:
\al{&\text{Im}\left(\Omega_{\mathrm{K3,PW}}\right) = -\text{Im}\left(\Omega_{\mathrm{K3,BHV}}\right) \,,\\
&\text{Re}\left(\Omega_{\mathrm{K3,PW}}\right) = J_{\mathrm{K3,BHV}}\,,\\
&J_{\mathrm{K3,PW}} = \text{Re}\left(\Omega_{\mathrm{K3,BHV}}\right)\,,\\
&t_{\mathrm{PW}} = - t_{\mathrm{BHV}}\,,\\
&\psi_{\mathrm{PW}} = \tilde{t}_{\mathrm{BHV}}\,,\\
&\tilde{t}_{\mathrm{PW}} = \psi_{\mathrm{BHV}}\,,
}
which is a variant of the usual Donaldson twist that Kovalev employed in the TCS construction, except here
some of the directions involved in the gluing are non-compact.

\subsection{Abelian BHV--PW Interpolation}\label{ssec:AbelianInterpolation}

In this section we turn to interpolating profiles between BHV and PW solutions. We again confine our analysis to abelian configurations. We will aim to give an interpolating profile between an abelian BHV solution on the left ($t<0$) of the tubular region and an abelian PW solution on the right ($t>0$). In what follows, we shall need to reference the asymptotic profile for the self-dual two-form
$\phi_{\mathrm{SD}}$ in the ``BHV region'' and the ``PW region.'' As we have already remarked, we can interchangeably work in terms of the Higgs field of these local systems, or can instead repackage this data in terms of a self-dual two-form. With this in mind, we let $\phi_{\mathrm{SD,BHV}}$ denote the profile of the self-dual two-form in the BHV region, and let $\phi_{\mathrm{SD,PW}}$ denote the profile of the self-dual two-form in the PW region.

Setting the unitary connection to zero and conjugating all the Higgs fields to the Cartan, our equations for the local $Spin(7)$ system
become simply
\begin{equation}
  d\phi_{\text{SD}} = 0\,.
\end{equation}
The main advantage is that now the system is linear which allows us to simply decompose $\phi_{\text{SD}} = \phi_{\text{SD,BHV}} + \phi_{\text{SD,PW}}$, where each of the two pieces are individually closed self-dual two-forms which satisfy the equations of their namesake throughout the interpolating region. In order to recover the local geometric gluing of the BHV and PW blocks, we demand that $\phi_{\text{SD,BHV}}$ vanishes as $t\rightarrow \infty$ and $\phi_{\text{SD,PW}}$ vanishes as $t\rightarrow -\infty$.
Ignoring Cartan factors for simplicity, we can write down a class of $\phi_{\text{SD,BHV}}$ solutions satisfying these constraints on $\mathbb{C}\times \mathbb{C}^{\ast} \simeq \mathbb{C} \times (\mathbb{R}\times S^1)$ with local coordinates $z=x+iy$ and $w=t+i\theta$  on the two factors as
\begin{equation}
  \phi_{\text{SD,BHV}}=g(z,w)\left[\tanh(w)-1\right]dz\wedge dw\, + h.c.,
\end{equation}
with $g(z,w)$ any holomorphic function in $w$ and $z$.\footnote{To avoid interfering with the boundary conditions we choose $g(z,w)$ to be finite as $t$ approaches infinity.} To further generalize this solution, we can consider again the tubular region where the topology of the four-manifold is the product of a Riemann surface, $C$, and a cylinder $\mathbb{C}^{\ast}$. Then we can write
\begin{equation}
    \phi_{\text{SD,BHV}}=\omega^{(1)}_C \wedge \rho^{(1)}(w) + h.c.,
\end{equation}
where $\omega^{(1)}$ is a global holomorphic one-form on $C$ and $\rho^{(1)}(w)$ is a meromorphic one-form on the cylinder with at least three simple poles. To see why, notice that after a change of coordinates from the cylinder to the complex projective line with coordinate $s=e^w\in\mathbb{P}^1$, our interpolation then requires that $\rho^{(1)}(s)$ is a section of $K_{\mathbb{P}^1}$ that is zero at $s=\infty$ and regular but non-zero at $s=0$. Because $\text{deg} \; K_{\mathbb{P}^1} =-2$, we must have three poles (counted with multiplicity) at some other points in $\mathbb{P}^1$ so in a local patch around $s=0$ we have
\begin{equation}
  \rho^{(1)}(w)=  \frac{-ds}{(s-s_a)(s-s_b)(s-s_c)}.
\end{equation}
 Taking $s_as_bs_c=1$, $ \rho^{(1)}$ is just $ds$ at $s=0$ and $0$ at $s=\infty$. Notice that in the $w$-coordinate system $\rho^{(1)}$ is
\begin{equation}
   \rho^{(1)}(w)= \frac{-e^wdw}{(e^w-e^{w_a})(e^w-e^{w_b})(e^w-e^{w_c})}
\end{equation}
which goes as $e^{-2w}dw$ for $t\rightarrow{}\infty$, which fits our gluing requirements. But, as $t\rightarrow{}-\infty$ it seems to asymptote as $e^wdw$ and not a non-zero constant. This is simply a feature of one-forms that one needs a suitable coordinate transformation to understand its asymptotic behavior, and in this case is in fact required for regularity at $s=\infty$. This is something we want for a healthy gluing procedure. It is important to pay attention to the fact that $\phi_{\text{SD,BHV}}$ ceases to be holomorphic at the locations of the simple poles. Rather than signaling a failure of $\phi_{\text{SD,BHV}}$ to solve the BPS equations, the presence of these poles is directly related to the presence of localized defects discussed in section \ref{sec:CAINANDABEL}.

On the other hand, because $\phi_{\text{SD,PW}}$ is constant along the $S^1$-factor, it can be presented as either a harmonic one-form or two-form on $C \times \mathbb{R}\times S^1$. For a local patch of $C$ diffeomorphic to $\mathbb{R}^2$, we can write it as a one-form $\phi_{\text{PW}} = df$ where $f$ is a solution to the (possibly singular) 3D Laplace equation on $\mathbb{R}^2\times \mathbb{R}_t$, while as a self-dual two-form we have:
\begin{equation}
  \phi_{\text{SD,PW}}=\partial_{z}fdz\wedge dw + \partial_{\Bar{z}}fd\Bar{z}\wedge d\Bar{w}+\frac{i}{2}\partial_t f (dz \wedge d\Bar{z}+dw \wedge d\Bar{w})\,.
\end{equation}
One ansatz for $f$ is to introduce coordinates $u\equiv t+ix$, $v \equiv t+iy$ and take advantage of the fact that real and imaginary parts of holomorphic functions are 2D harmonic. Then we can have
\begin{equation}
    \partial_u f= \mathrm{Re}\left[f_1(u)\frac{\tanh(u)+1}{2}\right], \; \; \; \partial_v f =\text{Re}\left[f_2(v)\frac{\coth(v)+1}{2}\right],
\end{equation}
where $f_1(u)$, $f_2(v)$ can be any holomorphic functions. Since the solution is periodic along $x$ and $y$, one can easily make this solution compact by appropriately quotienting $x$ and $y$ to include at least three singularities along both the $x$- and $y$-directions at $\{t=0\}$. The reason being is similar to $\phi_{\text{SD,BHV}}$ above where making, say, $x$ periodic means that $f_1(u)\frac{\tanh(u)+1}{2}du$ should be thought of as a section of the canonical bundle of $\mathbb{P}^1$, which after a conformal transformation to the $xt$-cylinder has a zero at $e^{t+ix}=0$ and is regular but non-zero at $e^{t+ix}=\infty$. Putting the pieces together, our local $Spin(7)$ solution, $\phi_{\text{SD,BHV}}+\phi_{\text{SD,PW}}$ is an explicit solution on $T^2\times \mathbb{P}^1$ with punctures at $\{s=0,1,\infty\}$, $\{t=0\}\cap \{ x = \frac{\pi}{2}+\pi n \}$, and $\{t=0\}\cap \{ y = n\pi\}$, where all of the punctures of the $Spin(7)$ system occur on Riemann surfaces which are topologically just copies of $T^2$.


\section{Conclusions} \label{sec:CONC6}

Higgs bundles are an important tool in
linking the geometry of extra dimensions in string theory
to low energy effective field theory. In this chapter we have developed a
detailed correspondence between a local $Spin(7)$ space given by a four-manifold of ADE singularities
and the corresponding partially twisted field theory localized on the four-manifold.
These systems engineer 3D $\mathcal{N} = 1$ theories (two real supercharges) and also generate
interfaces between 4D $\mathcal{N} = 1$ vacua. Focusing primarily on abelian configurations in which no gauge field
fluxes are switched on, we have shown that such 3D systems serve as interpolating profiles between
Higgs bundles used in 4D vacua. Additionally, we have developed the local model analog of the generalized connected sums
construction, showing that it is closely related to the twisted sums construction for $G_2$ spaces. In the remainder of this
section we discuss some potential areas for future investigation.

Much of our analysis has centered on the special class of Higgs bundles obtained from abelian Higgs field configurations.
There are more general ``fluxed'' configurations associated with T-brane vacua (see e.g. \cite{Aspinwall:1998he, Donagi:2003hh,Cecotti:2009zf,Cecotti:2010bp,Donagi:2011jy,Anderson:2013rka,
Collinucci:2014qfa,Cicoli:2015ylx,Heckman:2016ssk,Collinucci:2016hpz,Bena:2016oqr,
Marchesano:2016cqg,Anderson:2017rpr,Collinucci:2017bwv,Cicoli:2017shd,Marchesano:2017kke,
Heckman:2018pqx, Apruzzi:2018xkw, Cvetic:2018xaq, Collinucci:2018aho, Carta:2018qke, Marchesano:2019azf, Bena:2019rth, Barbosa:2019bgh, Hassler:2019eso}).
Recently T-brane configurations for $G_2$ backgrounds were investigated in \cite{Barbosa:2019bgh}
and it is natural to expect that these could be used as a starting point for generating T-brane configurations in local $Spin(7)$ systems.

One of the important applications of the local $Spin(7)$ system is that it engineers a broad class of 3D $\mathcal{N} = 1$ theories.
There are now many proposals for supersymmetric as well as non-supersymmetric dualities in such systems (see e.g. \cite{Aharony:2015mjs}).
In string theory, such dualities often arise from brane maneuvers in the extra-dimensional geometry. It would be interesting
to see whether the methods developed here could be adapted to study such proposed dualities.

Along these lines, one of the elements we have only lightly touched on is the structure of interactions amongst matter fields in
the resulting 3D $\mathcal{N} = 1$ theories. One reason is that from a 3D perspective, we expect strong quantum
corrections to such interaction terms. In the geometry, however, some of these interactions can be sequestered in the extra dimensions,
since they arise either from classical intersection geometry as in the case of Yukawa couplings for F-theory models, or from
non-perturbative instanton effects, as in the case of M-theory superpotentials. Determining robust estimates of the
resulting interaction terms would be most informative.

More generally, from the standpoint of effective field theory, we have explained how the local $Spin(7)$ equations
can be viewed as defining an interface between 4D vacua in which the Wilson coefficients of higher dimension
operators develop position dependent profiles. This raises an interesting possibility of tracking 4D dualities perturbed by different,
possibly ``dangerous irrelevant'' operators. A canonical example of this sort is the duality of reference \cite{Kutasov:1995ve}.
In this case again, we anticipate that geometric insights will likely
constrain possible behavior for the resulting IR physics.

We have also observed that some of the interpolating profiles obtained here are also part of another four-dimensional system, as captured by the
Kapustin-Witten equations. The natural setting for the appearance of this in type II string theory is
branes wrapped on a four-manifold $M$ in the cotangent space $T^{\ast} M$, a non-compact Calabi-Yau fourfold. It would be very interesting to
develop the corresponding spacetime interpretation, in line with our analysis of interpolating vacua presented here.

Lastly, all of our examples have focused on non-compact geometries. It would of course be interesting to see
how to build compact examples illustrating the same singularity structure. In contrast to the case of
$G_2$ spaces, $Spin(7)$ spaces are even-dimensional and there are many examples which directly descend from
quotients of Calabi-Yau fourfold geometries \cite{Joyce:1999nk}. Since there are relatively clear techniques for generating
the requisite geometric structures in elliptically fibered Calabi-Yau fourfolds, it would seem natural to track such
structures under a suitable quotient. Such compact examples would have applications to the study of 3D and 4D supersymmetric
vacua, as well as more ambitiously, to 4D ``$\mathcal{N} = 1/2$'' vacua \cite{Heckman:2018mxl, Heckman:2019dsj}.

This chapter has explored three-dimensional systems in the context of interpolating profiles between 4D vacua, but more specifically, for $\N=1$ theories living on a local $Spin(7)$ space. However, it would be interesting to also explore other classes of 3D systems generated by position dependent couplings on more general grounds. Such setups can provide access to quantum field theories with strong coupling features. Once again several geometric tools come in handy to shed light on the non-perturbative structure of these systems. In the following chapter we will focus on theories with time-reversal invariance, and we will see how the geometry of modular curves naturally emerges to characterize a large class of 3D interfaces.

\chapter{Geometric Approach to 3D Interfaces at Strong Coupling}\label{chapter7}
\section{Introduction} \label{sec:INTRO7}

As was illustrated in the previous chapter, insights from geometry and topology provide a non-trivial handle on many quantum systems,
even at strong coupling. In the context of high energy theory, this has typically
been applied in systems with supersymmetry. More generally,
however, one can hope that constraints on the topological structure of quantum fields
are enough to deduce many features of physics at long distance scales.

Indeed, there has recently been some progress in understanding 
some quantum field theories using constraints on the topological structure of
such systems. An example of this sort involves the effective field theory
associated with topological insulators \cite{Kane:2004bvs, Kane:2005zz, bernevig2006quantum, Moore:2006pjk, 2007Sci...318..766K, Fu:2007uya, PhysRevB.79.195322, Hasan:2010xy, 2011RvMP...83.1057Q, Hasan:2010hm, Ye:2017axd} in $3+1$ dimensions, which is one special type of symmetry-protected topological (SPT) phase of matter \cite{2010PhRvB..81f4439P, 2011PhRvB..83g5103F, 2011PhRvB..83g5102T, 2011PhRvB..83c5107C, 2011PhRvB..84p5139S, 2013PhRvB..87o5114C, 2008PhRvB..78s5424Q, 2008arXiv0810.2998E} with highly interesting surface behavior \cite{Vishwanath:2012tq, Chen:2013jha, Bonderson:2013pla, Wang:2013uky, 2014Sci...343..629W, Mross:2014gla, Wang:2014lca, Metlitski:2014xqa, Metlitski:2015bpa, Metlitski:2015eka, Fialkovsky:2019rum, Kurkov:2020jet}.
This phenomenon can be modeled in terms of the effective field theory
of a background $U(1)$ gauge theory with a position dependent $\theta$ angle \cite{Wilczek:1987mv, Qi:2008pi}.
Both $\theta = 0$ and $\theta = \pi$ preserve time-reversal symmetry, and demanding the system remain time-reversal invariant throughout
means that an interface between $\theta = 0$ and $\theta = \pi$ has trapped modes \cite{Jackiw:1975fn}. Indeed, this can
be explicitly verified by considering a 4D Dirac fermion with a  mass $m(x_\bot)$ which depends on a spatial direction of the
4D spacetime. A sign flip in $m$ leads to a trapped mode. There have been a
number of developments aimed at extending this analysis in various directions,
including new examples of dualities at weak coupling \cite{Seiberg:2016rsg}, as
well as possible strongly coupled phases for trapped edge modes \cite{Seiberg:2016gmd} and related dualities, see e.g.\ \cite{Hsin:2016blu, Karch:2016aux, Cordova:2017kue, Aitken:2017nfd, Gaiotto:2017tne, Benini:2018umh, Carl:2019bbf}.

In this chapter we study a similar class of questions but in which we allow the system
to approach a regime of ``strong coupling in the bulk.'' This also means that we allow the $U(1)$ to be dynamical, but we will
assume that degrees of freedom charged under it are still quite heavy. We can, of course, still require that far away from the interface we are
at very weak coupling, but even this assumption can in principle be relaxed (though that would of course be more difficult to realize experimentally but might be relevant for materials that have magnetic excitations such as pyrochlores \cite{Jaubert:2009fe}). Our aim will be to develop methods which apply in such situations as well.

The main theme running through our analysis will be to use methods from geometry to better understand the possible behavior of
localized modes. While much of our inspiration comes from the analysis of supersymmetric gauge theories in which these geometric structures descend from the extra-dimensional world of supersymmetric string compactifications, some aspects of our analysis
do not actually require the full machinery of these constructions. That being said, we will find it worthwhile to consider both low energy effective field theories in four dimensions, as well as compactification of six-dimensional superconformal field theories
as realized by string compactifications.

The first class of interfaces we study involve 4D $U(1)$ gauge theory with a complexified combination
of the gauge coupling $g$ and the theta angle:
\begin{equation}
\tau = \frac{4 \pi i}{g^2} + \frac{\theta}{2 \pi}.
\end{equation}
The main assumption we make is that our theory has a non-trivial set of duality transformations which act on this coupling as:
\begin{equation}
\tau \mapsto \frac{a \tau + b}{c \tau + d},
\end{equation}
for some $a,b,c,d$ integers such that $ad - bc = 1$. The most well-known case is that we just have a duality group $SL(2,\mathbb{Z}$) consisting
of all determinant one $2 \times 2$ matrices with integer entries, as associated with the famous electric-magnetic duality of Maxwell theory.
In systems with additional massive degrees of freedom, these duality groups can be smaller. Assuming this structure in the deep IR, we will be interested in the behavior of the 4D theory when $\tau(x_\bot)$ depends non-trivially on one of the spatial directions of the 4D spacetime.

In the case where the theory has an $SL(2,\mathbb{Z})$ duality group, there is a well-known correspondence between an equivalence class of $\tau$ and the geometry of a $T^2$ with complex structure $\tau$. One can think of this $T^2$ as the quotient $\mathbb{C} / \Lambda$ with $\Lambda = \omega^1 \mathbb{Z} \oplus \omega^2 \mathbb{Z}$ a two-dimensional lattice. In this case, the ratio $\omega^1 / \omega^2 = \tau$ dictates the ``shape'' of the $T^2$. In physical terms, $\Lambda$ is the lattice of electric and magnetic charges in the theory. Geometrically, we can replace $\tau(x_\bot)$ by a family of $T^2$'s which vary over a real line, building up a three-manifold with a boundary at $x_{\bot} \rightarrow \pm \infty$. Since there is a fixed choice of $T^2$ at both ends of the line, this $T^2$ comes with a distinguished marked point, and thus defines a one-dimensional family of elliptic curves.\footnote{An elliptic curve is a genus one curve with a marked point.}

We will be interested in a restricted class of 4D systems which enjoy time-reversal invariance in the bulk. This corresponds to a further condition of invariance of the physical theory under the mapping:
\begin{equation}
\tau \mapsto - \overline{\tau}.
\end{equation}
Geometrically, this corresponds to a further condition that the $j$-function of the elliptic curve is in fact a real number: $j \in \mathbb{R}$. This region splits into the familiar ``trivial phase'' with $\theta = 0$, the standard ``topological insulator phase'' with $\theta = \pi$ phase, and another ``strongly coupled phase'' in which $\vert \tau \vert = 1$. All other time-reversal invariant values of $\tau$ can be related to one of these three regions by an $SL(2,\mathbb{Z})$ transformation. As a point of nomenclature, we
note that this is somewhat of an abuse of terminology since in the topological insulator literature one views the $U(1)$ of the topological insulator as a global symmetry which is not broken (indeed it defines an SPT phase), and in which all excitations are gapped out. Part of the point of our analysis is to explore the effects of varying the gauge coupling as well as the theta angle. Hopefully the distinction will not be too distracting.

Viewed as a trajectory on the moduli space of elliptic curves, we thus see that an interface could a priori take two different routes between $\theta = 0$ and $\theta = \pi$. On the one hand, it could always remain at weak coupling. On the other hand, it could pass through a strongly coupled region. Asymptotically far away from the interface, both are a priori possible, but suggest very different possibilities for localized modes.  Singularities in this family of elliptic curves corresponds to the appearance of massless states. Since we are not assuming any supersymmetry, our knowledge of these states is somewhat limited, but we can, for example, deduce the electric and magnetic charge of states localized on the interface.

It can also happen that the duality group $\Gamma \subset SL(2,\mathbb{Z})$ is strictly smaller than that of the Maxwell theory. In this case, there are more possible phases, since the coset space $SL(2,\mathbb{Z}) / \Gamma$ is now non-trivial. Consequently, some values of $\tau$ related by an $SL(2,\mathbb{Z})$ duality transformation may now define different physical theories. The resulting moduli space of elliptic curves are specified by modular curves $X(\Gamma)$, and the geometry of these curves can be quite intricate. For our present purposes, we are interested in the subset of parameters which are time-reversal invariant. Thankfully, precisely this question has been studied in reference \cite{snowden2011real} which analyzes the real components of the modular curve, $X(\Gamma)_{\mathbb{R}}$. The key point for us is that $X(\Gamma)_{\mathbb{R}}$ consists of a collection of disjoint $S^1$'s. Each such $S^1$ itself breaks up into paths joined between ``cusps'' of the modular curve. These cusps are associated with the additional $SL(2,\mathbb{Z})$ images of the weak coupling point $\tau = i \infty$ which cannot be brought back to weak coupling via transformations in $\Gamma \subset SL(2,\mathbb{Z})$. Passing through such cusps is inevitable, and means that singularities in the family of elliptic curves are also dictated purely by topological considerations. For each such cusp, we can fix the associated electric and magnetic charge, thus indicating the corresponding charge of states localized on an interface.

We illustrate these general considerations with some concrete examples. As a first class, we consider some examples of 4D $\mathcal{N} = 2$ field theories in which the Seiberg-Witten curve has the topology of a $T^2$. As a second set of examples, we consider the compactification of a six-dimensional anti-chiral two-form on a family of elliptic curves. In this situation, we also present a general construction for realizing 4D $U(1)$ gauge theories with duality group given by the congruence subgroups $\Gamma_{0}(N), \Gamma_{1}(N),$ and $\Gamma(N)$.

As we have already mentioned, 3D interfaces appear in this geometric setting when the elliptic curve becomes singular. This raises the question
as to whether more singular transitions such as a change from a genus zero to a genus one curve could arise, and if so, what this would mean in
terms of the 4D effective field theory. Along these lines, we also consider a more general way to construct 3D interfaces from compactifying six-dimensional superconformal field theories on a three-manifold with boundaries. In this setting, we present explicit examples where the genus jumps as a function of $x_{\bot}$. By tracking the anomaly polynomial of the 4D theory before and after the jump, we deduce that the degrees of freedom on the two sides of a wall can be different. Such changes can be used to engineer more general examples of localized matter with a ``thickened interface.''

The rest of this chapter is organized as follows. We begin in section \ref{sec:DUALITY} with a geometric characterization of
3D interfaces of a $U(1)$ gauge theory
with duality group $SL(2,\mathbb{Z})$. In section \ref{sec:MOREDUAL} we generalize this to cases where the duality group is $\Gamma \subset SL(2,\mathbb{Z})$ a proper subgroup. Section \ref{sec:NTWO} presents some explicit constructions based on 4D $\mathcal{N} = 2$ theories, and
section \ref{sec:6DCOMPACTIFY} presents examples based on compactification of the theory of a six-dimensional anti-chiral two-form.
We generalize these constructions in section \ref{sec:6DGENERAL} by considering compactifications of six-dimensional superconformal field theories on three-manifolds with boundary. We conclude in section \ref{sec:CONC7}. Some additional details and examples are presented in the
Appendices.


\section{Time-Reversal Invariance and Duality} \label{sec:DUALITY}

In this section we review some elements of the ``standard'' case of a 4D $U(1)$ gauge theory which
has an interface between two time-reversal invariant phases with $\theta = 0$ and $\theta = \pi$.
We will be interested in developing a geometric characterization of this sort of system with an eye
towards generalizing to strongly coupled examples.

Throughout this chapter we will also confine our discussion to 4D theories on flat space $\mathbb{R}^{2,1} \times \mathbb{R}_{\bot}$.\footnote{Additionally, we will
ignore possible mixed gravitational/duality group anomalies which can appear on
some curved backgrounds \cite{Tachikawa:2017aux, Seiberg:2018ntt, Cordova:2019uob, Hsieh:2019iba, Cordova:2019jnf} as well as subtleties involving the spin-structure \cite{Rosenberg:2010ia, Metlitski:2013uqa, Metlitski:2015yqa}.
It would be interesting to extend the present considerations to these situations.}
We will, however, allow the coupling constants to depend on $x_{\bot}$, the local coordinate of $\mathbb{R}_{\bot}$.

The rest of this section is organized as follows. First, we introduce our conventions for time-reversal invariance, as well
$SL(2,\mathbb{Z})$ duality transformations. Using this, we identify different phases of parameter space which are time-reversal
invariant. Next, we study position dependent couplings which can generate an interface between these different phases.

\subsection{$U(1)$ Gauge Theory Revisited}

Consider an abelian gauge theory, with a possible coupling to some matter fields.
The corresponding Lagrangian density contains the terms:
\begin{equation}
\mathcal{L} = - \frac{1}{4 g^2} F_{\mu \nu} F^{\mu \nu} + \frac{\theta}{32 \pi^2} F_{\mu \nu} \widetilde{F}^{\mu \nu}  + \cdots \,,
\end{equation}
where the ``$\cdots$'' refers to contributions from all other matter fields. In terms of the electric and magnetic fields $\vec{E}$ and $\vec{B}$, we can also write this as:
\begin{equation}
\mathcal{L} = \frac{1}{2g^2} (\vec{E} \cdot \vec{E} - \vec{B} \cdot \vec{B}) - \frac{\theta}{8 \pi^2} \vec{E} \cdot \vec{B} + \cdots\,.
\end{equation}
It will be convenient to introduce the complexified coupling:
\begin{equation}
\tau = \frac{4 \pi i}{g^2} + \frac{\theta}{2 \pi}.
\end{equation}

Time reversal acts on the electric and magnetic fields as:
\begin{align}
\mathcal{T}: \quad \vec{E} \mapsto \vec{E} \,, \enspace \vec{B} \mapsto - \vec{B} \,.
\label{eq:tactemfields}
\end{align}
In terms of the original basis of fields, this has the effect of taking us to a new theory with the same gauge coupling, but with
$\theta_{\mathrm{new}} = -\theta_{\mathrm{old}}$. We can phrase this as a new choice of complexified gauge coupling:
\begin{equation}
\tau_{\mathrm{new}} = -\overline{\tau}_\mathrm{old}.
\end{equation}

We will be interested in values of the complexified coupling which can be identified with the old one via a duality transformation.
This takes us to a new basis of fields as well as dualized value of the coupling.
The most well-known situation is that our abelian gauge theory has an $SL(2,\mathbb{Z})$ duality group, which is the case for free Maxwell theory but also more interesting setups. We will shortly generalize this discussion to other duality groups.
Recall that the group $SL(2,\mathbb{Z})$ is defined as:
\begin{align}\label{generators}
SL(2,\mathbb{Z}) = \left\{ \begin{pmatrix} a & b \\ c & d \end{pmatrix}: a,b,c,d \in \mathbb{Z} \,, \enspace ad - bc = 1 \right\} \,.
\end{align}
Such duality transformations takes us to a new basis of electric and magnetic fields. Given a
state of electric charge $q_e$ and magnetic charge $q_m$, we introduce a two-component column vector which
transforms according to the rule:
\begin{equation}
\begin{pmatrix} q_e \\ q_m \end{pmatrix} \mapsto \begin{pmatrix} a & b \\ c & d \end{pmatrix} \begin{pmatrix} q_e \\ q_m \end{pmatrix} \,.
\end{equation}
For typographical purposes we shall also sometimes refer to this as a state having charge $(q_e,q_m)$, but we
stress that in our conventions this is to be viewed as a column vector, and not a row vector.
The Dirac pairing between two such charge vectors $\vec{q} \equiv q^{a}$ and $\vec{q^{\prime}} \equiv q^{\prime b}$ is:
\begin{equation}
\langle \vec{q} , \vec{q}^{\prime} \rangle = \epsilon_{ab} q^{a} q^{\prime b} = q_e q_m^{\prime} - q_{m} q_{e}^{\prime}.
\end{equation}
We can view a dyonic charge $(q_e , q_m)$ as coupling to a vector potential $A$ and its magnetic dual $A_D$ via the $SL(2,\mathbb{Z})$ invariant combination:
\begin{equation}
\epsilon_{ab} q^{a} \mathcal{A}^{b} = q_e A - q_m A_D,
\end{equation}
where we introduced the two-component vector $\mathcal{A}^{a}$ with
entries $\mathcal{A}^{1} = A_D$ and $\mathcal{A}^{2} = A$.
Under such a duality transformation, the complexified coupling also changes as:
\begin{equation}
\tau \mapsto \frac{a \tau + b}{c \tau + d}.
\end{equation}
Geometrically, the lattice of electric and magnetic charges can be written as:
\begin{equation}
\Lambda_{\tau} = \omega^1 \mathbb{Z} \oplus \omega^2 \mathbb{Z} = \tau \mathbb{Z} \oplus \mathbb{Z} \,
\end{equation}
where we can also view $\omega^{a}$ as a two-component column vector and the complex structure as $\tau = \omega^1 / \omega^2$.
Quotienting the complex plane $\mathbb{C}$ by this lattice results in an elliptic curve $E(\tau) = \mathbb{C} / \Lambda_{\tau}$. A pleasant feature of working with the elliptic curve is that $SL(2,\mathbb{Z})$ transformations leave the complex structure of the curve intact. This provides a geometric way to parameterize physically inequivalent $\tau$'s.

The group $SL(2,\mathbb{Z})$ is generated by the $T$ and $S$ transformations:
\begin{equation}
\begin{split}
T = \begin{pmatrix} 1 & 1 \\ 0 & 1 \end{pmatrix}:& \quad \tau \rightarrow \tau + 1 \,, \enspace \theta \rightarrow \theta + 2 \pi \,, \\
S = \begin{pmatrix} 0 & -1 \\ 1 & 0 \end{pmatrix}:& \quad \tau \rightarrow - \frac{1}{\tau} = - \frac{\overline{\tau}}{|\tau|^2} \,, \\
& \quad g^2 \rightarrow \Big( \Big( \frac{4 \pi}{g^2} \Big)^2 + \Big( \frac{\theta}{2 \pi} \Big)^2 \Big) g^2 \,, \enspace \theta \rightarrow - \Big( \Big( \frac{4 \pi}{g^2} \Big)^2 + \Big( \frac{\theta}{2 \pi} \Big)^2 \Big)^{-1} \theta \,
\end{split}
\end{equation}
observe that $\theta = -\pi$ can be mapped back to $\theta = +\pi$ under such a transformation. A priori, this gauge theory could be at strong or weak coupling, and have complicated interactions with other matter fields.

Assuming our theory enjoys an $SL(2,\mathbb{Z})$ duality group action, we need not work with the full set of
values of $\tau$, just the ones which are not identified by an $SL(2,\mathbb{Z})$ transformation. Implicit in this parameterization is that
when we label a theory, we allow ourselves to change to a dualized basis of fields.
Unitarity demands $\mathrm{Im} \tau  > 0$, so $\tau$ takes values in the upper half-plane $\mathbb{H}$. The quotient by $SL(2,\mathbb{Z})$
is known as the fundamental domain of $SL(2,\mathbb{Z})$, and we denote it as $Y = \mathbb{H} / SL(2,\mathbb{Z})$. Since we will also be interested in the very weakly coupled limit, we add on the ``point at infinity'' $\tau = i \infty$ as well as all of its $SL(2,\mathbb{Z})$ images (which are just rational numbers $a / c$ in the matrix presentation of line (\ref{generators})). Introducing the compactified upper half-plane:
\begin{equation}
\overline{\mathbb{H}} \equiv \mathbb{H} \cup \{ i \infty \} \cup \mathbb{Q},
\end{equation}
we can again consider the quotient space from an $SL(2,\mathbb{Z})$ action. This produces the compactified fundamental domain  which we denote as $X(\Gamma)$ with $\Gamma = SL(2,\mathbb{Z})$.

We will be interested in the space of couplings modulo such duality transformations. With this in mind, it is convenient to introduce an $SL(2,\mathbb{Z})$ invariant coordinate on the fundamental domain. This is simply the ``$j$-function'' of the parameter $\tau$.
The $j$-function is a modular form with $q$-expansion:
\begin{equation}
j = \frac{1}{q} + 744 + \cdots
\end{equation}
where $q = \exp(2 \pi i \tau)$. The $j$-function maps the fundamental domain $\overline{\mathbb{H}} / SL(2,\mathbb{Z})$ to the complex projective space $\mathbb{CP}^1$ with three distinguished points. This is the modular curve of the group $SL(2, \mathbb{Z})$. The three distinguished points are located at $\tau = i \infty, i, e^{2 \pi i/6}$, which are mapped to the points
\begin{align}
j (\tau) \underset{\tau \rightarrow i \infty}{\longrightarrow} \infty \,, \quad j (i) = 1728 \,, \quad j (e^{\pi i/3}) = 0 \,,
\end{align}
in the affine coordinate of $\mathbb{CP}^1$. For convenience we will use a rescaled version of the $j$-function defined by
\begin{align}
J (\tau) = \frac{j(\tau)}{1728} \,.
\end{align}

Having introduced a great deal of mathematical machinery, we now ask about which regions of our parameter space lead to a time-reversal invariant 4D theory. First of all, we can immediately identify a ``trivial phase'' with $\theta = 0$. This corresponds to the vertical line in the fundamental domain with $\tau \in i \mathbb{R}$. Additionally, we see that the region $\theta = \pi$ retains time-reversal invariance. We refer to this as the ``topological insulator'' phase. Using the $T$-generator of the $SL(2, \mathbb{Z})$ duality group one has
\begin{align}
(\theta = \pi) \overset{\mathcal{T}}{\longrightarrow} (\theta = - \pi) \overset{T}{\longrightarrow} (\theta = \pi) \,.
\end{align}
This means that utilizing the duality group, the value $\theta = \pi$ is also time-reversal invariant for arbitrary values of the gauge coupling $g$. In terms of the complex paremeter $\tau$, this region is given by $\tau \in \tfrac{1}{2} + i \mathbb{R}$. We will refer to a theory with $\theta = \pi$ as the ``topological insulator phase.'' This exhausts all possibilities for time-reversal invariance in regions of the moduli space that contain arbitrarily weak coupling, i.e.\ $g^2 \rightarrow 0$.

However, there is an additional phase that preserves time-reversal invariance at strong coupling.
In order to see that, assume $|\tau| = 1$ which means that we are at strong coupling. The $S$-generator of the $SL(2,\mathbb{Z})$ acts as
\begin{align}
\tau \rightarrow - \frac{\overline{\tau}}{|\tau|^2} = - \overline{\tau} \,,
\end{align}
i.e., exactly as $\mathcal{T}$! Therefore, there is a strongly coupled phase which preserves time-reversal
invariance for $|\tau| = 1$. We will refer to it as the ``strongly coupled phase''.

The time-reversal invariant subspace indicated above is mapped as follows to the modular curve $X(\Gamma ) \sim \mathbb{CP}^1$
\begin{equation}
\begin{split}
\text{Trivial}:& \quad \tau = i \alpha \enspace \text{with} \enspace \alpha \in [1, \infty) \,, \quad 1 < J \,, \\
\text{Topological Insulator}:& \quad \tau = \tfrac{1}{2} + i \alpha \enspace \text{with} \enspace \alpha \in [\tfrac{\sqrt{3}}{2}, \infty) \,, \quad J < 0 \,, \\
\text{Strongly Coupled}:& \quad \tau = e^{i \alpha} \enspace \text{with} \enspace \alpha \in [\pi/3 , \pi/2] \,, \quad 0 \leq J \leq 1 \,.
\end{split}
\label{eq:tauregions}
\end{equation}
So we find that the image under $J(\tau)$ of the time-reversal invariant values of $\tau$ is the real line in $\mathbb{C}$, which is compactified to a circle in $\mathbb{CP}^1$. Since $J$ is a one-to-one map from the fundamental domain we see that all real values of $J$ correspond to time-reversal invariant values of $\tau$. That is to say, the time-reversal invariant subspace of a $U(1)$ gauge theory with duality group $SL(2,\mathbb{Z})$ is given by the real subspace of the corresponding modular curve denoted $X(\Gamma)_{\mathbb{R}}$.
Note further, that all three distinguished points are contained in the time-reversal invariant subset of $X(\Gamma)_{\mathbb{R}}$, see figure \ref{fig:modcurve} for a depiction.

\begin{figure}[t!]
\centering
\includegraphics[width=0.7\textwidth]{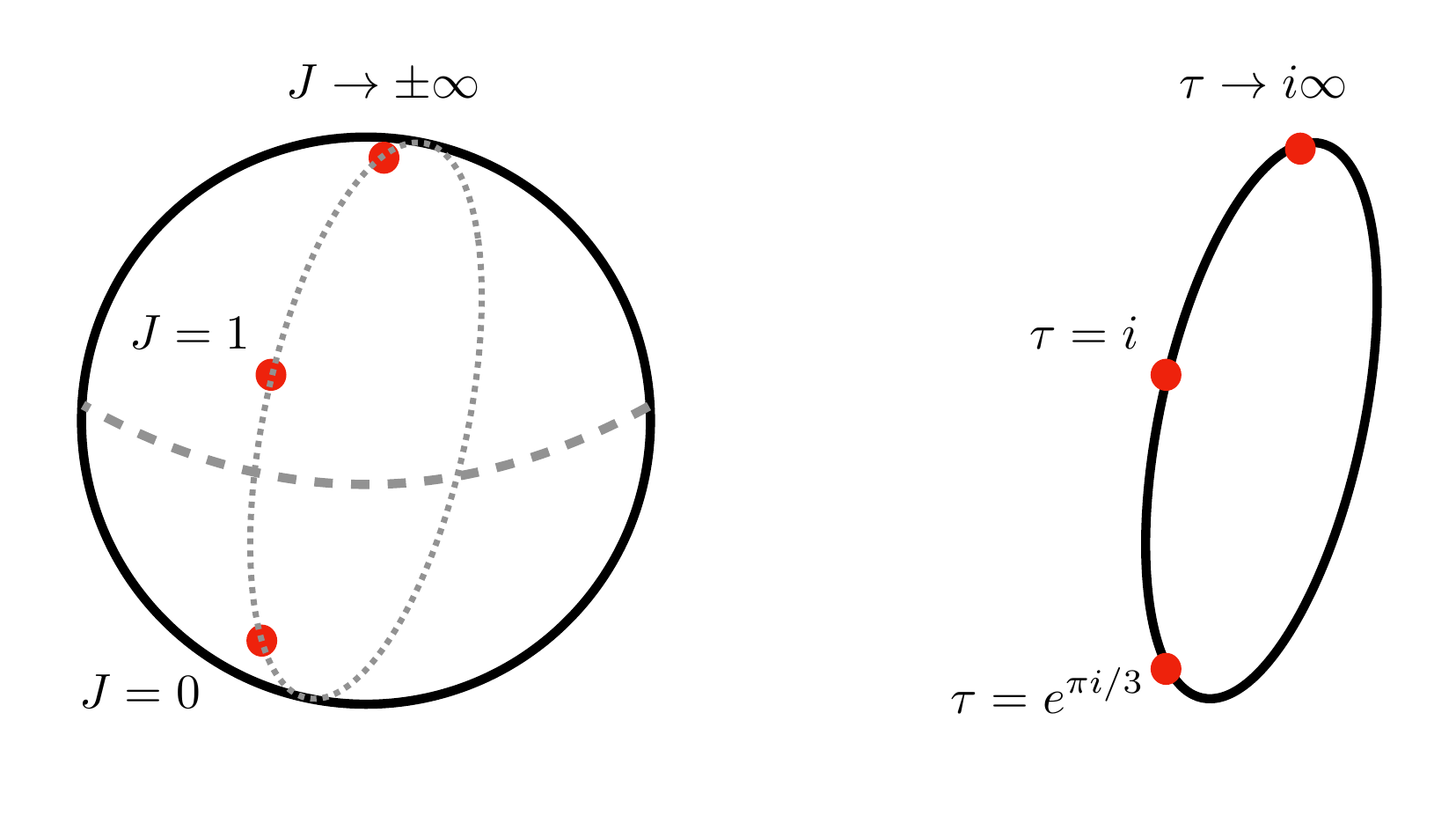}
\caption{Left: The image of the fundamental domain under $J$, with the marked points indicated as red dots. Right: The time-reversal subset $X(\Gamma)_{\mathbb{R}}$ of the modular curve $X (\Gamma)$ with $\Gamma = SL(2,\mathbb{Z})$.}
\label{fig:modcurve}
\end{figure}

We note that the above considerations have mainly focused on the structure of the effective Lagrangian. A priori, it could happen that time-reversal invariance is spontaneously broken, as happens in some gauge theory examples (see e.g. \cite{Gaiotto:2017tne}). Here we
assume that time-reversal invariance is preserved by the system and explore the geometric and physical consequences.

\subsection{Localized Matter and Real Elliptic Curves}

In the previous subsection we reviewed some general features of 4D $U(1)$ gauge theory for a fixed value of the coupling
$\tau$. We now consider more general configurations in which the parameter $\tau(x_\bot)$ is a non-trivial function of position
in the 4D spacetime $\mathbb{R}^{2,1} \times \mathbb{R}_{\bot}$. In particular, we would like to understand what happens when we have an interface between two different time-reversal invariant phases. We argue that
the geometry of real elliptic curves provides a helpful tool in analyzing these situations.

On general grounds, demanding time-reversal invariance between phases of the system with different
values of the parameters means that we should expect states to be localized at
the region of transition (see e.g. \cite{Seiberg:2016rsg}).
To this end, we now
allow $\tau(x_{\bot})$ to be a non-trivial function of the position coordinate in our
4D spacetime $\mathbb{R}^{2,1} \times \mathbb{R}_{\bot}$. For each point
$x_{\bot} \in \mathbb{R}_{\bot}$, we get a value of $\tau$, and can also think about a 4D Lorentz
invariant theory with that particular value of the coupling. Indeed, in an interval of $\mathbb{R}_{\bot}$ where $\tau(x_{\bot})$
is constant, we just have a 4D theory compactified on an interval, and so we can still speak of the action of the duality group on the 4D basis of fields. So, for sufficiently adiabatic variations of the coupling, we can still fruitfully apply our 4D Lorentz invariant analysis.
On the other hand, we will also be interested in regions where there is a sharp jump in the profile of the coupling (sharp compared to all other length scales in the system). In such situations, we can expect new phenomena to be localized in the region where a jump occurs.

To a large extent, demanding time-reversal invariance for the system leads to the prediction that there are localized states
trapped at such an interface. Our discussion follows reference \cite{Seiberg:2016rsg}. Observe that if nothing is localized at the interface, the shift in $\theta$ angle from $\pi$ to $0$ at $x_\bot = 0$ would break time-reversal invariance. This can be seen by considering the $\theta$ term on a geometry with boundary
\begin{align}
\underset{x_{\bot} < 0}{\int} \frac{\theta}{8 \pi^2} F \wedge F = \underset{x_{\bot} < 0}{\int} \frac{\pi}{8 \pi^2} d (A \wedge F) = \frac{1}{8 \pi} A \wedge F \big|_{x_\bot = 0} \,.
\end{align}
This induces a half-integer quantized Chern-Simons term at the boundary which breaks time-reversal invariance.
Therefore, there have to be degrees of freedom living at the interface to compensate the variation with respect to time-reversal. One weakly coupled solution to the problem is a localized charged 3D Dirac fermion which compensates this variation by its parity anomaly \cite{Redlich:1983dv, Niemi:1983rq, AlvarezGaume:1984nf, Witten:2015aba, Tachikawa:2016cha, Cordova:2017kue, Kurkov:2017cdz, Cordova:2019wpi}, a version of the anomaly inflow mechanism \cite{Callan:1984sa}. Other weakly coupled options were discussed in \cite{Seiberg:2016rsg}, and some strongly coupled options were considered in reference \cite{Seiberg:2016gmd}.

In terms of the geometry of the modular curve $X(\Gamma)$ for the duality group $\Gamma = SL(2,\mathbb{Z})$, these weakly coupled completions correspond to motion in $X(\Gamma)_{\mathbb{R}}$ through the point at $\tau = i \infty$. The geometry of $X(\Gamma)_{\mathbb{R}}$ suggests an alternative route which might connect these two phases. Indeed, we can instead contemplate passing down through the strong coupling phase to reach the same value of the parameters. Observe that along this route, we need not pass through a cusp at all. Instead, we can pass through the strong coupling region with values $\tau = i$ and $\tau = \exp(2 \pi i / 6)$ at the ``bottom'' of the fundamental domain. In this case, one might be tempted to say that there is nothing localized, since there is a smooth interpolating in the value of $\tau$ which completely bypasses the cusp.

We now argue that even along this other trajectory, there are localized states. The main reason is that if we demand time-reversal invariance
for the system, then in the limit where there is a sharp jump across the $\vert \tau \vert = 1$ region, there must also be \textit{something} localized in this region. The one loophole in this argument is that it could happen that time-reversal invariance is somehow broken in this region. This, however, would be in conflict with the fact that after compactifying our 4D spacetime on a very large circle $S^{1}$, we see that there is a non-trivial winding number associated with maps $S^1 \rightarrow X(\Gamma)_{\mathbb{R}}$. This instead indicates that the pair of jumps $(\theta = 0) \rightarrow (\theta = \pi)$ and  $(\theta = \pi) \rightarrow (\theta = 2 \pi)$ retains time-reversal invariance.

To better understand what is happening in this region, we now study the geometry of the elliptic curve associated with the parameter $\tau$. Because correlation functions of the physical theory will depend on duality covariant expressions built out of $\tau$,
possible singularities associated with localized states will in general be associated with singularities in the geometry of the elliptic curve.

We geometrize the above statements by defining an auxiliary elliptic curve $E$ with complex structure modulus identified with the complexified coupling constant $\tau$. Any elliptic curve can be represented as a hypersurface in the weighted projective space $\mathbb{CP}_{[2,3,1]}^{2}$ via the coordinates $x$, $y$, and $z$. This leads to the so-called Weierstrass form of the elliptic curve:
\begin{align}
y^2 = x^3 + f x z^4 + g z^6 \,,
\end{align}
with complex coefficients $f$ and $g$. Away from the point $[x,y,z] = [1,1,0]$ we can use the $\mathbb{C}^*$-rescaling in order to set $z$ to $1$ and one obtains the standard form
\begin{align}
y^2 = x^3 + f x + g \,.
\label{eq:Weierstrass}
\end{align}
In this form the elliptic curve is given by a branched double-cover, with three branch points at the roots of the right-hand side as well as a fourth root at infinity. For additional details on the geometry of elliptic curves, see Appendix \ref{app:ELLIPTIC}.

In terms of the parameter $\tau$, the coefficients $f$ and $g$ are associated with the Eisenstein series modular forms. We expect that
$f$ and $g$ depend non-trivially on the physical parameters of the system. This also holds for the discriminant:
\begin{equation}
\Delta = 4 f^3 + 27g^2.
\end{equation}
The $J$-function of the curve is given by the combination:
\begin{equation}
J = \frac{4 f^3}{4f^3 + 27g^2}.
\end{equation}

The appearance of this elliptic curve is quite familiar in a number of other contexts, including Seiberg-Witten theory, compactifications of 6D superconformal field theories on Riemann surfaces, as well as in the general approach to string vacua encapsulated by F-theory. In all of these cases, time-reversal invariance corresponds to a complex conjugation operation on the ``compactification coordinates'' $(x,y)$:
\begin{equation}
\mathcal{T}: (x,y) \mapsto (\overline{x}, \overline{y}).
\end{equation}
The special case of a time-reversal invariant Weierstrass model means we restrict to coefficients $f$ and $g$ which are real. Note that this is a strictly stronger condition than just demanding the $J$-function to be real. At least in supersymmetric settings, this is closely connected with the phase of BPS masses, and although we have less control in the non-supersymmetric setting, we expect a similar geometric condition to hold in this case as well. In section \ref{sec:NTWO} and Appendix \ref{app:FLAVA} we present some explicit $\mathcal{N} = 2$ examples illustrating these features, i.e., UV complete examples where $f$ and $g$ are purely real\footnote{Note that one could also consider models in which time-reversal invariance is restored in the deep IR, for which $f$ and $g$ can be complex numbers with correlated phases. In these cases, however, the mass parameters of the theory at high energies will break time reversal invariance in the UV.}.

Restricting $f$ and $g$ to be real means we are dealing with a real elliptic curve, namely the Weierstrass model makes sense over the real numbers. That being said, we will still view $x$ and $y$ as complex variables. This in turn leads to a constrained structure for the elliptic curve, especially as it moves through the different phases of $X(\Gamma)_{\mathbb{R}}$. To see this additional structure,
consider the factorization of the cubic in $x$:
\begin{equation}
x^3 + fx + g = \prod_{i = 1}^{3} (x - e_i),
\end{equation}
where the coefficients of the cubic are related to the roots as:
\begin{align}
0 & = e_1 + e_2 + e_3 \\
f & = e_1 e_2 + e_2 e_3 + e_3 e_1 \\
g & = -e_1 e_2 e_3\\
\Delta & = -\underset{i<j}{\prod}(e_i - e_j)^2.
\end{align}
The condition that $f$ and $g$ are real means that under complex conjugation, the roots $e_i$ must be permuted. There are two possibilities. Either all three roots are real, or one is real and the other two are complex conjugates. Without loss of generality, we can write these two cases as:
\begin{equation}
\begin{split}
\text{Case I}:& \quad e_1, e_2, e_3 \in \mathbb{R} \,, \\
\text{Case II}:& \quad e_1 \in \mathbb{R} \,, \enspace e_2 = \bar{e}_3 \,.
\end{split}
\end{equation}
Next, we want to relate the different configurations of the branch points to the time-reversal invariant values of $\tau$. The first comment is that from our explicit form of $f,g$ and $\Delta$, all of these quantities are real. In particular, the sign of the discriminant:
\begin{equation}
\Delta = - (e_1 - e_2)^2 (e_2 - e_3)^2 (e_3 - e_1)^2,
\end{equation}
tells us whether we are in Case I ($\Delta < 0$) or Case II ($\Delta > 0$). Since we also have:
\begin{equation}
J = \frac{4 f^3}{4 f^3 + 27 g^2} = \frac{4f^3}{\Delta},
\end{equation}
we conclude that when $f > 0$, we are in the regime of $0 \leq J \leq 1$, namely the strongly coupled phase.
If instead $f < 0$, then depending on the relative size of $4f^3$ and $27g^2$ we can get either sign of $\Delta$. Observe
that if $\Delta < 0$ and $f <0$ then, since $4f^3 + 27g^2 > 4f^3$ (recall $g^2$ is positive) we have $J > 1$, the ``trivial phase.'' If $\Delta >0$ and $f <0$ then we instead have $J <0$. Including the structure of the A- and B-cycles $\gamma_A$ and $\gamma_B$ of the elliptic curve, we see
there are three different phases of the time-reversal invariant contour specified by the following parameters:
\begin{itemize}
\item Trivial Phase: $J > 1 \Leftrightarrow \theta=0$ and $\tau = i \beta$ for $\beta>1$. There we have $\Delta < 0$, $f < 0$ and the roots $e_1 < e_3 < e_2$ are all real. The contours encircle $e_1$ to $e_3$ for $\gamma_B$ and $e_2$ to $e_3$ for $\gamma_A$.
\item Topological Insulator Phase: $J < 0 \Leftrightarrow \theta=\pi$. There we have $\Delta > 0$, $f < 0$ and the roots are such that $e_1 \in \mathbb{R}$, $e_2 = \bar{e}_3$, $\mathrm{Im}(e_2) > 0$. The contours encircle $e_1$ to $e_3$ for $\gamma_B$ and $e_2$ to $e_3$ for $\gamma_A$.
\item Strongly Coupled Phase: $0 \leq J \leq 1 \Leftrightarrow 0 \leq \theta \leq \pi$, $|\tau| = 1$. There we have $\Delta > 0$, $f \geq 0$ and the roots again satisfy $e_1 \in \mathbb{R}$, $e_2 = \bar{e}_3$, $\mathrm{Im}(e_2) > 0$. The contours encircle $e_1$ to $e_2$ for $\gamma_B$ and $e_1$ to $e_3$ for $\gamma_A$.
\end{itemize}
The different time-reversal invariant regions together with the signs of $f$, $g$, $\Delta$ are also indicated in figure \ref{fig:phasessigns}.
For some additional discussion, see Appendix \ref{app:ELLIPTIC}.

\begin{figure}[t!]
 \centering
 \includegraphics[width=.3\textwidth]{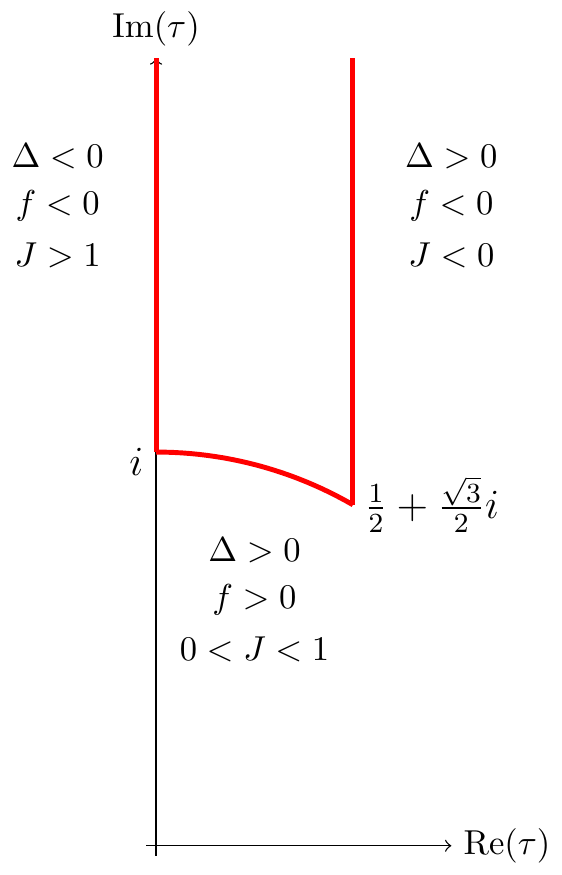}
 \caption{Values of the discriminant, $J(\tau)$ and $f$ for the elliptic curve as $\tau$ varies in its time-reversal invariant domain.}
 \label{fig:phasessigns}
\end{figure}

\begin{figure}[t!]
\centering
\includegraphics[width=\textwidth]{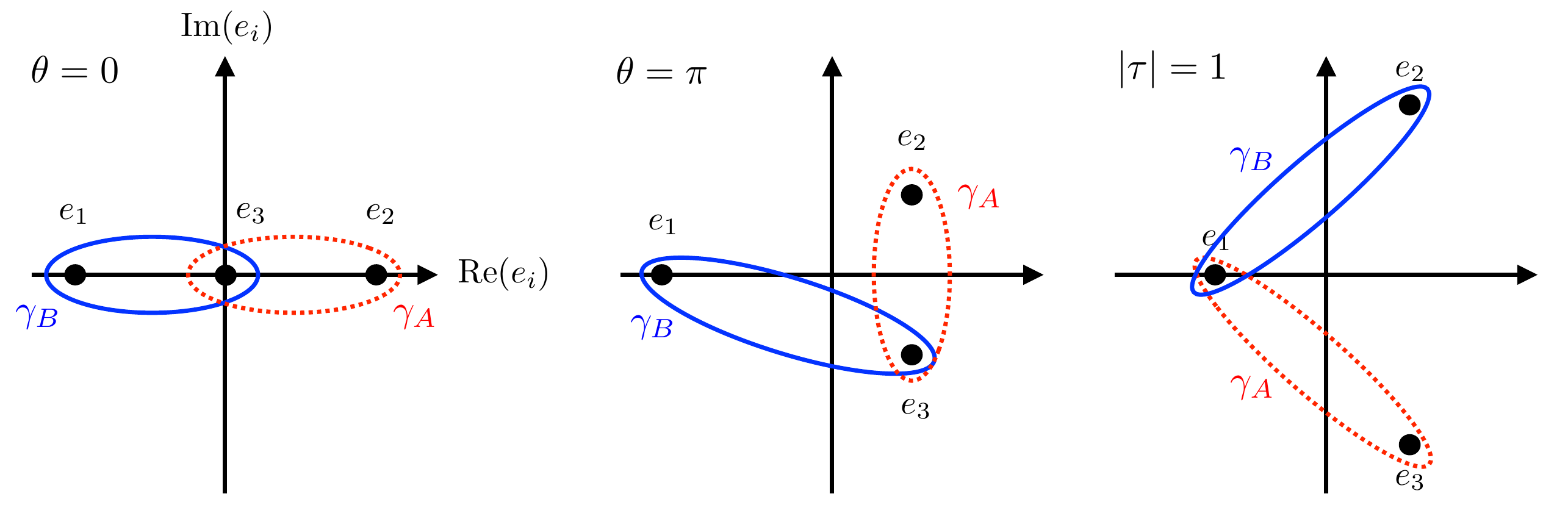}
\caption{Schematic view of the contours $\gamma_A$ (red dashed line) and $\gamma_B$ (blue solid line) for each phase.}
\label{fig:contours}
\end{figure}

Finally, we want to ensure that we can move between the three different time-reversal invariant regions by adjusting the three roots $e_i$. As already indicated above one can transition between the phase with $| \tau | = 1$ and the topological insulator phase $\theta = \pi$ by moving two roots in the imaginary direction. Collapsing two conjugate roots on the real axis and then separating them as real roots along the real axis leads to the transition between the topological insulator phase and the trivial phase with $\theta = 0$. The last transition seems to happen when two of the roots go off to infinity, see figures \ref{fig:Jreal} and \ref{fig:Jcomp}. However, this transition can also happen at finite values of the roots, when all three roots collapse at $0$.
This last transition is depicted in figure \ref{fig:transtrivstring}. We see that the discriminant vanishes in the transition between the trivial and topological insulator phase as well as in the transition between the trivial and the strongly coupled phase.

\begin{figure}[t!]
\centering
\includegraphics[width=\textwidth]{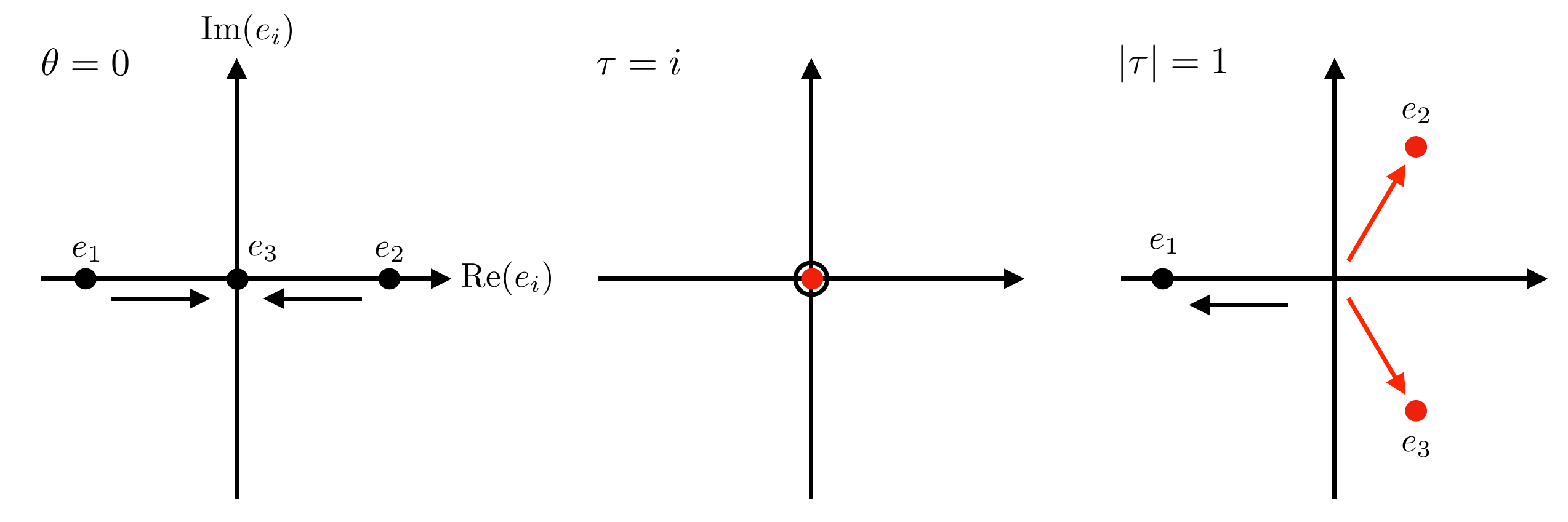}
\caption{Transition between the trivial and strongly coupled phase, with all three roots collapsing at $0$.}
\label{fig:transtrivstring}
\end{figure}

Our analysis in terms of the real elliptic curve reveals that passing through a singularity in the elliptic curve also occurs when we move along the ``alternative contour'' connecting $\theta = 0$ and $\theta = \pi$. We take this to mean that there is also localized dynamics trapped at such an interface, in accord with general expectations from time-reversal invariance.

\section{Other Duality Groups}\label{sec:MOREDUAL}

In the previous section we presented some geometric tools to study 3D interfaces in 4D $U(1)$ gauge theory in the special case
where the duality group is $SL(2,\mathbb{Z})$. In systems with interacting degrees of freedom, one often encounters $U(1)$ gauge theories
where the duality group $\Gamma$ is a subgroup of $SL(2,\mathbb{Z})$. A common situation where this arises is in the
case where the $U(1)$ gauge theory has a non-trivial spectrum of line operators, which one can think of as various heavy non-dynamical states.

Our aim in this section will be to study interfaces with these smaller duality groups. Compared with the case of
$SL(2,\mathbb{Z})$ duality, we find a significantly richer set of possible interfaces. This is simply because there are now many different physically distinct field configurations which can no longer be related by a duality transformation under the smaller group. As before, we shall assume that time-reversal invariance is preserved, and in particular is not spontaneously broken by the vacuum.

For now, we assume that we have a $U(1)$ gauge theory where the duality group $\Gamma \subset SL(2,\mathbb{Z})$ is a finite index subgroup
of $SL(2,\mathbb{Z})$. Starting from the original lattice of electric and magnetic charges $\Lambda$, we can consider the orbits swept out by the group action $\Gamma$. This results in a refinement in the lattice $\Lambda_{\mathrm{refined}} \subset \Lambda_{\mathrm{orig}}$. This new lattice of electric and magnetic charges specifies a different elliptic curve $E = \mathbb{C} / \Lambda_{\mathrm{refined}}$. This new elliptic curve is related to the other by an isogeny; The complex structure is actually unchanged under this refinement, but additional data is now being specified by this choice.

The space of physically distinct values of $\tau$ as captured by the fundamental domain $X(\Gamma) = \overline{\mathbb{H}} / \Gamma$ is consequently bigger. In fact, for general $\Gamma \subset SL(2,\mathbb{Z})$, the resulting modular curve can be considerably more complicated than that obtained in the special case of $SL(2,\mathbb{Z})$ where we have the geometry of a $\mathbb{CP}^1$ with a single cusp at $i \infty$. For example, the genus of this new modular curve can be greater than zero. Additionally, the set of cusps is always bigger. Recall that the space of cusps is specified by taking the quotient of $\{i \infty \} \cup \mathbb{Q}$ by the group action specified by $\Gamma$. In terms of the electric and magnetic charge of a state, these rational numbers are specified by the ratio $q_e / q_m$ so that the ``purely electric'' cusp is at
$i \infty$. Observe that the value of $\tau$ at a cusp indicates either zero gauge coupling (as in the case of $\tau = i \infty$) or ``infinite coupling'' (as in the case of $\tau \in \mathbb{Q}$).

This also translates to a bigger set of values for $\tau$ which can lead to time-reversal invariant phases.
As before, these are obtained by focusing on the points of $X(\Gamma)$ which are invariant under the anti-holomorphic involution:
\begin{equation}
c_0 : \tau \mapsto - \overline{\tau}.
\end{equation}
Here, to aid the reader interested in comparing with reference \cite{snowden2011real} we have used that paper's notation. This operation is, of course, nothing but time-reversal conjugation!

We refer to the real locus of the modular curve as $X(\Gamma)_{\mathbb{R}}$:
\begin{align}
X (\Gamma)_{\mathbb{R}} = \{ \tau \in X (\Gamma): \enspace c_0 (\tau) = \tau \} = \{ \tau \in \overline{\mathbb{H}}: \enspace c_0(\tau) = \gamma \tau \enspace \text{with} \enspace \gamma \in \Gamma \} \,.
\end{align}
Thankfully this space has actually been studied in great detail in reference \cite{snowden2011real} for the congruence subgroups $\Gamma(N), \Gamma_1(N), \Gamma_0(N) \subset SL(2,\mathbb{Z})$ (see Appendix \ref{app:CONG} for details on the congruence subgroups). The results there hold for general congruence subgroups of $SL(2,\mathbb{Z})$. The topology of $X(\Gamma)_{\mathbb{R}}$ is a disjoint union of circles. Each such circle contains at least one cusp, but some cusps of $X(\Gamma)$ do not belong to any real component.\footnote{For example let $\Gamma=\Gamma_0(N)$, then $N=16$ is the lowest $N$ for which there are non-real cusps, and in this case there is one real component that crosses four real cusps, and two additional $\mathcal{T}$-violating cusps on the genus zero curve $X_0(16)$.} We refer to the cusps which are members of $X(\Gamma)_{\mathbb{R}}$ as ``real cusps.'' We note that the point at infinity is always a real cusp, and it specifies a distinguished $S^1$. Observe also that there are $S^1$'s which only involve cusps at ``infinite coupling.'' These are intrinsically strongly coupled regions of parameter space which are in some sense ``cut off'' from weak coupling.

Let us now turn to the structure of interfaces between time-reversal invariant phases. To build an interface, we allow $\tau(x_\bot)$ to be a non-trivial function of position in the 4D spacetime $\mathbb{R}^{2,1} \times \mathbb{R}_{\bot}$. As we move along one of the $S^{1}$'s of $X(\Gamma)_{\mathbb{R}}$ we encounter a cusp of electric and magnetic charge $(q_e , q_m)$ associated with the rational number $q_e / q_m \in \mathbb{Q}$. From all that we have said, we expect that the condition of time-reversal invariance
enforces the appearance of localized degrees of freedom at such an interface.

To better understand this, suppose we have such an interface located at $x_\bot = 0$. We can first specialize to the case $\Gamma=SL(2,\mathbb{Z})$. In this case all cusps $q_e / q_m \in \mathbb{Q}\cup \{ i \infty \}$ are dual to each other so it is enough to consider the electric duality frame where $(q_e,q_m)=(1,0)$. Crossing such a cusp at $x_\bot = 0$ involves having $g^2\rightarrow 0$ as $|x_\bot|\rightarrow 0$ while $\theta=0$ for $x_\bot < 0$ and $\theta=\pi$ for $x_\bot > 0$. This induces a localized Chern-Simons theory at level-$\frac{1}{2}$ on the interface. As noted in reference \cite{Seiberg:2016rsg}, the states trapped at the interface could exhibit a wide range of phenomena, including a charged, massless 3D Dirac fermion, or a system with non-trivial topological order.\footnote{We use this language since one is often interested in situations where the Maxwell theory arises as the IR limit of a more complicated 4d gauge theory.} If we do act by an $SL(2,\mathbb{Z})$ transformation to transform the cusp to a more general choice $(q_e,q_m)$, then we have that the putative localized states are charged under a dualized gauge potential $A_{(q_e, q_m)}$. In terms of the vector potentials for the electric field strength $F_{\mu \nu}$ and its magnetic dual counterpart $\widetilde{F}_{\mu \nu}$, we can write this as:
\begin{equation}
A_{(q_e, q_m)} = q_e A- q_m A_D.
\end{equation}
In other words, we can speak of localized dyonic states of electric charge $q_e$ and magnetic charge $q_m$!
Suppose now that we have a theory with smaller duality group $\Gamma$ a proper subgroup of $SL(2,\mathbb{Z})$.
We assume that we can supplement this theory by adding additional degrees of freedom to it so that in this enlarged theory,
$SL(2,\mathbb{Z})$ is the resulting duality group. This in turn means that in this bigger theory we can ask about the effects of an $SL(2,\mathbb{Z})$ transformation. In the original theory with the smaller duality group,
then, we learn that there can be states trapped at an interface with different electric and magnetic charges. Summarizing, we see that
if we encounter a cusp $q_e / q_m \in \mathbb{Q}$ in the original theory, the
localized degrees of freedom can be viewed as carrying an electric and magnetic charge $(q_e , q_m)$.

In section \ref{sec:DUALITY} we noted that there can be additional singularities other than those located at the cusps,
as associated to degeneration in the elliptic curve near the points $\tau = i$ and $\tau = \exp(2 \pi i / 6)$. These points are distinguished in the sense that they are fixed under some of the elements of $SL(2,\mathbb{Z})$ and are referred to as ``elliptic points'' of order two ($\tau = i$) and three ($\tau = \exp(2 \pi i / 6)$). It turns out that for most finite index subgroups $\Gamma \subset SL(2,\mathbb{Z})$ there are no elliptic points, but in the few cases when they are present we can expect localized matter to also be present,
at least when the associated elliptic curve degenerates in approaching such a point of the real moduli space.
In such situations, we expect states with non-zero Dirac pairing to be simultaneously localized.

We can also deduce the relative spin-statistics of the excitations on neighboring interfaces,
which also lead to a quantization of the angular momentum induced by the electro-magnetic field between the interfaces.
Although not stated in these physical terms, reference \cite{snowden2011real}
computes the Dirac pairing between neighboring interfaces.
Focusing on the generic situation where our interfaces are generated by cusps, it turns out that the excitations localized on
neighboring interfaces always have a
non-vanishing Dirac pairing equal to $\pm 1$ or $\pm 2$:
\begin{equation}
\langle \vec{q} , \vec{q} \, '  \rangle \in \{\pm 1, \pm 2 \}.
\end{equation}
Recall that the Dirac pairing between dyons specifies an intrinsic angular momentum in the system. What this pairing indicates is that there is
an intrinsic \textit{spin} quantized in units of $\pm 1/2$ or $\pm 1$ associated with regions of the 4D bulk. This is an additional topological feature of our 4D bulk, as controlled by the dynamics of the interface! See figure \ref{fig:SpinChain} for a depiction.

\begin{figure}[t!]
\centering
\includegraphics[width=0.5\textwidth]{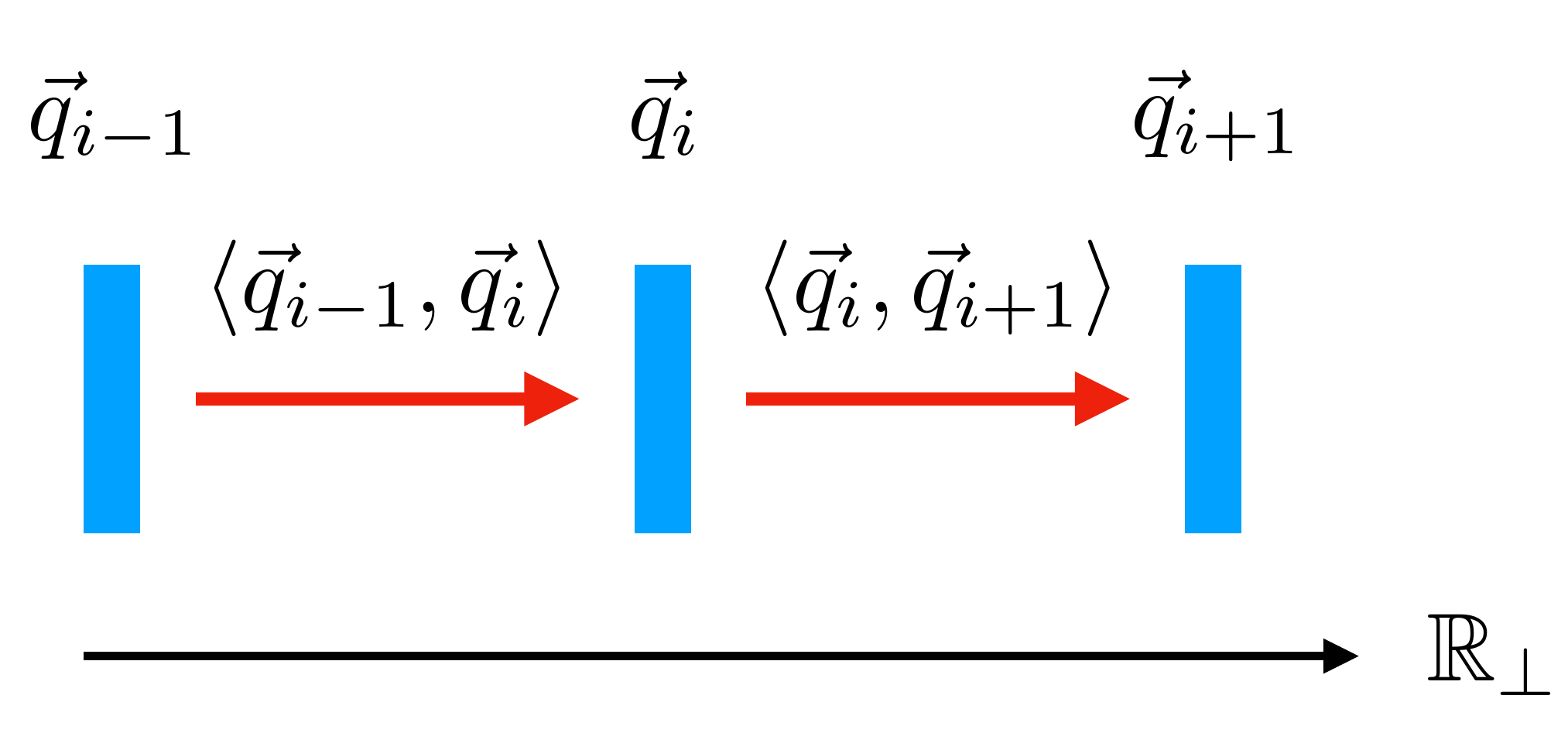}
\caption{Depiction of interfaces encountered in a trajectory through a component of $X(\Gamma)_{\mathbb{R}}$. Here,
each interface is associated with the $SL(2,\mathbb{Z})$ image of the cusp at weak coupling and therefore comes with excitations carrying an electric and magnetic charge which we denote as a two-component vector. States localized on neighboring walls have a non-zero Dirac
pairing, and this leads to a net angular momentum quantized in units of $\pm 1/2$ or $\pm 1$ between neighboring interfaces.}
\label{fig:SpinChain}
\end{figure}

In the remainder of this section we illustrate these general considerations by focusing on some specific choices of duality groups. In particular, we leverage the results of reference \cite{snowden2011real} to obtain explicit information on the structure of 3D interfaces
in these systems. We consider the three most well-known congruence subgroups $\Gamma(N), \Gamma_{1}(N),$ and $\Gamma_{0}(N)$ which also show up frequently in the study of modular curves:
\begin{equation}
\begin{split}
\Gamma_0 (N) &= \bigg\{ \gamma \in SL(2,\mathbb{Z}): \gamma = \left( \begin{array}{c c} * & * \\ 0 & * \end{array} \right) \, \text{mod} \, N\bigg\} \,, \\
\Gamma_1 (N) &= \bigg\{ \gamma \in SL(2,\mathbb{Z}): \gamma = \left( \begin{array}{c c} 1 & * \\ 0 & 1 \end{array} \right) \, \text{mod} \, N\bigg\} \,, \\
\Gamma (N) &= \bigg\{ \gamma \in SL(2,\mathbb{Z}): \gamma = \left( \begin{array}{c c} 1 & 0 \\ 0 & 1 \end{array} \right) \, \text{mod} \, N\bigg\} \,,
\end{split}
\end{equation}
where $*$ denotes an arbitrary integer entry. Clearly, these subgroups satisfy
\begin{align}
\Gamma (N) \subset \Gamma_1 (N) \subset \Gamma_0 (N) \subset SL(2,\mathbb{Z}) \,,
\end{align}
and each is a finite index subgroup of $SL(2,\mathbb{Z})$.

For each of these choices, there is a corresponding modular curve $X(\Gamma)$ which we denote by $X(N)$ for $\Gamma = \Gamma(N)$, $X_1(N)$ for $\Gamma = \Gamma_1(N)$ and $X_0(N)$ for $\Gamma = \Gamma_0(N)$.  Further it is clear that in each case $X(\Gamma)_{\mathbb{R}}$ is non-trivial since one can always choose the fundamental domain in a way that it contains (part of) the imaginary axis, which is invariant under $c_0$. This subset of $X (\Gamma)_{\mathbb{R}}$ is the region with $\theta = 0$. Moreover, it is clear that some remnant of the standard $T$ generator in $SL(2,\mathbb{Z})$ survives:
\begin{align}
T \in \Gamma_0 (N), \Gamma_1 (N) \,, \quad T^N \in \Gamma (N) \,,
\end{align}
which means that for $\Gamma_0(N)$ and $\Gamma_1(N)$ there are regions in $X (\Gamma)_{\mathbb{R}}$ which correspond to $\theta = \pi$. For $\Gamma (N)$ the non-trivial time-reversal invariant value of $\theta$ is given by $N \pi$. Note, that these two regions meet in the weakly coupled cusp situated at $\tau = i \infty$, which is also contained in the set $X(\Gamma)_\mathbb{R}$.

Since we have already explained the significance of the time-reversal invariant components of these modular curves, we now review the graphical rules developed in \cite{snowden2011real} which enumerate which ($\Gamma$-equivalence classes of) cusps are on a given real component. These graphs were arrived at by a group-theoretic analysis of each $\Gamma$ which assigns a solid dot to a cusp, on open dot to an elliptic point, with a single line connecting two cusps if their Dirac pairing is $\pm 1$, and a double line if their Dirac pairing is $\pm 2$ which reference \cite{snowden2011real} refers to as a ``weight''. Similar considerations hold for lines which connect an elliptic point to a cusp, but in this case the pairing is trajectory dependent. In these cases, the elliptic point connects to a cusp, once with weight one, and once with weight two. We take this to mean that there are states with mutually non-local charges localized at the elliptic point. This is a phenomenon which is
known to occur in 4D $\mathcal{N} = 2$ theories \cite{Argyres:1995jj}.

Each such line corresponds to a subset of points in $X(\Gamma)_{\mathbb{R}}$ satisfying:
\begin{align}
C_{\gamma}: \quad - \overline{\tau} = \gamma \tau \,.
\end{align}
for some conjugacy class $\gamma \in \Gamma$.
In general the subspace $X(\Gamma)_{\mathbb{R}}$ consists of the union of all these sets inside a single fundamental domain of the group $\Gamma$. For starters, we show the structure of this graph in figure \ref{fig:snowden2} in the case where $\Gamma = SL(2,\mathbb{Z})$.

\begin{figure}[t!]
  \centering
  \includegraphics[]{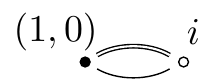}
  \caption{Real component for $X(1)$, namely the special case $\Gamma = SL(2,\mathbb{Z})$.
In the graph, cusps are denoted by solid dots and elliptic points are denoted by open dots.}%
  \label{fig:snowden2}%
\end{figure}

\begin{figure}[t!]
  \centering
  \includegraphics[]{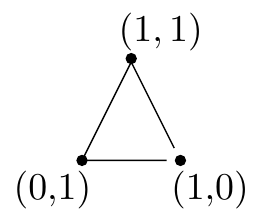}
  \includegraphics[]{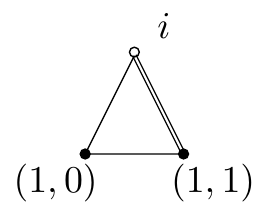}
 \caption{Real components of $X(2)$ (left) along with $X_0(2)$ and $X_1(2)$ (right).
On the right, the double line connecting $(1,1)$ to the elliptic point $\tau =i$
refers to the fact that if we follow a geodesic connecting
$(1,1)$ and $i$ we land on $(-1,1)$, and the Dirac pairing between $(1,1)$ and $(-1,1)$ is
$2$. Similarly, there is a single line connecting (1,0) and $i$ because the geodesic
through them lands on the cusp $(0,1)$, which has Dirac pairing $1$ with $(1,0)$.
In the graph, cusps are denoted by solid dots and elliptic points are denoted by open dots.
}
  \label{fig:snowden1}%
\end{figure}

As another example, consider the case of $X(2)$, for which there is one real component depicted in figure \ref{fig:snowden1} that (in a chosen duality frame) passes through the cusps $0, 1$, and $i \infty$. We represent this on the left side of figure \ref{fig:snowden1}. On the right side we depict the real component for $X_1(2)$ and $X_0(2)$ which passes through the cusps $1$ and $\infty$ and an order-2 elliptic point at $\tau=i$. Including $X(1)$, as shown in figure \ref{fig:snowden2}, we have actually exhausted all the cases where elliptic points can occur on a real component.

Having presented the general rules, we now summarize some of the important features of $X(\Gamma)_{\mathbb{R}}$ in the case of the aforementioned congruence subgroups. The statements we present amount to an adaptation of results given in \cite{snowden2011real}.

\subsubsection*{$\mathbf{X(N)}$}
Consider first the case where the duality group is $\Gamma = \Gamma(N) \subset SL(2,\mathbb{Z})$. In this case, the cusps are in the same $\Gamma(N)$-orbit if and only if $(a',b')\equiv \pm (a,b) \; \textnormal{mod} \; N$ and $\Gamma(N)$-equivalence classes of cusps are parametrized by pairs $\pm \frac{a}{b}$ of order-$N$ elements of $(\mathbb{Z}/N\mathbb{Z})^2$. To see the latter, note that we can reduce an element $(a,b)\in \mathbb{Z}^2$ modulo $N$ , which for $N>2$ is distinct from the modulo $N$ reduction of $(\pm a,b)$. Not every element of $(\mathbb{Z}/N\mathbb{Z})^2$ can be obtained from such a reduction though, since $\textnormal{gcd}(a,b)=1$. In particular $\textnormal{gcd}(a,b,N)=1$, which implies that at least either $a$ or $b$ must be an order-$N$ element of $\mathbb{Z}/N\mathbb{Z}$, making $(a,b)$ an order-$N$ element of $(\mathbb{Z}/N\mathbb{Z})^2$. The number of order-$N$ elements in $(\mathbb{Z}/N\mathbb{Z})^2$ is $N^2 \prod_{p|N}(1-1/p^2)$, where $p$ is a prime, but for $N>2$ we identify $(a,b) \, \textnormal{mod} \, N$ with $(-a,-b) \, \textnormal{mod} \, N$ since they represent the same cusp $\frac{a}{b}$, with similar considerations for the $-\frac{a}{b}$ cusp. Altogether we have
\begin{equation}
\textnormal{\# of cusps in fundamental domain} = \begin{cases}

\frac{1}{2} N^2 \underset{p|N}{\prod}(1-1/p^2) & N>2 \\
3 & N=2
\end{cases}
\end{equation}
for the total number of cusps.

Turning next to the real cusps and components, we characterize the cases by the power $r$ in $N=2^r N'$ with $\mathrm{gcd}(2,N') = 1$ and we quote the results mainly without proof. The case $r=0$ is perhaps the most complicated, we have $\phi(N)$ real cusps\footnote{This is the Euler totient function which expresses how many numbers $m<N$ are coprime to $N$, or equivalently, the order of the multiplicative group $(\mathbb{Z}/N\mathbb{Z})^\times$. It can be expressed as $N\prod_{p|N}(1-\frac{1}{p})$.} spread across $\psi(N)$ real components.\footnote{Borrowing notation from \cite{snowden2011real}, $\psi(N)$ is defined as the order of the group $(\mathbb{Z}/N\mathbb{Z})^\times/ \langle -1,2 \rangle $ which has no known closed form expression.} The neighborhood of a cusp $(a,b)$ (taken mod $N$)  is shown on the left-hand side of figure \ref{fig:snowden3}.

\begin{figure}[t!]
  \centering
  \includegraphics[]{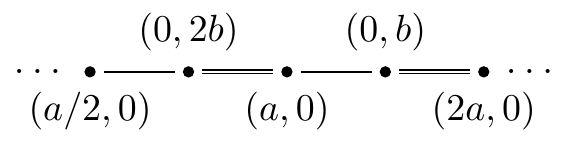}
  \includegraphics[]{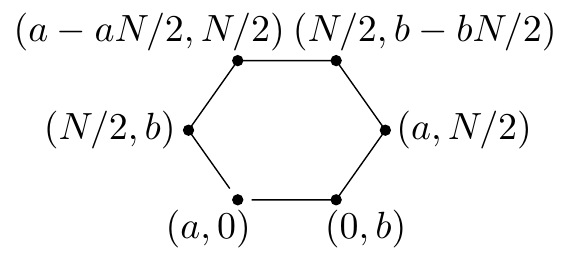}
  \includegraphics[]{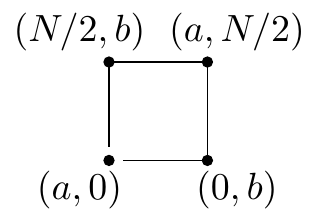}
 \caption{Real cusps/components (mod-$N$) for $r=0$ (left), $r=1$ (center), and $r\geq 2$ (right). Here $N=2^r N'$ for $N'$ odd. In all cases, $ab \equiv 1 \, \textnormal{mod} \, N$ and we take $\textnormal{gcd}(a,N) = 1$}%
  \label{fig:snowden3}%
\end{figure}

The case $r=1$ ($N>2$) has $3\phi(N)$ real cusps spread evenly across $\frac{1}{2}\phi(N)$ real components, i.e. six cusps per component whose charges (mod $N$) are shown in figure \ref{fig:snowden3}. While the $r \geq 2$ cases have $2\phi(N)$ real cusps spread evenly across $\frac{1}{2}\phi(N)$ real components, i.e. four cusps per component.

\subsubsection*{$\mathbf{X_1(N)}$}
Consider next the case of the modular curve $X_1(N)$ as specified by the duality group $\Gamma_1(N) \subset SL(2,\mathbb{Z})$. In this case, the cusps are in the same $\Gamma_1(N)$ orbit if and only if $(a,b)\equiv \pm (a+jb,b) \; \textnormal{mod} \; N$ for some integer $j$. Equivalence classes can be parametrized by first fixing $a \; \textnormal{mod} \; \textnormal{gcd}(b,N)$, then enumerating pairs $\pm \frac{a}{b}$ of order-$N$ elements of $(\mathbb{Z}/N\mathbb{Z})^2$ under this restriction. The number of cusps (see e.g. \cite{diamond2006first}), is
\begin{equation}
\textnormal{\# of cusps in fundamental domain} = \begin{cases}

2 & N=2 \\
3 & N=4  \\
\frac{1}{2} \underset{d|N}{\sum} \phi(d)\phi(N/d)& N=3 \;  \textnormal{or} \; N>4
\end{cases}
\end{equation}
where $d$ is any divisor. Just like the $X(N)$ curves, the properties of the real cusps and components depend on the exponent $r$ in $N=2^rN'$ (with $\mathrm{gcd}(2,N') = 1$, and in fact the $r=0$ case is exactly the same for $X_1(N)$ and $X(N)$. For the $r=1$ case, there are $2\phi (N)$ real cusps and $\psi(N/2)$ real components (making the number of cusps per component more irregular than for the $X(N)$ curves), while the $r\geq2$ case has $\frac{3}{2}\phi(N)$ real cusps and $\frac{1}{4}\phi(N)$ real components arranged as in figures \ref{fig:snowden4}, \ref{fig:snowden44} and \ref{fig:snowden444}. There is an exception to this classification for $N=4$. The real structure of this case
is displayed in figure \ref{fig:snowden5}.

\begin{figure}[t!]
  \centering
  \includegraphics[]{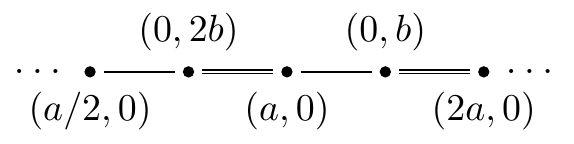}
 \caption{Real cusps/components (mod-$N$) for $X_1(N)$ ($N\neq 2,4$) for $r=0$. Here, $N=2^rN'$ with $\mathrm{gcd}(2,N') = 1$}%
  \label{fig:snowden4}%
\end{figure}

\begin{figure}[t!]
  \centering
  \includegraphics[width=50mm]{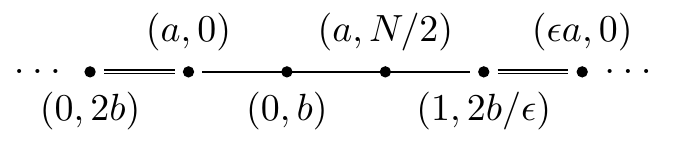}
 \caption{Real cusps/components (mod-$N$) for $X_1(N)$ ($N\neq 2,4$) for $r=1$. Here, $N=2^rN'$ with $\mathrm{gcd}(2,N') = 1$.
In the figure, $\epsilon \equiv 2+N/2$.}%
  \label{fig:snowden44}%
\end{figure}

\begin{figure}[t!]
  \centering
  \includegraphics[width=50mm]{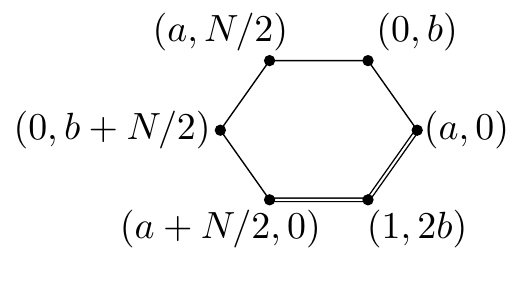}
 \caption{Real cusps/components (mod-$N$) for $X_1(N)$ ($N\neq 2,4$) for $r\geq 2$. Here, $N=2^rN'$ with $\mathrm{gcd}(2,N') = 1$.}%
  \label{fig:snowden444}%
\end{figure}

\begin{figure}[t!]
  \centering
  \includegraphics[]{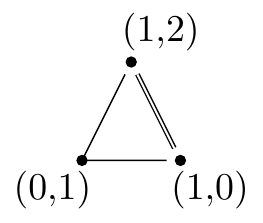}
 \caption{Real cusps/components for $X_1(4)$}%
  \label{fig:snowden5}%
\end{figure}

\subsubsection*{$\mathbf{X_0(N)}$}
Finally, consider the case of the modular curve $X_0(N)$ as associated with the duality group $\Gamma_0(N) \subset SL(2,\mathbb{Z})$.
The cusps in this case are in the same $\Gamma_0(N)$ orbit if and only if $(ya,b)\equiv \pm (a+jb,yb) \; \textnormal{mod} \; N$ for some integers $j$ and $y$ such that $\mathrm{gcd}(y,N)=1$. Conveniently, it turns out that equivalence class of cusps can be described simply as elements of $\mathbb{P}^1(\mathbb{Z}/N\mathbb{Z})$ and we can represent the mod-$N$ charges of cusps as $[a:b]$. The total number of cusps is then
\begin{equation}
\textnormal{\# of cusps in fundamental domain} =  \sum_{d|N} \phi(\textnormal{gcd}(d,N/d))
\end{equation}
for any $N$. For $r=0$ ($N$ odd), let $k$ be the number of distinct prime factors of $N$, then there are $2^{k-1}$ real components all of the form shown in figure \ref{fig:snowden6}.
\begin{figure}[t!]
  \centering
  \includegraphics[]{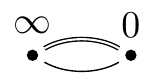}
  \caption{Real cusps/components for $X_0(N)$ when $N$ is odd.}%
  \label{fig:snowden6}%
\end{figure}

\begin{figure}[t!]
\centering
  \includegraphics[]{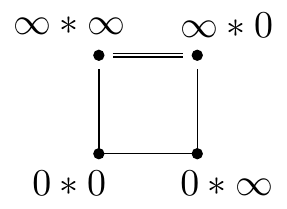}
  \includegraphics[]{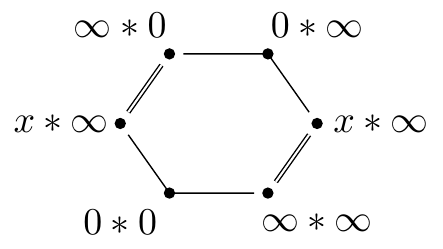}
 \caption{The real cusps/components for $X_0(N)$ when $r=1$ (left) and $r=2$ (right) where the $*$ notation refers to the decomposition $\mathbb{P}^1(\mathbb{Z}/N\mathbb{Z})=\mathbb{P}^1(\mathbb{Z}/2^r\mathbb{Z})\times \mathbb{P}^1(\mathbb{Z}/N'\mathbb{Z})$ since we do not want to conflate this with the parentheses notation $(\cdot, \cdot)$ used to label the electric and magnetic charges.
Here we define $x \equiv [1:2]$, viewed as an element of $\mathbb{P}^1(\mathbb{Z} / 2^r \mathbb{Z})$.}%
  \label{fig:snowden7}%
\end{figure}

The behavior for even $N$ is again governed by the number of distinct odd prime factors $k$. For $r = 1$, $r = 2$, and $r \geq 3$, there are respectively $2^{k+1}$, $3\cdot 2^k$ , and $2^{k+2}$ real cusps and $2^{k-1}$, $2^{k-1}$, and $2^k$ real components. See figures \ref{fig:snowden7} and \ref{fig:snowden8} for the corresponding real components of the modular curves.

\begin{figure}[t!]
\centering
  \includegraphics[]{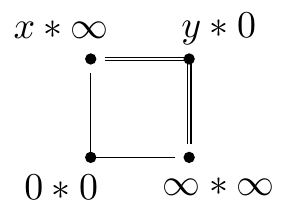}
  \includegraphics[]{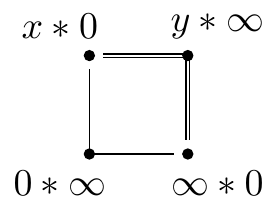}
 \caption{The real cusps/components for $X_0(N)$ when $r\geq 3$, where we have two flavors of components (an equal number of each).
The $*$ notation refers to the decomposition $\mathbb{P}^1(\mathbb{Z}/N\mathbb{Z})=\mathbb{P}^1(\mathbb{Z}/2^r\mathbb{Z})\times \mathbb{P}^1(\mathbb{Z}/N'\mathbb{Z})$ since we did not want to confuse with the parentheses $(,)$ for the electric and magnetic charges.
Here we defined $x \equiv [1:2]$ and $y \equiv [1:2^{r-1}]$ viewed as elements of $\mathbb{P}^1(\mathbb{Z} / 2^r \mathbb{Z})$.}%
  \label{fig:snowden8}%
\end{figure}

\section{$\mathcal{N} = 2$ Examples} \label{sec:NTWO}

To illustrate some of these general considerations, we now present some examples based on $\mathcal{N} = 2$ supersymmetry.
Recall that a helpful way to study such theories involves the geometry of the Seiberg-Witten curve \cite{Seiberg:1994rs, Seiberg:1994aj}.

We begin by considering a class of 4D $\mathcal{N} = 2$ superconformal field theories obtained from a D3-brane probing a stack of seven-branes with and ADE gauge group. This determines a flavor symmetry on the 4D worldvolume theory of the D3-brane \cite{Banks:1996nj,
Minahan:1996fg, Minahan:1996cj, Noguchi:1999xq}.
In these cases, there is a one-dimensional Coulomb branch, specified by a complex coordinate $u$, and mass parameters $m$ in the adjoint representation of the seven-brane gauge group. The Seiberg-Witten curves for this class of examples can all be written as:
\begin{equation}
y^2 = x^3 + f(u,m)x + g(u,m),
\end{equation}
where the $f$'s and $g$'s are polynomials in the Coulomb branch parameters and the $m$'s. These polynomials in the $m$'s are constructed
from Casimir invariants of the associated flavor symmetry. In the string compactification geometry, time-reversal invariance corresponds to a complex conjugation operation on the elliptic curve itself. We get a time-reversal invariant system by demanding the Weierstrass coefficients $f$ and $g$ are real. Observe that in a suitable basis of fields, we can simply demand that the $u$'s and $m$'s are all real. This corresponds to a situation in which any mass terms being switched on preserves time-reversal invariance
along the flow from the UV fixed point to the IR, namely where the Seiberg-Witten curve description is valid.

We obtain examples of interfaces by allowing position dependent mass terms
$m(x_{\bot})$. One can also contemplate giving a position dependent value to $u$, though in this case we need to consider
the spacetime dependence for a dynamical field. Switching on a $\mathcal{N} = 1$ superpotential deformation as well as possible supersymmetry breaking mass terms, we can also produce theories in the IR which only have a $U(1)$ gauge field remaining. This strategy was used, for example in \cite{Tachikawa:2016xvs} to analyze some examples of SPTs with non-abelian gauge dynamics.

Assuming we vary the mass parameters $m$ adiabatically, we can continue to use 4D $\mathcal{N} = 2$ supersymmetry to look for the appearance of localized states. In the F-theory realization of these systems as obtained from D3-branes probing a stack of seven-branes, this corresponds to moving the seven-branes around in the $\mathbb{R}_{\bot}$ direction of the 4D spacetime. In the vicinity of some of these seven-branes, however, we can continue to use a 4D analysis. In particular, the location of these seven-branes will occur at some locations $u = u_{\ast}$ in the original Coulomb branch parameter.

Now, the appearance of massless states occurs when the discriminant $\Delta$ vanishes to some order in the variable $(u - u_{\ast})$. In fact, for elliptically-fibered K3 spaces there is a Kodaira classification\footnote{Which also classifies possible codimension one singularities for higher-dimensional elliptically fibered Calabi-Yaus.} of possible singularities \cite{kodaira},
as controlled by the order of vanishing for:
\begin{align}
f & \sim (u - u_{\ast})^{\mathrm{ord}(f)}\\
g & \sim (u - u_{\ast})^{\mathrm{ord}(g)}\\
\Delta & \sim (u - u_{\ast})^{\mathrm{ord}(\Delta)}.
\end{align}
These tell us about the appearance of flavor enhancements, as well as the appearance of massless states, including the associated electric and magnetic charges. In Appendix \ref{app:FLAVA} we consider in detail the special case of $SU(2)$ gauge theory with four hypermultiplets in the fundamental representation of $SU(2)$. In particular, we calculate the periods and the appearance of massless states for a specific choice of mass parameters.

The case of a cusp corresponds to an $I_N$ singular fiber (associated with an $SU(N)$ flavor symmetry), in which $\mathrm{ord}(f) = \mathrm{ord}(g) = 0$, and $\mathrm{ord}(\Delta) = N$. Observe that in the vicinity of such a point, we have:
\begin{equation}
\tau \sim \frac{N}{2 \pi i} \log(u - u_{\ast}),
\end{equation}
indicating a jump of $\theta$ by $2 \pi N$ as we cross this sort of singularity.

The Kodaira classification also shows that we can expect mutually non-local states to be trapped at an interface. For example, a $III^{\ast}$ singular fiber (associated with an $E_7$ flavor symmetry) corresponds to the special case where $\mathrm{ord}(f) = 3$, $\mathrm{ord}(g) \geq 5$ and $\mathrm{ord}(\Delta) = 9$. In this case, we also note that the $J$-function has a well-defined limit, even though the elliptic curve becomes degenerate in this region. The specific value is $J = 1$, as associated with $\tau = i$.

We can also get trapped matter at the other elliptic point of $\Gamma = SL(2,\mathbb{Z})$, namely $\tau = \exp(2 \pi i / 6)$, as associated with $J = 0$. This occurs, for example, with a $II^{\ast}$ singularity (associated with an $E_8$ flavor symmetry), in which $\mathrm{ord}(f) \geq 4$, $\mathrm{ord}(g) = 5$, and $\mathrm{ord}(\Delta) = 10$. In the non-supersymmetric setting we have less analytic control over the ways in which $f,g$, and $\Delta$ might vanish.

Our discussion so far has focused on the case where the $U(1)$ gauge theory on the Coulomb branch enjoys an $SL(2,\mathbb{Z})$ duality group, as directly inherited from the F-theory realization of these systems.\footnote{Strictly speaking one should speak of the $\mathbb{Z} / 2 \mathbb{Z}$ extension of $SL(2,\mathbb{Z})$, as in reference \cite{Pantev:2016nze}. We will not dwell on this issue here.} We get examples with smaller duality groups by holding fixed some of the mass parameters of the system. For example, the ADE series of superconformal field theories just introduced can also be engineered by taking M5-branes wrapped on a $\mathbb{CP}^{1}$ with punctures \cite{Gaiotto:2009we}. These punctures dictate the behavior of mass parameters in the 4D effective field theory. In this formulation, the mapping class group of the curve determines the structure of the duality group. Doing so, we can engineer smaller duality groups. As an example, for $SU(2)$ gauge theory with four flavors, we have two M5-branes wrapped on a sphere with four punctures. In this case, taking some mass parameters held fixed to equal values can produce a smaller duality group such as $\Gamma_{0}(2)$.

We can also consider examples which have a smaller duality group right from the start. As an example of this sort, consider pure $\mathfrak{su}(2)$ gauge theory. Here, we have no mass parameters, so we will consider varying the Coulomb branch parameter $u$
as a function of $x_{\bot}$ with the implicit assumption that we have introduced a suitable $\mathcal{N} = 1$ superpotential deformation to generate jumps in the value of $\tau$ in a given interface region.

Consider first the limit where no superpotential deformation has been switched on.
Following \cite{Seiberg:1994rs, Seiberg:1994aj}, the $\mathcal{N} = 2$ vector multiplet contains a scalar field in the adjoint representation $\phi$. Non-zero values of this scalar move the theory onto the Coulomb branch. In the following we use the gauge invariant combination:
\begin{align}
u = \tfrac{1}{2} \text{tr} ( \phi^2 ) \,.
\end{align}

The Seiberg-Witten curve of the system is given by
\begin{align}
y^2 = (x - u) (x - \Lambda^2) (x + \Lambda^2) \,,
\end{align}
which can be brought to Weierstrass form by a coordinate transformation on $x$. The weakly coupled $U(1)$ gauge theory arises for $|u| \rightarrow \infty$ in which case the gauge coupling goes to zero. Other interesting limits are described by the limits $u \rightarrow \pm \Lambda^2$, which are at strong coupling. At these points one finds light magnetically charged states.

By moving around the moduli space parameterized by $u$ one finds the following monodromy actions in $SL(2,\mathbb{Z})$ on the auxiliary elliptic curve:
\begin{align}
\gamma_+ = \begin{pmatrix} 1 & 0 \\ -2 & 1 \end{pmatrix} \,, \quad  \gamma_- = \begin{pmatrix} -1 & 2 \\ -2 & 3 \end{pmatrix}   \,.
\end{align}
These do not generate the full $SL(2, \mathbb{Z})$ but instead a congruence subgroup given by $\Gamma(2)$.\footnote{Here we do not dwell on the distinctions between $\Gamma(2) \subset SL(2,\mathbb{Z})$ and $P\Gamma(2) \subset PSL(2,\mathbb{Z})$.}

Instead of using the usual Weierstrass form one can also describe the Seiberg-Witten curve in terms of a branched double cover of $\mathbb{CP}^1$, parameterized by the complex coordinate $z$. For a schematic description of the relation between the torus and the double cover of $\mathbb{CP}^1$, see figure \ref{fig:doublecov}.
\begin{figure}[t!]
\centering
\includegraphics[width=0.6\textwidth]{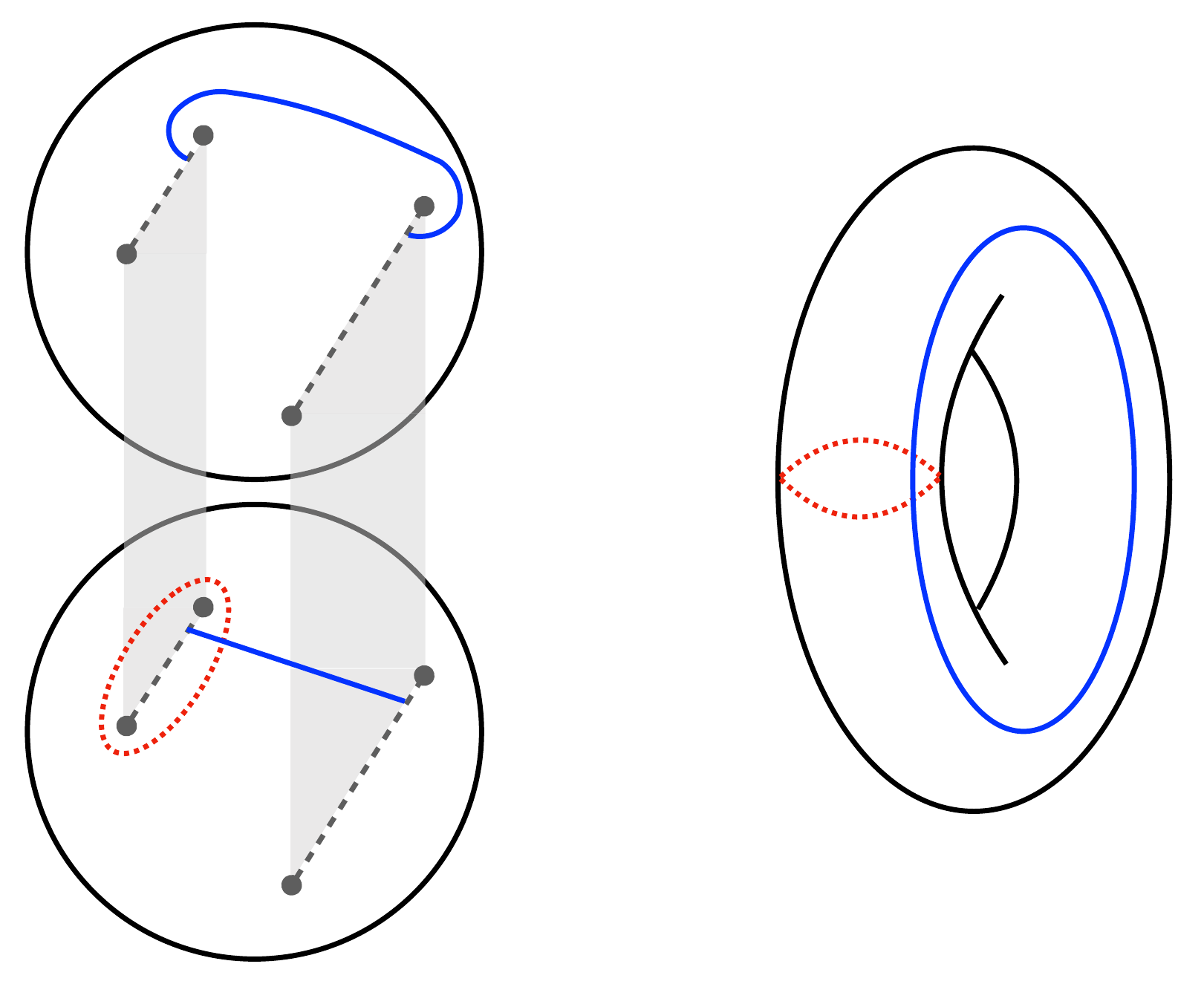}
\caption{Schematic description of the torus as double cover of $\mathbb{CP}^1$.}
\label{fig:doublecov}
\end{figure}
One possible parametrization is given in \cite{Tachikawa:2013kta} and reads as:
\begin{align}
\Lambda^2 z + \frac{\Lambda^2}{z} = x^2 - u \,.
\end{align}
In terms of these variables the Seiberg-Witten differential reads
\begin{align}
\lambda = x \frac{dz}{z} \,.
\end{align}
The UV curve is given by the $\mathbb{CP}^1$ in combination with the four branch points connected by two branch cuts.

The pure gauge theory describes an elliptic curve, with moduli space given by $X(2)$. The fundamental domain as well as its time-reversal invariant subset are depicted in figure \ref{fig:G2phases}.
\begin{figure}[t!]
\centering
\includegraphics[width=0.5\textwidth]{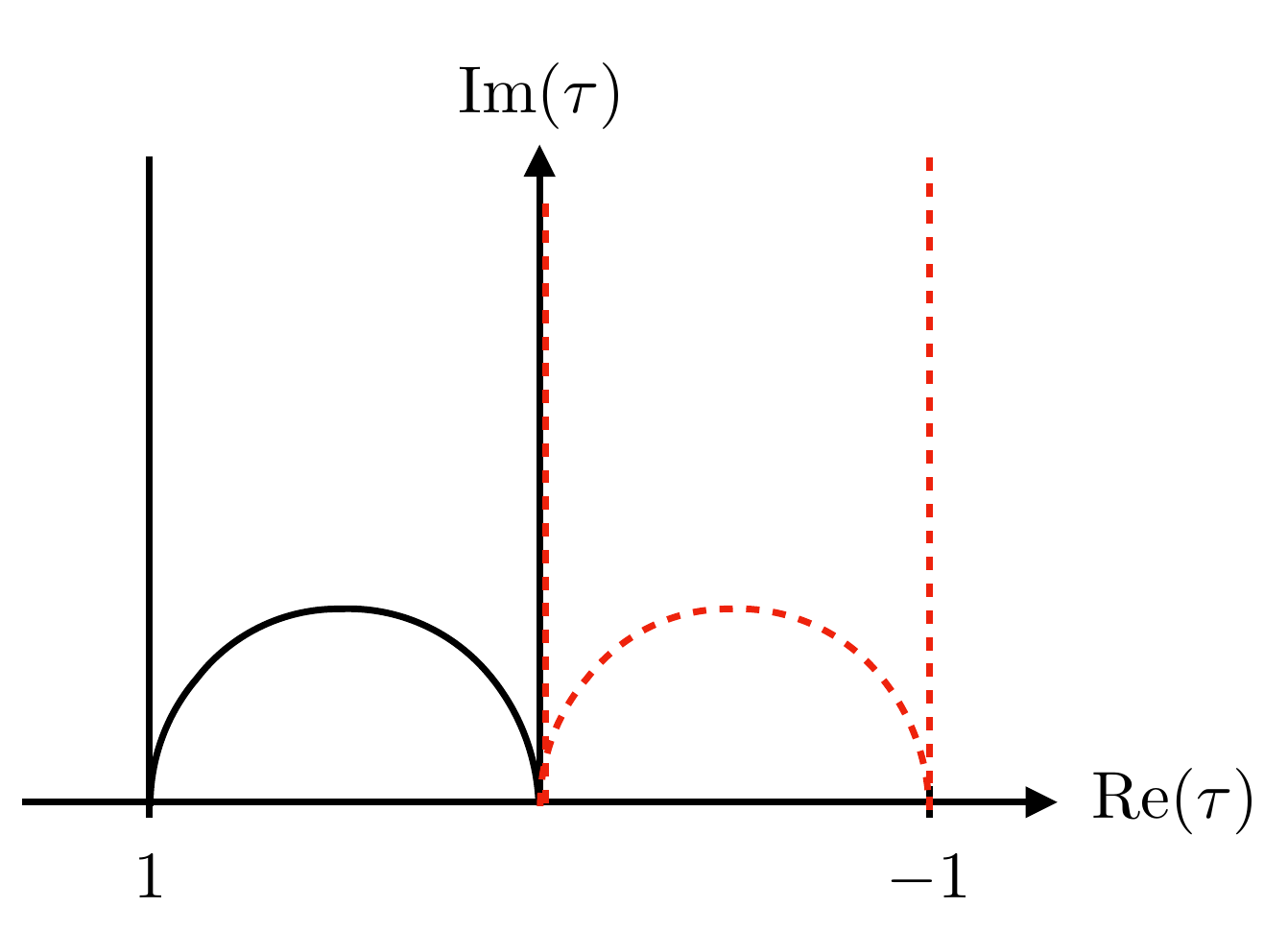}
\caption{Fundamental domain of $\Gamma (2)$ on the upper half plane as well as its time-reversal invariant subset $X(2)_{\mathbb{R}}$.}
\label{fig:G2phases}
\end{figure}
It contains three distinct cusps at $\tau \in \{ 0, 1, i \infty \}$ and is topologically a $\mathbb{CP}^1$ with the cusps marking three points. In this case the time-reversal invariant subset $X(2)_{\mathbb{R}}$ contains all three cusps.

Let us see what the three cusps correspond to in terms of data extracted from
the Seiberg-Witten curve. The equivalent of the $j$-function in the case
of $\Gamma = \Gamma(2)$ is its so-called Hauptmodul, defined by
\begin{align}
\lambda (\tau) = \bigg( \frac{\Theta_2 (\tau)}{\Theta_3 (\tau)} \bigg)^4 \,,
\end{align}
where the $\Theta$'s denote theta functions, the explicit form of which we will not need.
This yields a map $\lambda: X(2) \rightarrow \mathbb{CP}^1$. The values at the cusps are
\begin{align}
\lambda (0) = 1 \,, \quad \lambda (1) = \infty \,, \quad \lambda (i \infty) = 0 \,.
\end{align}
Taking the original form of the Seiberg-Witten curve, we expect cusps at the collision of two of the branch points, i.e.\
\begin{align}
u = \Lambda^2 \,, \quad u = - \Lambda^2 \,, \quad u \rightarrow \infty \,.
\end{align}
For the two strongly coupled cusps at $u = \pm \Lambda^2$, which are associated to $\tau = 0$ and $\tau = 1$, we know that we get either a massless monopole or dyon.

Next, we assume a suitable superpotential deformation has been switched on which produces a domain wall solution with multiple
kinks which passes through the different cusps. Our expectation is that the wall will now carry a charge as dictated by the sort of cusp encountered. The cusp at weak coupling corresponds to $u \rightarrow \infty$ and at first poses a puzzle. In the limit of large $u$ the theory becomes classical and one has the identification $a \sim \sqrt{u}$. Therefore, the $\mathfrak{su}(2)$ gauge algebra is broken to $U(1)$ at a very high scale and the supermultiplets containing the electrically charged $W$-bosons are very massive with
\begin{align}
m_{W} \sim a \rightarrow \infty \,.
\end{align}
Therefore, even though there is a cusp, one naively does not expect any light modes. That being said, building an interface that is very thin
relative to the mass scale, the corresponding energy scales are very high and the classical description in terms of a weakly coupled $\mathfrak{su}(2)$ gauge theory remains valid throughout the system. In this sense there actually are massless $W$ bosons and the $\mathfrak{su}(2)$
is restored.

Assuming the presence of light electric states of charge $q_e$ on the interfaces associated to the cusp at $\tau \rightarrow i \infty$, we can use coset representatives in order to investigate the other cusps at strong coupling. For this we choose
\begin{equation}
\begin{split}
\alpha_1 = \begin{pmatrix} 0 & -1 \\ 1 & 0 \end{pmatrix}:& \quad \tau = i \infty \enspace \mapsto \enspace \tau = 0 \,, \\
\alpha_2 = \begin{pmatrix} 1 & -1 \\ 1 & 0 \end{pmatrix}:& \quad \tau = i \infty \enspace \mapsto \enspace \tau = 1 \,.
\end{split}
\end{equation}
Then we can find the action on the charges of states as:
\begin{align}
\alpha_1 \begin{pmatrix} q \\ 0 \end{pmatrix} = \begin{pmatrix} 0 \\ q \end{pmatrix} \,, \quad \alpha_2 \begin{pmatrix} q \\ 0 \end{pmatrix} = \begin{pmatrix} q \\ q \end{pmatrix} \,,
\end{align}
which suggests the presence of massless purely magnetically charged and dyonic states, respectively. These are exactly the states associated to the monopole and dyon point for the pure gauge Seiberg-Witten theory! This can be precisely matched to the behavior of the elliptic $\lambda$-function in terms of the three branch points
\begin{align}
\lambda = \frac{2 \Lambda^2}{u + \Lambda^2} \,.
\end{align}
For $u \rightarrow \Lambda^2$, which is the monopole point one obtains $\lambda = 1$ which corresponds to $\tau = 0$. Similarly, for $u \rightarrow - \Lambda^2$, the dyon point, one has $\lambda \rightarrow  \infty$, i.e.\ $\tau = 1$.

\section{Examples via Compactification} \label{sec:6DCOMPACTIFY}

In this section we present a construction of 4D $U(1)$ gauge theories
with duality groups $\Gamma = \Gamma_{0}(N), \Gamma_{1}(N), \Gamma(N)$
by compactifying the theory of an anti-chiral two-form in six spacetime dimensions.
We view this theory as an edge mode coupled to a bulk 7D Chern-Simons theory.
This provides us with a geometric way to visualize much of the structure associated with
the spectrum of states and line operators in these 4D theories.

Using this, we can build 3D interfaces by just taking this 6D theory and compactifying on a three-manifold $M_{3}$ given by a family of elliptic curves fibered over the line $\mathbb{R}_{\bot}$ of the 4D spacetime $\mathbb{R}^{2,1} \times \mathbb{R}_{\bot}$. In this picture, singularities of the fibration indicate the locations of 3D interfaces.

This section is organized as follows. We begin by discussing the spectrum of charged states and line operators
for the different choices of duality groups. Much of this discussion follows what is presented in
reference \cite{Aharony:2013hda}. After this, we turn to the realization of this structure
via compactification of an anti-chiral two-form. In particular, we show that the level of the
associated 7D Chern-Simons theory provides a general way to control the set of possible duality groups.

\subsection{Line Operators and Charges \label{sec:lineops}}

A $U(1)$ gauge group is always specified together with a charge quantization condition. This quantization condition is not necessarily correlated with the presence of dynamical degrees of freedom with the corresponding charges. Instead it can be described by the set of genuine line operators.

For an abelian $U(1)$ gauge theory without any charged particles this defines a lattice of charges which are mutually local, i.e.\ they are consistent with the Dirac quantization condition, that enters in the definition of a general line operator. An electric line operator is given by
\begin{align}
\mathcal{O}^{(q_e , 0)}_{L} = \exp \Big( i q_e \underset{L}{\int} A \Big) \,,
\end{align}
where $A$ denotes the electric gauge field, and $L$ denotes a line in the 4D spacetime to integrate over.
The corresponding purely magnetically charged line operator can be given in terms of the dual gauge field $A_D$, and reads:
\begin{align}
\mathcal{O}^{(0, q_m)}_{L} = \exp \Big( - i q_m \underset{L}{\int} A_D \Big) \,.
\end{align}
In general, one can also define dyonic line operators $\mathcal{O}^{(q_e,q_m)}_{L}$, that carry both electric and magnetic charges.
For consistency, $q_e$ and $q_m$ have to be in the charge lattice defined by Dirac quantization. Moreover, these operators are charged with respect to global one-form symmetries \cite{Banks:2010zn, Gaiotto:2014kfa}. In the case of pure $U(1)$ gauge theory there are two global $U(1)$ one-form symmetries. The electric one-form symmetry acts by shifting $A$ by a flat $U(1)$ connection, the magnetic one acts accordingly on the dual gauge field $A_D$.

In the presence of dynamical charges the one-form symmetries are broken explicitly. However, if the dynamical charges only fill out a sublattice of the allowed charge lattice, discrete one-form symmetries remain. One example which will be relevant in the following is the case where the dynamical charges are of the form
\begin{align}
(q_e, q_m)_{\text{dyn}} = (N k, l) \,, \enspace \text{with} \enspace k,l \in \mathbb{Z} \,,
\end{align}
where without loss of generality we normalized the charges in a way that the full charge lattice is given by $\mathbb{Z} \times \mathbb{Z}$, i.e.\ integer charges. In this case the full magnetic one-form symmetry is broken. The electric one-form symmetry is only broken to a discrete subgroup, namely $\mathbb{Z} / N \mathbb{Z}$, with the charge carried by the line operators
\begin{align}
\mathcal{O}_L^{(r,0)} = \exp \Big( i r \underset{L}{\int} A \Big) \,, \enspace \text{with} \enspace r \in \{ 1, \dots, N - 1 \} \,.
\label{eq:ellinetors}
\end{align}
Note that line operators of the form discussed are objects in the theory which are also present at very low energies. The same is not necessarily true for dynamical charged particles, which can be integrated out below their mass scale.

On general grounds, the line operators transform non-trivially under duality, so to fully specify the action of the duality group we need to take this into account. To present explicit examples associated with different duality groups, we now turn to a 6D realization of these structures, starting first with $SL(2,\mathbb{Z})$.

\subsection{Geometrizing Duality}

One way of making this connection between line operators, charged states, and the congruence subgroups more apparent is to describe the $U(1)$ theory as a compactification of an anti-chiral two-form potential $B$ compactified on a torus, see e.g.\ \cite{Tachikawa:2013hya, Lawrie:2018jut, Eckhard:2019jgg, Garcia-Etxebarria:2019cnb}. At a classical level, we
can think of this as being specified by a three-form field strength $H$ subject to the condition:
\begin{equation} \label{6Dselfduality}
\ast_{6D} H = - H.
\end{equation}
The two-form potential couples to anti-chiral strings via integration of the pull-back of $B$ to the worldsheet of the string. It is well-known that the compactification of this theory on a $T^2$ produces a $U(1)$ gauge theory with complexified gauge coupling $\tau$ controlled by the complex structure of the $T^2$. Letting $\gamma_A$ and $\gamma_B$ denote the A- and B-cycles of this $T^2$, we observe that wrapping a string on the one-cycle $q_e \gamma_A + q_m \gamma_B$ results in a 4D point particle of electric and magnetic charge $(q_e,q_m)$. The celebrated S-duality of Maxwell theory corresponds to interchanging the A- and B-cycles of this
torus.

We would like to understand the structure of line operators and dynamical operators in the associated quantum theory.
To give a proper account, we of course need to quantize this 6D theory. This is somewhat subtle because the self-duality condition of equation \eqref{6Dselfduality} clashes with the condition that such fluxes should be quantized. As noted in \cite{Witten:1998wy, Belov:2006jd, Monnier:2017klz, Heckman:2017uxe}, the proper way to handle this sort of situation is to view the 6D theory as an edge mode coupled to a 7D Chern-Simons theory with three-form potential $C$ and action:
\begin{equation}
S_{7D} = \frac{k}{4 \pi i} \underset{M_7}{\int} C \wedge dC.
\end{equation}
with $M_7$ a seven-manifold with 6D boundary $M_{6} = \partial M_{7}$, e.g.\ \cite{Hsieh:2020jpj}. There are some subtleties in fully defining this 7D theory. For example, the analog of spin structure for a 3D Chern-Simons theory involves specifying a Wu structure (see e.g. \cite{Monnier:2017klz, Monnier:2018nfs}). Since we will primarily work on spaces with no metric curvature, most of these issues have little impact on the general statements we make. The boundary condition for the three-form potential is:
\begin{equation}
C \vert_{\partial M_7} = - \ast_{6D} C \vert_{\partial M_7}.
\end{equation}
This is the analog of the same condition one would impose for a bulk 3D Chern-Simons theory coupled to a chiral boson. In this bulk 7D theory we have a three-form potential, so our system couples to two-branes. Given a three-chain which ends on a two-cycle in the 6D spacetime, we obtain a two-dimensional string of the 6D theory. Much as in 3D Chern-Simons theory, the level $k \in \mathbb{Z}$ must be quantized. This is just to ensure that the phase factor $\exp(iS)$ remains well-defined under large gauge transformations of the three-form potential.

The analog of a line operator in this setting is specified by integrating the three-form potential over a three-chain.
Calling such a three-chain $\Sigma$, these operators take the form:
\begin{equation}
\mathcal{O}^{Q}_{\Sigma} = \exp \Big( i Q \underset{\Sigma}{\int} C \Big).
\end{equation}
If we were to quantize this theory with ``time'' indicated by the direction perpendicular to a 6D Euclidean slice, we
would obtain a non-trivial braid relation between these operators (see e.g. \cite{Witten:1998wy, DelZotto:2015isa}) given by:
\begin{equation}
\mathcal{O}^{Q}_{\Sigma} \mathcal{O}^{Q^{\prime}}_{\Sigma^{\prime}} = \exp \left(\frac{2 \pi i}{k} Q Q^{\prime} \Sigma \cdot \Sigma^{\prime} \right)\mathcal{O}^{Q^{\prime}}_{\Sigma^{\prime}} \mathcal{O}^{Q}_{\Sigma}.
\label{eq:corrfunc}
\end{equation}
In the case where the 6D slice is instead Lorentzian, this this fixes a Dirac pairing between strings
of the 6D theory \cite{Deser:1997se}. This Dirac pairing descends to the expected one in 4D.
Now, the important point for us is that we are interested in the spectrum of
line operators which commute, namely those which have integer valued Dirac pairing.
The main thing we will need to track is the level $k$ of the anti-chiral two-form $B$.

Let us now turn to the compactification of a level $k$ anti-chiral two-form on an elliptic curve $E$ with complex structure
$\tau$.  We will be interested in the
periods of the $B$-field on a two-cycle of the 6D spacetime $\mathbb{R}^{3,1} \times E$ of the form:
\begin{equation}
L \times (q_e \gamma_A + q_m \gamma_B).
\end{equation}
First of all, we see that the intersection pairing from the closed path on the elliptic curve amounts to the Dirac pairing which is invariant with respect to $SL(2, \mathbb{Z})$ transformations. Moreover, correlation functions are only sensitive to charges $(q_e,q_m)$ modulo $k \mathbb{Z}$. This naturally draws a connection to the classification of congruence subgroups acting in a particular way on operators specified by their electric and magnetic charges modulo $\mathbb{Z} / k \mathbb{Z} \times \mathbb{Z} / k \mathbb{Z}$, which we want to explain next.

First, let $k = N^2$ be a square of an integer $N$. Then one possible solution to the constraint that two genuine line operators have to commute is given by
\begin{align}
q_e \in N \mathbb{Z} + \tfrac{1}{N} m r \,, \quad q_m \in N \mathbb{Z} \,, \quad \text{with} \enspace r \in  \{ 0, 1, \dots, N-1 \}
\end{align}
which fills out a $\mathbb{Z} / N \mathbb{Z} \times \mathbb{Z} / N \mathbb{Z}$, a subset of $\mathbb{Z} / k \mathbb{Z} \times \mathbb{Z} / k \mathbb{Z}$.
\begin{figure}[t!]
\centering
\includegraphics[width=0.3\textwidth]{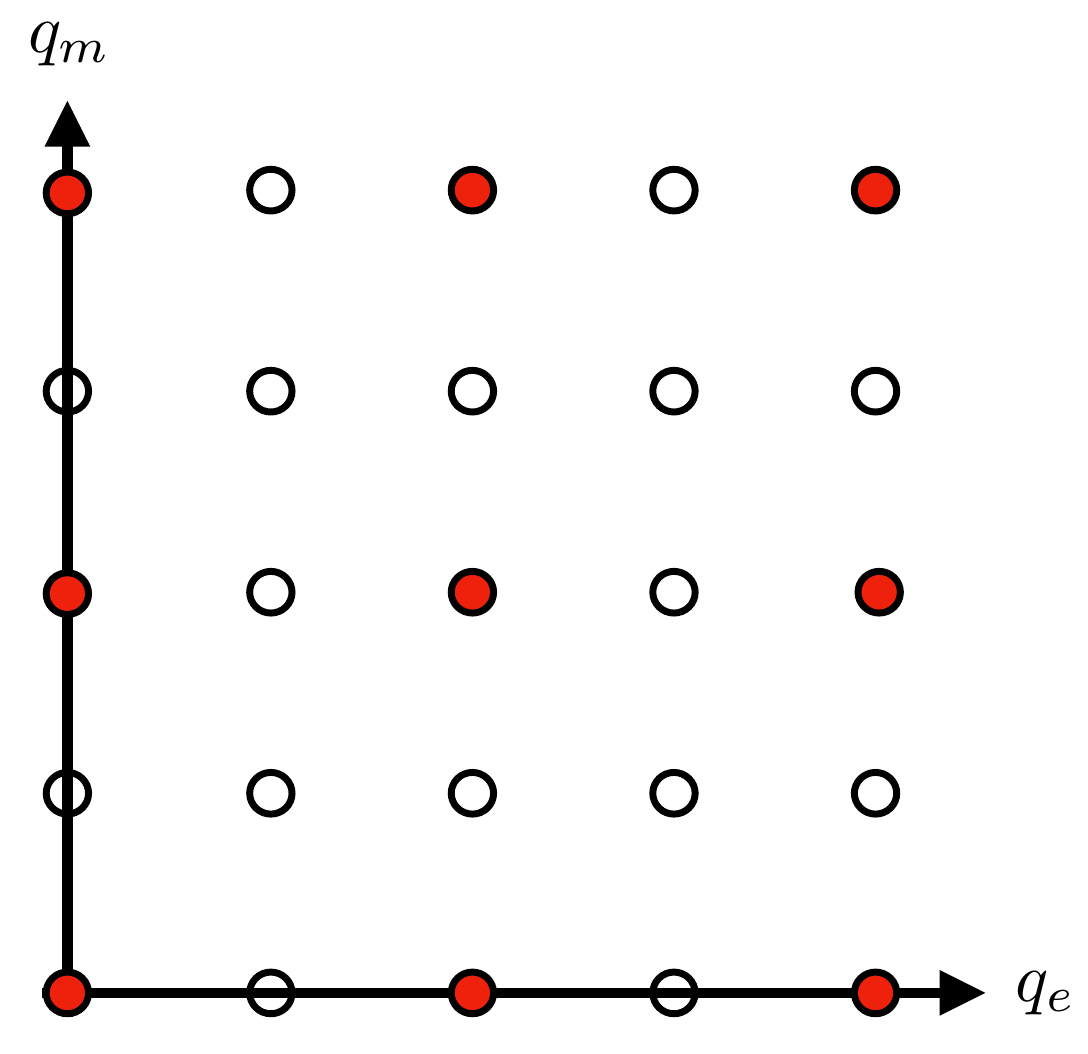}
\caption{Possible sublattice of commuting dynamical charges for $k = N^2 = 4$, corresponding to the case $\Gamma = \Gamma(2)$.}
\label{fig:chlatG4}
\end{figure}
Further demanding that $N$ times the charge has to be a trivial charge in $\mathbb{Z} / k \mathbb{Z} \times \mathbb{Z} / k \mathbb{Z}$ fixes $r$ to zero and one obtains the sublattice depicted in figure \ref{fig:chlatG4} for $k = N^2 = 4$. The charges of the genuine line operator are therefore labeled by elements of $\mathbb{Z} / N \mathbb{Z} \times \mathbb{Z} / N \mathbb{Z}$. Restricting the duality group to a subgroup keeping these operators invariant mod $k$ will lead to the congruence subgroup defined by $\Gamma (N)$.

For general $k$ such a sublattice is not accessible, but one always can define the charges to satisfy $q_e \in \mathbb{Z}$ and $q_m \in k \mathbb{Z}$, which naturally lead to a maximal set of charges with mutually local line operators.
\begin{figure}[t!]
\centering
\includegraphics[width=\textwidth]{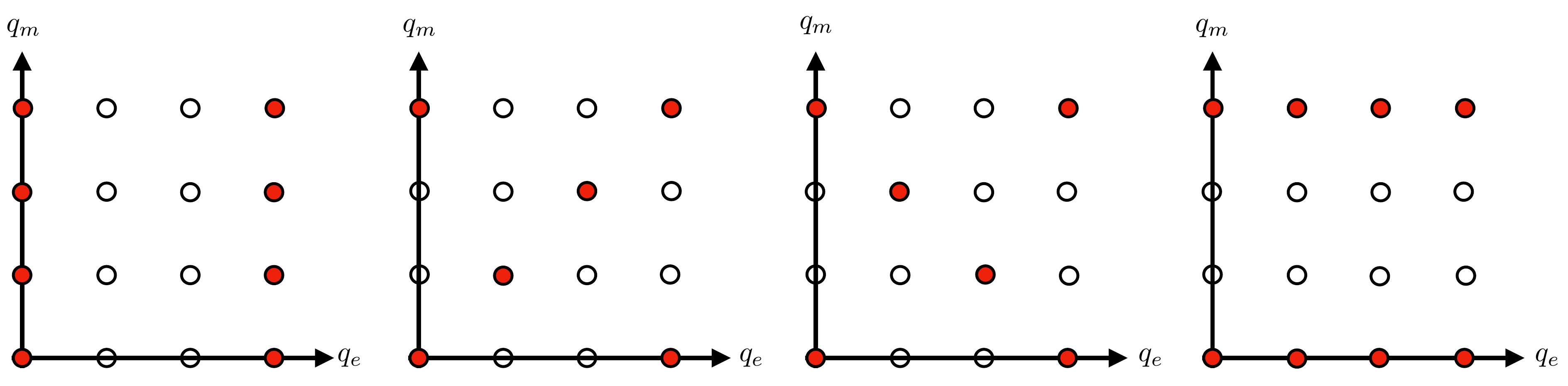}
\caption{Possible spectrum of genuine line operators for $k = 3$. Here, the duality group is taken to be either $\Gamma_0(3)$ or $\Gamma_1(3)$. In the case of $\Gamma_1(3)$, a torsional point (and its multiples) is fixed, while in the case of $\Gamma_0(3)$ only the zero element is fixed.}
\label{fig:chlatG1_3}
\end{figure}
Since the Dirac pairing is invariant with respect to the action of $SL(2, \mathbb{Z})$ one can also use the transformed spectrum of charges. In figure \ref{fig:chlatG1_3} we show the different possible choices for $k = 3$. Demanding invariance of the chosen spectrum of genuine line operators under the duality group then leads to the congruence subgroups $\Gamma_1 (k)$ and $\Gamma_0 (k)$, or a conjugate by a coset representative. In the case of $\Gamma_1 (k)$ one requires the invariance of each line operator individually. In the case of $\Gamma_0 (k)$ one allows an action on the line operators keeping the full spectrum fixed.

These congruence subgroups in connection with a specification of line operators also appear in the context of non-abelian gauge symmetries. There, the line operators specify the explicit realization of the gauge group as opposed to the gauge algebra \cite{Aharony:2013hda, Gaiotto:2014kfa, Garcia-Etxebarria:2019cnb}. In these cases the one-form symmetry is related to the center of the gauge group and mixed anomalies with time-reversal invariance can lead to interesting insights concerning the phase structure of four-dimensional theories as well as their possible interfaces \cite{Gaiotto:2017tne, Gaiotto:2017yup, Hsin:2018vcg}.

\subsection{The Jacobian Curve}

There is also a close connection with the Jacobian of the elliptic curve given as:
\begin{align}
\mathcal{J} (E) = H^1 (E, \mathbb{R}) / H^1 (E, \mathbb{Z}) \simeq \widetilde{E}
\end{align}
which itself is an elliptic curve with the origin defined as the vanishing gauge field. In physical terms, the Jacobian
specifies non-trivial flat fields on the torus $E$. In fact, the complex structure of this elliptic curve as specified by a parameter $\widetilde{\tau}$ is determined by the complex structure $\tau$ of the elliptic curve $E$; they are in fact the same.

With the basis of $H^1 (E, \mathbb{Z})$ given by $\{ \sigma_A, \sigma_B \}$ defining the lattice of $\widetilde{E}$, the relevant forms are given by $\alpha \sigma_A + \beta \sigma_B$, with $\alpha, \beta \in [0,1)$. Now we can specify the subset of $\mathcal{J}(E)$ which is trivial on the physical states, by which we mean that
\begin{align}
\underset{q_e \gamma_A + q_m \gamma_B}{\int}( \alpha \sigma_A + \beta \sigma_B ) \in \mathbb{Z} \,.
\end{align}
The structure specified by the level of the anti-chiral two-form
thus determines a corresponding level in the elliptic curve $\widetilde{E}$. This level structure is associated with the appearance of
torsional points in $\widetilde{E}$. Recall that these are obtained by viewing the curve $\widetilde{E} = \mathbb{C} / \widetilde{\Lambda}$ as a group. An $N$-torsional point $P$ in this group is one for which $N[P]$ is just the zero element of this additive group. In terms of the lattice $\widetilde{\Lambda} = \widetilde{\omega}^1 \mathbb{Z} \oplus \widetilde{\omega}^2 \mathbb{Z} \subset \mathbb{C}$, these $N$-torsion points can be written as:
\begin{equation}
\widetilde{E}(N) = a \frac{\widetilde{\omega}^1}{N} + b \frac{\widetilde{\omega}^2}{N} \,\,\, \text{for} \,\,\, a,b = 0,...,N-1.
\end{equation}

For the example above $(q_e,q_m) \in N \mathbb{Z} \times \mathbb{Z}$ this is given by the elements
\begin{align}
\big\{ \big( \tfrac{r}{N} + k \big) \sigma_A + l \sigma_B \big\} \,, \quad \text{with \ } k,l \in \mathbb{Z} \enspace \text{and} \enspace r \in \{0, 1, \dots, N-1 \} \,.
\end{align}
We see that up to lattice vectors this defines a set of $N$-torsion points on the Jacobian $\widetilde{E}$. In general, one can get the full set of $N$-torsion points by demanding that a dynamical state has charge $(q_e,q_m) \in N \mathbb{Z} \times N \mathbb{Z}$. An $SL(2,\mathbb{Z})$ action on the line operators can then be perceived as an action on the torsion points in the dual curve $\widetilde{E}$.

Invariance of (a subset of) the spectrum of line operators therefore restricts the duality group to a subgroup of $SL(2, \mathbb{Z})$.
One way to think about this is to start with the original lattice of electric and magnetic charges $\Lambda$, along with the corresponding elliptic curve $\widetilde{E}$. We can consider a non-zero holomorphic map to another complex torus $\widetilde{E^{\prime}}$ along with its corresponding defining lattice $\Lambda^{\prime}$. Such mappings are known as isogenies and in general correspond to either rescalings of the original lattice via the multiplication map $\Lambda \rightarrow N \Lambda$ or involve picking an order $N$ cyclic subgroup $C \subset \widetilde{E}[N] = \mathbb{Z} / N\mathbb{Z} \times \mathbb{Z} / N\mathbb{Z}$  and constructing a new lattice out of the cosets. All isogenies can be obtained from these two basic operations (see e.g. \cite{diamond2006first}), and they serve to define different lattices of electric and magnetic charges. We now turn to the three congruence subgroups $\Gamma(N), \Gamma_1(N)$, and $\Gamma_0(N)$, which are obtained as follows.

\subsubsection*{$\mathbf{\Gamma (N)}$}

For the congruence subgroup $\Gamma (N)$ the full set of line operators classified by the lattice $\mathbb{Z} / N \mathbb{Z} \times \mathbb{Z} / N \mathbb{Z}$ remains invariant. In terms of the Jacobian, that means that the full set of torsion points in $\widetilde{E}(N)$ is invariant up to lattice vectors. Specifically, the line operators are given by
\begin{align}
\mathcal{O}_{L}^{(r,s)} = \exp \Big( i \underset{L \times (r \gamma_A + s \gamma_B)}{\int} B\Big) = \exp \Big( i r \underset{L}{\int} A - i s \underset{L}{\int} A_D \Big) \,, \quad \text{with} \enspace r,s \in \mathbb{Z} / N \mathbb{Z}\,,
\end{align}
which are invariant under $\Gamma (N)$ up to the addition of a worldline of a dynamical particle. In the four-dimensional description this is a theory with dynamical electric and magnetic charges that are a multiple of $N$.

\subsubsection*{$\mathbf{\Gamma}_1 (N)$}

For the congruence subgroup $\Gamma_1(N)$ we fix an $N$-torsion point of $\widetilde{E}(N)$. This leads to the invariance of a full $\mathbb{Z} / N \mathbb{Z}$ subgroup of $\widetilde{E}(N)$ by the linearity of the modular transformation. With the help of an $SL(2, \mathbb{Z})$ element which is not in $\Gamma_1 (N)$ we can always map this torsion point to be $\tfrac{1}{N} \sigma_A$. We see that $\Gamma_1 (N)$ leaves invariant the line operators defined by
\begin{align}
\mathcal{O}^{(r,0)} =  \exp \Big( i \underset{L \times (r \gamma_A)}{\int} B \Big) = \exp \Big( i r \underset{L}{\int} A \Big) \,, \quad \text{with} \enspace r \in \mathbb{Z} / N \mathbb{Z}\,.
\label{eq:G1lines}
\end{align}
In the compactified theory this means that only dynamical electric charges which are a multiple of $N$ are present.
There can be other realizations of this choice which differ by the action of a coset representative.

\subsubsection*{$\mathbf{\Gamma}_0 (N)$}

Finally, in $\Gamma_0 (N)$ one has a set of elements generating a $\mathbb{Z} / N \mathbb{Z}$ subgroup of $\widetilde{E}(N)$ which stays invariant. The individual elements, however, can be transformed among each other. Again, we can use a coset representative in order to map the $\mathbb{Z} / N \mathbb{Z}$ subgroup to $\big\{ \tfrac{r}{N} \sigma_A \big\}$, which translates to the same line operators as in \eqref{eq:G1lines}. The transformation of the individual elements among each other defines an action on the line operators. For example if $\gamma \in \Gamma_0 (N)$ acts as
\begin{align}
\tfrac{r}{N} \sigma_A \mapsto \tfrac{r'}{N} \sigma_A \,,
\end{align}
up to lattice vectors, the induced action on the line operators reads
\begin{align}
\mathcal{O}_L^{(r, 0)} \rightarrow \mathcal{O}_L^{(r', 0)} \,.
\end{align}
In the four-dimensional effective action, we see that $\Gamma_0 (N)$, describes a theory with dynamical electric charges being a multiple of $N$ together with an action on the line operators $\mathcal{O}_{L}^{(r,0)}$.

\subsection{Generalization to Other Riemann Surfaces}
\label{subsec:genRiemann}

The generalization to higher-genus Riemann surfaces is straightforward from what we said above. Compactifying a 6D anti-selfdual tensor on a genus $g$ Riemann surface $C_g$ leads to $g$ abelian $U(1)$ gauge fields in four dimensions. Whereas the mapping class group of higher-genus realizations is highly complicated and these surfaces do not have a generic way to add points, the interpretation using the Jacobian is still applicable. The Jacobian of the Riemann surface is:
\begin{align}
\mathcal{J} (C_g) = H^1 (C_g, \mathbb{R}) / H^1 (C_g, \mathbb{Z}) \simeq \widetilde{T}^{2g} \,,
\end{align}
and on the torus $\widetilde{T}^{2g}$ we can define $N$-torsion elements as harmonic one-forms with
\begin{align}
N \sigma \in H^1 (\Sigma_g, \mathbb{Z}) \,,
\end{align}
which we denote by $\mathcal{J}_N (C_g)$. For the case of $C_g = E$ this lead to the identification of the congruence subgroups of $SL(2,\mathbb{Z})$ via the action on the torsion elements in $\widetilde{T}^2 = \widetilde{E}$.

For a general Riemann surface we can restrict the actions of the duality group, i.e.\ the mapping class group in such a way that the integral over a basis of one-cycles for all or a subset of torsion elements modulo $N$ has a well-defined behavior. It either remains fixed or it allows for an action on the set of torsion elements. Since now the set  of torsion elements in $\mathcal{J}_N(C_g)$ are defined by $( \mathbb{Z} / N \mathbb{Z} )^{2g}$ it is also conceivable that mixed version of the possibilities above are realized. For example, a certain $\mathbb{Z} / N \mathbb{Z}$ subgroup can be held fixed element by element and another subgroup might be held fixed up to an action on the individual elements. This leads to a generalization of congruence subgroups in the context of the mapping class groups of higher genus Riemann surfaces.

\section{More General Interfaces at Strong Coupling} \label{sec:6DGENERAL}

In the previous sections we used time-reversal invariance in 4D $U(1)$ gauge theories to produce examples of 3D interfaces
at strong coupling, and we also presented some explicit examples realizing these features.

A common theme in these constructions is the appearance of a six-dimensional field theory. In the case of the compactification of an anti-chiral
two-form, this is manifest from the start. In the case of our $\mathcal{N} = 2$ theories, this follows from the class $\mathcal{S}$ construction based on compactification of a 6D $\mathcal{N} = (2,0)$ superconformal field theory on a Riemann surface (see e.g. \cite{Witten:1997sc, Gaiotto:2009we}). In both these cases, the geometry of the interface can thus be understood in terms of compactification on a three-manifold with boundary, constructed from a family of Riemann surfaces fibered over the real line. Returning to the analysis of the previous sections, we have been considering singularities in the associated elliptic curve with real coefficients, deducing the appearance of localized matter from singular fibers. This method of construction relies heavily on the special features of time-reversal invariance, in tandem with the structure of congruence subgroups of $SL(2,\mathbb{Z})$.

In this section we present another method for generating interfaces at strong coupling. Instead of relying on the additional structure of time-reversal invariance we will instead consider compactification of higher-dimensional field theories on families of Riemann surfaces. The main theme here will be to identify the appearance of singularities in the associated fibers as a diagnostic for tracking the appearance of localized matter. We focus on the case of compactification of six-dimensional superconformal field theories on three-manifolds with boundary. There has recently been significant progress in understanding the construction and study of such 6D SCFTs (see e.g. \cite{Heckman:2013pva, Ohmori:2014kda, Heckman:2015bfa, Cordova:2015fha} and \cite{Heckman:2018jxk} for a recent review), and in particular the compactification of such theories to various lower-dimensional systems \cite{Morrison:2016nrt, Apruzzi:2016nfr, Razamat:2016dpl, Apruzzi:2018oge}. Notably, however, compactifications of 6D SCFTs on three-manifolds has mainly focused on the special case of $\mathcal{N} = (2,0)$ theories as in references \cite{Cecotti:2011iy, Dimofte:2011py}. From this perspective, the present study provides a general starting point for building 3D field theories associated with the degrees of freedom localized on an interface.

The main idea will be to first consider a 4D $\mathcal{N} = 1$ theory as obtained from compactification of a 6D SCFT on a Riemann surface. This sort of compactification involves a choice of background metric on the Riemann surface, and can also be supplemented by switching on various flavor symmetry fluxes. All of these choices lead to a wide range of possible 4D theories. In many cases, these compactifications are expected
to produce a 4D $\mathcal{N} = 1$ SCFT \cite{Morrison:2016nrt, Razamat:2016dpl, Apruzzi:2018oge}, but there are also situations where such a compactification instead leads to a trivial fixed point in the IR (either fully gapped or with just free fields) \cite{Apruzzi:2018oge}. Assuming we can switch on some choice of background fields in the 6D theory, the 4D theory inherits some of its symmetries as well their anomalies from the 6D theory.

To build a 3D interface, we can next consider a family of Riemann surfaces, each equipped with a set of flavor symmetry fluxes. Fibering over a real line $\mathbb{R}_{\bot}$ we can vary both the metric and the fluxes. In fact, by allowing for singular fibrations and gauge field configurations, we can allow both the genus and the Chern classes of these fluxes to jump as we move along $\mathbb{R}_{\bot}$. This is problematic when viewed as a motion inside the moduli space of genus $g$ Riemann surfaces with $n$ marked points (such as $\overline{\mathcal{M}}_{g,n}$, the Deligne-Mumford compactification of the moduli space), but is not problematic when viewed in terms of the geometry of the total space. Indeed, we can construct an interface by gluing together piecewise constant profiles for the metric and fluxes such that when interpreted as a 4D theory, the anomalies are bigger in an interior region. We view this as building an interface with non-zero thickness. In the singular limit where the interior region degenerates to zero thickness, we have a sharp interface.

The rest of this section is organized as follows. First, we set up the relevant mathematical bordism problem and show that there are no obstructions to constructing an interpolating profile of the sort needed to build a thick interface. We then illustrate these considerations with a few examples. We consider the special case of a 6D hypermultiplet compactified on a three-manifold with boundary, and then turn to the more general structure of compactifications of interacting 6D SCFTs.

\subsection{Cobordism Considerations}

To construct more general examples of 3D interfaces, we now discuss the general cobordism problem for our compactification.
Consider $Q$ a cobordism between two Riemann surfaces $C^L$ and $C^R$. A cobordism always has the structure of a fibration\footnote{To suit our needs, what we refer to as a cobordism here is actually a noncompact manifold gotten by deleting the boundary components of a cobordism (which is a compact manifold with boundary) so that $C^R$ and $C^L$ lie ``at infinity". The fibration structure is usually presented in the math literature as being over [0,1], but we use $\mathbb{R}_t$ for our physical purposes.} over $\mathbb{R}_{\bot}$ where the fiber may become singular, change its topology, and have multiple components. This is equivalent to the well-known statement that there always exists a smooth Morse function, $f$, on a cobordism with $f^{-1}(-\infty)=C^L$ and $f^{-1}(+\infty)=C^R$, which induces a codimension-one foliation which is singular at the critical points of $f$ \cite{milnor}. Further, we choose a metric on $Q$ that is in the conformal class of a metric that gives the same volume to each of the Morse fibers. We emphasize that while the fibers may become singular at given values of $x_\bot$, the smoothness of the compactification manifold $Q$ suggests we should be careful about our expectation of localized states since this is merely a coordinate singularity.

To understand what happens, first note that the second oriented cobordism group, $\Omega_2^{SO}$, is trivial for the reason that we can take any oriented three-manifold and cut out two disjoint oriented Riemann surfaces of any genus out of it. The fibration structure will depend on a choice of Morse function and will in general consist of several jumps in the genus of the fiber along with the possibility of the fiber being a disjoint union of Riemann surfaces. To eliminate certain pathologies, we will assume that this Morse function saturates the Morse inequalities from now on, and our choice of three-manifolds will force the fiber to always be connected.

As a warmup let us take our three-manifold to be an $S^3$. If we then cut out two $S^2$'s this is topologically $S^2\times \mathbb{R}_{\bot}$, so the fibration structure in this case is clear. If we instead cut out two tori, then the fibers of the fibration will jump in the following manner along $\mathbb{R}_{\bot}$:
\begin{equation}
  g=1 \quad \vert \quad g=0 \quad \vert \quad g=1.
\end{equation}
To generate thickened 3D interfaces, we will actually be interested in situations where the genus is bigger in the interior. The reason is that as a rule of thumb, compactifications of 6D SCFTs on higher genus spaces tend to produce 4D theories with more degrees of freedom. With this in mind, the typical situation of interest will be:
\begin{equation}
  g_L \quad \vert \quad g_{\text{mid}} \quad \vert \quad g_R, \,\,\, \text{with} \,\,\, g_{L},g_{R} < g_{\text{mid}}.
\end{equation}

Focusing on the case where the genus increases inside the interface, we accomplish this by cutting out Riemann surfaces with genera $g_{L,R}$ out of the suspension\footnote{Given a topological space X, the suspension is defined as $\Sigma X \coloneqq X\times [0,1]/\{ (x,0)\sim (y,0) \; \text{and} \; (x,1)\sim (y,1)\}$. This has the important property that $\Sigma S^2 \simeq_{Top.} S^3$ and we note that while normally $\Sigma$ is called the reduced suspension by mathematicians, we favor this symbol here for aesthetic purposes.} of a Riemann surface $\Sigma C_{g_{\mathrm{mid}}}$ such that $g_L,g_R < g_{\text{mid}}$. The 3D theory living on the interface can be equivalently studied as either the compactification of a 6D SCFT on $\Sigma C_{g_\mathrm{mid}} \; \; \textnormal{with lower genus ``punctures''}$ or (from the fibration point-of-view) as the compactification of the 4D theory associated to $C_{g_{\text{mid}}}$ on an interval with appropriate boundary conditions.

\begin{figure}[t!]
\centering
\includegraphics[width=0.24\textwidth]{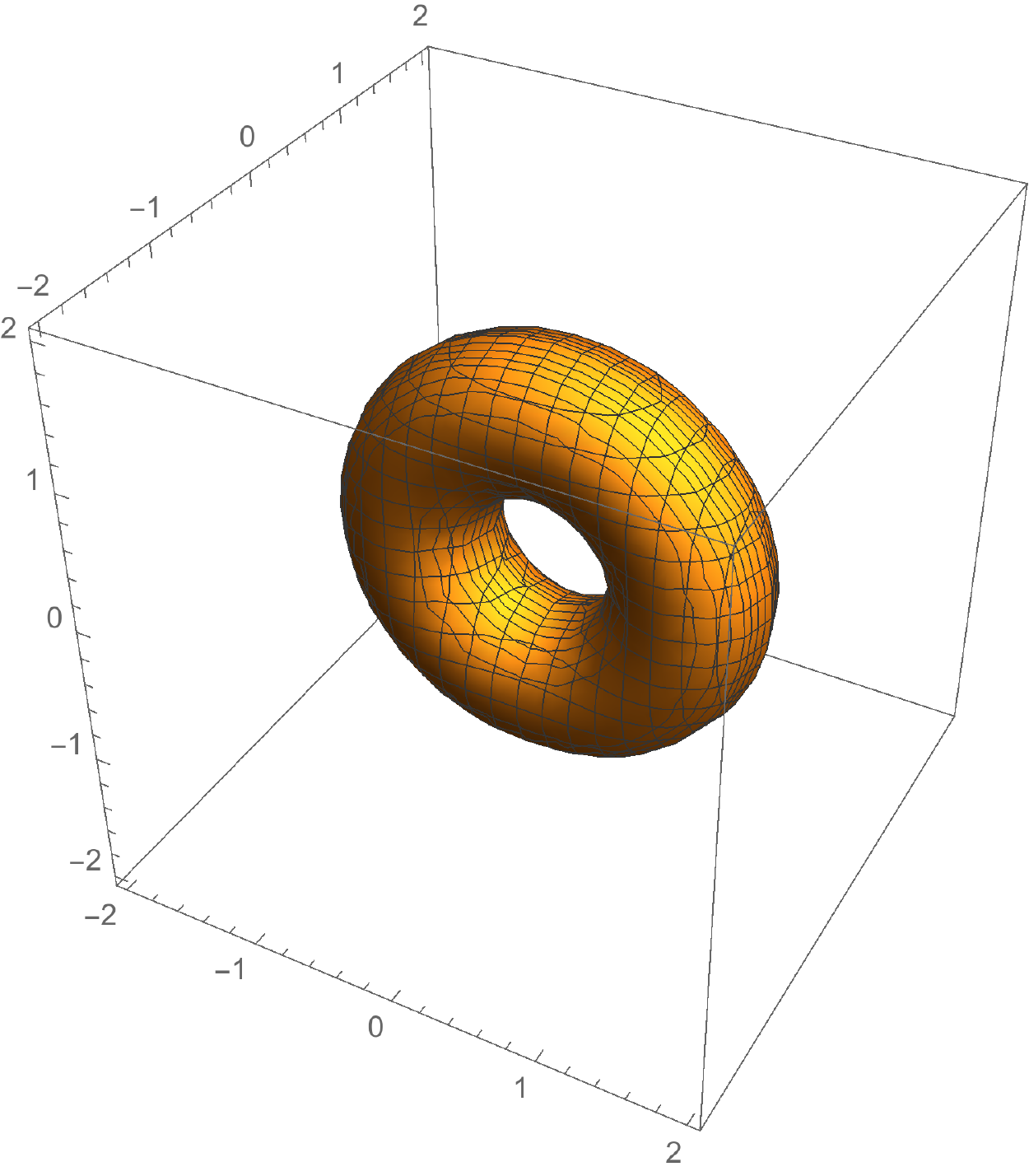}
\includegraphics[width=0.24\textwidth]{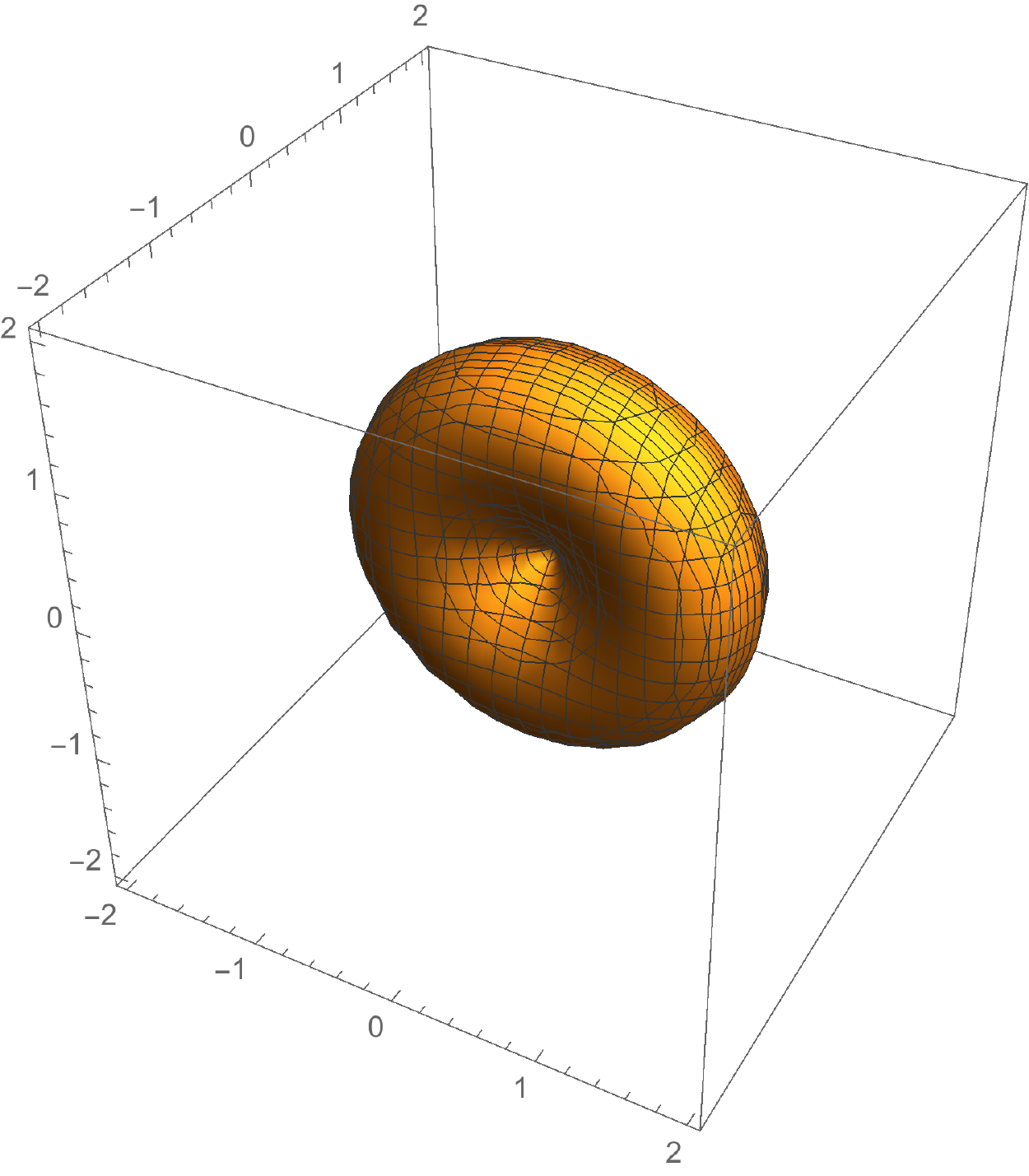}
\includegraphics[width=0.24\textwidth]{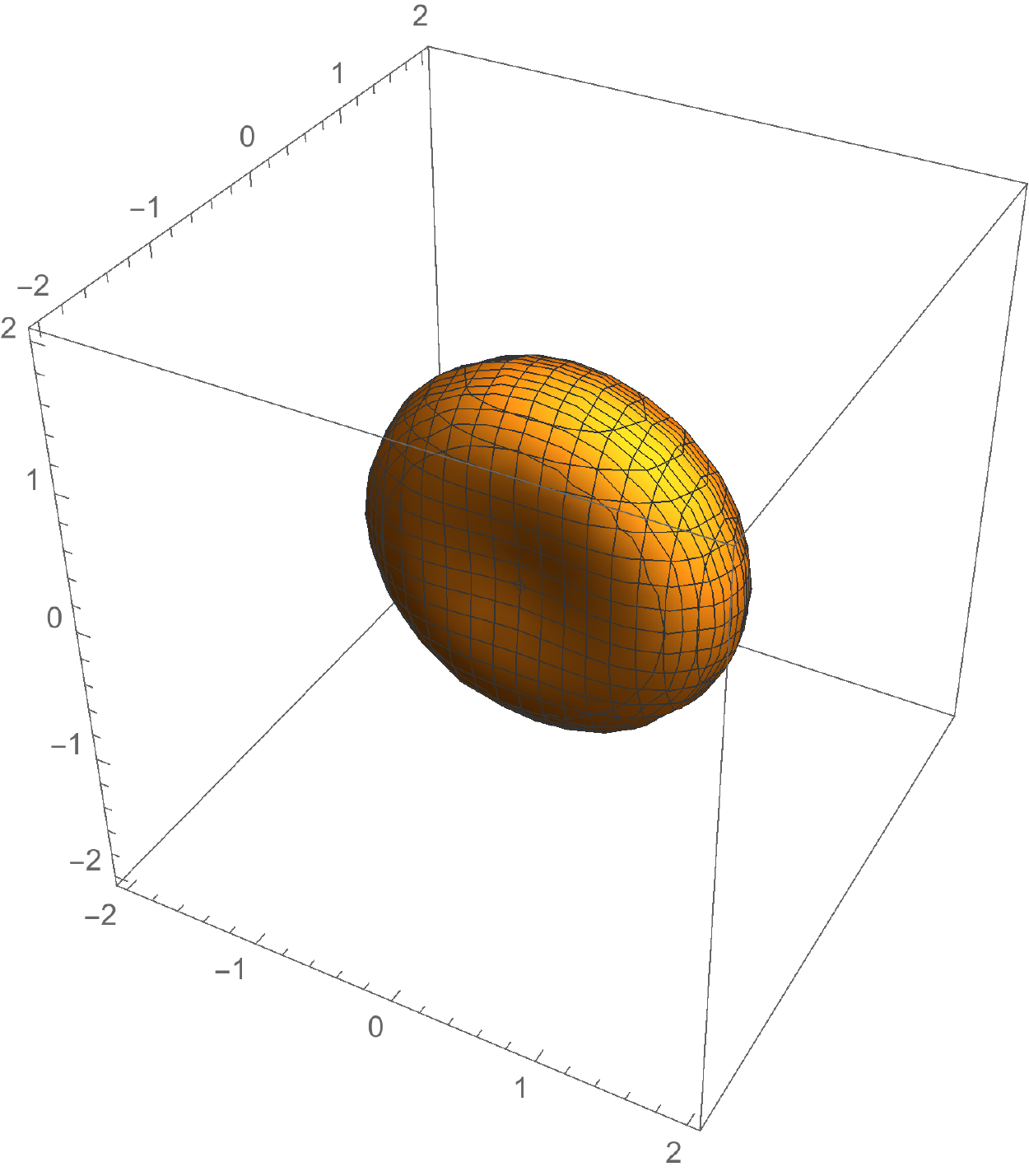}
\includegraphics[width=0.24\textwidth]{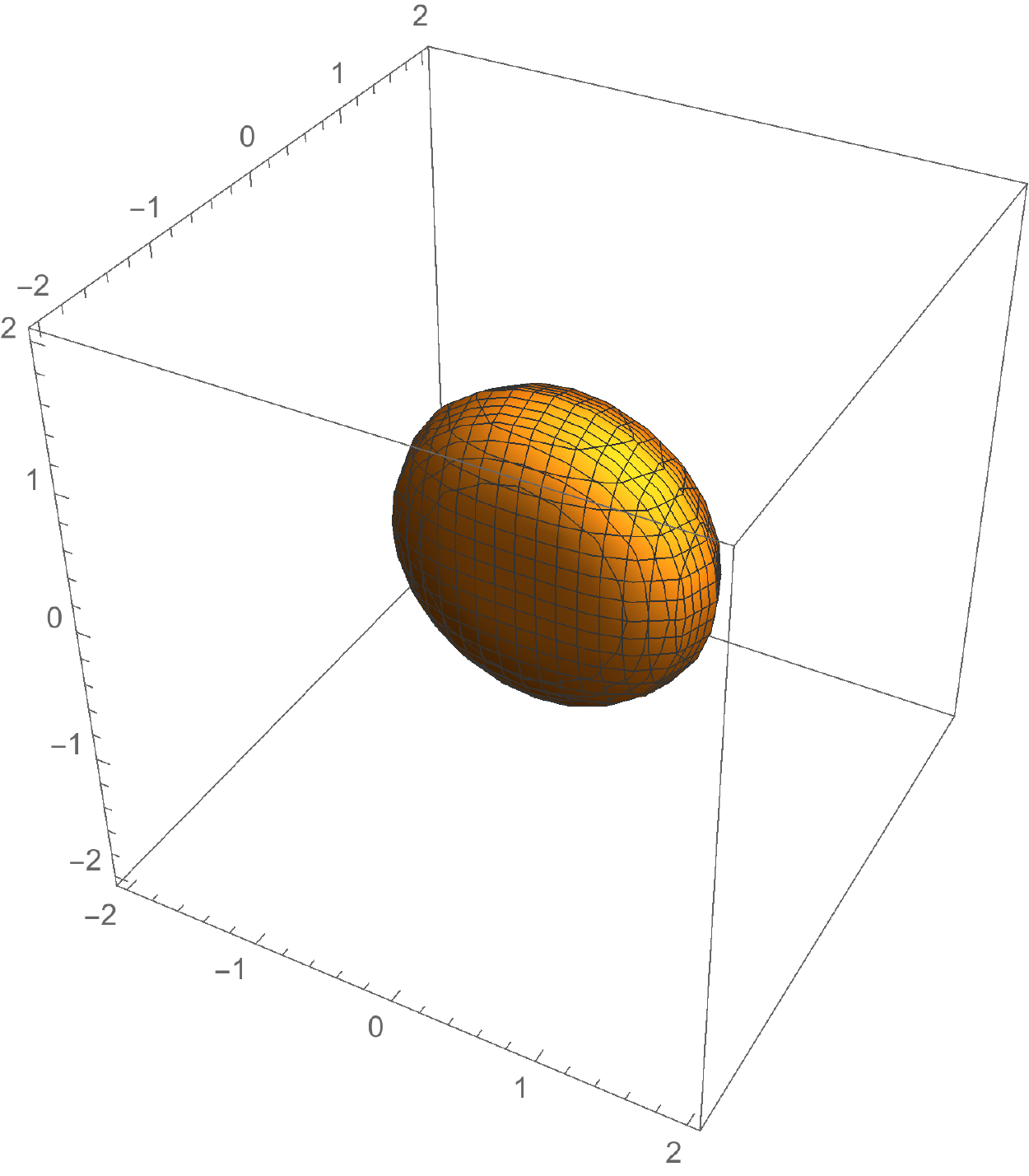}
\caption{Building a continuous family of Riemann surfaces with varying genus: we start on the left with a torus, which then fattens into a sphere. This construction can be extended to build more general interpolating profiles.}
\label{fig:interpolator}
\end{figure}

As an example of this parametrization of Riemann surfaces, we can define a family of tori given with parametrization variable $w$ as:
\begin{equation}
  Q = \{ (x^2+y^2+z^2+w^2+R^2-r^2)^2=4R^2(x^2+y^2+w^2) \},
\end{equation}\label{eq:interpolator}
where for $w=0$, $R$ and $r$ are the ``major'' and ``minor'' radii of the torus respectively. We then vary the parameter between $0$ and $R$, noting that at $w=R$ the Riemann surface described now turns into a two-sphere. This is illustrated in figure \ref{fig:interpolator} where we see a torus transform into a sphere as the parameter $w$ increases from $0$ to $R$. As a result, by compactifying on $Q$ with $w$ starting at $w<R$ (in the middle), and reaching $w=R$ as $|x_\bot|\rightarrow \infty$, we obtain families of 4D theories compactified on genus zero surfaces on the left and right, but compactified on a genus one surface in the middle, thus realizing two $S^2$'s cut out of $\Sigma T^2$. Note that once we transition to a genus zero Riemann surface, we can then consider further motion in the moduli space $\overline{\mathcal{M}}_{0,n}$. We can use this to also rotate the phases of ``mass parameters'' on the two sides of the thickened interface. Note that we can also extend this construction to produce interpolating profiles between different genus Riemann surfaces.

We can also consider interpolating profiles for flavor symmetry fluxes. The possibilities for the background gauge field that couples to the flavor current are: a non-trivial monodromy, a flux for an abelian portion, or a 't Hooft flux for a non-simply connected flavor group. We can build an interface that interpolates between any two pairs of monodromies since for the cobordism $Q=\Sigma C_g \backslash (C^R_{g_R}\sqcup C^L_{g_L})$, one is free to chose the monodromy around the cycles. Note also that these interfaces allow for the added possibility of monodromy associated only to the $\Sigma C_g$ cycles and not to either $C^R_{g_R}$ or $C^L_{g_L}$. For the flux cases, the relevant cobordism groups to look at are:
\begin{align}
 \Omega^{SO}_2(BU(1)) =\mathbb{Z} \; \; \; \; &(\textnormal{abelian flux})\\
 \Omega^{SO}_2(BG) = \pi_1 (G) \; \; \; \; &(\textnormal{'t Hooft flux})
\end{align}
where $G$ is the flavor group in question, and $BG$ denotes its classifying space. These express total abelian flavor and 't Hooft charge conversation and follow from an application of Stokes' theorem (along with the universal coefficient theorem for the 't Hooft case) to the cobordism with the assumptions $dF_{U(1)}=0$ and $\delta F_{\text{'t Hooft}}=0\in H^3(Q,\pi_1(G))$ (where here $\delta$ is the coboundary operator).

One can study more general interfaces by adding extra codimension-three defect operators with localized flux in the cobordism leading to the relation:
\begin{align}
\underset{C^L}{\int} c_1(F) & = \underset{C^R}{\int} c_1(F) + \text{monopoles} \\
\underset{C^L}{\int} w_2(F) & = \underset{C^R}{\int} w_2(F) + \text{twists}
\end{align}
where ``monopoles'' and ``twists'' refers to pointlike singular field configurations in the three-manifold.

\subsection{Hypermultiplet Example}

With these general considerations in place, we now turn to a concrete example of 6D hypermultiplets which, when suitably compactified, produces a 4D theory with a thickened 3D interface. This 6D theory arises from the theory of a single M5-brane probing an A-type singularity $\mathbb{C}^{2} / \mathbb{Z}_{k}$. Strictly speaking, this does not produce an interacting fixed point, but it will be adequate for the main ideas we wish to consider. In field theory terms, we have a theory of hypermultiplets in the bifundamental representation of $SU(k) \times SU(k)$.\footnote{The actual flavor symmetry in this case is $SU(2k)$} We will be interested in building an interpolating profile with modes trapped along a 3D interface. We review the case of a position dependent mass term for a Weyl fermion in Appendix \ref{app:WEYL}.

To begin, we consider the compactification of this theory on a genus $g$ Riemann surface $C$. We also consider switching on abelian fluxes in a subgroup $H \subset SU(2k)$ of the flavor symmetry. For ease of exposition, we concentrate on the case of a single $U(1)$ factor, and consider the mass spectrum for states of charge $\pm q$ under this $U(1)$ factor. We leave implicit the representation content under the commutant flavor symmetry. Letting $\mathcal{L}$ denote the line bundle associated with switching on this background flux, the zero mode content on the curve consists of 4D $\mathcal{N} = 1$ chiral multiplets of charge $+q$ and $-q$ under this $U(1)$. The 6D fermion obeys a Dirac equation of the form:
\begin{equation}
\Gamma_{6D} \cdot D_{6D} \Psi_{6D} = 0.
\end{equation}
We expand the 6D fermion in terms of a basis of 4D Weyl fermions and chiral modes on the curve $C$ via:
\begin{equation}
\Psi_{6D} = \sum_{a} \psi^{(a)}_{4D} \otimes \chi^{(a)}_{C}.
\end{equation}
The Dirac equation then takes the form:
\begin{equation}
(\gamma_{4D} \cdot D_{4D} + \gamma_{C} \cdot D_{C}) \sum_{a} \psi^{(a)}_{4D} \otimes \chi^{(a)}_{C} = 0.
\end{equation}
Consequently, the Dirac operator on $C$ controls the spectrum of zero modes and massive modes in the theory. More precisely,
in the expansion of $(\gamma_{C} \cdot D_{C})^{2}$, we see the appearance of the curvature in the spin connection and the gauge field flux.

The number of zero modes is controlled by the cohomology groups (see e.g. \cite{Beasley:2008dc}):
\begin{align}
\#_{+q} & = h^{0}(C, K_{C}^{1/2} \otimes \mathcal{L}^{+q})\\
\#_{-q} & = h^{0}(C, K_{C}^{1/2} \otimes \mathcal{L}^{-q}).
\end{align}
where here, $K_{C}$ denotes the canonical bundle and we need to specify a choice of spin structure, i.e. a choice of square root for $K_{C}$.

As an example, we can engineer a theory with no zero modes by considering the special case of $C$ a $\mathbb{CP}^1$ with $\mathcal{L} = \mathcal{O}$. We can view this as a situation in which all the modes of the 6D hypermultiplet have a Kaluza-Klein scale mass. As an example where we get a single chiral multiplet, we could consider switching on $\mathcal{L} = \mathcal{O}(1)$ on a $\mathbb{CP}^{1}$,
which includes a 4D Weyl fermion and a complex scalar, both of charge $+q$. Finally, we can also produce an example with a 4D Dirac fermion and its superpartners by compactifying on a $T^{2}$, with no fluxes switched on.

\subsection{Strongly Coupled Examples}

We now generalize the above considerations to consider compactifications of 6D SCFTs on three-manifolds with boundary. Our primary
interest will be in localizing states along a thickened 3D interface. To track the appearance of localized degrees of freedom, we consider the 4D anomaly polynomial obtained from compactification of a 6D theory on a curve $C$ with some background fluxes switched on. Recall that the general
form of the anomaly polynomial for a 6D SCFT takes the form:
\begin{align}
I_{8}  &  = \alpha c_{2}(R)^{2} + \beta c_{2}(R) p_{1}(T) + \gamma
p_{1}(T)^{2} + \delta p_{2}(T)\nonumber\\
&  + \sum_{i} \left[  \mu_{i} \, \mathrm{Tr} F_{i}^{4}
+ \, \mathrm{Tr} F_{i}^{2} \left(  \rho_{i}
p_{1}(T) + \sigma_{i} c_{2}(R) + \sum_{j} \eta_{ij} \, \mathrm{Tr} F_{j}^{2}
\right)  \right]  . \label{eq:anomalypoly}%
\end{align}
Here, $c_{2}(R)$ is the second Chern class of the $SU(2)_{R}$ symmetry,
$p_{1}(T)$ is the first Pontryagin class of the tangent bundle, $p_{2}(T)$ is
the second Pontryagin class of the tangent bundle, and $F_{i}$ is the field
strength of the $i$th symmetry, where the sum on $i$ and $j$ runs over the
global symmetries of the theory. In the case where we have sufficiently generic curvatures switched on, we can extract the anomalies of the 4D theory which are inherited from six dimensions by integrating this formal eight-form over a curve $C$ (see e.g. \cite{Benini:2009mz, Razamat:2016dpl, Apruzzi:2018oge}):
\begin{equation}
I_{6} = \underset{C}{\int} I_{8}.
\end{equation}
This, in tandem with $a$-maximization \cite{Intriligator:2003jj} makes it possible to extract the values of the conformal anomalies $a$ and $c$ (see e.g. \cite{Razamat:2016dpl, Apruzzi:2018oge}), which provides a crude ``count'' of the number of degrees of freedom in the 4D theory.

To generate examples of trapped matter, we can attempt to mimic our discussion of the 6D hypermultiplet. In particular, we can engineer examples where the anomalies split up as:
\begin{align}
  a_L \quad & \vert \quad a_{\text{mid}} \quad \vert \quad a_R, \,\,\, \text{with} \,\,\, a_{L},a_{R} < a_{\text{mid}}\\
  c_L \quad & \vert \quad c_{\text{mid}} \quad \vert \quad c_R, \,\,\, \text{with} \,\,\, c_{L},c_{R} < c_{\text{mid}}.
\end{align}
Of course, the anomalies provide only partial information on the structure of localized states, so a priori, it could happen that in each region, there are massless states present which are missing from the other regions. Though we cannot prove it in full generality, we expect that regions with higher $a$ and $c$ are typically the places which have more states as is expected by RG flow.

To illustrate this, consider the case of 6D SCFTs as generated by M5-branes probing an ADE singularity \cite{DelZotto:2014hpa}. In reference \cite{Ohmori:2014kda} the 6D anomalies for these theories were computed, and the anomalies of the 4D theories resulting from compactification were computed in \cite{Razamat:2016dpl, Apruzzi:2018oge}. For example, from compactification on a curve of genus $g \geq 1$ and in the absence of flavor symmetry fluxes, the values of $a$ and $c$ are both proportional to $(g-1)$. In the case of compactification on a genus one curve, one instead gets a 4D $\mathcal{N} = 2$ theory , and in the case of a genus zero curve (with no punctures), the resulting 4D system produces a trivial fixed point \cite{Apruzzi:2018oge}. When fluxes are switched on, the central charges become algebraic numbers, as determined by $a$-maximization. The general feature of $a$ and $c$ increasing with genus still holds in these cases \cite{Razamat:2016dpl, Apruzzi:2018oge}.

\subsection{Generating Thin Interfaces}

The construction we have provided generates a thickened 3D interface. This is simply because the ``middle region'' can also be thought of as compactification of a 6D theory on a Riemann surface which is then further compactified on an finite length interval. In the limit where the size of this interval collapses to zero size, this leads to a thin interface. What we would like to understand is whether the resulting construction still produces localized states.

Returning to the example of the 6D hypermultiplet, we can see some potential issues with such a procedure. For example, in the case of a 4D Dirac fermion with a position dependent mass, the appearance of a localized state in the thin wall limit relies on having a sign flip in the mass term, relating to the two time-reversal invariant values of $\theta$ at weak coupling. From the perspective of our compactification of a 6D anti-chiral two-form, this involves a bordism between two elliptic curves with different values of the complex structure moduli. In the example of a 6D hypermultiplet, we can arrange something similar since the spin connection and gauge field connection implicitly depend on the complex structure of the compactification curve. Working with curves with real coefficients, we can again enforce the appearance of a sign flip in the mass spectrum of Kaluza-Klein modes, thus ensuring that the trapped states ``in the middle'' do not disappear in the zero thickness limit. The same logic also applies in more general compactifications of 6D SCFTs. One reason is that a large number of such examples can be interpreted as 4D $\mathcal{N} = 1$ theories in which marginal couplings have been formally tuned to extremely large values \cite{Razamat:2019vfd}. From this perspective, we can impose a further condition that we restrict to time-reversal invariant values of these marginal couplings, thus providing a way to ``protect'' localized states in this more general setting.

\newpage

\section{Conclusions} \label{sec:CONC7}

Interfaces generated by position dependent couplings provide a general way to access non-perturbative structure in quantum field theories.
In this chapter we have investigated 3D interfaces generated from 4D theories at strong coupling. In the case of 4D $U(1)$ gauge theories we showed that the appearance of a finite index duality group $\Gamma \subset SL(2,\mathbb{Z})$, in tandem with the condition of time-reversal invariance leads to a rich phase structure for possible interfaces, as captured by the real component of a modular curve $X(\Gamma)_{\mathbb{R}}$. We have also seen that a more general starting point based on compactifications of 6D SCFTs on three-manifolds with boundary leads to a broad class of thickened 3D interfaces with states trapped in an interior region. In the remainder of this section we discuss some avenues for future investigation.

Throughout this chapter we have operated under the assumption that time-reversal invariance is preserved by the system, even as we vary
the parameters of the theory. Of course, this is not always the case, and in some cases there is good evidence that time-reversal invariance
is actually spontaneously broken (see for example \cite{Gaiotto:2017tne}). Given the strong constraints on the real component of a modular curve, it would be interesting to study these assumptions in more detail.

One of the outcomes of our analysis is the prediction that in some $U(1)$ gauge theories with duality group $\Gamma \subset SL(2,\mathbb{Z})$,
there are 3D interfaces which are inherently at strong coupling, namely, the resulting parameters are on a different component of $X(\Gamma)_{\mathbb{R}}$ from the one connected to the point of weak coupling. As a further generalization, it is natural to ask whether quantum transitions between these different phases could be activated by adding small time-reversal breaking couplings to the system. Calculating these transition rates would be very interesting in its own right, and would likely shed additional light on the non-perturbative structure of such theories.

The geometry of modular curves also suggests additional ways in which strong coupling phenomena may enter such setups. For example, for suitable duality groups, the modular curve $X(\Gamma)$ can have genus $g > 0$. This in turn means that there are one-cycles which can be traversed by a motion through parameter space. Compactifying our 4D theory on a circle, a non-zero winding number in moving through such a one-cycle of $X(\Gamma)$ suggests another way to produce features protected by topology.

It is also interesting to ask whether coupling such systems to gravity imposes any restrictions. At least in the context of F-theory constructions, there appear to be sharp constraints on the possible torsional structures which can be realized in UV complete models, see e.g.\ \cite{Aspinwall:1998xj, Hajouji:2019vxs}. More generally, Swampland type considerations suggest the possible existence of a sharp upper bound on the genus of the associated modular curves (perhaps they are always genus zero). Determining such bounds would be quite illuminating.

From a mathematical point of view, our study of the real components of the modular curve $X(\Gamma)$ has centered on a particular notion of conjugation given by $\tau \mapsto - \overline{\tau}$, which has a clear physical interpretation in terms of time-reversal. On the other hand, reference \cite{snowden2011real} considers another conjugation operation given by $\tau \mapsto 1 / \overline{\tau}$, and this choice also leads to a rather rich set of conjugation invariant components of the modular curve. This can be thought of
as the composition of time-reversal conjugation with an S-duality transformation. It would be very interesting to develop a physical interpretation of this case as well.

Much of our analysis has focused on the special case of 4D $U(1)$ gauge theory. When additional $U(1)$'s are present, there is again a
fundamental domain of possible couplings as swept out by a congruent subgroup of $Sp(2r, \mathbb{Z})$ acting on the Siegel upper half-space.
In this case, less is known about the analog of modular curves, let alone their real components, but it would nevertheless be interesting to study the phase structure of cusps in this setting.

The main thrust of our analysis has focused on formal aspects of 3D interfaces in 4D systems. One could envision applying these insights to specific concrete condensed matter systems. Additionally, in cases with additional $U(1)$ factors, one might consider scenarios in which a visible sector $U(1)$ kinetically mixes with a dark $U(1)$. The phenomenology of axionic domain walls leads to a rather rich set of signatures \cite{Sikivie:1984yz}, so it would be interesting to investigate the related class of questions for axionic domain walls charged under one of these hidden $U(1)$ factors.

Our analysis was inspired by string compactification considerations, though we have mainly focused on field-theory
considerations. In a related development, M-theory on non-compact $Spin(7)$ backgrounds can sometimes be interpreted as generating interpolating profiles between 4D M- and F-theory vacua \cite{Cvetic:2020piw}. It would be very interesting to study time-reversal invariant configurations engineered from this starting point.

In the same vein, we note that some of the techniques considered use supersymmetry only sparingly. It is therefore tempting to ask whether these considerations could be used to build non-supersymmetric brane configurations which are protected by topological structures. We leave an analysis of this exciting possibility for future work.

\titleformat{\chapter}[hang]{\large\center}{\thechapter}{0 pt}{}
\chapter*{CONCLUSIONS}
\addcontentsline{toc}{chapter}{CONCLUSIONS} 

%
%

In conclusion, there exists a of lot structure in quantum field theories and string theory. Through small deformations and RG flows we have observed intricate hierarchies which physically align well with known mathematical results. There are also many symmetries which add to the beauty of string theory. These symmetries have led to the discovery and understanding of new theories, as made evident for instance by the use of string junctions, orientifolds, and S-folds. 

In fact, part I of this thesis has established that there are still many symmetries and much structure to be explored even at strong coupling. For example, it is the geometrical aspects of string theory and quantum field theory that helped us establish the full structure of the nilpotent cone for various quiver theories in chapter \ref{chapter2}. By linking T-brane deformations of CFTs to nilpotent orbits of flavor symmetry algebras we were able to learn more about the networks of field theory fixed points and RG flows. It is also the geometry of S-folds and string junctions that allowed us to gain further intuition into non-perturbative effects of 4D and 6D SCFTs. In chapter \ref{chapter3}, this intuition led us to conjecture a prescription for how to define F-theory in the presence of S-folds even when there is discrete torsion. 

It would be rather interesting to see how those same principals could be applied to theories in other dimensions, or with different amount of supersymmetry. For instance, \cite{Kimura:2020hgw} recently explored T-brane deformations of 4D $\N=2$ SCFTs on S-folds, giving us further insight into 4D $\N=1$ SCFTs. A full understanding of all possible 4D $\N=1$ SCFTs could have profound consequences in phenomenology. In $\N=2$ theories, our analysis of S-folds has focused on rank one theories, but there are of course higher rank theories to consider. We should note that \cite{Martone:2021ixp}, began mapping rank two $\N=2$ SCFTs in four dimensions but the current catalogs reveals several gaps in our understanding. Of note is the existence of 4D SCFTs which do not yet have a stringy interpretation. Current techniques on S-folds and string junctions could fill this gap, and we leave an analysis of this possibility for future work.

While many symmetries of quantum field theories have been well explored, other symmetries still remain poorly understood. Here we have made some progress in better understanding one of them: Poisson-Lie T-duality. Its abelian counterpart is particularly fundamental to string theory. Thus it is important to figure out how much of a role Poisson-Lie T-duality has to play. Part II of this work has established that it is clearly a duality between conformal field theories. We have also established some of the ground work to show that it might be in fact a full duality of string theory. Yet there is no clear path to reaching an understanding of quantum corrections to PL T-duality in the string coupling $g_\mathrm{s}$, a point that definitely needs further investigating. For future considerations, there are however some attempts at non-perturbative generalizations of Poisson T-duality. Namely, U-duality which combines T and S dualities in M-theory. Some recent works have explored upgrading Poisson-Lie duality to the context of M-theory, using exceptional field theory. This is done by replacing the structure of Drinfeld doubles by an exceptional Drinfeld algebra. Hopefully, the rich geometrical structure found in Poisson-Lie T-duality, and further developed in chapters \ref{chapter4} and \ref{chapter5}, will lead to new insights into the mathematical tools necessary to develop a complete understanding of U-duality.

Ultimately, the goal of physics is to describe the inner workings of our universe. This work has also aimed especially in that direction by exploring dualities between systems with direct phenomenological relevance. The duality between F-theory on Calabi-Yau fourfolds, and M-theory on $G_2$ spaces could have very concrete applications. For instance, F-theory has many applications for dark matter as well as cosmology. Thus one could try to use $Spin(7)$ manifolds and the unification of Higgs bundle vacua we established to push known results of F-theory to the M-theory side. Moreover, many three-dimensional interfaces, either generated through special holonomy manifolds or time-reversal invariance conditions, are likely to have applications in condensed matter systems. For instance, a specific proposal for realizing QED-like systems at strong coupling was discussed in \cite{Pace:2020jiv} in the specific context of spin ice systems. This would provide an ideal setting for implementing a further study of the strong coupling phenomena indicated in chapter \ref{chapter7}. 

While string theory is yet to have made any testable predictions, it clearly has a lot of potential as a ``theory of everything". Its rich geometric structure and mathematical puzzles make it an interesting area of research in and of itself. Moreover, as we learn more about the intrinsic details of string theory and QFTs it becomes clear that any hope for a good description of our universe will require a better understanding of strong coupling effects between interacting strings. By continuing to study QFTs and strings, especially in strong coupling regimes, maybe some day we will be able to test whether or not string theory unifies all four fundamental forces in a way that is consistent with our reality.

\end{mainf}
\appendix
\newenvironment{appendixf}{}{}
\titleformat{\chapter}[hang]{\large\center}{\thechapter}{6 pt}{} 
\titleformat{\chapter}[hang]{\large\center}{CHAPTER \thechapter}{0 pt}{: }
\titlespacing*{\chapter}{0pt}{-29 pt}{6 pt} 
\begin{appendixf}

\addtocontents{toc}{\protect\setcounter{tocdepth}{1}} 
\clearpage
\chapter{Chapter 1 Appendix}
\addtocontents{toc}{\protect\setcounter{tocdepth}{-1}} 

\section{The Embedding Index \label{app:EMBED}}

\label{index} The embedding index $r$ here refers to that of a splitting of
the group $G=D_{4}$, or $E_{6,7,8}$ into irreducible representations (irreps) of $\SU(2)$.
There are two equivalent ways of computing this embedding index $r$. The first method is by
computing the sum of the indices of the $\SU(2)$ irreps divided by the index
of the representation of the group $G$ being split. That is, given a
representation $\rho(G)$ of $G$ and the branching $\rho(G) \rightarrow m_{1}
\mathbf{n}_{1} + m_{2} \mathbf{n}_{2} + \dots$ where $m_{(a)}$ are
multiplicities and $\mathbf{n}_{(a)}$ are $\SU(2)$ irreps, the embedding index
is given by:
\begin{align}
\label{rindex1}r  & = \frac{\sum_{(a)} m_{(a)} \cdot\mathrm{ind}%
(\mathbf{n}_{(a)})}{\mathrm{ind}(\rho(G))}.
\end{align}
For instance the splitting of $D_{4}$ according to the partition $[5,3]$ gives:
$\mathbf{28} \rightarrow3(\mathbf{3})+(\mathbf{5})+2(\mathbf{7})$ so that
\begin{align}
r=\frac{3\times4+ 20 +2 \times56}{12} = 12
\end{align}

As we can see, this definition of the embedding index is representation
independent. However it requires that we know the branching rule of splitting
of $G$ to $\SU(2)$ caused by the deformation of interest.

For this reason, we turn to the second method which makes use of the decorated Dynkin diagrams provided in
 \cite{Chacaltana:2012zy} for the exceptional groups.
Their labels specify a vector $v$ in the Cartan subalgebra which then yields
the projection matrix $\mathbb{P} = v \cdot C_{\mathfrak{g}}^{-1}$. $C_{\mathfrak{g}}$
is the Cartan matrix of the Lie algebra ${\mathfrak{g}}$, and $\mathbb{P}$ is the
projection matrix of the weights of ${\mathfrak{g}}$ into the $SU(2)_{D}$ nilpotent subalgebra.
As a result the decorated Dynkin diagrams can be directly used to obtain the branching rules and the embedding indices,
\begin{equation} \label{rindex}
r  = \frac{1}{2} \, \mathrm{Tr}(v\cdot C_{\mathfrak{g}}^{-1} \cdot v^{T})
\end{equation}
where the $\frac{1}{2}$ coefficient is simply a normalization factor.

Now, for $D_{4}$ we do not have the decorated Dynkin diagrams readily
available to us, so we need to compute them. We start with the 12 possible
partitions of $\SO(8)$ provided by \cite{rakotoarisoa2017bala}. Following
this procedure along with \cite{Haouzi:2016yyg} one can obtain the vectors
$v$ for $\SO(2k)$ in the same form as the ones provided by
\cite{Chacaltana:2012zy} for the exceptional groups. In summary the procedure
is as follows:

We begin by listing the possible partitions of $\SO(2k)$: $p_{i} = \{n_{l}\}$
where $i$ runs over the number of possible nilpotent deformations of $\SO(2k)$
and $n_{l}$ are integers summing to $2k$. The nilpotent deformation defines an
$\SU(2)$ subalgebra $[H,X]=2X$, $[H,X^{\dagger}]=-2X^{\dagger}$,
$[X,X^{\dagger}]=H$ where $X$ is the nilpotent orbit/deformation. $X$ is
directly constructed from the partitions: $X$ is a $2k \times2k$ matrix filled
on the first superdiagonal by the Jordan blocks corresponding to the $\SU(2)$
irreps defined by the partitions. Namely $\sqrt{j(j+1)-m(m+1)}$ where $-j \leq
m \leq j-1$. For instance, the $\SO(10)$ partition $\{7,3\}$ yields two Jordan
blocks. $X$ is zero everywhere except on the first super diagonal which is
given by the list $(\sqrt{6},\sqrt{10},\sqrt{12},\sqrt{12},\sqrt{10},\sqrt
{6},0,\sqrt{2},\sqrt{2})$ where for the first block (which defines the first 6
entries) we have $j=3$ and for the second block (which defines the last 2
entries) we have $j=1$.

Then the corresponding Cartan matrix $H$ is given by $[X,X^{\dagger}]=H$,
which is a diagonal matrix whose entries are then sorted in increasing order.
Furthermore, $\SO(2k)$ has $k$ Cartan matrices $H_{q}$ with $q=1,\cdots,k$.
The projection matrix (or just vector here) is $\alpha= \{\alpha_{i}\}$ given
by solving the linear equations:
\begin{align}
\label{H}\sum_{i=1}^{k} \alpha_{i} H_{i} = H
\end{align}
and the decorated Dynkin diagrams are given by the vector $v = \alpha\cdot C_{\SO(2k)}$.
Each partition yields a different $H$ and therefore a different set of
equations \eqref{H} and Dynkin labels $v$.

We should note that this analysis makes extensive use of the \texttt{LieArt} 
package of reference \cite{Feger:2012bs}.

\subsection*{$SO(8)$ Example}

To illustrate we work out an example with $\SO(8)$ in detail:

One partition of $\SO(8)$ is given by $[5,3]$. So the raising operator matrix
is:
\begin{align}
X =
\begin{pmatrix}
0 & 2 & 0 & 0 & 0 & 0 & 0 & 0\\
0 & 0 & \sqrt{6} & 0 & 0 & 0 & 0 & 0\\
0 & 0 & 0 & \sqrt{6} & 0 & 0 & 0 & 0\\
0 & 0 & 0 & 0 & 2 & 0 & 0 & 0\\
0 & 0 & 0 & 0 & 0 & 0 & 0 & 0\\
0 & 0 & 0 & 0 & 0 & 0 & \sqrt{2} & 0\\
0 & 0 & 0 & 0 & 0 & 0 & 0 & \sqrt{2}\\
0 & 0 & 0 & 0 & 0 & 0 & 0 & 0
\end{pmatrix}
\end{align}
and the corresponding Cartan matrix $H=[X,X^{\dagger}] = \text{diag}%
(4,2,2,0,0,-2,-2,-4)$ after sorting out the entries.

The 4 Cartans of $\SO(8)$ are given by:
\begin{align}
\label{cartansSO8}H_{1} & =\text{diag}(1,-1,0,0,0,0,1,-1)\\
H_{2} & =\text{diag}(0,1,-1,0,0,1,-1,0)\\
H_{3} & =\text{diag}(0,0,1,-1,1,-1,0,0)\\
H_{4} & =\text{diag}(0,0,1,1,-1,-1,0,0)
\end{align}
where we are using the mathematician's conventions to be consistent with the
use of the \texttt{LieArt} package.

The projection matrix $\alpha=(\alpha_{1}, \alpha_{2}, \alpha_{3}, \alpha
_{4})$ is then obtained by solving the equation:
\begin{align}
\label{projectionsolve}\alpha_{1} H_{1} + \alpha_{2} H_{2} + \alpha_{3} H_{3}
+ \alpha_{4} H_{4} = H
\end{align}
which yields:
\begin{align}
\label{projection}\alpha= (4,6,4,4).
\end{align}
Thus given the Cartan matrix:
\begin{align}
\label{CartanMatrix}C_{\SO(8)} =
\begin{pmatrix}
2 & -1 & 0 & 0\\
-1 & 2 & -1 & -1\\
0 & -1 & 2 & 0\\
0 & -1 & 0 & 2
\end{pmatrix}
\end{align}
the decorated Dynkin diagram specifies a vector $v = \alpha\cdot C_{\SO(8)}$ given by:
\begin{align}
\label{label}v = (2,0,2,2)
\end{align}

\begin{figure}
\centering
\includegraphics[scale=1]{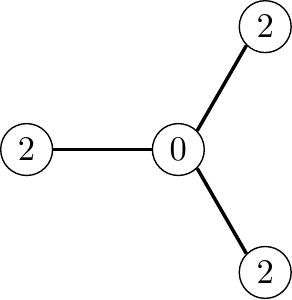}
\caption{Decorated Dynkin diagram for the $[5,3]$ partition of $\SO(8)$ }%
\label{DynkinLabel}%
\end{figure}

This procedure is repeated for every partition of $SO(2k)$ so as to obtain all
of the necessary decorated Dynkin diagrams and projection matrices.


\section{From 6D to 4D Conformal Matter \label{CMappendix}}

In this Appendix we collect some features of 6D conformal matter and its compactification on a $T^2$.
At long distances, this yields a 4D $\mathcal{N} = 2$ SCFT. Here, we review both the scaling dimensions
of Coulomb branch operators and the anomalies of these theories.

\subsection*{Coulomb Branch Operators}

In this subsection we calculate the scaling dimension of the operators parameterizing the Coulomb branch.
This data follows directly from the analysis of references \cite{Ohmori:2015pua, DelZotto:2015rca, Ohmori:2015pia}.
Our main task here is to extract from this analysis the corresponding scaling dimensions. References \cite{Ohmori:2015pua,
Ohmori:2015pia} implicitly give this information by showing that 4D $\mathcal{N} = 2$ $(G,G)$ conformal matter is actually a
compactification of a class $\mathcal{S}$ theory, specifying the corresponding Gaiotto curve as well. In reference \cite{DelZotto:2015rca}
the corresponding Seiberg-Witten curve is obtained via the mirror to the elliptically fibered Calabi-Yau threefold of
the F-theory background used to produce the 6D SCFT. Observe that F-theory compactified on a $T^2$ yields IIA on the same elliptic threefold,
and mirror symmetry takes us to type IIB. The advantage of the IIB presentation is that now the Coulomb branch is parameterized in
terms of the complex structure of this mirror geometry.

We opt to use the explicit Calabi-Yau geometries presented in reference \cite{DelZotto:2015rca}.
To aid comparison with the results of this reference, we refer to the theory of 6D conformal
matter with $(G,G)$ flavor symmetry given by $N$ M5-branes probing an ADE singularity as $\mathcal{T}(G,N)$.
In this chapter we focus exclusively on the case $N = 1$.

We now use the results of reference \cite{DelZotto:2015rca} on the associated
mirror geometries to compute  the scaling dimensions of the Coulomb branch for the
theories $\mathcal T(E_{6,7,8},1)$, on $T^2$. This method has been used before
for $\mathcal{N} = 2$ SCFTs, and is essentially adapted from the technique presented
in reference \cite{Argyres:1995xn}.

The IIB mirror geometry for $\mathcal T(E_{6},1)$ on $T^2$ is given by the following local Calabi-Yau threefold:
\be
 \begin{cases}
 &f=w^2+x_1^3+x_2^2\rho+\rho^2+(m_1+m_1^{'} y_1 )x_1x_2^2+(m_2+m_2^{'} y_1) x_1x_2 \nonumber \\ 
 & ~~~+ (m_3 +u_1 y_1+m_3^{'}y_1^2) x_2^2
 + (m_4+u_2 y_1+m_4^{'} y_1^2) x_1+(m_5+u_3 y_1+m_5^{'}y_1^2) x_2 \nonumber \\
 & ~~~+(m_6+u_4y_1+u_5y_1^2+m_6^{'}y_1^3)=0  \nonumber\\
& \rho=(1+y_1+y_2),\\
& x^2_2=\rho.
\end{cases}
\label{eq:E6mirrorcurve}
\ee
where $y_1$ is a $\mathbb C^*$ coordinate, $x_1,x_2,w,\rho$ are complex coordinates, $m_i$ are general mass parameters and $u_i$ are the coulomb branch operator vevs,
\begin{equation}
u_i \equiv \langle Z_i \rangle,
\end{equation}
$f$ is a homogeneous polynomial in the complex coordinates and it scales as follows:
\begin{equation}
f(\lambda^ax_1,\lambda^b x_2, \lambda^c \rho, \lambda^d w,y_1)= \lambda^e f(x_1,x_2,\rho,w,y_1).
\end{equation}
The holomorphic three-form is defined as follows
\begin{equation}
\Omega = \frac{dx_1 \wedge dx_2}{w}\wedge \frac{dy_1}{y_1}
\end{equation}
By fixing the scale of $\Omega(\lambda^ax_1,\lambda^b x_2, \lambda^c \rho, \lambda^d w,y_1) =\lambda \Omega(x_1,x_2,\rho,w,y_1)$ to the unity, i.e. $[\Omega]=1$, the first four monomials of $f$ uniquely fix the other scalings
\begin{equation}
[x_1]=a=4, \quad [x_2]=b=3, \quad [\rho]=c=[w]=d=6, \quad [f]=e=12.
\end{equation}
Recalling that $y_1$ does not scale since it is just a phase, we obtain the scaling dimension of the Coulomb branch parameters,
\begin{equation}
[u_1]=6, \quad [u_2]=8, \quad [u_3]=9, \quad [u_4]=[u_5]=12.
\end{equation}
This agrees with the scaling dimensions of the Coulomb branch operators for the class $\mathcal{S}$
trinions with two minimal and one maximal puncture in \cite{Chacaltana:2014jba}.

The IIB mirror Calabi-Yau for $\mathcal T(E_{7},1)$ on $T^2$ is described by
\be
 \begin{cases}
 &f = x_1^2+x_2^3\rho+\rho^3+(m_1+m_1^{'}y_1) x_2\rho^2+(m_2+u_1y_1+m_2^{'} y_1^2)\rho^2 + \nonumber\\
& ~~(m_3+u_2y_1+m_3^{'} y_1^2) x_2\rho +(m_4+u_3 y_1+m_4^{'} y_1^2)x_2^2 +(m_7+u_4 y_1+u_5y_1^2+m_4^{'} y_1^3)\rho +\nonumber\\
&(m_6+u_6 y_1+u_7y_1^2+m_5^{'} y_1^3)x_2 +(m_7+u_8y_1+u_9y_1^2+u_{10}y_1^3+m_7^{'}y_1^4)=0.   \nonumber\\
&\rho=(1+y_1+y_2).
 \end{cases}
 \ee
where again $y_1$ is a $\mathbb C^*$ coordinate, and $x_1,x_2,\rho$ are complex coordinates.
The homogeneous polynomial $f$ scales as follows:
\begin{equation} \label{eq:homo}
f(\lambda^ax_1,\lambda^b x_2, \lambda^c \rho,y_1)= \lambda^e f(x_1,x_2,y_1).
\end{equation}
The holomorphic three-form reads
\begin{equation} \label{eq:holo3form}
\Omega = \frac{dx_2 \wedge d\rho}{x_1}\wedge \frac{dy_1}{y_1}
\end{equation}
and we impose that it scales like $[\Omega]=1$. The first three monomials again fix the scaling of the complex coordinates and of $f$:
\begin{equation}
[x_1]=a=9, \quad [x_2]=b=4, \quad [\rho]=c=6, \quad [f]=e=18.
\end{equation}
By looking at the scaling of the other monomials involving the Coulomb branch vevs, the scaling dimensions of the Coulomb branch parameters are assigned
\begin{align}
& [u_1]=6, \quad [u_2]=8, \quad [u_3]=10, \quad [u_4]=[u_5]=12, \nonumber \\
& \quad [u_6]=[u_7]=14, \quad [u_8]=[u_9]=[u_{10}]=18.
\end{align}
This agrees with the scaling dimensions of the Coulomb branch operators for the class $\mathcal{S}$
trinions with two minimal and one maximal puncture in \cite{Chacaltana:2017boe}.

The IIB mirror Calabi-Yau for $\mathcal T(E_{8},1)$ on $T^2$ is described by
 \be
 \begin{cases}
 &f = x_1^2+x_2^3+\rho^5+(m_1+m_1^{'}y_1) x_2\rho^3+(uy_1^2)\rho^4+(m_2+u_1y_1+m_2^{'} y_1^2)x_2\rho^2+ \nonumber\\
& (m_3+u_2y_1+u_3y_1^2+m_3^{'} y_1^3) \rho^3 +(m_4+u_4 y_1+u_5y_1^2+m_4^{'} y_1^3) x_2\rho+\nonumber\\
&(m_5+u_6 y_1+u_7y_1^2+u_8y_1^3+m_5^{'} y_1^4) \rho^2+(m_6+u_9y_1+u_{10}y_1^2+u_{11}y_1^3+m_6^{'}y_1^4)x_2+ \nonumber\\
&(m_7+u_{12}y_1+u_{13}y_1^2+u_{14}y_1^3+u_{15}y_1^4+m_7^{'}y_1^5)\rho+  \nonumber\\
&((m_8+u_{16}y_1+u_{17}y_1^2+u_{18}y_1^3+u_{19}y_1^4+u_{20}y_1^5+m_8^{'}y_1^6)=0; \nonumber\\
&\rho=(1+y_1+y_2).
 \end{cases}
 \ee
where again $y_1$ is a $\mathbb C^*$ coordinate, and the $x_1,x_2,\rho$ are complex
coordinates. The homogeneous polynomial $f$ scales as in equation \eqref{eq:homo}.
The holomorphic three-form is analogous to the $E_7$ case, \eqref{eq:holo3form}.
By imposing $[\Omega]=1$, the first three monomials of $f$ fix the scaling of the coordinates,
\begin{equation}
[x_1]=a=15, \quad [x_2]=b=10, \quad [\rho]=c=6, \quad [f]=e=30.
\end{equation}
The other monomials involving the Coulomb branch vevs automatically assign the following scaling dimensions
\begin{align}
& [u]=6, \quad [u_1]=8, \quad [u_2]=[u_3]=12,\quad [u_4]=[u_5]=14, \quad [u_6]=[u_7]=[u_8]=18, \nonumber \\
&[u_9]=[u_{10}]=[u_{11}]=20, \quad [u_{12}]=[u_{13}]=[u_{14}]=[u_{15}]=24, \nonumber  \\
&[u_{16}]=[u_{17}]=[u_{18}]=[u_{19}]=[u_{20}]=30.
\end{align}
This agrees with the scaling dimensions of the Coulomb branch operators for the class $\mathcal{S}$ trinions with two minimal and one maximal puncture in \cite{Chcaltana:2018zag}.

Finally, for the $D_{k}$ conformal matter theories $\mathcal T(SO(2k),1)$ with $k>2$ on $T^2$ the scaling dimensions of the
Coulomb branch operators can be read off in a similar way from the curve (5.4) in \cite{Ohmori:2015pua}.

\subsection*{Anomaly Polynomials}

Given the importance of the UV anomalies we now review
how they were obtained in table \ref{CMUVscaling}. When studying an M5-brane
probing $D$- and $E$-type singularities we obtain 6D SCFTs also called ($G$,$G$)
6D conformal matter with anomaly polynomial:
\begin{align}
\label{eqn:I8CM}I_{8}=&\alpha c_{2}(R_{6D})^{2}+\beta c_{2} (R_{6D}%
)p_{1}(T)+\gamma p_{1}(T)^{2}+\delta p_{2}(T)+\kappa_{L} p_{1}(T) \frac{\,
\mathrm{Tr}(F_{L}^{2})}{4} \nonumber \\
& +\kappa_{R} p_{1}(T) \frac{\, \mathrm{Tr}(F_{R}%
^{2})}{4} + \dots
\end{align}
where the explicit expression for the 6D anomaly polynomial coefficients were
computed in \cite{Ohmori:2014kda}, and are listed in table~\ref{anoPolyCoeff}.
\begin{table}[h]
\centering
{\renewcommand{\arraystretch}{1.2}
\begin{tabular}
[c]{|c||c|c|c|c|}\hline
$(G,G)$ & $(D_k,D_k)$ & $(E_{6},E_{6})$ & $(E_{7},E_{7})$ & $(E_{8},E_{8})$\\\hline
$24 \alpha$ & $10k^{2}-57k+81$ & $319$ & $1670$ & $12489$\\\hline
$48 \beta$ & $-(2k^{2}-3k-9)$ & $-89$ & $-250$ & $-831$\\\hline
$\frac{5760}{7} \gamma$ & $k(2k-1)+1$ & $79$ & $134$ & $249$\\\hline
$\frac{5760}{4} \delta$ & $-\left( k(2k-1)+1\right) $ & $-79$ & $-134$ &
$-249$\\\hline
$24\kappa_{L}=24\kappa_{R} $ & $2k-2$ & $12$ & $18$ & $30$\\\hline
\end{tabular}
}\caption{Coefficients of 6D anomaly polynomial \eqref{eqn:I8CM}}%
\label{anoPolyCoeff}%
\end{table}

In order to obtain a 4D $\mathcal{N}=2$ SCFT, we compactify these theories on
$T^{2}$ and consider the general anomaly polynomial for a 4D theory
\begin{align}
\label{eqn:I6}I_{6} =& \frac{k_{RRR}}{6}c_{1}(R)^{3}-\frac{k_{R}}{24}%
p_{1}(T)c_{1}(R)+k_{RG_{L}G_{L}}\frac{\, \mathrm{Tr}(F^{2}_{G_{L}})}{4}%
c_{1}(R) \nonumber \\
&+k_{RG_{R}G_{R}}\frac{\, \mathrm{Tr}(F^{2}_{G_{R}})}{4}c_{1}(R)+
\dots\,,
\end{align}
where $R=R_{\mathrm{UV}}$ is the R-symmetry of the UV $\mathcal{N} = 2$ SCFT,
viewed as an $\mathcal{N} =1$ SCFT, $T$ is the
formal tangent bundle, $F$ is the field strength of $G_{L}$ or $G_{R}$ flavor
symmetries, and the dots indicate possible abelian flavor symmetries and mixed
contributions. Moreover we have the following relations
\begin{align}
\, \mathrm{Tr}(R^{3})=k_{RRR}, &  & \, \mathrm{Tr}(R)=k_{R}, &  & \,
\mathrm{Tr}(RF^{A}_{G_{L,R}}F^{B}_{G_{L,R}})=-\frac{k_{RG_{L,R}G_{L,R}}}%
{2}\delta^{AB} \,. &
\end{align}
From them, the definition of $R$, and
\begin{align}
\, \mathrm{Tr}\left( R_{\mathcal{N}=2}F^{A}_{G_{L,R}}F^{B}_{G_{L,R}}\right)
=-\frac{k_{L,R}}{2} \delta^{AB}%
\end{align}
we read off the anomalies
\begin{align}
a_{\mathrm{UV}}  & = \frac{9}{32}k_{RRR}-\frac{3}{32}k_{R}\\
c_{\mathrm{UV}}  & = \frac{9}{32}k_{RRR}-\frac{5}{32}k_{R}\\
k_{L}  & = 3k_{RG_{L}G_{L}}\\
k_{R}  & = 3k_{RG_{R}G_{R}}\,.
\end{align}
In terms of the 6D anomaly polynomial coefficients
\cite{Ohmori:2015pua,Ohmori:2015pia}, we finally identify
\begin{align}
a_{\mathrm{UV}}  & = 24\gamma-12\beta-18\delta\\
c_{\mathrm{UV}}  & = 64\gamma-12\beta-8\delta\\
k_{L}  & = 48\kappa_{L}\\
k_{R}  & = 48\kappa_{R}\,.
\end{align}
Once evaluated at the values of table \ref{anoPolyCoeff} the above equations
yield exactly the UV values of table \ref{CMUVscaling}, as expected.


\section{Accessing the Complete Tables \label{completeTables}}

Included with the \texttt{arXiv }submission of \cite{Apruzzi:2018xkw} is a set of \texttt{Mathematica}
scripts which can be used to access the full set of theories generated by
nilpotent deformations of the $\mathcal{N}=2$ theories considered in this
chapter. Indeed, due to the rather large size of the dataset it is impractical to list all
of our results in the format of a paper.

Instead we have written a \texttt{Mathematica} script which outputs the
complete list of all possible nilpotent deformations for the theories
described above. The necessary files are attached to \cite{Apruzzi:2018xkw}. To access
them, first proceed to the \texttt{arXiv} abstract of \cite{Apruzzi:2018xkw}. On the
right-hand side, there is a box with the title \textquotedblleft
Download.\textquotedblright\ Click on \textquotedblleft Other
formats\textquotedblright\ and then download the source files for the
\texttt{arXiv} submission.

\sloppy To access the full database, one simply needs to download the following six
files and store them in the same folder: \textquotedblleft
ProbeD3brane.m\textquotedblright, \textquotedblleft
ConformalMatter.m\textquotedblright, \textquotedblleft
ProbeD3braneFlavorK.m\textquotedblright, \textquotedblleft
ConformalMatterFlavorK.m\textquotedblright, \textquotedblleft
NilpotentDeformations.m\textquotedblright, \textquotedblleft
Results.nb\textquotedblright. Essentially, the first file contains all of the
information for nilpotent deformations of the probe D3-brane theories (with and without
flipper field deformations), except for the flavor central charge. The second
file stores all of the information for the nilpotent deformations of 4D conformal matter (with
and without flipper field deformations), except for the flavor central
charge. The next two files contain all of the information about the flavor
central charges for the Minahan-Nemeshansky and conformal matter theories
respectively. The file \textquotedblleft
NilpotentDeformations.m\textquotedblright\ does all of the formatting, and
finally the code \textquotedblleft Results.nb\textquotedblright\ loads the
previous three packages and outputs the results. Thus the only file the user
needs to run and worry about is the last one: \textquotedblleft
Results.nb\textquotedblright. When running this file the user is provided
with a list of options:

\begin{enumerate}
\item First one can choose between the four kinds of deformations:
probe D3-brane theories with plain mass deformations, probe D3-brane theories with
flipper field deformations, 4D conformal matter with plain mass deformations,
and 4D conformal matter with flipper field deformations.

\item Secondly one can choose between the $a_{\mathrm{IR}}$, $c_{\mathrm{IR}}$
anomalies and operator scaling dimensions or the tables with the flavor central charges.

\item Then the user should select the flavor groups: $D_{4}$, $E_{6}$, $E_{7}%
$, or $E_{8}$ for deformations of the probe D3-brane theories, and $(D_{4},D_{4})$,
$(E_{6},E_{6})$, $(E_{7},E_{7})$, or $(E_{8},E_{8})$ for deformations of 4D conformal matter.

\item If a probe D3-brane theory is selected then the user can
choose from two options:

\begin{enumerate}

\item select a single deformation by choosing the Bala-Carter label (or
partition of $D_{4}$) of the flavor group from the provided popup menu below.

\item select the whole table.
\end{enumerate}

\item If instead a 4D conformal matter theory is selected the user has three
options:

\begin{enumerate}

\item select a single deformation chosen by selecting the left and right
Bala-Carter labels (or partitions of $D_{4}$) for the breaking of the left and
right flavors.

\item select all of the deformations with a given left (or right) deformation,
by selecting a single Bala-Carter label (or partition of $D_{4}$).

\item select the whole table.
\end{enumerate}

\item The resulting table is then outputted. We also provide for the probe D3-brane theories
the branching rules from the adjoint of $G$ to the $\SU(2)$ irreps for the selected deformations.
\end{enumerate}

Finally, due to the form of the general equations used to compute the central
charges it is clear that all of our results are algebraic numbers. However not all are
rational. To differentiate the two in the tables we list the rational values
exactly (by keeping their rational form) while we only give numerical values
for the ones with irrational central charges.

For the convenience of the reader, in the following
subsections we list the explicit tables for all of the
nilpotent deformations of the probe D3-brane theory with $SO(8)$ flavor symmetry,
but only the rational theories for the other nilpotent networks.

As a point of notation, here we make reference to $K_{\mathrm{IR}}$ as well as $k_{\mathrm{IR}}$. 

\subsection*{Nilpotent Network for $SU(2)$ with Four Flavors}

\begin{table}[H]
  \centering
  \begin{adjustbox}{center}
\scalebox{1.0}{{\renewcommand{\arraystretch}{1.2} $\begin{array}{|c|c|c|c|c|c|c|}
 \hline
  \text{[B-C]} & \text{r} & a_{\text{IR}} & c_{\text{IR}} & t_* & \Delta _{\text{IR}}\text{(Z)} & \text{Min(}\Delta _{\text{IR}}\text{($\cO$'s))} \\
  \hline
  \left[1^8\right] & 0       & \frac{23}{24} & \frac{7}{6} &  \frac{2}{3} & \numprint{2.} & \numprint{2.} \\
  \left[2^2,1^4\right]       & 1  & \numprint{0.797038} & \numprint{0.955454} &  \numprint{0.506932} & \numprint{1.5208} & \numprint{1.4792}\\
  \left.\text{[3,}1^5\right] & 2  & \numprint{0.710272} & \numprint{0.846192} &  \numprint{0.434945} & \numprint{1.30483} & \numprint{1.69517} \\
  \left[2^4\text{]II}\right. & 2  & \numprint{0.710272} & \numprint{0.846192} &  \numprint{0.434945} & \numprint{1.30483} & \numprint{1.69517} \\
  \left[2^4\text{]I}\right.  & 2  & \numprint{0.710272} & \numprint{0.846192} &  \numprint{0.434945} & \numprint{1.30483} & \numprint{1.69517} \\
  \text{[3,}2^2\text{,1]}    & 3  & \numprint{0.651529} & \numprint{0.773372} &  \numprint{0.389898} & \numprint{1.16969} & \numprint{1.53788} \\
  \left[3^2,1^2\right]       & 4  & \numprint{0.607635} & \numprint{0.71946} &  \numprint{0.357838} & \numprint{1.07351} & \numprint{1.38973} \\
  \left[4^2\text{]I}\right.  & 10 & \{\numprint{0.452668},\numprint{0.473501}\} & \{\numprint{0.498618},\numprint{0.540284}\} &  \numprint{0.247886} & \numprint{1.} & \numprint{1.51269} \\
  \left[4^2\text{]II}\right. & 10 & \{\numprint{0.452668},\numprint{0.473501}\} & \{\numprint{0.498618},\numprint{0.540284}\} &  \numprint{0.247886} & \numprint{1.} & \numprint{1.51269} \\
  \left.\text{[5,}1^3\right] & 10 & \{\numprint{0.452668},\numprint{0.473501}\} & \{\numprint{0.498618},\numprint{0.540284}\} &  \numprint{0.247886} & \numprint{1.} & \numprint{1.51269} \\
  \text{[5,3]}               & 12 & \{\numprint{0.430022},\numprint{0.450856}\} & \{\numprint{0.467234},\numprint{0.508901}\} &  \numprint{0.227913} & \numprint{1.} & \numprint{1.63252} \\
  \text{[7,1]}               & 28 & \{\numprint{0.345121},\numprint{0.365954}\} & \{\numprint{0.348767},\numprint{0.390434}\} &  \numprint{0.151192} & \numprint{1.} & \numprint{1.63928} \\
 \hline
    \end{array}$}}
\end{adjustbox}

  \begin{adjustbox}{center}
\scalebox{0.9}{{\renewcommand{\arraystretch}{1.2} $\begin{array}{|c|c|c|c|}
 \hline
  \text{[B-C]} & \text{SU(2})_{D}\text{$\times $Residual}  & k_{\text{IR}}\text{ interact} & k_{\text{IR}}\text{+free} \\
  \hline
  \text{[$1^8$]} & \text{SO(8)}  & 4 & 4 \\
 \text{[$2^2,1^4$]} & \text{SU(2)}\times \text{SO(4)}\times \text{SU(2)} & \{\numprint{3.04159},\numprint{3.04159}\} &  \{\numprint{3.04159},\numprint{3.04159}\} \\
 \text{[$3,1^5$]} & \text{SU(2)}\times \text{SO(5)}  & \{\numprint{2.60967}\} & \{\numprint{2.60967}\} \\
 \text{[$2^4$]II} & \text{SU(2)}\times \text{Sp(4)}  & \{\numprint{2.60967}\} & \{\numprint{2.60967}\} \\
 \text{[$2^4$]I} & \text{SU(2)}\times \text{Sp(4)}  & \{\numprint{2.60967}\} & \{\numprint{2.60967}\} \\
 \text{[$3,2^2,1$]} & \text{SU(2)}\times \text{SU(2)}  & \{\numprint{2.33939}\} & \{\numprint{2.33939}\} \\
 \text{[$3^2,1^2$]} & \text{SU(2)}\times \text{U(1)}\times \text{U(1)}  & \{\numprint{3.22054},\numprint{1.07351}\} &  \{\numprint{3.22054},\numprint{1.07351}\} \\
 \text{[$5,1^3$]} & \text{SU(2)}\times \text{SU(2)} & \{\numprint{2.97463}\} & \{\numprint{2.97463}\} \\
 \text{[$4^2$]II} & \text{SU(2)}\times \text{SU(2)} & \{\numprint{2.97463}\} & \{\numprint{2.97463}\} \\
 \text{[$4^2$]I} & \text{SU(2)}\times \text{SU(2)}  & \{\numprint{2.97463}\} & \{\numprint{2.97463}\} \\
 \text{[$5,3$]} & \text{SU(2)}  & \{\} & \{\} \\
 \text{[$7,1$]} & \text{SU(2)}  & \{\} & \{\} \\
 \hline
    \end{array}$}}
\end{adjustbox}
  \caption{Plain nilpotent deformations of the probe D3-brane theory with $D_4$ flavor symmetry.
  The top table has the central charges $a_{\mathrm{IR}}$ and $c_{\mathrm{IR}}$ as well as scaling dimensions while the table below contains the information about the flavor central charges.}
  \label{MND4}
\end{table}

\begin{table}[H]
\centering
\begin{adjustbox}{center}
\scalebox{0.9}{{\renewcommand{\arraystretch}{1.2} $\begin{array}{|c|c|c|c|c|c|c|c|}
\hline
\text{[B-C]} & \text{r} & a_{\text{IR}} & c_{\text{IR}} &  t_* & \Delta _{\text{IR}}\text{(Z)} & \text{Min(}\Delta _{\text{IR}}\text{($\cO$'s))}\\
\hline
\left[1^8\right] & 0 & \left\{\frac{23}{24},\frac{37}{24}\right\} & \left\{\frac{7}{6},\frac{7}{3}\right\}  & \frac{2}{3} & \numprint{2.} & \numprint{2.} \\
\left[2^2,1^4\right] & 1       & \{ \numprint{0.962469},\numprint{1.3583}\} & \{ \numprint{1.26671},\numprint{2.05838}\} & \numprint{0.459126} & \numprint{1.37738} & \numprint{1.62262} \\
\left.\text{[3,}1^5\right] & 2 & \{ \numprint{0.80915}, \numprint{1.26748}\} & \{ \numprint{1.0197},\numprint{1.93637}\}  & \numprint{0.37588} &  \numprint{1.12764} & \numprint{1.87236} \\
\left[2^4\text{]II}\right. & 2 & \{ \numprint{0.80915}, \numprint{1.26748}\} & \{ \numprint{1.0197},\numprint{1.93637}\}  & \numprint{0.37588} &  \numprint{1.12764} & \numprint{1.87236} \\
\left[2^4\text{]I}\right. & 2  & \{ \numprint{0.80915}, \numprint{1.26748}\} & \{ \numprint{1.0197},\numprint{1.93637}\}  & \numprint{0.37588} &  \numprint{1.12764} & \numprint{1.87236} \\
\text{[3,}2^2\text{,1]} & 3    & \{ \numprint{0.727701},\numprint{1.20687}\} & \{ \numprint{0.911122},\numprint{1.86946}\}& \numprint{0.344209} & \numprint{1.03263} & \numprint{1.70921} \\
\rowcolor{LightCyan} \left[3^2,1^2\right] & 4 & \left\{\frac{7}{12},\frac{7}{6}\right\} & \left\{\frac{2}{3},\frac{11}{6}\right\} &  \frac{1}{3} & \numprint{1.} & \numprint{1.5} \\
\rowcolor{LightCyan} \left.\text{[5,}1^3\right] & 10 & \left\{\frac{11}{24},\frac{25}{24}\right\} & \left\{\frac{1}{2},\frac{5}{3}\right\} & \frac{2}{9} & \numprint{1.} & \numprint{1.66667} \\
\rowcolor{LightCyan} \left[4^2\text{]II}\right. & 10 & \left\{\frac{11}{24},\frac{25}{24}\right\} & \left\{\frac{1}{2},\frac{5}{3}\right\} & \frac{2}{9} & \numprint{1.} & \numprint{1.66667} \\
\rowcolor{LightCyan} \left[4^2\text{]I}\right. & 10 & \left\{\frac{11}{24},\frac{25}{24}\right\} & \left\{\frac{1}{2},\frac{5}{3}\right\} & \frac{2}{9} & \numprint{1.} & \numprint{1.66667} \\
\text{[5,3]} & 12 & \left\{\frac{6349}{13872},\frac{1769}{1734}\right\} & \left\{\frac{3523}{6936},\frac{5663}{3468}\right\} &  \frac{10}{51} & \numprint{1.} & \numprint{1.82353} \\
\rowcolor{LightCyan} \text{[7,1]} & 28 & \left\{\frac{43}{120},\frac{113}{120}\right\} & \left\{\frac{11}{30},\frac{23}{15}\right\} &  \frac{2}{15} & \numprint{1.} & \numprint{1.8} \\
\hline
\end{array}$}}
\end{adjustbox}
\begin{adjustbox}{center}
\scalebox{0.9}{{\renewcommand{\arraystretch}{1.2} $\begin{array}{|c|c|c|c|}
\hline
\text{[B-C]} & \text{SU(2})_{D}\text{$\times $Residual} & k_{\text{IR}}\text{ interact} & k_{\text{IR}}\text{+free} \\
\hline
\text{[$1^8$]} & \text{SO(8)}  & 4 & 16 \\
\text{[$2^2,1^4$]} & \text{SU(2)}\times \text{SO(4)}\times \text{SU(2)} &  \{\numprint{6.49049},\numprint{6.49049}\} &\{\numprint{10.4905},\numprint{10.4905}\} \\
\text{[$3,1^5$]} & \text{SU(2)}\times \text{SO(5)} & \{\numprint{3.74472}\} & \{\numprint{9.74472}\} \\
\text{[$2^4$]II} & \text{SU(2)}\times \text{Sp(4)} & \{\numprint{3.74472}\} & \{\numprint{9.74472}\} \\
\text{[$2^4$]I} & \text{SU(2)}\times \text{Sp(4)} & \{\numprint{3.74472}\} & \{\numprint{9.74472}\} \\
\text{[$3,2^2,1$]} & \text{SU(2)}\times \text{SU(2)}  & \{\numprint{2.48369}\} & \{\numprint{8.48369}\} \\
\text{[$3^2,1^2$]} & \text{SU(2)}\times \text{U(1)}\times \text{U(1)} & \{3,1\} & \{9,3\} \\
\text{[$5,1^3$]} & \text{SU(2)}\times \text{SU(2)}  & \left\{\frac{8}{3}\right\} &\left\{\frac{32}{3}\right\} \\
\text{[$4^2$]II} & \text{SU(2)}\times \text{SU(2)}  & \left\{\frac{8}{3}\right\} &\left\{\frac{32}{3}\right\} \\
\text{[$4^2$]I} & \text{SU(2)}\times \text{SU(2)}  & \left\{\frac{8}{3}\right\} &\left\{\frac{32}{3}\right\} \\
\text{[$5,3$]} & \text{SU(2)}  & \{\} & \{\} \\
\text{[$7,1$]} & \text{SU(2)}  & \{\} & \{\} \\
\hline
\end{array}$}}
\end{adjustbox}
\caption{Flipper field deformations of the probe D3-brane theory with $D_{4}$ flavor. The
top table has the central charges $a_{\mathrm{IR}}$ and $c_{\mathrm{IR}}$ as well
as scaling dimensions while the table below contains the information about the
flavor central charges. The cyan highlighted entries align with the $H_0$, $H_1$ and $H_2$ Argyres-Douglas theories,
as first noted in \cite{Maruyoshi:2016tqk, Maruyoshi:2016aim}. The other rational entry with partition [5,3] also aligns with \cite{Agarwal:2016pjo}}%
\label{MNMSD4}%
\end{table}

\subsection*{Tables of Rational Theories: Minahan-Nemeschansky Theories}

\begin{table}[H]
  \centering
  \begin{adjustbox}{center}
\scalebox{1}{\renewcommand{\arraystretch}{1.2} $\begin{array}{|c|c|c|c|c|c|c|}
 \hline
  \text{[B-C]} & \text{r} & a_{\text{IR}} & c_{\text{IR}} &  t_* & \Delta _{\text{IR}}\text{(Z)} & \text{Min(}\Delta _{\text{IR}}\text{($\cO$'s))} \\
  \hline
   0 & 0 & \frac{41}{24} & \frac{13}{6} &  \frac{2}{3} & \numprint{3.} & \numprint{2.} \\
   A_2+2A_1 & 6 & \frac{97}{96} & \frac{119}{96} &  \frac{1}{3} & \numprint{1.5} & \numprint{1.5} \\
 \hline
  \end{array}$}
\end{adjustbox}

  \begin{adjustbox}{center}
\scalebox{1}{{\renewcommand{\arraystretch}{1.2} $\begin{array}{|c|c|c|c|}
 \hline
  \text{[B-C]} & \text{SU(2})_{D}\text{$\times $Residual} &  k_{\text{IR}}\text{ interact} & k_{\text{IR}}\text{+free} \\
  \hline
  0 & E_6  & 6 & 6 \\
  \text{$A_2+2A_1$} & \text{SU(2)}\times \text{SU(2)}\times \text{U(1)} & \{18,18\} & \{18,18\} \\
  \hline
    \end{array}$}}
\end{adjustbox}
\caption{Plain nilpotent mass deformations of the Minahan-Nemeschansky theory with $E_6$ flavor. The top table has the central charges $a_{\mathrm{IR}}$ and $c_{\mathrm{IR}}$ as well as scaling dimensions while the table below contains the information about the flavor central charges.}
\label{MNE6}
\end{table}

\begin{table}[H]
  \centering
  \begin{adjustbox}{center}
\scalebox{1}{\renewcommand{\arraystretch}{1.2} $\begin{array}{|c|c|c|c|c|c|c|}
 \hline
  \text{[B-C]} & \text{r} & a_{\text{IR}} & c_{\text{IR}} &  t_* & \Delta _{\text{IR}}\text{(Z)} & \text{Min(}\Delta _{\text{IR}}\text{($\cO$'s))} \\
  \hline
    0 & 0 & \frac{59}{24} & \frac{19}{6} &  \frac{2}{3} & \numprint{4.} & \numprint{2.} \\
    A_1 & 1 & \frac{158}{75} & \frac{401}{150} &  \frac{8}{15} & \numprint{3.2} & \numprint{1.4} \\
    A_2+3A_1 & 7 & \frac{7150}{5043} & \frac{17785}{10086} &  \frac{40}{123} & \numprint{1.95122} & \numprint{1.53659} \\
    A_4+A_2 & 24 & \frac{478}{507} & \frac{1177}{1014} & \frac{8}{39} & \numprint{1.23077} & \numprint{1.46154} \\
    \text{($A_5$)'} & 35 & \frac{7075}{8664} & \frac{4345}{4332} &  \frac{10}{57} & \numprint{1.05263} & \numprint{1.42105} \\
    \text{($A_5$)$\texttt{"}$} & 35 & \frac{7075}{8664} & \frac{4345}{4332} &  \frac{10}{57} & \numprint{1.05263} & \numprint{1.42105} \\
    A_6 & 56 & \left\{\frac{3803}{5776},\frac{5885}{8664}\right\} & \left\{\frac{2253}{2888},\frac{890}{1083}\right\} &  \frac{8}{57} & \numprint{1.} & \numprint{1.52632} \\
    D_6(a_1) & 62 & \left\{\frac{253}{400},\frac{49}{75}\right\} & \left\{\frac{149}{200},\frac{59}{75}\right\} &  \frac{2}{15}  & \numprint{1.} & \numprint{1.4} \\
    E_7(a_3) & 111 & \left\{\frac{659}{1296},\frac{343}{648}\right\} & \left\{\frac{373}{648},\frac{50}{81}\right\} &  \frac{8}{81} & \numprint{1.} & \numprint{1.51852} \\
 \hline
\end{array}$}
\end{adjustbox}

  \begin{adjustbox}{center}
\scalebox{1}{{\renewcommand{\arraystretch}{1.2} $\begin{array}{|c|c|c|c|}
 \hline
  \text{[B-C]} & \text{SU(2})_{D}\text{$\times $Residual}  & k_{\text{IR}}\text{ interact} & k_{\text{IR}}\text{+free} \\
\hline
   0 & \text{E}_7  & 8 & 8 \\
 \text{$A_1$} & \text{SU(2)}\times \text{SO(12)} & \left\{\frac{32}{5}\right\} & \left\{\frac{32}{5}\right\} \\
 \text{$A_2+3A_1$} & \text{SU(2)}\times \text{G}_2 & \left\{\frac{320}{41}\right\} &   \left\{\frac{320}{41}\right\} \\
 \text{$A_4+A_2$} & \text{SU(2)}\times \text{SU(2)}  & \left\{\frac{480}{13}\right\} &   \left\{\frac{480}{13}\right\} \\
 \text{$A_5'$} & \text{SU(2)}\times \text{SU(2)}\times \text{SU(2)} &  \left\{\frac{40}{19},\frac{120}{19}\right\} & \left\{\frac{40}{19},\frac{120}{19}\right\} \\
 \text{$A_5''$} & \text{SU(2)}\times \text{G}_2 &  \left\{\frac{40}{19}\right\} & \left\{\frac{40}{19}\right\} \\
 \text{$A_6$} & \text{SU(2)}\times \text{SU(2)} &  \left\{\frac{224}{19}\right\} & \left\{\frac{224}{19}\right\} \\
 \text{$D_6(a_1)$} & \text{SU(2)}\times \text{SU(2)} & \left\{\frac{8}{5}\right\} & \left\{\frac{8}{5}\right\} \\
  \hline
    \end{array}$}}
\end{adjustbox}
\caption{Plain nilpotent mass deformations of the Minahan-Nemeschansky theory with $E_7$ flavor, only rational values. The top table has the central charges $a_{\mathrm{IR}}$ and $c_{\mathrm{IR}}$ as well as scaling dimensions while the table below contains the information about the flavor central charges.}
\label{MNE7}
\end{table}

\begin{table}[H]
  \centering
\begin{adjustbox}{center}
\scalebox{1}{\renewcommand{\arraystretch}{1.2} $\begin{array}{|c|c|c|c|c|c|c|}
 \hline
  \text{[B-C]} & \text{r} & a_{\text{IR}} & c_{\text{IR}} &  t_* & \Delta _{\text{IR}}\text{(Z)} & \text{Min(}\Delta _{\text{IR}}\text{($\cO$'s))} \\
  \hline
  0 & 0 & \frac{95}{24} & \frac{31}{6} &  \frac{2}{3} & 6. & 2. \\
  A_2+3A_1 & 7 & \frac{223}{96} & \frac{281}{96} &  \frac{1}{3} & 3. & 1.5 \\
  E_8(a_1) & 760 & \left\{\frac{5471}{13872},\frac{120}{289}\right\} & \left\{\frac{2897}{6936},\frac{531}{1156}\right\} & \frac{2}{51} & 1. & 1.58824 \\
 \hline
\end{array}$}
\end{adjustbox}

  \begin{adjustbox}{center}
\scalebox{1}{{\renewcommand{\arraystretch}{1.2} $\begin{array}{|c|c|c|c|}
 \hline
  \text{[B-C]} & \text{SU(2})_{D}\text{$\times $Residual} &  k_{\text{IR}}\text{ interact} & k_{\text{IR}}\text{+free} \\
\hline
  0 & \text{E}_8  & 12 & 12 \\
 \text{$A_2+3A_1$} & \text{SU(2)}\times \text{G}_2\times \text{SU(2)} & \{12,6\} & \{12,6\} \\
  \hline
    \end{array}$}}
\end{adjustbox}
\caption{Plain nilpotent mass deformations of the Minahan-Nemeschansky theory with $E_8$ flavor, only rational values. The top table has the central charges $a_{\mathrm{IR}}$ and $c_{\mathrm{IR}}$ as well as scaling dimensions while the table below contains the information about the flavor central charges.}
\label{MNE8}
\end{table}

\begin{table}[H]
\centering
\begin{adjustbox}{center}
\scalebox{0.9}{{\renewcommand{\arraystretch}{1.2} $\begin{array}{|c|c|c|c|c|c|c|c|}
\hline
\text{[B-C]} & \text{r} & a_{\text{IR}} & c_{\text{IR}} &  t_* & \Delta _{\text{IR}}\text{(Z)} & \text{Min(}\Delta _{\text{IR}}\text{($\cO$'s))}\\
\hline
0 & 0 & \left\{\frac{41}{24},\frac{10}{3}\right\} & \left\{\frac{13}{6},\frac{65}{12}\right\} &  \frac{2}{3} & \numprint{3.} & \numprint{2.} \\
\rowcolor{LightCyan}  D_4 & 28 & \left\{\frac{7}{12},\frac{53}{24}\right\} & \left\{\frac{2}{3},\frac{47}{12}\right\} &  \frac{1}{6} & \numprint{1.} & \numprint{1.5} \\
\rowcolor{LightCyan}  D_5 & 60 & \left\{\frac{11}{24},\frac{25}{12}\right\} & \left\{\frac{1}{2},\frac{15}{4}\right\} &  \frac{1}{9} & \numprint{1.} & \numprint{1.66667} \\
\rowcolor{LightCyan}  E_6 & 156 & \left\{\frac{43}{120},\frac{119}{60}\right\} & \left\{\frac{11}{30},\frac{217}{60}\right\}  & \frac{1}{15} & \numprint{1.} & \numprint{1.8} \\
\hline
\end{array}$}}
\end{adjustbox}
\par
\begin{adjustbox}{center}
\scalebox{0.9}{{\renewcommand{\arraystretch}{1.2} $\begin{array}{|c|c|c|c|}
\hline
\text{[B-C]} & \text{SU(2})_{D}\text{$\times $Residual} & k_{\text{IR}}\text{ interact} & k_{\text{IR}}\text{+free} \\
\hline
0 & \text{E}_6  & 6 & 30 \\
\text{$D_4$} & \text{SU(2)}\times \text{SU(3)}  & \{3\} & \{15\} \\
\text{$D_5$} & \text{SU(2)}\times \text{U(1)}  & \{6\} & \{24\} \\
\hline
\end{array}$}}
\end{adjustbox}
\caption{Flipper field deformations of the Minahan-Nemeschansky theory with $E_{6}$ flavor, only
rational values. The top table has the central charges $a_{\mathrm{IR}}$ and
$c_{\mathrm{IR}}$ as well as scaling dimensions while the table below contains
the information about the flavor central charges. The cyan highlighted
entries align with the $H_0$, $H_1$ and $H_2$ Argyres-Douglas theories,
as first noted in \cite{Maruyoshi:2016aim,Agarwal:2016pjo}.}%
\label{MN-MS-E6}%
\end{table}

\begin{table}[H]
\centering
\begin{adjustbox}{center}
\scalebox{0.9}{{\renewcommand{\arraystretch}{1.2} $\begin{array}{|c|c|c|c|c|c|c|c|}
\hline
\text{[B-C]} & \text{r} & a_{\text{IR}} & c_{\text{IR}} &  t_* & \Delta _{\text{IR}}\text{(Z)} & \text{Min(}\Delta _{\text{IR}}\text{($\cO$'s))}\\
\hline
0 & 0 & \left\{\frac{59}{24},\frac{251}{48}\right\} & \left\{\frac{19}{6},\frac{209}{24}\right\} &  \frac{2}{3} & \numprint{4.} & \numprint{2.}\\
A_2+3A_1 & 7 & \left\{\frac{12163}{8214},\frac{134899}{32856}\right\} & \left\{\frac{121465}{65712},\frac{466453}{65712}\right\} &  \frac{31}{111} & \numprint{1.67568} &
\numprint{1.74324} \\
\rowcolor{LightCyan}  E_6 & 156 & \left\{\frac{11}{24},\frac{155}{48}\right\} & \left\{\frac{1}{2},\frac{145}{24}\right\} &  \frac{2}{27} & \numprint{1.} & \numprint{1.66667} \\
\rowcolor{LightCyan} E_7 & 399 & \left\{\frac{43}{120},\frac{751}{240}\right\} & \left\{\frac{11}{30},\frac{709}{120}\right\}  & \frac{2}{45} & \numprint{1.} & \numprint{1.8} \\
\hline
\end{array}$}}
\end{adjustbox}
\par
\begin{adjustbox}{center}
\scalebox{0.9}{{\renewcommand{\arraystretch}{1.2} $\begin{array}{|c|c|c|c|}
\hline
\text{[B-C]} & \text{SU(2})_{D}\text{$\times $Residual}  & k_{\text{IR}}\text{ interact} & k_{\text{IR}}\text{+free} \\
\hline
0 & \text{E}_7  & 8 & 44 \\
\text{$A_2+3A_1$} & \text{SU(2)}\times \text{G}_2 & \left\{\frac{284}{37}\right\} &\left\{\frac{1246}{37}\right\} \\
\text{$E_6$} & \text{SU(2)}\times \text{SU(2)} & \left\{\frac{8}{3}\right\} & \left\{\frac{44}{3}\right\} \\
\hline
\end{array}$}}
\end{adjustbox}
\caption{Flipper field deformations of the Minahan-Nemeschansky theory with $E_{7}$ flavor, only
rational values. The top table has the central charges $a_{\mathrm{IR}}$ and
$c_{\mathrm{IR}}$ as well as scaling dimensions while the table below contains
the information about the flavor central charges. The cyan highlighted entries
align with the $H_0$ and $H_1$ Argyres-Douglas theories,
as first noted in \cite{Maruyoshi:2016aim}. Compared with reference \cite{Maruyoshi:2016aim},
we also find an additional flipper field deformation which yields the $H_1$ theory for the
$E_6$ Bala-Carter label, with embedding index $r = 156$. The other rational central charges are also in agreement with \cite{AgarwalMaruyoshiSong}.}%
\label{MN-MS-E7}%
\end{table}

\begin{table}[H]
\centering
\begin{adjustbox}{center}
\scalebox{0.9}{{\renewcommand{\arraystretch}{1.2} $\begin{array}{|c|c|c|c|c|c|c|c|}
\hline
\text{[B-C]} & \text{r} & a_{\text{IR}} & c_{\text{IR}} &  t_* & \Delta _{\text{IR}}\text{(Z)} & \text{Min(}\Delta _{\text{IR}}\text{($\cO$'s))}\\
\hline
0 & 0 & \left\{\frac{95}{24},\frac{73}{8}\right\} & \left\{\frac{31}{6},\frac{31}{2}\right\} &  \frac{2}{3} & \numprint{6.} & \numprint{2.} \\
A_3 & 10 & \left\{\frac{497803}{221952},\frac{529689}{73984}\right\} & \left\{\frac{635435}{221952},\frac{939321}{73984}\right\} &  \frac{53}{204} & \numprint{2.33824} &   \numprint{1.44118} \\
A_3+A_1 & 11 & \left\{\frac{139189}{60552},\frac{214667}{30276}\right\} & \left\{\frac{91127}{30276},\frac{95318}{7569}\right\} &  \frac{64}{261} &  \numprint{2.2069} &  \numprint{1.52874}  \\
E_7\left(a_5\right) & 39 & \left\{\frac{445}{324},\frac{2065}{324}\right\} & \left\{\frac{281}{162},\frac{1901}{162}\right\} & \frac{4}{27} & \numprint{1.33333} & \numprint{1.66667} \\
E_7\left(a_4\right) & 63 & \left\{\frac{1691}{1452},\frac{8951}{1452}\right\} & \left\{\frac{541}{363},\frac{4171}{363}\right\}  & \frac{4}{33} & \numprint{1.09091} & \numprint{1.54545} \\
\rowcolor{LightCyan}  E_8 & 1240 & \left\{\frac{43}{120},\frac{221}{40}\right\} & \left\{\frac{11}{30},\frac{107}{10}\right\}  & \frac{2}{75} & \numprint{1.} & \numprint{1.8} \\
\hline
\end{array}$}}
\end{adjustbox}
\begin{adjustbox}{center}
\scalebox{0.9}{{\renewcommand{\arraystretch}{1.2} $\begin{array}{|c|c|c|c|}
\hline
\text{[B-C]} & \text{SU(2})_{D}\text{$\times $Residual} & k_{\text{IR}}\text{ interact} & k_{\text{IR}}\text{+free} \\
\hline
0 & E_8  & 12 & 72 \\
\text{$A_3$} & \text{SU(2)}\times \text{SO(11)} & \{6\} & \{32\} \\
\text{$A_3+A_1$} & \text{SU(2)}\times \text{SO(7)}\times \text{SU(2)} &\left\{\frac{220}{29},\frac{421}{87}\right\} & \left\{\frac{800}{29},\frac{2161}{87}\right\} \\
\text{$E_7(a_5)$} & \text{SU(2)}\times \text{SU(2)} & \left\{\frac{29}{9}\right\} &\left\{\frac{137}{9}\right\} \\
\text{$E_7(a_4)$} & \text{SU(2)}\times \text{SU(2)} & \left\{\frac{31}{11}\right\} &\left\{\frac{152}{11}\right\} \\
\hline
\end{array}$}}
\end{adjustbox}
\caption{Flipper field deformations of the Minahan-Nemeschansky theory with $E_{8}$ flavor, only
rational values. The top table has the central charges $a_{\mathrm{IR}}$ and
$c_{\mathrm{IR}}$ as well as scaling dimensions while the table below contains
the information about the flavor central charges. The cyan highlighted entry
aligns with the $H_0$ Argyres-Douglas theory,
as first noted in \cite{Maruyoshi:2016aim}. The other rational central charges are also in agreement with \cite{AgarwalMaruyoshiSong}.}%
\label{MN-MS-E8}%
\end{table}

\subsection*{Tables of Rational Theories: Conformal Matter}

\begin{table}[H]
\centering
\begin{adjustbox}{left}
\scalebox{1.0}{{\renewcommand{\arraystretch}{1.2}
$\begin{array}{|c|c|c|c|c|c|c|c|}
\hline
\text{[B-C]}_L & \text{[B-C]}_R & r_L & r_R & r_L+r_R & a_{\text{IR}} & c_{\text{IR}} & t_* \\
\hline
0 & 0 & 0 & 0 & 0 & \frac{613}{24} & \frac{173}{6} &  \frac{2}{3} \\
2A_2+A_1 & 2A_2 & 9 & 8 & 17 & \frac{68050}{4107} & \frac{150715}{8214}  & \frac{40}{111} \\
A_5 & 2A_2+A_1 & 35 & 9 & 44 & \left\{\frac{316}{25},\frac{3817}{300}\right\} & \left\{\frac{346}{25},\frac{2101}{150}\right\}  & \frac{4}{15}\\
\hline
\end{array}$}} $\cdots$
\end{adjustbox}
\centering
\begin{adjustbox}{left}
\scalebox{1.0}{{\renewcommand{\arraystretch}{1.2}
$\begin{array}{|c|c|c|c|c|c|}
\hline
\text{[B-C]}_L & \text{[B-C]}_R & t_* & \text{Min(}\Delta _{\text{IR}}\text{(Z's))} & \text{Min(}\Delta_{\text{IR}}\left(\cO_L\text{'s))}\right. & \text{Min(}\Delta _{\text{IR}}\left(\cO_R\text{'s))}\right. \\
0 & 0 & \frac{2}{3} & \numprint{6.} & \numprint{2.} & \numprint{2.} \\
2A_2+A_1 & 2A_2 &\frac{40}{111} & \numprint{3.24324} & \numprint{1.10811} & \numprint{1.37838} \\
A_5 & 2A_2+A_1 & \frac{4}{15} & \numprint{2.4} & \numprint{1.} & \numprint{1.6} \\
\hline
\end{array}$ }}
\end{adjustbox}
\caption{Plain nilpotent mass deformations of ($E_{6}$, $E_{6}$) conformal matter, only rational values.}%
\label{CME6}%
\end{table}

\begin{table}[H]
\centering
\begin{adjustbox}{left}
\scalebox{1.0}{{\renewcommand{\arraystretch}{1.2} $\begin{array}{|c|c|c|c|c|c|c|c|}
\hline
\text{[B-C]}_L & \text{[B-C]}_R & r_L & r_R & r_L+r_R & a_{\text{IR}} & c_{\text{IR}} & t_*  \\
\hline
0 & 0 & 0 & 0 & 0 & \frac{817}{12} & \frac{221}{3} & \frac{2}{3} \\
D_4+A_1 & D_4+A_1 & 29 & 29 & 58 & \left\{\frac{314941}{8400},\frac{105097}{2800}\right\} & \left\{\frac{47843}{1200},\frac{15981}{400}\right\} & \frac{31}{105}  \\
D_5 & \text{(3}A_1\text{)''} & 60 & 3 & 63 & \left\{\frac{233959}{6272},\frac{235135}{6272}\right\} & \left\{\frac{247315}{6272},\frac{249667}{6272}\right\} & \frac{13}{42}  \\
D_5 & \text{(3}A_1\text{)'} & 60 & 3 & 63 & \left\{\frac{233959}{6272},\frac{235135}{6272}\right\} & \left\{\frac{247315}{6272},\frac{249667}{6272}\right\} &  \frac{13}{42}  \\
D_5+A_1 & 0 & 61 & 0 & 61 & \left\{\frac{63612}{1681},\frac{1022835}{26896}\right\} & \left\{\frac{538047}{13448},\frac{271545}{6724}\right\} & \frac{13}{41} \\
D_5+A_1 & D_4\left(a_1\right) & 61 & 12 & 73 & \left\{\frac{27729}{784},\frac{6969}{196}\right\} & \left\{\frac{3663}{98},\frac{14799}{392}\right\} & \frac{2}{7}  \\
D_5+A_1 & A_3+2A_1 & 61 & 12 & 73 & \left\{\frac{27729}{784},\frac{6969}{196}\right\} & \left\{\frac{3663}{98},\frac{14799}{392}\right\} & \frac{2}{7} \\
E_6\left(a_1\right) & A_3 & 84 & 10 & 94 & \left\{\frac{1583}{48},\frac{199}{6}\right\} & \left\{\frac{4177}{120},\frac{2111}{60}\right\}  & \frac{4}{15}\\
E_6 & A_3 & 156 & 10 & 166 & \left\{\frac{995}{36},\frac{1999}{72}\right\} & \left\{\frac{1049}{36},\frac{529}{18}\right\} &  \frac{2}{9} \\
E_7\left(a_1\right) & A_2 & 231 & 4 & 235 & \left\{\frac{2992009}{121104},\frac{187789}{7569}\right\} & \left\{\frac{1576001}{60552},\frac{198577}{7569}\right\} &
\frac{52}{261}\\
E_7\left(a_1\right)  & 4A_1 & 231 & 4 & 235 & \left\{\frac{2992009}{121104},\frac{187789}{7569}\right\} & \left\{\frac{1576001}{60552},\frac{198577}{7569}\right\} & \frac{52}{261} \\
\hline
\end{array}$}} $\cdots$
\end{adjustbox}
\begin{adjustbox}{left}
\scalebox{1.0}{{\renewcommand{\arraystretch}{1.2} $\begin{array}{|c|c|c|c|c|c|}
\hline
\text{[B-C]}_L & \text{[B-C]}_R &  t_* & \text{Min(}\Delta _{\text{IR}}\text{(Z's))} & \text{Min(}\Delta_{\text{IR}}\left(\cO_L\text{'s))}\right. & \text{Min(}\Delta _{\text{IR}}\left(\cO_R\text{'s))}\right. \\
\hline
0 & 0  & \frac{2}{3} & \numprint{6.} & \numprint{2.} & \numprint{2.} \\
D_4+A_1 & D_4+A_1 & \frac{31}{105}     & \numprint{2.65714} & \numprint{1.} & \numprint{1.} \\
D_5 & \text{(3}A_1\text{)''}  & \frac{13}{42} & \numprint{2.78571} & \numprint{1.} & \numprint{2.07143} \\
D_5 & \text{(3}A_1\text{)'}   & \frac{13}{42}  & \numprint{2.78571} & \numprint{1.} & \numprint{1.83929} \\
D_5+A_1 & 0 & \frac{13}{41}   & \numprint{2.85366} & \numprint{1.} & \numprint{2.5243}9 \\
D_5+A_1 & D_4\left(a_1\right) & \frac{2}{7} & \numprint{2.57143} & \numprint{1.} & \numprint{1.28571} \\
D_5+A_1 & A_3+2A_1    & \frac{2}{7} & \numprint{2.57143} & \numprint{1.} & \numprint{1.28571} \\
E_6\left(a_1\right) & A_3  & \frac{4}{15} & \numprint{2.4}     & \numprint{1.} & \numprint{1.4} \\
E_6 & A_3            & \frac{2}{9} & \numprint{2.}      & \numprint{1.} & \numprint{1.66667} \\
E_7\left(a_1\right)  & A_2  &  \frac{52}{261} & \numprint{1.7931} & \numprint{1.} & \numprint{2.10345} \\
E_7\left(a_1\right)  & 4A_1 & \frac{52}{261} & \numprint{1.7931} & \numprint{1.} & \numprint{2.25287} \\
\hline
\end{array}$}}
\end{adjustbox}
\caption{Plain nilpotent mass deformations of ($E_{7}$, $E_{7}$) conformal matter, only rational values.}%
\label{CME7}%
\end{table}\begin{table}[H]
\centering
\begin{adjustbox}{left}
\scalebox{0.85}{{\renewcommand{\arraystretch}{1.2} $\begin{array}{|c|c|c|c|c|c|c|c|}
\hline
\text{[B-C]}_L & \text{[B-C]}_R & r_L & r_R & r_L+r_R & a_{\text{IR}} & c_{\text{IR}} &  t_* \\
\hline
0 & 0 & 0 & 0 & 0 & \frac{1745}{8} & \frac{457}{2} &  \frac{2}{3} \\
A_3+A_2 & 3A_1 & 14 & 3 & 17 & \left\{\frac{2594465245}{14362032},\frac{325018777}{1795254}\right\} & \left\{\frac{1347452419}{7181016},\frac{676568695}{3590508}\right\}  & \frac{824}{1641}\\
D_5 & 0 & 60 & 0 & 60 & \left\{\frac{88198105}{591576},\frac{44209973}{295788}\right\} & \left\{\frac{11389690}{73947},\frac{45780601}{295788}\right\} & \frac{194}{471}\\
E_7\left(a_3\right) & 2A_2+A_1 & 111 & 9 & 120 & \left\{\frac{12055}{96},\frac{12091}{96}\right\} & \left\{\frac{12425}{96},\frac{12497}{96}\right\} & \frac{1}{3} \\
E_8\left(b_5\right) & D_6\left(a_1\right) & 160 & 62 & 222 & \left\{\frac{823817}{8112},\frac{103463}{1014}\right\} & \left\{\frac{422939}{4056},\frac{213413}{2028}\right\} &  \frac{10}{39} \\
D_7 & E_6\left(a_1\text{)+}A_1\right. & 182 & 85 & 267 & \left\{\frac{187823116685}{1971613488},\frac{47191962037}{492903372}\right\} &
\left\{\frac{96328408265}{985806744},\frac{12159142466}{123225843}\right\} & \frac{4588}{19227} \\
E_8\left(b_4\right) & A_2+A_1 & 232 & 5 & 237 & \left\{\frac{1832579}{17328},\frac{76553}{722}\right\} & \left\{\frac{943241}{8664},\frac{157989}{1444}\right\} & \frac{16}{57} \\
\hline
\end{array}$}} 
\end{adjustbox}
\begin{adjustbox}{left} $\cdots$
\scalebox{0.85}{{\renewcommand{\arraystretch}{1.2} $\begin{array}{|c|c|c|c|c|c|c|c|c|c|c|c|}
\hline
\text{[B-C]}_L & \text{[B-C]}_R & t_* & \text{Min(}\Delta _{\text{IR}}\text{(Z's))} & \text{Min(}\Delta_{\text{IR}}\left(\cO_L\text{'s))}\right. & \text{Min(}\Delta _{\text{IR}}\left(\cO_R\text{'s))}\right. \\
\hline
0 & 0 & \frac{2}{3} & \numprint{6.} & \numprint{2.} & \numprint{2.} \\
A_3+A_2 & 3A_1 & \frac{824}{1641} & \numprint{4.5192} & \numprint{1.} & \numprint{1.117} \\
D_5 & 0 & \frac{194}{471} & \numprint{3.70701} & \numprint{1.} & \numprint{2.38217} \\
E_7\left(a_3\right) & 2A_2+A_1 & \frac{1}{3} & \numprint{3.} & \numprint{1.} & \numprint{1.25} \\
E_8\left(b_5\right) & D_6\left(a_1\right) & \frac{10}{39} & \numprint{2.30769} & \numprint{1.} & \numprint{1.} \\
D_7 & E_6\left(a_1\text{)+}A_1\right. & \frac{4588}{19227} & \numprint{2.1476} & \numprint{1.} & \numprint{1.} \\
E_8\left(b_4\right) & A_2+A_1 & \frac{16}{57} & \numprint{2.52632} & \numprint{1.} & \numprint{1.73684} \\
\hline
\end{array}$}}
\end{adjustbox}
\caption{Plain nilpotent mass deformations of ($E_{8}$, $E_{8}$) conformal matter, only rational
values.}%
\label{CME8}%
\end{table}

\begin{table}[H]
\centering
\begin{adjustbox}{left}
\scalebox{0.85}{{\renewcommand{\arraystretch}{1.2} $\begin{array}{|c|c|c|c|c|c|c|c|}
\hline
\text{[B-C]}_L & \text{[B-C]}_R & r_L & r_R & r_L+r_R & a_{\text{IR}} & c_{\text{IR}} &  t_* \\
\hline
\left[1^8\right] & \left[1^8\right] & 0 & 0 & 0 & \left\{\frac{95}{24},\frac{41}{8}\right\} & \left\{\frac{31}{6},\frac{15}{2}\right\} & \frac{2}{3} \\
\text{[7,1]} & \left[4^2\text{]I}\right. & 28 & 10 & 38 & \left\{\frac{245399}{107736},\frac{87785}{35912}\right\} & \left\{\frac{95905}{26934},\frac{34961}{8978}\right\} &  \frac{34}{201} \\
\text{[7,1]} & \left[4^2\text{]II}\right. & 28 & 10 & 38 & \left\{\frac{245399}{107736},\frac{87785}{35912}\right\} & \left\{\frac{95905}{26934},\frac{34961}{8978}\right\} &  \frac{34}{201}\\
\text{[7,1]} & \left.\text{[5,}1^3\right] & 28 & 10 & 38 & \left\{\frac{245399}{107736},\frac{87785}{35912}\right\} & \left\{\frac{95905}{26934},\frac{34961}{8978}\right\} &  \frac{34}{201} \\
\hline
\end{array}$}} $\cdots$
\end{adjustbox}
\begin{adjustbox}{left}
\scalebox{0.85}{{\renewcommand{\arraystretch}{1.2} $\begin{array}{|c|c|c|c|c|c|c|c|c|}
\hline
\text{[B-C]}_L & \text{[B-C]}_R & t_* & \text{Min(}\Delta _{\text{IR}}\text{(Z's))} & \text{Min(}\Delta_{\text{IR}}\left(\cO_L\text{'s))}\right. & \text{Min(}\Delta _{\text{IR}}\left(\cO_R\text{'s))}\right. \\
\hline
\left[1^8\right] & \left[1^8\right]  & \frac{2}{3} & \numprint{6.} & \numprint{2.} & \numprint{2.}  \\
\text{[7,1]} & \left[4^2\text{]I}\right.  & \frac{34}{201} & \numprint{1.52239} & \numprint{1.47761} & \numprint{1.98507} \\
\text{[7,1]} & \left[4^2\text{]II}\right. & \frac{34}{201} & \numprint{1.52239} & \numprint{1.47761} & \numprint{1.98507} \\
\text{[7,1]} & \left.\text{[5,}1^3\right] & \frac{34}{201} & \numprint{1.52239} & \numprint{1.47761} & \numprint{1.98507} \\
\hline
\end{array}$}}
\end{adjustbox}
\caption{Flipper field deformations of ($D_{4}$, $D_{4}$) conformal matter, only rational values.}%
\label{MS-CM-D4}%
\end{table}

\begin{table}[H]
\centering
\begin{adjustbox}{left}
\scalebox{0.85}{{\renewcommand{\arraystretch}{1.2} $\begin{array}{|c|c|c|c|c|c|c|c|}
\hline
\text{[B-C]}_L & \text{[B-C]}_R & r_L & r_R & r_L+r_R & a_{\text{IR}} & c_{\text{IR}} & t_* \\
\hline
0 & 0 & 0 & 0 & 0 & \left\{\frac{613}{24},\frac{691}{24}\right\} & \left\{\frac{173}{6},\frac{106}{3}\right\} &  \frac{2}{3} \\
A_3+A_1 & A_1 & 11 & 1 & 12 & \left\{\frac{248983}{13872},\frac{144577}{6936}\right\} & \left\{\frac{137641}{6936},\frac{44453}{1734}\right\}  & \frac{20}{51} \\
D_4 & A_3+A_1 & 28 & 11 & 39 & \left\{\frac{5271}{400},\frac{3223}{200}\right\} & \left\{\frac{2893}{200},\frac{1017}{50}\right\}  & \frac{4}{15} \\
D_5\left(a_1\right) & A_3 & 30 & 10 & 40 & \left\{\frac{15737}{1200},\frac{9631}{600}\right\} & \left\{\frac{8651}{600},\frac{1522}{75}\right\} & \frac{4}{15} \\
D_5\left(a_1\right) & A_3+A_1 & 30 & 11 & 41 & \left\{\frac{1364659}{104976},\frac{836513}{52488}\right\} & \left\{\frac{749681}{52488},\frac{132256}{6561}\right\} & \frac{64}{243} \\
\hline
\end{array}$}} $\cdots$
\end{adjustbox}
\begin{adjustbox}{left}
\scalebox{0.85}{{\renewcommand{\arraystretch}{1.2} $\begin{array}{|c|c|c|c|c|c|c|c|c|}
\hline
\text{[B-C]}_L & \text{[B-C]}_R & t_* & \text{Min(}\Delta _{\text{IR}}\text{(Z's))} & \text{Min(}\Delta_{\text{IR}}\left(\cO_L\text{'s))}\right. & \text{Min(}\Delta _{\text{IR}}\left(\cO_R\text{'s))}\right. \\
\hline
0 & 0 & \frac{2}{3} & \numprint{6.} & \numprint{2.} & \numprint{2.} \\
A_3+A_1 & A_1 & \frac{20}{51} & \numprint{3.52941} & \numprint{1.} & \numprint{1.82353} \\
D_4 & A_3+A_1 & \frac{4}{15} & \numprint{2.4} & \numprint{1.} & \numprint{1.4} \\
D_5\left(a_1\right) & A_3 & \frac{4}{15} & \numprint{2.4} & \numprint{1.} & \numprint{1.4} \\
D_5\left(a_1\right) & A_3+A_1 & \frac{64}{243} & \numprint{2.37037} & \numprint{1.} & \numprint{1.41975} \\
\hline
\end{array}$}}
\end{adjustbox}
\caption{Flipper field deformations of ($E_{6}$, $E_{6}$) conformal matter, only rational values.}%
\label{MS-CM-E6}%
\end{table}

\begin{table}[H]
\centering
\begin{adjustbox}{left}
\scalebox{0.85}{{\renewcommand{\arraystretch}{1.2} $\begin{array}{|c|c|c|c|c|c|c|c|}
\hline
\text{[B-C]}_L & \text{[B-C]}_R & r_L & r_R & r_L+r_R & a_{\text{IR}} & c_{\text{IR}} &  t_*  \\
\hline
0 & 0 & 0 & 0 & 0 & \left\{\frac{817}{12},\frac{589}{8}\right\} & \left\{\frac{221}{3},\frac{339}{4}\right\} &  \frac{2}{3} \\
A_3+A_2+A_1 & 0 & 15 & 0 & 15 & \left\{\frac{1241}{24},\frac{453}{8}\right\} & \left\{\frac{661}{12},\frac{779}{12}\right\} &  \frac{4}{9} \\
A_5+A_1 & A_2+3A_1 & 36 & 7 & 43 & \left\{\frac{3931}{96},\frac{4417}{96}\right\} & \left\{\frac{4163}{96},\frac{5135}{96}\right\} &  \frac{1}{3} \\
E_6\left(a_1\right) & A_2+A_1 & 84 & 5 & 89 & \left\{\frac{235499}{6936},\frac{270757}{6936}\right\} & \left\{\frac{62449}{1734},\frac{40039}{867}\right\}  & \frac{14}{51} \\
E_6 & A_4+A_1 & 156 & 21 & 177 & \left\{\frac{44180297}{1642800},\frac{1096539}{34225}\right\} & \left\{\frac{23344541}{821400},\frac{2649843}{68450}\right\} & \frac{116}{555} \\
\hline
\end{array}$}} $\cdots$
\end{adjustbox}
\begin{adjustbox}{left}
\scalebox{0.85}{{\renewcommand{\arraystretch}{1.2} $\begin{array}{|c|c|c|c|c|c|c|c|c|}
\hline
\text{[B-C]}_L & \text{[B-C]}_R & t_* & \text{Min(}\Delta _{\text{IR}}\text{(Z's))} & \text{Min(}\Delta_{\text{IR}}\left(\cO_L\text{'s))}\right. & \text{Min(}\Delta _{\text{IR}}\left(\cO_R\text{'s))}\right. \\
\hline
0 & 0 & \frac{2}{3} & \numprint{6.} & \numprint{2.} & \numprint{2.} \\
A_3+A_2+A_1 & 0  & \frac{4}{9} & \numprint{4.} & \numprint{1.} & \numprint{2.33333} \\
A_5+A_1 & A_2+3A_1 & \frac{1}{3} & \numprint{3.} & \numprint{1.} & \numprint{1.5} \\
E_6\left(a_1\right) & A_2+A_1 & \frac{14}{51} & \numprint{2.47059} & \numprint{1.} & \numprint{1.76471} \\
E_6 & A_4+A_1 & \frac{116}{555} & \numprint{1.88108} & \numprint{1.} & \numprint{1.43243} \\
\hline
\end{array}$}}
\end{adjustbox}
\caption{Flipper field deformations of ($E_{7}$, $E_{7}$) conformal matter, only rational values.}%
\label{MS-CM-E7}%
\end{table}

\begin{table}[H]
\centering
\begin{adjustbox}{left}
\scalebox{0.85}{{\renewcommand{\arraystretch}{1.2} $\begin{array}{|c|c|c|c|c|c|c|c|}
\hline
\text{[B-C]}_L & \text{[B-C]}_R & r_L & r_R & r_L+r_R & a_{\text{IR}} & c_{\text{IR}} &  t_* \\
\hline
0 & 0 & 0 & 0 & 0 & \left\{\frac{1745}{8},\frac{5483}{24}\right\} & \left\{\frac{457}{2},\frac{1495}{6}\right\} &  \frac{2}{3} \\
A_4+A_2+A_1 & A_4+2A_1 & 25 & 22 & 47 & \left\{\frac{122989}{816},\frac{21657}{136}\right\} & \left\{\frac{63463}{408},\frac{2934}{17}\right\} &  \frac{20}{51} \\
D_5\left(a_1\right) & A_2+3A_1 & 30 & 7 & 37 & \left\{\frac{1200211}{7500},\frac{632293}{3750}\right\} & \left\{\frac{620893}{3750},\frac{342634}{1875}\right\}  & \frac{32}{75} \\
D_5\left(a_1\text{)+}A_2\right. & A_3+A_2+A_1 & 34 & 15 & 49 & \left\{\frac{122683}{816},\frac{64801}{408}\right\} & \left\{\frac{63361}{408},\frac{8785}{51}\right\} &   \frac{20}{51} \\
E_6\left(a_3\text{)+}A_1\right. & A_2+3A_1 & 37 & 7 & 44 & \left\{\frac{237476949}{1527752},\frac{187894873}{1145814}\right\} & \left\{\frac{30711077}{190969},\frac{407681569}{2291628}\right\}  & \frac{180}{437} \\
D_5+A_1 & D_5 & 61 & 60 & 121 & \left\{\frac{1760291}{14700},\frac{948133}{7350}\right\} & \left\{\frac{903893}{7350},\frac{519934}{3675}\right\}  & \frac{32}{105} \\
D_6\left(a_1\right) & D_5\left(a_1\right) & 62 & 30 & 92 & \left\{\frac{1553}{12},\frac{6655}{48}\right\} & \left\{\frac{6385}{48},\frac{7271}{48}\right\} & \frac{1}{3} \\
D_6 & D_4+A_1 & 110 & 29 & 139 & \left\{\frac{25707707}{218886},\frac{27750643}{218886}\right\} & \left\{\frac{52819073}{437772},\frac{60990817}{437772}\right\}  & \frac{172}{573} \\
E_8\left(b_5\right) & D_4\left(a_1\text{)+}A_1\right. & 160 & 13 & 173 & \left\{\frac{49357}{432},\frac{26681}{216}\right\} & \left\{\frac{25463}{216},\frac{7367}{54}\right\} &  \frac{8}{27} \\
\hline
\end{array}$}} $\cdots$
\end{adjustbox}
\begin{adjustbox}{left}
\scalebox{0.85}{{\renewcommand{\arraystretch}{1.2} $\begin{array}{|c|c|c|c|c|c|c|c|c|}
\hline
\text{[B-C]}_L & \text{[B-C]}_R & t_* & \text{Min(}\Delta _{\text{IR}}\text{(Z's))} & \text{Min(}\Delta_{\text{IR}}\left(\cO_L\text{'s))}\right. & \text{Min(}\Delta _{\text{IR}}\left(\cO_R\text{'s))}\right. \\
\hline
0 & 0 & \frac{2}{3} & \numprint{6.} & \numprint{2.} & \numprint{2.} \\
A_4+A_2+A_1 & A_4+2A_1 & \frac{20}{51} & \numprint{3.52941} & \numprint{1.} & \numprint{1.} \\
D_5\left(a_1\right) & A_2+3A_1 & \frac{32}{75} & \numprint{3.84} & \numprint{1.} & \numprint{1.08} \\
D_5\left(a_1\text{)+}A_2\right. & A_3+A_2+A_1 & \frac{20}{51} & \numprint{3.52941} & \numprint{1.} & \numprint{1.} \\
E_6\left(a_3\text{)+}A_1\right. & A_2+3A_1 & \frac{180}{437} & \numprint{3.70709} & \numprint{1.} & \numprint{1.14645} \\
D_5+A_1 & D_5 &  \frac{32}{105} & \numprint{2.74286} & \numprint{1.} & \numprint{1.} \\
D_6\left(a_1\right) & D_5\left(a_1\right) & \frac{1}{3} &  \numprint{3.} & \numprint{1.} & \numprint{1.} \\
D_6 & D_4+A_1 & \frac{172}{573} & \numprint{2.70157} & \numprint{1.} & \numprint{1.} \\
E_8\left(b_5\right) & D_4\left(a_1\text{)+}A_1\right. & \frac{8}{27} & \numprint{2.66667} & \numprint{1.} & \numprint{1.22222} \\
\hline
\end{array}$}}
\end{adjustbox}
\caption{Flipper field deformations of ($E_{8}$, $E_{8}$) conformal matter, only rational values.}%
\label{MS-CM-E8}%
\end{table}

\addtocontents{toc}{\protect\setcounter{tocdepth}{1}} 
\chapter{Chapter 2 Appendix}
\addtocontents{toc}{\protect\setcounter{tocdepth}{-1}}

\section{Partial Ordering for Nilpotent Orbits\label{sec:Nilpotents}}

In this Appendix, we review some aspects of nilpotent orbits of simple Lie algebras
and their partial ordering. We refer the interested reader to \cite{collingwood1993nilpotent} for further details.

The general linear group $GL(N,\mathbb{C})$ acts on its Lie algebra
$\mathfrak{gl}_{n}$ of all complex $n \times n$ matrices by conjugation; the
orbits are similarity classes of matrices. The theory of the Jordan
form gives a satisfactory parametrization of these classes and allows us to
regard two kinds of classes as distinguished: those represented by diagonal
matrices, and those represented by strictly upper triangular matrices, i.e., nilpotent matrices. There are only finitely many similarity classes of nilpotent
matrices, which are labeled by partitions of
of $n$. There is a similar parametrization of nilpotent orbits by partitions in any classical semisimple Lie algebra, with some additional restrictions imposed.

Semi-simple orbits are parametrized by points in a fundamental domain for the
action of the Weyl group on a Cartan subalgebra. In particular, there are
infinitely many semi-simple orbits.

\subsection{Weighted Dynkin Diagrams}

Associated to each nilpotent orbit is a unique (completely invariant) weighted
Dynkin diagram \cite{collingwood1993nilpotent}. In general, the Dynkin labels $\alpha_{i}(H)$, $1 \leq i \leq rank(G)$ of a weighted Dynkin diagram
are defined by the commutator relation:
\begin{align}
[H,X_{i}] = \alpha_{i}(H) X_{i},
\end{align}
where the $X_{i}$ are the raising operators corresponding to the positive
simple roots of $\mathfrak{g}$, and $H$ is directly constructed from the
partition $\mathbf{d}=[d_{1},\cdots, d_{n}]$ associated with the nilpotent orbit
as follows:
\begin{align}
H_{[d_{1},\cdots,d_{n}]} =
\begin{pmatrix}
D(d_{1}) & 0 & \cdots & 0\\
0 & D(d_{2}) & \cdots & 0\\
\vdots & \vdots & \ddots & \vdots\\
0 & 0 & \cdots & D(d_{k})
\end{pmatrix}
,
\end{align}
where
\begin{align}
D(d_{i}) =
\begin{pmatrix}
d_{i} -1 & 0 & 0 & \cdots & 0 & 0\\
0 & d_{i}-3 & 0 & \cdots & 0 & 0\\
0 & 0 & d_{i}-5 & \cdots & 0 & 0\\
\vdots & \vdots & \vdots & \ddots & \vdots & \vdots\\
0 & 0 & 0 & \cdots & -d_{i} + 3 & 0\\
0 & 0 & 0 & \cdots & 0 & -d_{i}+1
\end{pmatrix}
\end{align}
The nilpositive element $X$ in the $\{H, X, Y\}$ Jacobson-Morozov
standard triple is then given by:
\begin{align}
X_{[d_{1},\cdots,d_{n}]} =
\begin{pmatrix}
J^{+}(d_{1}) & 0 & \cdots & 0\\
0 & J^{+}(d_{2}) & \cdots & 0\\
\vdots & \vdots & \ddots & \vdots\\
0 & 0 & \cdots & J^{+}(d_{k})
\end{pmatrix}
,
\end{align}
where now
\begin{align}
J^{+}_{i,j}(d_{m})  &  = \delta_{i+1,j} \sqrt{i d_{m}-i^{2}} \nonumber \\
& =
\begin{pmatrix}
0 & \sqrt{d_{m} -1} & 0 & 0 & \cdots & 0 & 0\\
0 & 0 & \sqrt{2 d_{m}-4} & 0 & \cdots & 0 & 0\\
0 & 0 & 0 & \sqrt{3d_{m}-9} & \cdots & 0 & 0\\
\vdots & \vdots & \vdots & \ddots & \vdots & \vdots & \vdots\\
0 & 0 & 0 & \cdots & 0 & \sqrt{ 2 d_{m} -4} & 0\\
0 & 0 & 0 & \cdots & 0 & 0 & \sqrt{d_{m} -1}\\
0 & 0 & 0 & \cdots & 0 & 0 & 0
\end{pmatrix}
\end{align}
and similarly the nilnegative element $Y$ is given by:
\begin{align}
Y_{[d_{1},\cdots,d_{n}]} =
\begin{pmatrix}
J^{-}(d_{1}) & 0 & \cdots & 0\\
0 & J^{-}(d_{2}) & \cdots & 0\\
\vdots & \vdots & \ddots & \vdots\\
0 & 0 & \cdots & J^{-}(d_{k})
\end{pmatrix}
,
\end{align}
where $J^{-} = (J^{+})^{\dagger}$ so that $Y=X^{\dagger}$:
\begin{align}
J^{-}_{i,j}(d_{m}) = \delta_{j+1,i} \sqrt{j d_{m}-j^{2}}.
\end{align}

Direct matrix multiplication then gives the required commutation relations:
\begin{align}
[X,Y]  &  =H,\nonumber\\
[H,X]  &  =2X,\nonumber\\
[H,Y]  &  =-2Y.
\end{align}

This nilpositive matrix is similar to the nilpotent matrix
$X_{\mathcal{O}}$ we used to generate the partition in the first place.
Indeed, any two matrices with the same Jordan block decomposition (and therefore
corresponding to the same partition) are similar matrices and thus belong to
the same nilpotent orbit.

As a summary, the following are equivalent:

\begin{itemize}
\item A nilpotent orbit

\item A given Bala-Carter label

\item A corresponding set of simple roots generating the Levi subalgebra and
one or more positive roots ($X_{\alpha_{i}}$) for the distinguished orbits

\item A corresponding partition

\item An $\{H, X, Y\}$ Jacobson-Morozov standard triple, where $H$ is
explicitly built out of the partitions as described above and $X$ is similar
to the sum of the $X_{\alpha_{i}}$ specified in our brane diagrams.

\item A Weighted Dynkin diagram with weights $\alpha_{i}(H)$ given by the
relation $[H,X_{i}] = \alpha_{i}(H) X_{i}$ for $H$ defined above in the
standard Jacobson-Morozov triple and the $X_{i}$ being the positive simple roots.
\end{itemize}

Finally, we remark that the dimension of the orbit is given by:
\begin{align}
\mathrm{dim} (\mathcal{O}) = \mathrm{dim} (\mathfrak{g}) - \mathrm{dim}
(\mathfrak{g}_{0}) - \mathrm{dim} (\mathfrak{g}_{1}),
\end{align}
where
\begin{align}
\mathfrak{g}_{j} = \{Z \in\mathfrak{g} \, | \, [H,Z] = j Z \}.
\end{align}

\section{Review of Anomaly Polynomial Computations \label{app:ANOMALYPOLY}}

In this Appendix, we briefly review the computation of the anomaly polynomial $I_8$ for any 6D SCFT, as originally developed in \cite{Ohmori:2014kda}. For explicit step-by-step examples of anomaly polynomial computations, we refer
the interested reader to section 7.1 of \cite{Heckman:2018jxk}.

In a theory with a well-defined tensor branch and conventional matter,
the anomaly polynomial can be viewed as a sum of two terms: a 1-loop term and
a Green-Schwarz term,
\begin{equation}
I_{8} = I_{\text{1-loop}}+ I_{\text{GS}} .
\label{eq:Itot}
\end{equation}
The full anomaly polynomial of a 6D SCFT takes the form
\begin{align}
I_{8}  &  = \alpha c_{2}(R)^{2} + \beta c_{2}(R) p_{1}(T) + \gamma
p_{1}(T)^{2} + \delta p_{2}(T)\nonumber\\
&  + \sum_{i} \left[  \mu_{i} \, \mathrm{Tr} F_{i}^{4}
+ \, \mathrm{Tr} F_{i}^{2} \left(  \rho_{i}
p_{1}(T) + \sigma_{i} c_{2}(R) + \sum_{j} \eta_{ij} \, \mathrm{Tr} F_{j}^{2}
\right)  \right]  . \label{eq:anomalypolyAPP}%
\end{align}
Here, $c_{2}(R)$ is the second Chern class of the $SU(2)_{R}$ symmetry,
$p_{1}(T)$ is the first Pontryagin class of the tangent bundle, $p_{2}(T)$ is
the second Pontryagin class of the tangent bundle, and $F_{i}$ is the field
strength of the $i^{th}$ symmetry, where $i$ and $j$ run over the flavor symmetries of the theory.

The 1-loop term receives contributions from
free tensor multiplets, vector multiplets, and hypermultiplets:
\begin{align}
I_{\text{tensor}} = \frac{c_{2}(R)^{2}}{24} + \frac{c_{2}(R)p_{1}(T)}{48}  &
+ \frac{23 p_{1}(T)^{2} -116 p_{2}(T)}{5760},\label{eq:tensor}\\
I_{\text{vector}} = -\frac{\tr_{\text{adj}} F^{4} + 6 c_{2}(R) \tr_{\text{adj}%
} F^{2} + d_{G} c_{2}(R)^{2}}{24}  &  - \frac{\tr_{\text{adj}}F^{2}+d_{G}
c_{2}(R)p_{1}(T)}{48}\nonumber\\
&  - d_{G} \frac{7 p_{1}(T)^{2} - 4 p_{2}(T)}{5760},\label{eq:vector}\\
I_{\text{hyper}} = \frac{\tr_{\rho} F^{4} }{24} + \frac{\tr_{\rho}F^{2}
p_{1}(T)}{48}  &  + d_{\rho}\frac{7 p_{1}(T)^{2} - 4 p_{2}(T)}{5760}.
\label{eq:hyper}%
\end{align}
Here, $\tr_{\rho}$ is the trace in the representation $\rho$, $d_{\rho}$ is
the dimension of the representation $\rho$, and $d_{G}$ is the dimension of
the group $G$. In computing the anomaly polynomial, one should convert the
traces in general representations to the trace in a defining
representation. One may write
\begin{align}
\tr_{\rho}F^{4} = x_{\rho}\, \mathrm{Tr} F^{4} + y_{\rho}(\, \mathrm{Tr}
F^{2})^{2}\\
\tr_{\rho}F^{2} = \text{Ind}_{\rho}\, \mathrm{Tr} F^{2} ,
\end{align}
with $x_{\rho}$, $y_{\rho}$, and $\text{Ind}_{\rho}$ well-known constants in
group theory, which can be found in the Appendix of \cite{Ohmori:2014kda} or \cite{Heckman:2018jxk}. For the adjoint representation, $\text{Ind}_{\rho}$ is also known as
the dual Coxeter number, $h_{G}^{\vee}$. Note that the groups $SU(2)$,
$SU(3)$, $G_{2}$, $F_{4}$, $E_{6}$, $E_{7}$, and $E_{8}$ do not have an independent quartic Casimir $\, \mathrm{Tr} F^{4}$, so $x_\rho = 0$ for all representations of these groups.

The Green-Schwarz term takes the form
\begin{equation}
I_{\text{GS}} = \frac{1}{2} A^{ij} I_{i} I_{j},
\end{equation}
where $A^{ij}$ is a negative-definite matrix given by the inverse of the Dirac pairing on the string charge
lattice.
The term $I_i$ can be written as
\begin{equation}
I_{i} = a_{i} c_{2}(R) + b_{i} p_{1}(T) + \sum_{j} c_{ij} \, \mathrm{Tr}
F_{j}^{2}.
\label{eq:Ii}
\end{equation}
The coefficients $a_i$, $b_i$, and $c_{ij}$ are chosen so that the gauge anomalies $(\mathrm{Tr}
F_{i}^{2})^2$ and mixed gauge-gauge or gauge-global anomalies (e.g. $\mathrm{Tr}
F_{i}^{2} \mathrm{Tr} F_{j}^{2}$, $\mathrm{Tr}F_{i}^{2} c_2(R)$, $\mathrm{Tr} F_{i}^{2} p_1(T)$) vanish. In other words, these anomalies must precisely cancel between the Green-Schwarz term and the 1-loop term. In practice, one need not compute the individual $I_i$: one can simply complete the square with respect to the quadratic Casimir $\mathrm{Tr} F_{i}^{2}$ of each of the gauge groups in turn.
This is guaranteed to cancel out the gauge anomalies and mixed gauge anomalies, and what is left
is simply the total anomaly polynomial $I_8$.


\section{Catalogs of Short Quiver Theories}
\label{apdx:shortQuiverCatalogs}

In this Appendix we present explicit catalogs of ``kissing cases'' for $SO(8)$ and $SO(10)$ short quiver theories, each under a particular UV gauge group but varying UV length. For each case, we give the exact ``kissing case'', together with the ``preceding theory'' obtained from the nilpotent orbit but with a slightly longer quiver to illustrate how such collisions between the nilpotent deformations take place. As in \cite{Heckman:2018pqx}, we may compute the anomaly polynomial of the kissing theory directly, but we can also compute it via analytic continuation from a formal type IIA quiver. In most cases, this procedure gives the same result, but in some cases, there is an additional correction term, which we display in the right-hand columns of the following tables. This additional correction term can also be read off from the brane picture, as explained in section \ref{subsubsec:ShortAnomalyRules}.

\begin{longtable}[l]{|@{}c@{}|@{}c@{}|@{}c@{}|@{}c@{}|@{}c@{}|@{}c@{}|@{}c@{}|@{}}
\hline
$\mathcal{O}_L$ & $\mathcal{O}_R$ & Preceding Theory & Kissing Theory & $\#I_{n}$ & $\Delta \alpha$ & $\Delta \beta$\\ \hline \hline
$[7, 1]$ & $[7, 1]$ & $2  \,\,  {{\overset{\ksu(2)}2}}  \,\,   {\overset{\mathfrak{g_{2}}}3}  \,\, \underset{[SU(2)]}1 \,\, {\overset{\mathfrak{g_{2}}}3} \,\, {{\overset{\ksu(2)}2}}  \,\, {2}$ & ${2}  \,\,   \underset{[N_f = 3/2]}{{\overset{\ksu(2)}2}}  \,\,   \underset{[SU(2)]}{\overset{\ksu(2)}2}  \,\, \underset{[N_f = 3/2]}{{\overset{\ksu(2)}2}}  \,\, {2}$ & 2 & $\frac{1}{12}$ & $\frac{1}{24}$ \\ \hline \hline

$[7, 1]$ & $[4^2]$ &
$2  \,\,   {\overset{\ksu(2)}2}  \,\,   {{\overset{\mathfrak{g_{2}}}3}}  \,\, 1 \,\, \underset{[SU(2)]}{{\overset{\mathfrak{so}(7)}3}}\,\, {\overset{\ksu(2)}2} $
&  ${2}  \,\,   \underset{[N_f = 1/2]}{\overset{\ksu(2)}2}  \,\,   \underset{[SU(2)]}{{\overset{\ksu(3)}2}}  \,\, \underset{[N_f = 1]}{\overset{\ksu(2)}2}$ & 2 & $\frac{1}{24}$ & $\frac{1}{48}$ \\ \hline

$[7, 1]$ & $[5, 1^3]$  &
$2  \,\,   {\overset{\ksu(2)}2}  \,\,   {{\overset{\mathfrak{g_{2}}}3}}  \,\, 1 \,\, \underset{[SU(2)]}{{\overset{\mathfrak{so}(7)}3}}\,\, {\overset{\ksu(2)}2}$
& ${2}  \,\,   \underset{[N_f = 1/2]}{\overset{\ksu(2)}2}  \,\,   \underset{[SU(2)]}{{\overset{\ksu(3)}2}}  \,\, \underset{[N_f = 1]}{\overset{\ksu(2)}2}$ & 2 & 0 & 0 \\ \hline

$[7, 1]$ & $[5, 3]$  &
$2  \,\,   {\overset{\ksu(2)}2}  \,\,   {{\overset{\mathfrak{g_{2}}}3}}  \,\, \underset{[SU(2)]}{1} \,\, {\overset{\mathfrak{g_2}}3}\,\, {\overset{\ksu(2)}2}$
& ${2}  \,\,   \underset{[N_f = 3/2]}{{\overset{\ksu(2)}2}}  \,\,   {\overset{\ksu(2)}2}  \,\, {\overset{\ksu(2)}2}  \,\, [SU(2) \times SU(2)]$ & 2 & $\frac{1}{12}$ & $\frac{1}{24}$ \\ \hline \hline

$[4^2]$ & $[4^2]$&
$ {\overset{\ksu(2)}2}  \,\,   \underset{[SU(2)]}{{\overset{\kso(7)}3}}  \,\, 1  \,\, \underset{[SU(2)]}{{\overset{\kso(7)}3}}  \,\, {\overset{\ksu(2)}2}$
& $ \underset{[N_f = 1/2]}{\overset{\ksu(2)}2}  \,\,   \underset{[Sp(2)]}{{\overset{\mathfrak{g_{2}}}2}}  \,\, \underset{[N_f = 1/2]}{\overset{\ksu(2)}2}$ & 2 & 0 & 0 \\ \hline

$[5, 1^3]$ & $[4^2]$ &
$ {\overset{\ksu(2)}2}  \,\,   \underset{[SU(2)]}{{\overset{\kso(7)}3}}  \,\, 1  \,\, \underset{[SU(2)]}{{\overset{\kso(7)}3}}  \,\, {\overset{\ksu(2)}2}$
& $ \underset{[N_f = 1/2]}{\overset{\ksu(2)}2}  \,\,   \underset{[Sp(2)]}{{\overset{\mathfrak{g_{2}}}2}}  \,\, \underset{[N_f = 1/2]}{\overset{\ksu(2)}2}$ & 1 & 0 & 0 \\ \hline

$[5, 1^3]$ & $[5, 1^3]$  &
$ {\overset{\ksu(2)}2}  \,\,   \underset{[SU(2)]}{{\overset{\kso(7)}3}}  \,\, 1  \,\, \underset{[SU(2)]}{{\overset{\kso(7)}3}}  \,\, {\overset{\ksu(2)}2}$
& $ {\overset{\ksu(2)}2}  \,\,   \underset{[SU(4)]}{{\overset{\ksu(4)}2}}  \,\, {\overset{\ksu(2)}2}$ & 0 & 0 & 0 \\ \hline

$[5, 3]$ & $[4^2]$ &
$ {\overset{\ksu(2)}2}  \,\,  {{\overset{\mathfrak{g_{2}}}3}}  \,\, 1  \,\, \underset{[SU(2)]}{{\overset{\kso(7)}3}}  \,\, {\overset{\ksu(2)}2}$
&  $\underset{[N_f = 1]}{\overset{\ksu(2)}2}  \,\,   \underset{[SU(2)]}{{\overset{\ksu(3)}2}}  \,\, \underset{[N_f = 1]}{\overset{\ksu(2)}2}$ & 2 & $\frac{1}{24}$ & $\frac{1}{48}$ \\ \hline

$[5, 3]$ & $[5, 1^3]$ &
$ {\overset{\ksu(2)}2}  \,\,  {{\overset{\mathfrak{g_{2}}}3}}  \,\, 1  \,\, \underset{[SU(2)]}{{\overset{\kso(7)}3}}  \,\, {\overset{\ksu(2)}2}$
& $\underset{[N_f = 1]}{\overset{\ksu(2)}2}  \,\,   \underset{[SU(2)]}{{\overset{\ksu(3)}2}}  \,\, \underset{[N_f = 1]}{\overset{\ksu(2)}2}$ & 2 & 0 & 0 \\ \hline

$[5, 3]$ & $[5, 3]$ &
$ {\overset{\ksu(2)}2}  \,\,  {\overset{\mathfrak{g_{2}}}3}  \,\, \underset{[SU(2)]}{1}  \,\, {{\overset{\mathfrak{g_{2}}}3}}   \,\, {\overset{\ksu(2)}2}$
& $[SU(2) \times SU(2)] \,\,  {\overset{\ksu(2)}2}  \,\,
{\overset{\ksu(2)}2}  \,\, {\overset{\ksu(2)}2}   \,\,    [SU(2)]$ & 2 & $\frac{1}{12}$ & $\frac{1}{24}$ \\ \hline

$[7, 1]$ & $[2^2, 1^4]$ &
${2}  \,\,   {\overset{\ksu(2)}2}  \,\,   {{\overset{\mathfrak{g_{2}}}3}}  \,\, 1 \,\, \underset{[SU(2)^{\otimes 3}]}{{\overset{\mathfrak{so}(8)}3}}$
&$ 2 \,\,  {\overset{\ksu(2)}2}  \,\,
{\overset{\mathfrak{g_{2}}}2}  \,\,   [Sp(3)]$
& 2 & 0 & 0 \\ \hline

$[7, 1]$ & $[2^4]$ &
${2}  \,\,   {\overset{\ksu(2)}2}  \,\,   {{\overset{\mathfrak{g_{2}}}3}}  \,\, 1 \,\, \underset{[Sp(2)]}{{\overset{\mathfrak{so}(7)}3}}$
&${2}  \,\,  \underset{[N_f = 1/2]}{\overset{\ksu(2)}2}  \,\,   {\overset{\ksu(3)}2}  \,\,   [SU(4)]$
& 3 & $\frac{1}{24}$ & $\frac{1}{48}$ \\ \hline

$[7, 1]$ & $[3, 1^5]$ &
${2}  \,\,   {\overset{\ksu(2)}2}  \,\,   {{\overset{\mathfrak{g_{2}}}3}}  \,\, 1 \,\, \underset{[Sp(2)]}{{\overset{\mathfrak{so}(7)}3}}$
& ${2}  \,\,  \underset{[N_f = 1/2]}{\overset{\ksu(2)}2}  \,\,   {\overset{\ksu(3)}2}  \,\,   [SU(4)]$
& 4 & 0 & 0 \\ \hline

$[7, 1]$ & $[3,2^2,1]$ &
${2}  \,\,   {\overset{\ksu(2)}2}  \,\,   {{\overset{\mathfrak{g_{2}}}3}}  \,\, \underset{[SU(2)]}{1} \,\, \underset{[SU(2)]}{{\overset{\mathfrak{g_2}}3}}$
&  ${2}  \,\,  \underset{[N_f = 3/2]}{\overset{\ksu(2)}2}  \,\,   {\overset{\ksu(2)}2}  \,\,   [SU(2) \times SU(2)]$
& 4 & $\frac{1}{12}$ & $\frac{1}{24}$ \\ \hline

$[7, 1]$ & $[3^2, 1^2]$ &
${2}  \,\,   {\overset{\ksu(2)}2}  \,\,   {{\overset{\mathfrak{g_{2}}}3}}  \,\, \underset{[SU(3)]}{1} \,\, {\overset{\ksu(3)}3}$
&  ${2}  \,\,  \underset{[SU(3)]}{\overset{\ksu(2)}2}  \,\,   2$
& 4 & $\frac{1}{6}$ & $\frac{1}{12}$ \\ \hline \hline

$[4^2]$ & $[2^2, 1^4]$ &
$ {\overset{\ksu(2)}2}  \,\,   \underset{[SU(2)]}{\overset{\kso(7)}3} \,\, 1  \,\,  \underset{[SU(2)^{\otimes 3}]}{\overset{\kso(8)}3}   $
& $ {\overset{\ksu(2)}2}  \,\,   {\overset{\kso(7)}2}  \,\,  [Sp(3)\times Sp(1)]$
& 1 & 0 & 0 \\ \hline

$[4^2]$ & $[2^4]$ &
$ {\overset{\ksu(2)}2}  \,\,   \underset{[SU(2)]}{\overset{\kso(7)}3} \,\, 1  \,\,  \underset{[Sp(2)]}{\overset{\kso(7)}3}  $
&  $ \underset{[N_f = 1/2]}{\overset{\ksu(2)}2}  \,\,   {\overset{\mathfrak{g_{2}}}2}  \,\,  [Sp(3)]$
& 3 & 0 & 0 \\ \hline

$[4^2]$ & $[3, 1^5]$ &
$ {\overset{\ksu(2)}2}  \,\,   \underset{[SU(2)]}{\overset{\kso(7)}3}\,\, 1  \,\,  \underset{[Sp(2)]}{\overset{\kso(7)}3}   $
& $ \underset{[N_f = 1/2]}{\overset{\ksu(2)}2}  \,\,   {\overset{\mathfrak{g_{2}}}2}  \,\,  [Sp(3)]$
& 2 & 0 & 0 \\ \hline

$[4^2]$ & $[3, 2^2, 1]$ & $ {\overset{\ksu(2)}2}  \,\,   \underset{[SU(2)]}{\overset{\kso(7)}3} \,\, 1  \,\,  \underset{[SU(2)]}{\overset{\mathfrak{g_{2}}}3}   $
& $\underset{[N_f = 1]}{\overset{\ksu(2)}2}  \,\,   {\overset{\ksu(3)}2}  \,\,  [SU(4)]$ & 3 & $\frac{1}{24}$ & $\frac{1}{48}$ \\ \hline

$[4^2]$ & $[3^2, 1^2]$ & $ {\overset{\ksu(2)}2}  \,\,   \underset{[SU(2)]}{\overset{\kso(7)}3} \,\, \underset{[SU(2)]}{1}  \,\,  {\overset{\ksu(3)}3}  $ & $[SU(2)]  \,\,  {\overset{\ksu(2)}2}  \,\,   {\overset{\ksu(2)}2}  \,\,  [SU(2) \times SU(2)]$ & 4 & $\frac{1}{12}$ & $\frac{1}{24}$ \\ \hline

$[5, 1^3]$ & $[2^2, 1^4]$ &
$ {\overset{\ksu(2)}2}  \,\,   \underset{[SU(2)]}{\overset{\kso(7)}3} \,\, 1  \,\,  \underset{[SU(2)^{\otimes 3}]}{\overset{\kso(8)}3}   $
&$ {\overset{\ksu(2)}2}  \,\,   {\overset{\kso(7)}2}  \,\,  [Sp(3)\times Sp(1)]$
& 0 & 0 & 0 \\ \hline

$[5, 1^3]$ & $[2^4]$ &
$ {\overset{\ksu(2)}2}  \,\,   \underset{[SU(2)]}{\overset{\kso(7)}3} \,\, 1  \,\,  \underset{[Sp(2)]}{\overset{\kso(7)}3}   $
&$ \underset{[N_f = 1/2]}{\overset{\ksu(2)}2}  \,\,   {\overset{\mathfrak{g_{2}}}2}  \,\,  [Sp(3)]$
& 1 & 0 & 0 \\ \hline

$[5, 1^3]$ & $[3, 1^5]$ &
$ {\overset{\ksu(2)}2}  \,\,   \underset{[SU(2)]}{\overset{\kso(7)}3} \,\, 1  \,\,  \underset{[Sp(2)]}{\overset{\kso(7)}3}   $
&${\overset{\ksu(2)}2}  \,\,   {\overset{\ksu(4)}2}  \,\,  [SU(6)]$
& 0 & 0 & 0 \\ \hline

$[5, 1^3]$ & $[3, 2^2, 1]$ &
$ {\overset{\ksu(2)}2}  \,\,   \underset{[SU(2)]}{\overset{\kso(7)}3} \,\, 1  \,\,  \underset{[SU(2)]}{\overset{\mathfrak{g_{2}}}3}   $
& $ \underset{[N_f = 1]}{ \overset{\ksu(2)}2}  \,\,   {\overset{\ksu(3)}2}  \,\,  [SU(4)]$
& 2 & $0$ & $0$ \\ \hline

$[5, 1^3]$ & $[3^2, 1^2]$ &
$ {\overset{\ksu(2)}2}  \,\,   \underset{[SU(2)]}{\overset{\kso(7)}3} \,\, \underset{[SU(2)]}{1}  \,\,  {\overset{\ksu(3)}3}   $
& $[SU(2)]  \,\,  {\overset{\ksu(2)}2}  \,\,   {\overset{\ksu(2)}2}  \,\,  [SU(2) \times SU(2)]$
& 4 & 0 & 0 \\ \hline

$[5, 3]$ & $[2^2, 1^4]$ &
$ {\overset{\ksu(2)}2}  \,\,   {\overset{\mathfrak{g_{2}}}3} \,\, 1  \,\,  \underset{[SU(2)^{\otimes 3}]}{\overset{\kso(8)}3}   $
&$ \underset{[N_f = 1/2]}{\overset{\ksu(2)}2}  \,\,   {\overset{\mathfrak{g_{2}}}2}  \,\,  [Sp(3)]$ & 2 & 0 & 0 \\ \hline

$[5, 3]$ & $[2^4]$ &
$ {\overset{\ksu(2)}2}  \,\,   {\overset{\mathfrak{g_{2}}}3} \,\, 1  \,\,  \underset{[Sp(2)]}{\overset{\kso(7)}3}   $
& $  \underset{[N_f = 1]}{\overset{\ksu(2)}2}  \,\,   {\overset{\ksu(3)}2}  \,\,  [SU(4)]$
& 3 & $\frac{1}{24}$ & $\frac{1}{48}$ \\ \hline

$[5, 3]$ & $[3, 1^5]$ &
$ {\overset{\ksu(2)}2}  \,\,   {\overset{\mathfrak{g_{2}}}3} \,\, 1  \,\,  \underset{[Sp(2)]}{\overset{\kso(7)}3}   $
& $ \underset{[N_f = 1]}{\overset{\ksu(2)}2}  \,\,   {\overset{\ksu(3)}2}  \,\,  [SU(4)]$
& 4 & 0 & 0 \\ \hline

$[5, 3]$ & $[3, 2^2, 1]$ &
$ {\overset{\ksu(2)}2}  \,\,   {\overset{\mathfrak{g_{2}}}3} \,\, \underset{[SU(2)]}{1}  \,\,  \underset{[SU(2)]}{\overset{\mathfrak{g_{2}}}3}   $
& $[SU(2)]  \,\,  {\overset{\ksu(2)}2}  \,\,   {\overset{\ksu(2)}2}  \,\,  [SU(2) \times SU(2)]$
& 4 &  $\frac{1}{12}$ & $\frac{1}{24}$ \\ \hline

$[5, 3]$ & $[3^2, 1^2]$ &
$ {\overset{\ksu(2)}2}  \,\,   {\overset{\mathfrak{g_{2}}}3} \,\, \underset{[SU(3)]}{1}  \,\,  {\overset{\ksu(3)}3}   $
&  $[G_2]  \,\,  {\overset{\ksu(2)}2}  \,\,   2$
& 4 & $\frac{1}{6}$ & $\frac{1}{12}$ \\ \hline \hline

$[2^2, 1^4]$ & $[2^2, 1^4]$ &
$ \underset{[SU(2)^{\otimes 3}]}{\overset{\kso(8)}3} \,\, 1  \,\,  \underset{[SU(2)^{\otimes 3}]}{\overset{\kso(8)}3}   $
& $ {\overset{\kso(8)}2} \,\,  [Sp(2) \times Sp(2) \times Sp(2)]$ & 0 & 0 & 0 \\ \hline

$[2^4]$ & $[2^2, 1^4]$ &
$ \underset{[Sp(2)]}{\overset{\kso(7)}3} \,\, 1  \,\,  \underset{[SU(2)^{\otimes 3}]}{\overset{\kso(8)}3}   $
&$ {\overset{\kso(7)}2} \,\,  [Sp(4) \times Sp(1)]$ & 1 & 0 & 0 \\ \hline

$[3, 1^5]$    & $[2^2, 1^4]$ &
$ \underset{[Sp(2)]}{\overset{\kso(7)}3} \,\, 1  \,\,  \underset{[SU(2)^{\otimes 3}]}{\overset{\kso(8)}3}   $
& $ {\overset{\kso(7)}2} \,\,  [Sp(4) \times Sp(1)]$ & 0 & 0 & 0 \\ \hline

$[2^4]$ & $[2^4]$ &
$ \underset{[Sp(2)]}{\overset{\kso(7)}3} \,\, 1  \,\,  \underset{[Sp(2)]}{\overset{\kso(7)}3}   $
&$ {\overset{\mathfrak{g_{2}}}2} \,\,  [Sp(4)]$
& 4 & 0 & 0 \\ \hline

$[3, 1^5]$    & $[2^4]$ &
$ \underset{[Sp(2)]}{\overset{\kso(7)}3} \,\, 1  \,\,  \underset{[Sp(2)]}{\overset{\kso(7)}3}   $
&$ {\overset{\mathfrak{g_{2}}}2} \,\,  [Sp(4)]$
& 2 & 0 & 0 \\ \hline

$[3, 1^5]$    & $[3, 1^5]$ &
$ \underset{[Sp(2)]}{\overset{\kso(7)}3} \,\, 1  \,\,  \underset{[Sp(2)]}{\overset{\kso(7)}3}   $
&$ {\overset{\ksu(4)}2} \,\,  [SU(8)]$ & 0 & 0 & 0 \\ \hline

$[3, 2^2, 1]$ & $[2^2, 1^4]$ &
$ \underset{[SU(2)]}{\overset{\mathfrak{g_{2}}}3} \,\, 1  \,\,  \underset{[SU(2)^{\otimes 3}]}{\overset{\kso(8)}3}   $
&$ {\overset{\mathfrak{g_{2}}}2} \,\,  [Sp(4)]$ & 2 & 0 & 0 \\ \hline

$[3, 2^2, 1]$  & $[2^4]$  &
$ \underset{[SU(2)]}{\overset{\mathfrak{g_{2}}}3} \,\, 1  \,\, \underset{[Sp(2)]}{\overset{\kso(7)}3}    $
&$ {\overset{\ksu(3)}2} \,\, [SU(6)]$
& 4 & $\frac{1}{24}$ & $\frac{1}{48}$ \\ \hline

$[3, 2^2, 1]$  & $[3, 1^5]$ &
$ \underset{[SU(2)]}{\overset{\mathfrak{g_{2}}}3} \,\, 1  \,\, \underset{[Sp(2)]}{\overset{\kso(7)}3}   $
&$ {\overset{\ksu(3)}2} \,\, [SU(6)]$ & 4 & 0 & 0 \\ \hline

$[3, 2^2, 1]$ & $[3, 2^2, 1]$ &
$ \underset{[SU(2)]}{\overset{\mathfrak{g_{2}}}3} \,\, \underset{[SU(2)]}{1}  \,\, \underset{[SU(2)]}{\overset{\mathfrak{g_{2}}}3} $
&$ {\overset{\ksu(2)}2} \,\,  [SO(7)]$ & 6 & $\frac{1}{12}$ & $\frac{1}{24}$ \\ \hline

$[3^2, 1^2]$ & $[2^2, 1^4]$ &
$ {\overset{\ksu(3)}3} \,\, 1  \,\,  \underset{[SU(2)^{\otimes 3}]}{\overset{\kso(8)}3}   $
& $ {\overset{\ksu(3)}2} \,\, [SU(6)]$  & 4 & 0 & 0 \\ \hline

$[3^2, 1^2]$  & $[2^4]$  &
$ {\overset{\ksu(3)}3} \,\, \underset{[SU(2)]}{1}  \,\, \underset{[Sp(2)]}{\overset{\kso(7)}3}   $
&$ {\overset{\ksu(2)}2} \,\, [SO(7)]$
& 6 & $\frac{1}{12}$ & $\frac{1}{24}$ \\ \hline

$[3^2, 1^2]$  & $[3, 1^5]$ &
$ {\overset{\ksu(3)}3} \,\, \underset{[SU(2)]}{1}  \,\, \underset{[Sp(2)]}{\overset{\kso(7)}3}   $
&$ {\overset{\ksu(2)}2} \,\, [SO(7)]$ & 8 & 0 & 0 \\ \hline

$[3^2, 1^2]$ & $[3, 2^2, 1]$ &
$ {\overset{\ksu(3)}3} \,\, \underset{[SU(3)]}{1}  \,\,  \underset{[SU(2)]}{\overset{\mathfrak{g_{2}}}3}  $
&$ 2 \,\,  [SU(2) \subset Sp(2)_{R}]$ & 7 & $\frac{1}{6}$ & $\frac{1}{12}$ \\ \hline


\caption{A catalog for $SO(8)$ kissing short quiver cases, their preceding longer theory, and the relevant terms for anomaly matching. The $\mathcal{O}_{L,R}$ columns correspond to the left and right deformations. Here $\Delta \alpha = \alpha_{\textrm{formal}} - \alpha_{F}$, and likewise for $\Delta \beta$. The ``Preceding Theory" column gives the theory whose length is one longer than the kissing theory, under the same pair of nilpotent orbits. The ``Theory" column gives the actual deformed short quiver theory, while the $\#I_{n}$ columns stands for the number of anomaly of neutral hypermultiplets to be added to the F-theory quiver in order to match the coefficients $\gamma$ and $\delta$ of the formal quiver. The last entry indicates that there is an $SU(2) \subset Sp(2)_R$ flavor symmetry. By this, we mean that the IR theory ends up flowing to a theory with $\mathcal{N} = (2,0)$ supersymmetry, where the R-symmetry group is $Sp(2)_{R}$. Viewed as an $\mathcal{N} = (1,0)$ SCFT, there is an $SU(2)$ flavor symmetry and an $SU(2)_R$ R-symmetry.}
\label{table:SO8tangentialShortQuiver}
\end{longtable}

{\footnotesize
\begin{longtable}[l]{|@{}c@{}|@{}c@{}|@{}c@{}|@{}c@{}|@{}c@{}|@{}c@{}|@{}c@{}|@{}}
\hline
$\mathcal{O}_L$ & $\mathcal{O}_R$ & ``Preceding Theory" & ``Kissing Theory" & $\#I_{n}$ & $\Delta \alpha$ & $\Delta \beta$\\ \hline \hline
$[9, 1]$ & $[9, 1]$ & $2  \,\,   {\overset{\mathfrak{su}(2)}2}  \,\,   {{\overset{\mathfrak{g}_2}3}} \,\, 1 \,\, {\overset{\mathfrak{so}(8)}4} \,\, 1  {\,\,{\overset{\mathfrak{g}_2}3}}  \,\, {\overset{\mathfrak{su}(2)}2}  \,\, 2$ &  $2  \,\,   \underset{[N_f = 1/2]}{\overset{\mathfrak{su}(2)}2}  \,\,   \underset{[N_f = 1]}{{\overset{\mathfrak{su}(3)}2}} \,\, \underset{[N_f = 1]}{\overset{\mathfrak{su}(3)}2}  \,\, \underset{[N_f = 1/2]}{\overset{\mathfrak{su}(2)}2}  \,\, 2$  & 1 & 0 & 0 \\ \hline \hline
$[9, 1]$ & $[7, 1^3]$ & $2 \,\, {\overset{\mathfrak{su}(2)}2} \,\,  {\overset{\mathfrak{g}_2}3} \,\, 1 \,\, {\overset{\mathfrak{so}(8)}4} \,\ 1 \,\, \underset{[SU(2)]}{{\overset{\mathfrak{so}(7)}3}} \,\, {\overset{\mathfrak{su}(2)}2}$ & $2  \,\,   \underset{[N_f = 1/2]}{\overset{\mathfrak{su}(2)}2}  \,\,   {\overset{\mathfrak{su}(3)}2}  \underset{[SU(3)]}{\,\,{\overset{\mathfrak{su}(4)}2}}  \,\, {\overset{\mathfrak{su}(2)}2}$ & 0 & 0 & 0 \\ \hline

$[9, 1]$ & $[7, 3]$ & $2 \,\, {\overset{\mathfrak{su}(2)}2} \,\,  {\overset{\mathfrak{g}_2}3} \,\, 1 \,\, {\overset{\mathfrak{so}(8)}4} \,\ 1 \,\, {\overset{\mathfrak{g}_2}3} \,\, {\overset{\mathfrak{su}(2)}2}$ &  $2  \,\,   \underset{[N_f = 1/2]}{\overset{\mathfrak{su}(2)}2}  \,\,   \underset{[N_f = 1]}{{\overset{\mathfrak{su}(3)}2}}  \,\, \underset{[N_f = 1]}{{\overset{\mathfrak{su}(3)}2}}  \,\, \underset{[N_f = 1]}{\overset{\mathfrak{su}(2)}2}$ & 1 & 0 & 0 \\ \hline \hline
$[7, 1^3]$ & $[7, 1^3]$ & ${\overset{\mathfrak{su}(2)}2}  \,\,  \underset{[SU(2)]}{{\overset{\mathfrak{so}(7)}3}} \,\, 1 \,\, {\overset{\mathfrak{so}(8)}4} \,\ 1 \,\, \underset{[SU(2)]}{{\overset{\mathfrak{so}(7)}3}} \,\, {\overset{\mathfrak{su}(2)}2}$ & ${\overset{\mathfrak{su}(2)}2}  \,\,  \underset{[SU(2)]}{{\overset{\mathfrak{su}(4)}2}}  \underset{[SU(2)]}{\,\,{\overset{\mathfrak{su}(4)}2}}  \,\,{\overset{\mathfrak{su}(2)}2}$ &0 &0 &0 \\ \hline

$[7,3]$ & $[7, 1^3]$ & ${\overset{\mathfrak{su}(2)}2}  \,\,  {\overset{\mathfrak{g}_2}3} \,\, 1 \,\, {\overset{\mathfrak{so}(8)}4} \,\ 1 \,\, \underset{[SU(2)]}{{\overset{\mathfrak{so}(7)}3}} \,\, {\overset{\mathfrak{su}(2)}2}$ & $\underset{[N_f = 1]}{\overset{\mathfrak{su}(2)}2}  \,\,   {\overset{\mathfrak{su}(3)}2}  \,\,   \underset{[SU(3)]}{{\overset{\mathfrak{su}(4)}2}}  \,\,{\overset{\mathfrak{su}(2)}2}$ &0 &0 &0 \\ \hline

$[7, 3]$ & $[7, 3]$ & ${\overset{\mathfrak{su}(2)}2}  \,\,  {\overset{\mathfrak{g}_2}3} \,\, 1 \,\, {\overset{\mathfrak{so}(8)}4} \,\ 1 \,\, {\overset{\mathfrak{g}_2}3} \,\, {\overset{\mathfrak{su}(2)}2} $ & $\underset{[N_f = 1]}{\overset{\mathfrak{su}(2)}2}  \,\,  \underset{[N_f = 1]}{{\overset{\mathfrak{su}(3)}2}} \,\, \underset{[N_f = 1]}{{\overset{\mathfrak{su}(3)}2}} \,\,\underset{[N_f = 1]}{\overset{\mathfrak{su}(2)}2}$ &1 &0 &0 \\ \hline

$[9, 1]$ & $[4^2, 1^2]$ & $2 \,\, {\overset{\mathfrak{su}(2)}2} \,\,  {\overset{\mathfrak{g}_2}3} \,\, 1 \,\, \underset{[Sp(1)]}{\overset{\mathfrak{so}(9)}4} \,\ 1 \,\, {\overset{\mathfrak{su}(3)}3}$ & $2  \,\,   {\overset{\mathfrak{su}(2)}2}  \,\,   \underset{[Sp(2)]}{{\overset{\mathfrak{g_{2}}}2}}  \,\, \underset{[N_f = 1/2]}{\overset{\mathfrak{su}(2)}2} $&1 &0 &0 \\ \hline

$[9, 1]$ & $[5, 1^5]$ & $2 \,\, {\overset{\mathfrak{su}(2)}2} \,\,  {\overset{\mathfrak{g}_2}3} \,\, 1 \,\, {\overset{\mathfrak{so}(8)}4} \,\ 1 \,\,{\overset{\mathfrak{so}(7)}3} \,\, [Sp(2)]$ & $ 2 \,\,   \underset{[N_f = 1/2]}{\overset{\mathfrak{su}(2)}2}  \,\,   {\overset{\mathfrak{su}(3)}2}  \,\,{\overset{\mathfrak{su}(4)}2}  \,\,  [SU(5)]$ & 0 & 0 & 0 \\ \hline

$[9, 1]$ & $[5, 2^2, 1]$ & $2 \,\, {\overset{\mathfrak{su}(2)}2} \,\,  {\overset{\mathfrak{g}_2}3} \,\, 1 \,\, {\overset{\mathfrak{so}(8)}4} \,\ 1 \,\, \underset{[SU(2)]}{{\overset{\mathfrak{g}_2}3}}$  & $  2 \,\,   \underset{[N_f = 1/2]}{\overset{\mathfrak{su}(2)}2}  \,\,   \underset{[N_f = 1]}{{\overset{\mathfrak{su}(3)}2}}  \,\,{\overset{\mathfrak{su}(3)}2} \,\, [SU(3)]$ & 1 & 0 & 0 \\ \hline

$[9, 1]$ & $[5, 3, 1^2]$ & $2 \,\, {\overset{\mathfrak{su}(2)}2} \,\,  {\overset{\mathfrak{g}_2}3} \,\, 1 \,\, {\overset{\mathfrak{so}(8)}4} \,\ 1 \,\, {{\overset{\mathfrak{su}(3)}3}}$ & $  2  \,\,   \underset{[N_f = 1/2]}{\overset{\mathfrak{su}(2)}2}  \,\,   \underset{[SU(2)]}{{\overset{\mathfrak{su}(3)}2}}  \,\,\underset{[N_f = 1]}{\overset{\mathfrak{su}(2)}2}$ & 2 & 0 & 0 \\ \hline\hline

$[5^2]$ & $[5^2]$ & ${\overset{\mathfrak{su}(2)}2}  \,\,  {\overset{\mathfrak{so}(7)}3} \,\, \underset{[SO(4)]}{{\overset{\mathfrak{sp}(1)}1}}  \,\, {\overset{\mathfrak{so}(7)}3} \,\, {\overset{\mathfrak{su}(2)}2} $  & $[SU(2)]  \,\,  {\overset{\mathfrak{su}(2)}2}  \,\,   {{\overset{\mathfrak{su}(2)}2}}  \,\,{\overset{\mathfrak{su}(2)}2}  \,\,   [SU(2) \times SU(2)]$ & 2 & $\frac{1}{12}$ & $\frac{1}{24}$ \\ \hline

$[7, 1^3]$ & $[4^2, 1^2]$ & ${\overset{\mathfrak{su}(2)}2}  \,\,  \underset{[SU(2)]}{{\overset{\mathfrak{so}(7)}3}} \,\, 1 \,\, \underset{[Sp(1)]}{\overset{\mathfrak{so}(9)}4} \,\ 1 \,\, {{\overset{\mathfrak{su}(3)}3}} $  & ${\overset{\mathfrak{su}(2)}2}  \,\,   \underset{[Sp(2) \times Sp(1)]}{{\overset{\mathfrak{so}(7)}2}}  \,\,{\overset{\mathfrak{su}(2)}2} $ &0 &0 &0 \\ \hline

$[7, 1^3]$ & $[5, 1^5]$ & ${\overset{\mathfrak{su}(2)}2}  \,\,  \underset{[SU(2)]}{{\overset{\mathfrak{so}(7)}3}} \,\, 1 \,\, {\overset{\mathfrak{so}(8)}4} \,\ 1 \,\, {{\overset{\mathfrak{so}(7)}3}} \,\, [Sp(2)] $  & ${\overset{\mathfrak{su}(2)}2}  \,\,   \underset{[SU(2)]}{{\overset{\mathfrak{su}(4)}2}}  \,\,{\overset{\mathfrak{su}(4)}2} \,\, [SU(4)] $ &0 &0 &0 \\ \hline

$[7, 1^3]$ & $[5, 2^2, 1]$ & ${\overset{\mathfrak{su}(2)}2}  \,\,  \underset{[SU(2)]}{{\overset{\mathfrak{so}(7)}3}} \,\, 1 \,\, {\overset{\mathfrak{so}(8)}4} \,\ 1 \,\, {{\overset{\mathfrak{g}_2}3}} \,\, [SU(2)] $ & ${\overset{\mathfrak{su}(2)}2}  \,\,   \underset{[SU(3)]}{{\overset{\mathfrak{su}(4)}2}}  \,\,{\overset{\mathfrak{su}(3)}2} \,\, [SU(2)]$ &0 &0 & 0\\ \hline

$[7, 1^3]$ & $[5, 3, 1^2]$ & ${\overset{\mathfrak{su}(2)}2}  \,\,  \underset{[SU(2)]}{{\overset{\mathfrak{so}(7)}3}} \,\, 1 \,\, {\overset{\mathfrak{so}(8)}4} \,\ 1 \,\, {{\overset{\mathfrak{su}(3)}3}} $ & ${\overset{\mathfrak{su}(2)}2}  \,\,   \underset{[SU(4)]}{{\overset{\mathfrak{su}(4)}2}}  \,\,{\overset{\mathfrak{su}(2)}2} $& 0&0 &0 \\ \hline

$[7, 3]$ & $[4^2, 1^2]$& ${\overset{\mathfrak{su}(2)}2}  \,\,  {\overset{\mathfrak{g}_2}3} \,\, 1 \,\, \underset{[Sp(1)]}{\overset{\mathfrak{so}(9)}4} \,\ 1 \,\, {{\overset{\mathfrak{su}(3)}3}} $  & $\underset{[N_f = 1/2]}{\overset{\mathfrak{su}(2)}2}  \,\,   \underset{[Sp(2)]}{{\overset{\mathfrak{g_{2}}}2}}  \,\, \underset{[N_f = 1/2]}{\overset{\mathfrak{su}(2)}2}$&1 &0&0 \\ \hline

$[7, 3]$ & $[5, 1^5]$ & ${\overset{\mathfrak{su}(2)}2}  \,\, {\overset{\mathfrak{g}_2}3} \,\, 1 \,\, {\overset{\mathfrak{so}(8)}4} \,\ 1 \,\, \underset{[Sp(2)]}{{\overset{\mathfrak{so}(7)}3}} $  & $ \underset{[N_f = 1]}{\overset{\mathfrak{su}(2)}2}  \,\,   {\overset{\mathfrak{su}(3)}2}  \,\,{\overset{\mathfrak{su}(4)}2} \,\,[SU(5)]$ & 0&0 & 0\\ \hline

$[7, 3]$ & $[5, 2^2, 1]$  & ${\overset{\mathfrak{su}(2)}2}  \,\,  {\overset{\mathfrak{g}_2}3} \,\, 1 \,\, {\overset{\mathfrak{so}(8)}4} \,\ 1 \,\, \underset{[SU(2)]}{{\overset{\mathfrak{g}_2}3}} $ & $ \underset{[N_f = 1]}{\overset{\mathfrak{su}(2)}2}  \,\,  \underset{[N_f = 1]}{{\overset{\mathfrak{su}(3)}2}}  \,\,{\overset{\mathfrak{su}(3)}2} \,\, [SU(3)] $  &1 &0 &0 \\ \hline

$[7, 3]$ & $[5, 3, 1^2]$ & ${\overset{\mathfrak{su}(2)}2}  \,\,  {\overset{\mathfrak{g}_2}3} \,\, 1 \,\, {\overset{\mathfrak{so}(8)}4} \,\ 1 \,\, {{\overset{\mathfrak{su}(3)}3}} $ & $ \underset{[N_f = 1]}{\overset{\mathfrak{su}(2)}2}  \,\,   \underset{[SU(2)]}{{\overset{\mathfrak{su}(3)}2}}  \,\, \underset{[N_f = 1]}{\overset{\mathfrak{su}(2)}2}$  &3 & 0& 0\\ \hline \hline

$[4^2, 1^2]$  & $[4^2, 1^2]$ & ${\overset{\mathfrak{su}(3)}3}  \,\, 1 \,\,  \underset{[Sp(2)]}{{\overset{\mathfrak{so}(10)}4}} \,\, 1 \,\,{\overset{\mathfrak{su}(3)}3}$  & $[SU(3)] \,\, {\overset{\mathfrak{su}(3)}2}  \,\,   {\overset{\mathfrak{su}(3)}2} \,\,[SU(3)]$ & 1& 0 & 0\\ \hline

$[5, 1^5]$ & $[4^2, 1^2]$ & $\underset{[Sp(2)]}{\overset{\mathfrak{so}(7)}3}  \,\, 1 \,\,  \underset{[Sp(1)]}{\overset{\mathfrak{so}(9)}4} \,\, 1 \,\,{\overset{\mathfrak{su}(3)}3}$ & $[Sp(3) \times Sp(1)] \,\,{\overset{\mathfrak{so}(7)}2}  \,\,   {\overset{\mathfrak{su}(2)}2}   $ &0 &0 & 0\\ \hline

$[5, 1^5]$  &$[5, 1^5]$ & $\underset{[Sp(2)]}{\overset{\mathfrak{so}(7)}3}  \,\, 1 \,\,  {\overset{\mathfrak{so}(8)}4} \,\, 1 \,\,\underset{[Sp(2)]}{\overset{\mathfrak{so}(7)}3} $ & $[SU(4)] \,\, {\overset{\mathfrak{su}(4)}2}  \,\,   {\overset{\mathfrak{su}(4)}2} \,\,[SU(4)]$  &0 &0 &0 \\ \hline

$[5, 2^2, 1]$ & $[4^2, 1^2]$ & $\underset{[SU(2)]}{\overset{\mathfrak{g}_2}3}  \,\, 1 \,\,  \underset{[Sp(1)]}{\overset{\mathfrak{so}(9)}4} \,\, 1 \,\,{\overset{\mathfrak{su}(3)}3}$  &$[Sp(3)] \,\,{\overset{\mathfrak{g_{2}}}2}  \,\,   \underset{[N_f = 1/2]}{\overset{\mathfrak{su}(2)}2}   $& 1&0 &0 \\ \hline

$[5, 2^2, 1]$ & $[5, 1^5]$  &$\underset{[SU(2)]}{\overset{\mathfrak{g}_2}3}  \,\, 1 \,\,  {\overset{\mathfrak{so}(8)}4} \,\, 1 \,\,\underset{[Sp(2)]}{\overset{\mathfrak{so}(7)}3} $ &  $[SU(2)] \,\,{\overset{\mathfrak{su}(3)}2}  \,\,   {\overset{\mathfrak{su}(4)}2} \,\,[SU(5)]$ &0 &0 &0 \\ \hline

$[5, 2^2, 1]$ & $[5, 2^2, 1]$ & $\underset{[SU(2)]}{\overset{\mathfrak{g}_2}3}  \,\, 1 \,\,  {\overset{\mathfrak{so}(8)}4} \,\, 1 \,\,\underset{[SU(2)]}{\overset{\mathfrak{g}_2}3} $ & $[SU(3)] \,\, {\overset{\mathfrak{su}(3)}2}  \,\,   {\overset{\mathfrak{su}(3)}2} \,\,[SU(3)]$ &1 &0&0 \\ \hline

$[5, 3, 1^2]$ & $[4^2, 1^2]$  &  ${\overset{\mathfrak{su}(3)}3}  \,\, 1 \,\,  \underset{[Sp(1)]}{\overset{\mathfrak{so}(9)}4} \,\, 1 \,\,{\overset{\mathfrak{su}(3)}3}$ & $[SU(4)] \,\, {\overset{\mathfrak{su}(3)}2}  \,\,   \underset{[N_f = 1]}{\overset{\mathfrak{su}(2)}2} $ &2 &0 &0 \\ \hline

$[5, 3, 1^2]$ & $[5, 1^5]$   & ${\overset{\mathfrak{su}(3)}3}  \,\, 1 \,\,  {\overset{\mathfrak{so}(8)}4} \,\, 1 \,\,\underset{[Sp(2)]}{\overset{\mathfrak{so}(7)}3} $ & ${\overset{\mathfrak{su}(2)}2}  \,\,   {\overset{\mathfrak{su}(4)}2} \,\,[SU(6)]  $  &0 &0 & 0\\ \hline

$[5, 3, 1^2]$ & $[5, 2^2, 1]$   & ${\overset{\mathfrak{su}(3)}3}  \,\, 1 \,\,  {\overset{\mathfrak{so}(8)}4} \,\, 1 \,\,\underset{[SU(2)]}{\overset{\mathfrak{g}_2}3} $  & $\underset{[N_f = 1]}{\overset{\mathfrak{su}(2)}2}  \,\,   {\overset{\mathfrak{su}(3)}2} \,\, [SU(4)] $  &2 &0&0 \\ \hline

$[5, 3, 1^2]$ & $[5, 3, 1^2]$  & ${\overset{\mathfrak{su}(3)}3}  \,\, 1 \,\,  {\overset{\mathfrak{so}(8)}4} \,\, 1 \,\,{\overset{\mathfrak{su}(3)}3} $ & $[SU(2)] \,\,{\overset{\mathfrak{su}(2)}2}  \,\,   {\overset{\mathfrak{su}(2)}2}\,\,[SU(2) \times SU(2)]$   &4 &0 &0 \\ \hline

$[5^2]$ & $[2^4, 1^2]$      & ${\overset{\mathfrak{su}(2)}2}  \,\,  {\overset{\mathfrak{so}(7)}3} \,\, \underset{[N_f = 1]}{\overset{\mathfrak{sp}(1)}1} \,\, \underset{[N_s = 1]}{\overset{\mathfrak{so}(10)}3} \,\, [Sp(2)] $ & ${\overset{\mathfrak{su}(2)}2}  \,\,{\overset{\mathfrak{so}(7)}2} \,\, [Sp(3) \times Sp(1)]$  &1  & 0 &0  \\ \hline

$[5^2]$ & $[3, 2^2, 1^3]$     & ${\overset{\mathfrak{su}(2)}2}  \,\,  {\overset{\mathfrak{so}(7)}3} \,\, \underset{[SO(3)]}{\overset{\mathfrak{sp}(1)}1} \,\, \underset{[Sp(1)\times Sp(1)]}{\overset{\mathfrak{so}(9)}3} $ &  $\underset{[N_f = 1/2]}{\overset{\mathfrak{su}(2)}2}  \,\,{\overset{\mathfrak{g_{2}}}2} \,\, [Sp(3)]$  &2  &0 &  0\\ \hline

$[5^2]$ & $[3^2, 1^4]$      & ${\overset{\mathfrak{su}(2)}2}  \,\,  {\overset{\mathfrak{so}(7)}3} \,\, \underset{[SO(4)]}{\overset{\mathfrak{sp}(1)}1} \,\, \underset{[Sp(1) \times Sp(1)]}{\overset{\mathfrak{so}(8)}3} $ &  $\underset{[N_f = 1]}{\overset{\mathfrak{su}(2)}2}  \,\,   {\overset{\mathfrak{su}(3)}2} \,\, [SU(4)] $  &4  &0  & 0 \\ \hline

$[5^2]$ & $[3^2,2^2]$      &${\overset{\mathfrak{su}(2)}2}  \,\,  {\overset{\mathfrak{so}(7)}3} \,\, \underset{[SO(4)]}{\overset{\mathfrak{sp}(1)}1} \,\, \underset{[Sp(2)]}{\overset{\mathfrak{so}(7)}3} $&  $[SU(2)] \,\,{\overset{\mathfrak{su}(2)}2}  \,\,   {\overset{\mathfrak{su}(2)}2}\,\,[SU(2) \times SU(2)]$ & 4&$\frac{1}{12}$ &$\frac{1}{24}$  \\ \hline

$[5^2]$ & $[3^3, 1]$        &${\overset{\mathfrak{su}(2)}2}  \,\,  {\overset{\mathfrak{so}(7)}3} \,\, \underset{{[SO(5)]}}{\overset{\mathfrak{sp}(1)}1} \,\, {\overset{\mathfrak{g}_2}3} $  &  $[G_2] \,\, {\overset{\mathfrak{su}(2)}2} \,\,   2 $  &4 &$\frac{1}{6}$ &$\frac{1}{12}$   \\ \hline \hline

$[2^4, 1^2]$& $[2^4, 1^2]$& $[Sp(2)] \,\, \underset{[N_s = 1]}{\overset{\mathfrak{so}(10)}3}  \,\, {\overset{\mathfrak{sp}(1)}1} \,\,  \underset{[N_s = 1]}{\overset{\mathfrak{so}(10)}3} \,\, [Sp(2)] $ & $ {\overset{\mathfrak{so}(10)}2}  \,\,  [Sp(4) \times SU(2)]$ & 0 & 0  & 0 \\ \hline

$[3,2^2,1^3]$& $[2^4, 1^2]$& $ \underset{[Sp(1)\times Sp(1)]}{\overset{\mathfrak{so}(9)}3}  \,\, \underset{[N_f = 1/2]}{\overset{\mathfrak{sp}(1)}1} \,\,  \underset{[N_s = 1]}{\overset{\mathfrak{so}(10)}3} \,\, [Sp(2)]$ & $ {\overset{\mathfrak{so}(9)}2} \,\, [Sp(3) \times Sp(2)] $ & 0 & 0 & 0\\ \hline

$[3,2^2,1^3]$&$[3, 2^2, 1^3]$& $ \underset{[Sp(1)\times Sp(1)]}{\overset{\mathfrak{so}(9)}3}  \,\, \underset{[N_f = 1]}{\overset{\mathfrak{sp}(1)}1} \,\,  \underset{[Sp(1)\times Sp(1)]}{\overset{\mathfrak{so}(9)}3} $ & $ {\overset{\mathfrak{so}(8)}2} \,\, [Sp(2) \times Sp(2) \times Sp(2)]  $ & 0 & 0 & 0\\ \hline

$[3^2, 1^4]$& $[2^4, 1^2]$& $ \underset{[Sp(1) \times Sp(1)]}{\overset{\mathfrak{so}(8)}3}  \,\, \underset{[N_f = 1]}{\overset{\mathfrak{sp}(1)}1} \,\,  \underset{[N_s = 1]}{\overset{\mathfrak{so}(10)}3} \,\, [Sp(2)]$ & $ {\overset{\mathfrak{so}(8)}2}\,\, [Sp(2) \times Sp(2) \times Sp(2)] $ & 0 & 0 & 0\\ \hline

$[3^2, 1^4]$& $[3, 2^2, 1^3]$& $ \underset{[Sp(1) \times Sp(1)]}{\overset{\mathfrak{so}(8)}3}  \,\, \underset{[SO(3)]}{\overset{\mathfrak{sp}(1)}1} \,\,  \underset{[Sp(1) \times Sp(1)]}{\overset{\mathfrak{so}(9)}3} $ & $ {\overset{\mathfrak{so}(7)}2} \,\, [Sp(4) \times Sp(1) ]$ & 0 & 0 & 0\\ \hline

$[3^2, 1^4]$& $[3^2, 1^4]$& $ \underset{[Sp(1) \times Sp(1)]}{\overset{\mathfrak{so}(8)}3}  \,\, \underset{[SO(4)]}{\overset{\mathfrak{sp}(1)}1} \,\,  \underset{[Sp(1) \times Sp(1)]}{\overset{\mathfrak{so}(8)}3} $ & $ {\overset{\mathfrak{su}(4)}2} \,\, [SU(8)]$ & 0 & 0 & 0\\ \hline

$[3^2,2^2]$& $[2^4, 1^2]$& $ \underset{[Sp(1)]}{\overset{\mathfrak{so}(7)}3}  \,\, \underset{[N_f = 1]}{\overset{\mathfrak{sp}(1)}1} \,\,  \underset{[N_s = 1]}{\overset{\mathfrak{so}(10)}3} \,\, [Sp(2)] $ & $ {\overset{\mathfrak{so}(7)}2} \,\, [Sp(4) \times Sp(1) ] $ & 1 & 0 & 0\\ \hline

$[3^2,2^2]$& $[3, 2^2, 1^3]$& $ \underset{[Sp(1)]}{\overset{\mathfrak{so}(7)}3}  \,\, \underset{[SO(3)]}{\overset{\mathfrak{sp}(1)}1} \,\,  \underset{[Sp(1) \times Sp(1)]}{\overset{\mathfrak{so}(9)}3} $ & $ {\overset{\mathfrak{g_{2}}}2} \,\, [Sp(4)] $ &2& 0 & 0\\ \hline

$[3^2,2^2]$& $[3^2, 1^4]$& $ \underset{[Sp(1)]}{\overset{\mathfrak{so}(7)}3}  \,\, \underset{[SO(4)]}{\overset{\mathfrak{sp}(1)}1} \,\,  \underset{[Sp(1) \times Sp(1)]}{\overset{\mathfrak{so}(8)}3} $ & $ {\overset{\mathfrak{su}(3)}2} \,\, [SU(6)]$ & 4 & 0 & 0\\ \hline

$[3^2,2^2]$& $[3^2, 2^2]$& $ \underset{[Sp(1)]}{\overset{\mathfrak{so}(7)}3}  \,\, \underset{[SO(4)]}{\overset{\mathfrak{sp}(1)}1} \,\,  \underset{[Sp(1)]}{\overset{\mathfrak{so}(7)}3} $ & $ {\overset{\mathfrak{su}(2)}2} \,\, [SO(7)]$ & 6 & $\frac{1}{12}$ & $\frac{1}{24}$\\ \hline

$[3^3,  1]$& $[2^4, 1^2]$& $ {\overset{\mathfrak{g}_2}3}  \,\, \underset{{[SO(3)]}}{\overset{\mathfrak{sp}(1)}1} \,\,  \underset{[N_s = 1]}{\overset{\mathfrak{so}(10)}3} \,\, [Sp(2)] $ & $ {\overset{\mathfrak{g_{2}}}2} \,\, [Sp(4)]$ & 2 & 0 & 0\\ \hline

$[3^3,  1]$& $[3,2^2,1^3]$& $ {\overset{\mathfrak{g}_2}3}  \,\, \underset{{[SO(4)]}}{\overset{\mathfrak{sp}(1)}1} \,\,  \underset{[Sp(1) \times Sp(1)]}{\overset{\mathfrak{so}(9)}3} $ & $ {\overset{\mathfrak{su}(3)}2} \,\, [SU(6)]$ & 4 & 0 & 0\\ \hline

$[3^3,  1]$& $[3^2, 1^4]$& $ {\overset{\mathfrak{g}_2}3}  \,\, \underset{{[SO(5)]}}{\overset{\mathfrak{sp}(1)}1} \,\,  \underset{[Sp(1) \times Sp(1)]}{\overset{\mathfrak{so}(8)}3} $ & $ {\overset{\mathfrak{su}(2)}2} \,\, [SO(7)]$ & 8 & 0 & 0\\ \hline

$[3^3,  1]$& $[3^2, 2^2]$& $ {\overset{\mathfrak{g}_2}3}  \,\underset{{[SO(5)]}}{\overset{\mathfrak{sp}(1)}1} \,\,  \underset{[Sp(1)]}{\overset{\mathfrak{so}(7)}3} $ & $ 2 \,\, [SU(2) \subset Sp(2)_{R}]$ & 7 & $\frac{1}{6}$ & $\frac{1}{12}$\\ \hline


%

\caption{$SO(10)$ short quiver tangential cases, in parallel to table \ref{table:SO8tangentialShortQuiver}. See table \ref{table:SO8tangentialShortQuiver} for conventions and notation.}
\label{table:SO10TangentshortQuiver}
\end{longtable}
}

\section{Generators of $E_{6,7,8}$}

\label{app:ExpMat} In this section we list the generators $X_i$ and $Y_i$ for the exceptional
algebras $E_{6,7,8}$ in the basis used throughout this chapter. All
other generators can be obtained from appropriate commutators.

The six positive simple roots of $E_{6}$ are associated with:
\begin{align}
X_{1}  &  = E_{1,2}+E_{12,13}+E_{15,16}+E_{17,18}+E_{19,20}+E_{21,22}%
,\nonumber\\
X_{2}  &  = E_{4,6}+E_{5,8}+E_{7,9}+E_{19,21}+E_{20,22}+E_{23,24},\nonumber\\
X_{3}  &  = E_{2,3}+E_{10,12}+E_{11,15}+E_{14,17}+E_{20,23}+E_{22,24}%
,\nonumber\\
X_{4}  &  = E_{3,4}+E_{8,10}+E_{9,11}+E_{17,19}+E_{18,20}+E_{24,25}%
,\nonumber\\
X_{5}  &  = E_{4,5}+E_{6,8}+E_{11,14}+E_{15,17}+E_{16,18}+E_{25,26}%
,\nonumber\\
X_{6}  &  = E_{5,7}+E_{8,9}+E_{10,11}+E_{12,15}+E_{13,16}+E_{26,27}.
\end{align}
The corresponding negative roots are $Y_{i} = X_{i}^{T}$ and Cartans $H_{i} =
[X_{i},Y_{i}]$.

The seven positive simple roots of $E_{7}$ are taken to be:
\begin{align}
X_{1}  &  = E_{7,8}+E_{9,10}+E_{11,12}+E_{13,14}+E_{16,17}+E_{19,20}%
+E_{37,38}+E_{40,41}+E_{43,44}\nonumber \\ &+E_{45,46}+E_{47,48}+E_{49,50},\nonumber\\
X_{2}  &  = E_{5,6}+E_{7,9}+E_{8,10}+E_{22,25}+E_{24,28}+E_{26,30}%
+E_{27,31}+E_{29,33}+E_{32,35}\nonumber \\ &+E_{47,49}+E_{48,50}+E_{51,52},\nonumber\\
X_{3}  &  = E_{5,7}+E_{6,9}+E_{12,15}+E_{14,18}+E_{17,21}+E_{20,23}%
+E_{34,37}+E_{36,40}+E_{39,43}\nonumber \\ &+E_{42,45}+E_{48,51}+E_{50,52},\nonumber\\
X_{4}  &  = E_{4,5}+E_{9,11}+E_{10,12}+E_{18,22}+E_{21,24}+E_{23,26}%
+E_{31,34}+E_{33,36}+E_{35,39}\nonumber \\ &+E_{45,47}+E_{46,48}+E_{52,53},\nonumber\\
X_{5}  &  = E_{3,4}+E_{11,13}+E_{12,14}+E_{15,18}+E_{24,27}+E_{26,29}%
+E_{28,31}+E_{30,33}+E_{39,42}\nonumber \\ &+E_{43,45}+E_{44,46}+E_{53,54},\nonumber\\
X_{6}  &  = E_{2,3}+E_{13,16}+E_{14,17}+E_{18,21}+E_{22,24}+E_{25,28}%
+E_{29,32}+E_{33,35}+E_{36,39}\nonumber \\ &+E_{40,43}+E_{41,44}+E_{54,55},\nonumber\\
X_{7}  &  = E_{1,2}+E_{16,19}+E_{17,20}+E_{21,23}+E_{24,26}+E_{27,29}%
+E_{28,30}+E_{31,33}+E_{34,36}\nonumber \\ &+E_{37,40}+E_{38,41}+E_{55,56}.
\end{align}
Again corresponding negative roots are $Y_{i} = X_{i}^{T}$ and Cartans $H_{i}
= [X_{i},Y_{i}]$.

Finally, the eight positive simple roots of $E_{8}$ are taken to be:
\begin{align}
X_{1}  &  = E_{8,9}+E_{10,11}+E_{12,13}+E_{14,15}+E_{17,18}+E_{20,21}%
+E_{24,25}+E_{46,47}+E_{52,53}+E_{57,59}\nonumber \\ &+E_{58,60}+E_{63,65} +E_{64,66}+E_{68,71}+E_{69,72}+E_{70,73}+E_{75,78}+E_{76,79}+E_{77,80}%
+E_{82,85}\nonumber \\ &+E_{83,86}+E_{84,87}+E_{90,92}+E_{91,93}+E_{97,99}+E_{98,100}+E_{105,106}+E_{112,113}+E_{120,121}\nonumber \\ &+2 E_{121,129}
-E_{122,129}+E_{136,137}+E_{143,144}+E_{149,151}+E_{150,152}+E_{156,158}+E_{157,159}\nonumber \\ &+E_{162,165}+E_{163,166}+E_{164,167}
+E_{169,172}+E_{170,173}+E_{171,174}+E_{176,179}+E_{177,180}\nonumber \\ &+E_{178,181}+E_{183,185}+E_{184,186}+E_{189,191}+E_{190,192}
+E_{196,197}+E_{202,203}+E_{224,225}\nonumber \\ &+E_{228,229}+E_{231,232}+E_{234,235}+E_{236,237}+E_{238,239}+E_{240,241},\nonumber\\
X_{2}  &  = -E_{6,7}-E_{8,10}-E_{9,11}-E_{23,28}-E_{27,32}-E_{30,35}
-E_{31,36}-E_{33,39}-E_{34,40}-E_{37,43}\nonumber \\ &-E_{38,44}-E_{42,49}-E_{48,54}-E_{70,77}-E_{73,80}-E_{76,84}-E_{79,87}-E_{81,89}-E_{83,91}-E_{86,93}\nonumber \\ &-E_{88,95}-E_{90,98}-E_{92,100}-E_{94,102}-E_{97,105}-E_{99,106}-E_{101,108}-E_{107,114}+E_{115,128}\nonumber \\ &-E_{123,134}+2E_{128,134}-E_{135,142}-E_{141,148}-E_{143,150}-E_{144,152}-E_{147,155}-E_{149,157}\nonumber \\ &-E_{151,159}-E_{154,161}-E_{156,163}%
-E_{158,166}-E_{160,168}-E_{162,170}-E_{165,173}-E_{169,176}\nonumber \\ &-E_{172,179}-E_{195,201}-E_{200,207}-E_{205,211}-E_{206,212}%
-E_{209,215}-E_{210,216}-E_{213,218}\nonumber\\
&  -E_{214,219}-E_{217,222}-E_{221,226}-E_{238,240}-E_{239,241}-E_{242,243}%
,\nonumber\\
X_{3}  &  = -E_{6,8}-E_{7,10}-E_{13,16}-E_{15,19}-E_{18,22}-E_{21,26}%
-E_{25,29}-E_{41,46}-E_{45,52}-E_{50,57}\nonumber \\ &-E_{51,58}-E_{55,63}-E_{56,64}-E_{61,68}-E_{62,69}-E_{67,75}-E_{73,81}-E_{74,82}-E_{79,88}%
-E_{80,89}\nonumber \\ &-E_{86,94}-E_{87,95}-E_{92,101}-E_{93,102}-E_{99,107}-E_{100,108}-E_{106,114}-E_{112,120}+E_{113,122}\nonumber \\ &-E_{121,136}+2
E_{122,136}-E_{123,136}-E_{129,137}-E_{135,143}-E_{141,149}-E_{142,150}-E_{147,156}\nonumber \\ &-E_{148,157}-E_{154,162}-E_{155,163}%
-E_{160,169}-E_{161,170}-E_{167,175}-E_{168,176}-E_{174,182}\nonumber \\ &-E_{180,187}-E_{181,188}-E_{185,193}-E_{186,194}-E_{191,198}%
-E_{192,199}-E_{197,204}-E_{203,208}\nonumber \\ &-E_{220,224}-E_{223,228}-E_{227,231}-E_{230,234}-E_{233,236}-E_{239,242}-E_{241,243}%
,\nonumber\\
X_{4}  &  = E_{5,6}+E_{10,12}+E_{11,13}+E_{19,23}+E_{22,27}+E_{26,30}%
+E_{29,33}+E_{36,41}+E_{40,45}+E_{43,50}\nonumber \\ &+E_{44,51}+E_{49,55}+E_{54,61}+E_{64,70}+E_{66,73}+E_{69,76}+E_{72,79}+E_{75,83}+E_{78,86}+E_{82,90}\nonumber \\ &+E_{85,92}+E_{89,96}+E_{95,103}+E_{102,109}+E_{105,112}+E_{106,113}+E_{107,115}+E_{108,116}\nonumber \\ &+E_{114,123}-E_{122,135}+2
E_{123,135}-E_{124,135}-E_{128,135}+E_{133,141}+E_{134,142}+E_{136,143}\nonumber \\ &+E_{137,144}+E_{140,147}+E_{146,154}+E_{153,160}+E_{157,164}%
+E_{159,167}+E_{163,171}+E_{166,174}\nonumber \\ &+E_{170,177}+E_{173,180}+E_{176,183}+E_{179,185}+E_{188,195}+E_{194,200}+E_{198,205}%
+E_{199,206}\nonumber \\ &+E_{204,209}+E_{208,213}+E_{216,220}+E_{219,223}+E_{222,227}+E_{226,230}+E_{236,238}+E_{237,239}\nonumber \\ &+E_{243,244}%
,\nonumber\\
X_{5}  &  = -E_{4,5}-E_{12,14}-E_{13,15}-E_{16,19}-E_{27,31}-E_{30,34}%
-E_{32,36}-E_{33,37}-E_{35,40}-E_{39,43}\nonumber \\ &-E_{51,56}-E_{55,62}-E_{58,64}-E_{60,66}-E_{61,67}-E_{63,69}-E_{65,72}-E_{68,75}-E_{71,78}%
-E_{90,97}\nonumber \\ &-E_{92,99}-E_{96,104}-E_{98,105}-E_{100,106}-E_{101,107}-E_{103,110}-E_{108,114}-E_{109,117}\nonumber \\ &+E_{116,124}-E_{123,133}+2
E_{124,133}-E_{125,133}-E_{132,140}-E_{135,141}-E_{139,146}-E_{142,148}\nonumber \\ &-E_{143,149}-E_{144,151}-E_{145,153}-E_{150,157}%
-E_{152,159}-E_{171,178}-E_{174,181}-E_{177,184}\nonumber \\ &-E_{180,186}-E_{182,188}-E_{183,189}-E_{185,191}-E_{187,194}-E_{193,198}%
-E_{206,210}-E_{209,214}\nonumber \\ &-E_{212,216}-E_{213,217}-E_{215,219}-E_{218,222}-E_{230,233}-E_{234,236}-E_{235,237}-E_{244,245}%
,\nonumber\\
X_{6}  &  = E_{3,4}+E_{14,17}+E_{15,18}+E_{19,22}+E_{23,27}+E_{28,32}%
+E_{34,38}+E_{37,42}+E_{40,44}+E_{43,49}\nonumber \\ &+E_{45,51}+E_{50,55}+E_{52,58}+E_{53,60}+E_{57,63}+E_{59,65}+E_{67,74}+E_{75,82}+E_{78,85}%
+E_{83,90}\nonumber \\ &+E_{86,92}+E_{91,98}+E_{93,100}+E_{94,101}+E_{102,108}+E_{104,111}+E_{109,116}+E_{110,118}\nonumber \\ &+E_{117,125}-E_{124,132}+2
E_{125,132}-E_{126,132}+E_{131,139}+E_{133,140}+E_{138,145}+E_{141,147}\nonumber \\ &+E_{148,155}+E_{149,156}+E_{151,158}+E_{157,163}%
+E_{159,166}+E_{164,171}+E_{167,174}+E_{175,182}\nonumber\\
&  +E_{184,190}+E_{186,192}+E_{189,196}+E_{191,197}+E_{194,199}+E_{198,204}%
+E_{200,206}+E_{205,209}\nonumber \\ &+E_{207,212}+E_{211,215}+E_{217,221}+E_{222,226}+E_{227,230}+E_{231,234}+E_{232,235}+E_{245,246}%
,\nonumber\\
X_{7}  &  = -E_{2,3}-E_{17,20}-E_{18,21}-E_{22,26}-E_{27,30}-E_{31,34}%
-E_{32,35}-E_{36,40}-E_{41,45}-E_{42,48}\nonumber \\ &-E_{46,52}-E_{47,53}-E_{49,54}-E_{55,61}-E_{62,67}-E_{63,68}-E_{65,71}-E_{69,75}-E_{72,78}%
-E_{76,83}\nonumber \\ &-E_{79,86}-E_{84,91}-E_{87,93}-E_{88,94}-E_{95,102}-E_{103,109}-E_{110,117}-E_{111,119}\nonumber \\ &+E_{118,126}-E_{125,131}+2
E_{126,131}-E_{127,131}-E_{130,138}-E_{132,139}-E_{140,146}-E_{147,154}\nonumber \\ &-E_{155,161}-E_{156,162}-E_{158,165}-E_{163,170}%
-E_{166,173}-E_{171,177}-E_{174,180}-E_{178,184}\nonumber\\
&  -E_{181,186}-E_{182,187}-E_{188,194}-E_{195,200}-E_{196,202}-E_{197,203}%
-E_{201,207}-E_{204,208}\nonumber \\ &-E_{209,213}-E_{214,217}-E_{215,218}-E_{219,222}-E_{223,227}-E_{228,231}-E_{229,232}-E_{246,247}%
,\nonumber\\
X_{8}  &  = E_{1,2}+E_{20,24}+E_{21,25}+E_{26,29}+E_{30,33}+E_{34,37}%
+E_{35,39}+E_{38,42}+E_{40,43}+E_{44,49}\nonumber \\ &+E_{45,50}+E_{51,55}+E_{52,57}+E_{53,59}+E_{56,62}+E_{58,63}+E_{60,65}+E_{64,69}+E_{66,72}%
+E_{70,76}\nonumber \\ &+E_{73,79}+E_{77,84}+E_{80,87}+E_{81,88}+E_{89,95}+E_{96,103}+E_{104,110}+E_{111,118}\nonumber \\ &+E_{119,127}-E_{126,130}+2
E_{127,130}+E_{131,138}+E_{139,145}+E_{146,153}+E_{154,160}+E_{161,168}\nonumber \\ &+E_{162,169}+E_{165,172}+E_{170,176}+E_{173,179}%
+E_{177,183}+E_{180,185}+E_{184,189}+E_{186,191}\nonumber\\
&  +E_{187,193}+E_{190,196}+E_{192,197}+E_{194,198}+E_{199,204}+E_{200,205}%
+E_{206,209}+E_{207,211}\nonumber \\ &+E_{210,214}+E_{212,215}+E_{216,219}+E_{220,223}+E_{224,228}+E_{225,229}+E_{247,248}.
\end{align}
The corresponding negative roots are almost the transpose of these positive
roots:
\begin{align}
Y_{1}  &  = E_{9,8}+E_{11,10}+E_{13,12}+E_{15,14}+E_{18,17}+E_{21,20}%
+E_{25,24}+E_{47,46}+E_{53,52}+E_{59,57}\nonumber \\ &+E_{60,58}+E_{65,63}+E_{66,64}+E_{71,68}+E_{72,69}+E_{73,70}+E_{78,75}+E_{79,76}+E_{80,77}%
+E_{85,82}\nonumber \\ &+E_{86,83}+E_{87,84}+E_{92,90}+E_{93,91}+E_{99,97}+E_{100,98}+E_{106,105}+E_{113,112}+2 E_{121,120}\nonumber \\ &-E_{122,120}%
+E_{129,121}+E_{137,136}+E_{144,143}+E_{151,149}+E_{152,150}+E_{158,156}+E_{159,157}\nonumber \\ &+E_{165,162}+E_{166,163}+E_{167,164}%
+E_{172,169}+E_{173,170}+E_{174,171}+E_{179,176}+E_{180,177}\nonumber \\ &+E_{181,178}+E_{185,183}+E_{186,184}+E_{191,189}+E_{192,190}%
+E_{197,196}+E_{203,202}+E_{225,224}\nonumber \\ &+E_{229,228}+E_{232,231}+E_{235,234}+E_{237,236}+E_{239,238}+E_{241,240},\nonumber\\
Y_{2}  &  = -E_{7,6}-E_{10,8}-E_{11,9}-E_{28,23}-E_{32,27}-E_{35,30}%
-E_{36,31}-E_{39,33}-E_{40,34}-E_{43,37}\nonumber \\ &-E_{44,38}-E_{49,42}-E_{54,48}-E_{77,70}-E_{80,73}-E_{84,76}-E_{87,79}-E_{89,81}-E_{91,83}%
-E_{93,86}\nonumber \\ &-E_{95,88}-E_{98,90}-E_{100,92}-E_{102,94}-E_{105,97}-E_{106,99}-E_{108,101}-E_{114,107}-E_{123,115}\nonumber \\ &+2 E_{128,115}%
+E_{134,128}-E_{142,135}-E_{148,141}-E_{150,143}-E_{152,144}-E_{155,147}-E_{157,149}\nonumber \\ &-E_{159,151}-E_{161,154}-E_{163,156}%
-E_{166,158}-E_{168,160}-E_{170,162}-E_{173,165}-E_{176,169}\nonumber \\ &-E_{179,172}-E_{201,195}-E_{207,200}-E_{211,205}-E_{212,206}%
-E_{215,209}-E_{216,210}-E_{218,213}\nonumber \\ &-E_{219,214}-E_{222,217}-E_{226,221}-E_{240,238}-E_{241,239}-E_{243,242},\nonumber\\
Y_{3}  &  = -E_{8,6}-E_{10,7}-E_{16,13}-E_{19,15}-E_{22,18}-E_{26,21}%
-E_{29,25}-E_{46,41}-E_{52,45}-E_{57,50}\nonumber \\ &-E_{58,51}-E_{63,55}-E_{64,56}-E_{68,61}-E_{69,62}-E_{75,67}-E_{81,73}-E_{82,74}-E_{88,79}%
-E_{89,80}\nonumber \\ &-E_{94,86}-E_{95,87}-E_{101,92}-E_{102,93}-E_{107,99}-E_{108,100}-E_{114,106}-E_{120,112}-E_{121,113}\nonumber \\ &+2
E_{122,113}-E_{123,113}+E_{136,122}-E_{137,129}-E_{143,135}-E_{149,141}-E_{150,142}-E_{156,147}\nonumber \\ &-E_{157,148}-E_{162,154}-E_{163,155}%
-E_{169,160}-E_{170,161}-E_{175,167}-E_{176,168}-E_{182,174}\nonumber \\ &-E_{187,180}-E_{188,181}-E_{193,185}-E_{194,186}-E_{198,191}%
-E_{199,192}-E_{204,197}-E_{208,203}\nonumber \\ &-E_{224,220}-E_{228,223}-E_{231,227}-E_{234,230}-E_{236,233}-E_{242,239}-E_{243,241}%
,\nonumber\\
Y_{4}  &  = E_{6,5}+E_{12,10}+E_{13,11}+E_{23,19}+E_{27,22}+E_{30,26}%
+E_{33,29}+E_{41,36}+E_{45,40}+E_{50,43}\nonumber \\ &+E_{51,44}+E_{55,49}+E_{61,54}%
+E_{70,64}+E_{73,66}+E_{76,69}+E_{79,72}+E_{83,75}+E_{86,78}+E_{90,82}\nonumber \\ &%
+E_{92,85}+E_{96,89}+E_{103,95}+E_{109,102}+E_{112,105}+E_{113,106}+E_{115,107}+E_{116,108}\nonumber \\ &-E_{122,114}+2 E_{123,114}%
-E_{124,114}-E_{128,114}+E_{135,123}+E_{141,133}+E_{142,134}+E_{143,136}\nonumber \\ &+E_{144,137}+E_{147,140}+E_{154,146}+E_{160,153}+E_{164,157}%
+E_{167,159}+E_{171,163}+E_{174,166}\nonumber \\ &+E_{177,170}+E_{180,173}+E_{183,176}+E_{185,179}+E_{195,188}+E_{200,194}+E_{205,198}%
+E_{206,199}\nonumber \\ &+E_{209,204}+E_{213,208}+E_{220,216}+E_{223,219}+E_{227,222}+E_{230,226}+E_{238,236}+E_{239,237}\nonumber \\ &+E_{244,243}%
,\nonumber\\
Y_{5}  &  = -E_{5,4}-E_{14,12}-E_{15,13}-E_{19,16}-E_{31,27}-E_{34,30}%
-E_{36,32}-E_{37,33}-E_{40,35}-E_{43,39}\nonumber \\ &-E_{56,51}-E_{62,55}-E_{64,58}-E_{66,60}-E_{67,61}-E_{69,63}-E_{72,65}-E_{75,68}-E_{78,71}%
-E_{97,90}\nonumber \\ &-E_{99,92}-E_{104,96}-E_{105,98}-E_{106,100}-E_{107,101}-E_{110,103}-E_{114,108}-E_{117,109}\nonumber \\ &-E_{123,116}+2
E_{124,116}-E_{125,116}+E_{133,124}-E_{140,132}-E_{141,135}-E_{146,139}-E_{148,142}\nonumber \\ &-E_{149,143}-E_{151,144}-E_{153,145}-E_{157,150}%
-E_{159,152}-E_{178,171}-E_{181,174}-E_{184,177}\nonumber\\
&  -E_{186,180}-E_{188,182}-E_{189,183}-E_{191,185}-E_{194,187}-E_{198,193}%
-E_{210,206}-E_{214,209}\nonumber \\ &-E_{216,212}-E_{217,213}-E_{219,215}-E_{222,218}-E_{233,230}-E_{236,234}-E_{237,235}-E_{245,244}%
,\nonumber\\
Y_{6}  &  = E_{4,3}+E_{17,14}+E_{18,15}+E_{22,19}+E_{27,23}+E_{32,28}%
+E_{38,34}+E_{42,37}+E_{44,40}+E_{49,43}\nonumber \\ &+E_{51,45}+E_{55,50}+E_{58,52}+E_{60,53}+E_{63,57}+E_{65,59}+E_{74,67}+E_{82,75}+E_{85,78}%
+E_{90,83}\nonumber \\ &+E_{92,86}+E_{98,91}+E_{100,93}+E_{101,94}+E_{108,102}+E_{111,104}+E_{116,109}+E_{118,110}\nonumber \\ &-E_{124,117}+2
E_{125,117}-E_{126,117}+E_{132,125}+E_{139,131}+E_{140,133}+E_{145,138}+E_{147,141}\nonumber \\ &+E_{155,148}+E_{156,149}+E_{158,151}+E_{163,157}%
+E_{166,159}+E_{171,164}+E_{174,167}+E_{182,175}\nonumber\\
&  +E_{190,184}+E_{192,186}+E_{196,189}+E_{197,191}+E_{199,194}+E_{204,198}%
+E_{206,200}+E_{209,205}\nonumber \\ &+E_{212,207}+E_{215,211}+E_{221,217}+E_{226,222}+E_{230,227}+E_{234,231}+E_{235,232}+E_{246,245}%
,\nonumber\\
Y_{7}  &  = -E_{3,2}-E_{20,17}-E_{21,18}-E_{26,22}-E_{30,27}-E_{34,31}%
-E_{35,32}-E_{40,36}-E_{45,41}-E_{48,42}\nonumber \\ &-E_{52,46}-E_{53,47}-E_{54,49}-E_{61,55}-E_{67,62}-E_{68,63}-E_{71,65}-E_{75,69}-E_{78,72}%
-E_{83,76}\nonumber \\ &-E_{86,79}-E_{91,84}-E_{93,87}-E_{94,88}-E_{102,95}-E_{109,103}-E_{117,110}-E_{119,111}\nonumber \\ &-E_{125,118}+2
E_{126,118}-E_{127,118}+E_{131,126}-E_{138,130}-E_{139,132}-E_{146,140}-E_{154,147}\nonumber \\ &-E_{161,155}-E_{162,156}-E_{165,158}-E_{170,163}%
-E_{173,166}-E_{177,171}-E_{180,174}-E_{184,178}\nonumber\\
&  -E_{186,181}-E_{187,182}-E_{194,188}-E_{200,195}-E_{202,196}-E_{203,197}%
-E_{207,201}-E_{208,204}\nonumber \\ &-E_{213,209}-E_{217,214}-E_{218,215}-E_{222,219}-E_{227,223}-E_{231,228}-E_{232,229}-E_{247,246}%
,\nonumber\\
Y_{8}  &  = E_{2,1}+E_{24,20}+E_{25,21}+E_{29,26}+E_{33,30}+E_{37,34}%
+E_{39,35}+E_{42,38}+E_{43,40}+E_{49,44}\nonumber \\ &+E_{50,45}+E_{55,51}+E_{57,52}+E_{59,53}+E_{62,56}+E_{63,58}+E_{65,60}+E_{69,64}+E_{72,66}%
+E_{76,70}\nonumber \\ &+E_{79,73}+E_{84,77}+E_{87,80}+E_{88,81}+E_{95,89}+E_{103,96}+E_{110,104}+E_{118,111}\nonumber \\ &-E_{126,119}+2 E_{127,119}%
+E_{130,127}+E_{138,131}+E_{145,139}+E_{153,146}+E_{160,154}+E_{168,161}\nonumber \\ &+E_{169,162}+E_{172,165}+E_{176,170}+E_{179,173}%
+E_{183,177}+E_{185,180}+E_{189,184}+E_{191,186}\nonumber\\
&  +E_{193,187}+E_{196,190}+E_{197,192}+E_{198,194}+E_{204,199}+E_{205,200}%
+E_{209,206}+E_{211,207}\nonumber \\ &+E_{214,210}+E_{215,212}+E_{219,216}+E_{223,220}+E_{228,224}+E_{229,225}+E_{248,247}.
\end{align}


\addtocontents{toc}{\protect\setcounter{tocdepth}{1}}
\chapter{Chapter 3 Appendix}
\addtocontents{toc}{\protect\setcounter{tocdepth}{-1}}

\section{Brane Motions}\label{app:BRANE}

In this Appendix we present an illustrative example for how to rearrange various $[p,q]$ 7-branes so that S-fold projection
acts geometrically on the associated string junction states. This is best illustrated via pictures, so we mainly display the relevant figures
here. Our starting point is an $E_6$ stack written as $A^{5} B C^2 \sim A^{6}
X C \sim AAAC AAAC$ (see figure \ref{fig:E6move}), a $D_4$ stack written as $A^{4} BC \sim AAC AAC $ (see figure \ref{fig:D4move})
and an $H_2$ stack written as $A^{3} C \sim AC Y^2 \sim AC AC \sim DADA$ (see figure \ref{fig:H2move}).

\begin{figure}
\begin{center}
\includegraphics[scale = 0.6]{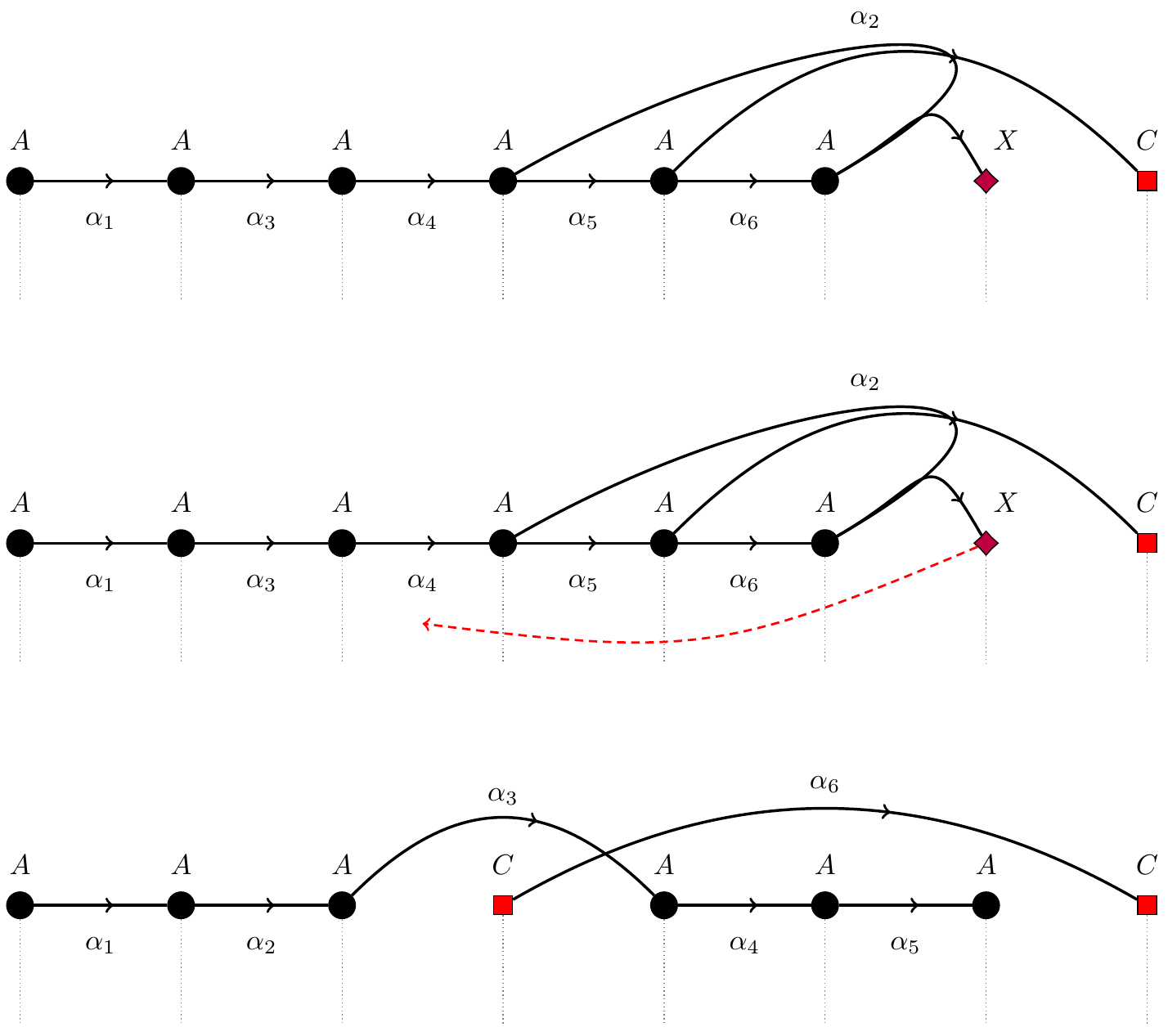}
\end{center}
\caption{Brane motion for $E_6$ 7-branes to a configuration which is $\mathbb{Z}_2$ symmetric, and thus amenable to a $\mathbb{Z}_2$ S-fold projection, i.e. an orientifold projection. In the figure we also indicate how the $X$-brane is moved to accomplish this rearrangement to the $\mathbb{Z}_2$ symmetric configuration $AAACAAAC$.}
\label{fig:E6move}
\end{figure}

\begin{figure}
\begin{center}
\includegraphics[scale = 0.6]{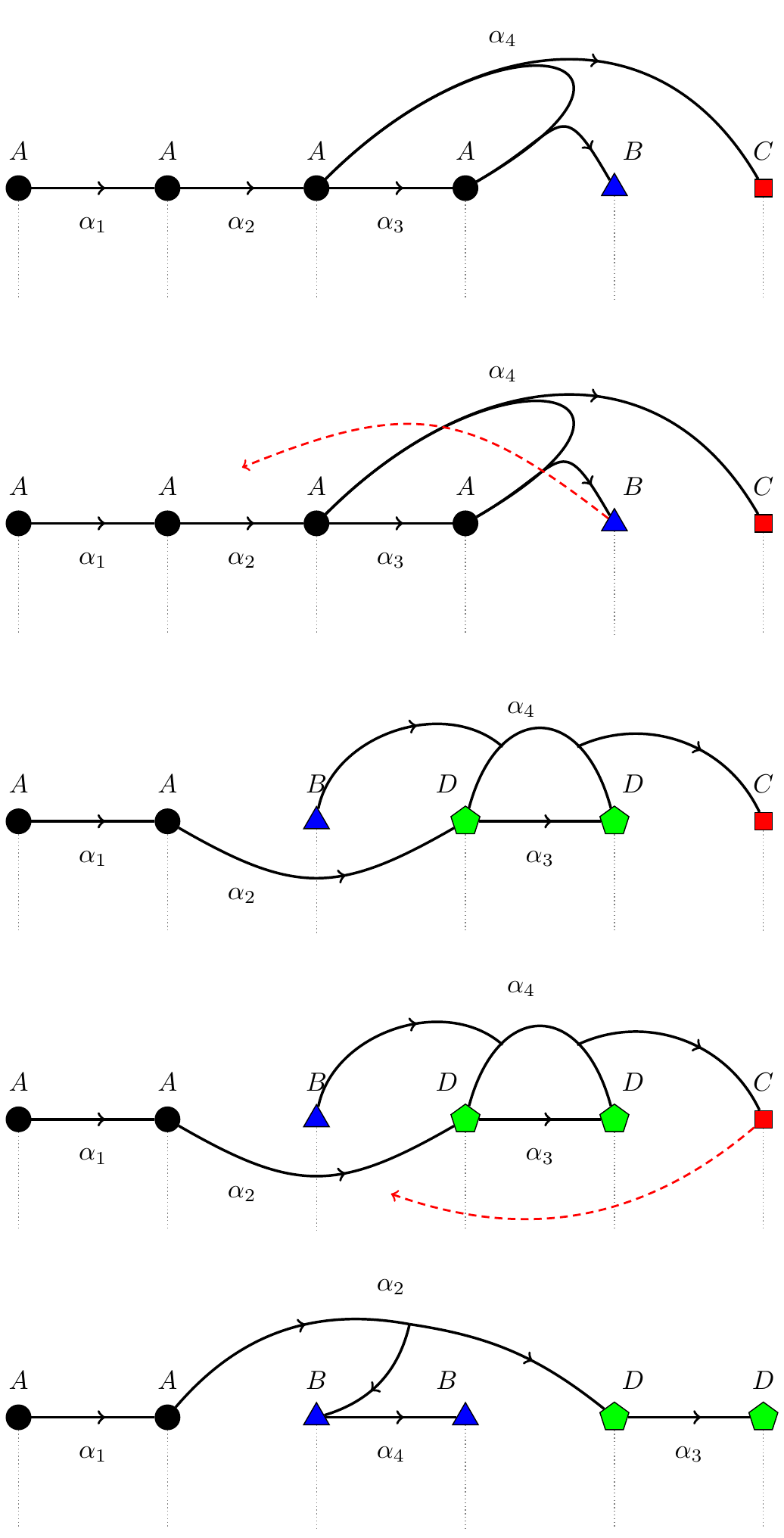}
\end{center}
\caption{Brane motion for $D_4$ 7-branes to a configuration which is $\mathbb{Z}_3$ symmetric, and thus amenable to a $\mathbb{Z}_3$ S-fold projection. In the figure we start with the presentation of this brane system as the bound state $A^{4} B C$, which we then split up into three stacks of branes which are permuted under the $\mathbb{Z}_3$ group action.}
\label{fig:D4move}
\end{figure}

\begin{figure}
\begin{center}
\includegraphics[scale = 0.6]{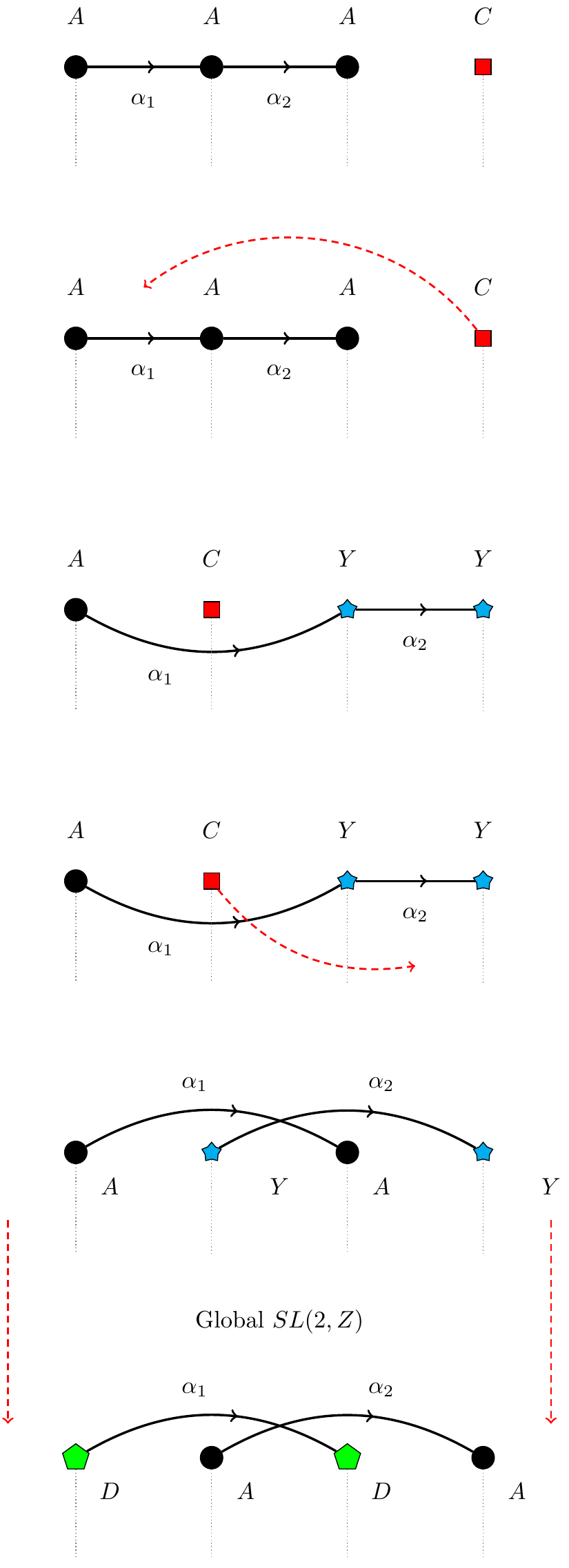}
\end{center}
\caption{Brane motion for $H_2$ 7-branes to the configuration $DADA$ which is $\mathbb{Z}_4$ symmetric, and thus amenable to an S-fold projection. In the last step of rearrangement we apply an $SL(2,\mathbb{Z})$ transformation as indicated by $Y$, with notation as in equation (\ref{monomove}).}
\label{fig:H2move}
\end{figure}

\section{Explicit $\mathbb{Z}_2$ Quotient of $E_6$ without Torsion}\label{app:E6SJ}
In this Appendix we give the explicit root system of $\mathfrak{e}_6$ and show how only $48$ roots survive the $\mathbb{Z}_2$ quotient (without torsion), corresponding exactly to the roots of an $\mathfrak{f}_4$ algebra. The roots of $E_6$, which are given in line \eqref{eq:E6roots} can be written as:
\begin{equation}
  \begin{split}
    &\pm\{(-1, 1, 0, 0, 0, 0, 0, 0), (0, 0, 0, -1, 0, 0, 0, 1), (0, -1, 1, 0, 0, 0, 0, 0), (0, 0, -1, 0, 1, 0, 0, 0),\\
    &(0, 0, 0, 0, -1, 1, 0, 0),(0, 0, 0, 0, 0, -1, 1, 0), (-1, 0, 1, 0, 0, 0, 0, 0), (0, 0, -1, -1, 1, 0, 0, 1),\\
    &(0, -1, 0, 0, 1, 0, 0, 0), (0, 0, -1, 0, 0, 1, 0, 0), (0, 0, 0, 0, -1, 0, 1, 0), (-1, 0, 0, 0, 1, 0, 0, 0),\\
    &(0, -1, 0, -1, 1, 0, 0, 1), (0, 0, -1, -1, 0, 1, 0, 1), (0, -1, 0, 0, 0, 1, 0, 0), (0, 0, -1, 0, 0, 0, 1, 0),\\
    &(-1, 0, 0, -1, 1, 0, 0, 1), (-1, 0, 0, 0, 0, 1, 0, 0), (0, -1, 0, -1, 0, 1, 0, 1), (0, 0, -1, -1, 0, 0, 1, 1), \\
    &(0, -1, 0, 0, 0, 0, 1, 0), (-1, 0, 0, -1, 0, 1, 0, 1), (-1, 0, 0, 0, 0, 0, 1, 0), (0, -1, -1, -1, 1, 1, 0, 1), \\
    &(0, -1, 0, -1, 0, 0, 1, 1), (-1, 0, -1, -1, 1, 1, 0, 1), (-1, 0, 0, -1, 0, 0, 1, 1), \\ 
    &(0, -1, -1, -1, 1, 0, 1, 1),(-1, -1, 0, -1, 1, 1, 0, 1),  (-1, 0, -1, -1, 1, 0, 1, 1), \\
    &(0, -1, -1, -1, 0, 1, 1, 1),(-1, -1, 0, -1, 1, 0, 1, 1), (-1, 0, -1, -1, 0, 1, 1, 1), \\
    &(-1, -1, 0, -1, 0, 1, 1, 1),(-1, -1, -1, -1, 1, 1, 1, 1), (-1, -1, -1, -2, 1, 1, 1, 2)\},
  \end{split}\label{eq:E6rootvectors}
\end{equation}
where the vectors follow the order of the branes of figure \ref{fig:E6Z2}. Namely, for instance, the highest root: $(1,1,1,2,-1,-1,-1,-2)$ corresponds to the string junction $(a_1+a_2+a_3+2c_1-a_4-a_5-a_6-2c_2)$. For the projection, we define the matrix
\begin{equation}
  Z = -\begin{pmatrix}
    0 & 0 & 0 & 0 & 1 & 0 & 0 & 0 \\
    0 & 0 & 0 & 0 & 0 & 1 & 0 & 0 \\
    0 & 0 & 0 & 0 & 0 & 0 & 1 & 0 \\
    0 & 0 & 0 & 0 & 0 & 0 & 0 & 1 \\
    1 & 0 & 0 & 0 & 0 & 0 & 0 & 0 \\
    0 & 1 & 0 & 0 & 0 & 0 & 0 & 0 \\
    0 & 0 & 1 & 0 & 0 & 0 & 0 & 0 \\
    0 & 0 & 0 & 1 & 0 & 0 & 0 & 0
  \end{pmatrix}.
\end{equation}
We then map every root $r$ in \eqref{eq:E6rootvectors} to $\frac12 (r + Z \cdot r)$. This results in the following $48$ roots:
\begin{equation}
  \begin{split}
    \pm\{
    &(0, 0, 0, -1, 0, 0, 0, 1), (0, 0, -1, 0, 0, 0, 1, 0), (0, -\frac12, \frac12, 0, 0, \frac12, -\frac12, 0),\\
    &(-\frac12, \frac12, 0, 0, \frac12, -\frac12, 0, 0),(0, 0, -1, -1, 0, 0, 1, 1), (0, -\frac12, -\frac12, 0, 0, \frac12, \frac12, 0),\\
    &(-\frac12, 0, \frac12, 0, \frac12, 0, -\frac12, 0),(0, -1, 0, 0, 0, 1, 0, 0), (0, -\frac12, -\frac12, -1, 0, \frac12, \frac12, 1),\\
    &(-\frac12, 0, -\frac12, 0, \frac12, 0, \frac12, 0),(0, -1, 0, -1, 0, 1, 0, 1), (-\frac12, 0, -\frac12, -1, \frac12, 0, \frac12, 1),\\
    &(-\frac12, -\frac12, 0, 0, \frac12, \frac12, 0, 0),(-1, 0, 0, 0, 1, 0, 0, 0), (0, -1, -1, -1, 0, 1, 1, 1),\\
    &(-\frac12, -\frac12, 0, -1, \frac12, \frac12, 0, 1),(-1, 0, 0, -1, 1, 0, 0, 1), (-\frac12, -\frac12, -1, -1, \frac12, \frac12, 1, 1),\\
    &(-1, 0, -1, -1, 1, 0, 1, 1),(-\frac12, -1, -\frac12, -1, \frac12, 1, \frac12, 1), (-1, -\frac12, -\frac12, -1, 1, \frac12, \frac12, 1),\\
    &(-1, -1, 0, -1, 1, 1, 0, 1), (-1, -1, -1, -1, 1, 1, 1, 1), (-1, -1, -1, -2, 1, 1, 1, 2)\}.
  \end{split}\label{eq:F4rootvectors}
\end{equation}
From there, we can extract the four simple roots of $F_4$:
\begin{equation}
\begin{split}
  \{&(\frac12, -\frac12, 0, 0, -\frac12, \frac12, 0, 0),(0, \frac12, -\frac12, 0, 0, -\frac12, \frac12, 0),(0, 0, 1, 0, 0, 0, -1, 0),\\&(0, 0, 0, 1, 0, 0, 0, -1)\},
  \end{split}
\end{equation}
corresponding exactly to the simple roots chosen in line \eqref{eq:F4simpleroots}.

\addtocontents{toc}{\protect\setcounter{tocdepth}{1}}
\chapter{Chapter 5 Appendix}
\addtocontents{toc}{\protect\setcounter{tocdepth}{-1}}

\section{Two-Loop \texorpdfstring{$\beta$}{beta}-Functions}\label{app:var2laction}
In this appendix we demonstrate how the two-loop $\beta$-functions arise from the variation of the two-loop low-energy effective target space action in the MT scheme. We follow the presentation of \cite{Metsaev:1987zx} closely but keep the one-loop $\beta$-functions in all steps instead of setting them to zero.

\subsection{Metric}
We begin by varying the two-loop low-energy effective target space action \eqref{eq:S2} with respect to the metric. As explained in section \ref{sec:oneloopdetail}, it is important that one does not vary with respect to the inverse metric, as this would introduce a wrong sign. For instance, the variation of the Riemann tensor is given by
\begin{equation}
  \delta R^e{}_{abc} = \frac{1}{2} g^{de}\left(\nabla_b\delta\left(\partial_a g_{dc}+\partial_c g_{da}-\partial_l g_{ca}\right)
  -\nabla_c\delta\left(\partial_a g_{db}+\partial_b g_{da}- \partial_d g_{ba}\right)\right),
\end{equation}
so that the variation of the first term in \eqref{eq:S2} reads
\begin{equation}\label{eqn:vargfirstterm}
  \begin{split}
    \delta \int \dd^D x \sqrt{g} e^{-2 \phi} \frac14  R_{efcd}R^{efcd} &= -\int \dd^D x \sqrt{g} e^{-2 \phi} R^{abcd} \nabla_a \nabla_c \delta g_{bd}\\
    &= -\int \dd^D x \sqrt{g} \nabla_a \nabla_c \left( e^{-2 \phi}R^{abcd} \right) \delta g_{bd}\\
    &= -\int \dd^D x \sqrt{g} \left(2R^{acdb}\Big(\nabla_c\partial_d\phi - 2\partial_c\phi\partial_d\phi\right) \\
    &\hspace{5em}  +4\nabla_c R^{c(ab)d}\partial_d\phi -\nabla_c\nabla_d R^{acdb} \Big)\delta g_{ab},
  \end{split}
\end{equation}
where we integrated by parts twice in the second line. Following \cite{Metsaev:1987zx}, we break down the metric variation of \eqref{eq:S2} into the three terms
\begin{align}
  P_{ab} &\equiv e^{2\phi} \frac{\delta}{\delta g^{ab}}\int \frac{1}{4} e^{-2\phi}R_{cdef}R^{cdef}, \label{Pmunu}\\
  Q_{ab} &\equiv e^{2\phi} \frac{\delta}{\delta g^{ab}}\int \left(-\frac{1}{8}\right)e^{-2\phi}R^{cdef}H_{cdh}H_{ef}{}^{h}, \label{Qmunu} \\
  O_{ab} &\equiv \frac{\delta}{\delta g^{ab}} \int \frac{1}{4}\left(\frac{1}{24}H_{cde}H^d{}_{fg}H^{fhe}H_h{}^{cg}-\frac{1}{8}H_{cde}H_f{}^{de}H^{cgh}H^f{}_{gh} \right), \label{Omunu}
\end{align}
where we have already partially treated the first term in \eqref{eqn:vargfirstterm}. We have to apply Bianchi identities to further simplify these three contributions. In particular, starting from the second Bianchi identity of the Riemann tensor, we derive the identity
\begin{equation}
  \nabla^c\nabla^d R_{acdb} = - \nabla^2R_{ab} +\frac12 \nabla_a\nabla_b R + R_{acdb}R^{cd} + R_{ac}  R^c{}_{b}.
\end{equation}
Applying it to the first term, we obtain
\begin{equation}
  \begin{split}
    P_{ab} = &-\frac12R_{acde}R_b{}^{cde} - \nabla^2R_{ab} + \frac12 \nabla_a\nabla_bR + R_{acdb}R^{cd} + R_{ac}  R^c{}_{b} \\
    &- 4\left(\nabla_{(a} R_{b)c}-\nabla_c R_{ab}\right)\partial^c \phi -2R_{acdb}\left(\nabla^c\partial^d\phi - 2\partial^c\phi\partial^d\phi\right).
  \end{split}
\end{equation}
Similarly, the second and third terms reduce to
\begin{align}
  Q_{ab} &= \frac{1}{2}R^{cdef}H_{acd}H_{bef}+\frac{3}{2}R^{cde}{}_{(a}H_{b)ge}H_{cd}{}^{g} + e^{2\phi} \nabla^c\nabla^d\left(e^{-2\phi}H_{ace}H_{bd}{}^{e}\right), \\
  O_{ab} &=-\frac{1}{16} (H^4)_{ab} + \frac{1}{16} H^2_{ac}(H^2)_b{}^{c} + \frac{1}{8}(H^2)^{cd}H_{ace}H_{bd}{}^{e}.
\end{align}

As suggested by \cite{Metsaev:1987zx}, we use the one-loop $\beta$-functions to remove all $\phi$-dependence. Consequentially, the variation of the action decomposed into terms containing $\bh^{(1)g}$ or $\bh^{(1)b}$ and terms without them. The latter form the two-loop $\beta$-function for the metric, $\bh^{(2)g}_{ab}$, while the former give rise to $\Kh^{(1)g}(\bh^{(1)g},\bh^{(1)B})$. During the computation we use the identities
\begin{equation}
  \begin{aligned}  
    & R_{acdb}\nabla^c\phi = 2 \nabla_{[c} \nabla_{b]}\nabla_a \phi = \nabla_{[d} \left(\bh^g_{b]a}-R_{b]a} + \frac{1}{4}H^2_{b]a} \right), \\
    &\nabla^c\nabla_a R_{cb} = \frac{1}{2} \nabla_a \nabla_b R + R_{ac}R^{c}{}_{b} + R_{acdb}R^{cd}, \\
    &\nabla_b H_{acd} = \nabla_c H_{bda}+\nabla_d H_{bac}+\nabla_a H_{bcd}\,,    
  \end{aligned}
\end{equation}
which eventually yield
\begin{equation}
  \begin{split}
    P_{ab} =& -\frac12 R_{acde}R_b{}^{cde} - \frac12 \nabla_e H_{acd}\nabla^e H_{b}{}^{cd} - \frac{1}{16}(H^2)_{a}{}^{c} H^2_{cb} - \frac14 H_{cea}H_{bd}{}^{e}(H^2)^{cd}  \\
    & + \frac12 R^{cdef}H_{acd}H_{bef} + \frac14 R_{acdb}(H^2)^{cd} + R_{(a}{}^{cde}H_{b)cf}H_{de}{}^{f} +\frac{1}{24} \nabla_a\nabla_bH^2 \\
    & - \nabla^2\bh^{(1)g}_{ab} + \nabla^c \nabla_{(a} \bh^{(1)g}_{b)c} - H^{cd}{}_{a}H_{edb}(\bh^{(1)g})_c{}^{e} - \frac14 H^2_{ac} (\bh^{(1)g})_{b}{}^{c} \\
    & - 2\left(\nabla_{(a}\bh^{(1)g}_{b)c} -\nabla_c\bh^{(1)g}_{ab}\right)\partial^c\phi \\
    & + 2H^{cd}{}_{a} \nabla_c \bh^{(1)b}_{db} + \frac12 H^{cd}{}_{a} \nabla_{b} \bh^{(1)b}_{cd} - \frac12 \nabla_{(a} H_{b)cd}(\bh^{(1)b})^{cd}, \\
    Q_{ab} =& \frac14 R_{(a}{}^{cde}H_{b)cf}H_{de}{}^{f}+\frac{1}{16}H_{ace}H_{bd}{}^{e}(H^2)^{cd}+\frac{1}{8}\nabla_e H_{acd}\nabla^e H_{b}{}^{cd}-\frac{1}{24}\nabla_a H_{cde}\nabla_b H^{cde} \\
    & +\frac14 H_{ace} H_{bd}{}^{e} (\bh^{(1)g})^{dc} - \nabla^c \bh_{d(a}^{(1)b} H_{b)c}{}^{d} + (\bh^{(1)b})_{ca}(\bh^{(1)b})^{c}{}_{b}, \\
    O_{ab} =& -\frac{1}{16} (H^4)_{ab} + \frac{1}{16} H^2_{ac}(H^2)_b{}^{c} + \frac{1}{8}(H^2)^{cd}H_{ace}H_{bd}{}^{e}\,.
  \end{split}
\end{equation}
Adding those terms back together gives rise to
\begin{equation}
  P_{ab} + Q_{ab} + O_{ab} = - \bh^{(2)g}_{ab} + \Kh^{(1)g}_{ab} ( \bh^{(1)g}, \bh^{(1)b} )\,.
\end{equation}
Now, we read off $\Kh^{(1)g}_{ab} ( \beta^g, \beta^B )$ in \eqref{eqn:Kh(1)g} and
\begin{equation}
  \bh^{(2)g}_{ab} =  \bb^{(2)g}_{ab} - \frac1{24} \nabla_a \nabla_b H^2\,.
\end{equation}

\subsection{\texorpdfstring{$B$}{B}-field}\label{app:varBfield}
Since the $B$-field only appears indirectly through $H=d B$ ($H_{abc} = 3\partial_{[a} B_{bc]}$),\footnote{Due to the total antisymmetrisation we can equivalently write $H_{abc} = 3\nabla_{[a} B_{bc]}$.}
we vary the action \eqref{eq:S2} with respect to $H$, apply the chain-rule and integrate by parts
\begin{equation}
  \delta \Sh^{(2)}= -3\int \dd^D x \sqrt{g} \nabla_c \frac{\delta \Lh^{(2)}}{\delta H_{cab}}\delta B_{ab}\,,
  \qquad \text{where} \qquad \Sh^{(2)} = \int \dd^D x \sqrt{g} \Lh^{(2)}\,.
\end{equation}
The result reads
\begin{equation}
  \begin{split}
  \frac{\delta \Sh^{(2)}}{\delta B^{ab}} =& \int \mathrm{d}^Dx\sqrt{g} \frac{1}{4} \nabla^e \left(e^{-2\phi}\left(R_{ab}{}^{cd}H_{cde}+R_{ea}{}^{cd}H_{cdb}+R_{be}{}^{cd}H_{cda}\right)\right) \\
       & -\frac{1}{8} \nabla^f  \left(e^{-2\phi}H_{acd}H_{be}{}^{c}H_f{}^{ed}\right) \\
  & +\frac{1}{8}\nabla^f \left(e^{-2\phi}\left(H_{abc}H_{def}H^{dec}+H_{fac}H_{deb}H^{dec}+H_{bfc}H_{dea}H^{dec}\right)\right) \\
  =& \int \mathrm{d}^Dx\sqrt{g} e^{-2\phi} \Big[
    -\frac{1}{2}\nabla^e \phi \left(R_{ab}{}^{cd}H_{cde}+R_{ea}{}^{cd}H_{cdb}+R_{be}{}^{cd}H_{cda}\right) \\
    & +\frac{1}{4}\nabla^f \phi H_{acd}H_{be}{}^{c}H_f{}^{ed} \\
    & -\frac{1}{4}\nabla^f \phi \left(H_{abc}H_{def}H^{dec}+H_{fac}H_{deb}H^{dec}+H_{bfc}H_{dea}H^{dec}\right) \\
    & +\frac{1}{4} \nabla^e \left(R_{ab}{}^{cd}H_{cde}+R_{ea}{}^{cd}H_{cdb}+R_{be}{}^{cd}H_{cda}\right) -\frac{1}{8} \nabla^f  \left(H_{acd}H_{be}{}^{c}H_f{}^{ed}\right) \\
    &+\frac{1}{8} \nabla^f \left(H_{abc}H_{def}H^{dec}+H_{fac}H_{deb}H^{dec}+H_{bfc}H_{dea}H^{dec}\right)
    \Big]\,.
  \end{split}
\end{equation}
Again all terms containing the dilaton can be eliminated in favour of $\bh^{(1)g}$ and $\bh^{(1)B}$, yielding
\begin{equation}
  \begin{split}
    \frac{\delta \Sh^{(2)}}{\delta B^{ab}} &= \int \mathrm{d}^Dx\sqrt{g} e^{-2\phi} \Big[
    -\frac{1}{2} R^{}_{[aecd}\nabla^e H^{cd}{}_{b]} - \frac{1}{4} \nabla^f  H^{}_{cd[a}H_{b]e}{}^{c}H_f{}^{ed} + \frac{1}{2} \nabla_c H^2_{d[b} H_{a]}{}^{dc}\\
    &+\frac{1}{8} H^2_{ec} \nabla^e H^c{}_{ab} + \frac{1}{8} H_{abc} H_{fde}\nabla^f H^{dec} -H_{[a|}{}^{cd}\nabla_c \bh^{(1)g}_{d |b]}- \frac{1}{2} R_{ab}{}^{cd} \bh^{(1)B}_{cd} \\
    &- \frac{1}{4} H_{abc} H^{dec} \bh^{(1)B}_{de} + \frac{1}{4} H_{acd}H_{be}{}^{c} (\bh^{(1)B})^{ed} - \frac{1}{2} H_{de[a|}H^{dec}\bh^{(1)B}_{c|b]}
    \Big]\,,
  \end{split}
\end{equation}
from which we read off $\Kh^{(1)B}_{ab}(\beta^g,\beta^B)$ in \eqref{eqn:Kh(1)B} and
\begin{equation}
  \bh^{(2)B\,ab} = \bb^{(2)B} - \frac1{48} H_{ab}{}^c \nabla_c H^2\,.
\end{equation}

\subsection{Dilaton}
Finally, for the dilaton, we begin with the two-loop $\beta$-function in the MT scheme, given in equation (6.10) of \cite{Metsaev:1987zx}, namely
\begin{equation}
  \bh^{(2)\phi} = \bb^{(2)\phi} - \frac1{48}\nabla^c\phi \nabla_cH^2\,.
\end{equation}
Combined with the $\beta$-function of the metric, it gives rise to\footnote{We make use of the Bianchi identity $\nabla^2 H^2 = 6 R^{ab} (H^2)_{ab} - 6 RHH + 2 \nabla_d H_{abc} \nabla^d H^{abc} + 6 H_{abc}\nabla^c\nabla_l H^{lab}$ in the step before last.}
\begin{align}
    \bh^{(2)d} =& \bh^{(2)\phi} - \frac14 g^{ab} \bh^{(2)g}_{ab} \nonumber \\
      =& -\frac1{16} \Big( R_{abcd} R^{abcd} + \frac1{24} H^4 + \frac13 \nabla_d H_{abc} \nabla^d H^{abc} -\frac18 H_{ab}^2(H^2)^{ab} \nonumber \\
      & -\frac32 RHH + R^{ab}H^2_{ab}  -\frac16\nabla^2H^2 + \frac13 \nabla^c\phi \nabla_c H^2 + 2H^2_{ab} \nabla^a\nabla^b \phi \Big) \nonumber \\
      =& -\frac1{16} \Big( R_{abcd} R^{abcd} + \frac1{24} H^4  -\frac18 H_{ab}^2(H^2)^{ab} -\frac12 RHH  - H_{abc}\nabla^c\nabla_l H^l{}_{ab} \nonumber \\
      & +\frac13 \nabla^c\phi \nabla_c H^2 +2 H^2_{ab} \nabla^a\nabla^b \phi \Big) \nonumber \\ \label{eqn:bh2dpure}
      =& -\frac1{16} \Big( R_{abcd} R^{abcd} + \frac1{24} H^4  -\frac18 H_{ab}^2(H^2)^{ab} -\frac12 RHH  + 2H^{abc}\nabla_c \bh^{(1)B}_{ab} \Big)\,,
\end{align}
where in the last step, we absorbed the terms involving $\phi$ into the one-loop $\beta$-function of the $B$-field. Moreover, the variation of the action \eqref{eq:S2}, with respect to the dilaton is given
\begin{equation}
  \begin{aligned}
    \frac{\delta S^{(2)}}{\delta d} = -2 \int \dd^D x e^{-2 d} \frac14 \Big[ & R_{abcd}R^{abcd} - \frac12 R^{abcd} H_{abe} H_{cd}{}^{e} \\ & + \frac1{24} H_{abc} H^b{}_{de} H^{dfc} H_f{}^{ae} - \frac18 H_{abc} H_d{}^{bc} H^{aef} H^d{}_{ef} \Big] \,. 
  \end{aligned}
\end{equation}
Combining it with \eqref{eqn:varS2}, we read off the value of $\Kh^{(1)d}(\beta^B)$ given in \eqref{eqn:Kh(1)d}.

\section{Transformation from HT to MT Scheme}\label{app:HTtoMT}
Starting from the two-loop $\beta$-function of the $B$-field in the HT scheme, we show the details of the scheme transformations required to obtain the corresponding $\beta$-function in the MT scheme. Our main motivation for this calculation is to have a cross check for \eqref{eqn:betaB2}, because it deviates by two signs from \cite{Metsaev:1987zx}. $\beta$-functions in both schemes are in general related by
\begin{equation}
  \bh^{\mathrm{MT}}_{ij} = \bh^{\mathrm{HT}}_{ij} - \Delta \bh_{ij} \label{eq:HTtoMT}\,,
\end{equation}
and the metric is shifted by \cite{Metsaev:1987zx}
\begin{equation}
  \Delta g^{(1)}_{ij} = \frac12 H_{ij}^2\,, \label{eq:scheme}
\end{equation}
while the $B$-field and the dilaton are not affected. Accordingly, the $B$-field $\beta$-function is shifted by
\begin{equation}
  \Delta \bh^{(2)B} = \Delta g^{(1)} \cdot \frac{\delta}{\delta g} \bh^{(1)B}\,. 
\end{equation}
Explicitly calculating the variation with respect to the metric on the right-hand side yields
\begin{equation}
  \delta_g \bh^{(1)B}_{ij} = \frac12 H_{ij}{}^k \nabla^l \delta g_{kl} + H_{[i}{}^{kl}\nabla_k \delta g_{j]l} - \frac14 g^{kl} H_{ij}{}^n\nabla_n \delta g_{kl} + \frac12 \delta g_{lk} \nabla^l H^k{}_{ij} - \delta g_{lk} \nabla^l\phi H^k{}_{ij}\,.
\end{equation}
and thus
\begin{equation}\label{eqn:Deltabh(2)B}
  \Delta \bh^{(2)B} = -\frac12 \nabla_k H^2_{l[j}H_{i]}{}^{kl} + \frac14 H^2_{lk}\nabla^l H^k{}_{ij} + H_{ij}{}^k \xi_k
\end{equation}
with
\begin{equation}
  \xi_k = \frac14\nabla^lH^2_{lk} -\frac18 \nabla_k H^2 - \frac12\nabla^l\phi H^2_{lk}\,.
\end{equation}
Since that last term in \eqref{eqn:Deltabh(2)B} just generates an infinitesimal diffeomorphism, we can drop it when computing $\bb^{(2)B\,\mathrm{MT}}_{ab}$ from the expression in the HT scheme,
\begin{equation}
  \bb^{(2)B\,\mathrm{HT}}_{ij} = \frac12\nabla^k H^{lm}{}_{[j}R_{i]klm} - \frac14 \nabla^l H^{km}{}_{[j}H_{i]kn}H_{lm}{}^n + \frac18 \nabla_k H_{lij} (H^2)^{kl}\,.
\end{equation}
The result
\begin{equation}
    \bb^{(2)B\,\mathrm{MT}}_{ij} = \frac12R_{[i|klm}\nabla^k H^{lm}{}_{|j]} +\frac14 \nabla^lH_{mn[i}H_{j]k}{}^mH_l{}^{kn} - \frac18H^2_{kl}\nabla^kH^l{}_{ij} + \frac12 H_{[i}{}^{kl}\nabla_k H^2_{j]l} \,.
\end{equation}
matches \eqref{eqn:betaB2} and confirms our result from appendix~\ref{app:varBfield}.

\addtocontents{toc}{\protect\setcounter{tocdepth}{1}}
\chapter{Chapter 6 Appendix}
\addtocontents{toc}{\protect\setcounter{tocdepth}{-1}}
\section{Proofs of Power Series Expansion}\label{app:Power}

In this Appendix we provide additional details on the power series expansions discussed in section \ref{sec:HIGGS}.

\subsection{BHV Power Series}
In the local coordinates given in \eqref{eq:SDtriplet}, and assuming a flat metric, the BHV equations become:
\begin{align}
  \begin{split}
  F_{t\theta} + F_{xy} = [\phi_\alpha,\phi_\beta],& \\
  F_{tx} + F_{y\theta} = 0,& \\
  F_{ty} - F_{x\theta} = 0,& \\
  D_x\phi_\alpha     + D_y\phi_\beta        =  0,& \\
  D_\theta\phi_\beta + D_t\phi_\alpha        = 0,& \\
  D_t\phi_\beta     - D_\theta\phi_\alpha    = 0,& \\
  D_x\phi_\beta     - D_y\phi_\alpha        = 0.&
  \end{split}
\end{align}
A power series expansion in $t$ then yields the following set of equations:
\begin{align}
  \begin{split}
    \displaystyle{\sum_{j=0}^{\infty}}\left(
    \begin{array}
      [c]{l}
      (j+1)A_\theta^{(j+1)}-\partial_\theta A_t^{(j)}+\partial_x A_y^{(j)}-\partial_y A_x^{(j)} \vspace{1mm} \\
      +\displaystyle{\sum_{n=0}^j}\left(\left[A_t^{(j-n)},A_\theta^{(n)}\right]+\left[A_x^{(j-n)},A_y^{(n)}\right]-\left[\phi_\alpha^{(j-n)},\phi_\beta^{(n)}\right]\right)
      \end{array}\right) t^j &= 0, \\
  \displaystyle{\sum_{j=0}^{\infty}}\left(
    \begin{array}
      [c]{l}(j+1)A_x^{(j+1)}-\partial_x A_t^{(j)}+\partial_y A_\theta^{(j)}-\partial_\theta A_y^{(j)} \vspace{1mm} \\
      +\displaystyle{\sum_{n=0}^j}\left(\left[A_t^{(j-n)},A_x^{(n)}\right]+\left[A_y^{(j-n)},A_\theta^{(n)}\right]\right)
      \end{array}\right) t^j &= 0, \\
  \displaystyle{\sum_{j=0}^{\infty}}\left(
    \begin{array}
      [c]{l}
      (j+1)A_y^{(j+1)}-\partial_y A_t^{(j)}-\partial_x A_\theta^{(j)}-\partial_\theta A_x^{(j)} \vspace{1mm} \\
      +\displaystyle{\sum_{n=0}^j}\left(\left[A_t^{(j-n)},A_y^{(n)}\right]-\left[A_x^{(j-n)},A_\theta^{(n)}\right]\right)
      \end{array}\right) t^j &= 0, \\
  \displaystyle{\sum_{j=0}^{\infty}}\left(
    \begin{array}
      [c]{l}
      \partial_x \phi_\alpha^{(j)}+\partial_y \phi_\beta^{(j)} 
      +\displaystyle{\sum_{n=0}^j}\left(\left[A_x^{(j-n)},\phi_\alpha^{(n)}\right]+\left[A_y^{(j-n)},\phi_\beta^{(n)}\right]\right)
      \end{array}\right) t^j &= 0, \\
  \displaystyle{\sum_{j=0}^{\infty}}\left(
    \begin{array}
      [c]{l}
      (j+1)\phi_\alpha^{(j+1)}+\partial_\theta \phi_\beta^{(j)} 
      +\displaystyle{\sum_{n=0}^j}\left(\left[A_\theta^{(j-n)},\phi_\beta^{(n)}\right]+\left[A_t^{(j-n)},\phi_\alpha^{(n)}\right]\right)
      \end{array}\right) t^j &= 0, \\
  \displaystyle{\sum_{j=0}^{\infty}}\left(
    \begin{array}
      [c]{l}
      (j+1)\phi_\beta^{(j+1)}-\partial_\theta \phi_\alpha^{(j)} 
      +\displaystyle{\sum_{n=0}^j}\left(\left[A_t^{(j-n)},\phi_\beta^{(n)}\right]-\left[A_\theta^{(j-n)},\phi_\alpha^{(n)}\right]\right)
      \end{array}\right) t^j &= 0, \\
  \displaystyle{\sum_{j=0}^{\infty}}\left(
    \begin{array}
      [c]{l}
      \partial_x \phi_\beta^{(j)}-\partial_y \phi_\alpha^{(j)} 
      +\displaystyle{\sum_{n=0}^j}\left(\left[A_x^{(j-n)},\phi_\beta^{(n)}\right]-\left[A_y^{(j-n)},\phi_\alpha^{(n)}\right]\right)
      \end{array}\right) t^j &= 0.
\end{split}
\end{align}
By taking the temporal gauge $A_t^{(j)} = 0$ we indeed obtain the differential equations \eqref{eq:difBHV} and recursion relations \eqref{eq:recBHV}.
To show that solving the zeroth order equations
\begin{align}
  \begin{split}
  & \mathcal{G}_{ab}^{(0)} = \partial_x \phi_\beta^{(0)}-\partial_y \phi_\alpha^{(0)}
  +\left[A_x^{(0)},\phi_\beta^{(0)}\right]-\left[A_y^{(0)},\phi_\alpha^{(0)}\right] = 0,\\
  & \mathcal{H}_{ab}^{(0)} = \partial_x \phi_\alpha^{(0)}+\partial_y \phi_\beta^{(0)}
  +\left[A_x^{(0)},\phi_\alpha^{(0)}\right]+\left[A_y^{(0)},\phi_\beta^{(0)}\right]= 0,
  \end{split}
\end{align}
leads to a solution at all orders in the power series expansion we substitute \eqref{eq:recBHV} into \eqref{eq:difBHV}. Explicitly we need to do the following computations.

\paragraph{-- The Commutators:}
Initially we have that:
\begin{equation}
  \left[A_x^{(k)},\phi_\beta^{(j-k)}\right] = \frac{k}{j}\left[A_x^{(k)},\phi_\beta^{(j-k)}\right] + \frac{j-k}{j}\left[A_x^{(k)},\phi_\beta^{(j-k)}\right]\,.
\end{equation}
Taking into account the summations we have that
\begin{equation}
\makebox[\textwidth]{
$\begin{aligned}
  &\sum_{k=0}^j\frac{k}{j}\left[A_x^{(k)},\phi_\beta^{(j-k)}\right] = \sum_{k=1}^j\frac{k}{j}\left[A_x^{(k)},\phi_\beta^{(j-k)}\right] \\
  &= \frac{1}{j} \sum_{k=1}^j \left[-\partial_y A_\theta^{(k-1)}+\partial_\theta A_y^{(k-1)}
     -\sum_{l=0}^{k-1}\left(\left[A_y^{(k-1-l)},A_\theta^{(l)}\right]\right),\phi_\beta^{(j-k)}\right] \\
  &= \frac{1}{j} \sum_{k=0}^{j-1} \left[-\partial_y A_\theta^{(k)}+\partial_\theta A_y^{(k)},\phi_\beta^{(j-k-1)}\right]
      -\frac{1}{j}\sum_{l=0}^{j-1}\sum_{m+n=j-l-1}\left[\left[A_y^{(l)},A_\theta^{(m)}\right],\phi_\beta^{(n)}\right],
  \end{aligned}$
}
\end{equation}
after substituting the recursion relation for $A_x^{(k)}$.
Similarly, by using the recursion relation for $\phi_\beta^{(k)}$ we find
\begin{equation}
\makebox[\textwidth]{
$\begin{aligned}
  &\sum_{k=0}^j \frac{j-k}{j}\left[A_x^{(k)},\phi_\beta^{(j-k)}\right]
    = \sum_{k=0}^{j-1} \frac{j-k}{j}\left[A_x^{(k)},\phi_\beta^{(j-k)}\right]
    = \sum_{k=1}^{j} \frac{k}{j}\left[A_x^{(j-k)},\phi_\beta^{(k)}\right] \\
  &=\frac{1}{j} \sum_{k=1}^j \left[A_x^{(j-k)}, \partial_\theta \phi_\alpha^{(k-1)}
    +\sum_{l=0}^{k-1}\left(\left[A_\theta^{(k-1-l)},\phi_\alpha^{(l)}\right]\right)\right] \\
  &=\frac{1}{j} \sum_{k=0}^{j-1} \left[A_x^{(j-k-1)}, \partial_\theta \phi_\alpha^{(k)}\right]
    + \frac{1}{j}\sum_{l=0}^{j-1}\sum_{m+n=j-l-1}\left[A_x^{(m)},\left[A_\theta^{(l)},\phi_\alpha^{(n)}\right]\right].
\end{aligned}$
}
\end{equation}
The computation for the other three commutators is identical. Together we have
\begin{align}
  \sum_{k=0}^j \left[A_x^{(k)},\phi_\beta^{(j-k)}\right] = &\frac{1}{j} \sum_{k=0}^{j-1} \left(
    \left[\partial_\theta A_y^{(k)}-\partial_y A_\theta^{(k)},\phi_\beta^{(j-k-1)}\right] + \left[A_x^{(j-k-1)}, \partial_\theta \phi_\alpha^{(k)}\right] \right) \nonumber \\
    -&\frac{1}{j}\sum_{l=0}^{j-1}\sum_{m+n=j-l-1} \left(
    \left[\left[A_y^{(l)},A_\theta^{(m)}\right],\phi_\beta^{(n)}\right] - \left[A_x^{(m)},\left[A_\theta^{(l)},\phi_\alpha^{(n)}\right]\right]
    \right), \\
  \sum_{k=0}^j \left[A_y^{(k)},\phi_\alpha^{(j-k)}\right] = -&\frac{1}{j} \sum_{k=0}^{j-1} \left(
    \left[\partial_\theta A_x^{(k)}-\partial_x A_\theta^{(k)},\phi_\alpha^{(j-k-1)}\right] + \left[A_y^{(j-k-1)}, \partial_\theta \phi_\beta^{(k)}\right] \right) \nonumber \\
    +&\frac{1}{j}\sum_{l=0}^{j-1}\sum_{m+n=j-l-1} \left(
    \left[\left[A_x^{(l)},A_\theta^{(m)}\right],\phi_\alpha^{(n)}\right] - \left[A_y^{(m)},\left[A_\theta^{(l)},\phi_\beta^{(n)}\right]\right]
    \right), \\
  \sum_{k=0}^j \left[A_x^{(k)},\phi_\alpha^{(j-k)}\right] = &\frac{1}{j} \sum_{k=0}^{j-1} \left(
    \left[\partial_\theta A_y^{(k)}-\partial_y A_\theta^{(k)},\phi_\alpha^{(j-k-1)}\right] - \left[A_x^{(j-k-1)}, \partial_\theta \phi_\beta^{(k)}\right] \right) \nonumber \\
    -&\frac{1}{j}\sum_{l=0}^{j-1}\sum_{m+n=j-l-1} \left(
    \left[\left[A_y^{(l)},A_\theta^{(m)}\right],\phi_\alpha^{(n)}\right] - \left[A_x^{(m)},\left[A_\theta^{(l)},\phi_\beta^{(n)}\right]\right]
    \right), \\
  \sum_{k=0}^j \left[A_y^{(k)},\phi_\beta^{(j-k)}\right] = -&\frac{1}{j} \sum_{k=0}^{j-1} \left(
    \left[\partial_\theta A_x^{(k)}-\partial_x A_\theta^{(k)},\phi_\beta^{(j-k-1)}\right] - \left[A_y^{(j-k-1)}, \partial_\theta \phi_\alpha^{(k)}\right] \right) \nonumber \\
    +&\frac{1}{j}\sum_{l=0}^{j-1}\sum_{m+n=j-l-1} \left(
    \left[\left[A_x^{(l)},A_\theta^{(m)}\right],\phi_\beta^{(n)}\right] + \left[A_y^{(m)},\left[A_\theta^{(l)},\phi_\alpha^{(n)}\right]\right]
     \right).
\end{align}

\paragraph{-- The Derivatives:}
Next, we have:
\begin{align}
  \partial_x \phi_\beta^{(j)}  =  \frac{1}{j} & \left\{\partial_x \partial_\theta \phi_\alpha^{(j-1)} + \sum_{n=0}^{j-1} \partial_x\left[A_\theta^{(j-1-n)},\phi_\alpha^{(n)}\right]\right\}, \\
  \partial_y \phi_\alpha^{(j)} = -\frac{1}{j} & \left\{\partial_y \partial_\theta \phi_\beta^{(j-1)} + \sum_{n=0}^{j-1} \partial_y\left[A_\theta^{(j-1-n)},\phi_\beta^{(n)}\right]\right\}, \\
  \partial_x \phi_\alpha^{(j)} = -\frac{1}{j} & \left\{\partial_x \partial_\theta \phi_\beta^{(j-1)} + \sum_{n=0}^{j-1} \partial_x\left[A_\theta^{(j-1-n)},\phi_\beta^{(n)}\right]\right\}, \\
  \partial_y \phi_\beta^{(j)}  =  \frac{1}{j} & \left\{\partial_y \partial_\theta \phi_\alpha^{(j-1)} + \sum_{n=0}^{j-1} \partial_y\left[A_\theta^{(j-1-n)},\phi_\alpha^{(n)}\right]\right\}.
\end{align}

\paragraph{-- Putting Everything Together:}
Finally, by summing all the pieces together and making use of the Jacobi identities we obtain:
\begin{align}
  (j+1) \mathcal{G}_{ab}^{j+1} &=  \partial_\theta \mathcal{H}_{ab}^{(j)} + \sum_{n=0}^{j}\left[A_\theta^{(j-n)},\mathcal{H}_{ab}^{(n)} \right], \\
  (j+1) \mathcal{H}_{ab}^{j+1} &= -\partial_\theta \mathcal{G}_{ab}^{(j)} - \sum_{n=0}^{j}\left[A_\theta^{(j-n)},\mathcal{G}_{ab}^{(n)} \right].
\end{align}
These expressions make obvious the inductive proof that if $\mathcal{G}_{ab}^{(0)} = \mathcal{H}_{ab}^{(0)} = 0$, which we assume, then it follows that $\mathcal{G}_{ab}^{(j)} = \mathcal{H}_{ab}^{(j)} = 0$ to all orders $j \geq 1$.

\subsection{Full Local $Spin(7)$ Expansion}
Similarly, we can write the local $Spin(7)$ equations as follows:
\begin{align}
  \begin{split}
  F_{t\theta} + F_{xy} = [\phi_\alpha,\phi_\beta],& \\
  F_{tx} + F_{y\theta} = [\phi_\gamma,\phi_\alpha],& \\
  F_{ty} - F_{x\theta} = [\phi_\gamma,\phi_\beta],& \\
  D_t\phi_\gamma     + D_x\phi_\alpha     + D_y\phi_\beta    =  0,& \\
  D_\theta\phi_\beta + D_t\phi_\alpha     - D_x\phi_\gamma    = 0,& \\
  D_t\phi_\beta     - D_\theta\phi_\alpha - D_y\phi_\gamma    = 0,& \\
  D_x\phi_\beta     - D_y\phi_\alpha     + D_\theta\phi_\gamma= 0.&
  \end{split}
\end{align}
Then a power series expansion in $t$ yields the following set of equations:
\begin{align}
  \begin{split}
  \displaystyle{\sum_{j=0}^{\infty}}\left(
    \begin{array}
      [c]{l}
      (j+1)A_\theta^{(j+1)}-\partial_\theta A_t^{(j)}+\partial_x A_y^{(j)}-\partial_y A_x^{(j)}  \vspace{1mm} \\
      + \displaystyle{\sum_{n=0}^j}\left(
      \left[A_t^{(j-n)},A_\theta^{(n)}\right] +\left[A_x^{(j-n)},A_y^{(n)}\right] -\left[\phi_\alpha^{(j-n)},\phi_\beta^{(n)}\right]\right)
    \end{array}
    \right)t^j &= 0, \\
  \displaystyle{\sum_{j=0}^{\infty}}\left(
    \begin{array}
      [c]{l}
      (j+1)A_x^{(j+1)}-\partial_x A_t^{(j)}+\partial_y A_\theta^{(j)}-\partial_\theta A_y^{(j)}  \vspace{1mm} \\
      + \displaystyle{\sum_{n=0}^j}\left(
      \left[A_t^{(j-n)},A_x^{(n)}\right]+\left[A_y^{(j-n)},A_\theta^{(n)}\right]-\left[\phi_\gamma^{(j-n)},\phi_\alpha^{(n)}\right]\right)
    \end{array}
    \right)t^j &= 0, \\
  \displaystyle{\sum_{j=0}^{\infty}}\left(
    \begin{array}
      [c]{l}
      (j+1)A_y^{(j+1)}-\partial_y A_t^{(j)}-\partial_x A_\theta^{(j)}-\partial_\theta A_x^{(j)}  \vspace{1mm} \\
      + \displaystyle{\sum_{n=0}^j}\left(
      \left[A_t^{(j-n)},A_y^{(n)}\right]-\left[A_x^{(j-n)},A_\theta^{(n)}\right]-\left[\phi_\gamma^{(j-n)},\phi_\beta^{(n)}\right]\right)
    \end{array}
    \right)t^j &= 0, \\
  \displaystyle{\sum_{j=0}^{\infty}}\left(
    \begin{array}
      [c]{l}
      (j+1)\phi_\gamma^{(j+1)}+\partial_x \phi_\alpha^{(j)}+\partial_y \phi_\beta^{(j)}  \vspace{1mm} \\
      + \displaystyle{\sum_{n=0}^j}\left(
      \left[A_t^{(j-n)},\phi_\gamma^{(n)}\right]+\left[A_x^{(j-n)},\phi_\alpha^{(n)}\right]+\left[A_y^{(j-n)},\phi_\beta^{(n)}\right]\right)
    \end{array}
    \right)t^j &= 0, \\
  \displaystyle{\sum_{j=0}^{\infty}}\left(
    \begin{array}
      [c]{l}
      (j+1)\phi_\alpha^{(j+1)}+\partial_\theta \phi_\beta^{(j)}-\partial_x \phi_\gamma^{(j)}  \vspace{1mm} \\
      + \displaystyle{\sum_{n=0}^j}\left(
      \left[A_\theta^{(j-n)},\phi_\beta^{(n)}\right]+\left[A_t^{(j-n)},\phi_\alpha^{(n)}\right]-\left[A_x^{(j-n)},\phi_\gamma^{(n)}\right]\right)
    \end{array}
    \right)t^j &= 0, \\
    \displaystyle{\sum_{j=0}^{\infty}}\left(
    \begin{array}
      [c]{l}
      (j+1)\phi_\beta^{(j+1)}-\partial_\theta \phi_\alpha^{(j)}-\partial_y \phi_\gamma^{(j)}  \vspace{1mm} \\
      + \displaystyle{\sum_{n=0}^j}\left(
      \left[A_t^{(j-n)},\phi_\beta^{(n)}\right]-\left[A_\theta^{(j-n)},\phi_\alpha^{(n)}\right]-\left[A_y^{(j-n)},\phi_\gamma^{(n)}\right]\right)
    \end{array}
    \right)t^j &= 0, \\
  \displaystyle{\sum_{j=0}^{\infty}}\left(
    \begin{array}
      [c]{l}
      \partial_x \phi_\beta^{(j)}-\partial_y \phi_\alpha^{(j)} + \partial_\theta \phi_\gamma^{(j)} \vspace{1mm}\\
      + \displaystyle{\sum_{n=0}^j}\left(
        \left[A_x^{(j-n)},\phi_\beta^{(n)}\right]-\left[A_y^{(j-n)},\phi_\alpha^{(n)}\right]+\left[A_\theta^{(j-n)},\phi_\gamma^{(n)}\right]\right)
    \end{array}
    \right)t^j &= 0.
  \end{split}
\end{align}
By taking the temporal gauge $A_t^{(j)} = 0$ we indeed obtain the differential equations \eqref{eq:difSpin7} and recursion relations \eqref{eq:recSpin7}. To show that solving the zeroth order equation
\begin{equation}
  D_x^{(0)} \phi_\beta^{(0)}-D_y^{(0)} \phi_\alpha^{(0)} + D_\theta^{(0)} \phi_\gamma^{(0)} = 0,
\end{equation}
leads to a solution at all orders in the power series expansion, we substitute \eqref{eq:recSpin7} into \eqref{eq:difSpin7}. Explicitly we need to do the following computations.

\paragraph{-- The Commutators:}
Using the same technique as before, the three commutators of interest are given by:
\begin{align}
    \sum_{k=0}^j\left[A_x^{(k)},\phi_\beta^{(j-k)}\right] &=\frac{1}{j} \displaystyle{\sum_{k=0}^{j-1}} \left(
      \begin{array}
        [c]{l}
        \left[-\partial_y A_\theta^{(k)}+\partial_\theta A_y^{(k)},\phi_\beta^{(j-k-1)}\right] \\
        + \left[A_x^{(j-k-1)}, \partial_\theta \phi_\alpha^{(k)}+\partial_y \phi_\gamma^{(k)}\right]
      \end{array}\right) \nonumber \\
      & -\frac{1}{j}\displaystyle{\sum_{l=0}^{j-1}}\displaystyle{\sum_{m+n=j-l-1}} \left(
      \begin{array}
        [c]{l}
        \left[\left[A_y^{(l)},A_\theta^{(m)}\right]-\left[\phi_\gamma^{(l)},\phi_\alpha^{(m)}\right],\phi_\beta^{(n)}\right] \\
        + \left[A_x^{(m)},-\left[A_\theta^{(l)},\phi_\alpha^{(n)}\right]-\left[A_y^{(l)},\phi_\gamma^{(n)}\right]\right]
      \end{array}\right), \\
  \sum_{k=0}^j \left[A_y^{(k)},\phi_\alpha^{(j-k)}\right] &= \frac{1}{j} \displaystyle{\sum_{k=0}^{j-1}} \left(
      \begin{array}
        [c]{l}
        \left[\partial_x A_\theta^{(k)}-\partial_\theta A_x^{(k)},\phi_\alpha^{(j-k-1)}\right] \\
        + \left[A_y^{(j-k-1)}, \partial_x \phi_\gamma^{(k)}-\partial_\theta \phi_\beta^{(k)}\right]
      \end{array}\right) \nonumber \\
      &-\frac{1}{j}\sum_{l=0}^{j-1}\sum_{m+n=j-l-1} \left(
      \begin{array}
        [c]{l}
        \left[-\left[A_x^{(l)},A_\theta^{(m)}\right]-\left[\phi_\gamma^{(l)},\phi_\beta^{(m)}\right],\phi_\alpha^{(n)}\right] \\
        + \left[A_y^{(m)},\left[A_\theta^{(l)},\phi_\beta^{(n)}\right]-\left[A_x^{(l)},\phi_\gamma^{(n)}\right]\right]
      \end{array}\right), \\
 \sum_{k=0}^j \left[A_\theta^{(k)},\phi_\gamma^{(j-k)}\right] &= \frac{1}{j} \displaystyle{\sum_{k=0}^{j-1}} \left(
     \begin{array}
       [c]{l}
       \left[\partial_y A_x^{(k)}-\partial_x A_y^{(k)},\phi_\gamma^{(j-k-1)}\right] \\
       + \left[A_\theta^{(j-k-1)}, -\partial_x \phi_\alpha^{(k)}-\partial_y \phi_\beta^{(k)}\right]
     \end{array}\right) \nonumber  \\
   &-\frac{1}{j}\displaystyle{\sum_{l=0}^{j-1}}\sum_{m+n=j-l-1} \left(
     \begin{array}
       [c]{l}
       \left[\left[A_x^{(l)},A_y^{(m)}\right]-\left[\phi_\alpha^{(l)},\phi_\beta^{(m)}\right],\phi_\gamma^{(n)}\right] \\
       + \left[A_\theta^{(m)},\left[A_x^{(l)},\phi_\alpha^{(n)}\right]+\left[A_y^{(l)},\phi_\beta^{(n)}\right]\right]
     \end{array}\right).
\end{align}
Then, making use of Jacobi's identities the sum of those commutators simplifies to:
\begin{equation}
\begin{split}
      &\displaystyle{\sum_{k=0}^j}\left(
       \left[A_x^{(k)},\phi_\beta^{(j-k)}\right]
        -\left[A_y^{(k)},\phi_\alpha^{(j-k)}\right]
        +\left[A_\theta^{(k)},\phi_\gamma^{(j-k)}\right]
        \right)
      = \\  & \qquad \frac{1}{j} \displaystyle{\sum_{k=0}^{j-1}}\left(
      \begin{array}
        [c]{l}
        \left[-\partial_y A_\theta^{(k)}+\partial_\theta A_y^{(k)},\phi_\beta^{(j-k-1)}\right] +\left[A_x^{(j-k-1)}, \partial_\theta \phi_\alpha^{(k)}+\partial_y \phi_\gamma^{(k)}\right] \vspace{1mm} \\
        -\left[\partial_x A_\theta^{(k)}-\partial_\theta A_x^{(k)},\phi_\alpha^{(j-k-1)}\right]-\left[A_y^{(j-k-1)}, \partial_x \phi_\gamma^{(k)}-\partial_\theta \phi_\beta^{(k)}\right] \vspace{1mm}  \\
        +\left[\partial_y A_x^{(k)}-\partial_x A_y^{(k)},\phi_\gamma^{(j-k-1)}\right]+\left[A_\theta^{(j-k-1)}, -\partial_x \phi_\alpha^{(k)}-\partial_y \phi_\beta^{(k)}\right]
        \end{array}\right).
\end{split}
\end{equation}

\paragraph{-- The Derivatives:}
Furthermore, the relevant derivatives are simply given by:
\begin{align}
  \partial_x \phi_\beta^{(j)} &= \frac{1}{j} \left\{ \partial_x \partial_y \phi_\gamma^{(j-1)} + \partial_x \partial_\theta \phi_\alpha^{(j-1)} +
  \sum_{n=0}^{j-1} \left(\partial_x\left[A_\theta^{(j-1-n)},\phi_\alpha^{(n)}\right]+\partial_x\left[A_y^{(j-1-n)},\phi_\gamma^{(n)}\right]\right)\right\}, \\
  \partial_y \phi_\alpha^{(j)} &= \frac{1}{j} \left\{ \partial_y \partial_x \phi_\gamma^{(j-1)} - \partial_y \partial_\theta \phi_\beta^{(j-1)} -
  \sum_{n=0}^{j-1} \left(\partial_y\left[A_\theta^{(j-1-n)},\phi_\beta^{(n)}\right]-\partial_y\left[A_x^{(j-1-n)},\phi_\gamma^{(n)}\right]\right)\right\}, \\
  \partial_\theta \phi_\gamma^{(j)} &= -\frac{1}{j} \left\{ \partial_\theta \partial_x \phi_\alpha^{(j-1)} + \partial_\theta \partial_y \phi_\beta^{(j-1)} +
  \sum_{n=0}^{j-1} \left(\partial_\theta \left[A_x^{(j-1-n)},\phi_\alpha^{(n)}\right]+\partial_\theta\left[A_y^{(j-1-n)},\phi_\beta^{(n)}\right]\right)\right\}.
\end{align}

\paragraph{-- Putting Everything Together:}
Summing up everything, we see that it all vanishes:
\begin{equation}
  \partial_x \phi_\beta^{(j)}-\partial_y \phi_\alpha^{(j)} + \partial_\theta \phi_\gamma^{(j)}
  +\sum_{n=0}^j\left(\left[A_x^{(j-n)},\phi_\beta^{(n)}\right]-\left[A_y^{(j-n)},\phi_\alpha^{(n)}\right]+\left[A_\theta^{(j-n)},\phi_\gamma^{(n)}\right]\right)
  = 0
\end{equation}
at all orders $j\geq 1$.

Therefore it is sufficient to solve the zeroth order differential equation
\begin{equation}
  D_x^{(0)} \phi_\beta^{(0)}-D_y^{(0)} \phi_\alpha^{(0)} + D_\theta^{(0)} \phi_\gamma^{(0)} = 0,
\end{equation}
and then one can simply propagate through the recursion equations \eqref{eq:recSpin7} to build up the higher order components.

\subsection{Abelian Case}
Finally, taking $A_i = 0$ gives some major simplifications. The local $Spin(7)$ recursion relations \eqref{eq:recSpin7} now become:
\begin{align}
  \begin{split}
  \phi_\gamma^{(j)} &= -\frac{1}{j} \left(\partial_x \phi_\alpha^{(j-1)} + \partial_y \phi_\beta^{(j-1)} \right),\\
  \phi_\alpha^{(j)} &= \frac{1}{j} \left(\partial_x \phi_\gamma^{(j-1)} -\partial_\theta \phi_\beta^{(j-1)} \right),\\
  \phi_\beta^{(j)}  &= \frac{1}{j} \left(\partial_\theta \phi_\alpha^{(j-1)}+\partial_y \phi_\gamma^{(j-1)} \right).
  \end{split}
\end{align}\label{eq:abelrecSpin7}
These can then be further expanded as:
\begin{align}
\begin{split}
  \phi_\gamma^{(j)} &= \frac{1}{(j+1)j(j-1)}\left(\partial_\theta^2+\partial_y^2+\partial_x^2\right)\left(\partial_x\phi_\alpha^{(j-2)}+\partial_y\phi_\beta^{(j-2)}\right),\\
  \phi_\alpha^{(j)} &= -\frac{1}{(j+1)j(j-1)}\left(\partial_\theta^2+\partial_y^2+\partial_x^2\right)\left(\partial_x\phi_\gamma^{(j-2)}-\partial_\theta\phi_\beta^{(j-2)}\right),\\
  \phi_\beta^{(j)}  &= -\frac{1}{(j+1)j(j-1)}\left(\partial_\theta^2+\partial_y^2+\partial_x^2\right)\left(\partial_\theta\phi_\alpha^{(j-2)}+\partial_y\phi_\gamma^{(j-2)}\right).
\end{split}
\end{align}

From there we note an obvious pattern,
\begin{align}
\begin{split}
  \phi_\gamma^{(j)} &= -\frac{1}{(j+1)j}\left(\partial_\theta^2+\partial_y^2+\partial_x^2\right)\phi_\gamma^{(j-1)},\\
  \phi_\alpha^{(j)} &= -\frac{1}{(j+1)j}\left(\partial_\theta^2+\partial_y^2+\partial_x^2\right)\phi_\alpha^{(j-1)},\\
  \phi_\beta^{(j)}  &= -\frac{1}{(j+1)j}\left(\partial_\theta^2+\partial_y^2+\partial_x^2\right)\phi_\beta^{(j-1)}.
\end{split}
\end{align}
yielding \eqref{eq:abel}.

\addtocontents{toc}{\protect\setcounter{tocdepth}{1}}
\chapter{Chapter 7 Appendix}
\addtocontents{toc}{\protect\setcounter{tocdepth}{-1}}
\section{Aspects of Elliptic Curves} \label{app:ELLIPTIC}

In this Appendix we review some aspects of the geometry of elliptic curves used in chapter~\ref{chapter7}. In
normal Weierstrass form, an elliptic curve can be presented as the hypersurface cut out by the equation:
\begin{align}
y^2 = x^3 + f x z^4 + g z^6 \,,
\end{align}
with complex coefficients $f$ and $g$ and $(x,y,z)$ inhomogeneous coordinates on the weighted projective space $\mathbb{CP}^{2}_{[2,3,1]}$. In the patch $z \neq 0$ one can rescale $z$ to $1$ via the $\mathbb{C}^*$ rescaling leading to the more standard form
\begin{align}\label{eq:WEIER}
y^2 = x^3 + f x + g \,,
\end{align}
which has to be supplemented by the ``point at infinity'' given by $[x,y,z] = [1,1,0]$.
Expressing the cubic equation according to its roots $e_i$ one can write
\begin{align}
y^2 = (x - e_1) (x - e_2) (x - e_3) \,,
\end{align}
and one has
\begin{align}
e_1 + e_2 + e_3 = 0 \,, \quad f = e_1 e_2 + e_2 e_3 + e_3 e_1 \,, \quad g = - e_1  e_2 e_3
\end{align}
The discriminant is given by:
\begin{align}
D_{\mathrm{disc}} = \prod_{i < j} (e_i - e_j)^2 = -( 4f^3 + 27 g^2) \equiv - \Delta \,.
\end{align}
In what follows we follow F-theory conventions and refer to $\Delta = 4 f^3 + 27 g^2$ as the discriminant.

We can define the modular $\lambda$ function knowing the position of the branch cuts $e_i$.
In the Weierstrass form, where one of the roots is at infinity it is given by:
\begin{align}
\lambda = \frac{e_3 - e_2}{e_1 - e_2} \,,
\end{align}
In terms of this, the $j$-function can be expressed as
\begin{align}
j (\tau) = \frac{256 (1 - \lambda - \lambda^2)^3}{\lambda^2 (1 - \lambda)^2} \,.
\end{align}

One can also work in terms of a presentation such as:
\begin{equation}
x^2 = P_4(z) = (z - z_1)(z - z_2)(z - z_3) (z - z_4)
\end{equation}
in which all four roots are at finite values. In this case, the modular $\lambda$
function is defined by the conformal cross ratio
\begin{align}
\lambda = \frac{(z_2 - z_3) (z_1 - z_4)}{(z_1 - z_3) (z_2 - z_4)} \,,
\end{align}
where the branch cuts are chosen between $z_2$ and $z_3$ and $z_1$ and $z_4$.
One can also consider the elliptic curve defined by the equation:
\begin{align}
x^2 = \frac{P_4 (z)}{(z - 1)^2 (z - q)^2} \,,
\end{align}
as is the case for the Seiberg-Witten curve with $N_f = 4$. In this case, we can clear denominators and perform blowups at $z = 1$ and $z = q$
to get an elliptic curve. In this case one can identify the branch points at the zeros of $P_4$ and plug
them into the formula for $\lambda$ which in turn can be used to compute $j (\tau)$.

Let us analyze the behavior of $j(\tau)$ in terms of the cross ratio $\lambda$. Clearly, $j (\tau)$ diverges for the three cases
\begin{align}
\lambda \rightarrow 0 \,, \quad \lambda \rightarrow 1 \,, \quad \lambda \rightarrow \infty \,.
\end{align}
In these limits the branch point at $\lambda$ collides with one of the other three branch points.

\subsection{Phase Structure for Real Elliptic Curves}

Having discussed the general structure of roots in an elliptic curve, we now specialize further, taking $f,g \in \mathbb{R}$.
In section \ref{sec:DUALITY} we argued that the time-reversal invariant components of the fundamental domain of $SL(2,\mathbb{Z})$
split up into three distinct phases based on singularities in the elliptic curve, as dictated by the vanishing of $f,g$ and $\Delta$.
Here we provide some complementary details.

Going back to the description in terms of the explicit branch
points we find that up to a
permutation of indices one has the following two possibilities.
\begin{equation}
\begin{split}
\text{Case I}:& \quad e_1, e_2, e_3 \in \mathbb{R} \,, \\
\text{Case II}:& \quad e_1 \in \mathbb{R} \,, \enspace e_2 = \bar{e}_3 \,.
\end{split}
\label{eq:rootcases}
\end{equation}
Next, we want to relate the different configurations of the branch
points to the regions of $\tau$ given in \eqref{eq:tauregions}
that describe the distinct time-reversal invariant phases of the
abelian gauge theory. For that we hold the root $e_1$ fixed at negative real value.

For Case I in \eqref{eq:rootcases} we can parametrize the two other roots as
\begin{align}
e_2 = - \tfrac{1}{2} e_1 + \delta \,, \quad e_3 = - \tfrac{1}{2} e_1 - \delta \,,
\end{align}
with $\delta \in \mathbb{R}$. In terms of the variable $\delta$ the Weierstrass coefficients and discriminant read
\begin{align}
f = - \tfrac{3}{4} e_1^2 - \delta^2 \,, \quad g = - e_1 \big( \tfrac{1}{4} e_1^2 - \delta^2 \big) \,, \quad \Delta = - \tfrac{1}{4} \delta^2 (9 e_1^2 - 4 \delta^2)^2 \,.
\end{align}
The discriminant vanishes for $\delta = 0$ and $\delta = \pm e_1$, and as expected these points are associated to the collision of two of the branch points. Note also that all the coefficients are invariant with respect to $\delta \rightarrow - \delta$, which corresponds to an exchange of $e_2$ and $e_3$.
\begin{figure}[t!]
\centering
\includegraphics[width=\textwidth]{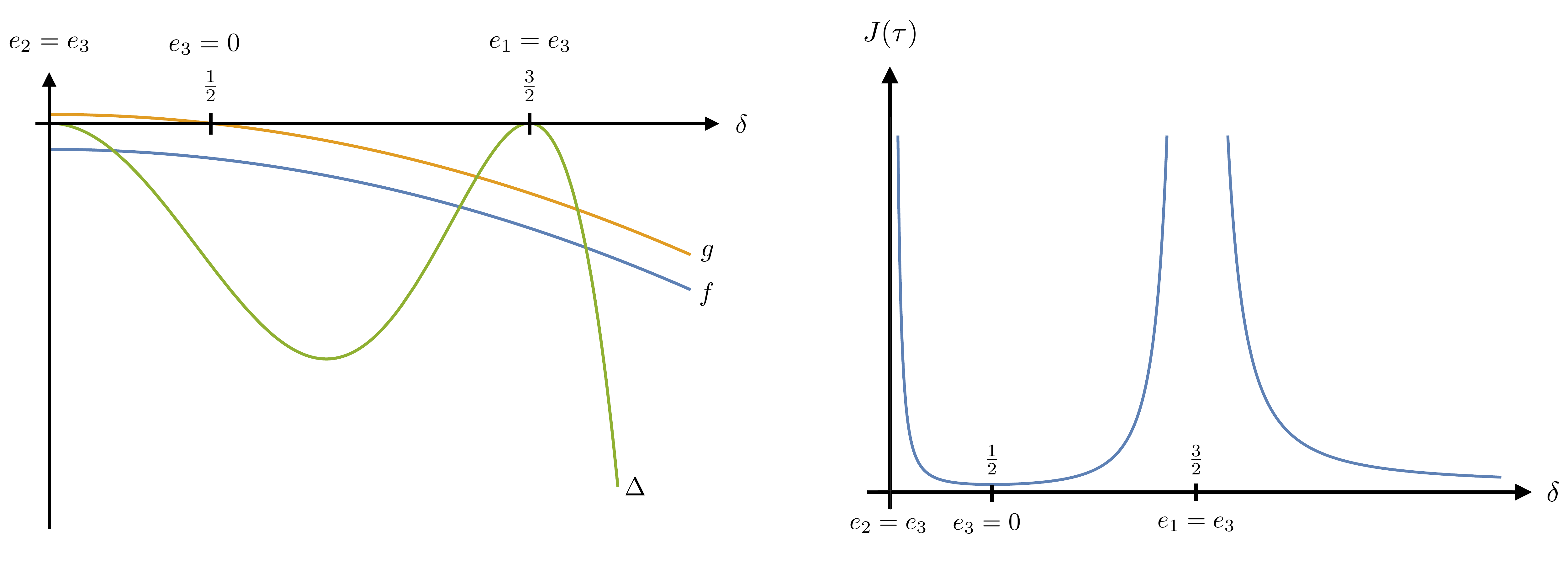}
\caption{The parameters $f$, $g$, and $\Delta$, as well as the $J$-function for all three branch points on the real axis (here: $e_1 = - 1$).}
\label{fig:Jreal}
\end{figure}
The $J$-function is then given by:
\begin{align}
J (\tau) = \frac{(3 e_1^2 + 4 \delta^2)^3}{4 \delta^2 (9 e_1^2 - 4 \delta^2)^2} \,.
\end{align}
Together with $f$, $g$, and $\Delta$ it is depicted in figure \ref{fig:Jreal}. We find that $J(\tau) \geq 1$, which means that all the configurations translate to the trivial phase with $\theta = 0$ and varying gauge coupling. At the collision of two branch points, which happens at $\delta = 0$ and $\delta = - \tfrac{3}{2} e_1$ the $J$-function diverges $J \rightarrow + \infty$. For the special values $\delta = - \tfrac{1}{2} e_1$ and $\delta \rightarrow \infty$ the $J$-function goes to $1$, which means that $\tau$ approaches the strong coupling point $\tau = i$.

For Case II in \eqref{eq:rootcases}, we use the following parametrization:
\begin{align}
e_2 = - \tfrac{1}{2} e_1 + i \tilde{\delta} \,, \quad e_3 = - \tfrac{1}{2} e_1 - i \tilde{\delta} \,,
\end{align}
with $\tilde{\delta} \in \mathbb{R}$. The Weierstrass coefficients and discriminant are given by
\begin{align}
f = - \tfrac{3}{4} e_1^2 + \tilde{\delta}^2 \,, \quad g = - e_1 \big( \tfrac{1}{4} e_1^2 + \tilde{\delta}^2 \big) \,, \quad \Delta = \tfrac{1}{4} \tilde{\delta}^2 (9 e_1^2 + 4 \tilde{\delta}^2)^2 \,.
\end{align}
The discriminant only vanishes at $\tilde{\delta} = 0$, when the two branch points collide on the real axis. Again, we find the symmetry $\tilde{\delta} \rightarrow - \tilde{\delta}$ which exchanges $e_2$ and $e_3$.
\begin{figure}[t!]
\centering
\includegraphics[width=\textwidth]{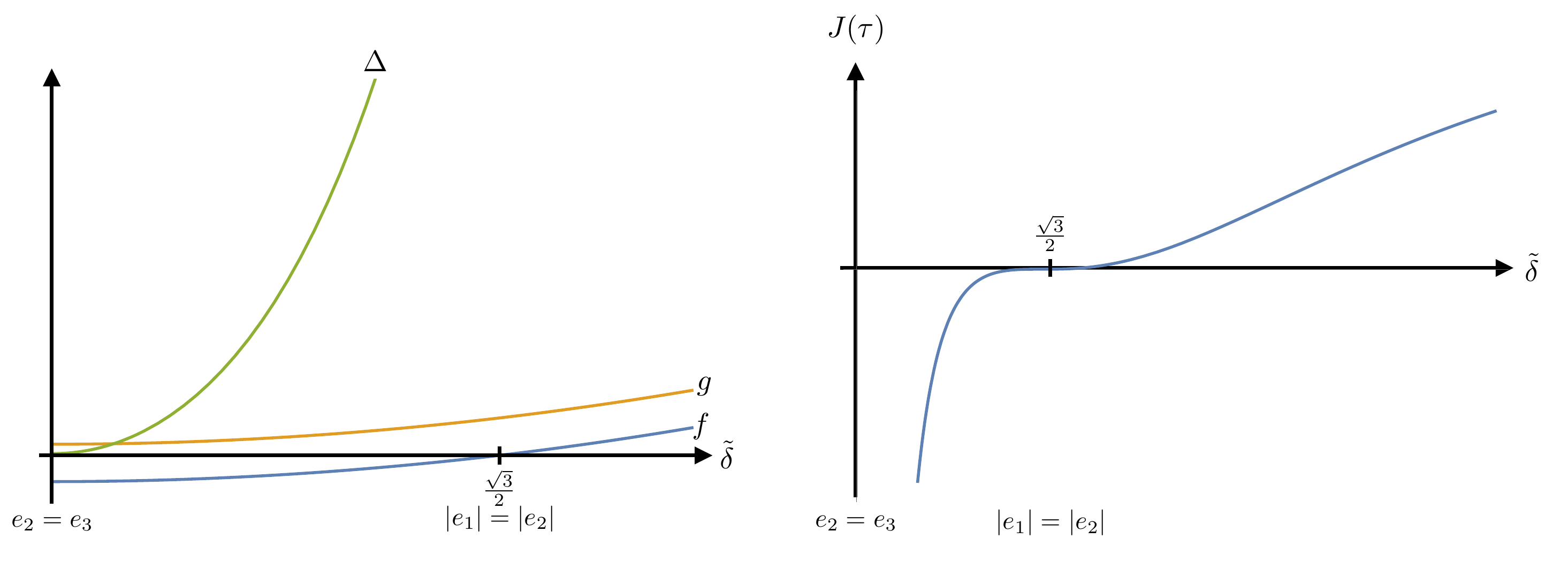}
\caption{The parameters $f$, $g$, and $\Delta$, as well as the $J$-function for two complex conjugate branch points (here: $e_1 = - 1$).}
\label{fig:Jcomp}
\end{figure}
The $J$-function is given by:
\begin{align}
J (\tau) = \frac{(4 \tilde{\delta}^2 - 3 e_1^2)^3}{4 \tilde{\delta}^2 (9 e_1^2 + 4 \tilde{\delta}^2)^2}
\end{align}
and is depicted in figure \ref{fig:Jcomp}. We find that $J(\tau) < 0$ for $\tilde{\delta} \in \big(- \tfrac{\sqrt{3}}{2} |e_1|, \tfrac{\sqrt{3}}{2} |e_1| \big)$, with $J (\tau) \rightarrow - \infty$ for $\tilde{\delta} \rightarrow 0$. This is the region where, $\theta = \pi$ and the gauge coupling varies. Finally, for $|\tilde{\delta}| > \tfrac{\sqrt{3}}{2} |e_1|$ one has $J (\tau) \in (0, 1)$ which indicates the strong coupling region with $|\tau| = 1$.

\begin{figure}[t!]
\centering
\includegraphics[width=\textwidth]{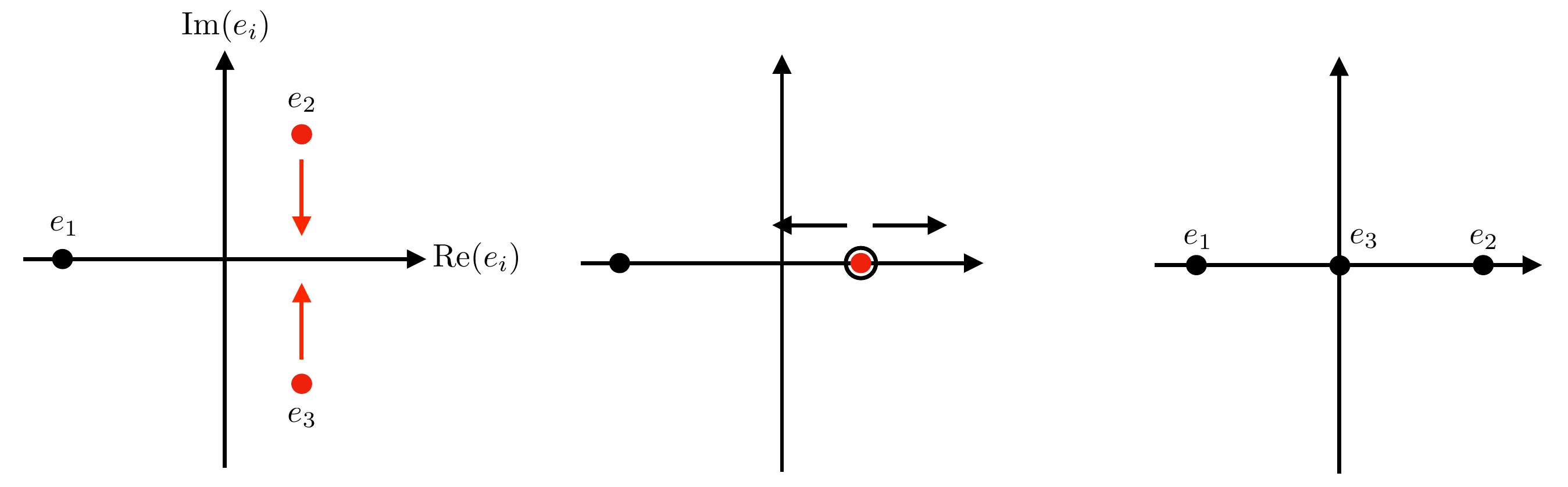}
\caption{The roots in the Weierstrass equation along the considered paths.}
\label{fig:rootpath}
\end{figure}
We see that by considering the configuration above, and depicted in figure \ref{fig:rootpath}, we can scan the full set of real $J (\tau)$ and therefore all the time-reversal invariant values of the complexified coupling constant $\tau$.

To summarize, the three different phases of the time-reversal invariant contour are specified by the following parameters:
\begin{itemize}
\item Trivial Phase: $J > 1 \Leftrightarrow \theta=0$ and $\tau = i \beta$ for $\beta>1$. There we have $\Delta < 0$, $f < 0$ and the roots $e_1 < e_3 < e_2$ are all real. The contours encircle $e_1$ to $e_3$ for $\gamma_B$ and $e_2$ to $e_3$ for $\gamma_A$.
\item Topological Insulator Phase: $J < 0 \Leftrightarrow \theta=\pi$. There we have $\Delta > 0$, $f < 0$ and the roots are such that $e_1 \in \mathbb{R}$, $e_2 = \bar{e}_3$, $\mathrm{Im}(e_2) > 0$. The contours encircle $e_1$ to $e_3$ for $\gamma_B$ and $e_2$ to $e_3$ for $\gamma_A$.
\item Strongly Coupled Phase: $0 \leq J \leq 1 \Leftrightarrow 0 \leq \theta \leq \pi$, $|\tau| = 1$. There we have $\Delta > 0$, $f \geq 0$ and the roots again satisfy $e_1 \in \mathbb{R}$, $e_2 = \bar{e}_3$, $\mathrm{Im}(e_2) > 0$. The contours encircle $e_1$ to $e_2$ for $\gamma_B$ and $e_1$ to $e_3$ for $\gamma_A$.
\end{itemize}
The different time-reversal invariant regions together with the signs of $f$, $g$, $\Delta$ are also indicated in figure \ref{fig:phasessigns}.

\section{Congruence Subgroups and Torsion Points} \label{app:CONG}

In section \ref{sec:MOREDUAL} we showed that compactifying the 6D theory of an anti-chiral two-form on an elliptic curve
can generate 4D $U(1)$ gauge theories with duality group given by a congruence subgroup $\Gamma \subset SL(2,\mathbb{Z})$.
In this Appendix we discuss in greater detail the relation between these congruence subgroups and torsion points.
As a point of notation, in the main text these torsion points are elements of $\widetilde{E}$, the Jacobian of the elliptic curve $E$
on which the 6D theory is compactified. To avoid cluttering the notation, we shall simply discuss an elliptic curve $E$ with torsion points.
The two characterizations are related by the Abel-Jacobi map, so we will not belabor this point in what follows.

We now consider the action of the congruence subgroups on the $N$-torsion points of an elliptic curve $E$,
denoted by $E(N)$, see e.g.\ \cite{diamond2006first}.
\begin{figure}[t!]
\centering
\includegraphics[width=0.7\textwidth]{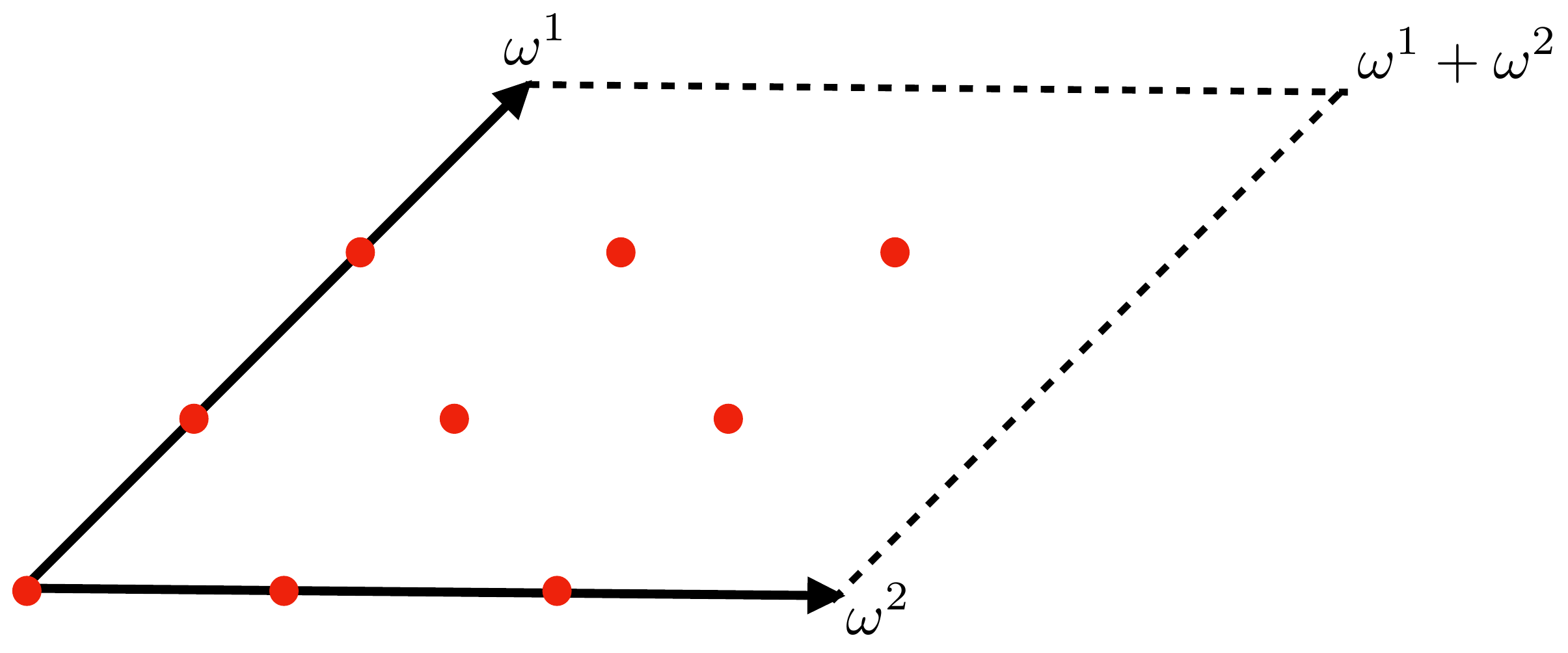}
\caption{Set of 3-torsion points $E(3)$ in the torus fundamental domain spanned by $\omega^1$ and $\omega^2$.}
\label{fig:3tors}
\end{figure}
When we describe $E$ as the quotient of the complex numbers $\mathbb{C}$ by a lattice $\Lambda = \omega^1 \mathbb{Z} \oplus \omega^2 \mathbb{Z}$, these torsion points are simply given by (see figure \ref{fig:3tors}):
\begin{align}
E(N) = \big\{ P \in E: \enspace P = \tfrac{m}{N} \omega^1 + \tfrac{n}{N} \omega^2 \,, \enspace m,n \in \{ 0, 1, \dots, N - 1 \} \big\} \,.
\end{align}
We see that the torsion points generate a subgroup of $E$ isomorphic to $\mathbb{Z} / N \mathbb{Z} \times \mathbb{Z} / N \mathbb{Z}$
with respect to the natural addition on the elliptic curve. An $N$-torsion point $P$ satisfies the condition:
\begin{align}
N P = P + P + \dots + P \in \Lambda \,,
\end{align}
i.e., the point $N P$ it is a lattice vector $k \omega^1 + l \omega^2$ with $k,l \in \mathbb{Z}$. This means that the full $N$-torsion subgroup is generated by two elements. We can choose $\omega^1 = \tau$ and $\omega^2 = 1$ on which a general $SL(2,\mathbb{Z})$ element acts as
\begin{align}
\begin{pmatrix} a & b \\ c & d \end{pmatrix} \begin{pmatrix} \tau \\ 1 \end{pmatrix} = \begin{pmatrix} a \tau + b \\ c \tau + d \end{pmatrix} \sim \begin{pmatrix} \tfrac{a \tau + b}{c \tau + d} \\ 1 \end{pmatrix}
\end{align}

The congruence subgroup $\Gamma (N)$ preserves two $N$-torsion points $P$ and $Q$ which generate the torsion subgroup $E(N)$ and have a Weil pairing given by $e_N (P,Q) = e^{2 \pi i / N}$. For two $N$-torsion points $P$ and $Q$ the Weil pairing is defined by
\begin{align}
e_N (P, Q) = e^{2 \pi i \det \alpha / N} \,,
\end{align}
where $\alpha$ is the matrix with entries in $\mathbb{Z}_N$, which maps $\big( \tfrac{1}{N} \omega^1, \tfrac{1}{N} \omega^2 \big)$ to $(P,Q)$ up to lattice vectors. Therefore, the subgroup $\Gamma (N)$ preserves all $N^2$ torsion points individually.

The congruence subgroup $\Gamma_1 (N)$ preserves a specific $N$-torsion point $P$ and consequently its multiples. This is, it fixes all elements in a $\mathbb{Z}_N$ subgroup of $E(N)$ individually. Note, that by an $SL(2,\mathbb{Z})$ transformation all such points can be mapped to e.g.\ $\tfrac{1}{N} \omega^2$. Conversely, starting from $\tfrac{1}{N} \omega^2$ we can generate all possible choices of the $N$-torsion element by the action of elements in $SL(2,\mathbb{Z}) / \Gamma_1(N)$, i.e.\ by the coset representatives.

Finally, the subgroup $\Gamma_0 (N)$ also preserves a $\mathbb{Z} / N \mathbb{Z}$ subgroup of $E(N)$, but it does not fix the individual elements, which can be mapped to one another in the process. As for $\Gamma_1(N)$ different choices of the $\mathbb{Z} / N \mathbb{Z}$ subgroup are related by a coset representative in $SL(2,\mathbb{Z}) / \Gamma_0(N)$.

Note that some of these congruence subgroups also appear in F-theory models with non-trivial Mordell-Weil torsion \cite{Aspinwall:1998xj, Hajouji:2019vxs}, see also \cite{Mayrhofer:2014opa, Kimura:2016crs, Baume:2017hxm, Cvetic:2017epq, Kimura:2018oze}. These models contain extra torsional sections, which can constrain the global realization of the gauge groups.

Let us illustrate the correspondence between $E(N)$ and the congruence subgroups for the case $N = 3$. We will use the description in terms of $\Lambda = \omega^1 \mathbb{Z} \oplus \omega^2 \mathbb{Z}$.

\subsection{$\Gamma(3)$}

The congruence subgroup $\Gamma(3)$ is generated by the elements
\begin{align}
\gamma_1 = \begin{pmatrix} 1 & 3 \\ 0 & 1 \end{pmatrix} \,, \quad \gamma_2 = \begin{pmatrix} -8 & 3 \\ -3 & 1 \end{pmatrix} \,, \quad \gamma_3 = \begin{pmatrix} 4 & -3 \\ 3 & -2 \end{pmatrix} \,.
\end{align}
A general point $P = (x,y)$ in $E$ is acted on by $SL(2, \mathbb{Z})$ as follows:
\begin{align}
\begin{pmatrix} x \\ y \end{pmatrix} \mapsto \begin{pmatrix} a & b \\ c & d \end{pmatrix} \begin{pmatrix} x \\ y  \end{pmatrix} = \begin{pmatrix} a x + b y \\ c x + d y \end{pmatrix} \,.
\end{align}
Furthermore, we use that the lattice $\Lambda$ is simply given by $\mathbb{Z} \oplus \mathbb{Z}$ and thus all points are understood modulo an integer. Since a point is invariant under the full group if it is invariant with respect to a set of generators, we check which points are invariant with respect to the action of $\gamma_1$, $\gamma_2$, and $\gamma_3$.

The first generator yields
\begin{align}
\begin{pmatrix} x \\ y \end{pmatrix} \mapsto \gamma_1 \begin{pmatrix} x \\ y \end{pmatrix} = \begin{pmatrix} x + 3 y \\ y \end{pmatrix} \sim \begin{pmatrix} x \\ y \end{pmatrix} \,,
\end{align}
which demands that $3 y$ is a lattice vector, or in other words $y \in \big\{ 0, \tfrac{1}{3}, \tfrac{2}{3} \big\}$. For the second generator one finds
\begin{align}
\begin{pmatrix} x \\ y \end{pmatrix} \mapsto \gamma_2 \begin{pmatrix} x \\ y \end{pmatrix} = \begin{pmatrix} - 8 x + 3 y \\ - 3 x + y \end{pmatrix} \,,
\end{align}
telling us that also $x \in \big\{ 0, \tfrac{1}{3}, \tfrac{2}{3} \big\}$. The last generator does not lead to any new constraints and one concludes that the set of invariant points is given by
\begin{align}
\big\{ \tfrac{m}{3} \omega^1 + \tfrac{n}{3} \omega^2 \,, \enspace m,n \in \{ 0, 1, 2\} \big\} = E(3) \,,
\end{align}
as desired.

\subsection{$\Gamma_1 (3)$}

The congruence subgroup $\Gamma_1 (3)$ is generated by the elements
\begin{align}
\widetilde{\gamma}_1 = \begin{pmatrix} 1 & 1 \\ 0 & 1 \end{pmatrix} \,, \quad \widetilde{\gamma}_2 = \begin{pmatrix} 1 & -1 \\ 3 & -2 \end{pmatrix} \,.
\end{align}
From the action of the two generators
\begin{align}
\begin{pmatrix} x \\ y \end{pmatrix} \mapsto \widetilde{\gamma}_1 \begin{pmatrix} x \\ y \end{pmatrix} = \begin{pmatrix} x + y \\ y \end{pmatrix} \,, \quad \begin{pmatrix} x \\ y \end{pmatrix} \mapsto \widetilde{\gamma}_2 \begin{pmatrix} x \\ y \end{pmatrix} = \begin{pmatrix} x - y \\ 3x - 2 y \end{pmatrix} \,,
\end{align}
one concludes that the only invariant points are given by
\begin{align}
\big\{ \tfrac{m}{3} \omega^1 \,, \enspace m \in \{ 0, 1, 2 \} \big\} \subset E(3) \,.
\end{align}
This fixes the elements of a $\mathbb{Z} / 3 \mathbb{Z}$ subgroup of the full torsion subset $E(3)$. Using a coset representative of $\Gamma_1 (3)$ with respect to $SL(2, \mathbb{Z})$, one can also generate different $\mathbb{Z} / 3 \mathbb{Z}$ subgroups which are preserved on the level of the individual elements.

\subsection{$\Gamma_0 (3)$}

The congruence subgroup $\Gamma_0 (3)$ is generated by the elements
\begin{align}
\gamma'_1 = \begin{pmatrix} 1 & 1 \\ 0 & 1 \end{pmatrix} \,, \quad \gamma'_2 = \begin{pmatrix} -1 & 1 \\ -3 & 2 \end{pmatrix} \,.
\end{align}
The action of the generators on points in $E$ is given by
\begin{equation}
\begin{split}
\begin{pmatrix} x \\ y \end{pmatrix} &\mapsto \gamma'_1 \begin{pmatrix} x \\ y \end{pmatrix} = \begin{pmatrix} x + y \\ y \end{pmatrix} \,, \\
\begin{pmatrix} x \\ y \end{pmatrix} &\mapsto \gamma'_2 \begin{pmatrix} x \\ y \end{pmatrix} = \begin{pmatrix} -x + y \\ - 3x + 2 y \end{pmatrix} \,,
\end{split}
\end{equation}
and no point beside the origin is kept fixed. However, the full set
\begin{align}
\big\{ \tfrac{m}{3} \omega^1 \,, \enspace m \in \{ 0, 1, 2 \} \big\} \subset E(N) \,.
\end{align}
is fixed under this group action. The individual elements are mapped to each other as follows
\begin{align}
\tfrac{0}{3} \omega^1 \mapsto \tfrac{0}{3} \omega^1 \,, \quad \tfrac{1}{3} \omega^1 \mapsto - \tfrac{1}{3} \omega^1 \,, \quad \tfrac{2}{3} \omega^1 \mapsto - \tfrac{2}{3} \omega^1 \,.
\end{align}
Again, we can use a coset representatives with respect to $SL(2, \mathbb{Z})$ in order to generate different $\mathbb{Z} / 3 \mathbb{Z}$ subgroups that are fixed by $\Gamma_0 (3)$ as a set but not element by element.

\section{4D $\mathcal{N} = 2$ Gauge Theory with Four Flavors} \label{app:FLAVA}

In this Appendix we discuss in greater detail some aspects of 4D $\mathcal{N} = 2$ gauge theory with gauge group $SU(2)$ and four hypermultiplets in the fundamental representation of $SU(2)$, as studied in reference \cite{Seiberg:1994aj}.
This theory leads to a 4D $\mathcal{N} = 2$ SCFT with flavor symmetry $SO(8)$. Our plan will be to first review some general aspects of
the $\mathcal{N} = 2$ curve in this setting. We then fix a choice of Coulomb branch parameter and vary the mass parameters of the theory
under the condition that the IR theory is time-reversal invariant, and that the mass parameters and Coulomb branch scalar vev preserve
time-reversal invariance.

\subsection{General $\mathcal{N} = 2$ Considerations}
We begin by stating some general considerations about $\mathcal{N} = 2$ theories.
For a state of charge $(q_e, q_m, q_f)$ under the electric, magnetic and flavor symmetry $U(1)$'s, this is controlled by the formula:
\begin{equation}
Z = q_e a - q_m a_D + \frac{1}{\sqrt{2}} \sum_{f = 1}^{\mathrm{dim} \mathcal{R}} q_f m^f, \,\,\, \text{with} \,\,\, M = \sqrt{2} \vert Z \vert.
\end{equation}
where here, $a$ denotes a coordinate on the Coulomb branch, $a_D = \partial \mathcal{F} / \partial a$ is a magnetic dual coordinate controlled by the derivative of $\mathcal{F}$, the $\mathcal{N} = 2$ prepotential, $\mathcal{R}$ denotes a representation of the flavor symmetry, and $M$ denotes the mass of the particle. Recall that in terms of the Seiberg-Witten geometry a massless state occurs whenever a one-cycle of the curve collapses. Following \cite{Minahan:1996cj, Minahan:1996fg}, we introduce a fixed representation $\mathcal{R}$ of the flavor symmetry and write the Seiberg-Witten one-form as:
\begin{equation}
\lambda_{\mathcal{R}} = (c_1 u + c_3) \frac{dx}{y} + c_2 \sum_{b} \frac{m_b y_b(u)}{x - x_b(u)}\frac{dx}{y}.
\end{equation}
for some coefficients $c_i$ which depend on the mass parameters. Introducing an A-cycle and a B-cycle
on the elliptic curve, the coordinates $a$ and $a_D$ can be written as:
\begin{equation}
a = \underset{\gamma_A}{\int} \lambda_{\mathcal{R}} \,\,\, \text{and} \,\,\, a_{D} = \underset{\gamma_B}{\int} \lambda_{\mathcal{R}},
\end{equation}
and the complex structure of the curve is encoded in the derivatives:
\begin{equation}
\tau = \frac{\partial a_{D}}{\partial a} = \frac{\partial a_{D} / \partial u}{\partial a / \partial u}.
\end{equation}

\subsubsection{Seiberg-Witten Curve}

Let us now turn to the Seiberg-Witten curve for the case of $SU(2)$ gauge theory with four flavors. This was originally considered in \cite{Seiberg:1994rs}, and was also presented in a different parametrization in reference \cite{Gaiotto:2009we}.

One way to present the Seiberg-Witten curve is by introducing the 6D SCFT with $\mathcal{N}=(2,0)$ of $A_1$-type, namely the one coming from
the worldvolume of two M5-branes. Wrapping the M5-branes on a $\mathbb{CP}^1$ with four marked points, the moduli space of $\mathcal{N} = 2$ vacua is controlled by the moduli space of the $SU(2)$ Hitchin system on this curve. At a generic point of the moduli space, we obtain a branched double cover of this genus zero curve, namely the ``IR curve'' or Seiberg-Witten curve as obtained from the spectral equation for the Higgs field:
\begin{align}
\lambda^2 - \phi_2 = 0 \,,
\end{align}
with Seiberg-Witten differential $\lambda = x dz/ z$ and $\phi_2$ the quadratic Casimir of the Hitchin system Higgs field
given by:
\begin{align}
\phi_2 =  \frac{P_4 (z)}{(z - 1)^2 (z - q)^2} \frac{dz^2}{z^2} \,.
\end{align}
In the above, $z$ is an affine coordinate on the $\mathbb{CP}^1$. Here, $q$ encodes the UV coupling constant $\tau_{\text{UV}}$ of the $SU(2)$ gauge theory via $q = e^{2 \pi i \tau_{\text{UV}}}$ and $P_4 (z)$ is a fourth order polynomial in $z$ whose coefficients determine the position of the four branch points on the $\mathbb{CP}^1$. Note that the differential on the lefthand side has double poles at $z = 0, 1, \infty$, and $q$. Clearing denominators, we can write this as a hypersurface equation inside $T^{\ast} \mathbb{CP}^{1}$ given by:
\begin{align}
x^2 (z - 1)^2 (z - q)^2 = P_4 (z) \,.
\end{align}
Since we have quadratic order terms on the left-hand side, we can blowup at these zeros, and instead consider the hypersurface equation:\
\begin{equation}
x^2 = P_4 (z),
\end{equation}
which we recognize as the equation of an elliptic curve. To pass to the Weierstrass form, we can use the general prescription given in Appendix \ref{app:ELLIPTIC} to first compute the conformal cross ratio in the roots of $P_4$, and from this extract the $J$-function for the elliptic curve. Next, apply a Moebius transformation on $z$
\begin{align}
z \rightarrow \frac{a z + b}{c z + d} \,, \quad dz \rightarrow \frac{a d - b c}{(c z + d)^2} \, dz = \frac{1}{(c z + d)^2} \, dz \,,
\end{align}
which can be understood as $x \rightarrow (c z + d)^{-2} x$ on the coordinate on the fiber of the cotangent bundle. This can be used to map three marked points to fixed positions, and recover the desired form of the Weierstrass model.

We now use the parametrization of the Seiberg-Witten curve in Weierstrass form as obtained from a D3-brane probe of an $SO(8)$ seven-brane.
From reference \cite{Noguchi:1999xq}, we have:
\begin{equation}
  f = u^2 + \widetilde{w}_4, \quad
  g = w_2 u^2 + w_4 u + w_6.
\end{equation}
The Casimir invariants are given by (equations (2.12)-(2.15) of \cite{Noguchi:1999xq}):
\begin{align}
  u_2 = -\sum_a m_a^2, \qquad & u_4 = \sum_{a<b} m_a^2 m_b^2, \nonumber \\
  u_6= -\sum_{a<b<c} m_a^2 m_b^2 m_c^2, \qquad & \widetilde{u}_4 = -2im_1m_2m_3m_4.
\end{align}
\begin{align}
  u_2=-3w_2, \qquad & u_4=\widetilde{w}_4+3 w_2^2, \nonumber \\
  u_6 = w_6 -w_2 \widetilde{w}_4 -w_2^3, \qquad & \widetilde{u}_4 = w_4.
\end{align}

To simplify we can set all the mass parameters equal to $m$ so that the computations only depend on two parameters.
Furthermore, the Coulomb branch is parameterized by $\widetilde{u}=i u$, and is taken to be real.

Thus,
\begin{equation}
  f = -\tilde{u}^2 + \frac{2}{3}m^4, \quad
  g = -\frac{4}{3}m^2 \widetilde{u}^2 + 2m^4 \widetilde{u} -\frac{20}{27}m^6.
\end{equation}
And the Seiberg-Witten differential is:
\begin{align}
  \lambda_{8_v} &= \frac{\sqrt{2}}{8\pi i} \left(2u \frac{dx}{y} + \sum_{a=1}^{4} \frac{m_a^2u + w_4/2}{x-m_a^2+w_2}\frac{dx}{y} \right) \nonumber \\
  &= \frac{\sqrt{2}}{8\pi i} \left(2u \frac{dx}{y} + 4m^2 \frac{u-im^2}{x+m^2/3}\frac{dx}{y} \right) \nonumber \\
  &= \frac{\sqrt{2}}{8\pi} \left(2 \widetilde{u} \frac{dx}{y} + 4m^2 (\widetilde{u}-m^2)\frac{dx}{y(x+m^2/3)} \right).
\end{align}
In figure \ref{fig:4flavNumeric} we then plot the result of those computations. We give the period integrals $a$ and $a_D$ across all three possible regions in which $\tau$ belongs to the real component of $X(\Gamma)_{\mathbb{R}}$ for $\Gamma = SL(2,\mathbb{Z})$. We note that as one moves around in the moduli space, the value of $\tau = \partial a_{D} / \partial a$ might move outside the fundamental domain. When this occurs, we perform a change in the ordering of roots $e_i$ appearing in the elliptic curve. This in turn leads to a jump in the values of the periods $a$ and $a_D$, as occurs by applying an $SL(2,\mathbb{Z})$ transformation. In our analysis, it proves convenient to use a slightly different convention from the rest of chapter~\ref{chapter7}. So, in this Appendix we take $e_1 > e_2 > e_3$ in the trivial phase, $e_2 \in \mathbb{R}$, Im$(e_1)>\mathrm{Im}(e_3)$ in the strongly coupled phase, and $e_1 \in \mathbb{R}$, Im$(e_3)>\mathrm{Im}(e_2)$ in the topological insulator phase.

\begin{figure}[t!]
\centering
\begin{overpic}[width=0.45\textwidth]{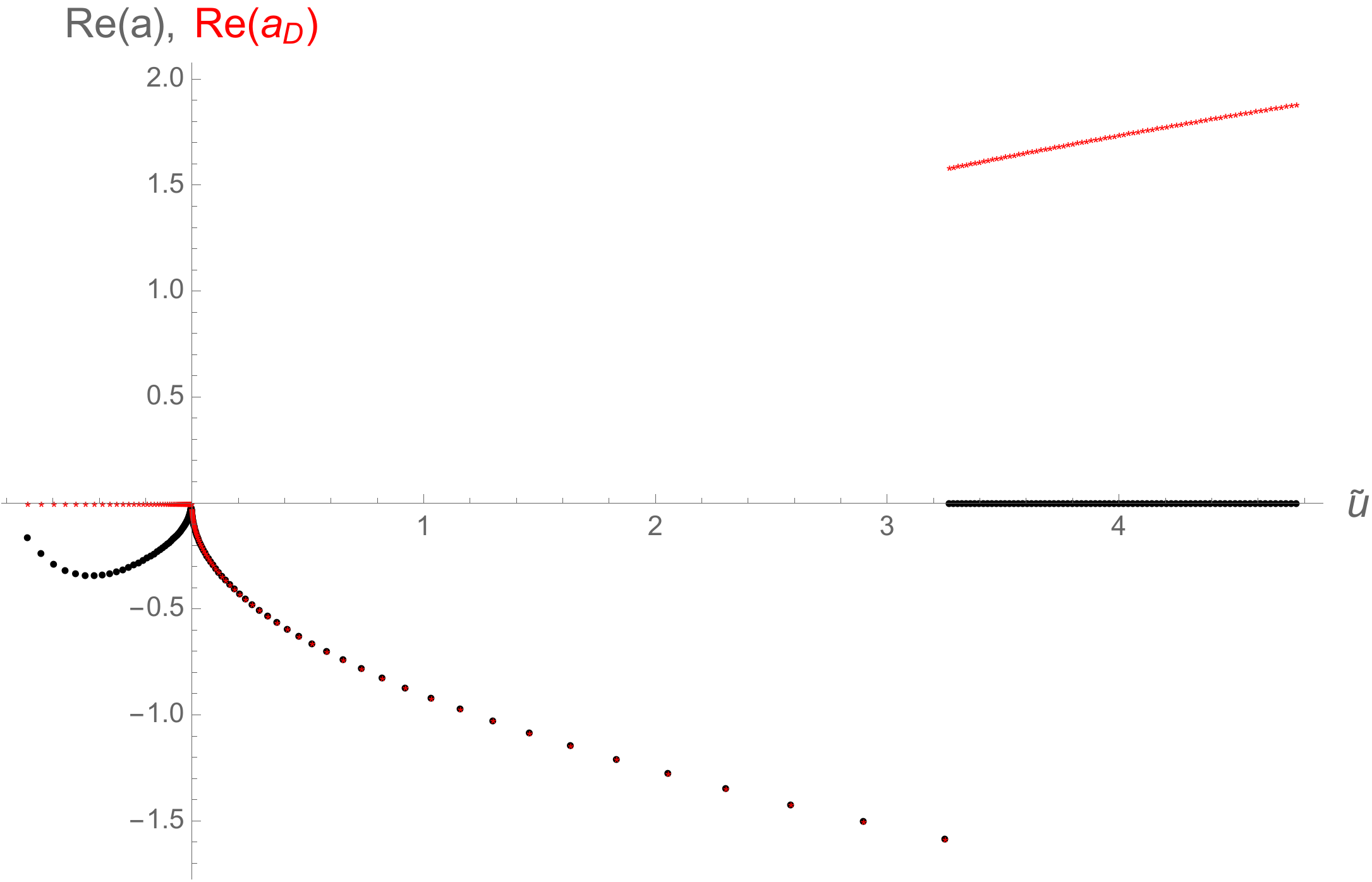}
  \put (1,40) {\tiny\textbf{$\theta=0$}}
  \put (40,40) {\tiny\textbf{$|\tau|=1$}}
  \put (80,40) {\tiny\textbf{$\theta=\pi$}}
\end{overpic}\hspace{.5cm}
\begin{overpic}[width=0.45\textwidth]{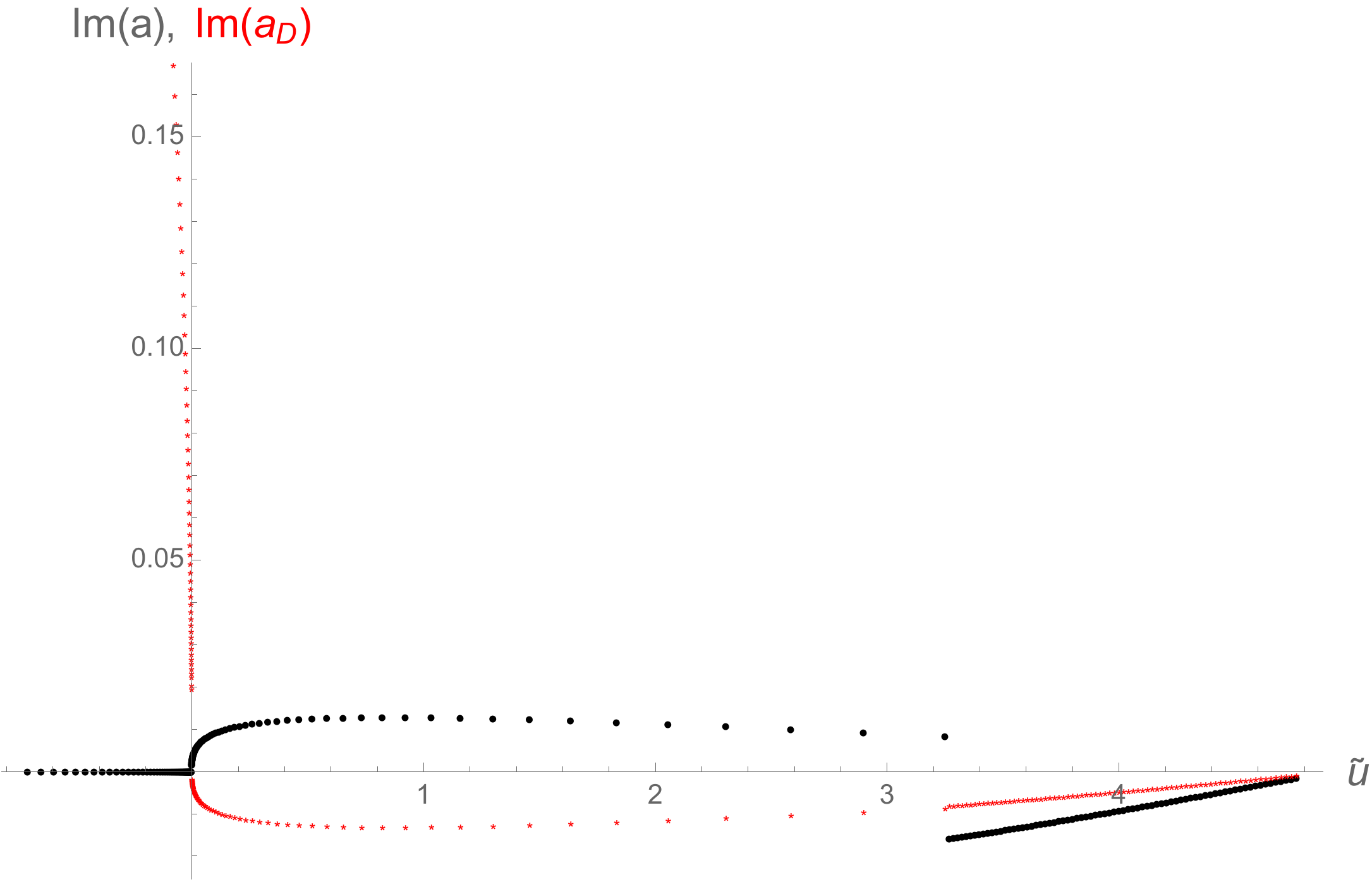}
  \put (1,40) {\tiny\textbf{$\theta=0$}}
  \put (40,40) {\tiny\textbf{$|\tau|=1$}}
  \put (80,40) {\tiny\textbf{$\theta=\pi$}}
\end{overpic}\vspace{.5cm}

\includegraphics[width=0.45\textwidth]{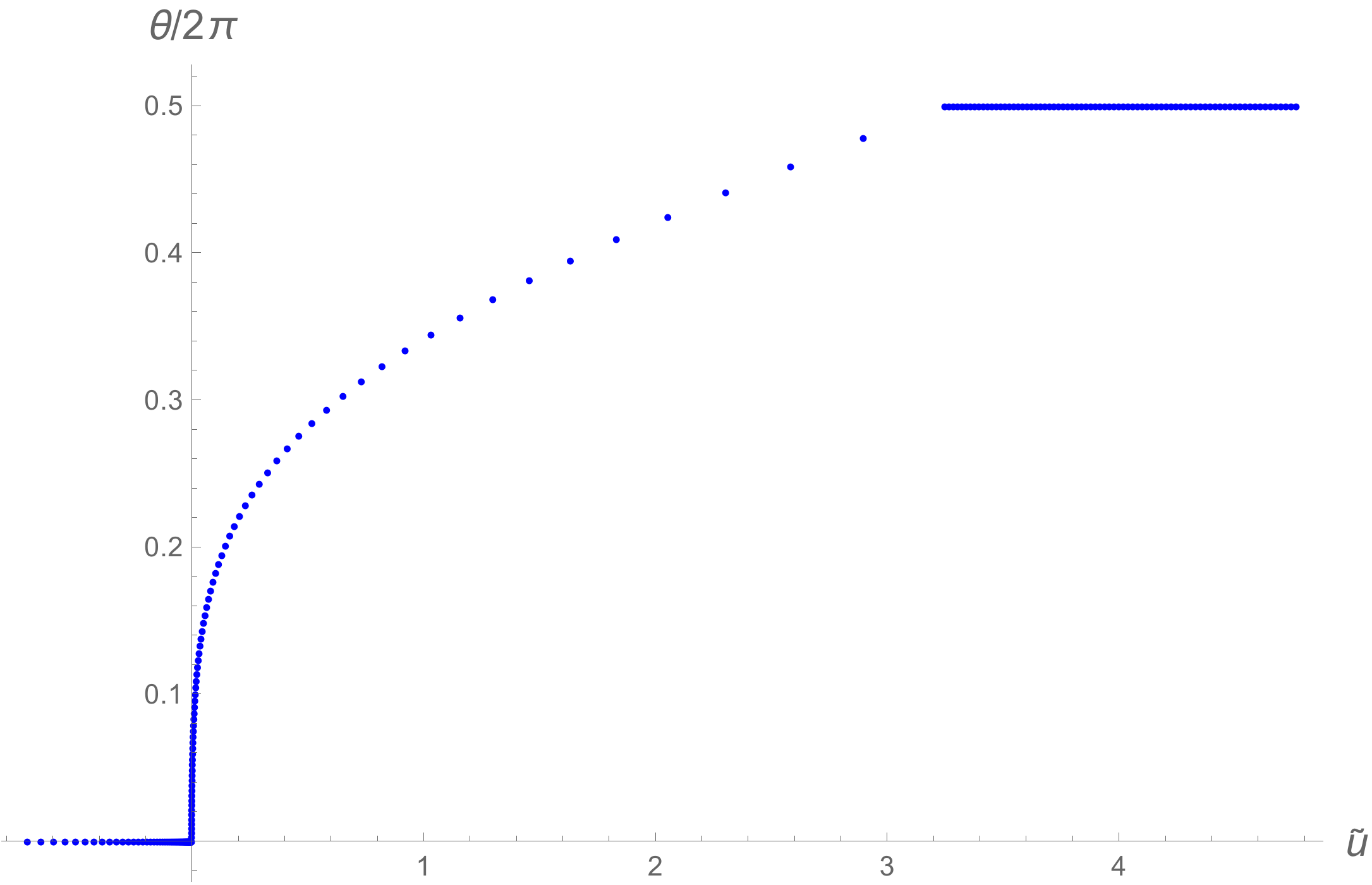} \hspace{.5cm}
\includegraphics[width=0.45\textwidth]{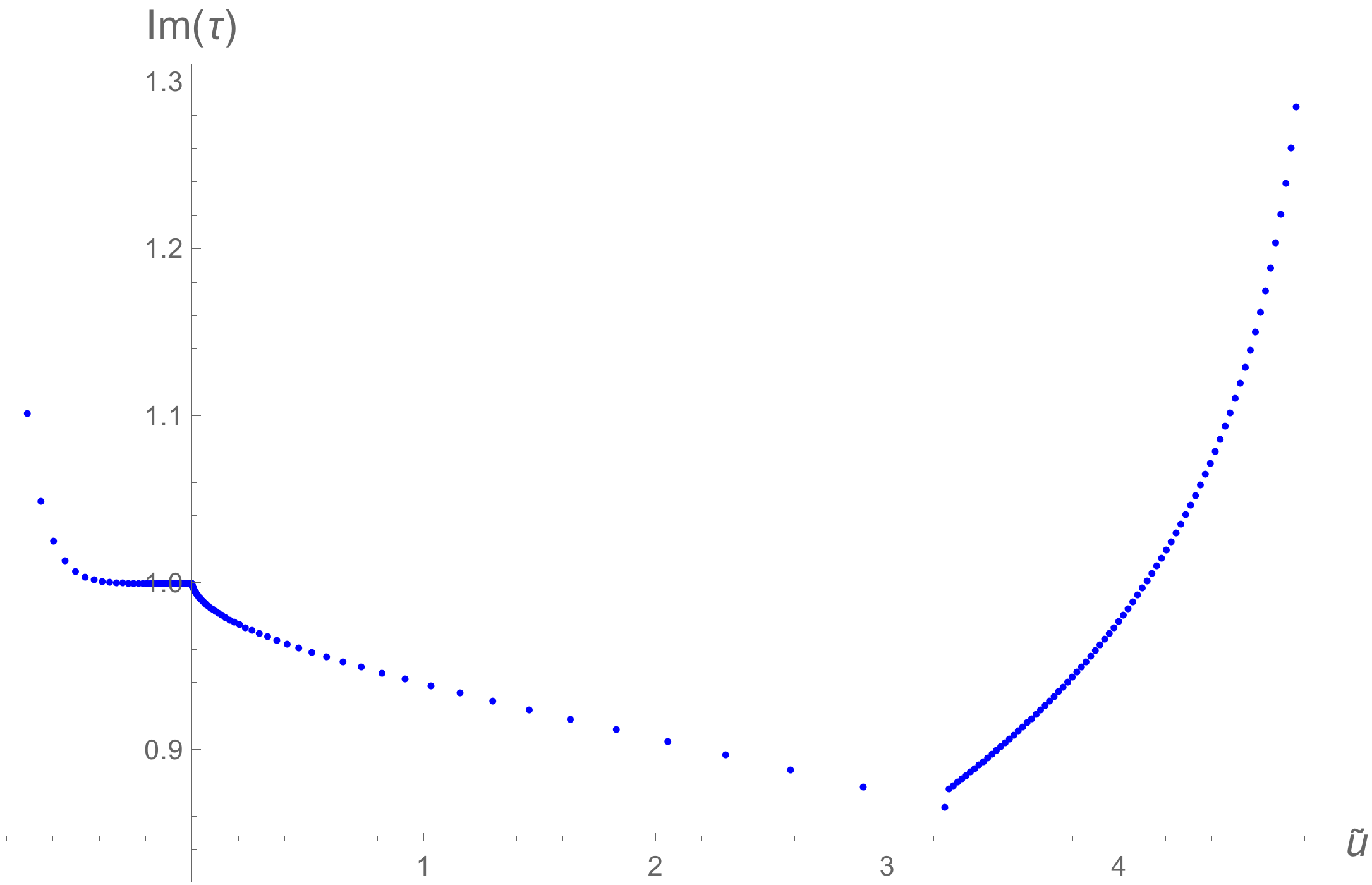}
\caption{The period integrals $a$ (black) and $a_D$ (red) plotted against the Coulomb branch parameter $\tilde{u}$ across the three different phases. The top panel gives the periods while the bottom shows the coupling $\tau$. The left-hand side gives the real part while the right-hand side shows the imaginary piece.
We start off in the trivial phase ($\theta=0$), then transition at $\tau=i$ into the strongly coupled phase $|\tau|=1$. The topological insulator phase ($\theta= \pi$) is then reached at $\tau=e^{\pi i/3}$. Finally, going to the weak coupling limit ($\tau = i \infty$) we can go back into the trivial ($\theta=0)$ phase. Note that the mass parameter $m$, while not plotted, also varies.}%
\label{fig:4flavNumeric}%
\end{figure}
In each of the different phases, we observe (from figure \ref{fig:4flavNumeric}) that:
\begin{itemize}
\item $\theta=0$:   $\Delta < 0$, $f < 0$ gives $a \in \mathbb{R}$, $a_D \in i\mathbb{R}$.
\item $|\tau|=1$: $\Delta > 0$, $f \geq 0$ gives $a_D = a^\dagger$.
\item $\theta=\pi$:  $\Delta > 0$, $f < 0$ gives $a \in i \mathbb{R}$, $\mathrm{Im}(a_D) = \mathrm{Im(a)}/2$.
\end{itemize}
Furthermore, both periods vanish at the transition point $\tau=i$, while only $a$ goes to zero at the weak coupling limit $\tau=i\infty$.

\subsection{Elliptic Integrals and Relations Between $a$ and $a_D$}\label{app:contour}
We now derive some of the reality conditions for contour integrals in the three different phases.
\begin{figure}[t!]
\centering
\includegraphics[width=0.7\textwidth]{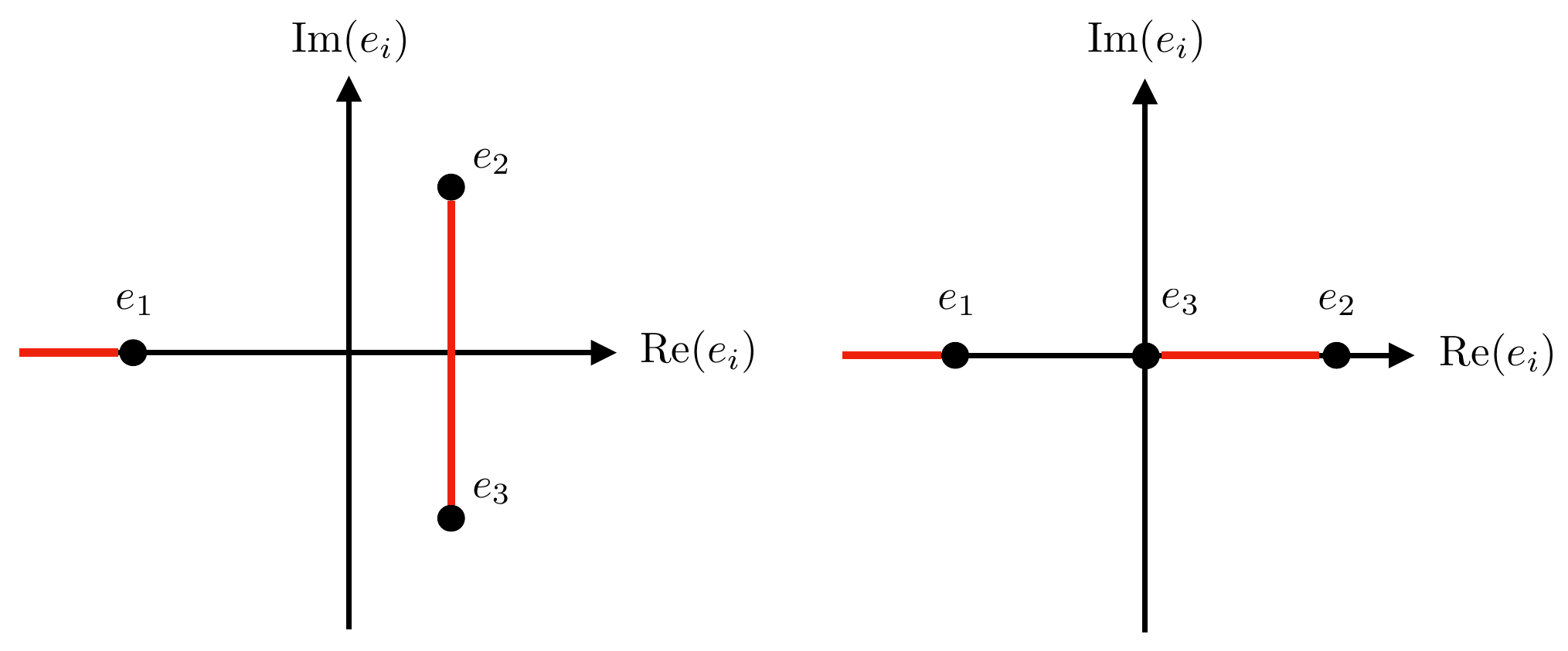}
\caption{Choice of branch cuts in the two cases \eqref{eq:rootcases}.}
\label{fig:branchcuts}
\end{figure}
We choose the distribution of branch cuts as depicted in figure \ref{fig:branchcuts} with contour integrals given in figure \ref{fig:contours}.
In order to prove the various relations between $a$ and $a_D$ we must first take a closer look at the elliptic integrals and fix some conventions about branch cuts.
We want to investigate the properties of the following integrals
\begin{equation}
\int_{e_a}^{e_b} \frac{dx}{y} \,, \quad \text{and} \quad  \int_{e_a}^{e_b} \frac{dx}{y(x-c)} \,,
\end{equation}
where $y$ on the chosen branch is given by $+ \sqrt{x^3 + f x + g}$.

\subsubsection{Proof that $I_A \in i \mathbb{R} $ and $I_B \in \mathbb{R}$ in Phase I (Trivial Phase)}
We fix the real roots such that $e_1 < e_3 < e_2$. Following the same notation as in \cite{Bilal:1997st, DelZotto:2016fju} we want to investigate the integrals:
\begin{align}
  I_A^{(1)} &= \int_{e_3}^{e_2} \frac{dx}{y}, \\
  I_A^{(3)} &= \int_{e_3}^{e_2} \frac{dx}{y(x-c)}, \\
  I_B^{(1)} &= \int_{e_1}^{e_3} \frac{dx}{y}, \\
  I_B^{(3)} &= \int_{e_1}^{e_3} \frac{dx}{y(x-c)},
\end{align}
where $y = \sqrt{(x-e_1)(x-e_2)(x-e_3)}$, so that $y$ is purely imaginary for $e_3 \leq x \leq e_2$, but $y$ is real for $e_1 \leq x \leq e_3$.

Therefore,
\begin{equation}
  I_A \in i \mathbb{R}, \quad I_B \in \mathbb{R}.
\end{equation}

\subsubsection{Proof that $I_A \in i \mathbb{R}$ and $\mathrm{Im}(I_B) = \mathrm{Im}(I_A)/2$ in Phase II (Topological Insulator)}
Let $e_2 = r+i\alpha$ so that $e_3 = r-i\alpha$ and $e_1=-2r$.
First of all we prove that $I_A \in i \mathbb{R}$ by noting that:
\begin{align}
  \begin{split}
    I_A^{(1)} &= \int_{e_2}^{e_3} \frac{dx}{y} \\
    &= \int_0^1 \frac{-2i\alpha}{y}dt, \quad x=(e_3-e_2)t+e_2 = -2i\alpha t + r + i\alpha \\
    &= \int_0^1 \frac{-2i\alpha}{\sqrt{4\alpha^2t\left(-3r+i\alpha(2t-1)\right)(t-1)}}dt \\
    &= \int_0^1 \frac{-i}{\sqrt{s(t)}}dt, \\
  \end{split}
\end{align}
where
\begin{equation}
  s(t) = t(t-1)\left(-3r+i\alpha(2t-1)\right).
\end{equation}
We observe that:
\begin{equation}
  s(t) = \overline{s(1-t)}.
\end{equation}
As a result,
\begin{align}
  \begin{split}
    I_A^{(1)} &= \int_0^1 \frac{-i dt}{\sqrt{s(t)}} \\
    &= \int_0^{\frac{1}{2}} \frac{-i dt}{\sqrt{s(t)}} + \int_{\frac{1}{2}}^{1} \frac{-i dt}{\sqrt{s(t)}} \\
    &= \int_0^{\frac{1}{2}} \frac{-i dt}{\sqrt{s(t)}} + \int_{\frac{1}{2}}^0 \frac{i dt'}{\sqrt{s(1-t')}} \\
    &= \int_0^{\frac{1}{2}} \frac{-i dt}{\sqrt{s(t)}} + \int_{\frac{1}{2}}^0 \frac{i dt'}{\sqrt{\overline{s(t')}}} \,, \\
    I_A^{(1)} &= -i\int_0^{\frac{1}{2}}dt\left(\frac{1}{\sqrt{s(t)}}+\frac{1}{\sqrt{\overline{s(t)}}}\right) \,, \\
  \end{split}
\end{align}
which implies  $I_A^{(1)} \in i \mathbb{R}$.

Furthermore, we have
\begin{equation}
  x(t) = -2i\alpha t + i\alpha + r = \overline{x(1-t)}.
\end{equation}
Thus the same reasoning applies to
\begin{equation}
  I_A^{(3)} = \int_{e_2}^{e_3} \frac{dx}{y(x-c)}.
\end{equation}
This concludes the proof that $I_A \in i \mathbb{R}$.

Next we note that
\begin{equation}
  I_B^{(1)} = \int_{e_1}^{e_3} \frac{dx}{y} = \overline{\int_{e_1}^{e_2} \frac{dx}{y}},
\end{equation}
which implies that
\begin{align}
  2i \mathrm{\, Im}I_B^{(1)} &= \int_{e_1}^{e_3} \frac{dx}{y} - \int_{e_1}^{e_2} \frac{dx}{y} \\
  &= \int_{e_1}^{e_3} \frac{dx}{y} + \int_{e_2}^{e_1} \frac{dx}{y} \\
  &= \int_{e_2}^{e_3} \frac{dx}{y} \\
  &= I_A^{(1)}.
\end{align}
And similarly we have $2i \mathrm{\, Im}I_B^{(3)} = I_A^{(3)}$. So that indeed, $\mathrm{Im}(I_B) = \mathrm{Im}(I_A)/2$

\subsubsection{Proof that $I_B= \bar{I}_A$ in Phase III (Strongly Coupled Phase)}
In this phase, we note that in order for $\tau$ to be in the fundamental domain, the roots are chosen so that $e_1 \in \mathbb{R}$, $e_2 = \bar{e}_3$, and the period integrals given by:
\begin{align}
  I_A^{(1)}= \int_{e_1}^{e_3} \frac{dx}{y}, && I_A^{(3)} = \int_{e_1}^{e_3} \frac{dx}{y(x-c)}, \nonumber \\
  I_B^{(1)}= \int_{e_1}^{e_2} \frac{dx}{y}, && I_B^{(3)} = \int_{e_1}^{e_2} \frac{dx}{y(x-c)}.
\end{align}
Therefore, $I_B^{(1)} = \overline{I_A^{(1)}}$ and $I_B^{(3)} = \overline{I_A^{(3)}}$.

\section{Localizing a 4D Weyl Fermion} \label{app:WEYL}

In this Appendix we consider the localization of a 4D Weyl fermion
$\chi_{\alpha}$ with a position dependent mass term on a thin wall. We will be
specifically interested in the case where the mass is non-zero outside some
finite size interval, but vanishes inside this interval. We take
\textquotedblleft particle physics conventions\textquotedblright\ and work in
signature $(+,-,-,-)$. We consider a position dependent mass term in the
spatial direction $x_{\bot}=x^{3}\equiv z$ given by:
\begin{equation}
m=m_{L}\Theta(-z)+m_{R}\Theta(z-h),
\end{equation}
where $\Theta$ denotes the Heaviside step function and $m_{L}=\left\vert
m_{L}\right\vert e^{i\phi_{L}}$ and $m_{R}=\left\vert m_{R}\right\vert
e^{i\phi_{R}}$ are non-zero complex numbers. The massless region runs from
$z=0$ to $z=h$, and would describe a thick interface. We will be interested in
the special case where $h\rightarrow0$. We will also need the derivative of
the mass term:%
\begin{equation}
\partial_{z}m=m_{R}\,\delta(z-h)-m_{L}\,\delta(z).
\end{equation}

Our 4D Weyl fermion satisfies the equation of motion:
\begin{equation}
i\left(  \overline{\sigma}^{\mu}\right)  ^{\dot{\alpha}\beta}\partial_{\mu}%
\chi_{\beta}=m(z)\left(  \chi^{\dagger}\right)  ^{\dot{\alpha}}\,,
\end{equation}
We will be interested in explicit solutions to this equation, so we write out
the form of the Dirac equation equation in terms of the two component
doublet:
\begin{equation}
\chi_{\beta}=%
\begin{pmatrix}
a\\
b
\end{pmatrix}
\quad\text{and}\quad\left(  \chi^{\dagger}\right)  ^{\dot{\alpha}}=%
\begin{pmatrix}
-b^{\dagger}\\
a^{\dagger}%
\end{pmatrix}
.
\end{equation}
From there, our Dirac equation can be simplified into a pair of differential
equations:
\begin{align}
(\partial_{\mathrm{4D}}^{2}+|m|^{2})a &  =i(\partial_{z}m^{\dagger}%
)b^{\dagger}\\
(\partial_{\mathrm{4D}}^{2}+|m|^{2})b &  =i(\partial_{z}m^{\dagger}%
)a^{\dagger}.
\end{align}
where the 4D$\ $D'Alembertian $\partial_{\mathrm{4D}}^{2}$ can be further
expanded as:%
\begin{equation}
\partial_{\mathrm{4D}}^{2}=\partial_{\mathrm{3D}}^{2}-\partial_{z}^{2},
\end{equation}
with $\partial_{\mathrm{3D}}^{2}$ the 3D D'Alembertian in the directions
transverse to the $z$-direction. We will mainly be interested in modes which
are exactly massless on a thin 3D slice, so we impose the condition that
$\partial_{\mathrm{3D}}^{2}$ annihilates all functions. We note that in the
case of a thick interface, this condition is not quite appropriate because we
really have a 4D Weyl fermion on an interval (in the interior region).

Focusing now on the case where $h\rightarrow0$, it is enough to consider
just the $z$-dependence of our solutions so we can now write our
differential equation as:%
\begin{align}
\left(  -\partial_{z}^{2}+|m|^{2}\right)  a &  =i(\partial_{z}m^{\dagger
})b^{\dagger}\\
\left(  -\partial_{z}^{2}+|m|^{2}\right)  b &  =i(\partial_{z}m^{\dagger
})a^{\dagger},
\end{align}

We now turn to the solutions of this differential equation. This is
essentially an exercise of the form found in introductory quantum mechanics
textbooks, but we include some general comments for completeness. In the thin
wall limit, the solution splits up into a piecewise smooth function. In the
$z<0$ region we have:
\begin{align}
z &  <0\\
a_{L} &  =A_{L}\exp(+\left\vert m_{L}\right\vert z)\\
b_{R} &  =B_{L}\exp(+\left\vert m_{L}\right\vert z).
\end{align}
for some as yet unfixed coefficients $A_{L}$ and $B_{L}$. Consider next the
solution in the region $z>0$. In this case we have:%
\begin{align}
z &  >0\\
a_{R} &  =A_{R}\exp(-|m_{R}|z)\\
b_{R} &  =B_{R}\exp(-|m_{R}|z).
\end{align}

Next, we need to match the form of our solutions across the three regions.
First, we impose continuity. This leads to the conditions:%
\begin{equation}
A_{L}=A_{R}=A\text{ \ \ and \ \ }B_{L}=B_{R}=B.
\end{equation}
Next, we integrate our differential equation across the interfaces. This
yields the conditions:%
\begin{align}
(\left\vert m_{R}\right\vert +\left\vert m_{L}\right\vert )A &  =i(m_{R}%
^{\dag}-m_{L}^{\dag})B^{\dag}\\
(\left\vert m_{R}\right\vert +\left\vert m_{L}\right\vert )B &  =i(m_{R}%
^{\dag}-m_{L}^{\dag})A^{\dag}.
\end{align}
so we get the condition:%
\begin{equation}
\left\vert m_{R}-m_{L}\right\vert ^{2}=\left\vert \left\vert m_{R}\right\vert
+\left\vert m_{L}\right\vert \right\vert ^{2}.
\end{equation}
To get a localized mode we therefore need to set $e^{i(\phi_{L}-\phi_{R})}%
=-1$, namely the mass term is rotated by a phase of exactly $\pi$ in passing
from the left to the right side of the thin interface. Note that
we also get a non-trivial constraint on the relative phases of $A$ and
$B$. Indeed, we have:
\begin{equation}
A=ie^{-i\phi_{R}}B^{\dag}.
\end{equation}
Consequently, we learn that out of the original two-dimensional complex
doublet of $\mathfrak{spin}(3,1)$, we only retain a single real doublet of
$\mathfrak{spin}(2,1)$ on the wall.

Returning to the more general setting where we have a thick interface, in this
case we should really include non-zero values of the three-momentum. We should
then consider a more general differential equation:%
\begin{align}
\left(  -\partial_{z}^{2}+\Delta\right)  a &  =i(\partial_{z}m^{\dagger
})b^{\dagger}\\
\left(  -\partial_{z}^{2}+\Delta\right)  b &  =i(\partial_{z}m^{\dagger
})a^{\dagger},
\end{align}
with:%
\begin{equation}
\Delta= \partial_{3D}^{2} + \left\vert m\right\vert ^{2} .
\end{equation}
In a thick interior region we have a standard
4D\ wave equation. Switching on specific phases for the mass
terms outside this region amounts to setting a boundary condition on the left $(z=0)$ and right
$(z=h)$ of the middle region. Note that this also leads to an oscillatory
behavior in the middle region. In the thin interface limit, the boundary
conditions on the left and right become correlated, and this imposes a further
condition on the zero modes (as we have seen).

\addtocontents{toc}{\protect\setcounter{tocdepth}{1}} 
\end{appendixf}

\clearpage
\phantomsection
\singlespacing
\newenvironment{bibliof}{}{}
\titleformat{\chapter}[hang]{\large\center}{\thechapter}{0 pt}{}
\titlespacing*{\chapter}{0pt}{-25 pt}{6 pt} 

\begin{bibliof}
\renewcommand\bibname{BIBLIOGRAPHY}
\addcontentsline{toc}{chapter}{BIBLIOGRAPHY}
\bibliographystyle{JHEPsorted}
\bibliography{refs} 

\end{bibliof}

\end{document}